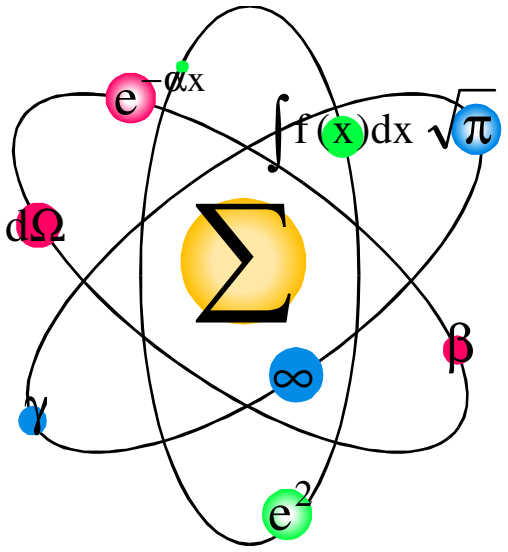

*С.Б. Дубовиченко*

# МЕТОДЫ РАСЧЕТА ЯДЕРНЫХ ХАРАКТЕРИСТИК

***Модели - методы - программы***

***Казахская Академия Труда и Социальных Отношений***

***С.Б. Дубовиченко***

# МЕТОДЫ РАСЧЕТА ЯДЕРНЫХ ХАРАКТЕРИСТИК

***Модели - методы - программы***

***Рецензенты:***

Директор Института проблем информатики и управления МОН РК, академик Международной инженерной академии РК, доктор физико - математических наук, профессор **Айдарханов М.Б.**,
директор астрофизического института МОН РК, академик Международной академии информатизации РК, доктор физико - математических наук, профессор **Чечин Л.М.,**
заведующий кафедрой КТПиП Каз.НТУ, академик и вице-президент Международной академии информатизации РК, доктор технических наук, профессор **Цеховой А.Ф.,**
заведующая кафедрой Информатики Каз.НУ, доктор физико - математических наук, профессор **Балакаева Г.Т.**

В книге изложены математические методы расчета ядерных сечений и фаз упругого рассеяния, энергии и характеристик связанных состояний в двух- и трехчастичных ядерных системах, когда потенциалы взаимодействия содержат не только центральную, но и тензорную компоненту. Приведены описания математических численных методов расчета и компьютерные программы на алгоритмическом языке "Бейсик" в среде компилятора "Turbo Basic" фирмы "Borland" для компьютеров типа IBM PC AT.

Для численных решений исходных уравнений Шредингера использован конечно - разностный и вариационный методы, а также метод Рунге - Кутта с автоматическим выбором шага по заданной точности результатов для фаз рассеяния и энергии связи. Приведено описание не стандартных методов решения системы уравнений Шредингера на связанные состояния и альтернативный методу Шмидта, метод решения обобщенной матричной задачи на собственные значения.

Разработанные программы позволяют определять волновые функции относительного движения ядерных фрагментов, нормированные на правильную асимптотику с учетом кулоновского взаимодействия. Приведены программы извлечения ядерных фаз (фазового анализа) из дифференциальных сечений упругого рассеяния.

Книга может быть использована в качестве учебника по численным математическим методам для студентов и аспирантов физических и математических специальностей высших учебных заведений.

*Автор выражает искреннюю благодарность*
*А.В. Джазаирову - Кахраманову и компании*
*«Tandem Translations» за оказание спонсорской*
*помощи при издании данной книги*

# ПРЕДИСЛОВИЕ

Множество задач теоретической ядерной физики, особенно в области легких атомных ядер, требует умения решать уравнение Шредингера или связанную систему уравнений такого типа. Результатом решения является волновая функция, которая описывает квантовое состояние некоторой системы ядерной частиц и, в принципе, содержит всю информацию о таком состоянии.

Существует довольно много различных математических методов решения дифференциальных уравнений или их систем второго порядка, которым является уравнение Шредингера. Однако, в математической литературе обычно приводятся довольно абстрактные методы решений таких уравнений, которые бывает достаточно сложно применить для решения конкретного уравнения, типа уравнения Шредингера. Проблему обычно составляет выбор наиболее оптимального математического метода, применимого для рассмотрения определенных задач, основанных на решениях уравнения Шредингера.

Именно решению этих проблем и посвящена данная книга, которая описывает некоторые математические методы, непосредственно применимые для нахождения волновых функций из решений уравнения Шредингера или систем таких уравнений в задачах ядерной физики. Рассматриваются математические численные и вариационные методы решений, применимые в задачах дискретного и непрерывного спектра состояний ядерных частиц и позволяющие получать конечные результаты с практически любой точностью.

На основе этих методов рассматривается возможность написания компьютерных программ на языке Basic для компилятора Turbo Basic фирмы Borland, которые реально позволяют решать все рассмотренные здесь задачи ядерной физики. К таким задачам относятся вариационные методы, применяемые в фазовом анализе при рассеянии ядерных частиц с разным спином, нахождение полных сечений фотоядерных процессов на легких ядрах, состояния взаимодействующих квантовых частиц, когда в ядерном потенциале присутствует тензорная компонента и т.д..

Автор надеется, что данная книга, в какой-то степени, сможет восполнить пробел, существующий в имеющейся литературе по описанию математических и численных методов и подходов, алгоритмов и компьютерных программ, используемых для решения определенного круга задач ядерной физики легких атомных ядер.

# СОДЕРЖАНИЕ

# ВВЕДЕНИЕ

В настоящее время не существует общей и законченной теории легких атомных ядер, и для анализа различных ядерных характеристик используются различные физические модели и методы. Хотя они часто позволяют получить хорошие результаты, нет целостной картины взаимодействий составных ядерных частиц в континууме и связанных состояниях, т.е. в непрерывном и дискретном спектре решений уравнения Шредингера, которое описывает такие системы [1,2].

Поэтому большой интерес представляет изучение возможностей потенциальной кластерной модели на основе межкластерных или межнуклонных взаимодействий с запрещенными состояниями. Такие потенциалы позволяют эффективно учитывать принцип Паули в межкластерных (нуклон - нуклонных) взаимодействиях, не требуя явной антисимметризации волновых функций системы, что заметно упрощает все компьютерные вычисления [3,4,5].

Ядерные межкластерные потенциалы, согласованные с фазами упругого рассеяния соответствующих ядерных частиц, должны быть способны воспроизвести свойства связанных состояний некоторых ядер в кластерной модели. Это можно рассматривать, как предпосылки к совместному описанию континуума и дискретного спектра на основе единых гамильтонианов в уравнении Шредингера, которое описывает данные системы .

Расчеты, проводимые на основе выбранных представлений, сравниваются с имеющимися экспериментальными данными, что позволяет сделать определенные выводы о качестве используемых физических моделей. И, тем самым, отобрать представления и подходы, которые приводят к лучшему согласию с экспериментом, а значит, максимально приближены к реальной ситуации, существующей в атомном ядре.

В начале 70г. в работах [6,7,8] впервые было показано, что фазы упругого рассеяния легких кластерных систем могут быть описаны на основе глубоких чисто притягивающих потенциалов Вудс - Саксоновского типа, которые содержат связанные запрещенные состояния (ЗС). Структура ЗС определяется перестановочной симметрией волновых функций (ВФ) системы относительно нуклонных перестановок. Поведение фаз рассеяния при нулевой энергии для таких взаимодействий подчиняется обобщенной теореме Левинсона [6-8,9,10,11]

$$\delta_L = \pi( N_L + M_L ) \quad ,$$

где $N_L$ и $M_L$ число запрещенных и разрешенных связанных со-стояний.

Фазы при больших энергиях стремятся к нулю все время оставаясь положительными. Такой подход, по - видимому, можно рассматривать, как альтернативу часто используемой концепции отталкивающего кора, который вводится для качественного учета принципа Паули без выполнения полной антисимметризации ВФ. Радиальная ВФ разрешенных состояний (РС) потенциалов с ЗС осциллирует на малых расстояниях, а не вымирает, как это было для взаимодействий с кором. Благодаря этому в рассмотрение включается внутренняя структура ядра, которая определяется поведением волновой функции системы в области малых расстояний.

В работах [9,12,13,14,15,16,17,18,19] были параметризованы межкластерные центральные гауссовы потенциалы взаимодействия, правильно воспроизводящие фазы упругого $^{4}$He$^{2}$H рассеяния при низких энергиях и содержащие запрещенные состояния. Показано, что на основе этих потенциалов в кластерной модели можно воспроизвести основные характеристики связанных состояний (СС) ядра $^{6}$Li, вероятность кластеризации которого в рассматриваемом канале сравнительно высока. Все состояния в такой системе оказываются чистыми по орбитальным схемам Юнга [6-19] и потенциалы, полученные из фаз рассеяния, можно непосредственно применять для описания характеристик основного состояния (ОС) ядра [20,21,22,23,24,25,26,27,28].

Для более легких кластерных систем вида N$^{2}$H, $^{2}$H$^{2}$H, p$^{3}$H, n$^{3}$He и т.д. в состояниях рассеяния с минимальным спином уже возможно смешивание по орбитальным симметриям и ситуация оказывается более сложной. В состояниях с минимальным спином, в непрерывном спектре разрешены две орбитальные симметрии с различными схемами Юнга, в то время, как связанным основным состояниям, по-видимому, соответствует только одна из этих схем [2,29,30,31,32,33,34,35,36,37,38,39,40,41]. Поэтому потенциалы, непосредственно полученные на основе экспериментальных фаз рассеяния, эффективно зависят от различных орбитальных схем и не могут в таком виде использоваться для описания характеристик основного состояния. Из таких взаимодействий, необходимо выделять чистую компоненту, применимую уже при анализе характеристик связанных состояний.

В более тяжелых ядерных системах N$^{6}$Li, N$^{7}$Li и $^{2}$H$^{6}$Li также реализуется подобная ситуация [2,42,43,44,45], когда в некоторых случаях различные состояния оказываются смешанными по схемам Юнга. В этих работ были впервые получены чистые по схемам Юнга потенциалы взаимодействия для перечисленных выше трех ядерных систем. Они, в основном, оказались способны правильно описывать как характеристики рассеяния, так и свойства связанных состояний соответствующих ядер.

Таким образом, большинство задач ядерной физики требуют знания волновой функции относительного движения частиц, которые участвуют в столкновениях (процессы рассеяния) или определяют связанное состояние ядра, т.е. являются внутренними фрагментами полной системы. Эту функцию можно найти из решений уравнения Шредингера для каждой конкретной физической задачи в дискретном или непрерывном спектре, если известен потенциал взаимодействия этих частиц.

Ядерный потенциал взаимодействия частиц (в задачах рассеяния или связанных состояниях) заведомо не известен, и определить его напрямую какими - либо способами не представляется возможным. Поэтому выбирается определенная форма его зависимости от расстояния (например, гауссова или экспоненциальная), и по некоторым ядерным характеристикам (обычно, это фазы ядерного рассеяния) фиксируются его параметры, так чтобы он описывал эти характеристики. В дальнейшем такой потенциал можно применять для расчетов любых других ядерных характеристик, например, энергий связи рассматриваемых ядер и свойств их связанных состояний или сечений различных реакций [2].

Практически весь круг, рассмотренных выше физических задач, требует умения решать уравнение Шредингера или связанную систему этих уравнений в случае тензорных ядерных сил с определенными начальными и асимптотическими условиями. В принципе, это чисто математическая задача из области математического моделирования физических процессов и систем. Существующие методы его решения [46,47,48,49,50,51,52,53] не всегда приводят к устойчивой численной схеме, а обычно используемые алгоритмы либо приводят к не высокой точности результатов, либо к переполнению в процессе работы компьютерных программ.

Решать уравнения Шредингера для связанных состояний и рассеяния можно, например, методом Рунге - Кутта или конечно - разностным методом [54,55]. Такие методы позволяют найти собственные, волновые функции и собственные энергии квантовой системы, если использовать предложенную нами комбинацию численных и вариационных методов и контролировать точность решения уравнения или системы связанных уравнений Шредингера методом невязок [56].

Описанию этих математических и численных методов, некоторых программных алгоритмов и самих компьютерных программ на языке Turbo Basic, непосредственно применяемых для решения уравнения Шредингера или системы таких уравнений в задачах ядерной физики и будет посвящена данная книга.

Перейдем теперь к непосредственному рассмотрению основных методов и подходов, используемых в потенциальной кластерной модели, где считается, что атомное ядро состоит из двух бесструктурных фрагментов, свойства которых совпадают или близки к свойствам соответствующих ядер в свободном состоянии. Поэтому для многих характеристик кластеров, например, зарядового радиуса, кулоновского формфактора, квадрупольного и магнитного моментов, других характеристик связанных фрагментов принимаются характеристики не взаимодействующих легких ядер $^4$He, $^3$H и $^2$H и т.д. Классическим образцом кластерного объекта являются ядра $^6$Li и $^7$Li в которых велика вероятность кластеризации в $^4$He$^2$H и $^4$He$^3$H каналах.

Полная волновая функция двухкластерной системы записывается в простом виде

$$\Psi = A\left( \varphi_1(x_1)\varphi_2(x_2)\Psi_{JM}(\vec{R}) \right) .$$

Здесь A - оператор антисимметризации волновых функций по всем возможным перестановкам нуклонов между разными кластерами, если волновые функции кластеров, зависящие от своих внутренних координат $x_i$ выбраны в правильном антисимметризованном виде и $\Psi_{JM}$ - функция относительного движения, которая разделяется на радиальную $\Phi_L(R)$ и спин - угловую $Y^{LS}_{JM}$ функции

$$\Psi_{JM}(\vec{R}) = \sum_L Y^{LS}_{JM}(\hat{R})\Phi_L(R) .$$

Спин - угловая часть волновой функции, определяемая в виде

$$Y^{LS}_{JM}(\hat{R}) = \sum_{m\sigma}\left(LmS\sigma|JM\right)Y_{Lm}(\hat{R})\chi_{s\sigma}(\sigma)$$

связывает орбитальную $Y_{Lm}$ и спиновую $\chi_{s\sigma}$ компоненты волновой функции ядерной системы.

Радиальная волновая функция относительного движения кластеров в ядре $\Phi_L(R)$ при заданном орбитальном моменте L зависит только от одной переменной R - радиус - вектора относительного движения фрагментов и является решением уравнения Шредингера

$$u'_L(r)+(k^2+V(r))u_L(r)=0 \quad , \quad \Phi_L=\frac{u_L}{r} ,$$

где V(r) - потенциал ядерного взаимодействия с учетом кулоновского и центробежного членов, $k^2 = 2\mu E/\hbar^2$ - волновое число относительного движения фрагментов, μ - приведенная масса ядра в рассматриваемом кластерном канале, E - энергия относительного движения в центре масс кластерной системы.

В том случае если ядерные ассоциации сильно обособлены роль эффектов антисимметризации, т.е. обменных процессов между кластерами оказывается малой и действием оператора A можно пренебречь. Однако, сказать заранее какова роль этих эффектов достаточно сложно. Вообще говоря, в каждом конкретном случае надо рассматривать точную антисимметризованную волновую функцию системы и только сравнивая ее с функцией без антисимметризации можно сделать на этот счет определенные выводы.

Процедура антисимметризации волновой функции обычно оказывается довольно сложной, поэтому часто используют приближенные способы учета принципа Паули. В частности, в течении многих лет в потенциал межкластерного взаимодействия вводили отталкивающий кор, который не позволяет кластерам слиться в некоторую общую нуклонную систему, обеспечивая тем самым явное разделение ядра на два фрагмента. Использование потенциалов с кором приводило к вымиранию волновой функции относительного движения кластеров на малых расстояниях.

В последствии появился другой класс ядерных глубоких чисто притягивающих потенциалов, содержащих запрещенные состояния, благодаря которым обеспечивается выполнение принципа Паули.

Автор выражает большую признательность Неудачину В.Г, Кукулину В.И., Краснопольскому В.М., Померанцеву В.Н. (Научно - исследовательский институт ядерной физики, МГУ им. М. Ломоносова, Москва), Часникову И.Я. и Босс Э.Г. (Физико - технический институт МОН РК, Алматы), Буртебаеву Н.Т. и Дуйсебаеву А.Д. (Институт ядерной физики Национального ядерного центра РК, Алматы), Страковскому И.И. и Парке В.С. (Ядерный центр Вашингтонского университета, США), Узикову Ю.Н. (Объединенный институт ядерных исследований, Дубна, РФ), Кобушкину А.П. (Институт теоретической физики им. Боголюбова, Украина) за многочисленные и исключительно полезные обсуждения затронутых в работе вопросов. А так же Джазаирову - Кахраманову А.В. (Казахский Национальный Университет им. аль - Фараби, Алматы) за большую помощь в проведении численных расчетов и обсужде-нии полученных результатов.

Кроме того, выражаю благодарность Гарсону М., Де Ягеру К., Николенко Д.М. и Гилману Р. за предоставленные новые экспериментальные данные по поляризациям и формфакторам дейтрона. Виринге Р. и Маклейдту Р. за волновые функции дейтрона для Аргонского и Боннского потенциалов.

Особую благодарность автор выражает академику МАИн РК, д.ф.-м.н., профессору Чечину Л.М. за постоянную помощь, огромную поддержку и неоценимое содействие настоящей работе.

# 1. МЕТОДЫ РЕШЕНИЯ УРАВНЕНИЯ ШРЕДИНГЕРА С ЦЕНТРАЛЬНЫМИ ПОТЕНЦИАЛАМИ В НЕПРЕРЫВНОМ СПЕКТРЕ

Множество задач ядерной физики могут быть рассмотрены при использовании только центральной части ядерных сил [57,58]. В таком случае имеется одно уравнение Шредингера или система не связанных уравнений (при учете спин - орбитального взаимодействия) и математическая задача решается достаточно просто. Учет тензорной компоненты ядерных сил приводит нас к системе связанных уравнений Шредингера [59,60], решение которой несколько сложнее, но вполне выполнимо описанными далее методами.

В этой главе мы приведем математические и вычислительные методы, используемые при решении уравнений Шредингера для центральных потенциалов при положительных собственных значениях и их применение для рассмотрения квантовой задачи рассеяния частиц и расчетов действительных фаз ядерного рассеяния.

## 1.1 Общие методы решения уравнения Шредингера

Здесь будет рассмотрена общая постановка задачи для решения уравнения Шредингера при положительных непрерывных собственных значениях и определены начальные и граничные условия, при которых решается такая задача, применительно к описанию физических процессов и состояний, а именно, для расчета ядерных фаз рассеяния.

### *1.1.1 Центральные действительные потенциалы*

Уравнение Шредингера для центральных сил взаимодействия между двумя ядерными частицами без учета спин - орбитального и тензорного потенциалов имеет следующий вид [57,58,61,62]

$$u''(r) + [\, k^2 - V_c(r) - V_{cul}(r) - L(L+1)/r^2 \,]u(r) = 0 \,, \qquad (1.1)$$

где r – скалярное относительное расстояние между частицами в Фм. (1 Фм - Ферми $= 10^{-15}$ м.),

u – решения уравнения, т.е. волновые функции (ВФ), а u'' – ее вторая производная,

$V_{cul}(r) = 2\mu/\hbar^2 \; Z_1Z_2/r$ - кулоновский потенциал, приведенный к размерности Фм$^{-2}$,

$\hbar$ - постоянная Планка = 1.055 $10^{-34}$ Дж. с,

$Z_1$ и $Z_2$ – заряды частиц в единицах элементарного заряда (1 э.з. - элементарный заряд = 1.60 $10^{-19}$ Кл),

константа $\hbar^2/M_N$ = 41.4686 или 41.47 МэВ $Фм^2$ (1 МэВ - мега-электронвольт = 1.60 x $10^{-13}$ Дж.),

$M_N$ - масса нуклона, равная 1 а.е.м.,

$V_{цб} = L(L+1)/r^2$ - центробежный потенциал, который зависит от величины орбитального момента относительного движения частиц L, величина $k^2 = 2\mu E/\hbar^2$ - волновое число относительного движения частиц в $Фм^{-2}$,

E – энергия частиц в МэВ,

$\mu = m_1m_2/(m_1+m_2)$ - приведенная масса двух частиц в а.е.м. (1 а.е.м. - атомная единица массы = 1.66 x $10^{-27}$ кг.),

$V_c(r)$ - центральная часть ядерного потенциала, равная $2\mu/\hbar^2\, V_n(r)$,

$V_n(r)$ - радиальная зависимость потенциала, часто принимаемая в виде $-V_0\exp(-\alpha r^2)$ или $-V_0\exp(-\alpha r)$,

$V_0$ - глубина потенциала в МэВ,

величина $\eta = \dfrac{\mu Z_1 Z_2}{\hbar^2 k}$ = 0.0344476 $Z_1Z_2/k$ - называется кулоновским параметром и кулоновский потенциал можно представить в виде

$$V_{cul}(r) = 2\eta k/r.$$

Если учитывается спин - орбитальное взаимодействие, то центральный потенциал принимает вид [58,62]

$$V_c(r) = 2\mu/\hbar^2\,[\,V_n(r) + V_{sl}(r)\,] \quad , \qquad V_{sl}(r) = -\,(\mathbf{sl})\,V_{0sl}\,F(r)$$

где F(r) - функциональная зависимость потенциала от взаимного расстояния между частицами, которая также может быть принята в виде гауссойды $\exp(-\alpha r^2)$ или экспоненты $\exp(-\alpha r)$.

Величина (**sl**) называется спин - орбитальным оператором и ее значения могут быть найдены из хорошо известного выражения [58]

$$(\mathbf{sl})\,u(r) = 1/2\,[J(J+1) - L(L+1) - s(s+1)]\,u(r) \quad ,$$

где J - полный момент системы, L - орбитальный момент, s - спин системы частиц. При учете спин - орбитального взаимодействия уравнение Шредингера разбивается на систему несвязанных уравнений, каждое из которых позволяет найти ВФ для конкретно-го полного момента.

Иногда в потенциал взаимодействия вводят кулоновский радиус $R_c$, и тогда кулоновская часть потенциала принимает несколько иной вид

$$V_{cul}(r) = \frac{2\mu}{\hbar^2} \begin{cases} \dfrac{Z_1 Z_2}{r} & r > R_c \\ Z_1 Z_2 \left(3 - \dfrac{r^2}{R_c^2}\right) \Big/ 2R_c & r < R_c \end{cases} . \qquad (1.2)$$

Уравнение (1.1) образуют задачу Коши с начальными условиями, которые выбираются из физических соображений. Первое начальное условие требует равенства нулю ВФ при $r = 0$. Поскольку ВФ отражает вероятность каких – то процессов или состояний квантовых частиц, то это условие означает ,что две частицы не могут полностью слиться и занимать один и тот же объем. Вторым условием задачи Коши должно быть заданием величины первой производной этой функции. Но из физических соображений нельзя определить величину этой производной, поэтому она берется равной некоторой константе, которая определяет амплитуду волновой функции. В численных расчетах обычно принимают $u' = 0.1$-$1$. Действительная амплитуда функции, которая используется для многочисленных физических расчетов, определяется из асимптотических условий, накладываемых на эту функцию при больших расстояниях $r \to R$, когда ядерный потенциал практически равен нулю.

Асимптотика волновой функции на больших расстояниях, когда $V_n(r \to R) = 0$ является решением уравнения (1.1) и может быть представлена следующим образом

$$u_L(r \to R) \longrightarrow F_L(kr) + tg(\delta_L) G_L(kr) \qquad (1.3)$$

или

$$u_L(r \to R) \longrightarrow Cos(\delta_L) F_L(kr) + Sin(\delta_L) G_L(kr) \quad ,$$

где $F_L$ и $G_L$ - кулоновские функции [63,64] рассеяния, которые являются частными решениями уравнения (1.1) без ядерной части потенциала, т.е. когда $V_c = 0$.

Сшивая численное решение $u(r)$ уравнения (1.1) на больших расстояниях (R порядка 10 - 20 Фм) с этой асимптотикой, можно найти амплитуду функции и фазы рассеяния $\delta_L$ для каждого L при заданной энергии взаимодействующих частиц.

Фазы рассеяния в конкретной системе ядерных частиц могут быть определены из фазового анализа экспериментальных данных по их упругому рассеянию (глава 6,7 данной работы). Далее, выполняется варьирование параметров ядерного потенциала заранее определенной формы в уравнении (1.1) и определяются те параметры, которые позволяют описать результаты фазового анализа.

Таким образом, задача описания процессов рассеяния ядерных частиц состоит именно в поиске параметров ядерного потенциала, которые описывают результаты фазового анализа, а, значит, экспериментальные данные по сечениям рассеяния.

Рассмотрим более подробно процедуру сшивки волновых функций с их асимптотикой. При r = R можно записать два равенства для самих ВФ и их производных [65]

$$Nu_L(R) = F_L(kR) + tg(\delta_L)G_L(kR) \quad ,$$
$$Nu'_L(R) = F'_L(kR) + tg(\delta_L)G'_L(kR) \quad ,$$

где N - нормировочный множитель. Можно рассматривать подобные выражения не для функции и производной, а только для функции, но в двух разных точках

$$Nu_L(R_1) = F_L(kR_1) + tg(\delta_L)G_L(kR_1) \quad ,$$
$$Nu_L(R_2) = F_L(kR_2) + tg(\delta_L)G_L(kR_2) \quad . \qquad (1.4)$$

Введем обозначения

$$F_1=F_L(kR_1)\ , \qquad F_2=F_L(kR_2)\ ,$$
$$G_1=G_L(kR_1)\ , \qquad G_2=G_L(kR_2)\ ,$$
$$u_1=u_L(R_1)\ , \qquad u_2=u_L(R_2)$$

и найдем величину N, например, из первого уравнения

$$N= [F_1+tg(\delta_L)G_1]/u_1 \quad .$$

Подставляя это выражение во второе уравнение, получим

$$tg(\delta_L) = (u_1F_2 - u_2F_1)/(u_2G_1 - u_1G_2) = A_L \quad . \qquad (1.5)$$

Тогда

$$\delta_L = Arctg(A_L) \quad .$$

Нормировка функции, для наших целей поиска фаз, значения не имеет. Но если нужна и нормированная ВФ, т.е. полная функция рассеяния, то лучше рассматривать второе уравнение из (1.3), записав его в виде (1.4) и выполнив действия аналогичные, приведенным выше. Для фаз рассеяния получается такое же выражение, а нормировка запишется в виде

$$N = [Cos(\delta_L)F_1 + Sin(\delta_L)G_1]/u_1$$

или

$$N = [Cos(\delta_L)F_2 + Sin(\delta_L)G_2]/u_2 \ .$$

Тем самым, мы полностью определяет поведение волновой функции, ее амплитуду и фазовый сдвиг, во всей области решений уравнения (1.1) от нуля до некоторого большого R, которое определяет асимптотику ВФ.

### *1.1.2 Центральные комплексные потенциалы*

Если в ядерных процессах открыт неупругий канал рассеяния или реакций, то нужно использовать комплексный потенциал взаимодействия, учитывающий убывание потока частиц из упругого канала [57].

Потенциал принимает теперь вид

$$V_c = V_r(r) + iV_m(r) \ , \qquad (1.6)$$

где $V_r(r)$ - действительная часть потенциала и $V_m(r)$ – его мнимая часть. Волновая функция также становится комплексной и может быть записана в форме

$$u(r) = x(r) + iy(r) \ . \qquad (1.7)$$

Тогда уравнение Шредингера (1.1) можно переписать в виде связанной системы уравнений

$$x''(r) + [\, k^2 - V_r(r) - V_{cul}(r) - L(L+1)/r^2 \,]x(r) = - V_m y(r) \ ,$$
$$y''(r) + [\, k^2 - V_r(r) - V_{cul}(r) - L(L+1)/r^2 \,]y(r) = V_m x(r) \ . \qquad (1.8)$$

С начальными условиями вида

$$x(r{=}0) = 0 \ , \qquad x'(r{=}0) = const \ ,$$

$y(r=0) = 0 \quad , \qquad y'(r=0) = const \; .$

В численных расчетах, величина константы (const) для производных волновых функций обычно задается на уровне 0.1-1. Асимптотика волновых функций представляется теперь следующим образом [57]

$$u(r) = H^+(r) + SH^-(r) = [F(r) + iG(r)] + S[F(r) - iG(r)] \; , \qquad (1.9)$$

где $H^+$ - функции Ганкеля, F и G - кулоновские функции и S - матрица рассеяния, которая имеет вид

$$S = e^{2i\delta} = S_1 + iS_2 = Cos(2\delta) + iSin(2\delta) \; .$$

При учете неупругих процессов сами фазы упругого рассеяния становятся комплексными и представляются следующим образом

$$\delta = \sigma + i\Delta \; ,$$

где σ и Δ - действительная и мнимая часть фазы. Тогда матрицу рассеяния можно переписать в виде

$$S = e^{2i\delta} = e^{-2\Delta}e^{2i\sigma} = \eta e^{2i\sigma} = \eta(S_1+iS_2) = \eta[Cos(2\sigma)+iSin(2\sigma)] \; , \qquad (1.10)$$

где $\eta = e^{-2\Delta}$ - параметр неупругости. Для определения фаз рассеяния и параметра неупругости запишем граничные условия для функций в двух точках

$$\frac{u_1}{u_2} = \frac{H_1^+ + SH_1^-}{H_2^+ + SH_2^-} \; , \qquad (1.11)$$

откуда легко найти

$$S = \frac{u_2 H_1^+ - u_1 H_2^+}{u_1 H_2^- - u_2 H_1^-} \; .$$

Подставляя выражения для функций Ганкеля, приведенные выше (1.9), и разделяя действительную и мнимую часть получим

$$S = \frac{C + iD}{A + iB} = K + iM \; , \qquad (1.12)$$

где

$$K = \frac{AC + BD}{A^2 + B^2} \quad , \qquad M = \frac{AD - BC}{A^2 + B^2} \tag{1.13}$$

и

$A = b - a$ , $B = - c - d$ ,
$C = a + b$ , $D = c - d$ ,
$a = x_2F_1 - x_1F_2$ , $b = y_1G_2 - y_2G_1$ ,
$c = y_2F_1 - y_1F_2$ , $d = x_1G_2 - x_2G_1$

Таким образом, все элементы S - матрицы выражаются через кулоновские функции и решения исходного уравнения Шредингера (1.8) с заданным ядерным потенциалом.

Сравнивая действительную и мнимую часть выражений (1.10) и (1.12) получим

$$S_1 = Cos(2\sigma) = K/\eta \quad ,$$
$$S_2 = Sin(2\sigma) = M/\eta \quad . \tag{1.14}$$

и

$$S^2 = \eta^2(S_1 + iS_2)^2 = \eta^2 \quad ,$$
$$S^2 = K^2 + M^2 \quad , \tag{1.15}$$

откуда находим

$$\eta^2 = K^2 + M^2$$

- параметр неупругости. Зная теперь эти величины, получим

$$A = tg(\sigma) = \frac{S_2}{1 + S_1} \quad . \tag{1.16}$$

Тогда

$$\sigma = Arctg(A) \quad . \tag{1.17}$$

Не трудно проверить, что когда $V_m = 0$ и уравнения (1.8) становятся независимыми, то $\eta = 1$, а результаты для фаз (1.5) и (1.16) будут совпадать.

Для определения нормировки ВФ используем выражения (1.9) и (1.7)

$$N(x+iy) = H^{+}(r)+SH^{-}(r) = [F(r)+iG(r)] + (S_1+iS_2)\ [F(r)-iG(r)]\ .$$

Откуда находим

$$N = \frac{Ax + By}{x^2 + y^2} + i\frac{Bx - Ay}{x^2 + y^2} \quad ,$$

где

$$A = (1 + S_1)F(r) + S_2G(r)\quad ,\qquad B = (1 - S_1)G(r) + S_2F(r)\quad .$$

В общем случае, нормировка ВФ может быть записана в виде

$$Nu(r) = (N_1 + iN_2)(x + iy) = N_1x - N_2y +i[N_1y + N_2x] = v + iw\ .$$

Здесь v и w - уже нормированные полные волновые функции рассеяния. Приравнивая действительную и мнимую части, будем иметь

$$N_1 = \frac{Ax + By}{x^2 + y^2}\ ,\qquad N_2 = \frac{Bx - Ay}{x^2 + y^2}$$

- общие выражения для определения нормировки ВФ рассеяния в случае комплексных потенциалов [66].

## 1.2 Численные методы решения уравнения Шредингера

Для численного решения уравнения Шредингера можно использовать конечно – разностный метод, представляя функцию и ее производную в виде центральных разностей или использовать известный метод Рунге - Кутта, который в некоторых случаях позволяет получить более высокую точность решения в каждой точке численной схемы.

### *1.2.1 Центральные действительные потенциалы*

Уравнение Шредингера для центральных ядерных сил (1.1) запишем в виде [57]

$$u'' + [k^2 - V(r)]u = 0\quad . \tag{1.18}$$

Для его решения можно использовать конечно - разностный метод, в котором вторая производная может быть представлена следующим образом [58]

$$u''(r) = [u(r + h) - 2u(r) + u(r - h)]/h^2 = [u(r_{i+1}) - 2u(r_i) + u(r_{i-1})]/h^2 , \qquad (1.19)$$

где h - шаг конечно - разностной сетки, для определения которого весь интервал значений r от нуля до некоторого R, делится на N частей

$$h = R/N .$$

Здесь R - верхний предел, на котором выполняется сшивка численного решения уравнения (1.18) с его асимптотикой. Тогда

$$r_i = hi \quad ,$$

где i меняется от 0 до N ($r_0 = 0$ и $r_N = R$). Выражение (1.19) теперь можно переписать в виде

$$u'' = [u_{i+1} - 2u_i + u_{i-1}]/h^2 \quad ,$$

а все уравнение перепишется

$$[u_{i+1} - 2u_i + u_{i-1}]/h^2 + [k^2 - V(r_i)]u_i = 0 \quad .$$

Откуда находим

$$u_{i+1} = [2 + h^2V(r_i) - h^2k^2]u_i - u_{i-1} \quad . \qquad (1.20)$$

Функция при r = 0 должна быть равна нулю, а на первом шаге может быть принята равной некоторой константе, которая определяет только нормировку функции, не сказываясь на ее поведении при различных r.

Отсюда находится ВФ на следующем шаге $u_2$ и этот процесс повторяется пока i не станет равно N - 1. Такая процедура позволяет найти весь массив значений ВФ во всех точках от нуля до R. Далее мы выполняем ее сшивку в двух точках, например, при $r_N = R$ и $r_{N-5} = R - 5h$, как описано в параграфе (1.1.1) . Вторая точка определяется экспериментальным путем в каждом конкретном случае и зависит от энергии частиц, но при малых энергиях обычно бывает достаточно отступить назад на 3 - 5 шагов [67].

### *1.2.2 Центральные комплексные потенциалы*

Если имеется система уравнений (1.8) для комплексного потенциала [67]

$$x''(r) + [\, k^2 - V_r(r) - V_{cul}(r) - L(L+1)/r^2\,]x(r) = -\,V_m y(r)$$
$$y''(r) + [\, k^2 - V_r(r) - V_{cul}(r) - L(L+1)/r^2\,]y(r) = V_m x(r) \qquad (1.21)$$

то, используя такое же представление производной в конечно - разностном виде

$$u'' = [u_{i+1} - 2u_i + u_{i-1}]/h^2$$

для функций x и y получим

$$x_{i+1} = [\, 2 - A_i h^2\,]x_i - x_{i-1} - h^2 V_m(r_i) y_i\,,$$
$$y_{i+1} = [2 - A_i h^2\,]y_i - y_{i-1} + h^2 V_m(r_i) x_i\,, \qquad (1.22)$$

где

$$A_i = k^2 - V_r(r_i) - V_{cul}(r_i) - L(L+1)/r_i^2$$

Задавая значения функций в двух первых точках

$$x_0 = 0\,, \qquad x_1 = \text{const}\,, \qquad y_0 = 0\,, \qquad y_1 = \text{const}$$

можно найти значения функций во всех остальных точках [67], как и в случае выражения (1.20). Процедура сшивки численной функции со своей асимптотикой в случае комплексных потенциалов описана в параграфе 1.1.2.

Более сложный случай решения системы вида (1.21), когда в потенциале присутствует тензорная компонента мы рассмотрим в следующей главе.

### *1.2.3 Метод Рунге – Кутта для центральных действительных потенциалов*

Рассмотрим теперь другой метод решения таких уравнений, который называется методом Рунге - Кутта четвертого порядка [68,69,70,71,72]. Стандартный метод решения одного дифференциального уравнения первого порядка

$$y' = f(x,y) \tag{1.23}$$

с начальным условием

$$y(x_0) = y_0$$

заключается в представлении решения на интервале от 0 до некоторого R в виде

$$y_{n+1} = y_n + \Delta y_n \ , \tag{1.24}$$

где n может меняться от 0 до N ($x_N = hN$), h - шаг решения, а $\Delta y_n$ находится из выражения

$$\Delta y_n = 1/6(k_1 + 2k_2 + 2k_3 + k_4) \ , \tag{1.25}$$

где

$$k_1 = hf(x_n,y_n) \ , \qquad k_2 = hf(x_n+h/2,\ y_n+k_1/2) \ ,$$
$$k_3 = hf(x_n+h/2,\ y_n+k_2/2) \ , \quad k_4 = hf(x_n+h,\ y_n+k_3) \ .$$

В случае системы двух дифференциальных уравнений первого порядка [68-72]

$$y' = f(x,y,z) \ ,$$
$$z' = g(x,y,z) \tag{1.26}$$

с начальными условиями

$$y(x_0) = y_0 \ , \qquad z(x_0) = z_0$$

решения находятся из выражений

$$y_{n+1} = y_n + \Delta y_n \ ,$$
$$z_{n+1} = z_n + \Delta z_n \ , \tag{1.27}$$

где

$$\Delta y_n = 1/6(k_1 + 2k_2 + 2k_3 + k_4) \ , \tag{1.28}$$
$$\Delta z_n = 1/6(m_1 + 2m_2 + 2m_3 + m_4)$$

и

$k_1 = hf(x_n,y_n,z_n)$ ,
$m_1 = hg(x_n,y_n,z_n)$ ,

$k_2 = hf(x_n+h/2,\ y_n+k_1/2,\ z_n+m_1/2)$ ,
$m_2 = hg(x_n+h/2,\ y_n+k_1/2,\ z_n+m_1/2)$ ,

$k_3 = hf(x_n+h/2,\ y_n+k_2/2,\ z_n+m_2/2)$ ,
$m_3 = hg(x_n+h/2,\ y_n+k_2/2,\ z_n+m_2/2)$ ,

$k_4 = hf(x_n+h,\ y_n+k_3,\ z_n+m_3)$ ,
$m_4 = hg(x_n+h,\ y_n+k_3,\ z_n+m_3)$ .

В случае одного дифференциального уравнения второго порядка вида (1.18)

$$y'' = g(x,y,y') \tag{1.29}$$

с начальными условиями

$y(0) = y_0$ , $y'(0) = y'_0$

используем замену

$z = y'$ .

Тогда получаем систему вида

$$\begin{aligned} y' &= z \ , \\ z' &= g(x,y,z) \end{aligned} \tag{1.30}$$

с начальными условиями

$y(0) = y_0$ , $z(0) = z_0$

решение которой при $f(x,y,z) = z$ (см. 1.26 и 1.27) может быть представлено следующим образом

$$\begin{aligned} \Delta y_n &= hz_n + 1/6h(m_1 + m_2 + m_3) \ , \\ \Delta z_n &= 1/6(m_1 + 2m_2 + 2m_3 + m_4) \end{aligned} \tag{1.31}$$

и

$k_1 = hz_n$ , $m_1 = hg(x_n,y_n,z_n)$ ,
$k_2 = h(z_n+m_1/2)$ , $m_2 = hg(x_n+h/2, y_n+k_1/2, z_n+m_1/2)$ ,
$k_3 = h(z_n+m_2/2)$ , $m_3 = hg(x_n+h/2, y_n+k_2/2, z_n+m_2/2)$ ,
$k_4 = h(z_n+m_3)$ , $m_4 = hg(x_n+h, y_n+k_3, z_n+m_3)$ .

В дальнейшем будет показано, что оба эти метода (конечно - разностный и Рунге - Кутта) для рассмотренного круга задач приводят примерно к одинаковым результатам и обеспечивают одинаковую точность.

### *1.2.4 Методы расчета кулоновских фаз*

Для практических расчетов характеристик ядерных реакций и процессов рассеяния во многих случаях необходимо знать и, как правило, с высокой точностью, численные значения кулоновских функций в заданной точке R и кулоновских фаз в широком диапазоне значений кулоновского параметра $\eta$.

В настоящее время известно достаточно много различных численных методов, применимых для нахождения этих величин, однако, только сравнительно недавно появились достаточно простые и надежные представления для кулоновских функций, а известные способы вычисления кулоновских фаз и сейчас обладают рядом недостатков, так что при их использовании необходимо соблюдать определенную осторожность.

Кулоновские фазы определяются через Г - функцию следующим образом [57,73]

$$\sigma_L = \arg\{\Gamma(L+1+i\eta)\} \qquad (1.32)$$

и удовлетворяют рекуррентному процессу

$$\sigma_L = \sigma_{L+1} - \mathrm{Arctg}\left(\frac{\eta}{L+1}\right) , \qquad (1.33)$$

где $\eta = \dfrac{\mu Z_1 Z_2}{\hbar^2 k}$ - кулоновский параметр, $\mu$ - приведенная масса двух частиц, k - волновое число относительного движения и $k^2$ = $2\mu E/\hbar^2$, E - энергия сталкивающихся частиц в центре масс.

Откуда сразу можно получить следующее очевидное выражение

$$\alpha_L = \sigma_L - \sigma_0 = \sum_{n=1}^{L} Arctg\left(\frac{\eta}{n}\right) , \qquad \alpha_0=0 \ . \qquad (1.34)$$

Наиболее естественное представление для кулоновских фаз получается на основе интегральной формулы для Г - функции [73,86]

$$\sigma_L = \mathrm{Arctg}(y/x) \ ,$$

где

$$y = \int_0^\infty \exp(-t) t^L \mathrm{Sin}(\eta \ln t) dt \ ,$$
$$x = \int_0^\infty \exp(-t) t^L \mathrm{Cos}(\eta \ln t) dt \qquad (1.35)$$

Однако непосредственное вычисление этих интегралов оказывается достаточно сложной задачей, так как подынтегральные функции являются быстро осциллирующими при $t \to 0$. Поэтому часто используются различного рода приближения и асимптотические разложения, например, такие, как представление фазы при L = 0 в виде [75]

$$\sigma_0 = -\eta + \frac{\eta}{2}\ln(\eta^2+16) + \frac{7}{2}\mathrm{arctg}(\eta/4) - [\mathrm{arctg}\eta + \mathrm{arctg}(\eta/2) + \mathrm{arctg}(\eta/3)] -$$

$$-\frac{\eta}{12(\eta^2+16)}[1 + \frac{1}{30}\frac{\eta^2-48}{(\eta^2+16)^2} + \frac{1}{105}\frac{\eta^4-160\eta^2+1280}{(\eta^2+16)^4} + ....]$$

или для L >> 1 [57]

$$\sigma_L = \alpha(L+1/2) + \eta(\ln\beta - 1) + \frac{1}{\beta}\left(-\frac{\mathrm{Sin}\alpha}{12} + \frac{\mathrm{Sin}3\alpha}{360\beta^2} - \frac{\mathrm{Sin}5\alpha}{1260\beta^4} + \frac{\mathrm{Sin}7\alpha}{1680\eta^6} - ...\right)$$

где

$$\alpha = \mathrm{arctg}\left(\frac{\eta}{L+1}\right) , \qquad \beta = \sqrt{\eta^2 + (L+1)^2} \ .$$

Используя эти формулы, все остальные фазы определяются из рекуррентных соотношений (1.33). Хотя оба представления обладают высокой скоростью счета на компьютере, фазы получаются с некоторой ошибкой, оценить которую можно только сравнив полученный результат с табличными данными или вычислениями по точным формулам. Кроме того, последняя формула верна только при L ≈ 100 и рекуррентный процесс вносит дополнительную ошибку в величину фаз.

Известны и другие представления кулоновских фаз, в частности, при $\eta >> 1$ [76]

$$\sigma_0 = \frac{\pi}{4} + \eta(\log\eta - 1) - \sum_{s=1}^{\infty} \frac{B_s}{2s(2s-1)\eta^{2s-1}} \quad , \qquad (1.36)$$

где $B_s$ - числа Бернулли [73]. Однако, подобное разложение хорошо работает только при $\eta \approx 100$. В работах [75,76] было показано, что можно получить восемь верных знаков с учетом только первого члена суммы (1.36) только при $\eta = 85$. При малых $\eta$ ряд сходится плохо и требует, кроме того задания или вычисления чисел Бернулли.

В работе [75] имеется и другое определение кулоновских фаз

$$\sigma_L = \eta\Psi(L+1) + \sum_{n=1}^{\infty}\left[\frac{\eta}{L+n} - \mathrm{arctg}\left(\frac{\eta}{L+n}\right)\right] \quad , \qquad (1.37)$$

которое можно получить из известной формы записи Г - функции [73]

$$\Gamma(z)=\Gamma(x+iy)=r\,\exp(i\phi)=r\,(\mathrm{Cos}\phi+i\mathrm{Sin}\phi) \quad ,$$

где

$$\phi = y\Psi(x) + \sum_{n=0}^{\infty}(\mathrm{tg}\omega_n - \omega_n) \quad , \qquad \omega_n = \mathrm{arctg}\left(\frac{y}{x+n}\right) \quad ,$$

а $\Psi(x)$ - логарифмическая производная Г - функции [73], которая для целого аргумента имеет вид

$$\Psi(L+1) = -C + 1 + 1/2 + ..... + 1/L \; \cdot$$

Здесь C = 0.5772156649..... - постоянная Эйлера [73]. Ряд (1.56)

будет сходиться тем быстрее, чем меньше $\eta$ и больше L. Эта формула охватывает противоположную представлению (1.36) область и при $1 < \eta < 50$ оба разложения имеют плохую сходимость.

Чтобы оценить остаточный член ряда (1.37) разложим арктангенс в ряд при $\eta/n << 1$, что всегда возможно при больших n или маленьких $\eta$. Тогда получим

$$\sigma_0 = -C\eta + \sum_{n=1}^{\infty}\left(\frac{\eta^3}{3n^3} - \frac{\eta^5}{5n^5} + \frac{\eta^7}{7n^7} - ...\right) \quad . \tag{1.38}$$

Отсюда видно, что остаток ряда будет иметь порядок величины $\eta^3/n^2$ [68]. Ряд (1.37) при $\eta > 1$ сходится сравнительно плохо, так как для получения, например, относительной точности $10^{-8}$ требуется учитывать десятки тысяч членов этого ряда.

Однако, такой ряд допускает существенное улучшение сходимости при $\eta \approx 1$ [84] после преобразования его к виду

$$\sigma_0 = -\eta C + \frac{1}{3}\eta^3 S + \sum_{n=1}^{\infty}\left[\frac{\eta}{n} - \mathrm{arctg}\left(\frac{\eta}{n}\right) - \frac{1}{3}\frac{\eta^3}{n^3}\right] \quad , \tag{1.39}$$

где $S = \sum_{k=1}^{\infty}\frac{1}{k^3} = 1.202056903...$ . Несложно найти, что остаточный член такого ряда равен $\eta^5/n^4$ и для получения восьми верных знаков требуется учитывать только несколько сотен членов при $\eta \approx 1$.

Приведенный выше ряд (1.39) допускает дополнительное улучшение сходимости, после преобразования его к виду [84]

$$\sigma_0 = -\eta C + \frac{1}{3}\eta^3 S - \frac{1}{5}\eta^5 D + \sum_{n=1}^{\infty}\left[\frac{\eta}{n} - \mathrm{arctg}\left(\frac{\eta}{n}\right) - \frac{1}{3}\frac{\eta^3}{n^3} + \frac{1}{5}\frac{\eta^5}{n^5}\right] \quad , \tag{1.40}$$

где $D = \sum_{k=1}^{\infty}\frac{1}{k^5} = 1.036927751...$ . Такой ряд сходится очень быстро и имеет остаточный член порядка $\eta^7/n^6$, так что для удовлетворения указанной выше точности требуется учитывать только несколько десятков членов.

Ниже приведена программа для вычисления кулоновских фаз, описанным выше методом разложения в ряд и на основе интегральных представлений (1.35).

```
REM ПРОГРАММА ВЫЧИСЛЕНИЯ КУЛОНОВСКИХ ФАЗ РАССЕЯНИЯ
DEFDBL A-Z: DEFINT I,J,K,L,N,M: M=4000:DIM V(M): CLS
EPS=1.0E-15: H=1: REM КУЛОНОВСКИЙ ПАРАМЕТР
REM *** РАСЧЕТ ФАЗ НА ОСНОВЕ РЯДОВ ***
C=0.577215665: A1=1.202056903/3: A2=1.036927755/5: F=0: S1=0
3 F=F+1: B=H/F-ATN(H/F): S1=S1+B: IF B<EPS GOTO 2: GOTO 3
2 D=0: S=0
4 D=D+1: A=H/D-ATN(H/D)-(H/D)^3/3+(H/D)^5/5: S=S+A
IF A<EPS GOTO 1: GOTO 4
1 FAZ=-C*H+A1*H^3-A2*H^5+S
FAZ1=-C*H+S1: PRINT "FAZ = ";FAZ; "    N = ";D
PRINT "FAZ1 = ";FAZ1;"  N = ";F
REM *** РАСЧЕТ ФАЗ НА ОСНОВЕ ИНТЕГРАЛЬНОГО ПРЕДСТАВЛЕНИЯ ***
NN=4000: E=1E-300: N=500: R1=0.1: R2=1: R3=40: HH=R1/N
H1=HH/NN: H2=(R2-R1)/NN: H3=(R3-R2)/NN: YY=0: FOR K=1 TO N
AA=(K-1)*HH: FOR I=0 TO NN: X=H1*I+AA+E
V(I)=EXP(-X)*SIN(H*LOG(X)): NEXT I: CALL SIM(NN,H1,V(),Y1)
YY=YY+Y1: NEXT K: FOR I=0 TO NN: X=H2*I+R1
V(I)=EXP(-X)*SIN(H*LOG(X)): NEXT: CALL SIM(NN,H2,V(),Y1)
YY=YY+Y1: FOR I=0 TO NN: X=H3*I+R2
V(I)=EXP(-X)*SIN(H*LOG(X)): NEXT
CALL SIM(NN,H3,V(),Y1): YY=YY+Y1: XX=0: FOR K=1 TO N
AA=(K-1)*HH: FOR I=0 TO NN: X=H1*I+AA+E
V(I)=EXP(-X)*COS(H*LOG(X)): NEXT I: CALL SIM(NN,H1,V(),X1)
XX=XX+X1: NEXT K: FOR I=0 TO NN: X=H2*I+R1
V(I)=EXP(-X)*COS(H*LOG(X)): NEXT: CALL SIM(NN,H2,V(),X1)
XX=XX+X1: FOR I=0 TO NN: X=H3*I+R2
V(I)=EXP(-X)*COS(H*LOG(X))
NEXT: CALL SIM(NN,H3,V(),X1): XX=XX+X1: AA=ATN(YY/XX)
PRINT "FAZA = ";AA: END
SUB SIM(N,H,V(5000),S)
A=0: B=0: FOR I=1 TO N-1 STEP 2: B=B+V(I): NEXT
FOR J=2 TO N-2 STEP 2: A=A+V(J): NEXT
S=H*(V(0)+V(N)+2*A+4*B)/3: END SUB
```

В таблице 1.1 приведены фазы, вычисленные по этой программе на основании формулы (1.40) [84]. Ошибка составляет примерно половину последнего знака.

Таблица 1.1 - Вычисление кулоновских фаз.

| $\eta$ | $\sigma_0$ | $\eta$ | $\sigma_0$ |
|---|---|---|---|
| 0.1 | -0.05732294 | 0.6 | -0.27274381 |
| 0.2 | -0.11230222 | 0.8 | -0.30422560 |
| 0.3 | -0.16282067 | 1.0 | -0.30164032 |
| 0.4 | -0.20715583 | 1.3 | -0.23921678 |
| 0.5 | -0.24405830 | 1.5 | -0.16293977 |

Отметим, что для получения одинаковой точности при расчетах по формулам (1.40) и (1.38), в последнем случае, обозначенном в программе $\sigma_1$, нужно учитывать примерно в семьсот раз больше членов ряда (при точности $10^{-15}$), что видно из приведенной распечатки результатов расчета при $\eta = 1$

$\sigma_0 =$ -0.30164032059 для N = 106
$\sigma_1 =$ -0.30164032060 для N = 69337

Расчет кулоновских фаз на основе интегрального представления, может быть выполнен, если разделить весь интервал интегрирования на несколько частей. Наиболее сильно подынтегральная функция меняется при малых t, поэтому делим интервал интегрирования на следующие части 0÷0.1, 0.1÷1 и 1÷40 (при t = 40 подынтегральная функция имеет порядок величины $10^{-17}$), а первую часть (0÷0.1) делим еще на N = 500 частей.

Вычисление интегралов по всем частям приводит нас к величине фазы -.30164031, что отличается от результата, полученного на основе рядов, только на единицу восьмого знака. Если первый интервал 0÷0.1 не делить на N частей, т.е. положить N = 1, то для фазы получается величина -.30163425 и ошибка составляет единицу пятого знака.

Отметим, что вычисление таких интегралов на компьютере Intel Pentium 200 MMX занимает несколько минут, а вычисление ряда (1.40) - доли секунды.

### *1.2.5 Методы расчета кулоновских функций*

Перейдем теперь к рассмотрению кулоновских функций рассеяния, регулярная $F_L(\eta,\rho)$ и нерегулярная $G_L(\eta,\rho)$ части которых являются линейно независимыми решениями радиального уравнения Шредингера (1.18) только с кулоновским потенциалом которое имеет вид [57,60]

$$\chi_L''(\rho) + \left(1 - \frac{2\eta}{\rho} - \frac{L(L+1)}{\rho^2}\right)\chi_L(\rho) = 0 \quad , \qquad (1.42)$$

где $\chi_L = F_L(\eta,\rho)$ или $G_L(\eta,\rho)$, $\rho = kr$, а $\eta = \dfrac{\mu Z_1 Z_2}{\hbar^2 k}$ - кулоновский параметр. Вронскианы этих функций имеют вид [57]

$$W_1 = F_L' G_L - F_L G_L' = 1 ,$$
$$W_2 = F_{L-1} G_L - F_L G_{L-1} = \frac{L}{\sqrt{\eta^2 + L^2}} . \qquad (1.43)$$

Рекуррентные соотношения записываются в форме

$$L[(L+1)^2 + \eta^2]^{1/2} u_{L+1} = (2L+1)\left[\eta + \frac{L(L+1)}{\rho}\right] u_L - (L+1)[L^2 + \eta^2]^{1/2} u_{L-1} \quad ,$$

$$(L+1)u_L' = \left[\frac{(L+1)^2}{\rho} + \eta\right] u_L - [(L+1)^2 + \eta^2]^{1/2} u_{L+1} \quad ,$$

$$L u_L' = [L^2 + \eta^2]^{1/2} u_{L-1} - \left[\frac{L^2}{\rho} + \eta\right] u_L \quad , \qquad (1.44)$$

где $u_L = F_L(\eta,\rho)$ или $G_L(\eta,\rho)$. Асимптотика при $\rho \to \infty$ может быть представлена в виде [74]

$$F_L = \mathrm{Sin}(\rho - \eta \ln 2\rho - \pi L/2 + \sigma_L) ,$$
$$G_L = \mathrm{Cos}(\rho - \eta \ln 2\rho - \pi L/2 + \sigma_L) . \qquad (1.45)$$

Имеется достаточно много методов и приближений для вычисления кулоновских функций [57,75,76,77,78,79,80,81].

Однако, только сравнительно недавно появилось быстро сходящееся представление, позволяющее получить их значения с высокой степенью точности и в широком диапазоне переменных с малыми затратами компьютерного времени [82]. Кулоновские функции в таком методе представляются в виде цепных дробей [83]

$$f_L = F_L' / F_L = b_0 + \cfrac{a_1}{b_1 + \cfrac{a_2}{b_2 + \cfrac{a_3}{b_3 + \dots}}} \quad , \qquad (1.46)$$

где

$b_0 = (L + 1)/\rho + \eta/(L + 1)$ ,
$b_n = [2(L + n) + 1][(L + n)(L + n + 1) + \eta\rho]$ ,
$a_1 = - \rho[(L + 1)^2 + \eta^2](L + 2)/(L + 1)$ ,
$a_n = - \rho^2[(L + n)^2 + \eta^2][(L + n)^2 - 1]$

и

$$P_L + iQ_L = \frac{G_L^{'} + iF_L^{'}}{G_L + iF_L} = \frac{i}{\rho}\left( b_0 + \cfrac{a_1}{b_1 + \cfrac{a_2}{b_2 + \cfrac{a_3}{b_3 + \dots}}} \right), \quad (1.47)$$

где

$$b_0 = \rho - \eta , \qquad b_n = 2(b_0 + in) , \quad (1.48)$$
$$a_n = - \eta^2 + n(n - 1) - L(L + 1) + i\eta(2n - 1) .$$

Такой метод расчета оказывается применим в области при $\rho \ge \eta + \sqrt{\eta^2 + L(L+1)}$ и легко позволяет получить высокую точность благодаря быстрой сходимости цепных дробей. Поскольку кулоновский параметр $\eta$ обычно порядка единицы, а L, как правило, не более 5-7, то метод дает хорошие результаты уже при $\rho > 5$ Фм. Именно в этой области необходимо знать кулоновские функции при численных расчетах ядерных функций рассеяния и реакций.

Используя (1.46-1.48) можно получить связь между кулоновскими функциями и их производными [84,85]

$$F_L^{'} = f_L F_L ,$$
$$G_L = (F_L^{'} - P_L F_L)/Q_L = (f_L - P_L)F_L/Q_L ,$$
$$G_L^{'} = P_L G_L - Q_L F_L = [P_L(f_L - P_L)/Q_L - Q_L]F_L .$$

Таким образом, задавая некоторое значение $F_L$ в точке $\rho$, находим все остальные функции и их производные с точностью до постоянного множителя, который определяется из вронскианов (1.43). Вычисления кулоновских функций по приведенным формулам и сравнение их с табличным материалом [86] показывает, что можно легко получить восемь - девять правильных знаков, если ρ удовлетворяет приведенному выше условию.

Ниже приведен текст компьютерной программы для вычисления кулоновских волновых функций рассеяния. Данная программа и все другие программы этой книги написаны на алгоритмическом языке Basic для компилятора Turbo Basic фирмы Borland [87].

Здесь G - кулоновский параметр, L - орбитальный момент данной парциальной волны, X - расстояние от центра, на котором вычисляются кулоновские функции, FF и GG - сами кулоновские функции, FP и GP - их производные, а W - Вронскиан, определяющий точность вычисления кулоновских функций (Первая формула в выражении (1.43)).

```
SUB CULFUN(G,X,L,FF,GG,FP,GP,W)
REM ***** ПРОГРАММА ВЫЧИСЛЕНИЯ КУЛОНОВСКИХ
ФУНКЦИЙ *****
Q=G:R=X: GK=Q*Q: GR=Q*R:RK=R*R: K=1: F0=1
B01=(L+1)/R+Q/(L+1):BK=(2*L+3)*((L+1)*(L+2)+GR)
AK= - R*((L+1)^2+GK)/(L+1)*(L+2):DK=1/BK: DEHK=AK*DK
S=B01+DEHK
1 K=K+1:AK= - RK*((L+K)^2 - 1)*((L+K)^2+GK)
BK=(2*L+2*K+1)*((L+K)*(L+K+1)+GR):DK=1/(DK*AK+BK)
IF DK>0 GOTO 3
2 F0= - F0
3 DEHK=(BK*DK - 1)*DEHK: S=S+DEHK
IF (ABS(DEHK) - 1E - 10)>0 GOTO 1:FL=S: K=1:RMG=R - Q
LL=L*(L+1):CK= - GK - LL:DK=Q: GKK=2*RMG
HK=2: AA1=GKK*GKK+HK*HK: PBK=GKK/AA1
RBK= - HK/AA1: OMEK=CK*PBK-DK*RBK
EPSK=CK*RBK+DK*PBK: PB=RMG+OMEK:QB=EPSK
4 K=K+1:CK= - GK - LL+K*(K - 1):DK=Q*(2*K - 1)
HK=2*K:FI=CK*PBK                                  -
DK*RBK+GKK:PSI=PBK*DK+RBK*CK+HK
AA2=FI*FI+PSI*PSI:PBK=FI/AA2: RBK= - PSI/AA2
VK=GKK*PBK - HK*RBK:WK=GKK*RBK+HK*PBK
OM=OMEK:EPK=EPSK:OMEK=VK*OM - WK*EPK - OM
EPSK=VK*EPK+WK*OM - EPK:PB=PB+OMEK
QB=QB+EPSK:IF (ABS(OMEK)+ABS(EPSK) - 1E - 10)>0 GOTO 4
PL= - QB/R: QL=PB/R:G0=(FL - PL)*F0/QL
G0P=(PL*(FL - PL)/QL - QL)*F0:F0P=FL*F0
ALFA=1/(SQR(ABS(F0P*G0 - F0*G0P)))
GG=ALFA*G0:GP=ALFA*G0P:FF=ALFA*F0
FP=ALFA*F0P:W=1 - FP*GG+FF*GP:END SUB
```

Результаты контрольного счета кулоновских функций для $\eta$ = 1 [88,89] и сравнение их с табличными данными [86] приведены в таблице 1.2.

Видно, что при $\eta = 1$ и $L = 0$ правильные результаты получаются уже для $\rho = kr = 1$. Величина вронскиана (1.43), представленного в виде $W_1$-1, при любых $\rho$ не превышает $10^{-15}$-$10^{-16}$.

Таблица 1.2 - Кулоновские функции.

| ***ρ*** | ***F₀*** (Наш расчет) | ***F₀*** *[86]* | ***F₀'*** (Наш расчет) | ***F₀'*** *[86]* |
|---|---|---|---|---|
| 1 | 0.22752621 | 0.22753 | 0.34873442 | 0.34873 |
| 5 | 0.68493741 | 0.68494 | -0.72364239 | -0.72364 |
| 10 | 0.47756082 | 0.47756 | 0.84114311 | 0.84114 |
| 15 | -0.97878958 | -0.97879 | 0.31950815 | 0.31951 |
| 20 | -0.32922554 | -0.32923 | -0.92214689 | -0.92215 |
| ***ρ*** | ***G₀*** | ***G₀*** *[86]* | ***G₀'*** | ***G₀'*** *[86]* |
| 1 | 2.0430972 | 2.0431 | -1.2635981 | -1.2636 |
| 5 | -0.89841436 | -0.89841 | -0.51080476 | -0.51080 |
| 10 | 0.94287424 | 0.94287 | -0.43325965 | -0.43326 |
| 15 | 0.34046374 | 0.34046 | 0.91053182 | 0.91053 |
| 20 | -0.97242840 | -0.97243 | 0.31370038 | 0.31370 |
| ***ρ*** | ***F₂*** | ***G₂*** | ***F'₂*** | ***G'₂*** |
| 1 | 1.47867E-02 | 1.26407E 01 | 4.70896E-02 | -2.73727E 01 |
| 5 | 1.18637E 00 | 3.82961E-01 | 1.54145E-01 | -7.93149E-01 |
| 10 | -9.63615E-01 | 4.81305E-01 | 4.24848E-01 | 8.25557E-01 |
| 15 | -2.27973E-01 | -1.01918E 00 | -9.33599E-01 | 2.12743E-01 |
| 20 | -1.01801E 00 | -1.62845E-01 | -1.55072E-01 | -9.57506E-01 |

Правильность вычисления функций при L = 2 легко проверить по рекуррентным формулам (1.44). Зная функции и их производные при L = 0, по второй формуле находим сами функции при L = 1, а затем, по третьей формуле, находим их производные для L = 1. Продолжая этот процесс легко найти все функции и их производные при любых L [84].

## 1.3 Программа расчета фаз рассеяния для центральных действительных потенциалов

Приведем теперь компьютерную программу для расчета действительных фаз упругого рассеяния конечно - разностным методом и методом Рунге - Кутта, которая демонстрирует хорошее совпадение результатов, полученных обоими способами.

Здесь приняты следующие обозначения: NN - Нижнее значение цикла по энергии, NV - Верхнее значение цикла по энергии, NH - Шаг цикла по энергии, EH - Шаг по энергии, EN - Нижнее значение по энергии, AM1 - Масса первой частицы в а.е.м., AM2 - Масса второй частицы в а.е.м., PM - Приведенная масса $\mu$, Z1 - Заряд первой частицы в единицах заряда "e", Z2 - Заряд второй частицы в единицах заряда "e", A1 - Константа $\hbar^2/M_N$ = 41.4686, где $M_N$ - масса нуклона в а.е.м, равная 1, AK1 - Константа при кулоновском потенциале 1.439975 $Z_1Z_2$ $2\mu/\hbar^2$, N - Число шагов при интегрировании уравнения Шредингера, H - Величина шага при интегрировании уравнения Шредингера, R00 - Расстояние, на котором выполняется сшивка численных функций с асимптотикой, V0 - Глубина ядерного потенциала, R0 - Радиус ядерного потенциала, RCU - Кулоновский радиус, L - Орбитальный момент, EL - Энергия частиц в лабораторной системе, ECM - Энергия частиц в системе центра масс, SK - Квадрат волнового числа $k^2$, SS - Волновое число k, G - Кулоновский параметр = 3.44476 $10^{-2}$ $Z_1Z_2$ $\mu$/k, F1(I) - Фаза рассеяния при заданной энергии, полученная конечно - разностным методом, F2(I) - Фаза рассеяния при заданной энергии, полученная методом Рунге - Кутта, ABS(F1(I) - F2(I)) - Разница фаз в градусах, полученных разными методами, ABS(F1(I) - F2(I))/ABS(F1(I))*100 - относительная разница фаз в процентах.

**REM ПРОГРАММА РАСЧЕТА ДЕЙСТВИТЕЛЬНЫХ ФАЗ РАССЕЯНИЯ**

```
REM *** ОПРЕДЕЛЕНИЕ МАССИВОВ И ПЕРЕМЕННЫХ ****
DEFDBL A - Z: DEFINT I,J,K,L,N,M: CLS
NN=4000: DIM EL(100), F1(100), ECM(100), F2(100)
DIM V1(NN), U(NN), V(NN), U1(NN)
PRINT "  EL   ECM   FKR   FRK  ERR - DEG   ERR - %"
REM ******** ЗАДАНИЕ НАЧАЛЬНЫХ ЗНАЧЕНИЙ *********
PI=3.14159265359: NN=1: NV=20: NH=1: EH=1: EN=0
AM1=2: AM2=4: Z1=1: Z2=2: A1=41.4686
 PM=AM1*AM2/(AM1+AM2)
B1=2*PM/A1:  AK1=1.439975*Z1*Z2*B1:  N=1000:  R00=20:
H=R00/N
V0=76.12: R0=.2: A2= - V0*B1: RCU=0: L=0
REM *********** НАЧАЛО ЦИКЛА ПО ЭНЕРГИИ **********
FOR  I=NN  TO  NV  STEP  NH:  EL(I)=EN+EH*I:
ECM(I)=EL(I)*PM/AM1
SK=ECM(I)*B1: SS=SQR(SK): G=3.44476E - 02*Z1*Z2*PM/SS
REM ***** ВЫЧИСЛЕНИЕ КУЛОНОВСКИХ ФУНКЦИЙ *****
X1=H*SS*(N - 4): X2=H*SS*(N): CALL CUL(G,X1,L0,F1,G1,W1)
CALL CUL(G,X2,L0,F2,G2,W2)
```

```
 REM *** ВЫЧИСЛЕНИЕ ФАЗ КОНЕЧНО - РАЗНОСТНЫМ
МЕТОДОМ ***
 CALL FUN(N,H,U1(),L,A2,AK1,SK,R0,RCU)
 D1=U1(N - 4): D2=U1(N): AF= - (F1 - F2*D1/D2)/(G1 - G2*D1/D2)
 FF=ATN(AF): IF FF>0 GOTO 90: FF=FF+PI
90 XN1=(COS(FF)*F2+SIN(FF)*G2)/D2: F1(I)=FF*180/PI
 REM *** ВЫЧИСЛЕНИЕ ФАЗ МЕТОДОМ РУНГЕ - КУТТА ***
 CALL FUNRK(V(),N,H): D1=V(N - 4): D2=V(N)
 AF= - (F1 - F2*D1/D2)/(G1 - G2*D1/D2): F33=ATN(AF)
 IF F33>0 GOTO 91: F33=F33+PI
91 XN2=(COS(F33)*F2+SIN(F33)*G2)/D2: F2(I)=F33*180/PI
 REM ***************** ПЕЧАТЬ ФАЗ *********************
 PRINT USING "+#.####^^^^ "; EL(I); ECM(I); F1(I); F2(I);
ABS(F1(I) - F2(I)); ABS(F1(I) - F2(I))/ABS(F1(I))*100: NEXT I:
STOP
```

**SUB CUL(G,X,L,F0,G0,W)**

```
 REM * ВЫЧИСЛЕНИЕ КУЛОНОВСКИХ ФУНКЦИЙ РАССЕЯ-
НИЯ *
 Q=G: R=X: F0=1: K=1: GK=Q*Q: GR=Q*R: RK=R*R
 B01=(L+1)/R+Q/(L+1): BK=(2*L+3)*((L+1)*(L+2)+GR)
 AK= - R*((L+1)^2+GK)/(L+1)*(L+2): DK=1/BK: DEHK=AK*DK
 S=B01+DEHK
1 K=K+1: AK= - RK*((L+K)^2 - 1)*((L+K)^2+GK)
 BK=(2*L+2*K+1)*((L+K)*(L+K+1)+GR): DK=1/(DK*AK+BK)
 IF DK>0 GOTO 3
2 F0= - F0
3 DEHK=(BK*DK - 1)*DEHK: S=S+DEHK
 IF (ABS(DEHK) - 1E - 06)>0 GOTO 1: FL=S: K=1: RMG=R - Q
 LL=L*(L+1): CK= - GK - LL: DK=Q: GKK=2*RMG: HK=2
 AA1=GKK*GKK+HK*HK: PBK=GKK/AA1: RBK= - HK/AA1
 OMEK=CK*PBK - DK*RBK: EPSK=CK*RBK+DK*PBK
 PB=RMG+OMEK: QB=EPSK
5 K=K+1: CK= - GK - LL+K*(K - 1): DK=Q*(2*K - 1): HK=2*K
 FI=CK*PBK - DK*RBK+GKK: PSI=PBK*DK+RBK*CK+HK
 AA2=FI*FI+PSI*PSI: PBK=FI/AA2: RBK= - PSI/AA2
 VK=GKK*PBK - HK*RBK: WK=GKK*RBK+HK*PBK
 OM=OMEK: EPK=EPSK: OMEK=VK*OM - WK*EPK - OM
 EPSK=VK*EPK+WK*OM - EPK: PB=PB+OMEK: QB=QB+EPSK
 IF (ABS(OMEK)+ABS(EPSK) - 1E - 06)>0 GOTO 5: PL= - QB/R
 QL=PB/R: G0=(FL - PL)*F0/QL: G0P=(PL*(FL - PL)/QL - QL)*F0
 F0P=FL*F0: ALFA=1/(SQR(ABS(F0P*G0 - F0*G0P))):
G0=ALFA*G0
 GP=ALFA*G0P: F0=ALFA*F0: FP=ALFA*F0P: W=1 -
FP*G0+F0*GP
```

```
 END SUB
 SUB FUN(N,H,U(5000),L,AV,AK,SK,R0,RCU)
 REM ***** РЕШЕНИЕ УРАВНЕНИЯ ШРЕДИНГЕРА КОНЕЧНО
- РАЗНОСТНЫМ МЕТОДОМ *****
 U(0)=0:  U(1)=0.001: HK=H*H: FOR K=1 TO N - 1: X=K*H
 Q1=AV*EXP( - R0*X*X)+L*(L+1)/(X*X): IF X>RCU GOTO 11
 Q1=Q1+(3 - (X/RCU)^2)*AK/(2*RCU): GOTO 22
11 Q1=Q1+AK/X
22 Q2= - Q1*HK - 2+SK*HK: U(K+1)= - Q2*U(K) - U(K - 1): NEXT
K
 END SUB
 SUB FUNRK(V(5000),N,H)
 REM ****** РЕШЕНИЕ УРАВНЕНИЯ ШРЕДИНГЕРА МЕТО-
ДОМ РУНГЕ - КУТТА ВО ВСЕЙ ОБЛАСТИ ПЕРЕМЕННЫХ
******
 VA1=0: REM VA1 - Значение функции в нуле
 PA1=1.0E - 05: REM PA1 - Значение функции на первом шаге
 FOR I=0 TO N - 1: X=H*I+1.0E - 05
 CALL RRUN(VB1,PB1,VA1,PA1,H,X)
 VA1=VB1:  PA1=PB1:  V(I+1)=VA1: NEXT: END SUB
 SUB RRUN(VB1,PB1,VA1,PA1,H,X)
 REM ***** РЕШЕНИЕ УРАВНЕНИЯ ШРЕДИНГЕРА МЕТОДОМ
РУНГЕ - КУТТА НА ОДНОМ ШАГЕ *****
 X0=X: Y1=VA1: CALL F(X0,Y1,FK1): FK1=FK1*H: FM1=H*PA1
 X0=X+H/2: Y2=VA1+FM1/2: CALL F(X0,Y2,FK2): FK2=FK2*H
 FM2=H*(PA1+FK1/2): Y3=VA1+FM2/2: CALL F(X0,Y3,FK3)
 FK3=FK3*H: FM3=H*(PA1+FK2/2): X0=X+H: Y4=VA1+FM3
 CALL F(X0,Y4,FK4): FK4=FK4*H: FM4=H*(PA1+FK3)
 PB1=PA1+(FK1+2*FK2+2*FK3+FK4)/6
 VB1=VA1+(FM1+2*FM2+2*FM3+FM4)/6: END SUB
  SUB F(X,Y,F)
 REM * ВЫЧИСЛЕНИЕ ФУНКЦИИ F(X,Y) В МЕТОДЕ РУНГЕ -
КУТТА *
 SHARED SK,A2,R0,AK1,L,RCU
 VC=A2*EXP(-R0*X^2): IF X>RCU GOTO 121
 VK=(3-(X/RCU)^2)*AK1/(2*RCU): GOTO 222
121 VK=AK1/X
222 F=-(SK-VK-VC-L*(L+1)/(X^2))*Y: END SUB
```

Для выполнения контрольного счета мы использовали конечно - разностный метод с классическим нуклон – нуклонным потенциалом Рейда [90]. В его работе приведен вид $^1P_1$ потенциала и расчетные фазы рассеяния.  Наши вычисления фаз с таким потенциалом даны в таблице 1.3 в сравнении с результатами Рейда.

Таблица 1.3 - Сравнение результатов для потенциала Рейда.

| E, МэВ | δ, рад., [90] | δ, рад., (Наш расчет) |
|---|---|---|
| 48 | -0.071 | -0.072 |
| 144 | -0.312 | -0.314 |
| 208 | -0.456 | -0.458 |
| 352 | -0.708 | -0.710 |

Из этих результатов видно совпадение обеих расчетов с точностью порядка нескольких тысячных радиана, что является прекрасным примером работоспособности использованных численных методов и точности работы написанной программы.

Сравним теперь точность с которой можно получить фазы рассеяния двумя рассмотренными методами – Рунге-Кутта и конечно – разностным. Приведем результаты расчетов фаз двумя этими методами (FKR - конечно - разностный и FRK - Рунге - Кутт) для S - фазы рассеяния в $^2H^4He$ системе (параметры потенциала получены нами в работах [13] $V_0 = 76.12$ МэВ, $\alpha = 0.2$ Фм$^{-2}$, $R_c = 0$ Фм., L = 0 и представляют собой альтернативный вариант подобных потенциалов впервые предложенных в работах [6-9]) и степень их совпадения в градусах (ERR - DEG) и процентах (ERR - %) [16]. Здесь EL, ECM - энергия в лабораторной системе и системе центра масс сталкивающихся частиц.

| EL | ECM | FKR | FRK | ERR - DEG | ERR - % |
|---|---|---|---|---|---|
| +1.0000E+00 | +6.6667E-01 | +1.5053E+02 | +1.5052E+02 | +8.4828E-03 | +5.6353E-03 |
| +2.0000E+00 | +1.3333E+00 | +1.2598E+02 | +1.2597E+02 | +1.2241E-02 | +9.7170E-03 |
| +3.0000E+00 | +2.0000E+00 | +1.0892E+02 | +1.0890E+02 | +1.4219E-02 | +1.3055E-02 |
| +4.0000E+00 | +2.6667E+00 | +9.5886E+01 | +9.5871E+01 | +1.5547E-02 | +1.6214E-02 |
| +5.0000E+00 | +3.3333E+00 | +8.5335E+01 | +8.5319E+01 | +1.6536E-02 | +1.9378E-02 |
| +6.0000E+00 | +4.0000E+00 | +7.6457E+01 | +7.6440E+01 | +1.7629E-02 | +2.3057E-02 |
| +7.0000E+00 | +4.6667E+00 | +6.8801E+01 | +6.8782E+01 | +1.8828E-02 | +2.7365E-02 |
| +8.0000E+00 | +5.3333E+00 | +6.2083E+01 | +6.2063E+01 | +1.9867E-02 | +3.2000E-02 |
| +9.0000E+00 | +6.0000E+00 | +5.6102E+01 | +5.6081E+01 | +2.0743E-02 | +3.6974E-02 |
| +1.0000E+01 | +6.6667E+00 | +5.0715E+01 | +5.0693E+01 | +2.1731E-02 | +4.2848E-02 |
| +1.1000E+01 | +7.3333E+00 | +4.5818E+01 | +4.5795E+01 | +2.2996E-02 | +5.0189E-02 |
| +1.2000E+01 | +8.0000E+00 | +4.1334E+01 | +4.1310E+01 | +2.4399E-02 | +5.9028E-02 |
| +1.3000E+01 | +8.6667E+00 | +3.7206E+01 | +3.7180E+01 | +2.5677E-02 | +6.9014E-02 |
| +1.4000E+01 | +9.3333E+00 | +3.3383E+01 | +3.3356E+01 | +2.6730E-02 | +8.0071E-02 |
| +1.5000E+01 | +1.0000E+01 | +2.9825E+01 | +2.9798E+01 | +2.7712E-02 | +9.2913E-02 |
| +1.6000E+01 | +1.0667E+01 | +2.6500E+01 | +2.6471E+01 | +2.8870E-02 | +1.0894E-01 |
| +1.7000E+01 | +1.1333E+01 | +2.3379E+01 | +2.3349E+01 | +3.0323E-02 | +1.2970E-01 |

На рисунке 1.1а непрерывной спадающей линией показана вычисленная конечно - разностным методом $S_1$ - фаза $^2H^4He$ рассеяния.

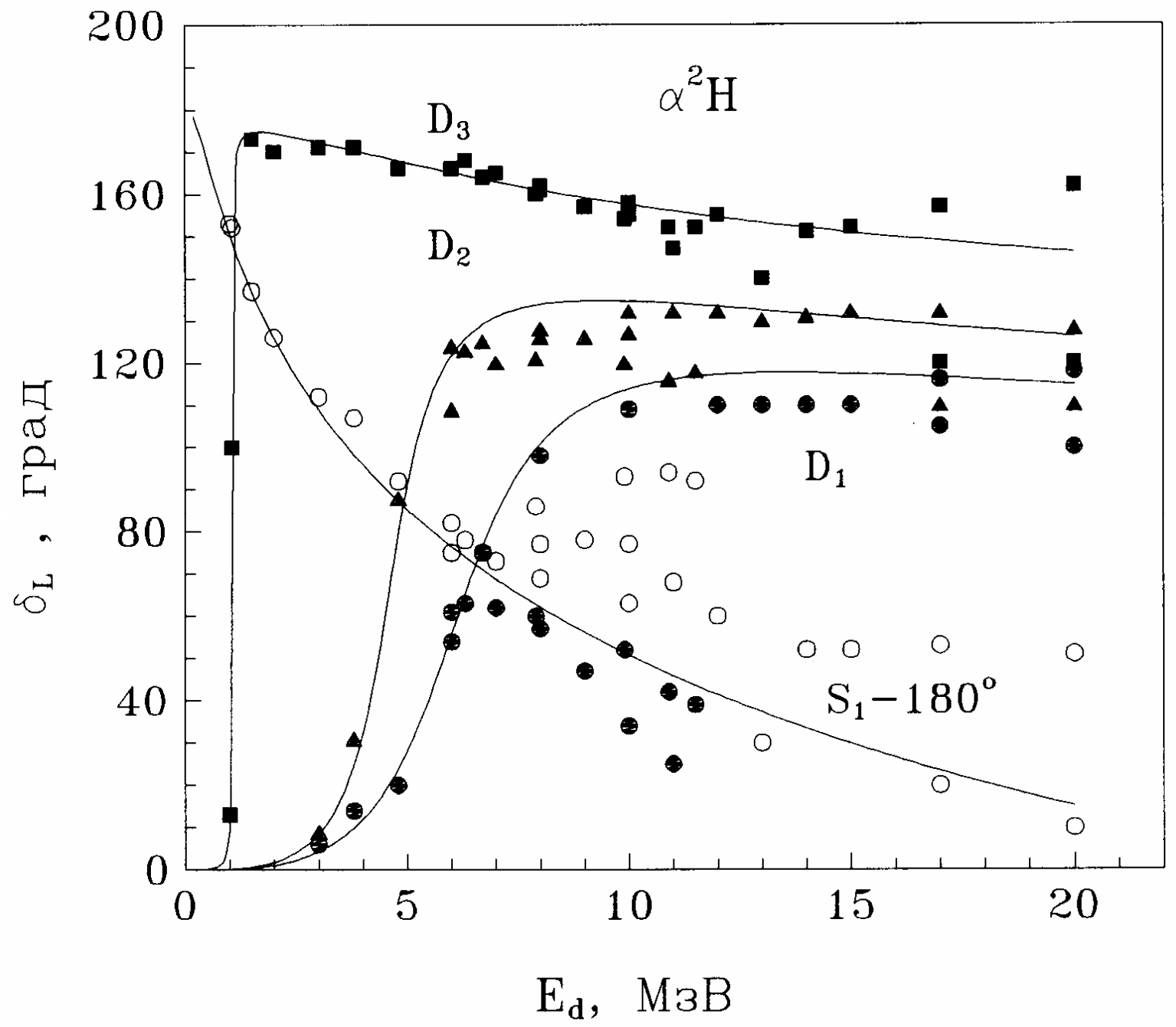

Кривые - расчеты по приведенной программе [16]. Точки, треугольники, кружки и квадраты - экспериментальные данные [91,92,93,94,95,96,97,98].

Рисунок 1.1а - Четные фазы упругого $^4He^2H$ рассеяния.

И в заключение этого параграфа приведем результаты расчетов $S_{1/2}$ - фазы для упругого $^3H^4He$ рассеяния при низких энергиях [16] двумя методами, показанные на рисунке 1.1б непрерывной линией вверху рисунка. Для параметров потенциала использованы следующие значения - $V_0$ = -67.5 МэВ, $\alpha$ = 0.15747 Фм$^{-2}$, $R_c$ = 3.095 Фм, L = 0 [13,16].

| EL | ECM | FKR | FRK | ERR-DEG | ERR-% |
|---|---|---|---|---|---|
| +1.0000E-01 | +4.2857E-02 | +1.8000E+02 | +1.8000E+02 | +3.0506E-07 | +1.6948E-07 |
| +1.1000E+00 | +4.7143E-01 | +1.7204E+02 | +1.7204E+02 | +1.2686E-03 | +7.3739E-04 |
| +2.1000E+00 | +9.0000E-01 | +1.6139E+02 | +1.6138E+02 | +2.8364E-03 | +1.7575E-03 |
| +3.1000E+00 | +1.3286E+00 | +1.5211E+02 | +1.5211E+02 | +4.0288E-03 | +2.6486E-03 |
| +4.1000E+00 | +1.7571E+00 | +1.4419E+02 | +1.4419E+02 | +4.9487E-03 | +3.4320E-03 |
| +5.1000E+00 | +2.1857E+00 | +1.3729E+02 | +1.3729E+02 | +5.6562E-03 | +4.1198E-03 |
| +6.1000E+00 | +2.6143E+00 | +1.3113E+02 | +1.3113E+02 | +6.2638E-03 | +4.7767E-03 |
| +7.1000E+00 | +3.0429E+00 | +1.2555E+02 | +1.2555E+02 | +6.8101E-03 | +5.4240E-03 |
| +8.1000E+00 | +3.4714E+00 | +1.2045E+02 | +1.2044E+02 | +7.2653E-03 | +6.0317E-03 |
| +9.1000E+00 | +3.9000E+00 | +1.1574E+02 | +1.1573E+02 | +7.6262E-03 | +6.5892E-03 |
| +1.0100E+01 | +4.3286E+00 | +1.1134E+02 | +1.1133E+02 | +7.9425E-03 | +7.1336E-03 |
| +1.1100E+01 | +4.7571E+00 | +1.0721E+02 | +1.0720E+02 | +8.2671E-03 | +7.7111E-03 |
| +1.2100E+01 | +5.1857E+00 | +1.0331E+02 | +1.0330E+02 | +8.6088E-03 | +8.3328E-03 |
| +1.3100E+01 | +5.6143E+00 | +9.9620E+01 | +9.9611E+01 | +8.9358E-03 | +8.9699E-03 |
| +1.4100E+01 | +6.0429E+00 | +9.6112E+01 | +9.6103E+01 | +9.2141E-03 | +9.5868E-03 |
| +1.5100E+01 | +6.4714E+00 | +9.2768E+01 | +9.2759E+01 | +9.4404E-03 | +1.0176E-02 |
| +1.6100E+01 | +6.9000E+00 | +8.9572E+01 | +8.9563E+01 | +9.6453E-03 | +1.0768E-02 |
| +1.7100E+01 | +7.3286E+00 | +8.6509E+01 | +8.6499E+01 | +9.8700E-03 | +1.1409E-02 |
| +1.8100E+01 | +7.7571E+00 | +8.3566E+01 | +8.3556E+01 | +1.0139E-02 | +1.2133E-02 |

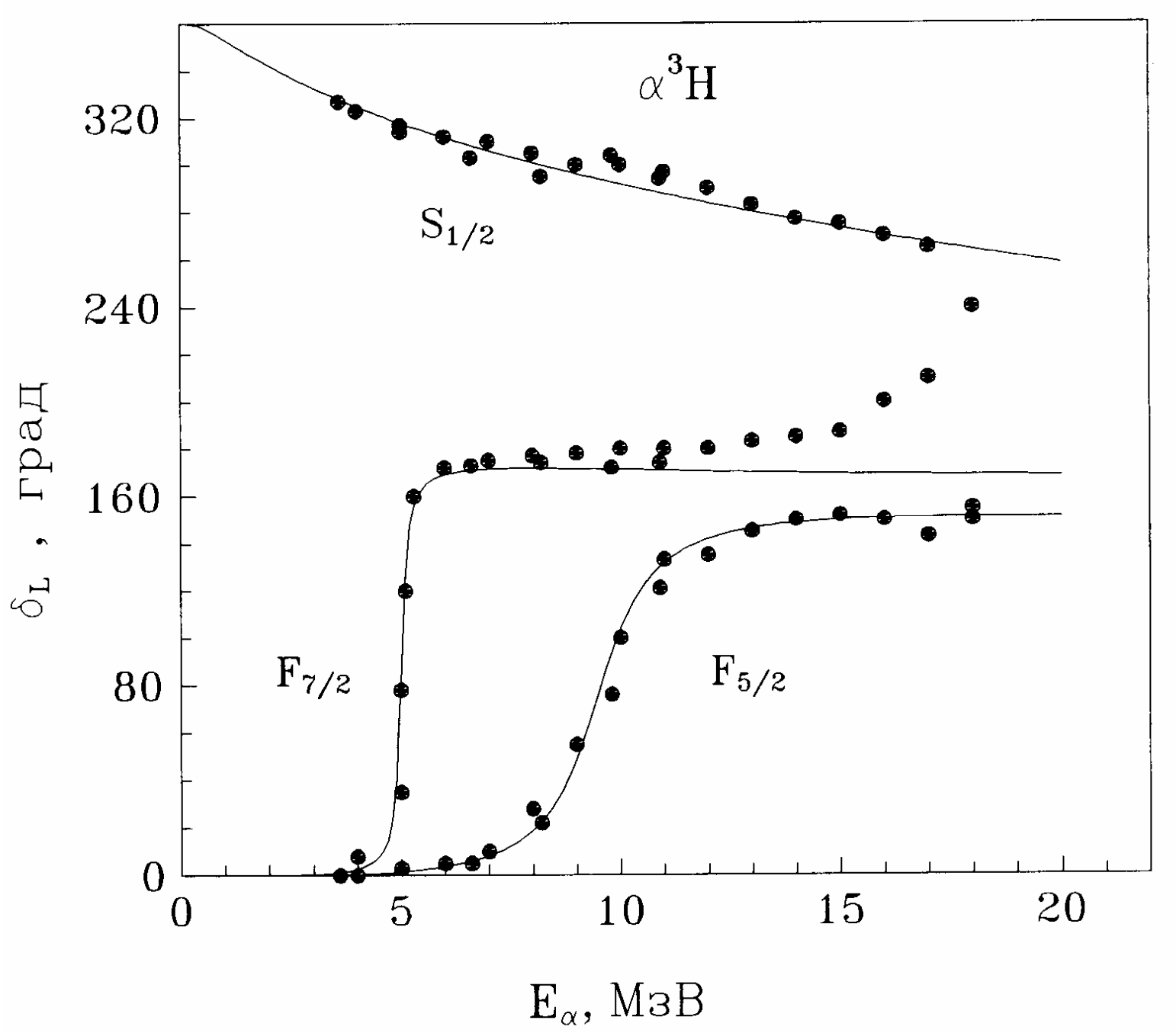

Кривые - расчеты по приведенной выше программе [16]. Точки и треугольники – экспериментальные данные из работ [99,100,101].
Рисунок 1.1б - Фазы упругого $^4He^3H$ рассеяния.

Все эти результаты показывают, что точность фаз ядерного рассеяния, получаемая из этих расчетов обоими методами находится на уровне сотых долей процента в области низких энергий и любой из этих способов может применяться для реальных расчетов ядерных фаз упругого рассеяния в любых кластерных системах с центральными силами.

## 1.5 Программа расчета фаз рассеяния для центральных комплексных потенциалов

Приведем теперь программу для расчета комплексных фаз упругого рассеяния конечно - разностным методом. Здесь приняты следующие обозначения: NN - Нижнее значение цикла по энергии, NV - Верхнее значение цикла по энергии, LN - Нижнее значение орбитального момента, LV - Верхнее значение орбитального момента, LH - Шаг по значениям орбитального момента, AM1 - Масса первой частицы в а.е.м., AM2 - Масса второй частицы в а.е.м., PM - Приведенная масса $\mu$, Z1 - Заряд первой частицы в единицах заряда "e", Z2 - Заряд второй частицы в единицах заряда "e", A1 - Константа $\hbar^2/M_N = 41.4686$, где $M_N$ - масса нуклона в а.е.м, равная 1, AK - Константа при кулоновском потенциале 1.439975 $Z_1Z_2\, 2\mu/\hbar^2$, N - Число шагов при интегрировании уравнения Шредингера, HH - Величина шага при интегрировании уравнения Шредингера, R00 - Расстояние, на котором выполняется сшивка численных функций с асимптотикой, VR1 - Глубина действительного потенциала, RRR - Радиус действительного потенциала, AR - Диффузность действительного потенциала Вудс - Саксоновского типа, VC1 - Глубина мнимой части потенциала, RRC - Радиус мнимой части потенциала, AC - Диффузность мнимой части потенциала Вудс - Саксоновского типа, RCU - Кулоновский радиус, L - Орбитальный момент, E1() - Энергия частиц в лабораторной системе, E() - Энергия частиц в системе центра масс, SK - Квадрат волнового числа $k^2$, SS - Волновое число k, GG - Кулоновский параметр = $3.44476\ 10^{-2} Z_1Z_2\, \mu/k$.

```
REM ** ВЫЧИСЛЕНИЕ КОМПЛЕКСНЫХ ФАЗ РАССЕЯНИЯ **
CLS: DEFDBL A-Z: DEFINT I,J,K,L,N,M: NN=4000: N=100
DIM E(N), DE(N), DEE(N), FAZA(N,15), E1(N), X(NN), Y(NN),
ETA(N,15), SIG(N,15), SEC(N), FAZ(N,15)
DIM V(NN),W(NN),FM(20),FR(20),FR1(20),FM1(20),ET(N)
REM ********************************************
A$="                    КОМПЛЕКСНЫЕ ФАЗЫ"
B$="  E(CM)      FAZR(EXP)    FAZR(TEOR)    FAZC(EXP)
FAZC(TEOR)  ETA(TEOR) "
SAVE=0: G$="C:\BASICA\FAZCOM\FAZALAL1.DAT"
PRINT "-------------------------------------------------------------":
PRINT B$
REM ************* ВИД ПОТЕНЦИАЛОВ ************
PI=4*ATN(1): NN=1: NV=1: LN=0: LV=10: LH=2: AM1=4: AM2=4
Z1=2: Z2=2: A1=41.4686: PM=AM1*AM2/(AM1+AM2):
B1=2*PM/A1
AK=1.439975*Z1*Z2*B1: N=2000: R00=20: HH=R00/N
REM *************** AL - AL 51.1 *******************
E1(1)=51.1
REM ************* ФАЗЫ AL - AL 51.1 ***************
FR(0)=291: FR(2)=245: FR(4)=163: FR(6)=28: FR(8)=4.2: FR(10)=0.5
```

```
 FM(0)=0.51: FM(2)=0.51: FM(4)=0.53: FM(6)=0.855: FM(8)=0.985
 FM(10)=0.998: FOR I=NN TO NV: E(I)=E1(I)*PM/AM1: NEXT I
 FOR I=LN TO LV STEP LH: FR1(I)=FR(I)*PI/180
 REM FM1(I)=FM(I)*PI/180
 NEXT I: REM *********  НАЧАЛЬНЫЕ ПАРАМЕТРЫ
********
 V22=122: A22=0.74: R22=1.81: V33=11: A33=0.74: R33=1.81
 RCU=1.81: VN2=-V22*B1: VN3=-V33*B1
 REM ***** ВЫЧИСЛЕНИЕ ФАЗ РАССЕЯНИЯ *******
 FOR JJ=NN TO NV: SK=E(JJ)*B1: SS=SQR(SK)
 GG=3.44476E-02*Z1*Z2*PM/SS: SIGMRR=0: SIGMAS=0
 FOR L=LN TO LV STEP LH
 CALL FUN (X(), Y(),  R22, VN2, A22, R33, VN3, A33, RCU, L, SK,
AK)
 RR1=HH*SS*(N-5): RR2=HH*SS*N: X1=X(N-5): X2=X(N)
 Y1=Y(N-5): Y2=Y(N)
 REM *********  КУЛОНОВСКИЕ ФУНКЦИИ  ********
 CALL CUL(GG,RR1,L,F1,G1,FP1,GP1)
 CALL CUL(GG,RR2,L,F2,G2,FP2,GP2)
 REM ***************  ПОИСК ФАЗ  ****************
 AA1=X2*F1-X1*F2: BB1=Y1*G2-Y2*G1: DD1=X1*G2-X2*G1
 CC1=Y2*F1-Y1*F2: AA=BB1-AA1: BB=-CC1-DD1: CC=AA1+BB1
 DD=CC1-DD1: DD0=AA^2+BB^2: SS1=(AA*CC+BB*DD)/DD0
 SS2=(AA*DD-BB*CC)/DD0: ETA(JJ,L)=SQR(SS1^2+SS2^2)
 SS22=SS2/ETA(JJ,L): SS11=SS1/ETA(JJ,L)
 SIG(JJ,L)=-LOG((ETA(JJ,L)))/2
 FAZ=SS22/(1+SS11): FAZ(JJ,L)=ATN(FAZ): IF FAZ(JJ,L)>0 GOTO
901
 FAZ(JJ,L)=FAZ(JJ,L)+PI
901 FAZA(JJ,L)=FAZ(JJ,L)*180/PI: IF SIG(JJ,L)>0 GOTO 911
 SIG(JJ,L)=SIG(JJ,L)+PI
911 SIG(JJ,L)=SIG(JJ,L)*180/PI: A=FAZ(JJ,L)
 SIGMAR = SIGMAR + (2*L+1)*(1 - (ETA(JJ,L))^2)
 SIGMAS = SIGMAS + (2*L+1)*(ETA(JJ,L))^2*(SIN(A))^2
PRINT USING " +#.###^^^^ "; L; FR(L); FAZA(JJ,L); FM(L);
SIG(JJ,L)
 ETA(JJ,L): NEXT L: SIGMAR=10*4*PI*SIGMAR/SK
 SIGMAS=10*4*PI*SIGMAS/SK
 PRINT "        SIGR - THEOR = ";SIGMAR;
 PRINT "        SIGS - THEOR = ";SIGMAS: NEXT JJ
 REM ******* ВЫЧИСЛЕНИЕ ДИФ. СЕЧЕНИЙ ********
 FOR J=NN TO NV: SK=E(J)*B1: SS=SQR(SK)
 GG=3.44476E-02*Z1*Z2*PM/SS: SIGMAR=0: SIGMAS=0
 FOR L=LN TO LV STEP LH: A=FR1(L): ET(L)=FM(L)
```

```
 SIGMAR = SIGMAR + (2*L+1)*(1 - (ET(JJ,L))^2)
 SIGMAS = SIGMAS + (2*L+1)*(ET(JJ,L))^2*(SIN(A))^2: NEXT L
 SIGMAR=10*4*PI*SIGMAR/SK: SIGMAS=10*4*PI*SIGMAS/SK
 PRINT "         SIGR - EXP = ";SIGMAR;
 PRINT "         SIGS - EXP = ";SIGMAS
 NEXT J: TMI=10: TMA=90: TH=1
 CALL SEC (FR1(), GG, SS, TMI, TMA, TH, SEC(), ET(), LN, LV,
LH, 1)
 FOR T=TMI TO TMA/3 STEP TH
 PRINT USING " ####.##  "; T; SEC(T); T+20; SEC(T+20); T+40;
SEC(T+40); T+60; SEC(T+60): NEXT
 REM **************   ЗАПИСЬ В ФАЙЛ   ************
 IF SAVE=0 THEN STOP: OPEN "O",1,G$
 PRINT#1, "ALPHA-ALPHA FOR LAB E=";: PRINT#1, E1(NN)
 FOR T=TMI TO TMA STEP TH: PRINT#1, USING " #.###^^^^
";T;SEC(T)
 NEXT: END
 SUB CUL(GG,RR2,L,F2,G2,FP2,GP2)
 Q=GG: R=RR2: F0=1: GK=Q*Q: GR=Q*R: RK=R*R
 B01=(L+1)/R+Q/(L+1): K=1: BK=(2*L+3)*((L+1)*(L+2)+GR)
 AK=-R*((L+1)^2+GK)/(L+1)*(L+2)
 DK=1/BK: DEHK=AK*DK: S=B01+DEHK
112 K=K+1: AK=-RK*((L+K)^2-1)*((L+K)^2+GK)
 BK=(2*L+2*K+1)*((L+K)*(L+K+1)+GR): DK=1/(DK*AK+BK)
 IF DK>0 GOTO 132: F0=-F0
132 DEHK=(BK*DK-1)*DEHK: S=S+DEHK
 IF (ABS(DEHK)-1E-10)>0 GOTO 112: FL=S: K=1: RMG=R-Q
 LL=L*(L+1): CK=-GK-LL: DK=Q: GKK=2*RMG: HK=2
 AA1=GKK*GKK+HK*HK: PBK=GKK/AA1: RBK=-HK/AA1
 OMEK=CK*PBK-DK*RBK: EPSK=CK*RBK+DK*PBK
 PB=RMG+OMEK: QB=EPSK
152 K=K+1: CK=-GK-LL+K*(K-1): DK=Q*(2*K-1): HK=2*K
 FI=CK*PBK-DK*RBK+GKK: PSI=PBK*DK+RBK*CK+HK
 AA2=FI*FI+PSI*PSI: PBK=FI/AA2: RBK=-PSI/AA2
 VK=GKK*PBK-HK*RBK: WK=GKK*RBK+HK*PBK: OM=OMEK
 EPK=EPSK: OMEK=VK*OM-WK*EPK-OM
 EPSK=VK*EPK+WK*OM-EPK
 PB=PB+OMEK: QB=QB+EPSK
 IF (ABS(OMEK)+ABS(EPSK)-1E-10)>0 GOTO 152: PL=-QB/R
 QL=PB/R: G0=(FL-PL)*F0/QL: G0P=(PL*(FL-PL)/QL-QL)*F0
 F0P=FL*F0: ALFA=1/(SQR(ABS(F0P*G0-F0*G0P))): G2=ALFA*G0
 GP2=ALFA*G0P: F2=ALFA*F0: FP2=ALFA*F0P
 W=1-FP2*G2+F2*GP2: END SUB
 SUB FUN(X(5000), Y(5000), R2, V2, A2, R3, V3, A3, RCU, L, SK,
```

```
AK)
 SHARED HH,N: HK=HH*HH: X(0)=0: X(1)=1E-3: Y(0)=0:
Y(1)=1E-3
 FOR K=1 TO N-1: R=K*HH: FR1=V2/(1+EXP((R-R2)/A2))
 FC1=V3/(1+EXP((R-R3)/A3)): FR=SK-FR1-L*(L+1)/R^2
 IF R>RCU GOTO 177: FR=FR-AK/(2*RCU)*(3-(R/RCU)^2): GOTO
188
177 FR=FR-AK/R
188 FC=FC1: F=2-FR*HK: G=FC*HK: X(K+1)=F*X(K)-X(K-1)-
G*Y(K)
 Y(K+1)=F*Y(K)-Y(K-1)+G*X(K): NEXT: END SUB
 SUB SEC (F(100), GG, SS, TMI, TMA, TH, S(100), E(100), LMI,
LMA, LH, NYS)
 SHARED PI: DIM S0(20),P(20)
 RECUL1=0: AIMCUL1=0: CALL CULFAZ(GG,S0())
 FOR TT=TMI TO TMA STEP TH: T=TT*PI/180: XP=COS(T)
 A=2/(1-XP): BB=-GG*A: ALO=GG*LOG(A)+2*S0(0)
 RECUL=BB*COS(ALO): AIMCUL=BB*SIN(ALO)
 IF NYS=0 GOTO 555
 PT=PI-T: X1P=COS(PT): A1=2/(1-X1P): BB1=-GG*A1
 ALO1=GG*LOG(A)+2*S0(0): RECUL1=BB1*COS(ALO1)
 AIMCUL1=BB1*SIN(ALO1)
555 RENUC=0: AIMNUC=0: RENUC1=0: AIMNUC1=0
 FOR L=LMI TO LMA STEP LH: AL=E(L)*COS(2*F(L))-1
 BE=E(L)*SIN(2*F(L)): LL=2*L+1: SL=2*S0(L)
 CALL POLLEG(XP,L,P())
 RENUC=RENUC+LL*(BE*COS(SL)+AL*SIN(SL))*P(L)
 AIMNUC=AIMNUC+LL*(BE*SIN(SL)-AL*COS(SL))*P(L)
 IF NYS=0 GOTO 556: CALL POLLEG(X1P,L,P())
 RENUC1=RENUC+LL*(BE*COS(SL)+AL*SIN(SL))*P(L)
 AIMNUC1=AIMNUC+LL*(BE*SIN(SL)-AL*COS(SL))*P(L)
556 NEXT L: RE=RECUL+RECUL1+RENUC+RENUC1
 AIM=AIMCUL+AIMCUL1+AIMNUC+AIMNUC1
 S(TT)=10*(RE^2+AIM^2)/4/SS^2: NEXT TT: END SUB
 SUB POLLEG(X,L,P(20))
 P(0)=1: P(1)=X: FOR I=2 TO L: P(I)=(2*I-1)*X/I*P(I-1)-(I-1)/I*P(I-2)
 NEXT: END SUB
 SUB CULFAZ(G,F(20))
 C=0.577215665: S=0: N=50: A1=1.202056903/3: A2=1.036927755/5
 FOR I=1 TO N: A=G/I-ATN(G/I)-(G/I)^3/3+(G/I)^5/5
 S=S+A: NEXT: FAZ=-C*G+A1*G^3-A2*G^5+S: F(0)=FAZ
 FOR I=1 TO 20: F(I)=F(I-1)+ATN(G/(I)): NEXT: END SUB
```

Приведем результаты контрольного счета по этой программе.

Рассматривался реальный физический процесс рассеяния в системе ядерных частиц $^4He^4He$ при энергии 51.1 МэВ с комплексным потенциалом. Экспериментальные данные по дифференциальным сечениям приведены в работе [102]. Там же выполнен анализ этих данных по оптической модели. В результате найдены параметры Вудс - Саксоновского потенциала вида

$$V(r) = \frac{V + iW}{\left[\exp\left(\frac{r - R}{a}\right) + 1\right]} + V_c(r) \quad ,$$

где $V_c(r)$ – кулоновский потенциал и

$V = -122 \pm 3$ МэВ , $W = -11 \pm 2$ МэВ ,
$R = 1.81$ Фм , $a = 0.74 \pm 0.03$ Фм , $R_c = R$ .

Такой потенциал приводит к фазам рассеяния и параметрам неупругости, которые приведены в таблице 1.4, вместе с нашими расчетами. В работе [102] приводится также полное экспериментальное сечение реакций $\sigma_r = 770 \pm 100$ мб при данной энергии.

Таблица 1.4 - Результаты фазового анализа.

| L | δ, град., [102] | δ, град. (Наш расчет) | η [102] | η (Наш расчет) |
|---|---|---|---|---|
| 0 | 111±4 | 1.123E+02 | 0.51±0.07 | 5.102E-01 |
| 2 | 65±4 | 6.655E+01 | 0.51±0.07 | 5.177E-01 |
| 4 | 163±4 | 1.649E+02 | 0.53±0.07 | 5.414E-01 |
| 6 | 28±3 | 2.935E+01 | 0.855±0.03 | 8.501E-01 |
| 8 | 4.2±0.6 | 4.422E+00 | 0.985±0.004 | 9.841E-01 |
| 10 | 0.5±0.1 | 7.464E -01 | 0.998±0.001 | 9.972E-01 |

Используя эти параметры потенциала, были выполнены расчеты фаз рассеяния, приведенные в таблице 1.9, и полного сечения реакций $\sigma_r$=766.1 мб. по описанной выше программе.

Видно, что практически все расчетные величины (исключая последнюю действительную фазу для L = 10), в пределах ошибок, совпадают с табличными данными [102]. Если использовать табличные фазы, то для сечения реакций получается величина $\sigma_r$ = 764.7 мб., которая также, как и предыдущая, хорошо согласуется с экспериментальными данными.

Таким образом, в случае только действительных центральных потенциалов, рассмотрены общие и численные методы решения одного уравнения Шредингера, продемонстрирована точность физических расчетов, которую позволяют получить такие методы и их полную применимость для нахождения фаз ядерного рассеяния.

Выполнены контрольные расчеты ядерных фаз рассеяния различных ядерных частиц конечно - разностным методом и методом Рунге - Кутта с реальными потенциалами взаимодействия и проведено их взаимное сравнение, которое демонстрирует совпадение результатов с точностью порядка нескольких сотых долей процента.

Рассмотрен случай, когда центральный потенциал содержит, как действительную, так и мнимую части, тогда уравнение Шредингера переходит в связанную систему уравнений. Изложены общие и численные методы решения такой системы с достаточно простыми начальными и асимптотическими условиями.

Для этого случая также приводятся контрольные расчеты ядерных фаз рассеяния и сравнение их с некоторыми результатами, полученными в других работах. Совпадение фаз наблюдается практически в пределах экспериментальных ошибок.

# 2 МЕТОДЫ РЕШЕНИЯ СИСТЕМЫ УРАВНЕНИЙ ШРЕДИНГЕРА ДЛЯ ПОТЕНЦИАЛОВ С ТЕНЗОРНОЙ КОМПОНЕНТОЙ В НЕПРЕРЫВНОМ СПЕКТРЕ

В этой главе приведены математические методы для решения системы уравнений Шредингера в случае наличия в действительном потенциале взаимодействия тензорной компоненты при положительных собственных значениях, которые являются энергией нуклон - нуклонной (NN) системы, т.е., так же, как и раньше, рассматриваются задачи ядерного рассеяния.

Описаны общие и вычислительные методы решения системы уравнений Шредингера для задачи рассеяния с тензорными силами, когда начальные и асимптотические условия записываются в сравнительно общем виде. В матричной форме дан весь основной математический аппарат для решения такой задачи. Приведен текст компьютерной программы, которая позволяет выполнять расчеты ядерных фаз рассеяния с тензорными потенциалами.

## 2.1 Общие методы решение системы уравнений Шредингера

Использование ядерных потенциалов с тензорной компонентой приводит нас к системе связанных уравнений Шредингера. Будем исходить в дальнейшем из обычных уравнений [59,60], которые учитывают действительные центральную и тензорную часть ядерных потенциалов

$$u''(r) + [\, k^2 - V_c(r) - V_{cul}(r)]u(r) = \sqrt{8}\, V_t(r)w(r) \quad ,$$

$$w''(r) + [\, k^2 - V_c(r) - 6/r^2 - V_{cul}(r) + 2V_t(r)\,]w(r) = \sqrt{8}\, V_t(r)u(r) \quad , \qquad (2.1)$$

где u(r) и w(r) – скалярные искомые волновые функции, а штрихи обозначают производные по r,

r – скалярное расстояние между частицами, измеряемое в Ферми (Фм): 1 Фм = $10^{-15}$ м,

$V_{cul}(r) = 2\mu/\hbar^2\; Z_1 Z_2/r$ - кулоновский потенциал,

$\hbar$ - постоянная Планка: 1.055 $10^{-34}$ Дж. с,

$Z_1$, $Z_2$ – заряды частиц в единицах элементарного заряда: 1 э.з. = 1.60 $10^{-19}$ Кл,

$\mu$ - приведенная масса двух частиц, равная $m_1m_2/(m_1+m_2)$ в атомных единица массы: 1 а.е.м. = 1.66 $10^{-27}$ кг,

Константа $\hbar^2/M_N$ = 41.4686 (или 41.47) МэВ Фм$^2$,

$M_N$ – средняя масса нуклона, равная 1 а.е.м.,

$k^2 = 2\mu E/\hbar^2$ - волновое число относительного движения частиц в $Фм^{-2}$,

E – энергия относительного движения частиц в мегаэлектронвольтах: 1 МэВ = 1.60 $10^{-13}$ Дж,

$V_c = 2\mu/\hbar^2\, V_{cn}(r)$ - центральная часть потенциала,

$V_t = 2\mu/\hbar^2\, V_{tn}(r)$ - тензорная часть потенциала взаимодействия,

$V_{cn}(r)$ и $V_{tn}(r)$ – радиальные части центрального и тензорного потенциалов, которые обычно берутся в виде гауссойды или экспоненты.

Решением системы (2.1) являются четыре волновые функции, получающиеся с двумя типами начальных условий вида [59,60]

$$\begin{aligned} &1)\ u_1(0)=0\ , \quad u'_1(0)=1\ , \quad w_1(0)=0\ , \quad w'_1(0)=0\ , \\ &2)\ u_2(0)=0\ , \quad u'_2(0)=0\ , \quad w_2(0)=0\ , \quad w'_2(0)=1\ , \end{aligned} \tag{2.2}$$

которые для состояний рассеяния ($k^2 > 0$) образуют линейно независимые комбинации, представляемые в виде [59,60]

$$\begin{aligned} u_\alpha &= C_{1\alpha} u_1 + C_{2\alpha} u_2 \longrightarrow Cos(\varepsilon)\,[F_0\, Cos(\delta_\alpha) + G_0\, Sin(\delta_\alpha)]\ , \\ w_\alpha &= C_{1\alpha} w_1 + C_{2\alpha} w_2 \longrightarrow Sin(\varepsilon)\,[F_2\, Cos(\delta_\alpha) + G_2\, Sin(\delta_\alpha)]\ , \\ u_\beta &= C_{1\beta} u_1 + C_{2\beta} u_2 \longrightarrow -Sin(\varepsilon)\,[F_0\, Cos(\delta_\beta) + G_0\, Sin(\delta_\beta)]\ , \\ w_\beta &= C_{1\beta} w_1 + C_{2\beta} w_2 \longrightarrow Cos(\varepsilon)\,[F_2\, Cos(\delta_\beta) + G_2\, Sin(\delta_\beta)]\ , \end{aligned} \tag{2.3}$$

где $F_L$ и $G_L$ - кулоновские функции рассеяния [59,60], $\delta$ - фазы рассеяния, $\varepsilon$ - параметр смешивания состояний с разными орбитальными моментами.

Пары функций $u_\alpha$ и $w_\alpha$ , и $u_\beta$ и $w_\beta$ являются наиболее общими решениями уравнений (2.1) и при больших расстояниях r порядка 10-20 Фм. стремятся к своим асимптотическим значениям, определяемым правой частью выражений (2.3). В случае равенства нулю тензорной части потенциала, параметр смешивания состояний с различным орбитальным моментом $\varepsilon$ становится равен нулю, уравнений (2.1) превращается в два не связанных уравнения и функции $u_\alpha$ и $w_\beta$ переходят в их решения $u_0$ и $w_2$, которые определяют волновые функции рассеяния частиц с относительным орбитальным моментом 0 и 2.

Если вынести в правой части $Cos(\delta)$, выражения (2.3) преобра-

зуются

$$u_{1\alpha} = C'_{1\alpha} u_1 + C'_{2\alpha} u_2 \longrightarrow Cos(\varepsilon)\,[F_0 + G_0\, tg(\delta_\alpha)] \;,$$
$$w_{1\alpha} = C'_{1\alpha} w_1 + C'_{2\alpha} w_2 \longrightarrow Sin(\varepsilon)\,[F_2 + G_2\, tg(\delta_\alpha)] \;,$$

$$u_{2\beta} = C'_{1\beta} u_1 + C'_{2\beta} u_2 \longrightarrow -Sin(\varepsilon)\,[F_0 + G_0\, tg(\delta_\beta)] \;,$$
$$w_{2\beta} = C'_{1\beta} w_1 + C'_{2\beta} w_2 \longrightarrow Cos(\varepsilon)\,[F_2 + G_2 tg(\delta_\beta)] \;,$$

где $C' = C/Cos(\delta)$ и $u_{i\alpha} = u_\alpha/Cos(\delta)$.

В случае нейтрон – протонной (np) задачи рассеяния, когда заряд одной из частиц равен нулю, кулоновские функции $F_L$ и $G_L$ превращаются в обычные сферические функции Бесселя.

Более компактно можно записать приведенные выше выражения для $u_{1\alpha}$ и $w_{1\alpha}$, и $u_{2\beta}$ и $w_{2\beta}$ в матричном виде [103]

$$V = XC' \longrightarrow FU + GU\sigma \;,$$

где

$$V = \begin{pmatrix} u_{1\alpha} & u_{2\beta} \\ w_{1\alpha} & w_{2\beta} \end{pmatrix}, \qquad X = \begin{pmatrix} u_1 & u_2 \\ w_1 & w_2 \end{pmatrix},$$

$$C' = \begin{pmatrix} C'_{1\alpha} & C'_{1\beta} \\ C'_{2\alpha} & C'_{2\beta} \end{pmatrix}, \qquad F = \begin{pmatrix} F_0 & 0 \\ 0 & F_2 \end{pmatrix}, \qquad G = \begin{pmatrix} G_0 & 0 \\ 0 & G_2 \end{pmatrix},$$

$$U = \begin{pmatrix} Cos\varepsilon & -Sin\varepsilon \\ Sin\varepsilon & Cos\varepsilon \end{pmatrix}, \qquad \sigma = \begin{pmatrix} tg\delta_\alpha & 0 \\ 0 & tg\delta_\beta \end{pmatrix} . \qquad (2.4)$$

Аналогичное уравнение можно написать и для производных ВФ:

$$V' = X'C' \longrightarrow F'U + G'U\sigma \;.$$

Исключая из этих уравнений C', после несложных преобразований, для K матрицы рассеяния, определяемой в виде $U\sigma U^{-1}$, окончательно будем иметь

$$K = U\sigma U^{-1} = -\,[\,X(X')^{-1}G' - G]^{-1}\,[X(X')^{-1}F' - F]\;.$$

Тем самым, K матрица рассеяния оказывается выраженной через кулоновские функции, решения исходных уравнений и их производные при некотором r = R.

Как известно, K матрица рассеяния в параметризации Блатта - Биденхарна выражается через фазы рассеяния и параметр смешивания следующим образом [59,60]

$$K = \begin{pmatrix} Cos^2\varepsilon tg\delta_\alpha + Sin^2\varepsilon tg\delta_\beta & Cos\varepsilon Sin\varepsilon(tg\delta_\alpha - tg\delta_\beta) \\ Cos\varepsilon Sin\varepsilon(tg\delta_\alpha - tg\delta_\beta) & Sin^2\varepsilon tg\delta_\alpha + Cos^2\varepsilon tg\delta_\beta \end{pmatrix} . \qquad (2.5)$$

Тогда, приравнивая соответствующие элементы, получим для матричных элементов K матрицы следующие выражения

$$K_{12} = K_{21} = 1/2\ (tg\delta_\alpha - tg\delta_\beta)\ Sin(2\varepsilon) \quad ,$$
$$K_{11} + K_{22} = tg\delta_\alpha + tg\delta_\beta \ ., \qquad (2.6)$$
$$K_{11} - K_{22} = (tg\delta_\alpha - tg\delta_\beta)\ Cos(2\varepsilon) \quad .$$

Откуда имеем

$$tg(2\varepsilon) = 2K_{12}/(K_{11}-K_{22}) \ , \quad tg\delta_\alpha = (A+B)/2 \ , \quad tg\delta_\beta = (A - B)/2 \ ,$$
$$A = K_{11} + K_{22} \ , \qquad B = (K_{11} - K_{22})/Cos(2\varepsilon) \ . \qquad (2.7)$$

Здесь

$$a = f\,(u_1w'_2 - u_2w'_1) \ , \quad b = f\,(u'_1u_2 - u_1u'_2), \qquad (2.8)$$
$$c = f\,(w_1w'_2 - w'_1w_2) \ , \quad d = f\,(u'_1w_2 - u'_2w_1) \ ,$$
$$f = (u'_1w'_2 - u'_2w'_1)^{-1} \ ,$$
$$A = aG'_0 - G_0 \ , \qquad B = bG'_2 \ ,$$
$$E = cG'_0 \ , \qquad D = dG'_2 - G_2 \ ,$$
$$F = PD \ , \qquad G = -\,PB \ ,$$
$$N = -\,PE \ , \qquad M = PA \ ,$$
$$P = -\,(AD - BE)^{-1} \ , \qquad R = aF'_0 - F_0 \ ,$$
$$S = bF'_2 \ , \qquad T = cF'_0 \ ,$$
$$Z = dF'_2 - F_2 \ , \qquad K_{11} = FR + GT \ ,$$
$$K_{12} = FS + GZ \ , \qquad K_{21} = NS + MZ \ ,$$
$$K_{22} = NR + MT \ .$$

Таким образом, получаются сравнительно простые выражения для определения фаз рассеяния $\delta_\alpha$ и $\delta_\beta$ и параметров смешивания $\varepsilon$, которые требуется определить для процессов рассеяния квантовых частиц, через значения численных волновых функций на асимптотике и известные кулоновские функции.

Для численных решений, производные и сами функции можно заменить на значения волновых функций в двух точках $R_1$ и $R_2$ - при этом вид полученных выражений не меняется. Надо только считать, что величины без штриха, например, находятся в первой точке, а со штрихом во второй. Расстояние между точками обычно выбирается равным 5-10 шагов решения численной схемы.

По определенным фазам легко можно найти и коэффициенты C' в любой из рассматриваемых точках

$$C' = X^{-1}(FU + GU\sigma) \ .$$

Расписывая это матричное выражение, имеем

$$C'_{1\alpha} = a(A + F) + b(E + H) \ , \quad C'_{2\alpha} = c(A + F) + d(E + H) \ ,$$
$$C'_{1\beta} = a(B + G) + b(D + K) \ , \quad C'_{2\beta} = c(B + G) + d(D + K) \ , \tag{2.9}$$

где

$$a = fw_2 \ , \quad b = - fu_2 \ , \quad c = - fw_1 \ , \ d = fu_1 \ , \quad f = (u_1w_2 - u_2w_1)^{-1} \ ,$$
$$A = F_0\mathrm{Cos}(\varepsilon) \ , \qquad B = - F_0\mathrm{Sin}(\varepsilon) \ ,$$
$$E = F_2\mathrm{Sin}(\varepsilon) \ , \qquad D = F_2\mathrm{Cos}(\varepsilon) \ ,$$
$$F = G_0\mathrm{Cos}(\varepsilon)\mathrm{tg}(\delta_\alpha) \ , \quad G = - G_0\mathrm{Sin}(\varepsilon)\mathrm{tg}(\delta_\beta) \ , \tag{2.10}$$
$$H = G_2\mathrm{Sin}(\varepsilon)\mathrm{tg}(\delta_\alpha) \ , \quad K = G_2\mathrm{Cos}(\varepsilon)\mathrm{tg}(\delta_\beta) \ .$$

В результате, можно получить полный вид ВФ во всей области при r<R. Радиус сшивки R обычно принимается равным 15 - 20 Фм. Для численного решения исходного уравнения можно использовать метод Рунге - Кутта четвертого порядка [68-72] с автоматическим выбором шага при заданной точности результатов по фазам и параметру смешивания. Относительная точность обычно задается на уровне 0.1%.

Фазы рассеяния для NN задачи обычно принято выражать в параметризации Сака, а не в используемом выше представлении Блатта - Биденхарна. Между этими представлениями фаз существует простая связь [59,60]

$$\theta_J^{J-1} + \theta_J^{J+1} = \delta_\alpha + \delta_\beta \ , \quad \mathrm{tg}(\theta_J^{J-1} - \theta_J^{J+1}) = \mathrm{Cos}(2\varepsilon)\mathrm{tg}(\delta_\alpha - \delta_\beta) \ ,$$

$$\mathrm{Sin}(2\hat{\varepsilon}) = \mathrm{Sin}(2\varepsilon)\mathrm{Sin}(\delta_\alpha - \delta_\beta) \ ,$$

где $\theta_J^{J\pm1}$, $\hat{\varepsilon}$ - фазы и параметр смешивания в параметризации Сака.

## 2.2 Численные методы решения системы уравнений Шредингера

Решения системы (2.1) связанных уравнений Шредингера вида [59]

$$u''(r) + [k^2 - V_c(r) - V_{cul}(r)]u(r) = \sqrt{8}\, V_t(r)w(r) \ ,$$

$$w''(r) + [k^2 - V_c(r) - 6/r^2 - V_{cul}(r) + 2\, V_t(r)\, ]w(r) = \sqrt{8}\, V_t(r)u(r) \qquad (2.11)$$

с начальными условиями

$$u_1(0)=0 \ , \quad u'_1(0)=1 \ , \quad w_1(0)=0 \ , \quad w'_1(0)=0 \ ,$$
$$u_2(0)=0 \ , \quad u'_2(0)=0 \ , \quad w_2(0)=0 \ , \quad w'_2(0)=1 \qquad (2.12)$$

образуют линейно независимые комбинации, представляемые в форме [59]

$$u_\alpha = C_{1\alpha}\, u_1 + C_{2\alpha}\, u_2 \ , \qquad w_\alpha = C_{1\alpha}\, w_1 + C_{2\alpha}\, w_2 \ ,$$
$$u_\beta = C_{1\beta}\, u_1 + C_{2\beta}\, u_2 \ , \qquad w_\beta = C_{1\beta}\, w_1 + C_{2\beta}\, w_2 \ .$$

Эта система может решаться методом Рунге - Кутта [68] с автоматическим выбором шага по заданной точности вычисления фаз.

Для нахождения решения системы двух уравнений второго порядка, перепишем уравнение (2.11) в следующем виде [67]

$$u'' + Ay = Bw \ ,$$
$$w'' + Cw = Bu \ , \qquad (2.13)$$

или

$$u'' = Bw - Au = F(x,u,w) \ ,$$
$$w'' = Bu - Cw = G(x,u,w) \ . \qquad (2.14)$$

Введем две новые переменные

$$y = u' \ , \qquad z = w' \ .$$

Тогда система перепишется в виде четырех уравнений первого порядка

$$u' = y \ , \qquad y' = F(x,u,w) \ ,$$

$$w' = z \ , \qquad z' = G(x,u,w) \ , \tag{2.15}$$

при

$$u_1(0) = 0 \ , \quad y_1(0) = const \ , \quad w_1(0) = 0 \ , \quad z_1(0) = 0 \ ,$$
$$u_2(0) = 0 \ , \quad y_2(0) = 0 \ , \quad w_2(0) = 0 \ , \quad z_2(0) = const \ . \tag{2.16}$$

Решение такой системы, записанной в общем виде

$$u' = f(x,y,z,u,w) \ , \qquad y' = g(x,y,z,u,w) \ ,$$
$$w' = d(x,y,z,u,w) \ , \qquad z' = s(x,y,z,u,w) \ , \tag{2.17}$$

можно представить в форме

$$y_{n+1} = y_n + \Delta y_n \ , \qquad z_{n+1} = z_n + \Delta z_n \ ,$$
$$u_{n+1} = u_n + \Delta u_n \ , \qquad w_{n+1} = w_n + \Delta w_n \ , \tag{2.18}$$

где

$$\Delta y_n = 1/6(k_1 + 2k_2 + 2k_3 + k_4) \ ,$$
$$\Delta z_n = 1/6(m_1 + 2m_2 + 2m_3 + m_4) \ ,$$
$$\Delta u_n = 1/6(v_1 + 2v_2 + 2v_3 + v_4) \ ,$$
$$\Delta w_n = 1/6(b_1 + 2b_2 + 2b_3 + b_4) \ , \tag{2.19}$$

и

$$k_1 = hg(x_n,y_n,z_n,u_n,w_n) \ , \qquad m_1 = hs(x_n,y_n,z_n,u_n,w_n) \ ,$$
$$v_1 = hf(x_n,y_n,z_n,u_n,w_n) \ , \qquad b_1 = hd(x_n,y_n,z_n,u_n,w_n) \ ,$$

$$k_2 = hg(x_n+h/2,\ y_n+k_1/2,\ z_n+m_1/2,\ u_n+v_1/2,\ w_n+b_1/2) \ ,$$
$$m_2 = hs(x_n+h/2,\ y_n+k_1/2,\ z_n+m_1/2,\ u_n+v_1/2,\ w_n+b_1/2) \ ,$$
$$v_2 = hf(x_n+h/2,\ y_n+k_1/2,\ z_n+m_1/2,\ u_n+v_1/2,\ w_n+b_1/2) \ ,$$
$$b_2 = hd(x_n+h/2,\ y_n+k_1/2,\ z_n+m_1/2,\ u_n+v_1/2,\ w_n+b_1/2) \ ,$$
$$(2.20)$$
$$k_3 = hg(x_n+h/2,\ y_n+k_2/2,\ z_n+m_2/2,\ u_n+v_2/2,\ w_n+b_2/2) \ ,$$
$$m_3 = hs(x_n+h/2,\ y_n+k_2/2,\ z_n+m_2/2,\ u_n+v_2/2,\ w_n+b_2/2) \ ,$$
$$v_3 = hf(x_n+h/2,\ y_n+k_2/2,\ z_n+m_2/2,\ u_n+v_2/2,\ w_n+b_2/2) \ ,$$
$$b_3 = hd(x_n+h/2,\ y_n+k_2/2,\ z_n+m_2/2,\ u_n+v_2/2,\ w_n+b_2/2) \ ,$$

$$k_4 = hg(x_n+h,\ y_n+k_3,\ z_n+m_3,\ u_n+v_3,\ w_n+b_3) \ ,$$
$$m_4 = hs(x_n+h,\ y_n+k_3,\ z_n+m_3,\ u_n+v_3,\ w_n+b_3) \ ,$$
$$v_4 = hf(x_n+h,\ y_n+k_3,\ z_n+m_3,\ u_n+v_3,\ w_n+b_3) \ ,$$

$b_4 = hd(x_n+h,\ y_n+k_3,\ z_n+m_3,\ u_n+v_3,\ w_n+b_3)$ .

Поскольку

$f(x,y,z,u,w) = y$ , $g(x,y,z,u,w) = F(x,u,w)$ ,
$d(x,y,z,u,w) = z$ , $s(x,y,z,u,w) = G(x,u,w)$ ,

то формулы (2.20) преобразуются к виду

$$k_1 = hF(x_n,u_n,w_n)\ ,\quad m_1 = hG(x_n,u_n,w_n)\ ,$$
$$v_1 = hy_n\ ,\quad b_1 = hz_n\ , \tag{2.21}$$

$$k_2 = hF(x_n+h/2,\ u_n+v_1/2,\ w_n+b_1/2)\ ,$$
$$m_2 = hG(x_n+h/2,\ u_n+v_1/2,\ w_n+b_1/2)\ ,$$
$$v_2 = h(y_n+k_1/2)\ ,\ b_2 = h(z_n+m_1/2)\ ,$$

$$k_3 = hF(x_n+h/2,\ u_n+v_2/2,\ w_n+b_2/2)\ ,$$
$$m_3 = hG(x_n+h/2,\ u_n+v_2/2,\ w_n+b_2/2)\ ,$$
$$v_3 = h(y_n+k_2/2)\ ,\ b_3 = h(z_n+m_2/2)\ ,$$

$$k_4 = hF(x_n+h,\ u_n+v_3,\ w_n+b_3)\ ,\quad m_4 = hG(x_n+h,\ u_n+v_3,\ w_n+b_3)\ ,$$
$$v_4 = h(y_n+k_3)\ ,\ b_4 = h(z_n+m_3)\ .$$

Тогда

$$\Delta u_n = 1/6h(6y_n + k_1 + k_2 + k_3)\ ,$$
$$\Delta w_n = 1/6h(6z_n + m_1 + m_2 + m_3) \tag{2.22}$$

и в формулах (2.20) вычислять нужно только два коэффициента k и m.

Ниже приведена программа вычисления ВФ и самих фаз рассеяния, использующая описанные выше методы. Начальные значения функций принимались равными

$VA1 = 0 = u_1(0)$
$PA1 = const = y_1(0)$
$WA1 = 0 = w_1(0)$
$QA1 = 0 = z_1(0)$
$VA2 = 0 = u_2(0)$
$PA2 = 0 = y_2(0)$
$WA2 = 0 = w_2(0)$
$QA2 = const = z_2(0)$

**SUB RRUN(VB1, WB1, VB2, WB2, PB1, QB1, PB2, QB2, VA1,**

```
WA1, VA2, WA2, PA1, QA1, PA2, QA2)
REM ***** ПРОГРАММА ВЫЧИСЛЕНИЯ ФУНКЦИЙ РАССЕЯ-
НИЯ МЕТОДОМ РУНГЕ - КУТТА *****
SHARED H,X
X0=X: CALL F(X0,VA1,WA1,FK1): CALL GG(X0,VA1,WA1,FM1)
CALL F(X0,VA2,WA2,SK1): CALL GG(X0,VA2,WA2,SM1)
FK1=FK1*H: FM1=FM1*H: SK1=SK1*H: SM1=SM1*H
X0=X0+H/2: V1=VA1+PA1*H/2: W1=WA1+QA1*H/2
V2=VA2+PA2*H/2: W2=WA2+QA2*H/2: CALL F(X0,V1,W1,FK2)
CALL GG(X0,V1,W1,FM2): CALL F(X0,V2,W2,SK2)
CALL GG(X0,V2,W2,SM2): FK2=FK2*H: FM2=FM2*H:
SK2=SK2*H
SM2=SM2*H: V1=VA1+PA1*H/2+FK1*H/4
 W1=WA1+QA1*H/2+FM1*H/4
V2=VA2+PA2*H/2+SK1*H/4: W2=WA2+QA2*H/2+SM1*H/4
CALL F(X0,V1,W1,FK3): CALL GG(X0,V1,W1,FM3)
CALL F(X0,V2,W2,SK3): CALL GG(X0,V2,W2,SM3)
FK3=FK3*H: FM3=FM3*H: SK3=SK3*H: SM3=SM3*H:
X0=X0+H/2
V1=VA1+PA1*H+FK2*H/2: W1=WA1+QA1*H+FM2*H/2
V2=VA2+PA2*H+SK2*H/2: W2=WA2+QA2*H+SM2*H/2
CALL F(X0,V1,W1,FK4): CALL GG(X0,V1,W1,FM4)
CALL F(X0,V2,W2,SK4): CALL GG(X0,V2,W2,SM4)
FK4=FK4*H: FM4=FM4*H: SK4=SK4*H: SM4=SM4*H
VB1=VA1+PA1*H+(FK1+FK2+FK3)*H/6
PB1=PA1+(FK1+2*FK2+2*FK3+FK4)/6
WB1=WA1+QA1*H+(FM1+FM2+FM3)*H/6
QB1=QA1+(FM1+2*FM2+2*FM3+FM4)/6
VB2=VA2+PA2*H+(SK1+SK2+SK3)*H/6
PB2=PA2+(SK1+2*SK2+2*SK3+SK4)/6
WB2=WA2+QA2*H+(SM1+SM2+SM3)*H/6
QB2=QA2+(SM1+2*SM2+2*SM3+SM4)/6: END SUB
SUB FAZ(DELA, DELB, EPS, C1A, C1B, C2A, C2B, VC1, WC1,
VC2, WC2, VB1, WB1, VB2, WB2, N)
REM *** ПРОГРАММА ВЫЧИСЛЕНИЯ ФАЗ РАССЕЯНИЯ ***
SHARED S,H,GGG
V11=VC1: V12=VB1: W11=WC1: W12=WB1: V21=VC2: V22=VB2
W21=WC2: W22=WB2: X1=H*S*(N - 3): X2=H*S*N: AL1=2
CALL CUL(GGG,X1,0,F01,G01): CALL CUL(GGG,X2,0,F02,G02)
CALL CUL(GGG,X1,2,F21,G21): CALL CUL(GGG,X2,2,F22,G22)
AP=V12*W22 - V22*W12: A=(V11*W22 - V21*W12)/AP
B=(V12*V21 - V22*V11)/AP: C=( - W12*W21+W22*W11)/AP
D=(V12*W21 - V22*W11)/AP: AA=A*G02 - G01: BB=B*G22
EE=C*G02: DD=D*G22 - G21: PP= - 1/(AA*DD - BB*EE)
```

```
FF=PP*DD: GG= - PP*BB: NN= - PP*EE: MM=PP*AA: RR=A*F02 -
F01
SS=B*F22: TT=C*F02: ZZ=D*F22 - F21: QK11=FF*RR+GG*TT
QK12=FF*SS+GG*ZZ: QK21=NN*RR+MM*TT
QK22=NN*SS+MM*ZZ
T2EPS=2*QK12/(QK11 - QK22): EPS=ATN(T2EPS)/2
AAA=QK11+QK22: BBB=(QK11 - QK22)/COS(2*EPS)
DELA=(AAA+BBB)/2: DELB=(AAA - BBB)/2: DELA=ATN(DELA)
DELB=ATN(DELB): A9=F02*COS(EPS)
F9=G02*COS(EPS)*TAN(DELA)
E9=F22*SIN(EPS): H9=G22*SIN(EPS)*TAN(DELA)
B9= - F02*SIN(EPS)
D9=F22*COS(EPS): G9= - G02*SIN(EPS)*TAN(DELB)
AK9=G22*COS(EPS)*TAN(DELB): FF1=1/(V12*W22 - V22*W12)
AA=FF1*W22: BB= - FF1*V22: CC= - FF1*W12: DD=FF1*V12
C1A=(AA*(A9+F9)+BB*(E9+H9))*COS(DELA)
C2A=(CC*(A9+F9)+DD*(E9+H9))*COS(DELA)
C1B=(AA*(B9+G9)+BB*(D9+AK9))*COS(DELB)
C2B=(CC*(B9+G9)+DD*(D9+AK9))*COS(DELB): END SUB
SUB F(X,Y,Z,F)
REM ***** ПОДПРОГРАММА ВЫЧИСЛЕНИЯ ПОТЕНЦИАЛОВ
В ПЕРВОМ УРАВНЕНИИ *****
SHARED SK,VCC,ACC,VTT,ATT,A5,A1,AKK
VC=VCC*EXP( - ACC*X^2): VC=VC+AKK/X
VT=VTT*EXP( - ATT*X^2)
UC=VC/A1: UT=VT/A1: F=UT*A5*Z - (SK - UC)*Y: END SUB
SUB GG(X,Y,Z,G)
REM ***** ПОДПРОГРАММА ВЫЧИСЛЕНИЯ ПОТЕНЦИАЛОВ
ВО ВТОРОМ УРАВНЕНИИ *****
SHARED SK,VCC,ACC,VTT,ATT,A5,A1,AKK
VC=VCC*EXP( - ACC*X^2): VC=VC+AKK/X
VT=VTT*EXP( - ATT*X^2)
UC=VC/A1: UT=VT/A1
G=UT*A5*Y - (SK - 6/X^2 - UC+2*UT+3*ULS)*Z: END SUB
```

Полностью программы расчета ядерных фаз упругого рассеяния при заданной энергии сталкивающихся частиц методом Рунге - Кутта, и некоторые полученные по ней результаты будут приведены в следующих параграфах.

## 2.3 Физические характеристики рассеяния при низких энергиях

Можно рассмотреть и низкоэнергетические характеристики процессов рассеяния в нуклон - нуклонной системе, такие как эф-фективный радиус и длина рассеяния, которые позволяют выполнять тестирование параметров ядерных потенциалов.

### *2.3.1 Центральные потенциалы*

Будем исходить их обычного уравнения Шредингера с L = 0 без кулоновского взаимодействия (т.е. при np рассеянии) для двух энергий и учитывать знак минус в ядерном потенциале [59]

$$u_1''(r) + [\, k_1^2 + V(r)]u_1(r) = 0 \,, \qquad (2.23)$$
$$u_2''(r) + [\, k_2^2 + V(r)]u_2(r) = 0 \,.$$

Если рассматривать триплетное состояние np системы, когда

$$u_g''(r) + [\, -\alpha^2 + V(r)]u_g(r) = 0 \,, \qquad (2.24)$$

то имеется связанное состояние, обозначенное индексом g. Волновые функции уравнений (1.72.) удовлетворяют условиям

$$u_1(0) = u_2(0) = u_g(0) = 0 \,.$$

Рассмотрим их асимптотику $\bar{u}(r)$, потребовав

$$\bar{u}_1(0) = \bar{u}_2(0) = \bar{u}_g(0) = 1$$

и

$$\bar{u}_i(r) = \frac{\mathrm{Sin}(k_i r + \delta_i)}{\mathrm{Sin}\delta_i} \,,$$
$$\bar{u}_g(r) = e^{-\alpha r} \,,$$

где i = 1,2. Эти ВФ удовлетворяют уравнениям (2.23) и (2.24) без ядерного потенциала. Комбинируя уравнения (2.23) можно найти

$$k_2\mathrm{Ctg}\delta_2 - k_1\mathrm{Ctg}\delta_1 = (k_2^2 - k_1^2)\int_0^\infty (\bar{u}_1\bar{u}_2 - u_1u_2)dr \,. \qquad (2.25)$$

Из уравнений (2.23) и (2.24) можно получить

$$k_2\mathrm{Ctg}\delta_2 + \alpha = (k_2^2 + \alpha^2)\int_0^\infty (\bar{u}_g\bar{u}_1 - u_gu_1)dr \,. \qquad (2.26)$$

Разложим величину $kCtg\delta$ в ряд

$$kCtg\delta_0 = - 1/a + 1/2\, r_0 k^2 \qquad (2.27)$$

и определим длину рассеяния в виде

$$\lim_{k^2 \to 0} [kCtg\delta_0] = -1/a \ ,$$

где введен индекс 0 для L = 0, причем $\delta_0$ относится к $k = k_0$. Устремив $k_2$ к нулю в уравнениях (2.25) и (2.26), и опустив индекс 2 получим

$$kCtg\delta_0 = - 1/a + 1/2\, k^2 \rho(0,E)$$

и

$$kCtg\delta_0 = - \alpha + 1/2\, (k^2 + \alpha^2)\rho(-\varepsilon,E) \ , \qquad (2.28)$$

где

$$\rho(0,E) = 2\int_0^\infty (\bar{u}_0 \bar{u} - u_0 u) dr \ ,$$

$$\rho(-\varepsilon,E) = 2\int_0^\infty (\bar{u}_g \bar{u} - u_g u) dr \ .$$

Устремив энергию E к нулю, будем иметь

$$1/a = \alpha \ - 1/2\alpha^2 \rho(-\varepsilon,0) \ ,$$

где

$$\rho(0,-\varepsilon) = 2\int_0^\infty (\bar{u}_g \bar{u}_0 - u_g u_0) dr \ .$$

Поскольку величины $\rho$ слабо зависят от энергии [59] можно аппроксимировать их следующим образом

$$r_0 = \rho(0,0) = 2\int_0^\infty (\bar{u}_0^2 - u_0^2) dr \ ,$$

$$\bar{u}_0 = 1 - r/a_t \ . \qquad (2.29)$$

Тогда получим

$$kCtg\delta_0 = -\alpha + 1/2(k^2+\alpha^2)r_0 \ , \qquad (2.30)$$

т.е.

$$1/a = \alpha - 1/2\,\alpha^2 r_0 \ ,$$

где величина $\alpha$ полностью зависит от энергии связи дейтрона (связанное состояние np системы). В зависимости от NN состояния рассматривают триплетные $a_t$ и синглетные $a_s$ длины рассеяния и эффективные радиусы $r_s$ и $r_t$.

Если учесть кулоновские силы, то можно получить аналогичные выражения [59]. Исходим из уравнения

$$u_l''(r) + [\,k_l^2 + V(r) - 1/(\rho r)]u_l(r) = 0 \ .$$

Его асимптотическое решение при V(r>R) = 0 имеет вид

$$\bar{u}(r) \to \left[\frac{2\pi\eta}{e^{2\pi\eta}-1}\right]^{1/2} \frac{Sin[kr-\eta\ln(2kr)+\delta_0+\sigma_0]}{Sin\delta_0} \ .$$

Тогда получим

$$K = \rho\,[\,-1/a + 1/2\,k^2 r_0] \ , \qquad (2.31)$$

где

$$K = \frac{\pi Ctg\delta_0}{e^{2\pi\eta}-1} + h(\eta) \ , \qquad h(\eta) = -\ln\eta - C + \eta^2\sum_{n=1}^{\infty}\frac{1}{n(n^2+\eta^2)} \ ,$$

$$-\rho/a = \lim_{k^2\to 0} K \ , \qquad \eta = \frac{\mu Z_1 Z_2}{k\hbar^2} \ , \qquad r_0 = 2\int_0^{\infty}(\bar{u}_0^2 - u_0^2)dr$$

и C = 0.57721... - постоянная Эйлера. Для вычисления эффективных радиусов нужно выполнить численное интегрирование разности волновых функций, которые при малых энергиях слабо меняют свою форму. Интеграл от слабо осциллирующих функций может быть вычислен методом Симпсона, который можно записать в виде [68-72]

$$\int_a^b f(x)dx = h/3[f(x_0)+2A+4B+f(x_{2m})] \ ,$$

где

$$A = \sum_{k=2}^{2m-2} f(x_k) \ - \text{суммирование по четным } k \ ,$$

$$B = \sum_{k=1}^{2m-1} f(x_k) \ - \text{суммирование по нечетным } k \ .$$

Весь интервал интегрирования разбивается на n = 2m частей (n должно быть четным числом), а шаг вычисляется следующим образом

$$h = (b - a)/n = (b - a)/2/m \ .$$

Тогда $x_i$ = ih и $f_i = f(x_i)$, где i меняется от 0 до 2m. Приведем теперь текст компьютерной программы на языке Turbo Basic для вычисления интеграла по методу Симпсону.

```
SUB SIMP(V(5000),N,H,S)
REM ***** ВЫЧИСЛЕНИЕ ИНТЕГРАЛА ПО СИМПСОНУ ***
A=0: B=0: FOR I=1 TO N-1 STEP 2: B=B+V(I): NEXT
FOR J=2 TO N-2 STEP 2: A=A+V(J): NEXT
S=H*(V(0)+V(N)+2*A+4*B)/3: END SUB
```

Чтобы использовать эту подпрограмму нужно предварительно вычислить массив значений подынтегральной функции F(x) в N точках (плюс одна точка при i = 0) с шагом H и передать его в подпрограмму оператором вызова

CALL SIMP(F(),N,H,SIM) .

Тогда значение переменной SIM будет равно величине интеграла.

### *2.3.2 Потенциалы с тензорной компонентой*

При учете тензорных потенциалов взаимодействия исходим из уравнений вида [59]

$$u''(r) + [\, k^2 - V_c(r) - V_{cul}(r)]u(r) = \sqrt{8}\, V_t(r)w(r) \ ,$$

$$(2.32)$$

$$w''(r) + [\, k^2 - V_c(r) - 6/r^2 - V_{cul}(r) + 2V_t(r)\,]w(r) = \sqrt{8}\, V_t(r)u(r) \ ,$$

с граничными условиями при $r \to \infty$

$$u_\alpha = \mathrm{Cos}(\varepsilon)\frac{\mathrm{Sin}(kr+\delta_\alpha)}{\mathrm{Sin}\delta_\alpha}$$
$$w_\alpha = \mathrm{Sin}(\varepsilon)\frac{\mathrm{Sin}(kr-\pi+\delta_\alpha)}{\mathrm{Sin}\delta_\alpha} ,$$

где $\delta_\alpha$ и $\varepsilon$ представляют сдвиг фазы и параметр смешивания состояний с разными орбитальными моментами. Если использовать кулоновские функции, то можно написать

$$u_\alpha = \mathrm{Cos}(\varepsilon)[\mathrm{Ctg}\delta_\alpha F_0(kr)+G_0(kr)]$$
$$w_\alpha = \mathrm{Sin}(\varepsilon)[\mathrm{Ctg}\delta_\alpha F_2(kr)+G_2(kr)] , \qquad (2.33)$$

где

$F_0(x) = \mathrm{Sin}(x)$ , $G_0(x) = \mathrm{Cos}(x)$ ,
$F_2(x) = (3/x^2 - 1)\mathrm{Sin}(x) - 3\mathrm{Cos}(x)/x$ ,
$G_2(x) = (3/x^2 - 1)\mathrm{Cos}(x) - 3\mathrm{Sin}(x)/x$ ,

при отсутствии кулоновского взаимодействия. Волновая функция (2.33) $w_\alpha$ расходится при $r = 0$ и мы будем использовать несколько другое представление [59]

$$\bar{u}_\alpha = u_\alpha = \mathrm{Cos}(\varepsilon)[\mathrm{Ctg}\delta_\alpha F_0(kr)+G_0(kr)]$$
$$\bar{w}_\alpha = w_\alpha - \frac{3\mathrm{Sin}\varepsilon}{(kr)^2} = \mathrm{Sin}(\varepsilon)\left[\mathrm{Ctg}\delta_\alpha F_2(kr)+G_2(kr)-\frac{3}{(kr)^2}\right] .$$

Эти функции удовлетворяют следующим измененным уравнениям

$$\left(\frac{d^2}{dr^2}+k^2\right)\bar{u}_\alpha = 0$$

$$\left(\frac{d^2}{dr^2}+k^2-\frac{6}{r^2}\right)\bar{w}_\alpha = -3\mathrm{Sin}\varepsilon/r^2 .$$

Выполняя процедуру, аналогичную описанной выше, получим

$$kCtg\delta_\alpha = - 1/a_t + k^2 r_{0t}/(2Cos\varepsilon) \quad , \qquad (2.34)$$

где

$$r_{0t} = 2\int_0^\infty [\bar{u}_\alpha^2 - u_\alpha^2 - w_\alpha^2]dr \ , \qquad \bar{u}_\alpha = 1 - r/a_t \qquad (2.35)$$

и

$$\lim_{k^2 \to 0} (kCtg\delta_\alpha) = -1/a_t$$

- формулы, которые полностью аналогичны, рассмотренному выше случаю центральных ядерных сил.

## 2.4 Программа расчета ядерных фаз рассеяния для потенциалов с тензорной компонентой

Программа для вычисления ядерных фаз рассеяния с учетом тензорных сил, приведенная ниже, написана на алгоритмическом языке “Basic” и использовалась для расчетов в среде компилятора “Turbo Basic” фирмы “Borland International Inc.” [103,104].

*Описание параметров программы*

ВХОДНЫЕ ПАРАМЕТРЫ - задание начальных условий работы программы для численного решения уравнения Шредингера и физические параметры, определяющие начальные условия задачи рассеяния:

AM1=2 - масса первой частицы,
AM2=4 - масса второй частицы,
Z1=1 - заряд первой частицы,
Z2=2 - заряд второй частицы,
AM=AM1+AM2 - сумма масс,
PM=AM1*AM2/AM - приведенная масса $\mu$,
A1=41.4686/(2*PM) - константа $\hbar^2/2\mu$,
PI=3.14159265 - число $\pi$,
VCC= - 71.979 - глубина центральной части потенциала в МэВ,
ACC=0.2 - параметр ширины центральной части потенциала в $Фм^{-2}$ (Ферми),
VTT= - 27. - глубина тензорной части потенциала в МэВ,
ATT=1.12 - параметр ширины тензорной части потенциала в $Фм^{-2}$,
EN=1 - нижний предел энергии в МэВ в лабораторной системе для расчета фаз,

NM=13 - число шагов по энергии,

EH=1 - величина шага по энергии в МэВ,

N=1000 - начальное число шагов при интегрировании системы уравнений,

H0=0.02 - начальная величина шага при интегрировании системы уравнений в Фм,

EPP=1D - 03 - относительная точность вычисления фаз рассеяния, т.е. 0.1%,

AKK=1.439975*Z1*Z2 - константа для кулоновского потенциала,

B1=2*PM/A1 - константа $\hbar^2/2m_N$, определяемая через массу нуклона и равная 41.4686 МэВ $Фм^2$,

A5=SQR(8) - константа $\sqrt{8}$,

S - волновое число k в системе центра масс, определяемое в виде S=SQR(SK), где SK=E1/A1 и E1 = E*PM/AM1 - или $k^2 = 2\mu E/\hbar^2$,

GGG=3.44477E-02*Z1*Z2*PM/S - кулоновский параметр, необходимый для нахождения кулоновский фаз.

*Описание блоков основной программы*

ИНТЕГРИРОВАНИЕ СИСТЕМЫ - блок программы, который обращается к подпрограмме RRUN, для решения системы начальных уравнений Шредингера по методу Рунге - Кутта при заданной точности результатов по фазам.

РАСЧЕТ ФАЗ РАССЕЯНИЯ - блок программы, вычисляющий физические фазы рассеяния на основе полученных при решении исходных уравнений функций рассеяния. Блок обращается к подпрограмме FAZ для расчета фаз рассеяния. Метод расчета фаз изложен выше в первой главе. В случае NN задачи, когда массы частиц равны, находятся фазы в параметризации Сака [103]. Для $^4He^2H$ рассеяния фазы определяются в параметризации Блатта - Биденхарна. Найденные фазы записываются в файл с заданным именем.

РАСЧЕТ ФУНКЦИЙ РАССЕЯНИЯ - блок расчета волновых функций рассеяния, нормированных на заданную асимптотику. Метод расчета изложен выше в первой главе. Вычисленные функции записываются в заданный числовой файл.

РАСЧЕТ ЭФФЕКТИВНОГО РАДИУСА И ДЛИНЫ РАССЕЯНИЯ - расчет эффективного радиуса и длины рассеяния по стандартным формулам теории ядерного рассеяния для NN потенциалов с тензорной компонентой [103]. Расчет проводится только при малых энергиях рассеяния < 0.01 МэВ

*Описание подпрограмм основной программы*

ПОДПРОГРАММА ВЫЧИСЛЕНИЯ ФУНКЦИЙ МЕТОДОМ РУНГЕ - КУТТА - подпрограмма для вычисления функций исходной системы уравнений методом Рунге - Кутта [103]. Метод был изложен в первой главе.

ПОДПРОГРАММА ВЫЧИСЛЕНИЯ ФАЗ РАССЕЯНИЯ - подпрограмма вычисления фаз рассеяния, изложенным выше методом (см. главу 1).

ПОДПРОГРАММА ВЫЧИСЛЕНИЯ ПОТЕНЦИАЛОВ В ПЕРВОМ УРАВНЕНИИ - подпрограмма для вычисления блока потенциалов в первом уравнении исходной системы.

ПОДПРОГРАММА ВЫЧИСЛЕНИЯ ПОТЕНЦИАЛОВ ВО ВТОРОМ УРАВНЕНИИ - подпрограмма для вычисления блока потенциалов во втором уравнении исходной системы.

ПОДПРОГРАММА ВЫЧИСЛЕНИЯ КУЛОНОВСКИХ ФУНКЦИЙ РАССЕЯНИЯ - подпрограмма вычисления кулоновских функций рассеяния. Метод вычисления основан на разложении функций по цепным дробях [103] и изложен в первой главе.

*Текст компьютерной программы*

Ниже приведена распечатка программы расчета ядерных фаз кластер - кластерного и нуклон - нуклонного рассеяния для потенциалов с тензорной компонентой.

```
REM РАСЧЕТ ФАЗ УПРУГОГО 4He2H РАССЕЯНИЯ ДЛЯ
ПОТЕНЦИАЛОВ С ТЕНЗОРНОЙ КОМПОНЕНТОЙ
DEFDBL A - Z:DEFINT K,L,N,M,I,J
DIM U1(4000), U2(4000), W1(4000), W2(4000), FAZ1(20), FAZ2(20),
E(20), EPSS(20)
GGG$="C:\WAVE.DAT":GG$="C:\FAZ.DAT"
A$=" E DELA DELB EPS"
REM ********** ВХОДНЫЕ ПАРАМЕТРЫ *********
AM1=2: AM2=4: Z1=1: Z2=2: AM=AM1+AM2: PM=AM1*AM2/AM
A1=41.4686/(2*PM): PI=3.14159265: VCC= - 71.979: ACC=0.2:
VTT=27.
ATT= - 1.12: EN=1: NM=10: EH=1: N=1000: H0=0.02: EPP=5D - 02
AKK=1.439975*Z1*Z2:B1=2*PM/A1:HK=H0^2:A5=SQR(8)
REM ********* ИНТЕГРИРОВАНИЕ СИСТЕМЫ *******
FOR I=0 TO NM: E=EN+I*EH: E1=E*PM/AM1: SK=E1/A1:
S=SQR(SK)
GGG=3.44477E - 02*Z1*Z2*PM/S: H=H0: N1=N: DB0=0: DB2=0
 EPSB=0
```

```
5 DA0=DB0: DA2=DB2: EPSA=EPSB: VA1=0: WA1=0: PA1=1E - 05
QA1=0: U1(1)=VA1: W1(1)=WA1: VA2=0: WA2=0: PA2=0
 QA2=1E - 05: U2(1)=VA2: W2(1)=WA2: KKK=1: FOR J=2 TO N1
IF J - 2>0 GOTO 3: X0=1E - 05: GOTO 4
3 X0=0
4 X=H*(J - 2)+X0
CALL RRUN(VB1, WB1, VB2, WB2, PB1, QB1, PB2, QB2, VA1,
WA1, VA2, WA2, PA1, QA1, PA2, QA2)
IF ABS(J - (N1 - 3))<1E - 01 GOTO 6:GOTO 10
6 WC1=WB1: VC1=VB1: VC2=VB2: WC2=WB2
10 VA1=VB1: WA1=WB1: VA2=VB2: WA2=WB2: IF N1>N GOTO
555
U1(J)=VA1: U2(J)=VA2: W1(J)=WA1: W2(J)=WA2
555 IF ABS(H0*KKK - H*J)>0.1*H GOTO 333
U1(KKK)=VA1: U2(KKK)=VA2: W1(KKK)=WA1: W2(KKK)=WA2
KKK=KKK+1
333 PA1=PB1: QA1=QB1: PA2=PB2: QA2=QB2: NEXT J
REM ************ РАСЧЕТ ФАЗ РАССЕЯНИЯ *********
CALL FAZ(DB0, DB2, EPSB, C1A, C1B, C2A, C2B, VC1, WC1,
VC2, WC2, VB1, WB1, VB2, WB2, N1)
H=0.5*H: N1=2*N1: KKK=2
IF ABS(DB0 - DA0)>ABS(EPP*DB0) GOTO 5
IF ABS(DB2 - DA2)>ABS(EPP*DB2) GOTO 5
IF ABS(EPSB - EPSA)>ABS(EPP*EPSB) GOTO 5
777 DELA=DB0:DELB=DB2:EPS=EPSB:FAZA=DELA*180/PI
FAZB=DELB*180/PI: EPSS=EPS*180/PI: IF AM1<>AM2 GOTO
3931
PRINT A$: PRINT USING " +#.####^^^^ ";E,FAZA,FAZB,EPSS
PRINT USING " +#.####^^^^ ";E,DELA,DELB,EPS
SIN2EPS=SIN(2*EPS)*SIN(DELA - DELB)
DELA1=(DELA+DELB+ATN(COS(2*EPS)*TAN(DELA - DELB)))/2
DELB1=DELA+DELB - DELA1
 TAN2EPS=SIN2EPS/SQR(1 - SIN2EPS^2)
EPS11=ATN(TAN2EPS)/2: IF DELA1>0 GOTO 111: DE-
LA1=DELA1+PI
111 FAZ1=DELA1*180/PI: FAZ2=DELB1*180/PI:
EPSS1=EPS11*180/PI
3931 E(I)=E: IF AM1=AM2 GOTO 3932: IF FAZA>0 GOTO 3211
FAZA=FAZA+180: 3211 IF FAZB>0 GOTO 3212: FAZB=FAZB+180
3212 FAZ1(I)=FAZA: FAZ2(I)=FAZB: IF EPSS<0 GOTO 3222
 EPSS=EPSS - 90
3222 EPSS(I)=EPSS: PRINT A$
PRINT USING " +#.####^^^^ ";E(I),DELA1,DELB1,EPS1
PRINT USING " +#.####^^^^ ";E(I),FAZ1(I),FAZ2(I),EPSS(I)
```

```
REM ********* РАСЧЕТ ФУНКЦИЙ РАССЕЯНИЯ ******
3932 FOR JJ=0 TO N: U1(JJ)=(C1A*U1(JJ)+C2A*U2(JJ))
W1(JJ)=(C1A*W1(JJ)+C2A*W2(JJ)): NEXT JJ: OPEN "O",1,GGG$
FOR JJ=0 TO N: X=H0*JJ: PRINT#1, USING " +#.####^^^^ ";X;U1(JJ);W1(JJ)
PRINT USING " +#.####^^^^ ";X;U1(JJ);W1(JJ): NEXT JJ: CLOSE
REM ******* РАСЧЕТ ЭФФЕКТИВНОГО РАДИУСА И ДЛИНЫ РАССЕЯНИЯ ********
IF E>.01 GOTO 155: FOR JJ=0 TO N: X=H0*JJ: FF=SIN(X*S+DELA)
U1(JJ)=(FF^2 - U1(JJ)^2) / SIN(DELA)^2 - W1(JJ)^2 / SIN(DELA)^2
NEXT JJ
A=0:  B=0: FOR II=1 TO N - 1 STEP 2: B=B+U1(II): NEXT II
FOR JJ=2 TO N - 2 STEP 2: A=A+U1(JJ): NEXT JJ
R0=2*H0*(U1(0)+U1(N)+2*A+4*B)/3: PRINT "R= ";
PRINT USING " +#.####^^^^ ";R0: AT1= -SIN(DELA)/COS(DELA)/S
PRINT: PRINT "AT=";: PRINT USING " +#.####^^^^";AT1
155 NEXT I: OPEN "O",1,GG$: FOR JJ=0 TO NM
PRINT#1, USING " +#.####^^^^ "; E(JJ); FAZ1(JJ); FAZ2(JJ); EPSS(JJ)
NEXT JJ:  STOP
SUB RRUN(VB1, WB1, VB2, WB2, PB1, QB1, PB2, QB2, VA1, WA1, VA2, WA2, PA1, QA1, PA2, QA2)
REM ПОДПРОГРАММА ВЫЧИСЛЕНИЯ ФУНКЦИЙ МЕТОДОМ РУНГЕ - КУТТА
SHARED H,X
X0=X: CALL F(X0,VA1,WA1,FK1): CALL F(X0,VA2,WA2,SK1)
CALL GG(X0,VA1,WA1,FM1): CALL GG(X0,VA2,WA2,SM1)
FK1=FK1*H: SK1=SK1*H: FM1=FM1*H: SM1=SM1*H: X0=X0+H/2
V1=VA1+PA1*H/2: W1=WA1+QA1*H/2: V2=VA2+PA2*H/2
W2=WA2+QA2*H/2: CALL F(X0,V1,W1,FK2): CALL F(X0,V2,W2,SK2)
CALL GG(X0,V1,W1,FM2): CALL GG(X0,V2,W2,SM2): FK2=FK2*H
SK2=SK2*H: FM2=FM2*H: SM2=SM2*H: V1=VA1+PA1*H/2+FK1*H/4
W1=WA1+QA1*H/2+FM1*H/4: V2=VA2+PA2*H/2+SK1*H/4
W2=WA2+QA2*H/2+SM1*H/4: CALL F(X0,V1,W1,FK3)
CALL F(X0,V2,W2,SK3): CALL GG(X0,V1,W1,FM3)
CALL GG(X0,V2,W2,SM3): FK3=FK3*H: SK3=SK3*H: FM3=FM3*H
SM3=SM3*H: X0=X0+H/2: V1=VA1+PA1*H+FK2*H/2
```

```
W1=WA1+QA1*H+FM2*H/2: V2=VA2+PA2*H+SK2*H/2
W2=WA2+QA2*H+SM2*H/2: CALL F(X0,V1,W1,FK4)
CALL F(X0,V2,W2,SK4): CALL GG(X0,V1,W1,FM4)
CALL GG(X0,V2,W2,SM4): FK4=FK4*H: SK4=SK4*H: FM4=FM4*H
SM4=SM4*H: VB1=VA1+PA1*H+(FK1+FK2+FK3)*H/6
VB2=VA2+PA2*H+(SK1+SK2+SK3)*H/6
PB1=PA1+(FK1+2*FK2+2*FK3+FK4)/6
PB2=PA2+(SK1+2*SK2+2*SK3+SK4)/6
WB1=WA1+QA1*H+(FM1+FM2+FM3)*H/6
WB2=WA2+QA2*H+(SM1+SM2+SM3)*H/6
QB1=QA1+(FM1+2*FM2+2*FM3+FM4)/6
QB2=QA2+(SM1+2*SM2+2*SM3+SM4)/6: END SUB
SUB FAZ(DELA, DELB, EPS, C1A, C1B, C2A, C2B, VC1, WC1, VC2, WC2, VB1, WB1, VB2, WB2, N)
REM ПОДПРОГРАММА ВЫЧИСЛЕНИЯ ФАЗ  РАССЕЯНИЯ
SHARED S,H,GGG
V11=VC1: V12=VB1: W11=WC1: W12=WB1: V21=VC2: V22=VB2
W21=WC2: W22=WB2: X1=H*S*(N - 3): X2=H*S*N: AL1=2
CALL CUL(GGG,X1,0,F01,G01): CALL CUL(GGG,X2,0,F02,G02)
CALL CUL(GGG,X1,2,F21,G21): CALL CUL(GGG,X2,2,F22,G22)
AP=V12*W22 - V22*W12: A=(V11*W22 - V21*W12)/AP
B=(V12*V21 - V22*V11)/AP: C=( - W12*W21+W22*W11)/AP
D=(V12*W21 - V22*W11)/AP: AA=A*G02 - G01: BB=B*G22
EE=C*G02: DD=D*G22 - G21: PP= - 1/(AA*DD - BB*EE)
FF=PP*DD: GG= - PP*BB: NN= - PP*EE: MM=PP*AA: RR=A*F02 - F01
SS=B*F22: TT=C*F02: ZZ=D*F22 - F21: QK11=FF*RR+GG*TT
QK12=FF*SS+GG*ZZ: QK21=NN*RR+MM*TT: QK22=NN*SS+MM*ZZ
T2EPS=2*QK12/(QK11 - QK22): EPS=ATN(T2EPS)/2: AAA=QK11+QK22
BBB=(QK11 - QK22)/COS(2*EPS): DELA=(AAA+BBB)/2
DELB=(AAA - BBB)/2: DELA=ATN(DELA): DELB=ATN(DELB)
A9=F02*COS(EPS): F9=G02*COS(EPS)*TAN(DELA)
 E9=F22*SIN(EPS):H9=G22*SIN(EPS)*TAN(DELA)
 B9= - F02*SIN(EPS): D9=F22*COS(EPS)
G9= - G02*SIN(EPS)*TAN(DELB): AK9=G22*COS(EPS)*TAN(DELB)
FF1=1/(V12*W22 - V22*W12): AA=FF1*W22: BB= - FF1*V22
CC= - FF1*W12: DD=FF1*V12
 C1A=(AA*(A9+F9)+BB*(E9+H9))*COS(DELA)
C2A=(CC*(A9+F9)+DD*(E9+H9))*COS(DELA)
C1B=(AA*(B9+G9)+BB*(D9+AK9))*COS(DELB)
```

```
C2B=(CC*(B9+G9)+DD*(D9+AK9))*COS(DELB): END SUB
SUB F(X,Y,Z,F)
REM ПОДПРОГРАММА ВЫЧИСЛЕНИЯ ПОТЕНЦИАЛОВ В
ПЕРВОМ УРАВНЕНИИ
SHARED SK,VCC,ACC,VTT,ATT,A5,A1,AKK
VC=VCC*EXP( - ACC*X^2): VC=VC+AKK/X
VT=VTT*EXP( - ATT*X^2)
UC=VC/A1: UT=VT/A1: F=UT*A5*Z - (SK - UC)*Y: END SUB
SUB GG(X,Y,Z,GG)
REM ПОДПРОГРАММА ВЫЧИСЛЕНИЯ ПОТЕНЦИАЛОВ ВО
ВТОРОМ УРАВНЕНИИ
SHARED SK,VCC,ACC,VTT,ATT,A5,A1,AKK
VC=VCC*EXP( - ACC*X^2): VC=VC+AKK/X
VT=VTT*EXP( - ATT*X^2):UC=VC/A1: UT=VT/A1
GG=UT*A5*Y - (SK - 6/X^2 - UC+2*UT+3*ULS)*Z:END SUB
SUB CUL(G,X1,L,F1,G1)
REM ПОДПРОГРАММА ВЫЧИСЛЕНИЯ КУЛОНОВСКИХ
ФУНКЦИЙ РАССЕЯНИЯ
Q=G: R=X1: F0=1: GK=Q*Q: GR=Q*R: RK=R*R:
B01=(L+1)/R+Q/(L+1)
K=1: BK=(2*L+3)*((L+1)*(L+2)+GR)
 AK= - R*((L+1)^2+GK)/(L+1)*(L+2)
DK=1/BK: DEHK=AK*DK: S=B01+DEHK
1 K=K+1: AK= - RK*((L+K)^2 - 1)*((L+K)^2+GK)
BK=(2*L+2*K+1)*((L+K)*(L+K+1)+GR): DK=1/(DK*AK+BK)
IF DK>0 GOTO 31
2 F0= - F0
31 DEHK=(BK*DK-1)*DEHK
S=S+DEHK: IF (ABS(DEHK) - 1E - 6)>0 GOTO 1
FL=S: K=1: RMG=R - Q: LL=L*(L+1): CK= - GK - LL: DK=Q
GKK=2*RMG:HK=2: AA1=GKK*GKK+HK*HK: PBK=GKK/AA1
 RBK= - HK/AA1:OMEK=CK*PBK - DK*RBK
 EPSK=CK*RBK+DK*PBK: PB=RMG+OMEK: QB=EPSK
51 K=K+1: CK= - GK - LL+K*(K - 1): DK=Q*(2*K - 1): HK=2*K
FI=CK*PBK - DK*RBK+GKK: PSI=PBK*DK+RBK*CK+HK
AA2=FI*FI+PSI*PSI: PBK=FI/AA2: RBK= - PSI/AA2
VK=GKK*PBK - HK*RBK:WK=GKK*RBK+HK*PBK: OM=OMEK
 EPK=EPSK:OMEK=VK*OM - WK*EPK - OM
 EPSK=VK*EPK+WK*OM - EPK:PB=PB+OMEK: QB=QB+EPSK
IF (ABS(OMEK)+ABS(EPSK) - 1E - 06)>0 GOTO 51: PL= - QB/R
QL=PB/R: G0=(FL - PL)*F0/QL: G0P=(PL*(FL - PL)/QL - QL)*F0
F0P=FL*F0: ALFA=1/(SQR(ABS(F0P*G0 - F0*G0P))):
G1=ALFA*G0
GP1=ALFA*G0P: F1=ALFA*F0: FP1=ALFA*F0P
```

W=1 - FP1*G1+F1*GP1:END SUB

Ниже приведены результаты контрольного счета по этой программе для случая классического потенциала Рейда с мягким кором [90]. Вычислительная точность в программе задавалась на уровне 0.5%, начальное число шагов 1000, а величина начального шага 0.02. Для определения фаз NN рассеяния, приведенных в работе Рейда, сшивка численной волновой функции с ее ассимптотикой выполнялась на расстояниях 20 фм. При расчетах, для получения заданной точности результатов, конечное число шагов доходило до 256 тысяч. Здесь E – энергия частиц, $\delta_{\alpha,\beta}$ - фазы рассеяния, $\varepsilon$ - параметр смешивания.

Таблица 2.1 - Сравнение результатов расчета ядерных фаз рассеяния.

| E, MeV | $\delta_\alpha$, rad [90] | $\delta_\alpha$, rad (Наш расчет) | $\delta_\beta$, rad [90] | $\delta_\beta$, rad (Наш расчет) | Sin(2ε), [90] | Sin(2ε), (Наш расчет) |
|---|---|---|---|---|---|---|
| 24 | 1.426 | 1.426 | -0.050 | -0.050 | 0.064 | 0.0638 |
| 48 | 1.105 | 1.105 | -0.115 | -0.116 | 0.081 | 0.0812 |
| 96 | 0.749 | 0.748 | -0.215 | -0.216 | 0.114 | 0.1140 |
| 144 | 0.521 | 0.520 | -0.281 | -0.282 | 0.152 | 0.1518 |
| 208 | 0.300 | 0.299 | -0.340 | -0.341 | 0.203 | 0.2030 |
| 304 | 0.057 | 0.056 | -0.403 | -0.404 | 0.269 | 0.2693 |

Как видно из этой таблицы отличие наших расчетов и результатов, приведенных в работе Рейда составляет величину порядка 0.001 радиана, что демонстрирует полную работоспособность, как описанных выше математических методов, правильность выбора численных способов решения уравнения Шредингера, так и работоспособность самой компьютерной программы.

Таким образом, разработанная нами программа для решения уравнения Шредингера, как для центральных, так и для тензорных потенциалов, позволяет получить хорошее совпадение всех результатов с полученными ранее, в других работах и, по-видимому, другими методами.

# 3. МЕТОДЫ РЕШЕНИЯ УРАВНЕНИЯ ШРЕДИНГЕРА С ЦЕНТРАЛЬНЫМИ ПОТЕНЦИАЛАМИ В ДИСКРЕТНОМ СПЕКТРЕ

Рассмотрим теперь случай, когда две ядерные частицы находятся в связанном состоянии, образуя атомное ядро, а его характеристики определяются некоторым центральными потенциалом взаимодействия этих частиц и их внутренними свойствами.

В этой главе описаны математические вариационные и численные методы решения задачи на связанные состояния для центральных действительных потенциалов, когда уравнение Шредингера имеет дискретный спектр своих решений. Каждое собственное значение такой задачи определяет энергию связи ядерных частиц, которая принимает только отрицательные значения.

## 3.1 Общие методы решения уравнения Шредингера

При нахождении волновых функций основных связанных и резонансных состояний в двухчастичной системе можно использовать вариационный [23,105,106] и конечно - разностный [13-19,42,1] методы решения радиального уравнения Шредингера.

Запишем еще раз уравнение Шредингера (1.1) с центральными ядерными силами для волновой функции системы двух частиц

$$\chi''(r) + [\, k^2 - V_c(r) - V_{cul}(r) - L(L+1)/r^2 \,]\chi(r) = 0 \quad . \qquad (3.1)$$

Решения этого уравнения для связанных состояний, т.е. при $k^2 < 0$, на бесконечности и в нуле подчиняются условиям

$$\chi_L(0) = \chi_L(\infty) = 0 \quad .$$

Однако уравнение (3.1), на расстояниях больших, чем радиус действия ядерных сил $R_0$, т.е. когда $V_c(r>R_0) = 0$, имеет аналитическое решение, называемое его асимптотикой. Поэтому условие на бесконечности можно заменить на требование неразрывности логарифмической производной на границе области взаимодействия, т.е. при $r = R_0$ [13-19,57,58]

$$\frac{\chi_L'(R_0)}{\chi_L(R_0)} = \frac{W_{\eta L}'(2kR_0)}{W_{\eta L}(2kR_0)} = f(\eta, L, Z) \quad , \qquad (3.2)$$

где $\chi(r)$ – скалярная волновая функция, а штрихи обозначают производные по r,

r – скалярное расстояние между частицами, измеряемое в Ферми (Фм): 1 Фм = $10^{-15}$ м,

$V_{cul}(r) = 2\mu/\hbar^2\ Z_1Z_2/r$ - кулоновский потенциал,

$\hbar$ - постоянная Планка: 1.055 $10^{-34}$ Дж. с,

$Z_1$, $Z_2$ – заряды частиц в единицах элементарного заряда: 1 э.з. = 1.60 $10^{-19}$ Кл,

$\mu$ - приведенная масса двух частиц, равная $m_1m_2/(m_1+m_2)$ в атомных единица массы: 1 а.е.м. = 1.66 $10^{-27}$ кг,

Константа $\hbar^2/M_N$ = 41.4686 (или 41.47) МэВ $Фм^2$,

$M_N$ – средняя масса нуклона, равная 1 а.е.м.,

$k^2 = 2\mu E/\hbar^2$ - волновое число относительного движения частиц в $Фм^{-2}$,

E – энергия относительного движения частиц в мегаэлектронвольтах: 1 МэВ = 1.60 $10^{-13}$ Дж,

$V_c = 2\mu/\hbar^2\, V_{cn}(r)$ - центральная часть потенциала,

$V_{cn}(r)$ – радиальная часть центрального потенциала, которая может быть представлена в виде гауссойды или экспоненты,

$W_{\eta L}(Z)$ - функция Уиттекера для связанных состояний, которая является решением уравнения (3.1) при $k^2<0$ без ядерного потенциала $V_c$. безразмерная переменная $Z = 2kR$,

$\eta = \dfrac{\mu Z_1 Z_2}{k\hbar^2}$ - кулоновский параметр,

L - орбитальный момент.

В том случае, когда в ядерном потенциале не учитывается кулоновское взаимодействие, асимптотика ВФ может быть представлена в наиболее простом виде

$$\chi(r>R_0) = e^{-kr}\ , \qquad \chi'(r>R_0) = -\,ke^{-kr}\ ,$$

где $k = \sqrt{|k^2|}$ и логарифмическая производная (3.2) будет просто равна -k.

Волновая функция связанного состояния любых частиц должна удовлетворять и условию

$$\int_0^\infty \chi^2(r)dr = 1\ ,$$

которое определяет ее нормировку.

## 3.2 Физические характеристики связанных состояний

К основным ядерным характеристикам связанных состояний можно отнести асимптотическую константу, среднеквадратичный радиус, кулоновские формфакторы, квадрупольный, октупольный и магнитные моменты, вероятности электромагнитных переходов между разными уровнями ядра, энергию связи в кластерном канале и т.д.

Асимптотическое поведение ВФ характеризуется асимптотической константой С, которую можно получить из сшивки волновой функции связанных состояний на расстояниях порядка 10 - 20 Фм с ее асимптотикой, определяемой функцией Уиттекера [107, 108, 109, 110, 111, 112]

$$R_{LJ} = \chi(r)/r = \frac{\sqrt{2k_0}}{r} C\, W_{\eta L}(2k_0 r) \quad , \tag{3.3}$$

где $\eta$ - кулоновский параметр, L - орбитальный момент и $k_0$ - волновое число, обусловленное энергией связи двухчастичной системы.

Для квадрупольного, магнитного, октупольного моментов и приведенной вероятности электрических и магнитных радиационных переходов можно написать выражения, зависящие от матричных элементов электрических операторов (более подробно об электрических операторах смотрите пятую часть) [22,42,113,114]

$$Q = \left(\frac{16\pi}{5}\right)^{1/2} \left\langle J_0 J_0 \left| Q_{20}(L) \right| J_0 J_0 \right\rangle \ ,$$

$$\mu = \left(\frac{4\pi}{3}\right)^{1/2} \left\langle J_0 J_0 \left| W_{10}(L)+W_{10}(S) \right| J_0 J_0 \right\rangle \ ,$$

$$\Omega = \left(\frac{4\pi}{3}\right)^{1/2} \left\langle J_0 J_0 \left| W_{30}(L)+W_{30}(S) \right| J_0 J_0 \right\rangle \ ,$$

$$B(M1) = \frac{1}{2J_i+1} \left| \left\langle J_f L_f \left\| W_1(L)+W_1(S) \right\| J_i L_i \right\rangle \right|^2 \ , \tag{3.4}$$

$$B(E2) = \frac{1}{2J_i+1} \left| \left\langle J_f L_f \left\| Q_2(L) \right\| J_i L_i \right\rangle \right|^2 \ .$$

Здесь i - начальное и f - конечное состояние системы, а значок 0 обозначает основное связанное состояние. Отсюда, например, в двухкластерной модели для основного состояния ядра $^7$Li с $J_0$=3/2$^-$ получаем [13-19,114]

$$Q = -\frac{2}{5} Y I_2 = -\frac{2}{5}\frac{34}{49} I_2 \ , \qquad Y = (Z_1 M_2^2 + Z_2 M_1^2)/M^2 \ ,$$

$$\frac{\mu}{\mu_0} = X + \mu_1 = \frac{17}{42} + \mu_1 \ , \qquad X = \frac{1}{M}\left(\frac{Z_1 M_2}{M_1} + \frac{Z_2 M_1}{M_2}\right) , \tag{3.5}$$

$$\frac{\Omega}{\mu} = \frac{3}{5}\frac{M_2^2}{M^2} I_2 = \frac{48}{245} I_2 \ ,$$

$$B(M1) = \frac{1}{4\pi}(2\mu_1 - X)^2 I_0^2 = \frac{1}{4\pi}(3\mu_1 - \mu)^2 I_0^2 \ ,$$

$$B(E2) = \frac{1}{4\pi} Y^2 I_2^2 \ .$$

Здесь первым кластером считается тритий или $^3$He, обладающий магнитным моментом и $\mu_0 = \frac{e\hbar}{2m_0 c}$ - ядерный магнетон.

Магнитный радиус ядра $^7$Li в кластерной $^3$H$^4$He модели определяется через зарядовые и магнитные радиусы фрагментов в следующем виде [115,116]

$$\mu R_m^2 = \frac{4}{21}\langle r_{te}^2 \rangle + \frac{3}{14}\langle r_{te}^2 \rangle + \mu_t \langle r_{tm}^2 \rangle + \left(\frac{209}{3430} + \frac{432}{1225}\mu_t\right) I_2 \ , \tag{3.6}$$

где $\mu$ - магнитный момент ядра, $\langle r_i \rangle$ - магнитные (m) и зарядовые (e) радиусы кластеров (в данном случае трития), а интегралы $I_k$ в (3.6) и (3.5) записываются в виде

$$I_k = \langle J_f L_f | R^k | J_i L_i \rangle \ .$$

Импульсное распределение кластеров в ядре, определяемое, как Фурье - образ волновой функции относительного движения фрагментов и нормированное на единицу при переданном импульсе q = 0 может быть записано [13-19]

$$P^2 = \sum_L P_L^2(q) \Big/ \sum_L P_L^2(0) \quad , \qquad P_L(q) = \int u_L j_L(qr) r dr \quad ,$$

где $j_L(x)$ - сферическая функция Бесселя, $u_L$ - радиальная ВФ СС ядра. Для расчетов продольных кулоновских формфакторов можно использовать, например, известное определение [3,117,118]

$$F(q) = 1/Z < \Psi_f \mid \Sigma\, (1/2+t_{zk}) \exp(iqr_k) \mid \Psi_i > \ .$$

Аналогично запишем формфакторы кластеров

$$F_{1,2}(q) = 1/Z_{1,2} < \Phi_{1,2} \mid \Sigma\, (1/2+t_{zn}) \exp(iq\rho_n) \mid \Phi_{1,2} > \ ,$$

где $t_{zk}$ - проекция изоспина k - й частицы. Используя векторные соотношения двухкластерной модели, для формфактора ядра находим [119]

$$F(q)=Z_1/Z\ F_1(q)<\Psi_f|\exp\ (iq_1R)|\Psi_i> + Z_2/Z\ F_2(q)<\Psi_f|\exp(iq_2R)|\Psi_i>$$

.

Здесь $q_1 = -\ qM_2/M$ и $q_2 = qM_1/M$ , а $F_{1,2}(q)$ - собственные формфакторы ассоциаций в свободном состоянии.

Разлагая плоские волны по функциям Бесселя и интегрируя по углам [120], квадрат кулоновского формфактора можно представить в виде [13-19]

$$F_J^2 = \frac{1}{Z^2} V_J^2 B_J \ , \tag{3.7}$$

$$B_J = (2J_f+1)(2J+1)(2L_i+1)(L_i 0 J 0 | L_f 0)^2 \begin{Bmatrix} L_i & S & J_i \\ J_f & J & L_f \end{Bmatrix}^2 \ ,$$

где $L_{i,f}$ и $J_{i,f}$ - орбитальные и полные моменты начального i и конечного f состояния ядра, J - мультипольность формфактора, S и Z - спин и заряд ядра, фигурная скобка - 6j символ Вигнера [120] и $V_J$ - структурный множитель, зависящий от характеристик фрагментов и их взаимного движения

$$V_J = Z_1 F_1 I_{2,J} + Z_2 F_2 I_{1,J} \ , \tag{3.8}$$

где $I_{k,J}$ - радиальные матричные элементы по функциям начального и конечного состояния от сферических функций Бесселя

$$I_{k,J} = < L_f J_f \mid j_J(g_k r) \mid L_i J_i > \quad . \qquad (3.9)$$

Здесь k = 1 для первого и 2 для второго кластера, $g_k = (M_k/M)q$, q - переданный импульс, $j_J(g_k r)$ - сферическая функция Бесселя.

Формфакторы $^3H$, $^3He$, $^4He$ кластеров представляются в виде следующей параметризации [13-19]

$$F_\alpha = [1 - (aq^2)^n]\exp(-bq^2) \ ,$$

где a = 0.09985 Фм$^2$, b = 0.46376 Фм$^2$ и n = 6 для $^4He$, a = 0.0785 Фм$^2$, b = 0.4075 Фм$^2$ и n = 5.46 для $^3H$ и a = 0.0872 Фм$^2$, b = 0.481 Фм$^2$ и n = 7.9 для $^3He$. Для дейтрона используется другая форма параметризации

$$F_d = \exp(-aq^2) + bq^2\exp(-cq^2) \ ,$$

с параметрами a = 0.49029 Фм$^2$, b = 0.01615 Фм$^2$ и c = 0.16075 Фм$^2$. Приведенные выше параметризации позволяют аппроксимировать формфакторы до 20 Фм$^{-2}$.

При вычислении неупругих формфакторов, когда конечное состояние лежит в непрерывном спектре собственных значений, волновые функции рассеяния при резонансных энергиях необходимо нормировать на асимптотику вида [118]

$$U_L = \exp(-\delta_L)\ [\ F_L \operatorname{Cos}(\delta_L) + G_L \operatorname{Sin}(\delta_L)] \ . \qquad (3.10)$$

Вероятность E2 - переходов и зарядовый радиус могут быть определены на основе кулоновских формфакторов [13-19,115,118] мультипольности CJ

$$B(E2) = \frac{225Z^2}{4\pi}\lim_{q\to 0}\left(\frac{F_{C2}^2(q)}{q^4}\right) \ , \qquad (3.11)$$

$$R_f^2 = 6\lim_{q\to 0}\left(\frac{1-F_{C0}(q)}{q^2}\right) \ . \qquad (3.12)$$

## 3.3 Вариационные методы решения уравнения Шредингера

Волновые функции в матричных элементах (3.9) для основных и резонансных состояний представимы в виде разложения по не ортогональному гауссовому базису вида [23]

$$R_L(r) = Nr^L \sum_i C_i \exp(-\alpha_i r^2) \quad , \tag{3.13}$$

где $\alpha_i$ и $C_i$ - вариационные параметры и коэффициенты разложения, которые находятся вариационным методом для связанных состояний или аппроксимацией гауссойдами численных волновых функций резонансных уровней [23,105].

Сами вариационные параметры $\alpha_i$ могут быть, например, получены из квадратурной сетки вида [23,105]

$$\alpha_i = \alpha_0 \, tg^2 \{ \pi (2i - 1) / 4N \} \ .$$

Для определения спектра собственных энергий и волновых функций в стандартном вариационном методе при разложении ВФ по ортогональному базису решается матричная задача на собственные значения [121]

$$\sum_i (H_{ij} - EI_{ij})C_i = 0 \quad ,$$

где H - симметричная матрица гамильтониана, I - единичная матрица, E - собственные значения и C - собственные вектора задачи. В данном случае, при не ортогональном базисе гауссойд, мы приходим к обобщенной задаче типа [122]

$$\sum_i (H_{ij} - EL_{ij})C_i = 0 \quad ,$$

где L - симметричная матрица интегралов перекрывания, которая не сводится к единичной матрице. При использовании ВФ вида (3.13) можно найти выражения для всех этих матричных элементов [13-19,23]

$$H_{ij} = T_{ij} + V_{ij} + < i \mid Z_1 Z_2/r \mid j > + < i \mid \hbar^2 L(L+1)/2\mu r^2 \mid j > \ ,$$

$$N_0 = [\, \Sigma \, C_i \, C_j \, L_{ij} \,]^{-1/2} \ ,$$

$$T_{ij} = -\frac{\hbar^2}{2\mu} \frac{\sqrt{\pi}(2L-1)!!}{2^{L+1}\alpha_{ij}^{L+1/2}} \left\{ L(2L+1) - L^2 - \frac{\alpha_i \alpha_j (2L+1)(2L+3)}{\alpha_{ij}^2} \right\} ,$$

$$V_{ij} = \int V(r) \, r^{2L+2} \exp(-\alpha_{ij} r^2) dr \ ,$$

$$L_{ij} = \frac{\sqrt{\pi}(2L+1)!!}{2^{L+2}\alpha_{ij}^{L+3/2}} \quad , \tag{3.14}$$

$$< i \mid Z_1 Z_2/r \mid j > = \frac{Z_1 Z_2 \; L!}{2\alpha_{ij}^{L+1}} \quad ,$$

$$< i \mid \hbar^2 L(L+1)/2\mu r^2 \mid j > = \frac{\sqrt{\pi}(2L-1)!!}{2^{L+1}\alpha_{ij}^{L+1/2}} \frac{L(L+1)\hbar^2}{2\mu} \quad ,$$

$$\alpha_{ij} = \alpha_i + \alpha_j \; .$$

В случае гауссова потенциала межкластерного взаимодействия

$$V(r) = V_0 \exp(-\beta r^2)$$

матричный элемент потенциала $V_{ij}$ определяется в аналитическом виде

$$V_{ij} = V_0 \frac{\sqrt{\pi}(2L+1)!!}{2^{L+2}(\alpha_{ij}+\beta)^{L+3/2}} \; .$$

Для приведенного выше вариационного разложения волновой функции, матричные элементы формфактора также вычисляются аналитически и, например, для ядра $^6$Li имеют вид [13-19]

$$I_{k,0}(C0) = \frac{\sqrt{\pi}}{4}\sum_{i,j} C_i C_j W_{ij} / \alpha_{ij}^{3/2} \quad , \tag{3.15}$$

$$I_{k,2}(C2) = \frac{\sqrt{\pi}}{16}\sum_{i,j} C_i C_j g_k^2 W_{ij} / \alpha_{ij}^{7/2} \quad .$$

Аналогичные выражения можно получить для форматоров ядра $^7$Li

$$J_0 = 3/2; \qquad I_{k,0}(C0) = \frac{\sqrt{\pi}}{8}\sum_{i,j} C_i C_j W_{ij}\left(3 - \frac{g_k^2}{2\alpha_{ij}}\right) / \alpha_{ij}^{5/2} ,$$

$$J_0 = 3/2; 1/2; \qquad I_{k,2}(C2) = \frac{\sqrt{\pi}}{16}\sum_{i,j} C_i C_j g_k^2 W_{ij} / \alpha_{ij}^{7/2} , \tag{3.16}$$

$$J_0 = 7/2; \qquad I_{k,2}(C2) = \frac{\sqrt{\pi}}{32}\sum_{i,j} C_i C_j g_k^2 W_{ij}\left(7 - \frac{g_k^2}{2\alpha_{ij}}\right)/\alpha_{ij}^{9/2},$$

$$J_0 = 7/2; \qquad I_{k,4}(C4) = \frac{\sqrt{\pi}}{64}\sum_{i,j} C_i C_j g_k^4 W_{ij}/\alpha_{ij}^{11/2},$$

где

$$W_{ij} = \exp\left(-\frac{g_k^2}{4\alpha_{ij}}\right), \qquad \alpha_{ij} = \alpha_i + \alpha_j, \qquad g_k = \frac{M_k}{M}q .$$

Используя разложение волновой функции по гауссойдам, для вероятности E2 переходов, имеем

$$B(E2) = \frac{1}{4\pi}W(M,Z)^2 R_0^4 B_2 , \tag{3.17}$$

$$Q = -\frac{2}{5}W(M,Z)R_0^2 ,$$

где

$$W(M,Z) = \frac{M_1^2 Z_2 + M_2^2 Z_1}{M^2} ,$$

$$R_0^2 = \frac{(2L+3)!!\sqrt{\pi}}{2^{L+3}}\sum_{i,j} C_i C_j \alpha_{ij}^{-(L+5/2)} ,$$

а величина $B_2$ была определена в выражении (3.7).

## 3.4 Методы решения обобщенной задачи на собственные значения

Будем и сходить из стандартного уравнения Шредингера в общем виде [61]

$$H\chi = E\chi ,$$

где H - гамильтониан системы и $\chi$ - волновые функции задачи. Разлагая эти волновые функции по некоторому вариационному базису

$$\chi = \sum_i C_i \phi_i \ ,$$

подставляя их в исходную систему, умножая ее слева на комплексно сопряженную $\phi^*_i$ базисную функцию и интегрируя по всем переменным, получим матричную систему вида [123]

$$(H - EL)C = 0 \ . \tag{3.18}$$

Если разложение ВФ выполнялось по ортогональному базису, матрица интегралов перекрывания L превращается в единичную матрицу I, и мы имеем стандартную задачу на собственные значения, для решения которой существует множество методов [68]. Если разложение выполнено по не ортогональному базису [23], то получаем обобщенную задачу на собственные значения [68].

Представляя, в таком случае, матрицу L в виде произведения нижней N и верхней V треугольных матриц [68]

$$L = NV$$

находим

$$HC = ENVC$$

или

$$H'C = EIC' \ , \tag{3.19}$$

где

$$H' = N^{-1}HV^{-1} \ , \qquad C' = VC \tag{3.20}$$

или

$$C = V^{-1}C'$$

и I - единичная матрица.

Тем самым, мы получаем стандартную матричную систему для задачи поиска собственных функций и значений [68,123] вида

$$(H' - EI)C' = 0 \ ,$$

которую можно решать известными методами [68] в общем матричном виде. Процедура перехода, от обобщенной к стандарт-ной задаче, называется ортогонализацией по Шмидту [124].

Вначале находим матрицы N и V, выполняя триангуляризацию симметричной матрицы L [68]. Затем находим обратные матрицы $N^{-1}$ и $V^{-1}$ и вычисляем элементы матрицы $H' = N^{-1}H\ V^{-1}$. Находим далее полную матрицу H' - EI и вычисляет ее детерминант /H' - EI/ при некоторой энергии E. Та энергия, которая приводит к нулю детерминанта является собственной энергией задачи, а соответствующие ей вектора C' - собственные вектора матричной системы. Зная C', не трудно найти и собственные вектора исходной задачи C, поскольку матрица $V^{-1}$ уже известна.

Однако в некоторых задачах при некоторых значениях вариационных параметров процедура нахождения обратных матриц оказывается неустойчивой и при работе компьютерной программы выдается переполнение. Далее мы рассмотрим не стандартный метод, решения обобщенной задачи на собственные значения, который не приводит к таким переполнениям в компьютерных программах.

Рассмотрим вначале общий случай матричного уравнения стандартного вида

$$Ax = b \quad , \tag{3.21}$$

где b и x - матрицы столбцы размерности N, а A - квадратная матрица размерности N×N. Матрицу A можно разложить на треугольные матрицы

$$A = BC \quad ,$$

где B - нижняя треугольная матрица и C - верхняя треугольная матрица, в главной диагонали которой стоят единицы. Нахождение нижней и верхней треугольных матриц выполняется по следующей схеме (метода Халецкого) [68]

$$b_{i1} = a_{i1} \quad , \tag{3.22}$$

$$b_{ij} = a_{ij} - \sum_{k=1}^{j-1} b_{ik} c_{kj} \quad ,$$

где $i \geq j > 1$ и

$$c_{1j} = a_{1j} / b_{11} \quad ,$$

$$c_{ij} = \frac{1}{b_{ii}} \left( a_{ij} - \sum_{k=1}^{i-1} b_{ik} c_{kj} \right) \quad ,$$

при $1 < i < j$. Такой метод позволяет определить и детерминант исходной матрицы A [68]

$$\det(A) = \det(B)\det(C) \; .$$

Известно, что детерминант треугольной матрицы равен произведению ее диагональных элементов, а поскольку

$$\det(C) = 1 \; ,$$

то

$$\det(A) = \det(B) = (b_{11}b_{22}....b_{nn}) \; .$$

Приведем теперь пример программы, которая реализует описанный выше метод разложения, и находит детерминант исходной матрицы.

```
 SUB TRIAN(A(20),B(20),C(20),DET,N)
 REM ***** РАЗЛОЖЕНИЕ МАТРИЦЫ "A" НА ДВЕ ТРЕУГОЛЬНЫЕ A=B*C И ВЫЧИСЛЕНИЕ ДЕТЕРМИНАНТА /A/ *****
 DIM AN(N,N)
 FOR I=1 TO N: C(I,I)=1: B(I,1)=A(I,1): C(1,I)=A(1,I)/B(1,1)
 NEXT I: FOR I=2 TO N: FOR J=2 TO N: S=0: IF J>I GOTO 551
 FOR K=1 TO I-1: S=S+B(I,K)*C(K,J): NEXT K: B(I,J)=A(I,J)-S
 GOTO 552
551 S=0: FOR K=1 TO I-1: S=S+B(I,K)*C(K,J): NEXT K
 C(I,J)=(A(I,J)-S)/B(I,I)
552 NEXT J: NEXT I
 REM - - - - - - - ВЫЧИСЛЕНИЕ ДЕТЕРМИНАНТА - - - - - -
 DET=1: FOR K=1 TO N: DET = DET *B(K,K): NEXT K
 REM - - - - - - - - - ВЫЧИСЛЕНИЕ НЕВЯЗОК - - - - - - - - - - -
 SS=0: FOR I=1 TO N: FOR J=1 TO N: S=0: FOR K=1 TO N
 S=S+B(I,K)*C(K,J): NEXT K: AN(I,J)=S-A(I,J): SS=SS+AN(I,J)
 NEXT J: NEXT I: PRINT "               N = A - B*C ": FOR I=1 TO N
 PRINT: FOR J=1 TO N: PRINT USING " +#.#####^^^^ ";AN(I,J);
 NEXT J: NEXT I: PRINT "            DET=";
 PRINT USING " +#.#####^^^^ "; DET;
 PRINT "            NEV=";: PRINT USING " +#.#####^^^^ ";SS
 END SUB
```

Для оценки точности решения т.е. точности разложения исходной матрицы на две треугольные, использован метод невязок

[68]. После разложения матрицы A на треугольные, вычисляется матрица невязок [68], как разность исходной матрицы A и

$$S = BC$$

где B и C найденные таким образом численные матрицы. Теперь берется разность по всем элементам с исходной матрицей A

$$AN = S - A$$

Матрица AN невязок дает отклонение приближенной величины BC, найденной численными методами, от истинного значения каждого элемента исходной матрицы A.

После разложения матрицы A на треугольные, решение матричной системы можно записать в виде

$$By = b \quad , \qquad Cx = y \quad ,$$

где сами решения находятся из следующих простых выражений [68]

$$y_1 = a_{1,n+1}/b_{11} \quad , \tag{3.23}$$

$$y_i = \frac{a_{i,n+1} - \sum_{k=1}^{i-1} b_{ik} y_k}{b_{ii}} \quad \text{при } i > 1$$

и

$$x_n = y_n \quad ,$$

$$x_i = y_i - \sum_{k=i+1}^{n} c_{ik} x_k \quad \text{при } i < n \quad ,$$

где $a_{i,n+1}$ - элементы матрицы - столбца b (здесь i меняется от 1 до N - размерности матрицы A). В такой задаче все треугольные матрицы и решения X определяются вполне однозначно.

Вернемся теперь к рассмотрению обобщенной матричной задачи на собственные значения и собственные вектора прежнего вида (3.18) [123]

$$(H - EL)C = 0 \quad . \tag{3.24}$$

Это однородная система уравнений и она имеет не тривиальные решение, только если ее детерминант

$$\det(H - EL) = 0 \quad . \tag{3.25}$$

Значения E, которые приводят к нулевому детерминанту, называются собственными значениями. Решения C системы при найденных собственных значениях являются собственными векторами исходной матрицы.

Для численных методов, реализуемых на компьютере, не обязательно разлагать матрицу L на треугольные и находить новую матрицу H' и новые вектора X', определяя обратные матрицы, как это было описано ранее. Мы можем сразу разлагать на треугольные недиагональную матрицу (H - EL) и численными методами искать энергии, которые приводят к нулю ее детерминанта.

Для этого сама матрица (H - EL) разлагается на две треугольные

$$A = H - EL = BC \quad ,$$

как описано выше, и вычисляется ее детерминант

$$\det(A) = \det(B) \quad ,$$

по нулю, которого ищутся собственные значения энергии E системы.

Здесь мы имеем довольно простую задачу поиска нуля некоторого функционала одной переменной

$$F(E) = 0 \quad ,$$

решение, которой не представляет большой сложности.

Такой метод показал свою полную работоспособность, как для контрольных задач, в качестве которых выбиралась нуклон – нуклонная система с классическим потенциалом Рейда, так для реальных расчетов физических характеристик связанных состояний кластеров в атомных ядрах [2].

Если функция F(E) непрерывная на отрезке [a,b] и $F(a)F(b) < 0$, т.е. такой отрезок включает корень функции, то для его поиска можно использовать метод половинного деления. Для поиска корня уравнения делим этот отрезок пополам $x = (b + a)/2$ и вычисляем F(x). Если $F(x) = 0$, то x и есть корень уравнения. Если F(x) не равно нулю, то выбираем ту половину [a,x] [x,b], на концах которой функция имеет разные знаки, например, $F(x)F(a) < 0$ и снова делим его пополам. Эта процедура выполняется до тех пор, пока значение функции в некоторой точке "y" не станет меньше заданного числа ε, которое определяет точность нахождения корня функции [68].

Приведем теперь текст компьютерной программы, реализующей описанный метод поиска корня функции. Здесь PN - нижний предел поиска корня, PV - верхний предел, H - шаг поиска, EP - точность поиска и X - значение корня.

```
 SUB NULFUN(PN,PV,H,EP,X)
 REM ******* ПОИСК НУЛЯ ФУНКЦИИ *******
 A2=PN: B2=PN+H: CALL FUN(A2,D12)
51 CALL FUN(B2,D11): IF D12*D11>0 GOTO 4
44 A3=A2: B3=B2
11 C3=(A3+B3)/2: IF ABS(A3-B3)<EP GOTO 151: CALL
FUN(C3,F2)
 IF D12*F2>0 GOTO 14: B3=C3: D11=F2: GOTO 15
14 A3=C3: D12=F2
15 IF ABS(F2)>EP GOTO 11
151 X=C3: GOTO 7
4 IF ABS(D11*D12)<EP GOTO 44: A2=A2+H
 B2=B2+H: D12=D11: IF B2-PV<0 GOTO 51
7 END SUB
 SUB FUN(X,F)
 F=COS(X)
 END SUB
```

Приведем результаты поиска корня функции Cos(x) при точности $10^{-15}$, в пределах 0 - 2 с шагом 0.1.

KOR = 1.570796326794897

$\pi/2$ = 1.570796326794897

FUN = -3.828317871046316E-016

Здесь первое число KOR - значение найденного корня, второе число – его истинное значение, которое известно и равно $\pi/2$, FUN - значение функции при найденном значении корня, которое должно быть меньше заданной величины, а именно, $10^{-15}$ .

После нахождения собственного значения (обычно это первое или второе собственное значение) решаем известную систему для поиска собственных векторов X, которая имеет вид

$$AX = BCX = (H - EL)X = 0 \quad .$$

Такая система линейных уравнений относительно N неизвестных X может быть решена при E, которое равно собственному значению. Равенство нулю ее детерминанта означает линейную зависимость одного из уравнений системы, т.е. ее ранг R меньше по-рядка системы N. Полагаем, что линейно зависимым является последнее N - е уравнение и, отбрасывая его, получаем систему N - 1 уравнений с N неизвестными [125]

$$a_{11}x_1 + a_{12}x_2 + a_{13}x_3 + \ldots\ldots + a_{1N}x_N = 0$$
$$a_{21}x_1 + a_{22}x_2 + a_{23}x_3 + \ldots\ldots + a_{2N}x_N = 0$$
$$\ldots\ldots\ldots\ldots\ldots\ldots\ldots\ldots\ldots\ldots\ldots\ldots\ldots\ldots\ldots\ldots\ldots$$
$$a_{N-11}x_1 + a_{N-12}x_2 + a_{N-13}x_3 + \ldots\ldots + a_{N-1N}x_N = 0$$

Принимаем $X_N = 1$, получаем систему N - 1 уравнений с N - 1 неизвестными и столбцом свободных членов, который получается из коэффициентов при N - ой неизвестной $a_{iN}$, где i меняется от 1 до N - 1.

В матричном виде это можно записать

$$A'X'=D \quad , \qquad (3.26)$$

где A' - матрица размерности N - 1, X' - решение системы, D - матрица свободных членов $-a_{iN}$. Решаем ее описанным выше методом, разлагая на две треугольные, и находим все X' при i = 1 ÷ N - 1.

Теперь нам известны все решение исходной системы

$$(H - EL)X = 0$$

при i = 1 ÷ N. Собственные вектора должны удовлетворять условию нормировки

$$N\sum_i X_i^2 = 1 \quad ,$$

откуда можно найти саму нормировку и окончательно определить собственные вектора.

Для оценки точности решения системы можно использовать метод невязок, т.е. вычислять матрицу

$$AN = (H - EL)X \quad ,$$

элементы которой должны быть близки к нулю при правильном определении всех X.

Приведем пример программы реализующей метод решения матричного уравнения A'X'=D после разложения матрицы на две треугольные A' = BC и нахождения всех решений уравнения (H - EL)X = 0. Здесь использовано обозначение матрицы L = AL() и X =

SV().

```
SUB SV(E,H(),AL(),C(), B(),Y(),D(),X(),SV(),N)
REM ***** ВЫЧИСЛЕНИЕ СОБСТВЕННЫХ ВЕКТОРОВ МАТ-
РИЧНОГО УРАВНЕНИЯ *****
DIM AN(N): REM ----- ВЕКТОРА СИСТЕМЫ N-1 УРАВНЕНИЙ --
Y(1)=D(1)/B(1,1): FOR I=2 TO N: S=0: FOR K=1 TO I-1
S=S+B(I,K)*Y(K): NEXT K: Y(I)=(D(I)-S)/B(I,I): NEXT I
X(N)=Y(N): FOR I=N-1 TO 1 STEP -1: S=0: FOR K=I+1 TO N
S=S+C(I,K)*X(K): NEXT K: X(I)=Y(I)-S: NEXT I
REM -- ВЕКТОРА ПОЛНОГО ОДНОРОДНОГО УРАВНЕНИЯ --
FOR I=1 TO N: SV(I)=X(I): NEXT I: N=N+1: SV(N)=1: S=0
FOR I=1 TO N: S=S+SV(I)^2: NEXT I: FOR I=1 TO N
SV(I)=SV(I)/SQR(S): NEXT I
REM ------------------ ВЫЧИСЛЕНИЕ НЕВЯЗОК ----------------
FOR I=1 TO N: S=0: SS=0: FOR J=1 TO N: S=S+H(I,J)*SV(J)
SS=SS+E*AL(I,J)*SV(J): NEXT J: AN(I)=S-SS: NEXT I
PRINT: PRINT "                    НЕВЯЗКИ       AN=H*X-E*L*X"
PRINT: FOR I=1 TO N: PRINT USING " +#.#####^^^^ ";AN(I);
NEXT I: END SUB
```

Если матрица A симметричная, можно применить и другой метод, позволяющий разложить ее на произведение транспонированных треугольных матриц вида (метод квадратного корня) [68]

$$A = T'T \quad ,$$

где

$$t_{11} = \sqrt{A_{11}} \quad , \quad t_{1j} = \frac{A_{1j}}{t_{11}} \quad ,$$

при j > 1 и

$$t_{ii} = \sqrt{A_{ii} - \sum_{k=1}^{i-1} t_{ki}^2} \quad \text{при} \quad 1 < i \le n \quad , \tag{3.27}$$

$$t_{ij} = \frac{A_{ij} - \sum_{k=1}^{i-1} t_{ki} t_{kj}}{t_{ii}} \quad \text{при} \quad i < j \quad .$$

Здесь T - верхняя треугольная матрица, а T' - нижняя, которая равна транспонированной матрице T.

Используя такое разложение, решение системы линейных уравнений Ax = b можно представить в виде

$$T'y = b \quad , \qquad Tx = y \quad ,$$

где

$$y_1 = b_1/t_{11} \quad , \tag{3.28}$$

$$y_i = \frac{b_i - \sum_{k=1}^{i-1} t_{ki} y_k}{t_{ii}} \quad \text{при } i > 1 \quad ,$$

где $b_i$ - элементы матрицы - столбца b и

$$x_n = y_n/t_{nn} \quad ,$$

$$x_i = \frac{y_i - \sum_{k=i+1}^{n} t_{ik} x_k}{t_{ii}}$$

при $i < n$ .

Это разложение позволяет легко находить определители симметричных матриц

$$A = T'T \quad ,$$

где T - верхняя треугольная матрица. Тогда

$$\det(A) = \det(T')\det(T) = [\det(T)]^2 = (t_{11}t_{22}....t_{nn})^2 \quad .$$

Использование такого способа при решении задачи на компьютере оказывается не всегда возможным. Из - за ошибок округления, при численных расчетах, под корнем может оказаться малая отрицательная величина (например, $-10^{-15}$), что приводит к аварийной остановке работы программы. Поэтому во всех расчетах мы использовали предыдущий метод решения такой задачи.

## 3.5 Вариационная программа решения уравнения Шредингера

Приведем вариант вариационной программы для поиска собственной энергии и ВФ с использованием независимого варьирования параметров в разложении ВФ по гауссойдам для $^2H^4He$ системы ядра $^6Li$ с реальным ядерным потенциалом. Обозначения параметров приведены в самой программе.

**REM ВАРИАЦИОННАЯ ПРОГРАММА ПОИСКА ЭНЕРГИИ СВЯЗИ**

```
CLS: DEFDBL A-Z: DEFINT I,L,J,N,M,K: NN=20: N=1000
DIM XP(NN), H(NN,NN), T(NN,NN), VN(NN,NN), VC(NN,NN),
AL1(NN,NN),VK(NN,NN)
DIM X(NN), Y(NN), B(NN,NN), C(NN,NN), D(NN), AD(NN,NN)
DIM SV(NN),AA(NN,NN), F1(N),FU(N),FF(N)
DIM AN(NN), AL(NN,NN), C0(5*NN), CW0(5*NN), CW(5*NN)
REM *********** НАЧАЛЬНЫЕ ЗНАЧЕНИЯ **********
FAIL$="C:\BASICA\SOB-ALD.DAT"
NSAVE=0     : REM =0 - НЕТ ЗАПИСИ В ФАЙЛ, =1 - ЗАПИСЬ
Z1=1: Z2=2                : REM МАССЫ И ЗАРЯДЫ КЛАСТЕРОВ
AM1=2 AM2=4
R01=1.96           : REM РАДИУСЫ КЛАСТЕРОВ
R02=1.67
AM=AM1+AM2              : REM ВХОДНЫЕ КОНСТАНТЫ
PM=AM1*AM2/AM
GK=3.44476E-02*Z1*Z2*PM
A11=20.7343
A22=1.439975*Z1*Z2
P1=3.14159265
NF=N   : REM ЧИСЛО ШАГОВ ВЫЧИСЛЕНИЯ ФУНКЦИИ
R00=25
HFF=R00/NF   : REM ШАГ ВЫЧИЛЕНИЯ ФУНКЦИИ
NP=10        : REM РАЗМЕРНОСТЬ БАЗИСА
NI=1         : REM  ЧИСЛО ИТЕРАЦИЙ
NV=1          : REM  =0 БЕЗ МИНИМИЗАЦИИ, =1 С МИНИМИЗА-
ЦИЕЙ ПО ЭНЕРГИИ
EP=1.0D-09    : REM ТОЧНОСТЬ ПОИСКА ЭНЕРГИИ
EPP=1.0D-15     : REM ТОЧНОСТЬ ПОИСКА НУЛЯ ДЕТЕРМИ-
НАНТА
HC=0.123      : REM ШАГ ПОИСКА НУЛЯ ДЕТЕРМИНАНТА
PNC=-1.6        : REM НИЖНЕЕ ЗНАЧЕНИЕ ЭНЕРГИИ ДЛЯ ПО-
ИСКА НУЛЯ ДЕТЕРМИНАНТА
PVC=3           : REM ВЕРХНЕЕ ЗНАЧЕНИЕ ЭНЕРГИИ ДЛЯ ПО-
ИСКА НУЛЯ ДЕТЕРМИНАНТА
PHN=0.123     : REM ШАГ ИЗМЕНЕНИЯ ПАРАМЕТРОВ АЛЬФА
REM ******* ПАРАМЕТРЫ ПОТЕНЦИАЛА *********
V0=-76.12: RN=0.2: LO=0: RC=0
REM ****** НАЧАЛЬНЫЕ ПАРАМЕТРЫ АЛЬФА ******
XP(1)=0.0108345: XP(2)=0.02535157: XP(3)=0.064665899
XP(4)=0.1456006: XP(5)=0.36370071: XP(6)=0.37917857
XP(7)=0.6401466:           XP(8)=16.7174:            XP(9)=0.95201183:
XP(10)=3.503041
```

```
REM ********** НАЧАЛЬНЫЕ КОНСТАНТЫ *********
C11=LO+1.5: C22=LO+0.5: PI=SQR(P1): C33=LO+1
N11=2*LO+3: S44=1: FOR K=1 TO N11 STEP 2
S44=S44*K: NEXT K: LK=LO*LO: S11=S44/(2*LO+3)
S22=S11/(2*LO+1): R1=1: FOR K=1 TO LO: R1=R1*K
NEXT K: B11=PI*S11/(2^(LO+2)): B22=B11*V0: B23=B11*V1
B33=LO*(LO+1)*PI*S22/(2^(LO+1)): B44=A22*R1/2
B55=PI/(2^(LO+1))
REM * ПОИСК ПАРАМЕТРОВ ВФ И ЭНЕРГИИ СВЯЗИ *
CALL VARMIN(ALA(),PHN,NP,NI,XP(),EP,B,NV,NN)
 REM ******* ЯДЕРНЫЕ ХАРАКТЕРИСТИКИ ********
 PRINT:  PRINT "------------------------ ЭНЕРГИЯ СВЯЗИ ------------
"
 PRINT:  PRINT "                        E = ";
 PRINT USING " +#.#####^^^^ ";B
 PRINT:  PRINT "- - - - - - - - - - - - -  ПАРАМЕТРЫ АЛЬФА  - - - - -
"
 PRINT:  FOR I=1 TO NP
 PRINT USING " +#.#####^^^^ ";XP(I);
 NEXT I:  PRINT
 REM ********* СОБСТВЕННЫЕ ВЕКТОРА ***********
 CALL SV(B,NP,XP())
 REM *********** НОРМИРОВКА ВЕКТОРОВ *********
 FOR I=1 TO NP:  FOR J=1 TO NP
 AL(I,J)=PI*S11/(2^(LO+2)*(XP(I)+XP(J))^C11)
 NEXT J:  NEXT I:  S=0
 FOR I=1 TO NP:  FOR J=1 TO NP
 S=S+SV(I)*SV(J)*AL(I,J)
 NEXT J:  NEXT I:  ANOR=1/SQR(S)
 PRINT:  PRINT
 PRINT "                        СОБСТВЕННЫЕ ВЕКТОРА"
 PRINT:  FOR IJK=1 TO NP
 SV(IJK)=ANOR*SV(IJK)
 PRINT USING " +#.#####^^^^ ";SV(IJK);
 NEXT IJK:  PRINT
 REM  ************** ВЫЧИСЛЕНИЕ ВФ *************
 INPUT FFFF: REM ЕСЛИ FFFF=0 ВФ НА ЭКРАН НЕ ВЫВО-
ДИТСЯ
 FOR I=0 TO NF:  R=HFF*I:  S=0
 FOR J=1 TO NP:  RRR=R^2*XP(J):  IF RRR>50 GOTO 9182
 S=S+SV(J)*EXP(-RRR)
9182  NEXT J:  FF(I)=R^(LO+1)*S:  NEXT I
 IF FFFF=0 GOTO 246:  PRINT " R              F(R)"
 FOR I=0 TO NF STEP NFF:  R=I*HFF
```

```
 PRINT USING " +#.##^^^^ ";R;
 PRINT USING "           +#.#####^^^^ ";FF(I):  NEXT I
246 REM ******** ПРОВЕРКА НОРМИРОВКИ *********
 FOR I=0 TO NF: R=I*HFF: F1(I)=FF(I)^2: NEXT I
 CALL SIM(F1(),NF,HFF,SIM): PRINT
 PRINT "                          NOR = ";
 PRINT USING " +#.########^^^^ ";SIM
 PRINT: REM ***** АСИМПТОТИЧЕСКИЕ КОНСТАНТЫ *****
 SKS=(ABS(B)*PM/A11): SS=SQR(ABS(SKS)): SQQ=SQR(2*SS)
 GGG=GK/SS: M1=NF/4: M3=NF/20: M2=NF/2+NF/4
 PRINT "          R              C0            CW0              CW"
 K=0: FOR I=M1 TO M2 STEP M3: K=K+1: R=I*HFF
 CALL ASIMP(R,SKS,GK,LO,I,C0,CW0,CW)
 C0(K)=C0: CW0(K)=CW0: CW(K)=CW
 PRINT USING "      +#.###^^^^  ";R,C0,CW0,CW: NEXT I
 REM ********** РАДИУС ЯДРА ****************
 SS=0: FOR I=1 TO NP: FOR J=1 TO NP
 SS=SS+SV(I)*SV(J)/(XP(I)+XP(J))^(LO+2.5)
 NEXT J: NEXT I: RR=PI*S44*SS/2^(LO+3): RRR=SQR(RR)
 RCH=AM1*R01^2/AM+AM2*R02^2/AM+AM1*AM2*RR/AM^2
 RCH=SQR(RCH): PRINT: PRINT "                      (R^2)^(1/2) =   ";
 PRINT USING " +#.#####^^^^ ";RCH
 IF NSAVE=0 GOTO 4567
 OPEN "O",1,FAIL$
  PRINT#1,"                          ЭНЕРГИЯ"
  PRINT#1,"                                          "
  PRINT#1,USING "                     +#.#####^^^^ ";B
  PRINT#1,"                                          "
  PRINT#1, "                 КОЭФФИЦИЕНТЫ АЛЬФА "
  PRINT#1,"                                          "
  FOR I=1 TO NP:  PRINT#1, USING " +#.#####^^^^ ";XP(I)
  NEXT I:  PRINT#1,"                                              "
  PRINT#1,"                                          "
  PRINT#1, "                 СОБСТВЕННЫЕ ВЕКТОРА    SV"
  PRINT#1,"                                          "
  FOR IJK=1 TO NP:  PRINT#1, USING " +#.#####^^^^ ";SV(IJK);
  NEXT IJK:  PRINT#1,"                                          "
  PRINT#1,"                                          "
  PRINT#1, "                  SV"
  FOR I=1 TO NP:  PRINT#1, USING " +#.#####^^^^ ";E2(I);
  NEXT I:  PRINT#1,"                                          "
  PRINT#1,"                                          "
  PRINT#1,"  R              C0            CW0            CW"
  PRINT#1,"                                          "
```

```
 KKK=0:  FOR I=M1 TO M2 STEP M3:  R=HFF*I
 KKK=KKK+1:  PRINT#1, USING " +#.##^^^^ ";R;
 PRINT#1,  USING  "          +#.#####^^^^  ";C0(KKK);  CW0(KKK);
CW(KKK)
 NEXT          I:                    PRINT#1,:                    PRINT#1,"
"
 PRINT#1, " МЕЖКЛАСТЕРНОЕ РАССТОЯНИЕ ";
 PRINT#1, USING " +#.#####^^^^ ";RRR
 PRINT#1,"                                            "
 PRINT#1, "РАДИУС ЯДРА   ";
 PRINT#1, USING " +#.#####^^^^ ";RCH
 PRINT#1,"                                            "
 PRINT#1,"   R              F(I)"
 PRINT#1,"                                            "
 FOR I=1 TO NF:  R=HFF*I:  PRINT#1, USING " +#.##^^^^ ";R;
 PRINT#1, USING "           +#.#####^^^^ ";FF(I):  NEXT I
 PRINT#1,"                                            "
 PRINT#1,"                                            "
 PRINT#1,"                                            " : CLOSE
4567 END
```

**SUB VARMIN(ALA(30), PHN, NP, NI, XP(30), EP, AMIN, NV, NNN)**

```
 DIM XPN(NNN)
 REM ************* ПОИСК МИНИМУМА ***********
 FOR I=1 TO NP: XPN(I)=XP(I): NEXT: NN=1
 PRINT USING " ### ";NN;
 PRINT USING " +###.#########   ";XPN(NN);
 PH=PHN: CALL DETNUL(XPN(),NP,ALA)
 B=ALA: IF NV=0 GOTO 3012
 PRINT USING "         +#.#######^^^^ ";ALA;
 PRINT: REM ----------------------------------------------------------------
 FOR IIN=1 TO NI: NN=0: GOTO 1119
1159 XPN(NN)=XPN(NN)-PH*XP(NN)
1119 NN=NN+1:  IN=0
2229 A=B: XPN(NN)=XPN(NN)+PH*XP(NN)
 IF XP(NN)<0 GOTO 1159: IN=IN+1
 REM -------------------------------------------------------------
 PRINT USING " ### ";NN;
 PRINT USING " +###.#########   ";XPN(NN);
 CALL DETNUL(XPN(),NP,ALA)
 B=ALA: PRINT USING "         +#.#######^^^^ ";ALA;
 PRINT: REM ----------------------------------------------------------------
 IF B<A GOTO 2229:  C=A:  XPN(NN)=XPN(NN)-PH*XP(NN)
 IF IN>1 GOTO 3339:  PH=-PH:  GOTO 5559
```

```
3339 IF ABS((C-B)/(B))<EP GOTO 4449:  PH=PH*0.5
5559 B=C:  GOTO 2229
4449 PH=PHN:  B=C:  IF NN<NP GOTO 1119:  PH=PHN*0.1
3012  AMIN=B:  NEXT IIN: FOR I=1 TO NP
 XP(I)=XPN(I): NEXT: END SUB
 SUB MAT(XP(20),NP)
 REM ****** ВЫИСЛЕНИЕ МАТРИЦ ***************
 SHARED B44, B23, B11, B33, A11, PM, B55, S22, S44, C22, LO,
S11,LK,RC,PI,C11,C33,B22
 SHARED T(),VC(),VN(),VK(),AL1(),H(),RN,RN1,F1()
 FOR KK=1 TO NP: FOR JJ=1 TO NP: AL=XP(KK)+XP(JJ)
 T(KK,JJ)=-B55*(LO*S11-LK*S22-
XP(KK)*XP(JJ)*S44/AL^2)/AL^C22
 SF=1: SS1=1: IF RC=0 GOTO 7654: PF=RC*SQR(AL)
 NFF=100: HF=PF/NFF: IF PF>3 GOTO 9765: FOR I=0 TO NFF
 X=HF*I: F1(I)=EXP(-X^2): NEXT I: CALL SIM(F1(),NFF,HF,SIM)
 SF=SIM*2/PI
9765 ALR=SQR(AL)*RC: ALR2=ALR^2: EX=EXP(-ALR2)
 SS=PI*(9*ALR-15/(2*ALR))*SF: SS1=(15*EX+SS)/(8*ALR2)
7654  VK(KK,JJ)=B44/AL^C33*SS1: VN(KK,JJ)=B22/(AL+RN)^C11
 VC(KK,JJ)=B33/AL^C22
 H(KK,JJ) = (A11/PM)*(T(KK,JJ) + VC(KK,JJ)) + VN(KK,JJ) +
VK(KK,JJ)
 AL1(KK,JJ)=B11/AL^C11: H(JJ,KK)=H(KK,JJ)
 AL1(JJ,KK)=AL1(KK,JJ)
 NEXT JJ: NEXT KK: END SUB
 SUB DETNUL(XP(20),NP,ALA)
 REM ******* ПОИСК НУЛЯ ДЕТЕРМИНАНТА *******
 SHARED EP,PNC,PVC,HC,EPP
 REM ----------------- ФОРМИРОВАНИЕ МАТРИЦЫ ---------
 CALL MAT(XP(),NP)
 REM --------------- ПОИСК НУЛЯ ДЕТЕРМИНАНТА --------
 A2=PNC: B2=PNC+HC: CALL DETER(A2,D12,NP)
51 CALL DETER(B2,D11,NP)
 REM --------------------------------------------------------------
 IF D12*D11>0 GOTO 4
44 A3=A2: B3=B2
11 C3=(A3+B3)/2: IF ABS(A3-B3)<EPP GOTO 151
 CALL DETER(C3,F2,NP): IF D12*F2>0 GOTO 14
 B3=C3: D11=F2: GOTO 15
14 A3=C3: D12=F2
15 IF ABS(F2)>EPP GOTO 11
151 ALA=C3: GOTO 7
 REM  - - - - - - - - - - - - - - - - - - - - - - - - - - - - - - - - - - - - - - -
```

```
4 IF ABS(D11*D12)<EPP GOTO 44
 A2=A2+HC: B2=B2+HC: D12=D11
 IF B2-PVC<0.1 GOTO 51
7 END SUB
 SUB DETER(AL,DET,NP)
 REM *** ВЫЧИСЛЕНИЕ ДЕРМИНАНТА МАТРИЦЫ ***
 SHARED H(),AL1(),AL(),B(),C()
 FOR I=1 TO NP: FOR J=1 TO NP
 AL(I,J)=(H(I,J)-AL*AL1(I,J)): B(I,J)=0: C(I,J)=0
 NEXT J: NEXT I: CALL TRIAN(AL(),B(),C(),DET,NP): END SUB
 SUB SV(AL,NP,XP(20))
 REM ******* СОБСТВЕННЫЕ ВЕКТОРА *************
 SHARED AL1(), C(), B(), AD(), AL(), Y(), AN(), D(), X(), SV(), H()
 REM ---------------- ФОРМИРОВАНИЕ МАТРИЦЫ ---------
 CALL MAT(XP(),NP)
 REM -------------------- ПОДГОТОВКА МАТРИЦЫ ----------
 FOR I=1 TO NP: FOR J=1 TO NP: AL(I,J)=(H(I,J)-AL*AL1(I,J))
 B(I,J)=0: C(I,J)=0: NEXT J: NEXT I
 FOR I=1 TO NP-1: FOR J=1 TO NP-1: AD(I,J)=AL(I,J)
 NEXT J: NEXT I: FOR I=1 TO NP-1: D(I)=-AL(I,NP)
 NEXT I: NP=NP-1: CALL TRIAN(AD(),B(),C(),DET,NP)
 REM - - - - - - - - - - - ВЫЧИСЛЕНИЕ ВЕКТОРОВ - - - - - - - -
 Y(1)=D(1)/B(1,1): FOR I=2 TO NP: S=0
 FOR K=1 TO I-1: S=S+B(I,K)*Y(K): NEXT K
 Y(I)=(D(I)-S)/B(I,I): NEXT I: X(NP)=Y(NP)
 FOR I=NP-1 TO 1 STEP -1: S=0: FOR K=I+1 TO NP
 S=S+C(I,K)*X(K): NEXT K: X(I)=Y(I)-S: NEXT I
 FOR I=1 TO NP: SV(I)=X(I): NEXT I: NP=NP+1: SV(NP)=1: S=0
 FOR I=1 TO NP: S=S+SV(I)^2: NEXT I: FOR I=1 TO NP
 SV(I)=SV(I)/SQR(S): NEXT I
 REM ------------------ ВЫЧИСЛЕНИЕ НЕВЯЗОК ----------------
 FOR I=1 TO NP: S=0: SS=0: FOR J=1 TO NP
 S=S+H(I,J)*SV(J): SS=SS+AL*AL1(I,J)*SV(J): NEXT J
 AN(I)=S-SS: NEXT I: PRINT
 PRINT "                    НЕВЯЗКИ        H*SV-LA*L*SV=0"
 PRINT: FOR I=1 TO NP
 PRINT USING " +#.#####^^^^ ";AN(I);: NEXT I: END SUB
 SUB TRIAN(AD(20),B(20),C(20),DET,NP)
 REM   РАЗЛОЖЕНИЕ МАТРИЦЫ НА ДВЕ ТРЕУГОЛЬНЫЕ
AD=B*C И ВЫЧИСЛЕНИЕ ДЕТЕРМИНАНТА
 SHARED AA()
 FOR I=1 TO NP: C(I,I)=1: B(I,1)=AD(I,1): C(1,I)=AD(1,I)/B(1,1)
 NEXT I: FOR I=2 TO NP: FOR J=2 TO NP: S=0
 IF J>I GOTO 551: FOR K=1 TO I-1: S=S+B(I,K)*C(K,J)
```

```
 NEXT K: B(I,J)=AD(I,J)-S:GOTO 552
551 S=0: FOR K=1 TO I-1: S=S+B(I,K)*C(K,J): NEXT K
 C(I,J)=(AD(I,J)-S)/B(I,I)
552 NEXT J: NEXT I: S=1: FOR K=1 TO NP: S=S*B(K,K)
 NEXT K: DET=S
 REM - - - - - - - - - ВЫЧИСЛЕНИЕ НЕВЯЗОК - - - - - - - - - - -
 SS=0: FOR I=1 TO NP: FOR J=1 TO NP: S=0
 FOR K=1 TO NP: S=S+B(I,K)*C(K,J): NEXT K
 AA(I,J)=S-AD(I,J): SS=SS+AA(I,J): NEXT J: NEXT I
 PRINT "                N = AD - B*C = 0"
 FOR I=1 TO NP: PRINT: FOR J=1 TO NP
 PRINT USING " +#.#####^^^^ ";AA(I,J);: NEXT J: NEXT I
 PRINT "             DET=";: PRINT USING " +#.#####^^^^ ";S;
 PRINT "                   NEV=";: PRINT USING " +#.#####^^^^ ";SS:
PRINT
 END SUB
 SUB WW(SK,L,GK,R,WH)
 REM ******** ФУНКЦИЯ УИТТЕКЕРА **************
 DIM F(2000)
 SS=SQR(ABS(SK)): AA=GK/SS: BB=L: NN=2000: HH=0.01
 ZZ=1+AA+BB: AAA=1/ZZ: NNN=30000: FOR I=1 TO NNN
 AAA=AAA*I/(ZZ+I): NEXT: GAM=AAA*NNN^ZZ
 RR=R: CC=2*RR*SS: FOR I=0 TO NN: TT=HH*I
 F(I)=TT^(AA+BB)*(1+TT/CC)^(BB-AA)*EXP(-TT): NEXT I
 CALL              SIM(F(),NN,HH,SIM):              WH=SIM*EXP(-
CC/2)/(CC^AA*GAM)
 END SUB
SUB SIM(V(5000),N,H,SIM)
 REM ********* ИНТЕГРАЛ ПО СИМПСОНУ *********
 A=0: B=0: FOR I=1 TO N-1 STEP 2: B=B+V(I)
 NEXT I: FOR J=2 TO N-2 STEP 2: A=A+V(J)
 NEXT J: SIM=H*(V(0)+V(N)+2*A+4*B)/3: END SUB
 SUB ASIMP(R,SK,GK,L,N,C0,CW0,CW)
 REM **** АСИМПТОТИЧЕСКАЯ КОНСТАНТА ******
 SHARED FF()
 SS=SQR(ABS(SK)): SQ=SQR(2*SS): GG=GK/SS
 CALL WW(SK,L,GK,R,WWW): CW=FF(N)/WWW/SQ
 C0=FF(N)/(EXP(-SS*R)*SQ): CW0=C0*(R*SS*2)^GG: END SUB
```

Дадим теперь результаты контрольного счета по этой программе при числе вариационных параметров N=10, который сравним с другими результатами, полученными конечно – разностным методом, для реальной физической системы $^{2}H^{4}He$ ядра $^{6}Li$.

Потенциал взаимодействия представляется в виде гауссойды с глубиной -76.12 МэВ и шириной 0.2 Фм$^{-2}$, кулоновский радиус и орбитальный момент основного состояния равны нулю. Асимптотические константы C0, CW0 и CW определяются из сшивки ВФ с обычной экспонентой exp(-kr), асимптотикой функции Уиттекера и точной функцией Уиттекера. Для радиусов кластеров использованы величины - $R_d$ = 1.96 Фм и $R_\alpha$ = 1.67 Фм [13-19].

------------------------------- ЭНЕРГИЯ СВЯЗИ ----------------------------

E = -1.4711E+00 - Энергия связи в МэВ.

- - - - - - - - - - - - - - - - ПАРАМЕТРЫ АЛЬФА - - - - - - - - - - - - - -

+9.50186E-03 +2.53516E-02 +6.46659E-02 +1.45601E-01 +3.18966E-01
+3.79179E-01 +6.40147E-01 +2.08299E+01 +8.34914E-01 +3.50304E+00

НЕВЯЗКИ (H-LA*L)*SV=0

+0.00000E+00 +9.76996E-15 +7.99361E-15 +1.49880E-14 +9.23393E-15
-9.85323E - 16 -1.24900E -15 -1.90820E -17 +9.15934E-16 -4.32966E- 13

СОБСТВЕННЫЕ ВЕКТОРА

+5.50487E-03 +5.13559E-02 +1.59658E-01 +2.67736E-01 +2.37404E-01
-9.26495E-01 -7.01996E-01 +9.98457E-03 +1.02114E-01 +2.05509E-02

NOR = +9.99999856E-01 - Нормировка ВФ.

АСИМПТОТИЧЕСКИЕ КОНСТАНТЫ

| R | C0 | CW0 | CW |
|---|---|---|---|
| +6.250E+00 | +1.991E+00 | +2.977E+00 | +3.225E+00 |
| +7.500E+00 | +1.905E+00 | +3.008E+00 | +3.223E+00 |
| +8.750E+00 | +1.837E+00 | +3.038E+00 | +3.228E+00 |
| +1.000E+01 | +1.781E+00 | +3.064E+00 | +3.235E+00 |
| +1.125E+01 | +1.723E+00 | +3.070E+00 | +3.224E+00 |
| +1.250E+01 | +1.663E+00 | +3.057E+00 | +3.198E+00 |
| +1.375E+01 | +1.612E+00 | +3.051E+00 | +3.179E+00 |
| +1.500E+01 | +1.582E+00 | +3.072E+00 | +3.192E+00 |
| +1.625E+01 | +1.569E+00 | +3.121E+00 | +3.235E+00 |
| +1.750E+01 | +1.562E+00 | +3.176E+00 | +3.284E+00 |
| +1.875E+01 | +1.542E+00 | +3.202E+00 | +3.304E+00 |

(R^2)^(1/2) = 2.614E+00 - Радиус ядра.

Из этих результатов видно, что элементы матрицы невязок не превышают величины $10^{-12}$.

Приведем подобные результаты для системы $^3H^4He$ ядра $^7Li$. Для па-раметров гауссового потенциала использовано - $V_0$ = 83.83 МэВ, $R_0$ = 0.15747 $Фм^{-2}$, L=1, $R_{cul}$ = 3.095 Фм.

-------------------------- ЭНЕРГИЯ СВЯЗИ ----------------------------

E = -2.4654E+00

- - - - - - - - - - - - - - ПАРАМЕТРЫ АЛЬФА - - - - - - - - - - - - - -

+3.48871E-02 +2.53516E-02 +6.46659E-02 +1.45601E-01 +3.63701E-01
+3.79179E-01 +7.82259E-01 +7.17400E-01 +9.52012E-01 +1.94769E+00

НЕВЯЗКИ H*SV-LA*L*SV=0

+8.88178E-15 +2.13163E-14 -6.30607E-14 -4.84057E-14 -1.56646E-14
-1.95260E-14 -5.44009E-15 -4.88498E-15 +2.92821E-15 -4.35187E-12

СОБСТВЕННЕ ВЕКТОРА

-3.58270E-04 -2.53526E-03 -3.65097E-02 -1.86901E-01 -2.79804E+00
+3.39896E+00 -5.11591E-01 +9.55986E-01 +4.50193E-02 +1.31924E-03

N = +1.00000000E+00

АСИМПТОТИЧЕСКИЕ КОНСТАНТЫ

| R | C0 | CW0 | CW |
|---|---|---|---|
| +6.250E+00 | -3.169E+00 | -4.984E+00 | -3.897E+00 |
| +7.500E+00 | -2.925E+00 | -4.825E+00 | -3.908E+00 |
| +8.750E+00 | -2.720E+00 | -4.672E+00 | -3.886E+00 |
| +1.000E+01 | -2.584E+00 | -4.594E+00 | -3.900E+00 |
| +1.125E+01 | -2.503E+00 | -4.590E+00 | -3.961E+00 |
| +1.250E+01 | -2.383E+00 | -4.492E+00 | -3.928E+00 |
| +1.375E+01 | -2.142E+00 | -4.140E+00 | -3.661E+00 |
| +1.500E+01 | -1.779E+00 | -3.518E+00 | -3.141E+00 |
| +1.625E+01 | -1.355E+00 | -2.736E+00 | -2.462E+00 |
| +1.750E+01 | -9.443E-01 | -1.944E+00 | -1.762E+00 |
| +1.875E+01 | -6.027E-01 | -1.263E+00 | -1.152E+00 |

(R^2)^(1/2) = 2.471E+00

И здесь элементы матрицы невязок не превышают величины $10^{-11}$, а нормировка N собственной функции равна единице с точностью до девятого знака. Из этих результатов видно, что в обеих случаях удается получить точность вычислений на уровне $10^{-11}$, асимптотика ВФ остается устойчивой по крайней мере в области 6-13 Фм, а нормировка функции определена с точностью порядка $10^{-6}$

## 3.6 Численные методы решения уравнения Шредингера

Уравнение Шредингера [57] для центральных потенциалов вида (3.1)

$$\chi''_L + [\,k^2 - V(r)\,]\,\chi_L = 0 \qquad (3.29)$$

с тем или иным граничным условием при $k^2 < 0$ образует краевую задачу типа Штурма - Лиувилля и при переходе к конечным разностям [58]

$$u'' = [u_{n+1} - 2u_n + u_{n-1}]/h^2$$

превращается в замкнутую систему линейных алгебраических уравнений [58,68]. Условие равенства нулю ее детерминанта позволяет определить энергию системы [126]

$$D_N = \begin{pmatrix} \theta_1 & 1 & 0 & . & . & . & 0 \\ \alpha_2 & \theta_2 & 1 & 0 & . & . & 0 \\ 0 & \alpha_3 & \theta_3 & 1 & 0 & . & 0 \\ . & . & . & . & . & . & . \\ . & . & . & . & . & . & . \\ 0 & . & 0 & 0 & \alpha_{N-1} & \theta_{N-1} & 1 \\ 0 & . & 0 & 0 & 0 & \alpha_N & \theta_N \end{pmatrix}, \qquad (3.30)$$

где N - число уравнений, $h = \Delta r/N$ - шаг конечно - разностной сетки, $\Delta r$ - интервал решения системы, и

$$\alpha_n = 1\ , \qquad \alpha_N = 2\ , \qquad \theta_n = k^2h^2 - 2 - V_nh^2\ ,$$
$$\theta_N = k^2h^2 - 2 - V_nh^2 + 2hf(\eta,L,Z_n)\ , \qquad Z_n = 2kr_n\ ,$$
$$r_n = nh\ , \qquad n = 1,2\ .....N\ , \qquad k = \sqrt{|k^2|}\ ,$$
$$f(\eta,L,Z_n) = -k - 2k\eta/Z_n - 2k(L-\eta)/Z_n^2\ , \qquad (3.31)$$

где $V_n = V(r_n)$ - потенциал взаимодействия кластеров в точке $r_n$. Такая форма записи граничных условий $f(\eta,L,Z_n)$ позволяет приближенно учитывать кулоновские взаимодействия, т.е. эффекты, которые дает учет функции Уиттекера.

Вид логарифмической производной ВФ во внешней области можно получить из интегрального представления функции Уиттекера [75]

$$f(\eta, L, Z) = -k - \frac{2k\eta}{Z} - \frac{2k(L-\eta)}{Z^2} S \quad , \tag{3.32}$$

где

$$S = \frac{\int\limits_0^\infty t^{L+\eta+1}(1+t/z)^{L-\eta-1} e^{-t} dt}{\int\limits_0^\infty t^{L+\eta}(1+t/z)^{L-\eta} e^{-t} dt} \quad .$$

Расчеты показывают, что величина S не превышает 1.05, и ее учет приводит к поправкам в энергию связи системы, порядка единицы четвертого знака после запятой.

Вычисление $D_N$ проводится по рекуррентным формулам вида [58]

$$D_{-1} = 0 \,,\, D_0 = 1 \quad , \quad D_n = \theta_n D_{n-1} - \alpha_n D_{n-2} \quad , \quad n = 1 \ldots N \ . \tag{3.33}$$

Для нахождения формы волновых функций связанных состояний используется другой рекуррентный процесс [58]

$$\chi_0 = 0,\ \chi_1 = \mathrm{const},\ \chi_n = \theta_{n-1}\chi_{n-1} + \alpha_{n-1}\chi_{n-2} \,,\ n = 2 \ldots N \tag{3.34}$$

Тем самым, при заданной энергии системы удается найти детерминант и волновую функцию связанного состояния. Энергия, приводящая к нулю детерминанта, считается собственной энергией системы, а волновая функция при этой энергии - собственной функцией задачи.

Приведенные выше выражения позволяют проводить расчеты многих характеристик связанных состояний атомных ядер в том случае, если межкластерные потенциалы имеют чисто центральный вид и не содержат слагаемых, зависящих от взаимной ориентации спинов частиц и их относительного расстояния.

Подобные ядерные силы называются тензорными и приводят к смешиванию орбитальных состояний с различным L. В частности, в дейтроне, такие силы смешивают орбитальные состояния S и D и приводят к появлению D компоненты в волновой функции системы. В следующем разделе мы перейдем к рассмотрению именно таких взаимодействий с изложением некоторых методов расчетов, применимые в данном случае.

### *3.6.1 Методы расчета Гамма функции*

Для нахождения численных значений Гамма функции можно использовать ее интегральное представление [73]

$$\Gamma(z)=\int_0^\infty t^{z-1}e^{-t}dt$$

при Re z >0. Или разложение вида

$$\Gamma(z)=\lim_{n\to\infty}\left(\frac{n!n^z}{z(z+1)...(z+n)}\right) \qquad (3.35)$$

при z не равном 0, - 1, - 2 ... . Существуют также полезные соотношения

$\Gamma(z+1)=z\Gamma(z)$ , $\Gamma(1)=\Gamma(2)=1$ , $\Gamma(n)=(n-1)!$

при n = 1,2,3.. . Выражение (3.35) можно использовать для численных расчетов значений Г - функции.

Ниже приведена программа вычисления Гамма - функции таким способом.

```
SUB GAMMA(Z,GF)
C=1:  N=100000
FOR I=1 TO N
C=C*I/(Z+I)
NEXT I
GF=C*N^Z/Z
END SUB
```

Здесь Z - значение переменной и GF - значение самой Гамма функции. В таблице 3.1 приведены результаты расчетов Гамма функции для различных N и Z и сравнение их с табличными данными [73].

Таблица 3.1 - Вычисление гамма - функции.

| | $N=10^5$ | $N=10^6$ | Точные |
|---|---|---|---|
| Z | Г(z) из программы | | Г(z) из таблиц |
| 1/3 | 2.67 89 32 | 2.67 89 38 | 2.67 89 38 |
| 0.5 | 1.77 24 47 | 1.77 24 53 | 1.77 24 54 |
| 1 | 0.99 99 90 | 0.99 99 99 | 1 |
| 2 | 0.99 99 70 | 0.99 99 97 | 1 |
| 3 | 1.99 98 80 | 1.99 99 88 | 2 |
| 4 | 5.99 94 00 | 5.99 99 40 | 6 |

Из этих результатов видно, что уже при числе членов выражения (3.35) удается получить точность порядка $5x10^{-4}$, даже при сравнительно больших значениях величины Z.

### *3.6.2 Методы расчета функций Уиттекера*

Функция Уиттекера является решение уравнения Шредингера без ядерного потенциала [86]

$$\frac{d^2W(\mu,\nu,z)}{dz^2}-\left(\frac{1}{4}-\frac{\nu}{z}-\frac{1/4-\mu^2}{z^2}\right)W(\mu,\nu,z)=0\ , \qquad (3.36)$$

которое можно привести к стандартному виду уравнения Шредингера

$$\frac{d^2\chi(k,L,r)}{dr^2}-\left(k^2+\frac{g}{r}+\frac{L(L+1)}{r^2}\right)\chi(k,L,r)=0\ ,$$

где $g=\dfrac{2\mu Z_1Z_2}{\hbar^2}=2k\eta$, $\eta=\dfrac{\mu Z_1Z_2}{k\hbar^2}$ - кулоновский параметр, $z=2kr$, $\nu=-\dfrac{g}{2k}=-\eta$ и $\mu = L+1/2$.

Для нахождения численных значений функции Уиттекера обычно используют ее интегральное представление

$$W(\mu,\nu,z)=\frac{z^{\nu}e^{-z/2}}{\Gamma(1/2-\nu+\mu)}\int t^{\mu-\nu-1/2}(1+t/z)^{\mu+\nu-1/2}e^{-t}dt\quad ,$$

которое можно привести к виду

$$W(L,\eta,z)=\frac{z^{-\eta}e^{-z/2}}{\Gamma(L+\eta+1)}\int t^{L+\eta}(1+t/z)^{L-\eta}e^{-t}dt\quad .$$

Легко видеть, что при L = 1 и $\eta$ = 1 приведенный интеграл превращается в $\Gamma(3)$, которая сокращается со знаменателем и остается

простое выражение [127]

$$W(1,1,z) = \frac{e^{-z/2}}{z},$$

которое можно использовать для контроля правильности вычислений функции Уиттекера при любых значениях z.

Ниже приведена программа для расчета функций Уиттекера, использующая ее интегральное представление [86].

```
SUB WH(V(5000),X,L,K,G,W)
REM ***** ВЫЧИСЛЕНИЕ ФУНКЦИИ УИТТЕКЕРА *****
A=G/K: B=L: H=0.005: N=6000: Z=1+A+B
C=1: M=100000: FOR I=1 TO M: C=C*I/(Z+I)
NEXT I: GF=C*M^Z/Z: Z=2*X*K: FOR I=0 TO N
T=H*I: V(I)=T^(A+B)*(1+T/Z)^(B-A)*EXP(-T): NEXT I
A=0:  B=0: FOR I=0 TO N STEP 2: A=A+V(I): NEXT I
FOR I=1 TO N STEP 2: B=B+V(I): NEXT I
S=H*(V(0)+4*A+2*B - V(N))/3
W=S*EXP( - Z/2)/(Z^A*GF): END SUB
```

Здесь $A = \frac{\mu Z_1 Z_2}{k\hbar^2} = G/k$ - кулоновский параметр, $G = \frac{\mu Z_1 Z_2}{\hbar^2}$, $k^2 = \frac{2\mu E}{\hbar^2} = \frac{2\mu E}{41.4686}$ - волновое число в Фм$^{-2}$, если E - энергия выражена в МэВ, X - расстояние от центра, равное r в Фм, Z=2kr - безразмерная переменная, L - орбитальный момент - 0,1,2 и т.д.

В таблице 3.2 приведены результаты расчета функции Уиттекера при $\eta = 1$ и $k = 1$ для разных орбитальных моментов и ее точные значения при $L = 1$ и $\eta = 1$.

Таблица 3.2 - Вычисление функции Уиттекера.

| r | L=0 | L=1 | L=1 (точные значения) | L=2 |
|---|---|---|---|---|
| 1 | 1.020287E-01 | 1.839506E-01 | 1.839397E-01 | 5.518741E-01 |
| 5 | 5.684637E-04 | 6.738346E-04 | 6.737947E-04 | 9.434065E-04 |
| 10 | 2.071580E-06 | 2.270131E-06 | 2.269996E-06 | 2.724267E-06 |
| 15 | 9.577254E-09 | 1.019735E-08 | 1.019674E-08 | 1.155746E-08 |
| 20 | 4.912900E-11 | 5.153190E-11 | 5.152884E-11 | 5.668736E-11 |

Из этой таблицы видно, при наблюдается совпадение с точными значениями функции на уровне порядка $5x10^{-4}$, т.е. не хуже 0.05 %.

### 3.7 Численная программа решения уравнения Шредингера

Приведем теперь текст конечно - разностной программы для вычисления волновых функций и энергий связанных состояний, основанной на описанном выше методе [13-19]. Здесь использованы следующие обозначения: AM1 - Масса первой частицы в а.е.м., AM2 - Масса второй частицы в а.е.м., Z1 - Заряд первой частицы в единицах заряда "e", Z2 - Заряд второй частицы в единицах заряда "e", PM - Приведенная масса $\mu$, A1 - Константа $\hbar^2/M_N = 41.4686$, где $M_N$ - масса нуклона в а.е.м., равная 1, AK1 - Константа при кулоновском потенциале $1.439975 Z_1 Z_2\ 2\mu/\hbar^2$, N - Число шагов при интегрировании уравнения Шредингера, H - Величина шага при интегрировании уравнения Шредингера, V0 - Глубина ядерного потенциала, R0 - Радиус ядерного потенциала, RCU - Кулоновский радиус, L - Орбитальным момент, SKS - Квадрат волнового числа, ESS - Энергия связи, GGG - Кулоновский параметр $3.44476\ 10^{-2}\ Z_1 Z_2\ \mu/k$, RK1, RK2 - Радиусы частиц, SKN - Нижний предел поиска энергии связанного состояния, HC - Начальный шаг при поиске энергии, SKV - Верхний предел поиска энергии связанного состояния, EP - Точность вычисления энергии связи.

```
REM ПРОГРАММА РАСЧЕТА ВФ И ЭНЕРГИИ СВЯЗИ
DEFDBL A - Z:DEFINT K,L,J,M,N,I:NN=4000:DIM U(NN),V(NN)
REM ******** НАЧАЛЬНЫЕ ПАРАМЕТРЫ ************
RAD=1:ASSIMP=1:WFUN=0
REM **********************************************
 Z1=2: Z2=1: Z=Z1+Z2: AM1=4: AM2=2: AM=AM1+AM2
 RK1=1.67: RK2=1.96: PI=3.1415926535899
 PM=AM1*AM2/(AM1+AM2): A1=41.4686: B1=2*PM/A1
 AK1=1.439975*Z1*Z2*B1: GK=3.44476E - 02*Z1*Z2*PM
 REM **********************************************
 N=1000: H=0.02: SKN= - 2: HC=0.1: SKV=1: SKN=SKN*B1
 SKV=SKV*B1: HC=HC*B1: EP=1.0E - 10
 REM **********************************************
 V0=76.12: R0=.2: V1=0: R1=0: A2= - V0*B1: A33=V1*B1: RCU=0:
L=0
REM ************ ПОИСК ЭНЕРГИИ **************
 CALL MIN(EP, B1, PM, SKN, SKV, HC, H, N, L, A2, R0, AK1,
RCU, GK, ESS, SKS): PRINT: PRINT " E=";: PRINT USING
"+#.####^^^^ ";ESS: PRINT
 REM ************ ПОИСК ФУНКЦИИ **************
```

```
 CALL FUN(U(),H,N,A2,R0,A33,R1,L,RCU,AK1,SKS)
 REM * * * * * * * * НОРМИРОВКА ВФ * * * * * * * * *
 FOR I1=0 TO N: V(I1)=U(I1)*U(I1): NEXT I1
 CALL SIM(V(),H,N,HN): HN1=1/SQR(HN): FOR I1=0 TO N
 U(I1)=U(I1)*HN1: NEXT I1
 REM * * * АСИМПТОТИЧЕСКИЕ КОНСТАНТЫ * * *
 SS=SQR(ABS(SKS)): SQQ=SQR(2*SS): GGG=GK/SS
 IF ASSIMP=0 GOTO 9191: PRINT " R C0 CW0 CW"
 FOR I=N/4 TO N STEP N/10: R=I*H
 CALL WW(SKS,L,GK,R,WWW,V())
 CW=U(I)/WWW/SQQ: C0=U(I)/EXP( - SS*R)/SQQ
 CW0=C0*(2*R*SS)^GGG
 PRINT USING " +#.####^^^^ ";R,C0,CW0,CW: NEXT I
9191 IF WFUN=0 GOTO 2233: PRINT: PRINT " R U"
 PRINT: FOR I1=0 TO N STEP 40: X=H*I1
 PRINT USING " +#.####^^^^ ";X,U(I1): NEXT I1
2233 REM * * * * * * * * РАДИУС * * * * * * * * * * * * * *
 IF RAD=0 GOTO 7733: FOR I1=0 TO N: X=I1*H:
V(I1)=X^2*U(I1)^2
 NEXT I1: CALL SIM(V(),H,N,RKV): AM=AM1+AM2
 RK=AM1 / AM * RK1 ^ 2 + AM2 / AM * RK2 ^ 2 + AM1 * AM2 /
AM ^ 2 * RKV
RZ=Z1 / Z * RK1 ^ 2 + Z2 / Z * RK2 ^ 2 + (Z1 * AM2 ^ 2 + Z2 * AM1
^ 2) / AM ^ 2 / Z * RKV
 PRINT: PRINT " (RM^2)^1/2= ";: PRINT USING " #.###^^^^
";SQR(RK);
 PRINT " (RZ^2)^1/2= ";: PRINT USING " #.###^^^^ ";SQR(RZ)
7733 PRINT
 REM * * * * ИМПУЛЬСНЫЕ РАСПРЕДЕЛЕНИЯ * * * * * *
 IF IMPULS=0 GOTO 8822: PRINT " Q P^2/P0^2
SQR(P^2/P0^2)"
 PRINT: FOR IL=0 TO NP: Q=HP*IL+.000001: FOR I1=1 TO N
 X=I1*H: V(I1)=(SIN(Q*X))*U(I1)/Q: NEXT I1
 CALL SIM(V(),H,N,S): IF Q>0.01 GOTO 963: QQ=S^2
963 AIMP=S^2/QQ: PRINT USING " +#.####^^^^ ";Q;
AIMP;SQR(AIMP)
 NEXT IL
8822 REM *********** ФОРМФАКТОР ***************
 IF FORM=0 GOTO 6655: PRINT
 PRINT "     QK        FC0K        Q        FC0"
 FOR II=0 TO NFOR: QK(II)=QH*II+QN: Q=SQR(QK(II))
 G1=AM1*Q/AM: G2=AM2*Q/AM
 FK1=(1 - (FS11*QK(II))^6)*EXP( - FS12*QK(II))
 FK2 = ABS(EXP( - FS21*QK(II)) + FS22*QK(II)*EXP( -
```

```
FS23*QK(II)))
 REM * * * * * * * * * * * * * * * * * * * * * * * * * * * * * * * *
 FOR K=1 TO N: R=H*K: RQ1=R*G1: V(K)=U(K)^2*SIN(RQ1)/RQ1
 NEXT K: CALL SIM(V(),H,N,SIM): AI1=SIM: F2=Z2*FK2*AI1
 REM * * * * * * * * * * * * * * * * * * * * * * * * * * * * * * * * *
 FOR K=1 TO N: R=H*K: RQ2=R*G2: V(K)=U(K)^2*SIN(RQ2)/RQ2
 NEXT K: CALL SIM(V(),H,N,SIM): AI2=SIM: F1=Z1*FK1*AI2
 REM * * * * * * * * * * * * * * * *  * * * * * * * * * * * * * *
 FOR1(II)=(F1+F2)/(Z1+Z2): FKB(II)=FOR1(II)^2
 IF QK(II)>0.01 GOTO 123: RA=6*(1-FOR1(II))/QK(II)
 RA=(SQR(RA)): PRINT "                              RF=";
 PRINT USING "#.####^^^^";RA: PRINT
123 PRINT USING "  +#.####^^^^"; QK(II); FKB(II); Q; FOR1(II)
 NEXT II: PRINT " КОНЕЦ ПРОГРАММЫ":  END
```

**SUB MIN(EP, B1, PM, SKN, SKV, HC, H, N3, L, A22, R0, AK1, RCU, GK, E, SKS)**

```
 REM ***** ПОИСК ЭНЕРГИИ СВЯЗАННЫХ СОСТОЯНИЙ ****
 SHARED A33,R1
 DK=SKN:                                                    CALL
DET(DK,GK,N3,A22,R0,L,AK1,RCU,H,DD,A33,R1)
 D12=DD: B2=A2+HC
51                          DK=B2:                          CALL
DET(DK,GK,N3,A22,R0,L,AK1,RCU,H,DD,A33,R1)
 D11=DD: IF D12*D11>0 GOTO 4
3 A3=A2: B3=B2
11 C3=(A3+B3)/2: IF (ABS(A3 - B3))<1D - 15 GOTO 151: DK=C3
 CALL DET(DK,GK,N3,A22,R0,L,AK1,RCU,H,DD,A33,R1)
 F2=DD: IF D12*F2>0 GOTO 14: B3=C3: D11=F2: GOTO 155
14 A3=C3: D12=F2
155 IF ABS(F2)>EP GOTO 11
151 CO=C3: GOTO 7
4 REM IF ABS(D11*D12)<1D - 15 GOTO 3: A2=A2+HC
 B2=B2+HC: D12=D11: IF B2 - SKV<0.1 GOTO 51
7 E=CO/B1: SKS=CO: END SUB
```

**SUB DET(DK,GK,N,A2,R0,L,AK,RCU,H,DD,A3,R1)**

```
 REM ****** ВЫЧИСЛЕНИЕ ДЕТЕРМИНАНТА ********
 HK=H^2: S1=SQR(ABS(DK)): G2=GK/S1: D1=0: D=1
 FOR  II=1  TO  N:  X=II*H:  A=A2*EXP(  -  X*X*R0)+A3*EXP(  -
X*X*R1)
 F=A+L*(L+1)/(X*X): IF X>RCU GOTO 67
 F=F+AK/(2*RCU)*(3 - (X/RCU)^2): GOTO 66
67 F=F+AK/X
66 IF II=N GOTO 111: D2=D1: D1=D: OM=DK*HK - F*HK - 2
 D=OM*D1 - D2: NEXT II
```

```
111 Z=2*X*S1: OM=DK*HK - F*HK - 2
 W= - S1 - 2*S1*G2/Z - 2*S1*(L - G2)/(Z*Z)
 OM=OM+2*H*W: DD=OM*D - 2*D1: END SUB
 SUB WW(SK,L,GK,R,WH,V(4000))
 REM ********** ФУНКЦИЯ УИТТЕКЕРА ************
 SS=SQR(ABS(SK)): AA=GK/SS: BB=L: NN=2000: HH=0.01
 ZZ=1+AA+BB: AAA=1/ZZ: NNN=30000: FOR I2=1 TO NNN
 AAA=AAA*I2/(ZZ+I2): NEXT I2: GAM=AAA*NNN^ZZ
 RR=R: CC=2*RR*SS: FOR I=0 TO NN: TT=HH*I
 V(I)=TT^(AA+BB)*(1+TT/CC)^(BB - AA)*EXP( - TT): NEXT I
 CALL SIM(V(),HH,NN,SIM): WH=SIM*EXP( -
CC/2)/(CC^AA*GAM)
 END SUB
 SUB FUN(U(4000),H,N,A2,R0,A3,R1,L,RCU,AK,SK)
 REM ************ ПОИСК ФУНКЦИИ **************
 U(0)=0: U(1)=0.001: HK=H*H: FOR K=1 TO N - 1: X=K*H
 Q1=A2 * EXP( -R0 * X * X) + A3 * EXP( - R1 * X * X) + L * (L + 1)
/ (X * X)
IF X>RCU GOTO 1571
Q1=Q1+(3 - (X/RCU)^2)*AK/(2*RCU)
GOTO 1581
1571 Q1=Q1+AK/X
1581 Q2= - Q1*HK - 2+SK*HK: U(K+1)= - Q2*U(K) - U(K - 1)
 NEXT K: END SUB
 SUB SIM(V(4000),H,N,SIM)
 REM ******** ИНТЕГРАЛ ПО СИМПСОНУ ***********
 A=0:  B=0: FOR II=1 TO N - 1 STEP 2: B=B+V(II): NEXT II
 FOR JJ=2 TO N - 2 STEP 2: A=A+V(JJ): NEXT JJ
 SIM=H*(V(0)+V(N)+2*A+4*B)/3: END SUB
```

Для контроля работы программы рассмотрим известный центральный нуклон – нуклонный потенциал простого гауссова типа

$$V(r)= -V_0 \exp(-kr^2)$$

с параметрами $V_0$ = 46.8 МэВ, k = 0.2669 Фм$^{-2}$. Известно [128], что он приводит к энергии связи дейтрона, равной -2.222 МэВ. Выполним теперь расчеты такой энергии по нашей программе конечно – разностным методом с разным числом шагов N и разной величиной шага H. Величина R = NH, показывает до какого расстояния вычисляется волновая функция, т.е. это расстояние считается уже асимптотикой, где ядерный потенциал практически равен нулю. Результаты контрольных расчетов приведены в таблице 3.3.

Таблица 3.3 - Сходимость конечно – разностного метода при вычислении дискретного собственного значения, т.е. энергии связи дейтрона.

| N | H, Фм | R, Фм. | E, Мэв |
|---|---|---|---|
| 200 | 0.1 | 20 | -2.2264 |
| 200 | 0.05 | 10 | -2.2230 |
| 500 | 0.01 | 5 | -2.2203 |
| 500 | 0.02 | 10 | -2.2220 |
| 1000 | 0.01 | 10 | -2.2219 |
| 1000 | 0.02 | 20 | -2.2220 |

Видно, что уже при 500 шагов и величине шага 0.02 Фм. наш результат с точностью до третьего знака после запятой совпадает с известным результатом для этого потенциала.

# 4. МЕТОДЫ РЕШЕНИЯ СИСТЕМЫ УРАВНЕНИЙ ШРЕДИНГЕРА В ДИСКРЕТНОМ СПЕКТРЕ ДЛЯ ПОТЕНЦИАЛОВ С ТЕНЗОРНОЙ КОМПОНЕНТОЙ

В этой главе мы рассмотрим математические методы решения связанной системы уравнений Шредингера для дискретного спектра собственных значений, которые имеют отрицательные величины. Эта задача предназначена для рассмотрения связанных состояния ядерных частиц для потенциалов их взаимодействия с центральными и тензорными силами.

Для решения такой задачи предложена комбинация численных и вариационных методов для нахождения отрицательных собственных значений, т.е. энергий связи, которая позволяет определять их с большой точностью, контролируемой на основе метода невязок.

## 4.1 Общие методы решение уравнения Шредингера

Для расчетов энергии и волновых функций связанных состояний ядерной системы с тензорными потенциалами использовался такой же метод, как в работах [59,60,90]. Исходим из обычных уравнений Шредингера для тензорных потенциалов

$$u''(r) + [\,k^2 - V_c(r) - V_{cul}(r)]u(r) = \sqrt{8}\,V_t(r)w(r)\ ,$$
$$w''(r) + [\,k^2 - V_c(r) - 6/r^2 - V_{cul}(r) + 2\,V_t(r)\,]w(r) = \sqrt{8}\,V_t(r)u(r)\ . \qquad (4.1)$$

Решением этой системы являются четыре волновые функции, получающиеся с начальными условиями типа (3.2), которые образуют линейно независимые комбинации, представляемые в виде (для S и D орбитальных состояний)

$$\chi_0 = C_1u_1 + C_2\,u_2 = \exp(-kr)\ , \qquad (4.2)$$
$$\chi_2 = C_1w_1 + C_2w_2 = [1 + 3/kr + 3/(kr)^2]\exp(-kr)\ ,$$

или с учетом кулоновских сил

$$\chi_0 = C_1u_1 + C_2\,u_2 = W_{\eta,0}(2kr)\ , \qquad (4.3)$$
$$\chi_2 = C_1w_1 + C_2w_2 = W_{\eta,2}(2kr)\ ,$$

где k - волновое число, определяемое энергией связи ядра в рассматриваемом канале, $\eta$ - кулоновский параметр и - функция Уиттекера.

Волновые функции связанных состояний нормированы на единицу следующим образом

$$\int [\chi_0^2+\chi_2^2]dr=1 \quad ,$$

а интеграл от квадрата волновой функции D состояния определяет ее вес, обычно выражаемый в процентах.

Орбитальные состояния системы при наличии тензорных потенциалов смешиваются, так что сохраняется только полный момент, который определяется векторной суммой орбитального и спинового моментов [57]

$$\mathbf{J} = \mathbf{S} + \mathbf{L}$$

Откуда для орбитального момента можно получить выражение

$$/J - S/ \le L \le /J + S/$$

В частности для дейтрона, полный момент равен единице, спин также единица, а орбитальный момент может принимать значения 0 и 2.

## 4.2 Физические результаты для связанных состояний

Изложенный метод можно использовать для рассмотрения кластерной системы $^4He^2H$, когда в потенциале взаимодействия присутствует тензорная компонента, например, гауссова вида [129]

$$V(r) = V_c(r) + V_t(r)\,S_{12} \; , \qquad S_{12} = [6(Sn)^2 - 2S^2] \; ,$$
$$V_c(r) = -V_0\exp(-\alpha r^2) \; , \qquad V_t(r) = -V_1\exp(-\beta r^2) \; .$$

Здесь S - полный спин системы, n - единичный вектор, совпадающий по направлению с вектором межкластерного расстояния, $S_{12}$ - тензорный оператор.

Под тензорным потенциалом, в рассматриваемой системе, следует понимать взаимодействие, оператор которого зависит от взаимной ориентации полного спина системы и межкластерного расстояния. Математическая форма записи такого оператора полностью совпадает с оператором двухнуклонной задачи, поэтому и потенциал, по аналогии, будем называть тензорным [22,130,131].

В проведенных расчетах, квадрупольный момент $^6$Li в $^4$He$^2$H кластерной модели, вычислялся с учетом момента дейтрона $Q_d$ следующим образом [13-19,22]

$$Q = Q_d + Q_0 \quad ,$$

$$Q_0 = \sqrt{\frac{16\pi}{5}} C_{\alpha d} \sum_{LL'} I_{LL'} R_{LL'} \quad , \tag{4.4}$$

где

$$C_{\alpha d} = \frac{Z_\alpha M_d^2 + Z_d M_\alpha^2}{M^2} \quad ,$$

$$R_{LL'} = \langle \chi_L | r^2 | \chi_{L'} \rangle \quad , \quad \Phi_L = \chi_L / r \quad ,$$

$$I_{LL'} = (-1)^{J+L'+S} \sqrt{\frac{5(2L+1)(2J+1)}{4\pi}} (L020|L'0)(JJ20|JJ) \begin{Bmatrix} L & S & J \\ J & 2 & L' \end{Bmatrix} \ .$$

Здесь $\chi_L$ - радиальные ВФ СС, L и L’ - могут принимать значения 0 и 2, Z и M - заряды и массы кластеров и ядра.

В конечном итоге для величины $Q_0$ получаем [13-19,22]

$$Q_0 = \frac{4\sqrt{2}}{15} \int r^2 (\chi_0 \chi_2 - \frac{1}{\sqrt{8}} \chi_2^2) dr \ . \tag{4.5}$$

Импульсное распределение кластеров, нормированное на единицу при переданном импульсе q = 0, определялось в виде [13-19,22]

$$P^2 = \sum_L P_L^2(q) \quad , \qquad P_L(q) = \int \chi_L j_L(qr) dr \ . \tag{4.6}$$

Здесь q - переданный импульс, $j_L$ - сферическая функция Бесселя, L = 0,2.

Магнитный момент ядра в двухкластерной системе в случае, когда только один из кластеров имеет магнитный момент $\mu_d$ и спин 1 может быть представлен [13-19,22]

$$\mu = \mu_d J + \frac{1}{2(j+1)} (B_{\alpha d} - \mu_d)(\hat{J} + \hat{L} - \hat{S}) P_D \quad , \qquad \hat{A} = A(A+1) \quad ,$$

где $P_D$ - величина примеси D состояния, L=2 и

$$B_{\alpha d} = \frac{1}{M}\left(\frac{Z_\alpha M_d}{M_\alpha} + \frac{Z_d M_\alpha}{M_d}\right) .$$

Магнитный момент дейтрона равен $0.857\mu_0$, а ядра $^6$Li несколько меньше $0.822\ \mu_0$. Поэтому для получения, в рассматриваемой модели, правильного момента ядра необходимо допустить примерно 6.5% примеси D состояния.

При расчетах кулоновских формфакторов использовалось выражение [13-19]

$$F^2 = \frac{1}{Z^2}\sum_J V_J^2 \quad , \qquad V_J = Z_1 F_1 I_{2,J} + Z_2 F_2 I_{1,J} \quad ,$$

где интегралы от радиальных функций СС представляются в виде

$$I_{k,0} = \int(\chi_0^2 + \chi_2^2) j_0(g_k r)dr \ , \ I_{k,2} = 2\int \chi_2(\chi_0 - \frac{1}{\sqrt{8}}\chi_2) j_2(g_k r)dr \ . \qquad (4.7)$$

Здесь k = 1 или 2 и обозначает $^2$H или $^4$He, $g_k=(M_k/M)q$ , J - мультипольность формфактора, равная 0 или 2, $j_J$ - сферическая функция Бесселя, q - переданный импульс.

Для вычисления асимптотических констант $C_L^0$ , $C_L^{W0}$ и $C_L^W$ использовались известные выражения [107,114]

$$\Phi_L = \frac{\sqrt{2k}}{r} C_L^0 A_L \exp(-kr) \ , \ A_0 = 1 \ , \ A_2 = [1 + 3/kr + 3/(kr)^2] \ ,$$

$$\Phi_L = \frac{\sqrt{2k}}{r} C_L^{W0} A_L W^0_{-\gamma,l+1/2}(2kr) \ ,$$

$$\Phi_L = \frac{\sqrt{2k}}{r} C_L^W W_{-\gamma,l+1/2}(2kr) \ , \qquad (4.8)$$

где k - волновое число, определяемое энергией связи ядра в рассматриваемом канале, $\eta$ - кулоновский параметр, $W_{-\gamma,l+1/2}(2kr)$ - функция Уиттекера и

$$W^{0}_{-\gamma,l+1/2}(2kr) = (2kr)^{-\gamma}\exp(-kr) \qquad (4.9)$$

- ее асимптотика. Для нуклон - нуклонной системы рассматривается и отношение

$$\eta = C^{W}{}_{2}/C^{W}{}_{0} \ .$$

Радиус ядра вычислялся двумя способами. В кластерной модели его можно представить в виде [13-19]

$$R_r^2 = \frac{M_d}{M}R_d^2 + \frac{M_\alpha}{M}R_\alpha^2 + \frac{M_d M_\alpha}{M^2}R_{\alpha d}^2 \ ,$$

$$R_{\alpha d}^2 = \int r^2(\chi_0^2 + \chi_2^2)dr \ . \qquad (4.10)$$

А из кулоновского формфактора он определяется следующим образом

$$R_f^2 = 6\lim_{q\to 0}\left(\frac{1 - F_{C0}(q)}{q^2}\right) \ .$$

Для радиусов кластеров использовались значения $R_\alpha$=1.67 Фм, $R_d$=1.96 Фм.

В случае нуклон - нуклонной задачи с тензорными силами несколько меняются формулы для формфакторов, которые принимают следующий вид [132, 133, 134, 135, 136, 137, 138, 139, 140, 141, 142, 143, 144, 145, 146]

$$\frac{d\sigma}{d\Omega} = \left(\frac{d\sigma_M}{d\Omega}\right)\left[A + B\,tg^2\left(\frac{\theta}{2}\right)\right] \ ,$$

$$A = G_0^2 + G_2^2 + \frac{2}{3}\eta(1+\eta)G_M^2 \ ,$$

$$B = \frac{4}{3}\eta(1+\eta)^2 G_M^2 \ ,$$

$$G_0 = 2G_E C_E \ ,$$

$$G_2 = 2G_E C_Q \ ,$$

$$G_M = \frac{M_d}{M_P}(2G_{M0}C_S + G_E C_L) \ ,$$

$$\eta = \frac{(\hbar c q)^2}{4M_d^2} = 0.002767 q^2 ,$$

$$2G_E = G_{Ep} + G_{En}, \qquad 2G_{M0} = G_{Mp} + G_{Mn} ,$$

$$C_E = \int (u^2 + w^2) j_0(x) dr ,$$

$$C_Q = 2\int w \left( u - \frac{w}{\sqrt{8}} \right) j_2(x) dr ,$$

$$C_L = \frac{3}{2} \int w^2 (j_0(x) + j_2(x)) dr ,$$

$$x = \frac{qr}{2} ,$$

$$C_S = \int \left( u^2 - \frac{w^2}{2} \right) j_0(x) dr + \frac{1}{\sqrt{2}} \int w \left( u + \frac{w}{\sqrt{2}} \right) j_2(x) dr .$$

Здесь u(r) и w(r) - волновые функции связанного состояния для орбитальных моментов 0 и 2, а $j_i$ - функции Бесселя i - го порядка.

Для масс нуклонов использовались значения $M_p$=938.28 МэВ и $M_n$=939.57 МэВ [147], масса дейтрона принималась равной 1875.63 МэВ. Зарядовый формфактор нейтрона считался равным нулю, а в качестве зарядового формфактора протона использовалась параметризация [147]

$$G_{Ep} = \frac{1}{(1 + 0.054844 q^2)^2} .$$

Здесь переданный импульс q измеряется в $Фм^{-1}$. Магнитные формфакторы нуклонов находились на основе “масштабного закона” [147]

$$G_{Mp} = \mu_p G_{Ep}, \qquad G_{Mn} = \mu_n G_{Ep} ,$$

а в качестве магнитных моментов нуклонов использовались следующие величины

$$\mu_p = 2.7928\, \mu_0, \qquad \mu_n = -\,1.9131\, \mu_0 .$$

## 4.3 Численные методы решения системы уравнений Шредингера

Для нахождения энергий и ВФ связанных состояний системы с тензорным потенциалом использовалась комбинация численных и вариационных методов.

При некоторой заданной энергии связанного состояния (которая не является собственным значением задачи) численным методом находилась ВФ системы (4.37). Для этого использовался обычный метод Рунге - Кутта, описанный выше. Затем система уравнений представлялась в конечно - разностном виде [59]

$$u_{i+1} - 2u_i + u_{i-1} + h^2[\, k^2 - V_c - V_{cul}]u_i = h^2 \sqrt{8}\, V_t\, w_i \;,$$
$$w_{i+1} - 2w_i + w_{i-1} + h^2[\, k^2 - V_c - 6/r^2 - V_{cul} + 2\, V_t\,]w_i = h^2 \sqrt{8}\, V_t u_i \qquad (4.11)$$

или

$$u_{i+1} + h^2[\, - 2/h^2 + k^2 - V_c - V_{cul}]u_i + u_{i-1} - h^2 \sqrt{8}\, V_t\, w_i = 0$$
$$w_{i+1} + h^2[\, - 2/h^2 + k^2 - V_c - 6/r^2 - V_{cul} + 2V_t]w_i + w_{i-1} - h^2 \sqrt{8}\, V_t u_i = 0$$

и полученная численная ВФ подставлялась в эту систему уравнений. Левая часть этих уравнений будет равна нулю только в случае, когда энергия и ВФ являются собственными решениями такой задачи. При произвольной энергии и найденной по ней ВФ левая часть будет отлична от нуля, и можно говорить о методе невязок [148], который позволяет оценить степень точности нахождения собственных функций и собственных значений.

Из уравнений

$$N_{si} = u_{i+1} + h^2[\, - 2/h^2 + k^2 - V_c - V_{cul}]u_i + u_{i-1} - h^2 \sqrt{8}\, V_t\, w_i \;,$$
$$N_{ti} = w_{i+1} + h^2[\, - 2/h^2 + k^2 - V_c - 6/r^2 - V_{cul} + 2V_t]w_i + w_{i-1} - h^2 \sqrt{8}\, V_t u_i \qquad (4.12)$$

вычислялась сумма невязок в каждой точке численной схемы

$$N_s = \sum_i N_{si} \;, \qquad\qquad N_t = \sum_i N_{ti} \;.$$

Варьируя энергию связи (или $k^2$), проводилась минимизация значений всех невязок

$$\delta[/N_s(k^2)/ + /N_t(k^2)/] = 0 \;. \qquad (4.13)$$

Энергия, дающая минимум невязок, считалась собственной энергией $k_0^2$, а функции u и w, приводящие к этому минимуму - собственными функциями задачи, т.е. ВФ связанного состояния системы.

Этот метод прекрасно показал свою работоспособность, как для контрольных задач, в качестве которых выбиралась нуклон – нуклонная система с классическим потенциалом Рейда, так для реальных расчетов физических характеристик связанных состояний кластеров в атомных ядрах [2].

## 4.4 Численная программа решения уравнения Шредингера

Программа для вычисления ядерных характеристик дейтрона и связанных состояний в $^4He^2H$ системе, приведенная ниже, написана на алгоритмическом языке “Basic” и использовалась для расчетов в среде компилятора “Turbo Basic” фирмы “Borland International Inc.” [149].

*Описание параметров программы*

НАЧАЛЬНЫЕ УСЛОВИЯ - задание входных начальных условий необходимых для решения системы уравнений и физических параметров:

AM1=2 - масса первой частицы,

AM2=4 - масса второй частицы,

Z1=1 - заряд первой частицы,

Z2=2 - заряд второй частицы,

AM=AM1+AM2 - сумма масс,

PM=AM1*AM2/AM - приведенная масса $\mu$,

A1=41.4686/(2*PM) - константа $\hbar^2/\mu$,

AKK=1.439975*Z1*Z2 - константа для кулоновского потенциала,

GK=0.0344476*Z1*Z2*PM - кулоновский параметр,

AKK=AKK/A1 - константа $\hbar^2/2m_N$ , определяемая через массу нуклона,

PI=3.14159265 - число $\pi$,

A5=SQR(8) - константа $\sqrt{8}$ ,

VC0= - 72.266 - глубина центральной части потенциала в МэВ,

RNC=0.2 - параметр ширины центральной части потенциала в $Фм^{-2}$ (Ферми),

VT0= - 27 - глубина тензорной части потенциала в МэВ,

RNT=1.12 - параметр ширины тензорной части потенциала в $Фм^{-2}$,

EP5=1D - 7 - абсолютная точность вычисления энергии,

PH5= - 1D - 5 - шаг по энергии, с которым ведется поиск энер-гии связи,

AL0= - 1.4735 - начальное значение энергии в МэВ с которого начинаются вычисления,

MIN=1E30 - условное число для поиска минимума невязок,

MINE=MIN - условное число для поиска энергии,

N=1000 - начальное число шагов для интегрирования системы уравнений,

H0=0.02 - начальный шаг интегрирования системы в Фм.

*Описание блоков основной программы*

РЕШЕНИЕ СИСТЕМЫ - блок решения исходных уравнений. Происходит обращение к подпрограмме RRUN, которая решает систему уравнений Шредингера с тензорными потенциалами методом Рунге - Кутта, как описано выше в тексте второй главы.

НОРМИРОВКА ФУНКЦИИ - проводится нормировка найденной функции на, приведенные в тексте второй главы, граничные условия.

ОПРЕДЕЛЕНИЕ ВЕСА D ВОЛНЫ - вычисляются веса функций с орбитальным моментом 0 и 2, как описано в тексте второй главы.

ВЫЧИСЛЕНИЕ РАДИУСА - определяется радиус кластерной системы на основе приведенных в тексте второй главы формул.

ВЫЧИСЛЕНИЕ КВАДРУПОЛЬНОГО МОМЕНТА - вычисляется квадрупольный момент системы ядерных кластеров по формулам второй главы.

ВЫЧИСЛЕНИЕ АСИМПТОТИЧЕСКОЙ КОНСТАНТЫ - находятся три асимптотические константы, определяющие поведение волновых функций на больших расстояниях.

ВЫЧИСЛЕНИЕ ИМПУЛЬСНОГО РАСПРЕДЕЛЕНИЯ - вычисляются импульсные распределения кластеров в ядре, как описано в тексте второй главы.

ВЫЧИСЛЕНИЕ ЗАРЯДОВОГО ФОРМФАКТОРА - вычисление зарядового кулоновского упругого формфактора ядра на основе кластерной модели.

ВЫЧИСЛЕНИЕ РАДИУСА И КВАДРУПОЛЬНОГО МОМЕНТА ИЗ ФОРМФАКТОРА - поиск зарядового радиуса и квадрупольного момента из найденных кулоновских формфакторов.

*Описание подпрограмм для основной программы*

ПОДПРОГРАММА ИНТЕГРИРОВАНИЯ СИСТЕМЫ УРАВНЕНИЙ МЕТОДОМ РУНГЕ - КУТТА - подпрограмма интегрирования исходных уравнений методом Рунге - Кутта с автоматическим выбором шага (этот блок приведен в основной программе) по заданной точности для энергии связи.

ПОДПРОГРАММА ВЫЧИСЛЕНИЯ ПОТЕНЦИАЛА В ПЕРВОМ УРАВНЕНИИ - вычисление потенциалов в первом уравнении для использования в подпрограмме RRUN.

ПОДПРОГРАММА ВЫЧИСЛЕНИЯ ПОТЕНЦИАЛА ВО ВТОРОМ УРАВНЕНИИ - вычисление потенциалов во втором уравнении для использования в подпрограмме RRUN.

ПОДПРОГРАММА ВЫЧИСЛЕНИЯ НЕВЯЗОК - подпрограмма вычисления невязок в исходной систем уравнений, как описано в тексте второй главы.

ПОДПРОГРАММА ВЫЧИСЛЕНИЯ ФУНКЦИИ УИТТЕКЕРА - вычисление функций Уиттекера на основе интегрального представления [73].

ПОДПРОГРАММА ИНТЕГРИРОВАНИЯ - вычисление интегралов методом Симпсона.

*Текст компьютерной программы*

Ниже приведена распечатка программы расчета энергий и других ядерных характеристик для связанных состояний в кластер - кластерной и нуклон - нуклонной системе для потенциалов с тензорной компонентой.

```
REM ПРОГРАММА ДЛЯ НАХОЖДЕНИЯ ХАРАКТЕРИСТИК СВЯЗАННОГО ДВУХЧАСТИЧНОГО СОСТОЯНИЯ С ТЕНЗОРНЫМИ СИЛАМИ
DEFDBL A - Z: DEFINT K,J,L,I,N,M: NN=4000
DIM V1(NN), W1(NN), V(NN), W(NN), FE(50), FC(50), FQ(50), FCK(50), QK(50),Q(50)
FW$="C: \ WAVE.DAT": F$="C: \FORM.DAT"
P$="C: \IMPUL.DAT": A$="            E            DELA            DELB            EPS"
REM *********** НАЧАЛЬНЫЕ УСЛОВИЯ ************
AM1=2: AM2=4: Z1=1: Z2=2: AM=AM1+AM2: PM=AM1*AM2/AM
A1=41.4686/(2*PM): AKK=1.439975*Z1*Z2: GK=0.0344476*Z1*Z2*PM
AKK=AKK/A1: PI=3.14159265: A5=SQR(8): VC0= - 71.979: RNC=0.2
VT0= - 27: RNT=1.12: EP5=1D - 7: PH5= - 1D - 5: AL0= - 1.4735
MIN=1E30: MINE=MIN: AL00=AL0*PH5: YSCH=0: KK=2: N=1000
H0=0.02: N0=KK*N: H00=H0/KK: H=H00: N1=N0: HK=H^2
REM ************** РЕШЕНИЕ СИСТЕМЫ ***********
60 AL0=AL0+AL00: SK=AL0/A1: S=SQR(ABS(SK)): SSV=S: SQ=SSV
```

```
PRINT " E= "; AL0;
5 VA1=0: WA1=0: PA1=1E - 01: QA1=0: VA2=0: WA2=0: PA2=0
QA2=1E - 01: KKK=1: FOR J=0 TO N1: IF J>0 GOTO 3: X0=1D – 07:
GOTO 4
3 X0=0
4 X=H*(J)+X0
CALL RRUN(VB1, WB1, VB2, WB2, PB1, QB1, PB2, QB2, VA1,
WA1, VA2, WA2, PA1, QA1, PA2, QA2)
VA1=VB1: WA1=WB1: VA2=VB2: WA2=WB2: PA1=PB1:
QA1=QB1
PA2=PB2: QA2=QB2: IF H0*KKK<>H*J GOTO 777: V(KKK)=VA2
W(KKK)=WA2: V1(KKK)=VA1: W1(KKK)=WA1: KKK=KKK+1
777 NEXT J: H=0.5*H: N1=2*N1: IF N1=<N0 GOTO 5
REM *********** НОРМИРОВКА ФУНКЦИИ *********
HF=H0: NF=N: X=H0*(NF): AA=EXP( - SSV*X)
BB=AA*(1+3/SSV/X+3/SSV^2/X^2)
C2=(BB - AA*W1(NF) / V1(NF))/ (W(NF) - V(NF) * W1(NF) /
V1(NF))
C1=(AA - C2*V(NF))/V1(NF): FOR I=0 TO NF: X=H0*(I)
V(I)=C1*V1(I)+C2*V(I): W(I)=C1*W1(I)+C2*W(I): W1(I)=0: NEXT
I
FOR I=0 TO NF: V1(I)=W(I)^2+V(I)^2: NEXT I
CALL SIMP(V1(),NF,HF,VV): NOR=1/SQR(VV): FOR I=0 TO NF
X=HF*I: V(I)=V(I)*NOR: W(I)=W(I)*NOR
 REM PRINT USING " +#.##^^^^ ";X;V(I);W(I): NEXT I
REM ********** ВЫЧИСЛЕНИЕ НЕВЯЗОК ***********
GOSUB 1000
REM ***********************************************
H=H00: N1=N0: F00=ABS(SS)+ABS(SD): IF F00=<MIN GOTO 61
YSCH=YSCH+1: AL0=AL0 - AL00: AL00= - AL00/2: GOTO 60
61 MIN=F00: AL0M=AL0: IF ABS(AL0 - MINE)<ABS(EP5) GOTO
71
MINE=AL0: GOTO 60
71 PRINT "E = ";: PRINT USING " +#.#####^^^^ ";AL0M;
PRINT USING " +#.###^^^^ ";MIN
REM ****** ОПРЕДЕЛЕНИЕ ВЕСА D ВОЛНЫ ********
NF=N: FOR I=0 TO NF: V1(I)=W(I)^2+V(I)^2: NEXT I
CALL SIMP(V1(),NF,HF,VV): NOR=1/SQR(VV): FOR I=0 TO NF
X=HF*I: V(I)=V(I)*NOR: W(I)=W(I)*NOR: NEXT I: FOR I=0 TO NF
V1(I)=V(I)^2: NEXT I: CALL SIMP(V1(),NF,HF,UU): FOR I=0 TO
NF
V1(I)=W(I)^2: NEXT I: CALL SIMP(V1(),NF,HF,WW)
PRINT "VAWE D = ";WW*100;" VAWE S = ";UU*100
REM ************ ВЫЧИСЛЕНИЕ РАДИУСА *********
```

```
FOR I=0 TO NF: X=HF*I: V1(I)=X^2*(V(I)^2+W(I)^2): NEXT I
CALL SIMP(V1(),NF,HF,RR): PRINT " R = ";
RCH=2/6*(1.96)^2+4/6*(1.67)^2+8*RR/36: RCH=SQR(RCH)
PRINT USING " +#.#####^^^^ ";RCH
REM ******* КВАДРУПОЛЬНЫЙ МОМЕНТ **********
FOR I=0 TO NF: X=HF*I: V1(I)=X^2*(V(I)*W(I) - W(I)^2/SQR(8))
 NEXT I
CALL SIMP(V1(),NF,HF,RR): QQQ=4*SQR(2)*RR/15
PRINT " Q = ";: PRINT USING " +#.#####^^^^ ";QQQ
REM ***** АСИМПТОТИЧЕСКАЯ КОНСТАНТА *****
PRINT " R C0 ETA0 CW ETA - W";: MM=NF/4: MMM=NF/10:
KK=0
FOR IJ=MM TO NF STEP MMM: KK=KK+1: X=HF*IJ
AA=SQR(2*SQ)*EXP( - SQ*X): C0=V(IJ)/AA
 BB=AA*(1+3/X/SQ+3/X^2/SQ^2)
C2=W(IJ)/BB: L=0: CALL WH(X,L,SK,GK,WH0)
AA=SQR(2*SQ)*WH0
CW0=V(IJ)/AA: L=2: CALL WH(X,L,SK,GK,WH2)
 BB=SQR(2*SQ)*WH2
CW2=W(IJ)/BB: PRINT USING " +#.##^^^^ ";X;
PRINT USING " +#.#####^^^^";C0;C2/C0;CW0;CW2/CW0: NEXT IJ
WFSAVE=0: IF WFSAVE=0 GOTO 4443: OPEN "O",1, FW$
FOR I=0 TO N: X=I*H0: PRINT#1, USING " +#.###^^^^ ";
X;V(I);W(I)
NEXT I: CLOSE
4443 REM **** ИМПУЛЬСНОЕ РАСПРЕДЕЛЕНИЯ *****
IMPULS=0: IF IMPULS=0 GOTO 4444
PRINT "          Q                 P1^2/P10^2                P2^2/P20^2
(P1^2+P2^2)/P20^2"
HP=0.1: NP=20: FOR IL=0 TO NP: Q=HP*IL+1D - 05: V1(0)=0
FOR I1=1 TO NF: X=I1*HF: V1(I1)=SIN(Q*X)*V(I1)/Q: NEXT I1
CALL SIMP(V1(),NF,HF,S): FE(IL)=S^2: FOR I1=1 TO NF: X=I1*HF
XX=X*Q: J2=(3/XX^3 - 1/XX)*SIN(XX) - 3/XX^2*COS(XX)
V1(I1)=X*J2*W(I1): NEXT I1
CALL SIMP(V1(),NF,HF,S): FC(IL)=S^2: IF Q>0.01 GOTO 9632
QQ2=FE(IL)+FC(IL)
9632 FQ(IL)=(FE(IL)+FC(IL))/QQ2: FC(IL)=FC(IL)/QQ2
 FE(IL)=FE(IL)/QQ2
PRINT USING " +#.###^^^^ ";Q;FE(IL);FC(IL);FQ(IL): NEXT IL
IMPSAVE=0: IF IMPSAVE=0 GOTO 4444: OPEN "O",1,P$
 FOR I=0 TO NP: Q=I*HP
 PRINT#1, USING " +#.###^^^^ ";Q;FE(I);FC(I);FQ(I): NEXT I
4444 REM *************** ФОРМФАКТОР ***************
FORM=0: IF FORM=0 GOTO 5151
```

```
PRINT "    Q      FC0^2        FC2^2        FC0^2+FC2^2"
HQ=.5: NFOR=50: NQ=.001: FOR II=0 TO NFOR:
QK(II)=HQ*II+NQ
Q(II)=SQR(QK(II)): G1=AM1*Q(II)/AM: G2=AM2*Q(II)/AM
FK1=ABS(EXP(-0.49029*QK(II))+0.01615*QK(II)*EXP(-0.16075*
QK(II)))
FK2=ABS(1-(0.09986*QK(II))^6)*EXP(-0.46376*QK(II))
FOR K=1 TO NF: R=HF*K: XX=R*G1
V1(K)=(V(K)^2+W(K)^2)*SIN(XX)/XX
J2=(3/XX^3 - 1/XX)*SIN(XX) - 3/XX^2*COS(XX)
W1(K)=J2*(W(K)*(V(K) - W(K)/SQR(8))): NEXT K
CALL SIMP(V1(),NF,HF,SIM): F2=Z2*FK2*SIM
CALL SIMP(W1(),NF,HF,SIM): F22=2*Z2*FK2*SIM: V1(0)=0:
W1(0)=0
FOR K=1 TO NF: R=HF*K: XX=R*G2
V1(K)=(V(K)^2+W(K)^2)*SIN(XX)/XX
J2=(3/XX^3 - 1/XX)*SIN(XX) - 3/XX^2*COS(XX)
W1(K)=J2*(W(K)*(V(K) - W(K)/SQR(8))): NEXT K
CALL SIMP(V1(),NF,HF,SIM): F1=Z1*FK1*SIM
 CALL SIMP(W1(),NF,HF,SIM)
F11=2*Z1*FK1*SIM: FC(II)=(F1+F2)/(Z1+Z2): FCK(II)=FC(II)^2
FE(II)=(F11+F22)/(Z1+Z2): FQ(II)=FE(II)^2
REM ****** ВЫЧИСЛЕНИЕ РАДИУСА И КВАДРУПОЛЬНОГО
МОМЕНТА ИЗ ФОРМФАКТОРА ********
IF QK(II) - 0.1>0 GOTO 123: RA=6*(1 - FC(II))/QK(II):
RA=(SQR(RA))
QF=9*SQR(2)*FE(II)/QK(II): PRINT "    RF        QF";
PRINT USING "#.###^^^^";RA;QF
123 PRINT USING " +#.###^^^^"; Q(II); FCK(II); FQ(II);
FCK(II)+FQ(II)
NEXT II: FORSAVE=0: IF FORSAVE=0 GOTO 5151
OPEN "O",1,F$: FOR I=0 TO NFOR
PRINT#1, USING "#.###^^^^ "; Q(I); FCK(I); FQ(I); FCK(I)+FQ(I)
NEXT I: CLOSE
5151 STOP
```

**SUB RRUN(VB1, WB1, VB2, WB2, PB1, QB1, PB2, QB2, VA1, WA1, VA2, WA2, PA1, QA1, PA2, QA2)**

```
REM ПОДПРОГРАММА ИНТЕГРИРОВАНИЯ СИСТЕМЫ
УРАВНЕНИЙ МЕТОДОМ РУНГЕ - КУТТА
SHARED H,X
X0=X: CALL F(X0,VA1,WA1,FK1): CALL F(X0,VA2,WA2,SK1)
CALL GG(X0,VA1,WA1,FM1): CALL GG(X0,VA2,WA2,SM1)
FK1=FK1*H: SK1=SK1*H: FM1=FM1*H: SM1=SM1*H:
X0=X0+H/2
```

```
V1=VA1+PA1*H/2: W1=WA1+QA1*H/2: V2=VA2+PA2*H/2
W2=WA2+QA2*H/2: CALL F(X0,V1,W1,FK2): CALL
F(X0,V2,W2,SK2)
CALL GG(X0,V1,W1,FM2): CALL GG(X0,V2,W2,SM2)
FK2=FK2*H: SK2=SK2*H: FM2=FM2*H: SM2=SM2*H
V1=VA1+PA1*H/2+FK1*H/4: W1=WA1+QA1*H/2+FM1*H/4
V2=VA2+PA2*H/2+SK1*H/4: W2=WA2+QA2*H/2+SM1*H/4
CALL F(X0,V1,W1,FK3): CALL F(X0,V2,W2,SK3)
CALL GG(X0,V1,W1,FM3): CALL GG(X0,V2,W2,SM3)
FK3=FK3*H: SK3=SK3*H: FM3=FM3*H: SM3=SM3*H:
X0=X0+H/2
V1=VA1+PA1*H+FK2*H/2: W1=WA1+QA1*H+FM2*H/2
V2=VA2+PA2*H+SK2*H/2: W2=WA2+QA2*H+SM2*H/2
CALL F(X0,V1,W1,FK4): CALL F(X0,V2,W2,SK4)
CALL GG(X0,V1,W1,FM4): CALL GG(X0,V2,W2,SM4)
FK4=FK4*H: SK4=SK4*H: FM4=FM4*H: SM4=SM4*H
VB1=VA1+PA1*H+(FK1+FK2+FK3)*H/6
VB2=VA2+PA2*H+(SK1+SK2+SK3)*H/6
PB1=PA1+(FK1+2*FK2+2*FK3+FK4)/6
PB2=PA2+(SK1+2*SK2+2*SK3+SK4)/6
WB1=WA1+QA1*H+(FM1+FM2+FM3)*H/6
WB2=WA2+QA2*H+(SM1+SM2+SM3)*H/6
QB1=QA1+(FM1+2*FM2+2*FM3+FM4)/6
QB2=QA2+(SM1+2*SM2+2*SM3+SM4)/6: END SUB
SUB F(X,Y,Z,F)
REM ПОДПРОГРАММА ВЫЧИСЛЕНИЯ ПОТЕНЦИАЛА В ПЕР-
ВОМ УРАВНЕНИИ
SHARED SK,A1,A5,VC0,RNC,VT0,RNT,AKK
X2=X^2: VC=VC0*EXP( - RNC*X2): VT=VT0*EXP( - RNT*X2)
UC=VC/A1: UT=VT/A1: F=UT*A5*Z - (SK - AKK/X - UC)*Y: END
SUB
SUB GG(X,Y,Z,GG)
REM ПОДПРОГРАММА ВЫЧИСЛЕНИЯ ПОТЕНЦИАЛА ВО
ВТОРОМ УРАВНЕНИИ
SHARED SK,A1,A5,VC0,RNC,VT0,RNT,AKK
X2=X^2: VC=VC0*EXP( - RNC*X2): VT=VT0*EXP( - RNT*X2)
UC=VC/A1: UT=VT/A1
GG=UT*A5*Y - (SK - 6/X^2 - AKK/X - UC+2*UT)*Z:END SUB
1000 REM
REM ПОДПРОГРАММА ВЫЧИСЛЕНИЯ НЕВЯЗОК
HFK=HF^2: SS=0: SD=0: FOR KK=1 TO NF: X=HF*KK: X2=X^2
VCCC=VC0*EXP( - RNC*X2): VVTT=VT0*EXP( - RNT*X2)
VVCC=VVCC/A1: VVTT=VVTT/A1: A=SK - VVCC - AKK/X
C=A - 6/X^2+2*VVTT: B=SQR(8)*VVTT
```

```
SS=SS+ABS(V(KK+1)-B*HFK*W(KK)+(HFK*A-2)* V(KK)+V(KK -
1))
SD=SD+ABS(W(KK+1)-B*HFK*V(KK)+(HFK*C-2)*
W(KK)+W(KK- 1))
NEXT KK: RETURN
```
**SUB WH(X,L,SK,GK,WH)**
```
REM ПОДПРОГРАММА ВЫЧИСЛЕНИЯ ФУНКЦИИ УИТТЕКЕ-
РА
DIM V(1000)
SS=SQR(ABS(SK)): AA=GK/SS: BB=L: H=0.02: N=1000:
ZZ=1+AA+BB
AAA=1/ZZ: NNN=30000: FOR I=1 TO NNN: AAA=AAA*I/(ZZ+I)
NEXT I: GAM=AAA*NNN^ZZ: CC=2*X*SS: FOR I=0 TO N
TT=H*I: V(I)=TT^(AA+BB)*(1+TT/CC)^(BB - AA)*EXP( - TT)
NEXT I: CALL SIMP(V(),N,H,SIM)
WH=SIM*EXP( - CC/2)/(CC^AA*GAM):END SUB
```
**SUB SIMP(V(),N,H,SIM)**
```
REM ***** ИНТЕГРИРОВАНИЕ ПО СИМПСОНУ *****
A=0:  B=0: FOR I=2 TO N-2 STEP 2: A=A+V(I): NEXT I
FOR I=1 TO N-1 STEP 2: B=B+V(I): NEXT I
SIM=H*(V(0)+2*A2+4*B2+V(N))/3: END SUB
```

Программа тестировалась на нуклон - нуклонном потенциале Рейда [90] и сравнение результатов, полученных в его работе другими методами, с найденными по разработанной нами программе, приведены в таблице 4.1.

Таблица 4.1 - Сравнение характеристик дейтрона и np рассеяния для нуклон - нуклонного взаимодействия Рейда.

| Характеристики дейтрона | Расчет Рейда (RSCA) [90] | Наш расчет для потенциала Рейда |
|---|---|---|
| $E_d$, МэВ | 2.22464 | 2.22458 |
| $Q_d$, Фм$^2$ | 0.2762 | 0.2757 |
| $P_d$ , % | 6.217 | 6.217 |
| $A_S$ | 0.87758 | 0.875(2) |
| $\eta=A_D/A_S$ | 0.02596 | 0.0260(2) |
| $a_t$, Фм | 5.390 | 5.390 |
| $r_t$, Фм | 1.720 | 1.723 |
| $a_s$, Фм | -17.1 | -17.12 |
| $r_s$, Фм | 2.80 | 2.810 |
| $R_d$, Фм | 1.956 | 1.951 |

Из этих результатов видно, что совпадение наших и предыдущих расчетов по энергии связанного состояния дейтрона имеет величину порядка нескольких тысячных процента. Совпадение результатов по другим характеристикам находится примерно на таком же уровне.

# 5. МЕТОДЫ ВАРИАЦИОННОЙ ТРЕХТЕЛЬНОЙ МОДЕЛИ

В этой главе рассмотрены математические методы решения трехчастичной вариационной задачи на связанные состояния с разложением волновой функции по не ортогональному гауссовому базису. Приведена математическая модель решения такой задачи с использованием не стандартного метода, который приводит к исключительно устойчивому алгоритму решения обобщенной матричной задачи на собственные значения.

Для примера рассмотрена трехтельная модель ядра $^7$Li и ее возможности по описанию некоторых характеристик связанного состояния $^4$He$^2$Hn кластеров.

## 5.1 Общие методы трехтельной модели

В работах [23-25] были подробно рассмотрены возможности трехтельной модели ядра $^6$Li и показана ее способность правильно описывать почти все наблюдаемые характеристики, включая электромагнитные формфакторы, если выполнить антисимметризацию волновой функции [150,151,152,153,154]. Исключение составляет только квадрупольный момент ядра, который во всех расчетах получается положительным, в то время как экспериментальные измерения дают отрицательную величину. Учет антисимметризации волновой функции позволил существенно улучшить качество описания поперечных формфакторов [150], но мало изменил другие характеристики ядра.

Этот результат может объяснить определенные успехи простых двухкластерных моделей легких ядер с запрещенными состояниями, в частности $^2$H$^4$He и $^3$H$^4$He моделей ядер $^6$Li и $^7$Li, в которых получается хорошее описание многих экспериментальных характеристик, но плохо воспроизводятся поперечные формфакторы при больших переданных импульсах [155,156].

Антисимметризация волновой функции, выполненная в [150], затрагивает в основном внутреннюю область ядра и заметно изменяет волновую функцию на малых расстояниях, которые определяют поведение высокоимпульсной компоненты формфакторов. Область больших расстояний при этом меняется мало, что не приводит к существенным изменениям других расчетных характеристик, зависящих в основном от поведения волновой функции периферийной области ядра. Поэтому, вполне можно предположить, что проведение антисимметризации волновой функции в двухкластерных моделях $^7$Li или $^6$Li с тензорными силами, которые позволяют передать квадрупольный момент ядра $^6$Li [157,158] может заметно улучшить описание поперечных формфакторов при больших пере-данных импульсах.

Ядро $^{7}$Li, не смотря на вполне успешное описание многих его характеристик на основе простой двухкластерной системы [17-19,38,11,29-36,159], можно рассматривать в трехтельной $n^{2}H^{4}He$ модели, которая имеет больше возможностей и, в частности, позволяет выделять различные двухчастичные каналы.

Рассмотрим такую модель более подробно и будем считать, что в основании треугольника из трех частиц находятся $^{2}$Hn кластеры (частицы 23) с радиус - вектором относительного расстояния $r = r_{23}$ и орбитальным моментом относительного движения $\lambda$, которые находятся в дублетном спиновом состоянии. Ядро $^{4}$He (частица 1) находится в вершине треугольника и его положение относительно центра масс двухкластерной системы определяется вектором $R = R_{(23),1}$ и моментом l. Полный орбитальный момент системы равен $\mathbf{L} = \mathbf{l} + \boldsymbol{\lambda}$, а полный спин $\mathbf{S} = \mathbf{S}_3 + \mathbf{S}_2$, ($S_1 = 0$) может принимать значения 1/2 и 3/2, т.е. система $n^{2}H$ может находиться в дублетном и квартетном состояниях, первое из которых соответствует основному состоянию ядра трития при $\lambda$=0.

Полный момент основного состояния ядра $^{7}$Li равен $3/2^{-}$ и, поскольку $\mathbf{J} = \mathbf{S} + \mathbf{L}$, может быть получен из комбинации L = 1 и S = 1/2, которая приводи к J = 1/2 и 3/2 с отрицательной четностью. Основному состоянию ядра соответствует момент 3/2, а первому возбужденному состоянию при энергии 0.478 МэВ момент 1/2.

Полный орбитальный момент системы $L = l + \lambda$, равный единице может быть получен из комбинации $l = 1$ и $\lambda = 0$, которая позволяет рассматривать систему $^{2}$Hn, как связанное состояние ядра трития в дублетном спиновом состоянии [160].

В качестве парных межкластерных потенциалов выбирались взаимодействия гауссовой формы с отталкивающим кором, позволяющие правильно передавать соответствующие фазы рассеяния. В паре частиц (13) используется чистый по схемам Юнга $n^{4}He$ потенциал для S - волны ($l_{13} = 0$) с параметрами, описывающими экспериментальную фазу [161], как показано на рисунке 5.1.

В паре (12) использован $P_0$ - потенциал $^{2}H^{4}He$ взаимодействия ($l_{12} = 1$), параметры которого уточнялись по трехтельной энергии, поскольку $P_0$ – фазы, показанные на рисунке 5.2 и полученные в разных работах [91-98] имеют большую неоднозначность.

В паре частиц (23) взято чистое по орбитальным симметриям $n^{2}H$ дублетное S - взаимодействие ($l_{23} = \lambda = 0$) с отталкиванием, параметры которого фиксированы по характеристикам связанного состояния ядра трития, а фазы показаны на рисунке 5.3 в сравнении с извлеченными из эксперимента чистыми фазами [98].

В каждой паре частиц использован только один потенциал для определенной парциальной волны и спинового состояния. Это представляется вполне оправданным, если потенциалы для остальных парциальных волн (в каждой паре) вносят меньший вклад и приводят только к небольшим поправкам к расчетным характеристикам ядра.

В отличие от трехчастичной модели ядра $^6$Li [23], где потенциалы в NN и N$^4$He системах хорошо определены по экспериментальным фазам рассеяния, имеющим сравнительно малые ошибки, параметры $^2$H$^4$He взаимодействия имеют заметную неопределенность из - за различия результатов разных фазовых анализов.

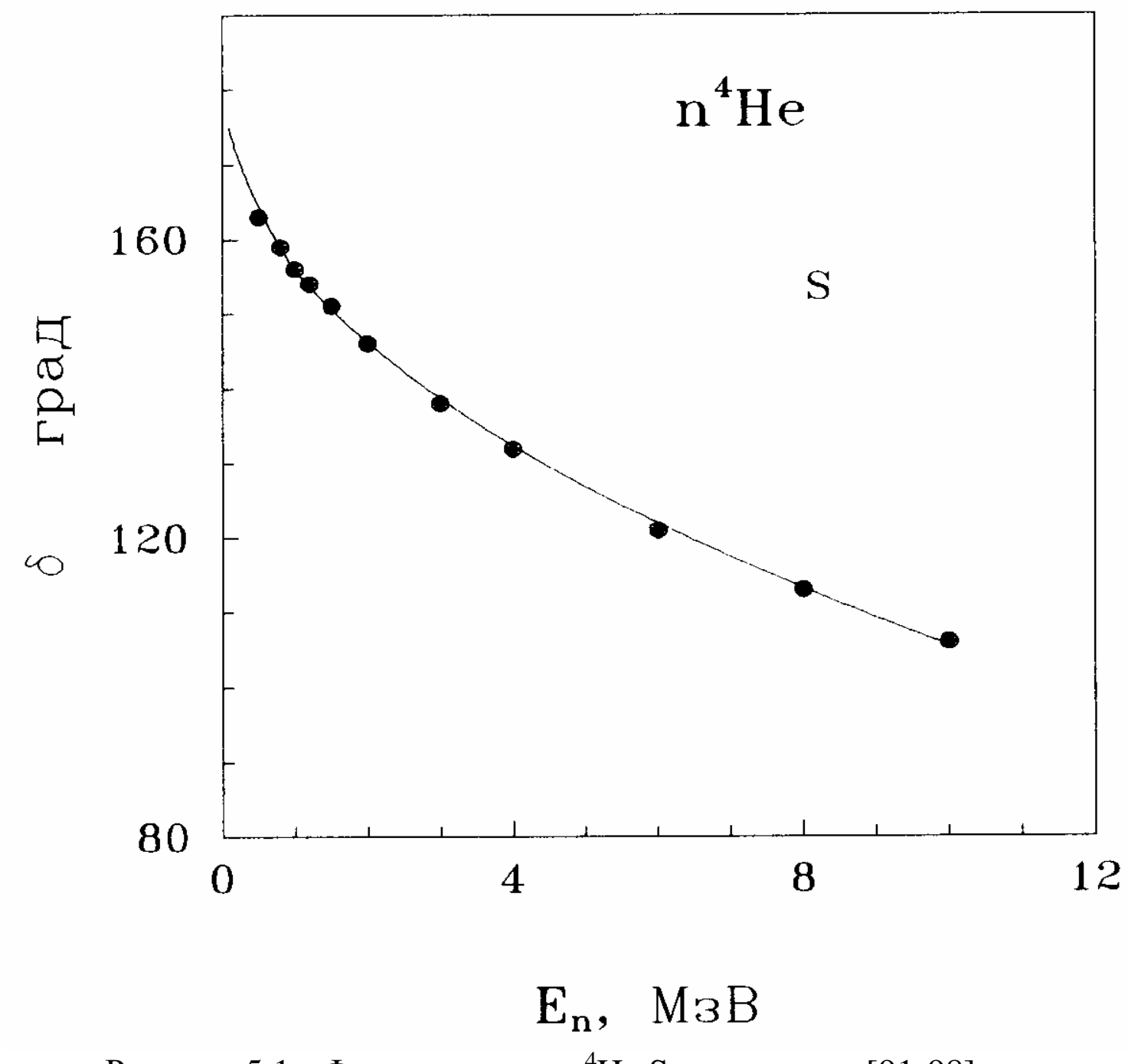

Рисунок 5.1 - Фазы упругого n$^4$He S - рассеяния [91-98] при низких энергиях для потенциала из таблицы 5.1.

Поэтому представляется интересным выяснить - можно ли в трехтельной $^4$He$^2$Hn модели согласовать, в пределах имеющихся экспериментальных неоднозначностей по фазам $^2$H$^4$He рассеяния, параметры этих двухкластерных потенциалов с энергией связи ядра $^7$Li в трехтельном канале.

Парные межкластерные потенциалы взаимодействия принимались в следующем виде

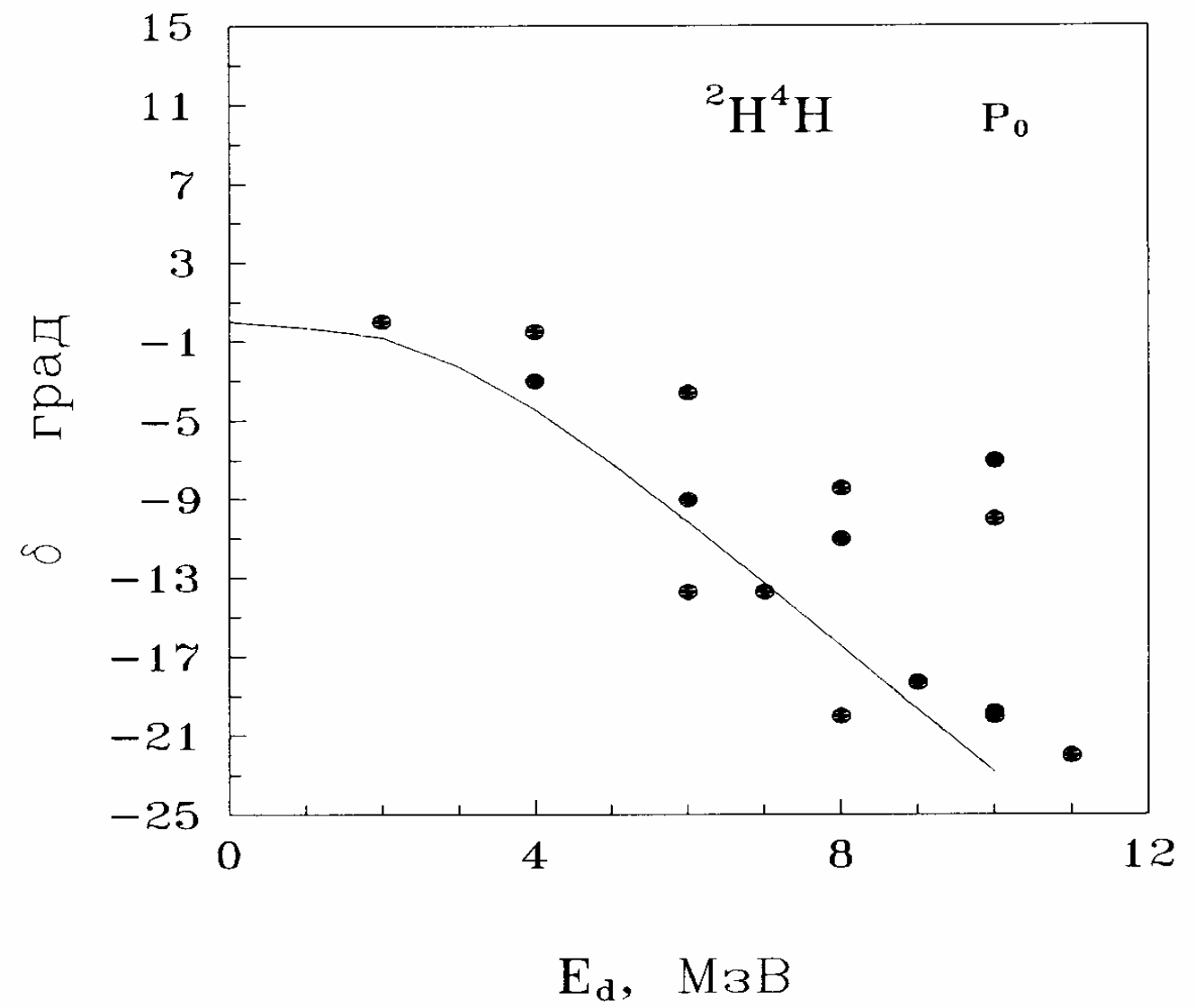

Рисунок 5.2 - Фазы $^4$He$^2$H упругого P - рассеяния [91-98] при низких энергиях для потенциала из таблицы 5.1.

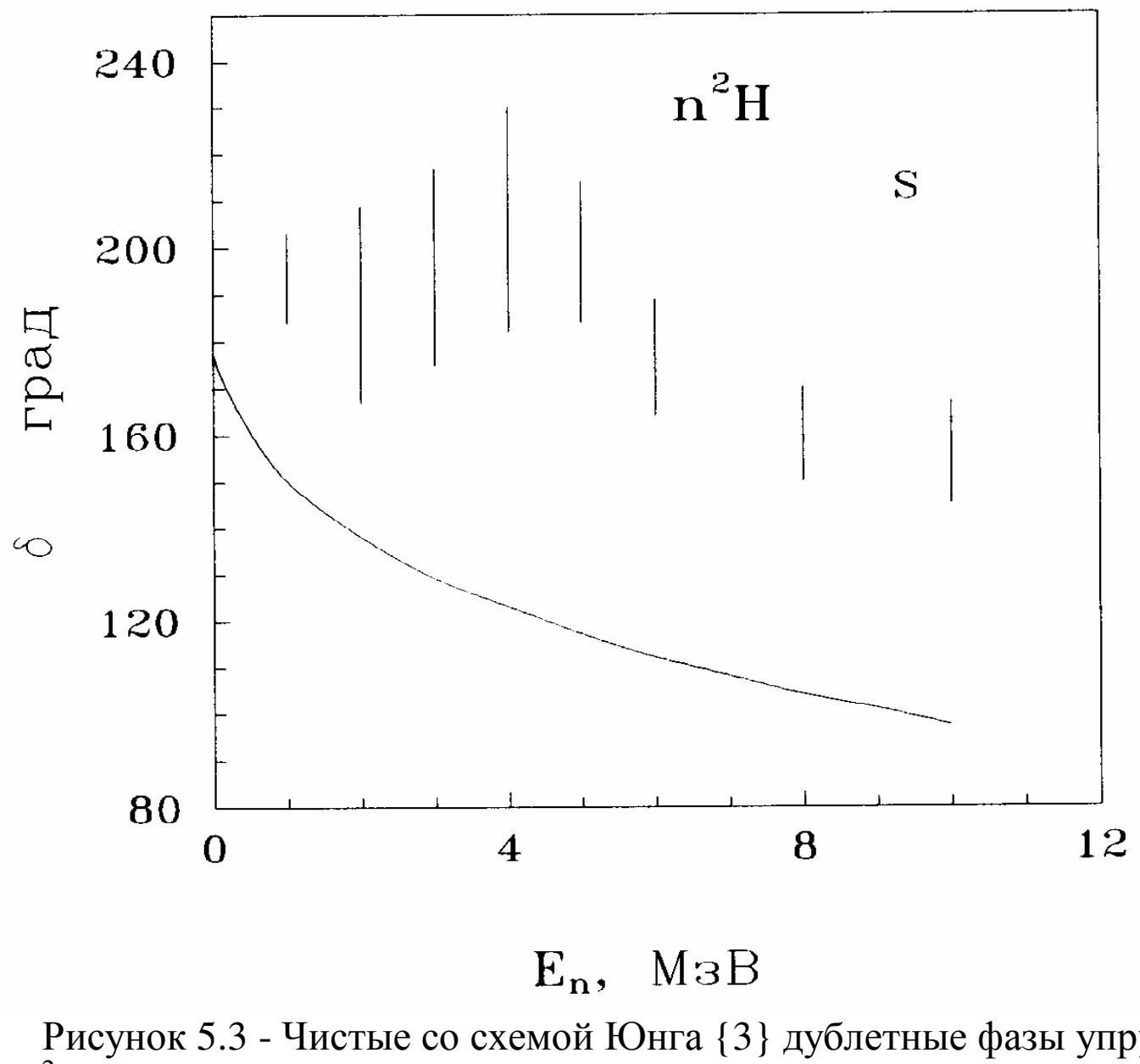

Рисунок 5.3 - Чистые со схемой Юнга {3} дублетные фазы упругого n$^2$H S - рассеяния [98] при низких энергиях для потенциала из таблицы 5.1.

$$V(r)=V_1 \exp(-\gamma r^2) + V_2 \exp(-\delta r^2) , \qquad (5.1)$$

а их параметры даны в таблице 5.1.

Таблица 5.1 - Параметры парных межкластерных потенциалов.

| Система | L | $V_1$, МэВ | $\gamma$, Фм$^{-2}$ | $V_2$, МэВ | $\delta$, Фм$^{-2}$ |
|---|---|---|---|---|---|
| $^2H^4He$ | 1 | -10.0 | 0.1 | 72.0 | 0.2 |
| $n^4He$ | 0 | -115.5 | 0.16 | 500 | 1.0 |
| $n^2H$ | 0 | -78.78 | 0.3 | 200 | 2 |

Потенциал основного состояния $n^2H$ системы дает энергию связи -6.25 МэВ, асимптотическую константу $C_0$=2.0(1) в хорошем соответствии с экспериментом [162, 163, 164, 165, 166, 167, 168, 169, 170, 171, 172, 173, 174, 175] и среднеквадратичный радиус 2.12 Фм, который несколько больше известной величины 1.70(3) Фм [176, 177, 178, 179, 180, 181, 182, 183, 184, 185, 186]. Здесь не удается полностью согласовать $n^2H$ потенциал со всеми наблюдаемыми [98] - расчетная фаза лежит заметно ниже извлеченных из эксперимента чистых по схеме Юнга {3} фаз, а радиус ядра завышен.

Последнее вполне объяснимо, поскольку дейтрон имеет радиус больше, чем тритий и не может находиться внутри него без деформаций, т.е. дейтронному кластеру в такой системе нельзя полностью сопоставлять характеристики свободного дейтрона [98]. Для того, чтобы получить правильный радиус трития необходимо деформировать дейтрон, уменьшив его радиус примерно на 27% и принять 1.42 Фм, что приводят к расчетному зарядовому радиусу трития 1.70 Фм в хорошем согласии с экспериментом.

Для нахождения энергии трехкластерной системы использовался неортогональный вариационный метод [23]. Полная трехтельная волновая функция имеет вид

$$\Psi(r,R) = \sum_{l,\lambda} \Phi_{l,\lambda}(r,R) Y_{LS}^{JM}(\hat{r},\hat{R}) \quad ,$$

где угловая часть записывается

$$Y_{LS}^{JM}(\hat{r},\hat{R}) = \sum_{M_S M_L} \langle LM_L SM_S | JM \rangle Y_{LM_L}(\hat{r},\hat{R}) \chi_{SM_S}(\sigma) \; .$$

Здесь L – орбитальный момент, S – спин, J – полный момент системы частиц, M – их проекции, $Y_{LS}^{JM}$ – спин – угловая функция, $\Phi_{l,\lambda}$ - радиальная волновая функция, r и R – скалярные расстояния между частицами, r и R со шляпкой – углы между направлениями векторов **r** и **R** и осью z, $Y_{LM}$ – сферическая функция, $\chi_{SM}$ – спиновая функция системы, зависящая от спина σ, угловые скобки обозначают коэффициенты Клебша - Гордона.

Радиальная волновая функция представляется в форме разложения по гауссойдам так же, как в трехтельной модели ядра $^6$Li [23]

$$\Phi_{l,\lambda}(r,R) = Nr^{\lambda}R^{l}\sum_i C_i \exp(-\alpha_i r^2 - \beta_i R^2) = N\sum_i C_i\Phi_i \cdot \qquad (5.2)$$

Здесь выражение

$$\Phi_i = r^{\lambda}R^{l}\exp(-\alpha_i r^2 - \beta_i R^2)$$

называется базисной функцией. Исходное радиальное уравнения Шредингера системы трех частиц запишем в форме [59,60]

$$( H - E )\, \Phi_{l,\lambda} = 0 \quad , \qquad (5.3)$$

где

$$H = T + V \;, \qquad T = T_1 + T_2 = -\frac{\hbar^2}{2\mu}\Delta_r - \frac{\hbar^2}{2\mu_0}\Delta_R \;,$$

$$V = V_{12} + V_{23} + V_{13} \;,$$

$$\mu = \frac{m_2 m_3}{m_{23}} \;, \quad \mu_0 = \frac{m_1 m_{23}}{m} \;, \quad m_{23} = m_2 + m_3 \;,$$

$$m = m_1 + m_2 + m_3$$

Здесь m и μ – массы и приведенные массы частиц, Δ - оператор Лапласа, $h^2$ – постоянная Планка, T и V – операторы кинетической и потенциальной энергии, H – гамильтониан и E – энергия системы.

Подставляя разложение (2.52) в уравнение (2.53), домножая слева это уравнение на базисную функцию $\Phi_j$ , и интегрируя по всем переменным, приводим (2.53) к матричному виду

$$\sum_i ( H_{ij} - E\, L_{ij})\, C_i = 0$$

или

$$KC = 0 \quad , \qquad (5.4)$$

где матрица K определяется в виде

$$K = H - EL \ .$$

В этих выражениях H - матрица гамильтониана, L - матрица интегралов перекрывания, которая при использовании ортогонального базиса переходит в единичную матрицу I. Отметим, что матрица K не диагональна по энергии и вместо обычной задачи на собственные значения мы имеем обобщенный вариант этой задачи. Поскольку уравнение (2.54) однородное, оно будет иметь не тривиальные решения только тогда, когда детерминант матрица K равен нулю. Условие равенства нулю ее детерминанта позволяет найти все собственные значения E системы (при заранее заданных параметрах $\alpha_i$ и $\beta_i$), а по ним все собственные вектора C, а значит и саму радиальную функцию $\Phi_{l,\lambda}$ в выражении (2.52).

Для решений обобщенной задачи на собственные значения матрица интегралов перекрывания L обычно разлагается на верхнюю V и нижнюю N треугольные матрицы, найти которые можно методом Халецкого. Затем, определяют обратные им матрицы и, с их помощью, новую матрицу гамильтониана, которая приводит к стандартной, диагональной задаче на собственные значения, как было подробно описано в третьей главе. Такая процедура обычно называется ортогонализацией по Шмидту.

Но если в двухтельной задаче с одним вариационным параметром такой метод оказывается сравнительно устойчив, то в трехтельной системе, при некоторых значениях двух вариационных параметров, метод нахождения обратным матриц приводит к существенной неустойчивости и переполнению при работе компьютерной программы.

Поэтому, мы предложили не стандартный метод вычисления детерминанта для обобщенной задачи на собственные значения, который заключается в разложении всей недиагональной по E матрицы K на треугольные, а не только матрицы интегралов перекрывания, как это делается обычно. Теперь, как и прежде, ищется ноль детерминанта нижней треугольной матрицы, который равен произведению ее диагональных элементов, зависящих от E. Предложенный метод позволил получить устойчивый алгоритм решения этой задачи, не приводящий к переполнению при работе компьютерных программ, поскольку уже не нужно определять обратные к V и N матрицы.

При каждом значении вариационных параметров $\alpha_i$ и $\beta_i$ находим некоторую энергию системы (которая дает ноль детерминанта), а затем, варьируя эти параметры, проводим поиск минимума этой энергии. Затем увеличивается размерность базиса N и повторяем все вычисления, до тех пор пока величина собственного значения, т.е. энергии связи $E_N$, на очередном шаге N не станет отличаться от предыдущего значения $E_{N-1}$ на величину $\varepsilon$, которая обычно задается на уровне 0.5-1.0%. В соответствии с теоремой Хилерааса – Ундгейма [60] эта минимальная энергия и будет реальной энергией связи в такой трехчастичной системе, т.е. энергией связи атомного ядра $^7$Li.

Матричные элементы гамильтониана системы и интегралов перекрывания, вычисленные по базисным функциям $\Phi_i$, имеют вид [187, 188]

$$T_{ij} = \frac{\pi}{16} N^2 \frac{(2l+1)!!(2\lambda+1)!!}{2^{l+\lambda}} \frac{\hbar^2}{m_N} \alpha_{ij}^{-\lambda-1/2} \beta_{ij}^{-l-1/2} G_{ij} \quad ,$$

$$G_{ij} = \frac{B_{ij}(\alpha,\lambda)}{\mu\beta_{ij}} + \frac{B_{ij}(\beta,l)}{\mu_0 \alpha_{ij}} \quad ,$$

$$B_{ij}(\delta,\nu) = \frac{\nu^2}{2\nu+1} + \frac{\delta_i \delta_j}{\delta_{ij}^2}(2\nu+3) - \nu \quad ,$$

$$L_{ij} = \frac{\pi}{16} N^2 \frac{(2l+1)!!(2\lambda+1)!!}{2^{l+\lambda}} \alpha_{ij}^{-\lambda-3/2} \beta_{ij}^{-l-3/2} \quad ,$$

$$(V_{23})_{ij} = \frac{\pi}{16} N^2 \frac{(2l+1)!!(2\lambda+1)!!}{2^{l+\lambda}} V_{23}(r)(\alpha_{ij}+\gamma)^{-\lambda-3/2} \beta_{ij}^{-l-3/2} ,$$

$$N = \left( \sum_{ij} C_i C_j L_{ij} \right)^{-1/2} ,$$

$$[(V_{цб})_R]_{ij} = \frac{\pi}{16} N^2 l(l+1) \frac{(2l-1)!!(2\lambda+1)!!}{2^{l+\lambda}} \frac{\hbar^2}{\mu_0} \alpha_{ij}^{-\lambda-3/2} \beta_{ij}^{-l-1/2} \quad ,$$

$$[(V_{цб})_r]_{ij} = \frac{\pi}{16} N^2 \lambda(\lambda+1) \frac{(2l+1)!!(2\lambda-1)!!}{2^{l+\lambda}} \frac{\hbar^2}{\mu} \alpha_{ij}^{-\lambda-1/2} \beta_{ij}^{-l-3/2} \quad ,$$

$$[(V_k)_r]_{ij} = Z_2 Z_3 \frac{\pi}{16} N^2 \frac{2}{\sqrt{\pi}} \frac{(2l+1)!!}{2^l} \frac{\lambda!}{\alpha_{ij}^{\lambda+1} \beta_{ij}^{l+3/2}} \quad ,$$

$$\alpha_{ij} = \alpha_i + \alpha_j \quad , \qquad \beta_{ij} = \beta_i + \beta_j \quad .$$

Далее, например, при значениях l = 1 и $\lambda$ = 0 (набор моментов для $^4$He$^2$Hn системы в ядре $^7$Li) имеем

$$(V_{12})_{ij} = \frac{\pi}{32} N^2 \frac{3}{A^{3/2}(\beta_{ij}+\gamma)} \left[ \frac{a^2\gamma^2}{A} + 1 \right] V_{12}(r_{12}) \quad ,$$

$$A = \alpha_{ij}\beta_{ij} + \gamma(\alpha_{ij} + a^2\beta_{ij}) \quad , \qquad a = m_3/m_{23} \quad .$$

В случае l = 0 и λ = 0 (для ядра $^3H$ или $^3He$) для этой части потенциала будем иметь

$$(V_{12})_{ij} = \frac{\pi}{16} N^2 \left[ \alpha_{ij}\beta_{ij} + \gamma(\alpha_{ij} + a^2\beta_{ij}) \right]^{-3/2} V_{12}(r_{12}) \quad .$$

Здесь величина γ является параметром ширины гауссового потенциала между частицами с номерами 1 и 2.

Среднеквадратичный массовый радиус ядра в такой модели представляется в виде [187, 189]

$$< r^2 >_m = m_1/m < r^2 >_{m1} + m_2/m < r^2 >_{m2} + m_3/m < r^2 >_{m3} + A/m \tag{5.5}$$

$$A = \frac{\pi}{16} N^2 \frac{(2l+1)!!(2\lambda+1)!!}{2^{l+\lambda+1}} \sum_{i,j} C_i C_j \alpha_{ij}^{-\lambda-3/2} \beta_{ij}^{-l-3/2} \left( \frac{2\lambda+3}{\alpha_{ij}} \mu + \frac{2l+3}{\beta_{ij}} \mu_0 \right)$$

где

$$\mu = \frac{m_2 m_3}{m_{23}}; \quad \mu_0 = \frac{m_1 m_{23}}{m}; \quad m_{23} = m_2 + m_3; \quad m = m_1 + m_2 + m_3$$

Квадрупольный момент ядра с учетом момента дейтрона записывается [22,157]

$$Q = Q_d - \frac{2}{5} B \quad , \tag{5.6}$$

$$B = \frac{\pi}{16} N^2 \frac{(2l+1)!!(2\lambda+1)!!}{2^{l+\lambda+1}} \sum_{i,j} C_i C_j \alpha_{ij}^{-\lambda-3/2} \beta_{ij}^{-l-3/2} \left( \frac{2l+3}{\beta_{ij}} C + \frac{2\lambda+3}{\alpha_{ij}} D \right) +$$

$$+ N^2 E \sum_{i,j} C_i C_j \frac{(\lambda+1)!(l+1)!}{2\alpha_{ij}^{\lambda+2} \beta_{ij}^{l+2}} \quad ,$$

$$C = \frac{Z_1 m_{23}^2 + Z_{23} m_1^2}{m^2} \ , \quad D = \frac{Z_2 m_3^2 + Z_3 m_2^2}{m_{23}^2} \ ,$$

$$E = \frac{m_1}{m m_{23}} (Z_3 m_2 - Z_2 m_3) \ ,$$

$$Z_{23} = Z_1 + Z_2 \ .$$

Среднеквадратичный зарядовый радиус имеет вид [23]

$$< r^2 >_z = Z_1 / Z < r^2 >_{z1} + Z_2 / Z < r^2 >_{z2} + Z_3 / Z < r^2 >_{z3} + B / Z \cdot \qquad (5.7)$$

В качестве зарядовых радиусов кластеров принимались величины $<r>_{mn}$ = 0.8 Фм, $<r>_{zn}$ = 0 Фм, $<r>_{md}$ =$<r>_{zd}$= 1.96 Фм, $<r>_{m\alpha}$= $<r>_{z\alpha}$=1.67 Фм, а квадрупольный момент дейтрона 2.86 мб. [162-186]. Экспериментальное значение квадрупольного момента $^7$Li, равное -36.6(3) мб, приведено в работах [190, 191].

Для нахождения волновой функции относительного движения и вероятности двухкластерной $^3$H$^4$He конфигурации использовалась волновая функция ядра $^3$H в n$^2$H модели в виде простого разложения по гауссойдам

$$\varphi(r) = N_0 \sum_j B_j \exp(-\chi_j r^2) \ . \qquad (5.8)$$

Здесь параметры $\chi_j$ и коэффициенты разложения $B_j$ находились на основе n$^2$H потенциала основного состояния, приведенного в таблице.

При использовании волновых функций $^3$H вида (2.58) функция относительного движения двух кластеров в $^3$H$^4$He канале ядра $^7$Li при l=1 и λ=0 может быть представлена в виде [192]

$$X(R) = \int \Phi(r,R) \varphi(r) r^2 dr = \frac{\sqrt{\pi}}{4} R \sum_i C_i \exp(-\beta_i R^2) \sum_j B_j (\alpha_i + \chi_j)^{-3/2} \qquad (5.9)$$

Интеграл от квадрата модуля этой волновой функции дает вероятность двухкластерной конфигурации в трехтельной модели [192]

$$P = \frac{3\pi^{3/2}}{16 \cdot 8} \sum_{i,k} C_i C_k \beta_{ik}^{-5/2} \sum_{j,n} B_j B_n [(\alpha_i + \chi_j)(\alpha_k + \chi_n)]^{-3/2} \, . \quad (5.10)$$

Получив волновую функцию двухчастичной системы (2.59), можно использовать ее для расчетов среднеквадратичных радиусов $^7$Li в обычной двухкластерной модели, которые записываются [23]

$$< r^2 >_m = m_1 / m < r^2 >_{m1} + m_2 / m < r^2 >_{m2} + \frac{m_1 m_2}{m^2} R_{1,2}^2 \quad ,$$

$$(5.11)$$

$$< r^2 >_z = Z_1 / Z < r^2 >_{z1} + Z_2 / Z < r^2 >_{z2} + \frac{(Z_1 m_2^2 + Z_2 m_1^2)}{Z m^2} R_{1,2}^2 \quad .$$

Здесь первая масса и заряд относятся, например, к $^4$He, вторая к $^3$H, m и Z - полная масса и заряд $^7$Li, а $R^2$ матричный элемент вида

$$R_{1,2}^2 = \langle X(R) | R^2 | X(R) \rangle \quad ,$$

который определяет относительное межкластерное расстояние для выделяемой пары частиц.

При поиске энергии связи ядра в трехтельной модели начальные значения вариационных параметров $\alpha_i$ и $\beta_i$ находились из линейной сетки вида [23]

$$\alpha_i = i/30 \, , \qquad \beta_i = 2\alpha_i \quad .$$

Затем проводилось независимое варьирование каждого из них так, чтобы минимизировать энергию системы с точностью до $10^{-3}$ МэВ, т.е. параметры изменяются до тех пор, пока изменение энергии не станет меньше этой заданной величины.

## 5.2 Вариационная программа трехтельной модели

Приведем теперь текст компьютерной программы на языке Turbo Basic, которая предназначена для расчета энергии трехтельной системы на основе описанных выше методов.

**REM ПРОГРАММА ДЛЯ РАСЧЕТА ЭНЕРГИИ N²H⁴He СИСТЕМЫ**

```
DEFDBL A-Z: DEFINT J,K,L,N,I
DIM XP(30), H(30,30), T(30,30), VN(30,30), VC(30,30), L1(30,30),
```

```
FF(500), FU(500)
 DIM          A(30,30),X(30),Y(30),B(30,30),C(30,30),D(30),Y1(30,30),
SV(30),A1(30)
 DIM  P(30),  Q(30),  AA(30,30),  AAA(30,30),  LA(30),  N(30,30),
NO(30,30), LLL(30,30)
 DIM V(30,30), OV(30,30), E(30,30), E1(30,30), VK(30,30), C1(30,30),
E2(40), L(30,30)
 DIM C0(30),CW(30),PH5(20)
 ALD$="C:\BASICA\TRITON\ALDN1.DAT"
 ALD1$="C:\BASICA\TRITON\TRIT\ALDN.DAT"
 ALD2$="C:\BASICA\SOB\SOB-ND2.DAT"
 Z1=2: Z2=1: Z3=0: M1=4: M2=2: M3=1
 Z=Z1+Z2+Z3: M23=M3+M2: PM23=M3*M2/M23
 AM0=M1+M2+M3: PM0=M1*M23/AM0
 NF=200: NFF=4: NS=1: HFF=0.1: PN=.5: PH=.05: NI=0: NNP=1
 NV=0: NSM=0: EP=1D-04: EPP=1D-04: NITER=2: PH=1: B=1E30
 HC=1: PNC=-10: PVC=10: A11=41.4686:A12=1.439975*Z1*Z2
A13=1.439975*Z1*Z3:A23=1.439975*Z3*Z2:P1=3.14159265:PI=SQR
(P1)
REM ---------------------------------------------------------------------------
REM 1 - AL; 2 - D; 3 - N;    L  -  AL-D - 1, AL-N -0, D-N - 0
REM N-D
V231=-78.78      : REM J= 1  ; L=0
R231=.3
V232=200.      : REM J= 1  ; L=0
 R232=2
 REM AL-N
 V131=-115.5    : REM J=1/2 ; L=0
 R131=.16
 V132=500     : REM J=1/2 ; L=0
 R132=1
 REM AL-D
 V121=-10.    : REM J=1/2 ; L=1
 R121=.1
 V122=72      : REM J=1/2 ; L=1
 R122=.2
 REM - - - - - - - - - - - - - - - - - - - - - - - - - - - - - - - - - - - - - - - - - -
 NP=9: FOR I=1 TO NP: XP(I)=(I^1)/30: XP(I+NP)=XP(I)*2: NEXT I
 REM GOTO 5454: OPEN "I",1,ALD1$: FOR I=1 TO NP
 INPUT1, XP(I),XP(I+NP),SV(I): NEXT I: CLOSE
5454 REM ВАРИАЦИОННАЯ ТОЧНОСТЬ: EP=1D-04
 REM ТОЧНОСТЬ ВЫЧИСЛЕНИЯ ДЕТЕРМИН: EPP=1D-04
 NITER=2: PH=.091234: NP2=2*NP: FOR ITER=1 TO NITER
 N55=0: PH5=PH/ITER
```

```
50 FMIN=B: FOR IJK=1 TO NP2: PH5(IJK)=XP(IJK)*PH5
60 XP(IJK)=XP(IJK)+PH5(IJK): IF XP(IJK)<0 GOTO 619
 PRINT ITER;IJK;: PRINT USING " +#.#####^^^^ ";XP(IJK);
 ZYS=0: GOSUB 5000: F=LA: IF PH=0 GOTO 7654: C=B
 B=F: IF F-C<0 GOTO 60
619 XP(IJK)=XP(IJK)-PH5(IJK): NEXT IJK: N55=N55+1
 PRINT "**********************************************"
 PRINT "                    E = ";: ENER=F
 PRINT USING "+#.####^^^^ ";F: FOR II=1 TO NP
 PRINT USING "+#.####^^^^ ";XP(II);: NEXT II
 PRINT: FOR II=NP+1 TO NP2
 PRINT USING "+#.####^^^^ ";XP(II);: NEXT II
 PRINT: IF  N55=1 GOTO 5522: PH5=-PH5/2: GOTO 6611
5522 PH5=-PH5
6611 IF ABS(FMIN-F)>EP GOTO 50: NEXT ITER
7654 PRINT "******************************************"
  PRINT "                    E= ";:  ENER=F
  PRINT USING "+#.#####^^^^ ";F:  PRINT "                    AL-
FA"
  FOR I=1 TO NP2:  PRINT USING " +#.#####^^^^ ";XP(I);:  NEXT I
899  REM - - - - - - - -  - - - - - - - - - - - - - - - - - - - - - - - - - - - - - - - -
  LLA=F:  ZYS=1:  GOSUB 5000
  REM - - - - - - - -  - - - - - - - - - - - - - - - - - - - - - - - - - - - - - - - -
  CN=-1.5:  BN=-2.5:  S=0:  FOR I=1 TO NP:  FOR J=1 TO NP
  AL=XP(I)+XP(J):                        BT=XP(I+NP)+XP(J+NP):
L(I,J)=AL^CN*BT^BN
  S=S+SV(I)*SV(J)*L(I,J):        NEXT     J:       NEXT     I:
ANOR=SQR(2^5/3/P1/S)
  PRINT "                    SV":  FOR IJK=1 TO NP
  SV(IJK)=ANOR*SV(IJK):     PRINT   USING   "   +#.#####^^^^
";SV(IJK);
  NEXT  IJK:   REM  ***********   ПРОВЕРКА  НОРМИРОВАКИ
**********
  SSS=0:  FOR JJ=1 TO NP:  FOR KK=1 TO NP:  AL=XP(JJ)+XP(KK)
  BT=XP(JJ+NP)+XP(KK+NP):  FOR I=0 TO NF:  R=HFF*I
  RRR=R^2*AL
  IF RRR>100 GOTO 9182:  A=R^2*EXP(-RRR)
9182 FF(I)=A:  NEXT I:  AA=0;  BB=0:  FOR I=1 TO NF-1 STEP 2
  BB=BB+FF(I):  NEXT I:  FOR J=2 TO NF-2 STEP 2:  AA=AA+FF(J)
  NEXT J:  SIM1=HFF*(FF(0)+FF(NF)+2*AA+4*BB)/3:  FOR I=0 TO
NF
  R=HFF*I:  RRR=R^2*BT:  IF RRR>100 GOTO 9183
 A=R^4*EXP(-RRR)
9183 FF(I)=A:  NEXT I: AA=0;  BB=0: FOR I=1 TO NF-1 STEP 2
```

```
 BB=BB+FF(I): NEXT I: FOR J=2 TO NF-2 STEP 2: AA=AA+FF(J)
 NEXT J: SIM2=HFF*(FF(0)+FF(NF)+2*AA+4*BB)/3
 SS=SV(JJ)*SV(KK)*SIM1*SIM2: SSS=SSS+SS: NEXT KK: NEXT
JJ
 PRINT: PRINT "    NORM =";: PRINT USING " +#.########^^^^
";SSS;
6622 REM GOTO 6319
PRINT "                    INPUT 1 FOR PRINT ASSIMPTOTIC CON-
STANTS"
INPUT A:IF A<>1 GOTO 6319
REM   * * * * * * * РАСЧЕТ ВФ В N-6LI КАНАЛЕ * * * * * * * * *
NPI=9:OPEN "I",1,ALD2$: FOR I=1 TO NPI: INPUT#1, E2(I),C0(I)
 REM PRINT USING " +#.####^^^^ ";E2(I); C0(I); XP(I); XP(I+NP);
SV(I)
 NEXT I: CLOSE: FOR K=1 TO NF: R=HFF*K: SS=0
 FOR I=1 TO NP: FOR J=1 TO NPI: BIK=XP(I+NP): DIJ=XP(I)+E2(J)
 SS=SS+C0(J)*SV(I)*DIJ^(-1.5)*EXP(-BIK*R^2): NEXT J: NEXT I
 FF(K)=PI/4*SS*R^2: FU(K)=FF(K)^2*R^2: NEXT K
 FU(0)=0: AA=0: BB=0: FOR I=1 TO NF-1 STEP 2: BB=BB+FU(I)
 NEXT I: FOR J=2 TO NF-2 STEP 2: AA=AA+FU(J): NEXT J
 PP=HFF*(FU(0)+FU(NF)+2*AA+4*BB)/3: PRINT SQR(PP)
 RRM=4/7*(1.67)^2+3/7*(1.7)^2+12/49*PP
 RRZ=2/3*(1.67)^2+1/3*(1.7)^2+(2*9+1*16)/49/3*PP
 PRINT 12/49;(2*9+16)/49/3
 PRINT "P(ALT)= ";PP;"; RM(ALT)= ";SQR(RRM);"; RZ(ALT)= ";
SQR(RRZ)
 SKS=2*PM0*2.47/A11: SS=SQR(ABS(SKS)): SQQ=SQR(2.*SS)
 GK=3.44476E-02*Z1*(Z2+Z3)*PM0: GGG=GK/SS
 PRINT "      R        C0         CW0        CW"
 FOR I=NF/5 TO NF/2 STEP NF/30: R=I*HFF
 CALL WW(SKS,1,GK,R,NF,HFF,WWW)
 CW=FF(I)/WWW/SQQ: C0=FF(I)/(EXP(-SS*R)*SQQ)
 CW0=C0*(R*SS*2)^GGG:   PRINT   USING   "   +#.###^^^^
";R,C0,CW0,CW
 NEXT I
6319 REM ************** НОРМИРОВКА *****************
 CN=-1.5: BN=-2.5: SS=0: FOR I=1 TO NP: FOR J=1 TO NP
 AL=XP(I)+XP(J): BT=XP(I+NP)+XP(J+NP): L(I,J)=AL^CN*BT^BN
 SS=SS+SV(I)*SV(J)*L(I,J):     NEXT     J:     NEXT     I:
SSS=SQR(3*P1/2^5*SS)
 PRINT "     N= ";: PRINT SSS
 REM ***************  РАДИУС МАССЫ *****************
 S=0: FOR I=1 TO NP: FOR J=1 TO NP: AL=XP(I)+XP(J)
 BT=XP(I+NP)+XP(J+NP)
```

```
 S=S+SV(I)*SV(J)*BT^BN*AL^CN*(3*PM23/AL+5*PM0/BT):
NEXT J
 NEXT I: RM=P1*3*S/2^6
 RRR=M1/AM0*(1.67)^2 + M2/AM0*(1.)^2 + M3/AM0*(.8)^2 +
1/AM0*RM
 RS=SQR(RRR): REM PRINT: PRINT "  RM = ";
 PRINT USING " +#.#####^^^^ ";RS;: REM PRINT "              RMM =
";
 REM PRINT USING " +#.#####^^^^ ";RM
 REM ******************* РАДИУС ЗАРЯДА **************
 AA=(Z1*M23^2+Z2*M1^2+Z3*M1^2)/AM0^2
 BB=(Z2*M3^2+Z3*M2^2)/M23^2:        CC=-M1/AM0/M23*(Z2*M3-
Z3*M2)
 S=0: SS=0: SSS=0: FOR I=1 TO NP: FOR J=1 TO NP
 AL=XP(I)+XP(J): BT=XP(I+NP)+XP(J+NP)
 S=S+SV(I)*SV(J)*AA*5/(BT^(3.5)*AL^(1.5))
 SS=SS+SV(I)*SV(J)*BB*3/(BT^(2.5)*AL^(2.5))
 SSS=SSS+SV(I)*SV(J)*CC/(BT^3*AL^2): NEXT J: NEXT I
 RM=3/2^6*P1*(S+SS)+SSS
 RRR=Z1/Z*(1.67)^2+Z2/Z*(1.42)^2+Z3/Z*(0.)^2+RM/Z:
RZ=SQR(RRR)
 PRINT "  RZ = ";: PRINT USING " +#.#####^^^^ ";RZ;
 PRINT "              RZM = ";: PRINT USING " +#.#####^^^^ ";RM
 REM ********** КВАДРУПОЛЬНЫЙ МОМЕНТ *************
 AA=(Z1*M23^2+(Z2+Z3)*M1^2)/AM0^2
 BB=(Z2*M3^2+Z3*M2^2)/M23^2:        CC=-M1/AM0/M23*(Z2*M3-
Z3*M2)
 S=0: SS=0: SSS=0: FOR I=1 TO NP: FOR J=1 TO NP
 AL=XP(I)+XP(J): BT=XP(I+NP)+XP(J+NP)
 S=S+SV(I)*SV(J)*AA*5/(BT^(3.5)*AL^(1.5))
 SS=SS+SV(I)*SV(J)*BB*3/(BT^(2.5)*AL^(2.5))
 SSS=SSS+SV(I)*SV(J)*CC/(BT^3*AL^2): NEXT J: NEXT I
 RM=-2/5*(P1/16*3/4*(S+SS)+SSS)*10: QQ=RM+2.86
 PRINT "  Q  = ";: PRINT USING " +#.#####^^^^ ";QQ
 REM ****** ВЕРОЯТНОСТЬ N-6LI КОНФИГУРАЦИИ *******
 NPI=9: OPEN "I",1,ALD2$: FOR I=1 TO NPI: INPUT#1, E2(I),C0(I)
 REM PRINT USING " +#.####^^^^ ";E2(I); C0(I); XP(I); XP(I+NP);
SV(I)
 NEXT I: CLOSE: SS=0: FOR I=1 TO NP: FOR K=1 TO NP
 S=0: FOR J=1 TO NPI: FOR N=1 TO NPI: DIJ=XP(I)+E2(J)
 DKN=XP(K)+E2(N): S=S+C0(J)*C0(N)*DIJ^(-1.5)*DKN^(-1.5)
 NEXT N: NEXT J: BIK=XP(I+NP)+XP(K+NP)
 SS=SS+SV(I)*SV(K)*S*BIK^(-2.5): NEXT K: NEXT I
 PP=3*P1^(1.5)/16/8*SS: PRINT "PP= ";PP
```

```
REM *************** COULUMB ENERGY  ****************
SS11=0: CN=-1.5: BN=-2.5: S=0: SS=0: SSS=0: SSDD=0: SSSS=0
S1=0: S3=0: S2=0: FOR I=1 TO NP: FOR J=1 TO NP:
AL=XP(I)+XP(J)
BT=XP(I+NP)+XP(J+NP): VK12=2*A12/PI/AL^(3/2)/BT^2*P1/16
VK13=2*A13/PI/AL^(3/2)/BT^2*P1/16:
VK23=3*A23/PI/AL/BT^(5/2)*P1/16
VK=(VK12+VK13+VK23): S=S+SV(I)*SV(J)*VK
S1=S1+SV(I)*SV(J)*VK12
S2=S2+SV(I)*SV(J)*VK13: S3=S3+SV(I)*SV(J)*VK23
VCB=A11*(AL*BT)^(-1.5)/PM0*P1/16
SS=SS+SV(I)*SV(J)*VCB: AL1=XP(I)*XP(J)
BT1= XP(I+NP)*XP(J+NP)
H1=3/2*3*A11*AL1/PM23*(AL*BT)^BN
H2=3/2*A11/PM0/(AL*BT)^(1.5)*(5*BT1/BT^2-2/3)
TTT=(H1+H2)*P1/16: SSS=SSS+SV(I)*SV(J)*TTT
L1=3/2/BT*(AL*BT)^CN*P1/16: SSDD=SSDD+SV(I)*SV(J)*L1
AA=AL*BT+R121*(AL+(M3/M23)^2*BT)
BB=(M3/M23)^2*R121^2:
VN12=3/2*V121*(AA^CN)*(BB/AA+1)/(BT+R121)
AA=AL*BT+R122*(AL+(M3/M23)^2*BT)
BB=(M3/M23)^2*R122^2
VN12=VN12+3/2*V122*(AA^CN)*(BB/AA+1)/(BT+R122)
AA=AL*BT+R131*(AL+(M2/M23)^2*BT):
BB=(M2/M23)^2*R131^2
VN13=3/2*V131*(AA^CN)*(BB/AA+1)/(BT+R131)
AA=AL*BT+R132*(AL+(M2/M23)^2*BT):
BB=(M2/M23)^2*R132^2
VN13=VN13+3/2*V132*(AA^CN)*(BB/AA+1)/(BT+R132)
VN231=3/2*V231*BT^BN*(AL+R231)^CN
VN232=3/2*V232*BT^BN*(AL+R232)^CN: VN23=VN231+VN232
SSSS=SSSS+SV(I)*SV(J)*(VN12+VN13+VN23)*P1/16: NEXT J
NEXT I
SS11=SSS+S+SS+SSSS: PRINT
PRINT "COUL. ENERGY=  ";"VK";S;"12";S1;"13";S2;"23";S3
PRINT "CENTROB. ENERGY=  ";SS
PRINT "KINETICH. ENERGY=  ";SSS
PRINT "M.E. OT E* L1=  ";SSDD
PRINT "POTENS. ENERGY=  ";SSSS
PRINT "POLNAY ENERGY=  ";SS11
REM **************************************************
OPEN "O",1,ALD$
PRINT#1,"                    ЭНЕРГИЯ"
PRINT#1,"                                        "
```

```
 PRINT#1,USING "                         +#.#####^^^^ ";ENER
 PRINT#1,"                                                "
 PRINT#1, "        ПАРАМЕТРЫ АЛЬФА  И КОЕФ. РАЗЛОЖЕНИЯ"
 PRINT#1,"                                                "
 FOR I=1 TO NP
 PRINT#1, USING " +#.#####^^^^ ";XP(I);XP(I+NP);SV(I)
 NEXT I: PRINT#1,"                                          "
 PRINT#1,"                                                "
 REM  PRINT#1, "                 КОЭФФИЦИЕНТЫ РАЗЛОЖЕНИЯ
SV"
 REM  PRINT#1,"                                           "
 REM  FOR IJK=1 TO NP
 REM  PRINT#1, USING " +#.#####^^^^ ";SV(IJK);
 REM             NEXT        IJK:        REM              PRINT#1,"
"
 REM  PRINT#1,"                                          "
 PRINT#1,"                                                "
 PRINT#1, "    НЕВЯЗКИ ПО НАЙДЕННЫМ КОЭФФИЦИЕНТАМ
H*SV-LA*L*SV=0:                                        PRINT#1,"
"
 FOR IJK=1 TO NP: PRINT#1, USING " +#.#####^^^^ ";E2(IJK);
 NEXT IJK:  PRINT#1,"                                    "
 GOTO 2345:  PRINT#1,"                                    "
  PRINT#1," R                 C0                 CW"
  PRINT#1,"                                               "
  KKK=0:  FOR I=MM TO NF STEP MMM:  R=HFF*I
  KKK=KKK+1:  PRINT#1, USING " +#.##^^^^ ";R;
  PRINT#1, USING "              +#.#####^^^^ ";C0(KKK); CW0(KKK);
CW(KKK)
  NEXT I
2345  PRINT#1,
  PRINT#1,"                                               "
  PRINT#1, "РАДИУС МАССЫ   ";
  PRINT#1, USING " +#.#####^^^^ ";RS
  PRINT#1,"                                               "
  PRINT#1, "ЗАРЯДОВЫЙ РАДИУС   ";
  PRINT#1, USING " +#.#####^^^^ ";RZ
  PRINT#1,"                                               "
  PRINT#1, "КВАДРУПОЛЬНИЙ МОМЕНТ   ";
  PRINT#1, USING " +#.#####^^^^ ";QQ
  PRINT#1,"                                               ":  CLOSE
  END
```

**4000 REM - - - - - - - - -  ГАММА ФУНКЦИЯ  - - - - - - - - - - - - - -**

```
 A1(0)=1:              A1(1)=0.5772156649015328б:              A1(2)=-
```

```
0.65587807152025388
 A1(3)=-0.04200263503409523: A1(4)=0.1665861138229148
 A1(5)=-0.04219773455554433: A1(6)=-0.00962197152787697
 A1(7)=0.00721894324666309: A1(8)=-0.00116516759185906
 A1(9)=-0.00021524167411495: A1(10)=0.00012805028238811
 A1(11)=-0.00002013485478078: A1(12)=-0.00000125049348214
 A1(13)=0.0000013302723198: A1(14)=-0.0000020563384169
 A1(15)=0.00000000611609510: A1(16)=0.00000000500200764
 A1(17)=-0.00000000118127457: A1(18)=0.0000000001044267
 A1(19)=0.00000000000778226: A1(20)=-0.0000000000069681
 Z=ET+LO: S=0: FOR I=0 TO 20: S=S+A1(I)*Z^I: NEXT I
 GA=1/S: RETURN
```

**5000 REM - - - - - - - - ПОИСК НУЛЯ ДЕТЕРМИНАТА - - - - -**

```
 CN=-1.5: BN=-2.5: FOR KK=1 TO NP: FOR JJ=KK TO NP
 AL=XP(KK)+XP(JJ): AL1=XP(KK)*XP(JJ)
 BT=(XP(KK+NP)+XP(JJ+NP)): BT1= XP(KK+NP)*XP(JJ+NP)
 H1=3/2*3*A11*AL1/PM23*(AL*BT)^BN
 H2=3/2*A11/PM0/(AL*BT)^(1.5)*(5*BT1/BT^2-2/3):
T(KK,JJ)=H1+H2
 L1(KK,JJ)=3/2/BT*(AL*BT)^CN:
AA=AL*BT+R121*(AL+(M3/M23)^2*BT)
 BB=(M3/M23)^2*R121^2
 VN12(KK,JJ)=3/2*V121*(AA^CN)*(BB/AA+1)/(BT+R121)
 AA=AL*BT+R122*(AL+(M3/M23)^2*BT):
BB=(M3/M23)^2*R122^2
VN12(KK,JJ)=VN12(KK,JJ)+3/2*V122*(AA^CN)*(BB/AA+1)/(BT+R
122)
 AA=AL*BT+R131*(AL+(M2/M23)^2*BT):
BB=(M2/M23)^2*R131^2
 VN13(KK,JJ)=3/2*V131*(AA^CN)*(BB/AA+1)/(BT+R131)
 AA=AL*BT+R132*(AL+(M2/M23)^2*BT):
BB=(M2/M23)^2*R132^2
VN13(KK,JJ)=VN13(KK,JJ)+3/2*V132*(AA^CN)*(BB/AA+1)/(BT+R
132)
 VN231(KK,JJ)=3/2*V231*BT^BN*(AL+R231)^CN:     REM     L=1,
LAM=0
 VN232(KK,JJ)=3/2*V232*BT^BN*(AL+R232)^CN:     REM     L=1,
LAM=0
 VN23(KK,JJ)=VN231(KK,JJ)+VN232(KK,JJ)
 VK12=2*A12/PI/AL^(3/2)/BT^2: VK13=2*A13/PI/AL^(3/2)/BT^2
 VK23=3*A23/PI/AL/BT^(5/2): VCB=A11*(AL*BT)^CN/PM0
H(KK,JJ)=T(KK,JJ)+VN23(KK,JJ)+VN12(KK,JJ)+VN13(KK,JJ)+VC
B+VK12+VK13+VK23: H(JJ,KK)=H(KK,JJ): L1(JJ,KK)=L1(KK,JJ)
 NEXT JJ: NEXT KK: IF ZYS=1 THEN GOSUB 1000
```

```
 IF ZYS=1 THEN RETURN
 REM - - - - - - - - - - - - - - - - - - - - - - - - - - - - - - - - - - - - - - - - - -
 I1=0
100 A2=PNC: B2=PNC+HC
 REM ------------------------------------------------------------------------------
 LLA=A2: GOSUB 2000: D12=S
 REM ------------------------------------------------------------------------------
51 LLA=B2: GOSUB 2000: D11=S
 REM ------------------------------------------------------------------------------
 IF D12*D11>0 GOTO 4
44 I1=I1+1: A3=A2: B3=B2
11 C3=(A3+B3)/2: IF ABS(A3-B3)<EPP GOTO 151
 REM ------------------------------------------------------------------------------
 LLA=C3: GOSUB 2000: F2=S
 REM ------------------------------------------------------------------------------
 IF D12*F2>0 GOTO 14: B3=C3: D11=F2: GOTO 15
14 A3=C3: D12=F2
15 IF ABS(F2)>EPP GOTO 11
151 CO=C3
 REM - - - - - - - - - - - - - - - - - - - - - - - - - - - - - - - - - - - - - - - - -
 LA=CO
 REM  - - - - - - - - - - - - - - - - - - - - - - - - - - - - - - - - - - - - - - - - -
 IF NSM=0 GOTO 2002: PNC=CO+ABS(0.1*CO)
2002 IF I1<NS GOTO 100: GOTO 7
4 IF ABS(D11*D12)<1.D-30 GOTO 44: A2=A2+HC: B2=B2+HC
 D12=D11: IF B2-PVC<0.1 GOTO 51: YS=PVC: GOTO 8
7 YS=NS
8 PRINT USING "  +#.##^^^^ ";F2;: FOR I=1 TO NS
 PRINT USING "     +#.#####^^^^ ";LA: NEXT I: RETURN
```

**2000 REM ПОИСК ДЕТЕРМИНАНТА МАТРИЦЫ**

```
 FOR I=1 TO NP: FOR J=1 TO NP: LLL(I,J)=(H(I,J)-LLA*L1(I,J))
 B(I,J)=0: C(I,J)=0: NEXT J: NEXT I: GOTO 234: PRINT
 FOR II=1 TO NP: FOR KK=1 TO NP
 PRINT  USING  "  +#.###^^^^  ";LLL(II,KK);:  NEXT  KK:  PRINT:
NEXT II
234 REM - - - - - - - - - - - - - - - - - - - - - - - - - - - - - - - - - - - - - - - -
 FOR I=1 TO NP: C(I,I)=1: B(I,1)=LLL(I,1): C(1,I)=LLL(1,I)/B(1,1)
 NEXT I: FOR I=2 TO NP: FOR J=2 TO NP: S=0: IF J>I GOTO 1
 FOR K=1 TO I-1: S=S+B(I,K)*C(K,J): NEXT K
 B(I,J)=LLL(I,J)-S: GOTO 2
1 S=0: FOR K=1 TO I-1: S=S+B(I,K)*C(K,J): NEXT K
 C(I,J)=(LLL(I,J)-S)/B(I,I)
2 NEXT J: NEXT I
 REM - - - - - - - - - - - - - - - - - - - - - - - - - - - - - - - - - - - - - - - - -
```

```
 SS=0: FOR I=1 TO NP: FOR J=1 TO NP: S=0: FOR K=1 TO NP
 S=S+B(I,K)*C(K,J): NEXT K: AAA(I,J)=S-LLL(I,J):
SS=SS+AAA(I,J)
 NEXT J: NEXT I: GOTO 678: PRINT "          -¦-¦¬¦¬  N=LLL-B*C
=0"
 FOR I=1 TO NP: PRINT: FOR J=1 TO NP
 PRINT USING " +#.#####^^^^ ";AAA(I,J);: NEXT J: NEXT I: PRINT
678 S=1: FOR K=1 TO NP: S=S*B(K,K): NEXT K: GOTO 991
 PRINT "      E=";: PRINT USING " +#.#####^^^^ ";LLA;
 PRINT "      DET=";: PRINT USING " +#.#####^^^^ ";S;
 PRINT "      NEV=";: PRINT USING " +#.#####^^^^ ";SS
991 RETURN
```

**1000 REM ВЫЧИСЛЕНИЕ СОБСТВЕННЫХ ВЕКТОРОВ**

```
 FOR I=1 TO NP: FOR J=1 TO NP: LLL(I,J)=(H(I,J)-LLA*L1(I,J))
 B(I,J)=0: C(I,J)=0: NEXT J: NEXT I: FOR I=1 TO NP-1
 FOR J=1 TO NP-1
 I1=I: J1=J: AD(I,J)=LLL(I1,J1): NEXT J: NEXT I: I1=1: I2=NP-1
 J=NP: FOR I=I1 TO I2: D(I)=-LLL(I,J): NEXT I: NP=NP-1: GOSUB
3000
 REM - - - - - - - - - - - - - - - - - - - - - - - - - - - - - - - - - - - - - - - - -
 Y(1)=D(1)/B(1,1): FOR I=2 TO NP: S=0: FOR K=1 TO I-1
 S=S+B(I,K)*Y(K)
 NEXT K: Y(I)=(D(I)-S)/B(I,I): NEXT I: X(NP)=Y(NP)
 FOR I=NP-1 TO 1 STEP -1: S=0: FOR K=I+1 TO NP:
S=S+C(I,K)*X(K)
 NEXT K: X(I)=Y(I)-S: NEXT I: FOR I=1 TO NP: SV(I)=X(I)
 NEXT I: NP=NP+1: SV(NP)=1: S=0: FOR I=1 TO NP: S=S+SV(I)^2
 NEXT I: SS=0: FOR I=1 TO NP: SV(I)=SV(I)/SQR(S): NEXT I
 AN=1: PRINT "                 H*SV-LA*L*SV=0: FOR I=1 TO NP
 S=0: SS=0: FOR J=1 TO NP: SV(J)=SV(J)*AN: S=S+H(I,J)*SV(J)
 SS=SS+LLA*L1(I,J)*SV(J): NEXT J: E2(I)=S-SS: NEXT I
 FOR I=1 TO NP: PRINT USING " +#.#####^^^^ ";E2(I);: NEXT I
 PRINT: RETURN
```

**3000 REM РАЗЛОЖЕНИЕ МАТРИЦЫ НА ТРЕУГОЛЬНЫЕ**

```
 FOR I=1 TO NP: C(I,I)=1: B(I,1)=AD(I,1): C(1,I)=AD(1,I)/B(1,1)
 NEXT I: FOR I=2 TO NP: FOR J=2 TO NP: S=0: IF J>I GOTO 551
 FOR K=1 TO I-1: S=S+B(I,K)*C(K,J): NEXT K: B(I,J)=AD(I,J)-S
 GOTO 552
551 S=0: FOR K=1 TO I-1: S=S+B(I,K)*C(K,J): NEXT K
 C(I,J)=(AD(I,J)-S)/B(I,I)
552 NEXT J: NEXT I
 REM - - - - - - - - - - - - - - - - - - - - - - - - - - - - - - - - - - - - - - - - -
 SS=0: FOR I=1 TO NP: FOR J=1 TO NP: S=0: FOR K=1 TO NP
 S=S+B(I,K)*C(K,J): NEXT K: AAA(I,J)=S-AD(I,J): SS=SS+AAA(I,J)
```

```
 NEXT J: NEXT I: GOTO 578
 PRINT "                    NEV = AD - B*C =0": FOR I=1 TO NP
 FOR J=1 TO NP: PRINT USING " +#.#####^^^^ ";AAA(I,J);: NEXT J
 NEXT I
578 S=1: FOR K=1 TO NP: S=S*B(K,K): NEXT K: GOTO 9753
 PRINT "          DET=";: PRINT USING " +#.#####^^^^ ";S;
 PRINT "          NEV=";: PRINT USING " +#.#####^^^^ ";SS
9753 PRINT: RETURN
```

**SUB WW(SK,L,GK,R,N,H,WH)**

```
 DIM V(500)
 SS=SQR(ABS(SK));  AA=GK/SS;  BB=L;  NN=500;  HH=.02
 ZZ=1+AA+BB;  AAA=1/ZZ;  NNN=2000: FOR I2=1 TO NNN
 AAA=AAA*I2/(ZZ+I2): NEXT I2: GAM=AAA*NNN^ZZ
 RR=R;  CC=RR*SS*2: FOR I=0 TO NN: TT=HH*I
 V(I)=TT^(AA+BB)*(1+TT/CC)^(BB-AA)*EXP(-TT): NEXT I
 A=0;  B=0: FOR II=1 TO NN-1 STEP 2: B=B+V(II): NEXT II
 FOR JJ=2 TO NN-2 STEP 2: A=A+V(JJ): NEXT JJ
 SIM=HH*(V(0)+V(NN)+2*A+4*B)/3
 WH=SIM*EXP(-CC/2)/(CC^AA*GAM): END SUB
```

Для проверки метода расчета и компьютерной программы рассматривалась модельная задача для трех частиц, взаимодействующих в потенциале Афнана - Танга [193] с усреднением триплетных и синглетных состояний. Для энергии такой системы в [193] получено -7.74 МэВ, а в работах [194,195], где использовался неортогональный вариационный метод с изменением параметров волновой функции на основе тангенциальной сетки, найдено -7.76 МэВ. Здесь, при независимом варьировании всех параметров и размерности базиса N=5, получено -7.83 МэВ, т.е. энергия изменилась примерно на 1% относительно результатов [193,194].

## 5.3 Физические результаты трехтельных расчетов

Приведем теперь распечатки результатов расчета трехтельной энергии связи ядра $^7$Li при N=9, 11 и реальными ядерными потенциалами, описанными выше.

**E= -8.678E+00 МэВ (N=9)**

ALFA

+1.75758E-01 +5.98566E-02 +2.41148E-01 +1.04891E-01 +1.78012E-01
+1.67172E-01 +2.23184E-01 +7.05920E-01 +1.21380E+00

BETA

+7.57214E-02 +6.43507E-02 +1.75199E-01 +2.80843E-01 +6.60368E-01

+8.13208E-01 +4.77234E-01 +3.71087E-01 +9.66861E-02

(H-LA*L)SV=0

-5.68434E-14 +2.27374E-13 +2.84217E-14 -8.52651E-14 -7.10543E-15
+1.77636E-15 -3.01981E-14 -1.42109E-14 +1.68490E-03

SV

-4.41044E-02 -1.20457E-02 -1.31495E-01 +6.08814E-02 -7.79223E-01
+3.40193E-01 +6.20480E-01 -1.05539E-01 +5.58986E-02

NORM = 1.00000000E+00

RM = 2.62E+00 Фм, RZ = 2.38E+00 Фм, Q = -3.45E+01 мб.

CENTRIFUGAL ENERGY= 1.926781715339702 МэВ
KINETIC ENERGY= 15.63147355849597 МэВ
POTENSIAL ENERGY= -27.01378556133874 МэВ
TOTAL ENERGY= -8.678288314964876 МэВ

**E= -8.713E+00 (N=11)**

ALFA

+3.09996E-02 +8.90401E-02 +2.57654E-01 +1.39035E-01 +2.02704E-01
+1.38880E-01 +2.21892E-01 +2.40877E-01 +1.24262E+00 +2.48192E-01
+9.59501E-01

BETA

+3.53874E-02 +7.40006E-02 +5.08743E-02 +2.92684E-01 +1.70156E-01
+4.59418E-01 +2.75673E-01 +7.54029E-01 +8.35727E-02 +6.74060E-01
+4.50788E-01

(H-LA*L)SV=0

+0.00000E+00 -2.27374E-13 +3.41061E-13 -7.10543E-15 +0.00000E+00
+2.84217E-14 +2.84217E-14 +2.35367E-14 -2.13163E-14 -2.66454E-15
-2.71238E-03

SV

+8.61351E-04 +2.56716E-02 +1.37998E-02 -2.73826E-01 +2.02813E-01
+2.09360E-01 -5.46337E-02 +5.40009E-01 -4.51367E-02 -6.87519E-01
+8.04027E-02

NORM = +1.00000000E+00

RM = +2.64E+00 RZ = +2.39E+00 Q = -3.555E+01

COUL. ENERGY= 0.7728061981754414 12
CENTROB. ENERGY= 1.910695580740683
KINETICH. ENERGY= 15.49351258590982
POTENS. ENERGY= -26.88993422313928

TOTAL ENERGY= -8.71291985831334

Результаты расчета вариационной энергии ядра $^7$Li, полученные изложенным методом, с использованием потенциалов из предыдущей таблицы и в зависимости от размерности вариационного базиса даны в таблице 5.2.

Таблица 5.2 -Результаты вычисления трехтельной энергии.

| N | 3 | 5 | 7 | 9 | 10 | 11 |
|---|---|---|---|---|---|---|
| E($^7$Li), МэВ | -7.68 | -8.63 | -8.66 | -8.678 | -8.706 | -8.713 |

Экспериментальная трехтельная энергия ядра в этом канале составляет -8.725 МэВ [190]. Из таблицы видно, что при размерности N=9-11 энергия системы практически сходится, и дальнейшее увеличение базиса может привести, по - видимому, к изменению энергии на величину порядка 0.01 - 0.02 МэВ. Как уже говорилось, параметры взаимодействия в $^2$H$^4$He системе из - за различия разных экспериментальных данных имеют большую неоднозначность. Однако теперь становится ясно, что в пределах этой неопределенности можно найти параметры, позволяющие правильно воспроизвести энергию связи ядра $^7$Li.

С полученными волновыми функциями для массового и зарядового трехтельных среднеквадратичных радиусов найдено 2.78 Фм и 2.51 Фм соответственно, что несколько больше эксперимента [162-186], где для зарядового радиуса получены величины 2.39(3) и 2.35(10) Фм.

Однако, здесь использовался n$^2$H потенциал, приводящий к завышенному радиусу трития, что повлияло и на радиус самого ядра $^7$Li. Тем самым, дейтронный кластер нужно деформировать, как в ядре трития, так и в $^7$Li, поскольку в свободном состоянии дейтрон очень "рыхлая" система. Для того чтобы получить правильный зарядовый радиус ядра 2.39 Фм необходимо уменьшить радиус дейтронного кластера и принят его равным 1.42 Фм, т.е. уменьшить его на те же 27%, как это было для ядра трития. Для массового радиуса, в таком случае, находим 2.64 Фм. Тем самым получаем, что дейтронный кластер примерно одинаково деформирован, как в тритии, так и в ядре $^7$Li, что хорошо согласуется с $^3$H$^4$He моделью этого ядра.

Другой возможной причиной завышения зарядового радиуса ядра $^7$Li без деформаций дейтрона может быть отсутствие учета, в разных парах частиц, потенциалов для других парциальных волн. Например, наряду с P волной в $^2$H$^4$He системе можно учитывать S взаимодействие, а в n$^4$He канале - P волну.

Но поскольку радиус ядра, без деформации дейтрона, завышен всего на 4 - 5%, а квадрупольный момент оказывается равен -35.5 мб, что меньше экспериментальной величины только на 3 - 4%, можно, по - видимому, считать, что учет дополнительных парциальных волн в парных потенциалах приведет только к небольшим поправкам для полученных величин.

Найденная вероятность двухчастичного $^3H^4He$ канала 98.1%, вполне объясняет успешное использование простой двухкластерной модели, позволяющей получить хорошие результаты для многих характеристик ядра $^7Li$ [162-186, 91].

Для двухчастичных радиусов на основе (3.59) найдено 2.68 Фм и 2.63 Фм соответственно, что несколько больше экспериментальной величины и результатов, получаемых в двухкластерной модели с запрещенными состояниями [162-186,91]. Кулоновская энергия ядра, которая представляется в виде среднего от кулоновского матричного элемента, оказалась равна 0.77 МэВ.

Для энергии связи ядра $^7Be$, если рассматривать его с теми же параметрами волновой функции, но учесть кулоновское взаимодействие между частицами (13) и (23) найдено -7.15 МэВ, что только на 1% отличается от экспериментальной величины -7.08 МэВ [190].

Таким образом, видно, что описанные методы и компьютерная программа позволяет воспроизвести известные ранее результаты, что говорит об их работоспособности. А рассмотренная трехкластерная модель ядра $^7Li$ позволяет получить новые результаты, хорошо передающие экспериментальные данные по основным характеристикам ядра $^7Li$ и приводящие к большой вероятности $^3H^4He$ канала. Именно использованные парные взаимодействия дают наибольший вклад в рассмотренные характеристики, а учет деформаций дейтрона приводит к правильному зарядовому радиусу ядра [187].

# 6. МЕТОДЫ РАСЧЕТА СЕЧЕНИЙ ЯДЕРНОГО РАССЕЯНИЯ

В этой главе рассматриваются математические численные методы расчета дифференциальных сечений упругого рассеяния ядерных частиц с различным спином. В таких задачах не требуется решения уравнения Шредингера, но все эти методы использованы в следующей главе при рассмотрении фазового анализа, основанного на многопараметрическом вариационном методе.

Когда известны дифференциальные сечения, определяемые экспериментальным путем, почти всегда можно найти определенный набор ядерных фаз (т.е. некоторых параметров), который способен, с той или иной точностью, передать форму этих сечений. Качество описания экспериментальных данных на основе некоторой функции (функционала нескольких переменных) можно оценить по методу $\chi^2$ - квадрат, который записывается в виде [57]

$$\chi^2 = \frac{1}{N}\sum_{i=1}^{N}\left[\frac{\sigma_i^t(\theta)-\sigma_i^e(\theta)}{\Delta\sigma_i^e(\theta)}\right]^2 ,$$

где $\sigma_e$ и $\sigma_t$ – экспериментальное и теоретическое, т.е. расчетное при некоторых заданных значениях фаз $\delta_{S,L}{}^J$ рассеяния сечение упругого рассеяния ядерных частиц для i – го угла рассеяния.

## 6.1 Система частиц с нулевым полным спином

Наиболее простые формулы для сечений ядерного рассеяния получаются в случае рассеяния частиц со спином ноль, поскольку отсутствует спин - орбитальное расщепление фаз. Если частицы не тождественны, например, $^4He^{16}O$ или $^4He^{12}C$, то сечение определяется наиболее просто и записывается, как квадрат модуля амплитуды рассеяния [57]

$$\frac{d\sigma(\theta)}{d\Omega} = |f(\theta)|^2 , \qquad (6.1)$$

где сама амплитуда представляется в виде суммы кулоновской (c) и ядерной (N) амплитуд

$$f(\theta) = f_c(\theta) + f_N(\theta) , \qquad (6.2)$$

которые выражаются через ядерные $\delta_L \to \delta_L + i\Delta_L$ и кулоновские $\sigma_L$ фазы рассеяния

$$f_c(\theta) = -\left(\frac{\eta}{2k\mathrm{Sin}^2(\theta/2)}\right)\exp\{i\eta\ln[\mathrm{Sin}^{-2}(\theta/2)] + 2i\sigma_0\} \quad ,$$

$$f_N(\theta) = \frac{1}{2ik}\sum_L (2L+1)\exp(2i\sigma_L)[S_L - 1]P_L(\mathrm{Cos}\theta) \quad . \qquad (6.3)$$

Здесь $S_L(k) = \eta_L(k)\exp[2i\delta_L(k)]$ - матрица рассеяния, которая может быть представлена в виде

$$\frac{(S_L - 1)}{2i} = \eta_L \mathrm{Sin}\delta_L \exp(i\delta_L) \quad ,$$

а $\eta_L(k) = \exp[-2\Delta_L(k)]$ - параметр неупругости, зависящий от мнимой части ядерной фазы $\mathrm{Im}\ \delta_L = \Delta_L(k)$, $P_L(x)$ - полиномы Лежандра

$$P_n(x) = \frac{1}{2^n n!}\frac{d^n}{dx^n}(x^2 - 1)^n \quad ,$$

$\eta$ - кулоновский параметр, $\mu$ - приведенная масса, k - волновое число относительного движения частиц - $k^2 = 2\mu E/\hbar^2$, E - энергия сталкивающихся частиц в системе центра масс.

Из (6.3) легко получить полное сечение упругого рассеяния при $f_c = 0$. Взяв квадрат модуля от $f_N$, интегрируя по углам и учитывая ортогональность полиномов Лежандра ($x = \mathrm{Cos}\theta$) [57]

$$\int_0^{2\pi}\int_{-1}^{1} P_L(x)P_{L'}(x)dxd\varphi = \frac{4\pi\delta_{LL'}}{2L+1}$$

будем иметь

$$\sigma_s = \frac{\pi}{k^2}\sum_L\left[(2L+1)\left(|1 - S_L|^2\right)\right] = \frac{4\pi}{k^2}\sum_L (2L+1)\eta_L^2 \mathrm{Sin}^2\delta_L \quad . \qquad (6.4)$$

Полное сечение реакций или неупругих процессов можно представить в виде [57,102]

$$\sigma_r = \frac{\pi}{k^2}\sum_L\left[(2L+1)\left(1 - |S_L|^2\right)\right]$$

и при подстановке выражения для матрицы рассеяния полу-

чим

$$\sigma_r = \frac{\pi}{k^2}\sum_L (2L+1)(1-\eta_L^2) \quad . \qquad (6.5)$$

Для нахождения полиномов Лежандра можно использовать начальные значения и рекуррентное соотношение [73]

$P_0(x) = 1$ ,  $P_1(x) = x$ ,
$P_L(x) = (2L - 1)x/LP_{L-1}(x) - (L - 1)/LP_{L-2}(x)$ ,

где значения L начинаются с 2. Кулоновскую амплитуду рассеяния (6.3), используя выражение

$A = Sin^{-2}(\theta/2) = 2/[1 - Cos(\theta)]$ ,

можно записать в виде

$$f_c = - \eta A/2k\ [Cos(B) + iSin(B)] \quad , \qquad (6.6)$$

где $B = 2\sigma_0+\eta \ln A$. Ядерная амплитуда (6.3) может быть представлена в следующей форме

$$f_n = \frac{1}{2k}\sum_L \hat{L}\left\{ [\beta Cos(2\sigma_L)+\alpha Sin(2\sigma_L)]+i[\beta Sin(2\sigma_L)-\alpha Cos(2\sigma_L)] \right\} P_L(x) \qquad (6.7)$$

где $x = Cos(\theta)$, $\hat{L} = 2L+1$, а $\alpha = \eta_L Cos(2\delta_L)-1$ и $\beta = \eta_L Sin(2\delta_L)$ зависят только от ядерных фаз, параметра неупругости и орбитального момента.

Кулоновские фазы рассеяния выражаются через Гамма - функцию [84]

$$\sigma_L = \arg\{\Gamma(L+1+i\eta)\}$$

и удовлетворяют рекуррентному процессу

$$\sigma_L = \sigma_{L+1} - Arctg\left(\frac{\eta}{L+1}\right) \quad .$$

Откуда сразу можно получить следующее выражение для кулоновских фаз

$$\alpha_L = \sigma_L - \sigma_{L-1} = \sum_{n=1}^{L} \mathrm{Arctg}\left(\frac{\eta}{n}\right), \qquad \alpha_0 = 0 . \qquad (6.8)$$

Величина $\alpha_L$ используется в преобразованных выражениях (6.3), если вынести общий множитель $\exp(2i\sigma_0)$. Тогда $\sigma_L \to \alpha_L$ с $\alpha_0 = 0$, что избавляет от необходимости вычислять кулоновские фазы в явном виде, а кулоновская амплитуда принимает вид

$$f_c(\theta) = -\left(\frac{\eta}{2k\mathrm{Sin}^2(\theta/2)}\right)\exp\{i\eta\ln[\mathrm{Sin}^{-2}(\theta/2)]\} \ .$$

В случае рассеяния тождественных бозонов, например, при рассеянии ядер $^4He^4He$, формула сечения (6.1) преобразуется к виду [196, 197,198]

$$\frac{d\sigma(\theta)}{d\Omega} = |f(\theta) + f(\pi - \theta)|^2 , \qquad (6.9)$$

где $f(\theta)$ определено формулами (6.2) и (6.3). Такая запись позволяет учитывать эффекты, которые дает симметризация волновых функций системы тождественных частиц.

Поскольку $\mathrm{Cos}(\pi - \theta) = - \mathrm{Cos}\theta = - x$ и $P_L(-x) = (-1)^L P_L(x)$, то величина ядерной амплитуды просто удваивается. Суммирование в (6.3) идет только по четным L, поскольку нечетные парциальные волны не дают вклада в суммарное сечение.

В кулоновской амплитуде выполняется преобразование величины A к виду

$$A = \mathrm{Sin}^{-2}[(\pi-\theta)/2] = 2/[1 - \mathrm{Cos}(\pi-\theta)] = 2/[1 + \mathrm{Cos}(\theta)]$$

Формулы (6.1) - (6.3) можно использовать для расчета синглетных сечений (при полном спине S = 0) в системах, когда обе частицы имеют спин 1/2, например, для np, $^3H^3He$ или $N^3He$, $N^3H$ и т.д.

Приведем пример программы для вычисления сечений упругого рассеяния тождественных и не тождественных бозонов с нулевым спином. Здесь приняты следующие обозначения: NN=0 - Нижнее значение энергии, NV=0 - Верхнее значение энергии, LN=0 - Нижнее значение момента, LV=12 - Верхнее значение момента, LH=2 - Шаг по значениям момента, TMI=10 - Нижнее значение угла рассеяния, TMA=90 - Верхнее значение угла рассеяния, TH=1 - Шаг по углу, AM1, AM2, Z1, Z2 - Массы и заряды частиц,

E1(0) - Лабораторная энергия частиц, FR(0) - Реальная часть фазы рассеяния, FM(0) - Мнимая часть фазы рассеяния [199].

**REM CROSS SECTION CALCULATE FOR COMPLEX PHASE SHIFTS OF SYSTEM WITH 0 SPIN**

```
CLS: DEFDBL A-Z: DEFINT I,J,K,L,N,M:  N=200
DIM E(N), DE(N), E1(N), ETA(N), SEC(N), SECE(N), FM(N/10), FR(N/10)
REM *********************************************
ISAVE=0:          REM =0 - NO SAVE, =1 - SAVE IN FILE
G$="C:\BASICA\SEC\SECALAL.DAT"
REM ************ INPUT PARAMETERS ************
PI=4*ATN(1): NN=0: NV=0: LN=0: LV=12: LH=2: TMI=10: TMA=90
TH=1: AM1=4: AM2=4: Z1=2: Z2=2: A1=41.4686
PM=AM1*AM2/(AM1+AM2)
B1=2*PM/A1
REM ********* ENERGY IN LAB SYSTEM ***********
E1(0)=53.4: E1(6)=119.86
REM ************* PHASE SHIFTS ******************
REM *********** FOR AL-AL ON E=53.4 *************
FR(0)=-75.2: FR(2)=47.9: FR(4)=137.9: FR(6)=27.5: FR(8)=2: FR(10)=0
FM(0)=12.1: FM(2)=22.1: FM(4)=16.3: FM(6)=3.2: FM(8)=0: FM(10)=0
REM ********** FOR AL-AL ON E=119.9 **************
GOTO 111
FR(0)=-161.5: FR(2)=-16: FR(4)=130.3: FR(6)=93.8: FR(8)=26: FR(10)=7
FR(12)=1.7: FM(0)=15.7: FM(2)=15.6: FM(4)=13.8
FM(6)=26.6: FM(8)=18.3: FM(10)=3.7: FM(12)=0.
111 REM ******** ПЕРЕВОД ФАЗ В РАДИАНЫ ********
FOR L=LN TO LV STEP LH: FM(L)=FM(L)*PI/180
FR(L)=FR(L)*PI/180
NEXT
REM ********** TRANSFORM TO C.M. ***************
FOR I=NN TO NV: E(I)=E1(I)*PM/AM1: NEXT I
REM ***********  TOTAL CROSS SECTION  ***********
FOR J=NN TO NV: SK=E(J)*B1: SS=SQR(SK)
GG=3.44476E-02*Z1*Z2*PM/SS: SIGMAR=0: SIGMAS=0
FOR L=LN TO LV STEP LH: A=FR(L): ETA(L)=EXP(-2*FM(L))
SIGMAR=SIGMAR+(2*L+1)*(1-(ETA(L))^2)
SIGMAS=SIGMAS+(2*L+1)* (ETA(L))^2 *(SIN(A))^2
NEXT L: SIGMA=10*4*PI*SIGMA/SK
```

```
 PRINT "                     SIGR -TOT = ";
 PRINT USING " ####.### ";SIGMAR
 PRINT "                     SIGS -TOT = ";
 PRINT USING " ####.### ";SIGMAS: NEXT J: PRINT
 REM ****** DIFFERENTIAL CROSS SECTION *******
 CALL SEC (FR(), GG, SS, TMI, TMA, TH, SEC(), ETA(), LN,
LV,LH, IUSLOV)
 FOR T=TMI TO TMA/3 STEP TH
 PRINT USING "####.##  "; T; SEC(T); T+20; SEC(T+20); T+40;
SEC(T+40); T+60; SEC(T+60): NEXT
 REM ************* SAVE IN FILE  *************
 IF ISAVE=0 GOTO 221
 OPEN "O",1,G$: PRINT#1, "             ALPHA - ALPHA FOR LAB
E=";
 PRINT#1, E1(NN): FOR T=TMI TO TMA STEP TH
 PRINT#1, USING " #.###^^^^ ";T;SEC(T): NEXT
221 END
 SUB SEC (F(100), GG, SS, TMI, TMA, TH, S(100),E (100), LMI,
LMA, LH, NYS)
 REM ------------------- CROSS SECTIONS -------------------------
 SHARED PI: DIM S0(20),P(20)
 RECUL1=0: AIMCUL1=0: CALL CULFAZ(GG,S0())
 FOR TT=TMI TO TMA STEP TH: T=TT*PI/180: X=COS(T): A=2/(1-
X)
 S0=2*S0(0): BB=-GG*A: ALO=GG*LOG(A)+S0
 RECUL=BB*COS(ALO)
 AIMCUL=BB*SIN(ALO): IF NYS=0 GOTO 555
 A1=2/(1+X): BB1=-GG*A1
 ALO1=GG*LOG(A1)+S0: RECUL1=BB1*COS(ALO1)
 AIMCUL1=BB1*SIN(ALO1)
555 RENUC=0: AIMNUC=0: FOR L=LMI TO LMA STEP LH
 AL=E(L)*COS(2*F(L))-1: BE=E(L)*SIN(2*F(L)): LL=2*L+1
 SL=2*S0(L)
 CALL POLLEG(X,L,P())
 RENUC=RENUC+LL*(BE*COS(SL)+AL*SIN(SL))*P(L)
 AIMNUC=AIMNUC+LL*(BE*SIN(SL)-AL*COS(SL))*P(L):  NEXT
L
 IF NYS=0 GOTO 556: AIMNUC=2*AIMNUC: RENUC=2*RENUC
556 RE=RECUL+RECUL1+RENUC
 AIM=AIMCUL+AIMCUL1+AIMNUC
 S(TT)=10*(RE^2+AIM^2)/4/SS^2: NEXT TT: END SUB
 SUB POLLEG(X,L,P(20))
 P(0)=1: P(1)=X: FOR I=2 TO L
 P(I)=(2*I-1)*X/I*P(I-1)-(I-1)/I*P(I-2): NEXT: END SUB
```

```
SUB CULFAZ(G,F(20))
 REM ---------------- COULOMB PHASE SHIFTS -----------------
 C=0.577215665: S=0: N=50: A1=1.202056903/3: A2=1.036927755/5
 FOR I=1 TO N: A=G/I-ATN(G/I)-(G/I)^3/3+(G/I)^5/5
 S=S+A: NEXT: FAZ=-C*G+A1*G^3-A2*G^5+S: F(0)=FAZ
 FOR I=1 TO 20: F(I)=F(I-1)+ATN(G/(I)): NEXT: END SUB
```

Контрольный счет по этой программе выполнен для сечений упругого рассеяния в $^4He^4He$ системе при энергии 29.5 МэВ с данными из работы [200]. В работе приведены экспериментальные сечения $\sigma_e$ вместе со своими ошибками, показанные в таблице 6.1 в зависимости от угла рассеяния θ и результаты фазового анализа, которые даны в таблице 6.2 для E = 29.5 МэВ.

Таблица 6.1 - Сечения рассеяния

| | E = 29.5 МэВ | | | E = 18.0 МэВ | | |
|---|---|---|---|---|---|---|
| θ, град | $\sigma_e$ , мб/ст [200] | $\Delta\sigma_e$ , мб/ст [200] | $\sigma_t$ , мб/ст [200] | $\sigma_e$, мб/ст [200] | $\Delta\sigma_e$ , мб/ст [200] | $\sigma_t$ , мб/ст [200] |
| 22 | 1523 | 11.9 | 1521.64 | 961.9 | 3.6 | 950.59 |
| 30 | 422.6 | 5.7 | 425.14 | 557.6 | 3.1 | 559.08 |
| 40 | 44.68 | 0.25 | 44.51 | 219.2 | 1.5 | 221.06 |
| 50 | 135.58 | 0.7 | 134.30 | 79.81 | 0.6 | 81.03 |
| 60 | 107.83 | 0.8 | 108.36 | 55.79 | 0.3 | 55.52 |
| 70 | 33.77 | 0.2 | 33.70 | 97.20 | 0.6 | 96.96 |
| 80 | 124.20 | 1.0 | 121.70 | 165.18 | 0.78 | 165.32 |
| 90 | 211.20 | 1.1 | 211.67 | 199.16 | 0.84 | 199.22 |

Таблица 6.2 - Фазы рассеяния

| E, МэВ | $\delta_0$, град | $\delta_2$, град | $\delta_4$, град | $\delta_6$, град | $\delta_8$, град |
|---|---|---|---|---|---|
| 29.5 | -29.12±0.17 | 86.90±0.13 | 121.19±0.17 | 2.20±0.11 | 0.11±0.08 |

Ниже приведены подробные результаты расчета сечений $\sigma_t$ при 29.5 МэВ с точными значениями углов рассеяния и табличными фазами, вычислением $\chi_i^2$ на каждую точку (экспериментальные ошибки из [200]) и среднего $\chi^2$ по всем точкам по нашей программе.

$\chi^2 = 1.086$ $\sigma_s = 1044.66$ мб

| θ | $\sigma_t$ | $\sigma_e$ | $\chi_i^2$ | θ | $\sigma_t$ | $\sigma_e$ | $\chi_i^2$ |
|---|---|---|---|---|---|---|---|
| 22.04 | 1514.27 | 1523.00 | 0.54 | 56.09 | 140.40 | 139.86 | 0.58 |
| 24.05 | 1167.29 | 1164.00 | 0.10 | 58.10 | 126.34 | 127.10 | 0.84 |
| 26.05 | 869.26 | 885.90 | 3.53 | 60.10 | 107.33 | 107.83 | 0.39 |
| 28.05 | 620.34 | 616.10 | 0.35 | 62.10 | 86.08 | 86.66 | 0.52 |
| 30.06 | 419.86 | 422.60 | 0.23 | 64.10 | 65.55 | 66.12 | 0.70 |
| 32.06 | 268.10 | 270.00 | 0.22 | 66.10 | 48.53 | 48.43 | 0.04 |
| 34.06 | 160.37 | 160.20 | 0.00 | 68.11 | 37.33 | 37.43 | 0.10 |
| 36.07 | 91.26 | 91.50 | 0.02 | 70.11 | 33.71 | 33.77 | 0.11 |
| 38.07 | 55.13 | 55.53 | 0.29 | 72.11 | 38.34 | 38.34 | 0.00 |
| 40.07 | 44.49 | 44.68 | 0.61 | 74.11 | 51.05 | 50.74 | 0.33 |
| 42.08 | 52.11 | 52.96 | 2.96 | 76.11 | 70.77 | 70.82 | 0.00 |
| 44.08 | 70.83 | 71.74 | 1.53 | 78.11 | 95.64 | 95.55 | 0.01 |
| 46.08 | 94.23 | 95.44 | 2.13 | 80.11 | 123.24 | 124.20 | 0.89 |
| 48.08 | 116.95 | 118.46 | 3.42 | 82.11 | 150.84 | 153.40 | 5.83 |
| 50.09 | 134.96 | 135.58 | 0.84 | 84.11 | 175.70 | 177.50 | 3.06 |
| 52.09 | 145.40 | 145.62 | 0.16 | 86.11 | 195.31 | 197.00 | 2.71 |
| 54.09 | 147.16 | 147.60 | 0.36 | 88.11 | 207.72 | 209.74 | 4.40 |

В работе [200] для среднего $\chi^2$ была получена величина 0.68. Однако нельзя напрямую сравнивать ее, с приведенным выше значением, поскольку в [200] использовались несколько другие методы расчета $\chi^2$, которые учитывают некоторый весовой множитель. Если учесть весовые множители из [200] и поделить среднее $\chi^2$ на степени свободы, как это делается в работе [200], то можно получить величину 0.6, вполне согласующуюся с результатами [200].

## 6.2 Система частиц с полным спином 1/2

Рассмотрим теперь рассеяние в системе частиц с полным спином 1/2, т.е. одна частица имеет нулевой, а вторая полуцелый спин и учтем спин - орбитальное расщепление фаз. Такое рассеяние имеет место в ядерных системах типа $N^4He$, $^3H^4He$ и т.д.

Сечение упругого рассеяния таких частиц представляется в виде [57]

$$\frac{d\sigma(\theta)}{d\Omega} = |A(\theta)|^2 + |B(\theta)|^2 \ , \qquad (6.10)$$

где

$$A(\theta) = f_c(\theta) + \frac{1}{2ik}\sum_{L=0}^{\infty}\{(L+1)S_L^+ + LS_L^- - (2L+1)\}\exp(2i\sigma_L)P_L(\cos\theta) \ ,$$

$$B(\theta) = \frac{1}{2ik}\sum_{L=0}^{\infty}(S_L^+ - S_L^-)\exp(2i\sigma_L)P_L^1(\cos\theta) \ . \qquad (6.11)$$

Здесь $S_L^{\pm} = \eta_L^{\pm}\exp(2i\delta_L^{\pm})$ - матрица рассеяния, $\eta^{\pm}_L$ - параметры неупругости, а знаки "±" соответствуют полному моменту системы J = L ± 1/2, $f_c$ - кулоновская амплитуда, описанная выше (6.2), $P_n^m(x)$ - присоединенные полиномы или функции Лежандра [73]

$$P_n^m(x) = (1 - x^2)^{m/2}\frac{d^m P_n(x)}{dx^m} \quad ,$$

которые можно вычислять по рекуррентным формулам вид

$$P_{L+1}^1(x) = \frac{(2L+1)x}{L}P_L^1(x) - \frac{L+1}{L}P_{L-1}^1(x)$$

с начальными значениями

$$P_0^1(x) = 0 \quad , \qquad P_1^1(x) = (1-x^2)^{1/2} \quad , \qquad P_2^1(x) = 3xP_1^1(x) \quad .$$

Через величины A и B можно выразить и поляризацию в упругом рассеянии таких частиц [57]

$$P(\theta) = \frac{2\,\mathrm{Im}(AB^*)}{|A|^2 + |B|^2} \quad . \tag{6.12}$$

Расписывая выражение (6.11) для B получим

$$\mathrm{Re}\,B = \frac{1}{2k}\sum_{L=0}^{\infty}[a\mathrm{Sin}(2\sigma_L) + b\mathrm{Cos}(2\sigma_L)]P_L^1(x) \quad , \tag{6.13}$$

$$\mathrm{Im}\,B = \frac{1}{2k}\sum_{L=0}^{\infty}[b\mathrm{Sin}(2\sigma_L) - a\mathrm{Cos}(2\sigma_L)]P_L^1(x) \quad ,$$

где

$$\begin{aligned} a &= \eta_L^+\mathrm{Cos}(2\delta_L^+) - \eta_L^-\mathrm{Cos}(2\delta_L^-) \\ b &= \eta_L^+\mathrm{Sin}(2\delta_L^+) - \eta_L^-\mathrm{Sin}(2\delta_L^-) \end{aligned} \quad .$$

Аналогичным способом для амплитуды A можно найти следующие выражения [201]

$$\mathrm{Re}\,A = \mathrm{Re}\,f_c + \frac{1}{2k}\sum_{L=0}^{\infty}[c\,\mathrm{Sin}(2\sigma_L) + d\,\mathrm{Cos}(2\sigma_L)]P_L(x) \ ,$$

$$\mathrm{Im}\,A = \mathrm{Im}\,f_c + \frac{1}{2k}\sum_{L=0}^{\infty}[d\,\mathrm{Sin}(2\sigma_L) - c\,\mathrm{Cos}(2\sigma_L)]P_L(x) \ , \qquad (6.14)$$

где

$$c = (L+1)\eta_L^+\mathrm{Cos}(2\delta_L^+) + L\eta_L^-\mathrm{Cos}(2\delta_L^-) - (2L+1)$$

.

$$d = (L+1)\eta_L^+\mathrm{Sin}(2\delta_L^+) + L\eta_L^-\mathrm{Sin}(2\delta_L^-)$$

Для вычисления поляризации можно использовать следующие соотношения

$B = \mathrm{Re}B + i\mathrm{Im}B = C + iD$ , $A = \mathrm{Re}A + i\mathrm{Im}A = G + iF$ ,
$AB^* = CG + FD + i(FC - GD)$ ,
$\mathrm{Im}(AB^*) = FC - GD = \mathrm{Re}B\ \mathrm{Im}A - \mathrm{Re}A\ \mathrm{Im}B$ ,

которые непосредственно используются для численных компьютерных расчетов.

Для полного сечения реакций существует выражение, аналогичное системе с нулевым спином [57]

$$\sigma_r = \frac{\pi}{k^2}\sum_L \left[ (L+1)\left(1 - \left|S_L^+\right|^2\right) + L\left(1 - \left|S_L^-\right|^2\right)\right] \ , \qquad (6.15)$$

а для полного сечения упругого рассеяния

$$\sigma_s = \frac{\pi}{k^2}\sum_L \left[ (L+1)\left|1 - S_L^+\right|^2 + L\left|1 - S_L^-\right|^2 \right] \ . \qquad (6.16)$$

Расписать их можно точно так же, как в предыдущем случае, например

$$\sigma_r = \frac{\pi}{k^2}\sum_L \left\{(L+1)\left[1 - \left(\eta_L^+\right)^2\right] + L\left[1 - \left(\eta_L^-\right)^2\right]\right\} \ . \qquad (6.17)$$

Приведем текст компьютерной программы для расчета полных и дифференциальных сечений упругого рассеяния частиц с полуцелым спином [201]. Используемые здесь обозначения практически совпадают с обозначениями в предыдущей программе.

```
 REM CROSS SECTION CALCULATE OF COMPLEX PHASE
SHIFTS FOR SYSTEM WITH 1/2 SPIN
 CLS: DEFDBL A-Z: DEFINT I,J,K,L,N,M: N=200
 DIM E(N), DE(N), E1(N), ETA(N), SEC(N), SECE(N), FM(N/10),
FR(N/10), POL(N)
 REM ************************************************
 ISAVE=0: REM =0 - NO SAVE, =1 - SAVE IN FILE
 G$="C:\BASICA\SEC\AL-N1.DAT"
 REM ************ INPUT PARAMETERS *************
 PI=4*ATN(1): NN=0: NV=0: LN=0: LV=2: LH=1: TMI=10:
TMA=170
 TH=2: AM1=1: AM2=4: Z1=1: Z2=2: A1=41.4686
 PM=AM1*AM2/(AM1+AM2)
 B1=2*PM/A1
 REM ********* ENERGY IN LAB. SYSTEM ***********
 E1(0)=9.954
 REM **************** PHASE SHIFTS ***************
 REM ************ FOR P-AL ON E=9.954 ************
 FP(0)=111: FP(1)=103: FP(2)=-2: FM(0)=111: FM(1)=52: FM(2)=-4
111 FOR L=LN TO LV STEP LH: FM(L)=FM(L)*PI/180
 FP(L)=FP(L)*PI/180: NEXT
 REM ************ TRANSFORM TO C.M. *************
 FOR I=NN TO NV: E(I)=E1(I)*PM/AM1: NEXT I
 REM ********** TOTAL CROSS SECTION ***********
 FOR J=NN TO NV: SK=E(J)*B1: SS=SQR(SK)
 GG=3.44476E-02*Z1*Z2*PM/SS: SIGMAR=0: SIGMAS=0
 FOR L=LN TO LV STEP LH: AP=FP(L): AM=FM(L): ETAP(L)=1
 ETAM(L)=1
 SIGMAR = SIGMAR + ((L + 1)*(1 - (ETAP(L))^2) + L*(1 -
(ETAM(L))^2))
 SIGMAS = SIGMAS + ((L + 1)*(ETAP(L))^2*(SIN(AP))^2 +
L*(ETAM(L))^2*(SIN(AM))^2): NEXT L
 SIGMAR=10*4*PI*SIGMAR/SK: SIGMAS=10*4*PI*SIGMAS/SK
 PRINT "                SIGMR-TOT=";
 PRINT USING " ####.### ";SIGMAR
 PRINT "                SIGMS-TOT=";
 PRINT USING " ####.### ";SIGMAS: NEXT J: PRINT
 REM ******* DIFFERENTIAL CROSS SECTION *******
 CALL SEC (FP(), GG, SS, TMI, TMA, TH, SEC(), FM(), LN, LV, LH,
POL())
 FOR T=TMI TO (TMA/3.3) STEP TH
 PRINT USING " ### ";T;: PRINT USING " ####.##      ";SEC(T);
```

```
 PRINT T+42;: PRINT USING " ####.##       ";SEC(T+42);
 PRINT T+84;: PRINT USING " ####.##       ";SEC(T+84);
 PRINT T+126;: PRINT USING " ####.##       ";SEC(T+126): NEXT
 PRINT "-----------------------------------------------------------------"
 FOR T=TMI TO (TMA/3.3) STEP TH
 PRINT T;: PRINT USING "+###.###       ";POL(T);
 PRINT T+42;: PRINT USING "+###.###       ";POL(T+42);
 PRINT T+84;: PRINT USING "+###.###       ";POL(T+84);
 PRINT T+126;: PRINT USING "+###.###       ";POL(T+126): NEXT
 REM ***************   SAVE IN FILE   ***************
 IF ISAVE=0 GOTO 221
 OPEN "O",1,G$: PRINT#1, "            P - ALPHA FOR LAB E=";
 PRINT#1, E1(NN): FOR T=TMI TO TMA STEP TH
 PRINT#1, USING "  +#.###^^^^  ";T;SEC(T);POL(T): NEXT
221 END
```

**SUB SEC (FP(100), GG, SS, TMI, TMA, TH, S(100), FM(100), LMI, LMA, LH, POL(100))**

```
 REM ----------- CROSS SECTION CALCULATION ------------
 SHARED PI: DIM S0(20),P(20),PP(20)
 CALL CULFAZ(GG,S0()): FOR TT=TMI TO TMA STEP TH
 T=TT*PI/180: X=COS(T): A=2/(1-X): S0=2*S0(0): BB=-GG*A
 ALO=GG*LOG(A)+S0: REC=BB*COS(ALO): AMC=BB*SIN(ALO)
 REA=0: AMA=0: REB=0: AMB=0: FOR L=LMI TO LMA STEP LH
 FP=2*FP(L): FM=2*FM(L): A=COS(FP)-COS(FM): B=SIN(FP)-SIN(FM)
 SL=2*S0(L): CALL FUNLEG(X,L,PP())
 REB=REB+(B*COS(SL)+A*SIN(SL))*PP(L)
 AMB=AMB+(B*SIN(SL)-A*COS(SL))*PP(L): LL=2*L+1: JJ=L+1
 A=JJ*COS(FP)+L*COS(FM)-LL: B=JJ*SIN(FP)+L*SIN(FM)
 CALL POLLEG(X,L,P()): REA=REA+(B*COS(SL)+A*SIN(SL))*P(L)
 AMA=AMA+(B*SIN(SL)-A*COS(SL))*P(L): NEXT L
 REA=REC+REA: AMA=AMC+AMA: RE=REA^2+AMA^2
 AM=REB^2+AMB^2: S(TT)=10*(RE+AM)/4/SS^2
 POL(TT)=2*(REB*AMA-REA*AMB)/(RE+AM): NEXT TT: END SUB
```

**SUB POLLEG(X,L,P(20))**

```
 P(0)=1: P(1)=X: FOR I=2 TO L: P(I)=(2*I-1)*X/I*P(I-1)-(I-1)/I*P(I-2)
 NEXT: END SUB
```

**SUB FUNLEG(X,L,P(20))**

```
 P(0)=0: P(1)=SQR(ABS(1-X^2)): P(2)=3*X*P(1): FOR I=2 TO L
 P(I+1)=(2*I+1)*X/I*P(I)-(I+1)/I*P(I-1): NEXT: END SUB
```

**SUB CULFAZ(G,F(20))**

```
REM ---------------- COULOMB PHASE SHIFTS ----------------
C=0.577215665: S=0: N=50: A1=1.202056903/3: A2=1.036927755/5
FOR I=1 TO N: A=G/I-ATN(G/I)-(G/I)^3/3+(G/I)^5/5: S=S+A: NEXT
FAZ=-C*G+A1*G^3-A2*G^5+S: F(0)=FAZ: FOR I=1 TO 20
F(I)=F(I-1)+ATN(G/(I)): NEXT: END SUB
```

Приведем теперь результаты контрольного счета сечений и поляризаций для $p^4He$ системы при энергии $E_p$ = 9.954 МэВ. Дифференциальные сечения $\sigma_e$ (Таблица 6.3) измерялись в работе [202], там же проведен фазовый анализ (Таблица 6.4). С найденными фазами в [202] вычислялись сечения $\sigma_t$ и поляризации $P_t(\theta)$, которые вместе с экспериментальными сечениями $\sigma_e$ также даны в таблице 6.3 [202]. Ошибки экспериментальных сечений не превышают 2%. При столь малых энергиях все фазы рассеяния число действительные и параметры неупругости равны единице.

Таблица 6.3 - Сечения рассеяния

| $\theta$, град | $\sigma_e$ , Мб/ст [202] экспер. | $\sigma_t$ , мб/ст [202] расчет | $\sigma_0$ , мб/ст Наш расчет | $P_t(\theta)$, % [202] расчет | $P_0,(\theta)$, % Наш расчет |
|---|---|---|---|---|---|
| 25.1 | 371 | 372 | 370.65 | -11 | -10.73 |
| 30.89 | 339 | 335 | 333.31 | -16 | -15.50 |
| 35.07 | 305 | 311 | 309.72 | -19 | -18.95 |
| 49.03 | 232 | 233 | 232.33 | -32 | -31.54 |
| 54.7 | 205 | 202 | 201.05 | -37 | -37.44 |
| 60.0 | 176 | 173 | 172.57 | -44 | -43.49 |
| 70.1 | 124 | 123 | 122.27 | -57 | -56.62 |
| 80.0 | 82 | 81 | 80.70 | -71 | -70.91 |
| 90.0 | 49.2 | 49.1 | 48.91 | -82 | -82.18 |
| 94.07 | 39.1 | 39.4 | 39.24 | -82 | -82.30 |
| 102.17 | 26.2 | 25.9 | 25.79 | -61 | -61.28 |
| 106.9 | 22.0 | 21.5 | 21.44 | -29 | -28.71 |
| 109.9 | 21.0 | 20.0 | 19.96 | -1 | -0.96 |
| 120.6 | 23.0 | 22.0 | 21.98 | 86 | 85.85 |
| 122.8 | 24.5 | 23.7 | 23.63 | 94 | 93.82 |
| 130.13 | 31.9 | 31.6 | 31.56 | 99 | 98.82 |
| 130.90 | 33.2 | 32.6 | 32.57 | 98 | 98.05 |
| 134.87 | 37.8 | 38.3 | 38.21 | 92 | 91.98 |
| 140.8 | 47.3 | 47.7 | 47.59 | 80 | 79.51 |
| 145.0 | 54.0 | 54.7 | 54.56 | 70 | 69.95 |
| 149.4 | 61.6 | 62.0 | 61.84 | 60 | 60.01 |
| 154.9 | 70.4 | 70.6 | 70.48 | 48 | 48.08 |
| 160.0 | 78.4 | 77.8 | 77.60 | 38 | 37.58 |
| 164.4 | 84.9 | 83 | 82.78 | 29 | 28.91 |

Таблица 6.4 - Фазы рассеяния

| $\delta_L$, град | $S_{1/2}$ | $P_{1/2}$ | $P_{3/2}$ | $D_{3/2}$ | $D_{5/2}$ |
|---|---|---|---|---|---|
| E = 9.954 МэВ | 111 | 52 | 103 | -4 | -2 |

Результаты вычисления сечений $\sigma_0$ и поляризаций $P_0$ по приведенной выше программе с табличными фазами для E= 9.954 МэВ также показаны в таблице 6.3 и приводят к величине $\chi^2$ для сечений равной 0.96 при 2% ошибках экспериментальных сечений.

Видно, что отклонение расчетных сечений от результатов расчета из работы [202] для первых трех точек около 1.0 ÷ 1.5 мб/ст, по остальным точкам менее 1.0 мб/ст, а отклонения поляризаций около 0.5%, что связано, по - видимому, с округлением численных результатов в работе [202].

Некоторые другие результаты для этой системы частиц мы приведем в следующей главе, где будут рассмотрены методы фазового анализа экспериментальных сечений рассеяния.

### 6.3 Система частиц с единичным полным спином

В случае системы частиц, когда одна из них имеет нулевой спин, а вторая равный 1, например, для $^2H^4He$ системы, формулы дифференциальных сечений с учетом только спин - орбитального взаимодействия записываются в виде [57]

$$\frac{d\sigma(\theta)}{d\Omega} = \frac{1}{3}\left[|A|^2 + 2\left(|B|^2 + |C|^2 + |D|^2 + |E|^2\right)\right] , \qquad (6.18)$$

где амплитуды рассеяния

$$A = f_c(\theta) + \frac{1}{2ik}\sum_{L=0}\{(L+1)\alpha_L^+ + L\alpha_L^-\}\exp(2i\sigma_L)P_L(\mathrm{Cos}\theta) ,$$

$$B = f_c(\theta) + \frac{1}{4ik}\sum_{L=0}\{(L+2)\alpha_L^+ + (2L+1)\alpha_L^0 + (L-1)\alpha_L^-\}\exp(2i\sigma_L)P_L(\mathrm{Cos}\theta) ,$$

$$C = \frac{1}{2ik\sqrt{2}}\sum_{L=1}\{\alpha_L^+ - \alpha_L^-\}\exp(2i\sigma_L)P_L^1(\mathrm{Cos}\theta) , \qquad (6.19)$$

$$D = \frac{1}{2ik\sqrt{2}} \sum_{L=1} \frac{1}{L(L+1)} \{L(L+2)\alpha_L^+ - (2L+1)\alpha_L^0 -$$

$$- (L-1)(L+1)\alpha_L^- \} \exp(2i\sigma_L) P_L^1(\mathrm{Cos}\theta) \ ,$$

$$E = \frac{1}{4ik} \sum_{L=2} \frac{1}{L(L+1)} \{L\alpha_L^+ - (2L+1)\alpha_L^0 + (L+1)\alpha_L^- \} \exp(2i\sigma_L) P_L^2(\mathrm{Cos}\theta) \ .$$

Здесь определена величина $\alpha^J = (S^J - 1)$ для каждого состояния с полным моментом $J = L \pm 1$ ($\alpha^+$ и $\alpha^-$) и $J = L$ ($\alpha^0$). Отметим, что существует и другая форма записи выражений для сечений, представленная через производные полиномов Лежандра [73].

Функция Лежандра $P^1{}_L(x)$ были рассмотрены выше, а $P^2{}_L(x)$ вычисляются по рекуррентным формулам типа

$$P_{L+1}^2(x) = \frac{(2L+1)x}{L-1} P_L^2(x) - \frac{L+2}{L-1} P_{L-1}^2(x)$$

с начальными значениями

$$P_0^2(x) = P_1^2(x) = 0 \ , \quad P_2^2 = 3(1-x^2) \ , \quad P_3^2(x) = 5xP_2^2(x) .$$

Полное сечение реакций и неупругих процессов записывается теперь в виде [57]

$$\sigma_r = \frac{\pi}{3k^2} \sum_L \left[ (2L+3)\left(1-\left|S_L^+\right|^2\right) + (2L+1)\left(1-\left|S_L^0\right|^2\right) + (2L-1)\left(1-\left|S_L^-\right|^2\right) \right] \ . \qquad (6.20)$$

Для векторной поляризации в такой системе существует выражение [57]

$$P = \frac{\sqrt{8}}{3} \frac{\mathrm{Im}(AC^* + BD^* + DE^*)}{d\sigma / d\Omega} \ . \qquad (6.21)$$

Кроме векторных поляризаций имеются и тензорные поляризации, которые выражаются через амплитуды рассеяния следующим образом [57]

$$t_{20} = \frac{1}{\sqrt{2}}\left(1 - \frac{|A|^2 + 2|D|^2}{d\sigma / d\Omega}\right) \quad ,$$

$$t_{21} = -\sqrt{\frac{2}{3}}\left(\frac{\mathrm{Re}(AC^* - BD^* + DE^*)}{d\sigma / d\Omega}\right) \quad , \qquad (6.22)$$

$$t_{22} = \frac{1}{\sqrt{3}}\left(\frac{2\,\mathrm{Re}(BE^*) - |C|^2}{d\sigma / d\Omega}\right) \quad .$$

Причем

$t_{11} = (3/2)^{1/2}P$ .

Расписывая выражения (6.19) и вынося за знак модуля общий делитель 2k, будем иметь

$$A = \mathrm{Re}\, f_c + i\, \mathrm{Im}\, f_c +$$

$$+\sum_L \{[A_2 \mathrm{Cos}(2\sigma_L) + A_1 \mathrm{Sin}(2\sigma_L)] + i[A_2 \mathrm{Sin}(2\sigma_L) - A_1 \mathrm{Cos}(2\sigma_L)]\} P_L^m(x) \quad ,$$

где m = 0 для A и B ($P^0{}_L = P_L$),

$$A_1 = (L+1)\alpha_1^+ + L\alpha_1^-$$
$$A_2 = (L+1)\alpha_2^+ + L\alpha_2^-$$

и

$$\alpha_L^k = \alpha_{1,L}^k + i\alpha_{2,L}^k = [\eta_L^k \mathrm{Cos}(2\delta_L^k) - 1] + i\eta_L^k \mathrm{Sin}(2\delta_L^k) \quad .$$

где k = "+", "-" и "0". Для всех остальных амплитуд получаются аналогичные выражения с заменой $A_i$ (i = 1,2) на следующие величины

$$B_i = \frac{1}{2}\left[(L+2)\alpha_i^+ + (2L+1)\alpha_i^0 + (L-1)\alpha_i^-\right] \quad ,$$

$$C_i = \frac{1}{\sqrt{2}}\left[\alpha_i^+ - \alpha_i^-\right] \quad ,$$

$$D_i = \frac{1}{\sqrt{2}\,L(L+1)}\left[L(L+2)\alpha_i^+ - (2L+1)\alpha_i^0 - (L^2-1)\alpha_i^-\right] \quad ,$$

$$E_i = \frac{1}{2L(L+1)}\left[L\alpha_i^+ - (2L+1)\alpha_i^0 + (L+1)\alpha_i^-\right] \quad .$$

Для степени функции Лежандра в амплитудах C и D принимаем m = 1, а для амплитуды E величина m = 2. Кулоновские амплитуды учитываются только в амплитудах A и B.

Приведем теперь программу для расчетов дифференциальных сечений в системе частиц с единичным спином [203]. Обозначения в этой программе практически совпадают с обозначениями для системы со спином 1/2.

**REM CALCULATE OF CROSS SECTION ON COMPLEX PHASE SHIFTS FOR SYSTEM WITH 1 SPIN**

```
CLS: DEFDBL A-Z: DEFINT I,J,K,L,N,M: N=200
DIM E(N),DE(N),E1(N),EP(N/10),EM(N/10),E0(N/10)
DIM SEC(N), SECE(N), FM(N/10), F0(N/10), FP(N/10)
REM ***********************************************
ISAVE=1: REM =0 - NO SAVE, =1 - SAVE IN FILE
G$="C:\BASICA\SEC\AL-d1.DAT"
REM ************* INPUT PARAMETERS ***********
PI=4*ATN(1): NN=1: NV=1: LN=0: LV=2: LH=1: TMI=10: TMA=170
TH=2: AM1=2: AM2=4: Z1=1: Z2=2: A1=41.4686
PM=AM1*AM2/(AM1+AM2): B1=2*PM/A1
REM ******** ENERGY IN LAB. SYSTEM ***********
E1(0)=4.3: E1(1)=4
REM **** ECSPERIMENTAL CROSS SECTION 4.0 *****
SECE(30)=220: SECE(40)=160: SECE(50)=120
SECE(60)=80: SECE(70)=60: SECE(80)=50
SECE(90)=45: SECE(100)=55: SECE(110)=70
SECE(120)=100: SECE(130)=120: SECE(140)=180
REM ************** PHASE SHIFTS ***************
REM ************* FOR D-AL ON E=4 ***************
FP(0)=99.3: FP(1)=6.6: FP(2)=172.9: F0(0)=99.3: F0(1)=1.2: F0(2)=53.2
FM(0)=99.3: FM(1)=-0.5: FM(2)=18.4: FOR L=LN TO LV STEP LH
FM(L)=FM(L)*PI/180: FP(L)=FP(L)*PI/180: F0(L)=F0(L)*PI/180: NEXT
```

```
 REM *********** TRANSFORM TO C.M. *************
 FOR I=NN TO NV: E(I)=E1(I)*PM/AM1: NEXT I
 REM ************  TOTAL CROSS SECTION  *********
 FOR J=NN TO NV: SK=E(J)*B1: SS=SQR(SK)
 GG=3.44476E-02*Z1*Z2*PM/SS: SIGMAR=0: SIGMAS=0
 FOR L=LN TO LV STEP LH: AP=FP(L): AM=FM(L): A0=F0(L)
 EP(L)=1: EM(L)=1: E0(L)=1
 SIGMAR = SIGMAR + (2 * L + 3) * (1 - (EP(L))^2) + (2 * L + 1) * (1
- (E0(L))^2) + (2 * L-1) * (1 - EM(L))^2
 SIGMAS = SIGMAS + (2 * L + 3) * (EP(L))^2 * (SIN(AP))^2 + (2 * L
+ 1) * (E0(L))^2 * (SIN(A0))^2 + (2 * L - 1) * (EM(L))^2 *
(SIN(AM))^2
 NEXT L
 SIGMAR=10*4*PI*SIGMAR/SK/3:                                 SIG-
MAS=10*4*PI*SIGMAS/SK/3
 PRINT "                    SIGMR-TOT=";: PRINT USING " ####.###
";SIGMAR
 PRINT "                    SIGMS-TOT=";: PRINT USING " ####.###
";SIGMAS
 NEXT J: PRINT
 REM ******  DIFFERENTIAL CROSS SECTION  *******
 CALL SEC (SS,GG,SEC()): FOR T=TMI TO (TMA/3.3) STEP TH
 PRINT USING " ### ";T;: PRINT USING " ####.##      ";SEC(T);
 PRINT T+42;: PRINT USING " ####.##      ";SEC(T+42);
 PRINT T+84;: PRINT USING " ####.##      ";SEC(T+84);
 PRINT T+126;: PRINT USING " ####.##      ";SEC(T+126): NEXT
 IF ISAVE=0 GOTO 221: OPEN "O",1,G$
 PRINT#1, "            P - ALPHA FOR LAB E=";
 PRINT#1, E1(NN): FOR T=TMI TO TMA STEP TH
 PRINT#1, USING "  +#.###^^^^  ";T;SEC(T): NEXT
221 END
```

**SUB SEC (SS, GG, S(100))**

```
 SHARED FP(),EP(),F0(),E0(),FM(),EM()
 SHARED PI,TMI,TMA,TH,LN,LV,LH
 DIM S0(20),P(20),P1(20),P2(20): CALL CULFAZ(GG,S0())
 FOR TT=TMI TO TMA STEP TH: T=TT*PI/180: X=COS(T)
 A=2/(1-X): S0=2*S0(0): BB=-GG*A: AL=GG*LOG(A)+S0
 RECUL=BB*COS(AL): AMCUL=BB*SIN(AL)
 REA=0: AMA=0: REB=0: AMB=0: REC=0
 AMC=0: RED=0: AMD=0: REE=0: AME=0
 FOR L=LN TO LV STEP LH: CALL POLLEG(X,L,P())
 FP=2*FP(L): FM=2*FM(L): F0=2*F0(L): SL=2*S0(L)
 C=COS(SL): S=SIN(SL): AL1P=EP(L)*COS(FP)-1
 AL2P=EP(L)*SIN(FP): AL1M=EM(L)*COS(FM)-1
```

```
AL2M=EM(L)*SIN(FM): AL10=E0(L)*COS(F0)-1
AL20=E0(L)*SIN(F0): A1=(L+1)*AL1P+L*AL1M
A2=(L+1)*AL2P+L*AL2M: REA=REA+(A2*C+A1*S)*P(L)
AMA=AMA+(A2*S-A1*C)*P(L)
B1=(L+2)*AL1P+(2*L+1)*AL10+(L-1)*AL1M
B2=(L+2)*AL2P+(2*L+1)*AL20+(L-1)*AL2M
REB=REB+(B2*C+B1*S)*P(L)/2: AMB=AMB+(B2*S-B1*C)*P(L)/2
NEXT L: FOR L=1 TO LV STEP LH: CALL FUNLEG1(X,L,P1())
FP=2*FP(L): FM=2*FM(L): F0=2*F0(L): SL=2*S0(L)
C=COS(SL): S=SIN(SL): AL1P=EP(L)*COS(FP)-1
AL2P=EP(L)*SIN(FP): AL1M=EM(L)*COS(FM)-1
AL2M=EM(L)*SIN(FM): AL10=E0(L)*COS(F0)-1
AL20=E0(L)*SIN(F0): C1=AL1P-AL1M: C2=AL2P-AL2M
BB=1/(SQR(2)): REC=REC+(C2*C+C1*S)*P1(L)*BB
AMC=AMC+(C2*S-C1*C)*P1(L)*BB: AA=1/(SQR(2)*L*(L+1))
D1=L*(L+2)*AL1P-(2*L+1)*AL10-(L^2-1)*AL1M
D2=L*(L+2)*AL2P-(2*L+1)*AL20-(L^2-1)*AL2M
RED=RED+(D2*C+D1*S)*P1(L)*AA
AMD=AMD+(D2*S-D1*C)*P1(L)*AA
NEXT L: FOR L=2 TO LV STEP LH: CALL FUNLEG2(X,L,P2())
FP=2*FP(L): FM=2*FM(L): F0=2*F0(L): SL=2*S0(L): C=COS(SL)
S=SIN(SL): AL1P=EP(L)*COS(FP)-1: AL2P=EP(L)*SIN(FP)
AL1M=EM(L)*COS(FM)-1
AL2M=EM(L)*SIN(FM): AL10=E0(L)*COS(F0)-1
AL20=E0(L)*SIN(F0)
CC=1/(2*L*(L+1)): E1=L*AL1P-(2*L+1)*AL10+(L+1)*AL1M
E2=L*AL2P-(2*L+1)*AL20+(L+1)*AL2M
REE=REE+(E2*C+E1*S)*P2(L)*CC
AME=AME+(E2*S-E1*C)*P2(L)*CC
NEXT L: REA=RECUL+REA: AMA=AMCUL+AMA
REB=RECUL+REB
AMB=AMCUL+AMB: A=REA^2+AMA^2: B=REB^2+AMB^2
C=REC^2+AMC^2: D=RED^2+AMD^2: E=REE^2+AME^2
SEC=(A+2*(B+C+D+E))/3: S(TT)=10*SEC/4/SS^2
POL(TT) = 2 * SQR(2) / 3 * (AMA * REC - REA * AMC + AMB *
RED - REB * AMD + AMD * REE - RED * AME) / SEC
T20(TT) = 1/SQR(2)*(1 - (AA + 2*DD)/SEC)
T22(TT) = 1/SQR(3) * (2 * (REB * REE + AMB * AME) - CC) / SEC
T21(TT) = - SQR(2 / 3) * (REA * REC + AMA * AMC - REB * RED -
AMB * AMD + RED * REE + AMD * AME) / SEC: NEXT TT: END
SUB
```

**SUB POLLEG(X,L,P(20))**

```
P(0)=1: P(1)=X: FOR I=2 TO L: P(I)=(2*I-1)*X/I*P(I-1)-(I-1)/I*P(I-2)
NEXT: END SUB
```

```
SUB FUNLEG1(X,L,P(20))
P(0)=0: P(1)=SQR(ABS(1-X^2)): FOR I=2 TO L
P(I)=(2*I-1)*X/(I-1)*P(I-1)-I/(I-1)*P(I-2): NEXT: END SUB
SUB FUNLEG2(X,L,P(20))
P(0)=0: P(1)=0: P(2)=3*ABS(1-X^2): FOR I=2 TO L
P(I+1)=(2*I+1)*X/(I-1)*P(I)-(I+2)/(I-1)*P(I-1): NEXT: END SUB
SUB CULFAZ(G,F(20))
C=0.577215665: S=0: N=50: A1=1.202056903/3: A2=1.036927755/5
FOR I=1 TO N: A=G/I-ATN(G/I)-(G/I)^3/3+(G/I)^5/5: S=S+A: NEXT
FAZ=-C*G+A1*G^3-A2*G^5+S: F(0)=FAZ: FOR I=1 TO 20
F(I)=F(I-1)+ATN(G/(I)): NEXT: END SUB
```

Дадим результаты контрольного счета по этой программе для сечений упругого рассеяния в $^2H^4He$ системе при энергии налетающего дейтрона 4 МэВ [92]. В работе [92] выполнен и фазовый анализ сечений, результаты которого показаны в таблице 6.5. Параметр смешивания при этой энергии оказывается равен $-2^0$ и смешивание парциальных волн практически отсутствует.

Результаты по экспериментальным сечениям $\sigma_e$ в работе [92] приведены только на рисунке, поэтому в нашей таблице 6.6, в скобках, показана точность, с которой их удается определить из этих рисунков. Результаты расчета сечений $\sigma_t$ по приведенной выше программе с табличными фазами (Таблица 6.5) также приведены в таблице 6.6.

Таблица 6.5 - Фазы рассеяния.

| $\delta_L$, град | $S_1$ | $P_0$ | $P_1$ | $P_2$ | $D_1$ | $D_2$ | $D_3$ |
|---|---|---|---|---|---|---|---|
| $E_\alpha$ = 4.0 МэВ | 99.3 | -0.5 | 1.2 | 6.6 | 18.4 | 14.7 | 171.6 |

Таблица 6.6 - Сечения рассеяния.

| θ, град | $\sigma_t$, мб/ст | $\sigma_e$, мб/ст [92] | θ, град | $\sigma_t$, мб/ст | $\sigma_e$, мб/ст [92] |
|---|---|---|---|---|---|
| 12 | 4009.3 | | 92 | 47.17 | |
| 14 | 1982.5 | | 94 | 48.16 | |
| 16 | 1096.3 | | 96 | 49.46 | |
| 18 | 678.10 | | 98 | 51.08 | |
| 20 | 469.14 | 450(50) | 100 | 53.04 | 50(5) |
| 22 | 359.72 | | 102 | 55.35 | |
| 24 | 299.69 | | 104 | 58.02 | |
| 26 | 264.83 | | 106 | 61.08 | |
| 28 | 242.89 | | 108 | 64.56 | |
| 30 | 227.56 | 220(10) | 110 | 68.48 | 70(5) |
| 32 | 215.48 | | 112 | 72.88 | |
| 34 | 204.90 | | 114 | 77.76 | |
| 36 | 194.93 | | 116 | 83.18 | |
| 38 | 185.12 | | 118 | 89.14 | |
| 40 | 175.30 | 170(10) | 120 | 95.66 | 100(10) |
| 42 | 165.44 | | 122 | 102.76 | |
| 44 | 155.58 | | 124 | 110.45 | |
| 46 | 145.80 | | 126 | 118.72 | |
| 48 | 136.19 | | 128 | 127.58 | |
| 50 | 126.84 | 125(10) | 130 | 137.01 | 125(10) |
| 52 | 117.83 | | 132 | 146.98 | |
| 54 | 109.25 | | 134 | 157.47 | |
| 56 | 101.15 | | 136 | 168.42 | |
| 58 | 93.58 | | 138 | 179.80 | |
| 60 | 86.58 | 85(5) | 140 | 191.54 | 185(10) |
| 62 | 80.16 | | 142 | 203.57 | |
| 64 | 74.36 | | 144 | 215.82 | |
| 66 | 69.15 | | 146 | 228.20 | |
| 68 | 64.54 | | 148 | 240.63 | |
| 70 | 60.51 | 60(5) | 150 | 253.01 | 250(10) |
| 72 | 57.05 | | 152 | 265.24 | |
| 74 | 54.11 | | 154 | 277.21 | |
| 76 | 51.69 | | 156 | 288.83 | |
| 78 | 49.74 | | 158 | 299.99 | |
| 80 | 48.24 | 50(5) | 160 | 310.59 | 300(10) |
| 82 | 47.15 | | 162 | 320.53 | |
| 84 | 46.46 | | 164 | 329.71 | |
| 86 | 46.13 | | 166 | 338.04 | |
| 88 | 46.15 | | 168 | 345.44 | |
| 90 | 46.50 | 45(5) | 170 | 351.83 | 350(10) |

Рассмотрим еще один пример контрольного счета сечений по фазам из работы [91], где были измерены дифференциальные сечения $\sigma_e$ и поляризации упругого $^2H^4He$ рассеяния при энергиях 3 - 11.5 МэВ. Возьмем энергию $E_d$ = 6.34 МэВ, при которой уже заметны неупругие процессы, но еще не проявляется сильное смешивание по орбитальным моментам, т.е. параметр смешивания $\varepsilon_1$ близок к нулю [91]. Фазы, полученные в работах [204,205,206] при энергии 6.24 МэВ приведены в таблице 6.7, а результаты расчетов $\sigma_t$, полученными по нашей программе с этими табличными фазами показаны далее на распечатке.

Для полного сечения реакций $\sigma_r$ с такими фазами получается величина 336 мб, а в работе [91] приведено экспериментальное значение 380(10) мб. Отметим, что в работе [207] для сечения реакций получено 290(20) мб, а в работах [204-206] приводится величина 200(10) мб. В работе [207] для параметра смешивания приводится значение равное $0\pm5^0$ при энергии дейтрона до 6.2 - 6.3 МэВ. При большей энергии происходит резкий скачек параметра смешивания и уже при 6.5 МэВ он равен $-35^0 \div -45^0$ ($\pm10^0$), а при 7 МэВ его значение уменьшается до $-80^0\pm15^0$.

Таблица 6.7 - Фазы рассеяния.

| $\delta_L$, град | $S_1$ | $P_0$ | $P_1$ | $P_2$ | $D_1$ | $D_2$ | $D_3$ | $F_2$ | $F_3$ | $F_4$ |
|---|---|---|---|---|---|---|---|---|---|---|
| Re δ | 78.0 | -21.5 | 2.6 | 10.5 | 63.0 | 123.1 | 168.4 | 3.3 | 5.0 | 2.8 |
| Im δ | 0.9 | 2.7 | 3.5 | 4.2 | 4.7 | 2.4 | 5.5 | 0.2 | 2.1 | 2.5 |

$\chi^2 = 3.85$ $\sigma_r =$ 336.0 мб. $\sigma_s = 1236.7$ мб.

| θ | $\sigma_e$ | $\sigma_t$ | $\chi^2_i$ | θ | $\sigma_e$ | $\sigma_t$ | $\chi^2_i$ |
|---|---|---|---|---|---|---|---|
| 22.48 | 450.50 | 413.62 | 7.45 | 109.71 | 63.80 | 60.70 | 2.62 |
| 26.20 | 340.20 | 332.53 | 0.56 | 115.09 | 65.80 | 64.00 | 0.83 |
| 29.91 | 263.10 | 274.86 | 2.22 | 120.00 | 69.30 | 69.23 | 0.00 |
| 33.60 | 214.30 | 227.44 | 4.18 | 125.09 | 79.60 | 77.73 | 0.61 |
| 37.28 | 178.60 | 186.28 | 2.05 | 129.71 | 89.20 | 88.68 | 0.04 |
| 44.57 | 117.30 | 120.14 | 0.65 | 130.00 | 92.40 | 89.48 | 1.11 |
| 51.78 | 77.20 | 75.08 | 0.84 | 134.08 | 103.70 | 102.04 | 0.28 |
| 58.87 | 53.00 | 49.66 | 4.41 | 135.00 | 107.70 | 105.22 | 0.59 |
| 65.84 | 43.30 | 39.63 | 7.99 | 138.22 | 119.60 | 117.30 | 0.41 |
| 72.67 | 42.60 | 39.52 | 5.82 | 140.00 | 129.10 | 124.57 | 1.37 |
| 79.34 | 47.00 | 44.14 | 4.11 | 145.00 | 156.10 | 146.87 | 3.88 |
| 85.84 | 52.60 | 49.69 | 3.40 | 150.00 | 184.90 | 171.05 | 6.24 |
| 92.13 | 58.50 | 54.10 | 6.29 | 155.00 | 217.50 | 195.67 | 11.20 |
| 98.22 | 60.50 | 56.91 | 3.92 | 160.00 | 247.30 | 219.07 | 14.48 |
| 104.08 | 62.00 | 58.73 | 3.10 | 165.00 | 270.70 | 239.55 | 14.72 |

В наших расчетах экспериментальная ошибка принималась равной 3% [91].

Ниже приведены расчетные и экспериментальные [204-206] поляризации для энергии 6.34 МэВ. В качестве $T_{11}(\theta)$ использована формула для $P(\theta)$ из [57] без перевода ее в $T_{11}$.

*Поляризации $T_{11}$*

| θ | $T_e$ | $T_t$ | θ | $T_e$ | $T_t$ |
|---|---|---|---|---|---|
| 22.48 | -0.200 | -0.220 | 109.71 | -0.162 | -0.130 |
| 26.20 | -0.225 | -0.233 | 115.09 | -0.100 | -0.073 |
| 29.91 | -0.240 | -0.239 | 120.00 | +0.000 | +0.010 |
| 33.60 | -0.264 | -0.242 | 125.09 | +0.116 | +0.108 |
| 37.28 | -0.263 | -0.242 | 129.71 | +0.219 | +0.189 |
| 44.57 | -0.262 | -0.228 | 130.00 | +0.216 | +0.193 |
| 51.78 | -0.209 | -0.168 | 134.08 | +0.289 | +0.246 |
| 58.87 | -0.075 | -0.031 | 135.00 | +0.306 | +0.256 |
| 65.84 | +0.108 | +0.132 | 138.22 | +0.322 | +0.280 |
| 72.67 | +0.204 | +0.193 | 140.00 | +0.336 | +0.288 |
| 79.34 | +0.169 | +0.143 | 145.00 | +0.329 | +0.294 |
| 85.84 | +0.070 | +0.051 | 150.00 | +0.310 | +0.279 |
| 92.13 | -0.038 | -0.039 | 155.00 | +0.272 | +0.249 |
| 98.22 | -0.127 | -0.109 | 160.00 | +0.227 | +0.208 |
| 104.08 | -0.179 | -0.142 | 165.00 | +0.168 | +0.161 |

*Поляризации $T_{20}$*

| θ | $T_e$ | $T_t$ | θ | $T_e$ | $T_t$ |
|---|---|---|---|---|---|
| 22.48 | +0.165 | +0.157 | 109.71 | +0.121 | +0.283 |
| 26.20 | +0.123 | +0.120 | 115.09 | -0.025 | +0.148 |
| 29.91 | +0.088 | +0.080 | 120.00 | -0.145 | +0.042 |
| 33.60 | +0.054 | +0.040 | 125.09 | -0.201 | -0.028 |
| 37.28 | +0.012 | -0.001 | 129.71 | -0.183 | -0.050 |
| 44.57 | -0.070 | -0.086 | 130.00 | -0.185 | -0.050 |
| 51.78 | -0.132 | -0.153 | 134.08 | -0.158 | -0.039 |
| 58.87 | -0.086 | -0.123 | 135.00 | -0.141 | -0.034 |
| 65.84 | +0.103 | +0.080 | 138.22 | -0.118 | -0.008 |
| 72.67 | +0.346 | +0.349 | 140.00 | -0.067 | +0.009 |
| 79.34 | +0.502 | +0.525 | 145.00 | -0.013 | +0.062 |
| 85.84 | +0.540 | +0.594 | 150.00 | +0.054 | +0.118 |
| 92.13 | +0.492 | +0.586 | 155.00 | +0.089 | +0.168 |
| 98.22 | +0.399 | +0.523 | 160.00 | +0.156 | +0.211 |
| 104.08 | +0.271 | +0.416 | 165.00 | +0.189 | +0.245 |

*Поляризации $T_{22}$*

| θ | $T_e$ | $T_t$ | θ | $T_e$ | $T_t$ |
|---|---|---|---|---|---|
| 22.48 | -0.040 | -0.048 | 109.71 | +0.350 | +0.268 |
| 26.20 | -0.068 | -0.068 | 115.09 | +0.180 | +0.102 |
| 29.91 | -0.090 | -0.090 | 120.00 | +0.013 | -0.038 |
| 33.60 | -0.110 | -0.115 | 125.09 | -0.108 | -0.147 |
| 37.28 | -0.173 | -0.143 | 129.71 | -0.171 | -0.204 |
| 44.57 | -0.234 | -0.202 | 130.00 | -0.175 | -0.206 |
| 51.78 | -0.268 | -0.235 | 134.08 | -0.206 | -0.222 |
| 58.87 | -0.182 | -0.149 | 135.00 | -0.197 | -0.223 |
| 65.84 | +0.096 | +0.119 | 138.22 | -0.198 | -0.216 |
| 72.67 | +0.411 | +0.420 | 140.00 | -0.188 | -0.208 |
| 79.34 | +0.646 | +0.596 | 145.00 | -0.164 | -0.175 |
| 85.84 | +0.705 | +0.651 | 150.00 | -0.127 | -0.136 |
| 92.13 | +0.693 | +0.628 | 155.00 | -0.099 | -0.098 |
| 98.22 | +0.623 | +0.549 | 160.00 | -0.066 | -0.064 |
| 104.08 | +0.503 | +0.425 | 165.00 | -0.028 | -0.036 |

*Поляризации $T_{21}$*

| θ | $T_e$ | $T_t$ | θ | $T_e$ | $T_t$ |
|---|---|---|---|---|---|
| 22.48 | -0.045 | -0.051 | 109.71 | -0.068 | -0.114 |
| 26.20 | -0.059 | -0.055 | 115.09 | -0.039 | -0.102 |
| 29.91 | -0.060 | -0.057 | 120.00 | +0.003 | -0.078 |
| 33.60 | -0.058 | -0.058 | 125.09 | +0.045 | -0.045 |
| 37.28 | -0.064 | -0.059 | 129.71 | +0.070 | -0.015 |
| 44.57 | -0.058 | -0.057 | 130.00 | +0.066 | -0.013 |
| 51.78 | -0.045 | -0.045 | 134.08 | +0.890 | +0.010 |
| 58.87 | -0.017 | -0.016 | 135.00 | +0.078 | +0.015 |
| 65.84 | +0.007 | +0.017 | 138.22 | +0.091 | +0.028 |
| 72.67 | +0.001 | +0.024 | 140.00 | +0.090 | +0.034 |
| 79.34 | -0.019 | +0.004 | 145.00 | +0.085 | +0.045 |
| 85.84 | -0.042 | -0.028 | 150.00 | +0.084 | +0.049 |
| 92.13 | -0.064 | -0.061 | 155.00 | +0.072 | +0.048 |
| 98.22 | -0.082 | -0.090 | 160.00 | +0.053 | +0.042 |
| 104.08 | -0.081 | -0.109 | 165.00 | +0.048 | +0.034 |

## 6.4 Система частиц с единичным спином и тензорными силами

В системе $^2H^4He$ возможно и смешивание состояний с различным орбитальным моментом, когда мы учитываем наличие тензорной компоненты ядерных сил. В частности имеется смешивание состояния с J = 1, которое возможно при L = 0 и 2. В таком случае вводят параметр смешивания $\varepsilon_J$ и формулы для сечений несколько усложняются [207]

$$\frac{d\sigma(\theta)}{d\Omega} = \frac{1}{3k^2}\left[|A|^2 + 2|B|^2 + |C|^2 + |D|^2 + |E|^2\right] , \qquad (6.23)$$

где k - волновое число, θ - угол рассеяния в системе центра масс, x = Cos(θ), а амплитуды рассеяния имеют вид

$$A = f_c + \sum_L [\exp(\frac{1}{2} i\alpha_L)/2i]\{\exp(\frac{1}{2} i\alpha_L) P_L(x)[L_4 U^{L+1}_{L,L} + LU^{L-1}_{L,L} - L_2] -$$

$$- \exp(\frac{1}{2} i\alpha_{L+2}) P_{L+2}(x)\sqrt{L_1 L_4}\, U^{L+1}_{L,L+2} - \exp(\frac{1}{2} i\alpha_{L-2}) P_{L-2}(x)\sqrt{LL_3}\, U^{L-1}_{L,L-2}\} \;, \quad (6.24)$$

$$B = f_c + \sum_L [\exp(\frac{1}{2} i\alpha_L)/4i]\{\exp(\frac{1}{2} i\alpha_L) P_L(x)[L_1 U^{L+1}_{L,L} + L_2 U^{L}_{L,L} + L_3 U^{L-1}_{L,L} - 2L_2] +$$

$$+ \exp(\frac{1}{2} i\alpha_{L+2}) P_{L+2}(x)\sqrt{L_1 L_4}\, U^{L+1}_{L,L+2} + \exp(\frac{1}{2} i\alpha_{L-2}) P_{L-2}(x)\sqrt{LL_3}\, U^{L-1}_{L,L-2}\} \;,$$

$$C = \sum_L [\exp(\frac{1}{2} i\alpha_L)/2i]\{\exp(\frac{1}{2} i\alpha_L) P^1_L(x)[U^{L+1}_{L,L} - U^{L-1}_{L,L-2}] +$$

$$+ \exp(\frac{1}{2} i\alpha_{L+2}) P^1_{L+2}(x)\sqrt{L_4/L_1}\, U^{L+1}_{L,L+2} - \exp(\frac{1}{2} i\alpha_{L-2}) P^1_{L-2}(x)\sqrt{L/L_3}\, U^{L-1}_{L,L-2}\} \;,$$

$$D = \sum_L [\exp(\frac{1}{2} i\alpha_L)/2i]\{\exp(\frac{1}{2} i\alpha_L) \frac{P^1_L(x)}{LL_4}[LL_1 U^{L+1}_{L,L} - L_2 U^{L}_{L,L} - (L^2-1) U^{L-1}_{L,L-2}] -$$

$$- \exp(\frac{1}{2} i\alpha_{L+2}) P^1_{L+2}(x)\sqrt{L_4/L_1}\, U^{L+1}_{L,L+2} + \exp(\frac{1}{2} i\alpha_{L-2}) P^1_{L-2}(x)\sqrt{L/L_3}\, U^{L-1}_{L,L-2}\} \;,$$

$$E = \sum_L [\exp(\frac{1}{2} i\alpha_L)/2i\sqrt{2}]\{\exp(\frac{1}{2} i\alpha_L) \frac{P^2_L(x)}{LL_4}[LU^{L+1}_{L,L} - L_2 U^{L-1}_{L,L-2} + L_4 U^{L-1}_{L,L}] +$$

$$+ \exp(\frac{1}{2} i\alpha_{L+2}) P^2_{L+2}(x)[L_4 L_1]^{-1/2} U^{L+1}_{L,L+2} + \exp(\frac{1}{2} i\alpha_{L-2}) P^2_{L-2}(x)[LL_3]^{-1/2} U^{L-1}_{L,L-2}\} \;,$$

где

$L_1$ = L+2 ,  $L_2$ = 2L+1 ,  $L_3$ = L-1 ,  $L_4$ = L+1

и

$$U^J_{L,L} = \exp(2i\delta^J_L) \to \eta^J_L \mathrm{Cos}(2\delta^J_L) + i\eta^J_L \mathrm{Sin}(2\delta^J_L)$$

- элементы матрицы рассеяния для состояний без смешивания, а

$$U^1_{0,0} = \mathrm{Cos}^2\varepsilon_1 \exp(2i\delta^1_\alpha) + \mathrm{Sin}^2\varepsilon_1 \exp(2i\delta^1_\beta) \;, \quad (6.25)$$

$$U^1_{2,2} = \mathrm{Sin}^2\varepsilon_1 \exp(2i\delta^1_\alpha) + \mathrm{Cos}^2\varepsilon_1 \exp(2i\delta^1_\beta) \quad ,$$

$$U^1_{0,2} = U^1_{2,0} = \frac{1}{2}\mathrm{Sin}(2\varepsilon_1)\left[\exp(2i\delta^1_\alpha) - \exp(2i\delta^1_\beta)\right]$$

- элементы матрицы рассеяния для смешанных состояния с J = 1 и L = 0,2 [208]. Аналогично записываются и другие матричные элементы состояний со смешиванием, например, при J = 2 и L=1,3. Кулоновская амплитуда $f_c$ имеет вид (6.3) с вынесенной фазой $\sigma_0$, а величины $\alpha_L$ определены выражением (6.8).

Все фазы $\delta_L$ считаются комплексными, т.е. все экспоненты могут быть представлены в виде

$$\exp(2i\delta_L) \to \eta_L \exp[2i(\mathrm{Re}\delta_L)] \quad ,$$

где $\eta_L$ - параметр неупругости.

Приведем текст компьютерной программы для расчета сечений упругого рассеяния в системе частиц с полным спином 1. Здесь использованы те же обозначения, что и в предыдущей программе для $^4He^4He$ рассеяния [203].

```
REM CALCULATE OF CROSS SECTION ON COMPLEX PHASE SHIFTS FOR SYSTEM WITH 1 SPIN AND TENSOR FORCE
CLS: DEFDBL A-Z: DEFINT I,J,K,L,N,M: N=200
DIM E(N), DE(N), E1(N), EP(N/10), EM(N/10), E0(N/10), TT(N), DEE(N), T11E(N), T21E(N)
DIM SEC(N), SECE(N), FM(N/10), F0(N/10), FP(N/10), POL(N), T20(N), T22(N), T21(N): DIM T22E(N),T20E(N),T(N)
REM ********************************************
ISAVE=1: REM =0 - NO SAVE, =1 - SAVE IN FILE
G$="C:\BASICA\SEC\AL-d.DAT"
REM ************** INPUT PARAMETERS ***********
PI=4*ATN(1): NN=0: NV=0: LN=0: LV=3: LH=1: TMI=10: TMA=170
TH=2: AM1=2: AM2=4: Z1=1: Z2=2: A1=41.4686
PM=AM1*AM2/(AM1+AM2): B1=2*PM/A1
REM ********* ENERGY IN LAB. SYSTEM ***********
IXI=1: REM IF =1 - XI^2 WILL BE CALCULATE, =0 NO CALCULATE
E1(0)=11.
REM **** ECSPERIMENTAL CROSS SECTION 11. ******
SECE(1)=00:    DEE(1)=00:    SECE(2)=00:    DEE(2)=00
SECE(3)=184.0: DEE(3)=7.38:  SECE(4)=139.0: DEE(4)=5.55
```

```
SECE(5)=98.4:   DEE(5)=3.93:  SECE(6)=67.2:   DEE(6)=1.34
SECE(7)=43.5:   DEE(7)=0.87:  SECE(8)=26.9:   DEE(8)=0.54
SECE(9)=15.8:   DEE(9)=0.32:  SECE(10)=10.1: DEE(10)=0.2
SECE(11)=8.49: DEE(11)=0.17: SECE(12)=10.2: DEE(12)=0.2
SECE(13)=14.3: DEE(13)=0.29: SECE(14)=20.2: DEE(14)=0.4
SECE(15)=26.9: DEE(15)=0.54: SECE(16)=34.1: DEE(16)=0.68
SECE(17)=41.3: DEE(17)=0.83: SECE(18)=48.0: DEE(18)=0.96
SECE(19)=53.9: DEE(19)=1.08: SECE(20)=58.2: DEE(20)=1.16
SECE(21)=61.1: DEE(21)=1.22: SECE(22)=62.2: DEE(22)=1.24
SECE(23)=62.8: DEE(23)=1.26: SECE(24)=62.0: DEE(24)=1.24
SECE(25)=59.8: DEE(25)=1.2:   SECE(26)=59.7: DEE(26)=1.19
SECE(27)=56.9: DEE(27)=1.14: SECE(28)=54.2: DEE(28)=1.08
SECE(29)=54.3: DEE(29)=1.09: SECE(30)=49.9: DEE(30)=1.
SECE(31)=46.9: DEE(31)=0.94: SECE(32)=44.5: DEE(32)=0.89
SECE(33)=41.7: DEE(33)=0.83: SECE(34)=39.1: DEE(34)=0.78
SECE(35)=38.3: DEE(35)=0.77: SECE(36)=34.4: DEE(36)=0.69
SECE(37)=33.1: DEE(37)=0.66: SECE(38)=30.0: DEE(38)=0.6
SECE(39)=29.6: DEE(39)=0.59: SECE(40)=33.8: DEE(40)=1.38
SECE(41)=41.9: DEE(41)=1.72: SECE(42)=52.1: DEE(42)=2.08
SECE(43)=66.4: DEE(43)=2.66: SECE(44)=79.7: DEE(44)=3.22
SECE(45)=92.9: DEE(45)=3.7
T(1)=18.75:   T11E(1)=.0038:   T20E(1)=.0339:   T21E(1)=-.008:   T22E(1)=-.0039
T(2)=26.20:   T11E(2)=.0214:   T20E(2)=.0045:   T21E(2)= .0015:  T22E(2)=-.0231
T(3)=29.91:   T11E(3)=.0207:   T20E(3)=.0008:   T21E(3)=-.0029:  T22E(3)=-.0291
T(4)=33.60:   T11E(4)=.0192:   T20E(4)=-.0282:  T21E(4)=-.0021:  T22E(4)=-.0574
T(5)=37.28:   T11E(5)=.0139:   T20E(5)=-.0417:  T21E(5)=.0023:   T22E(5)=-.0732
T(6)=40.94:   T11E(6)=-.0088:  T20E(6)=-.0723:  T21E(6)=-.0023:  T22E(6)=-.1003
T(7)=44.57:   T11E(7)=-.0435:  T20E(7)=-.1258:  T21E(7)=-.0002:  T22E(7)=-.1533
T(8)=48.19:   T11E(8)=-.0859:  T20E(8)=-.2020:  T21E(8)=.0068:   T22E(8)=-.2307
T(9)=51.78:   T11E(9)=-.1495:  T20E(9)=-.3045:  T21E(9)=.0162:   T22E(9)=-.3212
T(10)=55.34: T11E(10)=-.1762: T20E(10)=-.4107: T21E(10)=.0032:  T22E(10)=-.3571
T(11)=58.87: T11E(11)=-.0537: T20E(11)=-.3390: T21E(11)=-.0383: T22E(11)=-.2526
T(12)=62.37:  T11E(12)= .1129: T20E(12)=-.1243: T21E(12)=-.0765: T22E(12)=-.0023
T(13)=65.84:  T11E(13)= .2033: T20E(13)= .0344: T21E(13)=-.0960: T22E(13)=.1272
T(14)=69.28:  T11E(14)= .2183: T20E(14)= .1248: T21E(14)=-.1073: T22E(14)=.2095
T(15)=72.67:  T11E(15)= .1967: T20E(15)= .1611: T21E(15)=-.1186: T22E(15)=.2569
T(16)=76.03:  T11E(16)= .1736: T20E(16)= .1962: T21E(16)=-.1192: T22E(16)=.2687
T(17)=79.34:  T11E(17)= .1393: T20E(17)= .1965: T21E(17)=-.1194: T22E(17)=.2772
T(18)=82.61:  T11E(18)= .1043: T20E(18)= .2133: T21E(18)=-.1266: T22E(18)=.2954
T(19)=85.84:  T11E(19)= .0591: T20E(19)= .2077: T21E(19)=-.1225: T22E(19)=.3038
T(20)=89.01:  T11E(20)= .0086: T20E(20)= .2087: T21E(20)=-.1152: T22E(20)=.3003
T(21)=92.13:  T11E(21)=-.0296: T20E(21)= .2036: T21E(21)=-.1252: T22E(21)=.3145
T(22)=95.20:  T11E(22)=-.0784: T20E(22)= .2029: T21E(22)=-.1222: T22E(22)=.3126
T(23)=98.22:  T11E(23)=-.1214: T20E(23)= .2009: T21E(23)=-.1226: T22E(23)=.3183
T(24)=101.18: T11E(24)=-.1649: T20E(24)=.1923: T21E(24)=-.1082:  T22E(24)=.3105
T(25)=104.08: T11E(25)=-.2100: T20E(25)=.1880: T21E(25)=-.1148:  T22E(25)=.3159
T(26)=105.00: T11E(26)=-.2266: T20E(26)=.1840: T21E(26)=-.1168: T22E(26)=.3041
T(27)=106.92: T11E(27)=-.2599: T20E(27)=.1812: T21E(27)=-.1103: T22E(27)=.3048
T(28)=109.71: T11E(28)=-.2957: T20E(28)=.1797: T21E(28)=-.1161: T22E(28)=.3089
T(29)=110.00: T11E(29)=-.2908: T20E(29)=.1726: T21E(29)=-.1059: T22E(29)=.2961
T(30)=112.43: T11E(30)=-.3305: T20E(30)=.1625: T21E(30)=-.1023: T22E(30)=.2907
T(31)=115.00: T11E(31)=-.3517: T20E(31)=.1476: T21E(31)=-.1007: T22E(31)=.2705
T(32)=115.09: T11E(32)=-.3584: T20E(32)=.1625: T21E(32)=-.1059: T22E(32)=.2744
T(33)=117.68: T11E(33)=-.3672: T20E(33)=.1325: T21E(33)=-.0952: T22E(33)=.2431
T(34)=120.00: T11E(34)=-.3712: T20E(34)=.1312: T21E(34)=-.0892: T22E(34)=.2049
```

```
T(35)=120.21: T11E(35)=-.3739: T20E(35)=.1432: T21E(35)=-.0849: T22E(35)=.2181
T(36)=122.68: T11E(36)=-.3400: T20E(36)=.1337: T21E(36)=-.0702: T22E(36)=.1760
T(37)=125.00: T11E(37)=-.3091: T20E(37)=.0961: T21E(37)=-.0602: T22E(37)=.1189
T(38)=130.00: T11E(38)=-.1259: T20E(38)=.0766: T21E(38)=-.0269: T22E(38)=-.0048
T(39)=135.00: T11E(39)= .1203: T20E(39)=.0671: T21E(39)= .0033: T22E(39)=-.1141
T(40)=140.43: T11E(40)= .3461: T20E(40)=.0727: T21E(40)= .0256: T22E(40)=-.1774
T(41)=145.00: T11E(41)= .4402: T20E(41)=.0937: T21E(41)= .0212: T22E(41)=-.1777
T(42)=150.00: T11E(42)= .4200: T20E(42)=.1014: T21E(42)= .0183: T22E(42)=-.1519
T(43)=155.00: T11E(43)= .4046: T20E(43)=.1133: T21E(43)= .0211: T22E(43)=-.1097
T(44)=160.00: T11E(44)= .3200: T20E(44)=.1202: T21E(44)= .0125: T22E(44)=-.0615
T(45)=165.00: T11E(45)= .2479: T20E(45)=.1160: T21E(45)=-.0009: T22E(45)=-.0420
 NT=45
 REM ***************** PHASE SHIFTS **************
 REM *********** FOR D-AL ON E=10.9 ***************
 EPS(1)=-90
 FP(0)=93.6:    FPI(0)=23.4: FP(1)=-0.5:     FPI(1)=12.2
 FP(2)=152.2:   FPI(2)=13.8: FP(3)=12.4:     FPI(3)=3.1
 F0(1)=-6.0:      F0I(1)=4.: F0(2)=116.:     F0I(2)=5.6
 F0(3)=13.9:     F0I(3)=4.6: F0(4)=0.00:     F0I(4)=0.
 FM(1)=-34.6:   FMI(1)=12.4: FM(2)=42.6:    FMI(2)=1.3
 FM(3)=.8:        FMI(3)=9.1
111 FOR L=LN TO LV STEP LH: FM(L)=FM(L)*PI/180
 FP(L)=FP(L)*PI/180: F0(L)=F0(L)*PI/180: FMI(L)=FMI(L)*PI/180
 FPI(L)=FPI(L)*PI/180: F0I(L)=F0I(L)*PI/180: EP(L)=EXP(-2*FPI(L))
 EM(L)=EXP(-2*FMI(L)):                          E0(L)=EXP(-2*F0I(L)):
EPS(L)=EPS(L)*PI/180
 NEXT
 REM *********** TRANSFORM TO C.M. *************
 FOR I=NN TO NV: E(I)=E1(I)*PM/AM1: NEXT I
 REM ***********  TOTAL CROSS SECTION  **********
 FOR J=NN TO NV: SK=E(J)*B1: SS=SQR(SK)
 GG=3.44476E-02*Z1*Z2*PM/SS: SER=0: SES=0
 FOR L=LN TO LV STEP LH
 AP=FP(L): AM=FM(L): A0=F0(L): L1=2*L+3: L2=2*L+1: L3=2*L-1
 SER=SER+L1*(1-EP(L)^2)+L2*(1-E0(L)^2)+L3*(1-EM(L)^2)
SES=SES+L1*EP(L)^2*SIN(AP)^2+L2*E0(L)^2*SIN(A0)^2+L3*EM(
L)^2*SIN(AM)^2: NEXT L: SIGR=10*PI*SER/SK/3
 SIGS=10*4*PI*SES/SK/3
 PRINT "            SIGMR-TOT=";: PRINT USING " ####.### ";SIGR;
 PRINT  "            SIGMS-TOT=";:  PRINT  USING  "  ####.###  ";SIGS:
NEXT J
 REM *******  DIFFERENTIAL CROSS SECTION  *******
 CALL SEC(SS,GG,SEC(),POL(),T20(),T22(),T21(),N)
 REM ******************** XI^2 *******************
 IF IXI=0 GOTO 211: S1=0: FOR M=1 TO NT: IF M<3 GOTO 234
 DE(M)=( (SEC(M)-SECE(M) )/( DEE(M) ) )^2: S1=S1+DE(M)
```

```
234 NEXT M: XI=S1/NT: PRINT "                    XISEC=";XI
 REM PRINT "  T    SEC-EXP    SEC-TEOR    DE": FOR I=1 TO
(NT+1)/2
 PRINT USING " ###.## ";T(I);: PRINT USING " ###.##
";SECE(I);SEC(I);DE(I);
 PRINT USING "     ###.## ";T(I+23);
 PRINT USING " ###.## ";SECE(I+23);SEC(I+23);DE(I+23): NEXT
211 S1=0: FOR M=1 TO NT
 DE(M)=( (POL(M)-T11E(M))/(0.05*T11E(M)) )^2
 S1=S1+DE(M): NEXT M: XI=S1/NT: INPUT "      T11";Z
 IF Z=0 GOTO 321
 PRINT "------------------------- T11 -------------------------"
 PRINT "                      XIT11=";XI
 FOR I=1 TO (NT+1)/2: PRINT USING "###.## ";T(I);
 PRINT USING "+#.##^^^^ ";T11E(I);POL(I);DE(I);
 PRINT USING "   ###.## ";T(I+23);
 PRINT USING "+#.##^^^^ "; T11E(I+23); POL(I+23); DE(I+23):
NEXT
321 S1=0: FOR M=1 TO NT
 DE(M)=( (T20(M)-T20E(M))/(0.05*T20E(M)) )^2
 S1=S1+DE(M): NEXT M: XI=S1/NT: INPUT "      T20";Z
 IF Z=0 GOTO 322: PRINT "------------------------- T20 --------------------
-----"
 PRINT "                      XIT20=";XI: FOR I=1 TO (NT+1)/2
 PRINT USING "###.## ";T(I);
 PRINT USING "+#.##^^^^ ";T20E(I);T20(I);DE(I);
 PRINT USING "   ###.## ";T(I+23);
 PRINT USING "+#.##^^^^ ";T20E(I+23);T20(I+23);DE(I+23): NEXT
322 S1=0: FOR M=1 TO NT
 DE(M)=( (T22(M)-T22E(M))/(0.05*T22E(M)) )^2
 S1=S1+DE(M): NEXT M: XI=S1/NT: INPUT "      T22";Z
 IF Z=0 GOTO 212: PRINT "------------------------- T22 --------------------
-----"
 PRINT "                      XIT22=";XI: FOR I=1 TO (NT+1)/2
 PRINT USING "###.## ";T(I);
 PRINT USING "+#.##^^^^ ";T22E(I);T22(I);DE(I);
 PRINT USING "   ###.## ";T(I+23);
 PRINT USING "+#.##^^^^ ";T22E(I+23);T22(I+23);DE(I+23): NEXT
212 REM *************  SAVE IN FILE  *************
 IF ISAVE=0 GOTO 221: OPEN "O",1,G$
 PRINT#1, "          D - ALPHA FOR LAB E=";: PRINT#1, E1(NN)
 PRINT#1, " T        SEC        POL        T20        T21        T22"
 FOR T=1 TO NT
 PRINT#1, USING "  +#.###^^^^  "; T(T); SEC(T); POL(T); T20(T);
```

```
T21(T); T22(T): NEXT
221 END
```

**SUB SEC (SS, GG, S(100), POL(100), T20(100), T22(100), T21(100))**

```
 SHARED FP(),EP(),F0(),E0(),FM(),EM(),EPS(),TT()
 SHARED PI,TMI,TMA,TH,LN,LV,LH,NT,T()
 DIM S0(20),P(20),P1(20),P2(20)
 LVV=LV+2: CALL CULFAZ(GG,LVV,S0()): FOR I=1 TO NT
 T=T(I)*PI/180
 X=COS(T): CALL CULAMP(X,GG,S0(),RECUL,AMCUL)
 CALL POLLEG(X,LVV,P()): CALL FUNLEG1(X,LVV,P1())
 CALL FUNLEG2(X,LVV,P2()): REA=0: AMA=0: REB=0: AMB=0
 REC=0: AMC=0: RED=0: AMD=0: REE=0: AME=0
 FOR L=LN TO LV STEP LH: FP=2*FP(L):   FM=2*FM(L): F0=2*F0(L)
  SL=2*S0(L): C0=COS(SL):  S0=SIN(SL): EP=EP(L): E0=E0(L)
  EM=EM(L): CP=COS(S0(L)+S0(L+2)): SP=SIN(S0(L)+S0(L+2))
 AL1M=EM*COS(FM)-1:  AL2M=EM*SIN(FM)
 AL10=E0*COS(F0)-1:  AL20=E0*SIN(F0)
 CO=COS(EPS(L+1))^2: SI=SIN(EPS(L+1))^2
 AL1P=EP*CO*COS(FP)+EM(L+2)*SI*COS(2*FM(L+2))-1
 AL2P=EP*CO*SIN(FP)+EM(L+2)*SI*SIN(2*FM(L+2))
 SI=1/2*SIN(2*EPS(L+1))
 AP1=SI*( EP*COS(FP)-  EM(L+2)*COS(2*FM(L+2)) )
 AP2=SI*( EP*SIN(FP)-EM(L+2)*SIN(2*FM(L+2)) )
 IF L<2 GOTO 1121: CO=COS(EPS(L-1))^2: SI=SIN(EPS(L-1))^2
 AL1M=EP(L-2)*SI*COS(2*FP(L-2))+EM*CO*COS(FM)-1
 AL2M=EP(L-2)*SI*SIN(2*FP(L-2))+EM*CO*SIN(FM)
 SI=1/2*SIN(2*EPS(L-1))
 AM1=SI*(EP(L-2)*COS(2*FP(L-2))-EM*COS(FM))
 AM2=SI*(EP(L-2)*SIN(2*FP(L-2))-EM*SIN(FM))
1121 REP=SQR((L+1)*(L+2))*(AP2*CP+AP1*SP)*P(L+2)
 AMP=SQR((L+1)*(L+2))*(AP2*SP-AP1*CP)*P(L+2)
 REMM=0: AMM=0
 IF L<2 GOTO 1122: CM=COS(S0(L)+S0(L-2)): SM=SIN(S0(L)+S0(L-2))
 REMM=SQR(L*(L-1))*(AM2*CM+AM1*SM)*P(L-2)
 AMM=SQR(L*(L-1))*(AM2*SM-AM1*CM)*P(L-2)
1122 A1=(L+1)*AL1P+L*AL1M: A2=(L+1)*AL2P+L*AL2M
 REA=REA+(A2*C0+A1*S0)*P(L)+REP+REMM
 AMA=AMA+(A2*S0-A1*C0)*P(L)+AMP+AMM
 B1=(L+2)*AL1P+(2*L+1)*AL10+(L-1)*AL1M
 B2=(L+2)*AL2P+(2*L+1)*AL20+(L-1)*AL2M
 REB=REB+((B2*C0+B1*S0)*P(L)+(-REP-REMM))/2
```

```
 AMB=AMB+((B2*S0-B1*C0)*P(L)+(-AMP-AMM))/2
 IF L<1 GOTO 2111
 REP=SQR((L+1)/(L+2))*(AP2*CP+AP1*SP)*P1(L+2)
 AMP=SQR((L+1)/(L+2))*(AP2*SP-AP1*CP)*P1(L+2)
 REMM=0: AMM=0: IF L<2 GOTO 123
 CM=COS(S0(L)+S0(L-2)): SM=SIN(S0(L)+S0(L-2))
 REMM=SQR(L/(L-1))*(AM2*CM+AM1*SM)*P1(L-2)
 AMM=SQR(L/(L-1))*(AM2*SM-AM1*CM)*P1(L-2)
123 C1=AL1P-AL1M: C2=AL2P-AL2M: CC1=1/SQR(2)
 REC=REC+((C2*C0+C1*S0)*P1(L)-REP+REMM)*CC1
 AMC=AMC+((C2*S0-C1*C0)*P1(L)-AMP+AMM)*CC1
 DD1=1/(L*(L+1)):              D1=L*(L+2)*AL1P-(2*L+1)*AL10-(L^2-
1)*AL1M
 D2=L*(L+2)*AL2P-(2*L+1)*AL20-(L^2-1)*AL2M
 RED = RED + ((D2*C0 + D1*S0)*P1(L)*DD1 + REP - REMM)*CC1
 AMD = AMD + ((D2*S0 - D1*C0)*P1(L)*DD1 + AMP - AMM)*CC1
2111 IF L<2 GOTO 1222
 REP=1/SQR((L+1)*(L+2))*(AP2*CP+AP1*SP)*P2(L+2)
 AMP=1/SQR((L+1)*(L+2))*(AP2*SP-AP1*CP)*P2(L+2)
 REMM=0: AMM=0: IF L<2 GOTO 124
 CM=COS(S0(L)+S0(L-2)): SM=SIN(S0(L)+S0(L-2))
 REMM=1/SQR(L*(L-1))*(AM2*CM+AM1*SM)*P2(L-2)
 AMM=1/SQR(L*(L-1))*(AM2*SM-AM1*CM)*P2(L-2)
124 EE1=1/SQR(2): EE2=1/(L*(L+1))
 E1=L*AL1P-(2*L+1)*AL10+(L+1)*AL1M
 E2=L*AL2P-(2*L+1)*AL20+(L+1)*AL2M
 REE=REE+((E2*C0+E1*S0)*P2(L)*EE2+(-REP-REMM))/2
 AME=AME+((E2*S0-E1*C0)*P2(L)*EE2+(-AMP-AMM))/2
1222 NEXT L: REA=RECUL+REA:  AMA=AMCUL+AMA
 REB=RECUL+REB:  AMB=AMCUL+AMB
 REA=REA/2/SS:              AMA=AMA/2/SS:        REB=REB/2/SS:
AMB=AMB/2/SS
 REC=REC/2/SS:              AMC=AMC/2/SS:        RED=RED/2/SS:
AMD=AMD/2/SS
 REE=REE/2/SS:  AME=AME/2/SS: AA=REA^2+AMA^2
  BB=REB^2+AMB^2
 CC=REC^2+AMC^2: DD=RED^2+AMD^2:  EE=REE^2+AME^2
  SEC=(AA+2*BB+2*CC+2*DD+2*EE)/3: S(I)=10*SEC
 POL(I) = 2*SQR(2)/3*(AMA*REC - REA*AMC + AMB*RED -
REB*AMD + AMD*REE - RED*AME)/SEC
 T20(I)=1/SQR(2)*(1-(AA+2*DD)/SEC)
 T22(I)=1/SQR(3)*(2*(REB*REE+AMB*AME)-CC)/SEC
 T21(I) = - SQR(2/3)*(REA*REC + AMA*AMC - REB*RED -
AMB*AMD +  RED*REE + AMD*AME)/SEC: NEXT
```

```
END SUB
SUB CULAMP(X,GG,S0(20),RECUL,AMCUL)
A=2/(1-X): S0=2*S0(0): BB=-GG*A: AL=GG*LOG(A)+S0
RECUL=BB*COS(AL): AMCUL=BB*SIN(AL): END SUB
SUB POLLEG(X,L,P(20))
P(0)=1: P(1)=X: FOR I=2 TO L: P(I)=(2*I-1)*X/I*P(I-1)-(I-1)/I*P(I-2)
NEXT: END SUB
SUB FUNLEG1(X,L,P(20))
P(0)=0: P(1)=SQR(ABS(1-X^2)): FOR I=2 TO L
P(I)=(2*I-1)*X/(I-1)*P(I-1)-I/(I-1)*P(I-2): NEXT: END SUB
SUB FUNLEG2(X,L,P(20))
P(0)=0: P(1)=0: P(2)=3*ABS(1-X^2): FOR I=3 TO L
P(I)=(2*I-1)*X/(I-2)*P(I-1)-(I+1)/(I-2)*P(I-2): NEXT: END SUB
SUB CULFAZ(G,L,F(20))
C=0.577215665: S=0: N=50: A1=1.202056903/3: A2=1.036927755/5
FOR I=1 TO N: A=G/I-ATN(G/I)-(G/I)^3/3+(G/I)^5/5: S=S+A
NEXT: FAZ=-C*G+A1*G^3-A2*G^5+S: F(0)=FAZ: FOR I=1 TO L
F(I)=F(I-1)+ATN(G/(I)): NEXT
END SUB
```

Приведем теперь результаты контрольного счета дифференциальных сечений в $^2H^4He$ системе при энергии 11 МэВ. Экспериментальные данные по сечениям $\sigma_e$ и их ошибки для каждого угла взяты из работ [204-206], фазы и параметр смешивания, равный $-90^0$ при энергии 10.9 МэВ, взяты из [91] и приведены в таблице 6.8.

Таблица 6.8 - Фазы рассеяния.

| $\delta_L$, град | $S_1$ | $P_0$ | $P_1$ | $P_2$ | $D_1$ | $D_2$ | $D_3$ | $F_2$ | $F_3$ | $F_4$ |
|---|---|---|---|---|---|---|---|---|---|---|
| Re $\delta$ | 93.2 | -34.6 | -6.0 | -0.5 | 42.6 | 116.0 | 152.2 | 0.8 | 13.9 | 12.4 |
| Im $\delta$ | 23.4 | 12.4 | 4.0 | 12.2 | 1.3 | 5.6 | 13.8 | 9.1 | 4.6 | 3.1 |

С такими фазами получаем следующие результаты для дифференциальных сечений $\sigma_t$ упругого рассеяния

$\chi^2$ = 6.93 $\sigma_r$ = 427.35 мб. $\sigma_s$ = 528.95 мб.

| $\theta$ | $\sigma_e$ | $\sigma_t$ | $\chi^2_i$ | $\theta$ | $\sigma_e$ | $\sigma_t$ | $\chi^2_i$ |
|---|---|---|---|---|---|---|---|
| 18.75 | 0.00 | 366.15 | 0.00 | 98.22 | 62.80 | 57.14 | 20.15 |
| 26.20 | 0.00 | 233.01 | 0.00 | 101.18 | 62.00 | 56.47 | 19.87 |
| 29.91 | 184.00 | 183.24 | 0.01 | 104.08 | 59.80 | 54.96 | 16.29 |
| 33.60 | 139.00 | 139.33 | 0.00 | 105.00 | 59.70 | 54.32 | 20.47 |
| 37.28 | 98.40 | 101.41 | 0.59 | 106.92 | 56.90 | 52.76 | 13.21 |
| 40.94 | 67.20 | 70.10 | 4.69 | 109.71 | 54.20 | 50.04 | 14.84 |
| 44.57 | 43.50 | 45.68 | 6.29 | 110.00 | 54.30 | 49.73 | 17.57 |
| 48.19 | 26.90 | 27.86 | 3.14 | 112.43 | 49.90 | 47.01 | 8.37 |
| 51.78 | 15.80 | 16.21 | 1.68 | 115.00 | 46.90 | 43.94 | 9.92 |
| 55.34 | 10.10 | 9.95 | 0.57 | 115.09 | 44.50 | 43.83 | 0.57 |
| 58.87 | 8.49 | 8.11 | 5.02 | 117.68 | 41.70 | 40.69 | 1.47 |
| 62.37 | 10.20 | 9.70 | 6.29 | 120.00 | 39.10 | 37.98 | 2.07 |
| 65.84 | 14.30 | 13.75 | 3.53 | 120.21 | 38.30 | 37.74 | 0.53 |
| 69.28 | 20.20 | 19.40 | 3.99 | 122.68 | 34.40 | 35.10 | 1.04 |
| 72.67 | 26.90 | 25.84 | 3.85 | 125.00 | 33.10 | 32.97 | 0.04 |
| 76.03 | 34.10 | 32.48 | 5.65 | 130.00 | 30.00 | 29.98 | 0.00 |
| 79.34 | 41.30 | 38.81 | 9.01 | 135.00 | 29.60 | 29.85 | 0.18 |
| 82.61 | 48.00 | 44.47 | 13.50 | 140.43 | 33.80 | 33.51 | 0.04 |
| 85.84 | 53.90 | 49.23 | 18.72 | 145.00 | 41.90 | 39.68 | 1.67 |
| 89.01 | 58.20 | 52.90 | 20.90 | 150.00 | 52.10 | 49.21 | 1.93 |
| 92.13 | 61.10 | 55.43 | 21.59 | 155.00 | 66.40 | 60.75 | 4.52 |
| 95.20 | 62.20 | 56.83 | 18.76 | 160.00 | 79.70 | 73.03 | 4.29 |

Для полного сечения реакций при этой энергии в [91] была получена экспериментальная величина 430(10) мб, которая хорошо согласуется с нашими расчетами. Расчетные дифференциальные сечения с такими фазами полностью соответствуют результатам работы [91], где с найденными наборами фаз получено некоторое снижение сечения относительно эксперимента в области углов $80^0$ - $110^0$.

Можно несколько изменить значения фаз (таблица 6.9) из работы [91] и попытаться улучшить качество описания экспериментальных данных. Варьирование фаз приводит нас к следующим результатам.

Таблица 6.9 - Фазы рассеяния.

| $\delta_L$, град | $S_1$ | $P_0$ | $P_1$ | $P_2$ | $D_1$ | $D_2$ | $D_3$ | $F_2$ | $F_3$ | $F_4$ |
|---|---|---|---|---|---|---|---|---|---|---|
| Re δ | 97.1 | -34.5 | -6.2 | -0.4 | 42.6 | 116.6 | 151.6 | 0.9 | 14.0 | 12.4 |
| Im δ | 19.8 | 23.4 | 2.3 | 11.5 | 0.0 | 5.2 | 12.7 | 5.2 | 3.3 | 3.3 |

С такими фазами, для дифференциальных сечений, получается уменьшение $\chi^2$ примерно в 15 раз, хотя сами фазы, за исключением мнимой части $P_0$, существенно не изменились

$\chi^2$ = 0.437 $\sigma_r$ = 380.39 мб. $\sigma_s$ = 560.84 мб.

| θ | $\sigma_e$ | $\sigma_t$ | $\chi^2_i$ | θ | $\sigma_e$ | $\sigma_t$ | $\chi^2$ |
|---|---|---|---|---|---|---|---|
| 18.75 | 0.00 | 348.37 | 0.00 | 98.22 | 62.80 | 62.47 | 0.07 |
| 26.20 | 0.00 | 221.21 | 0.00 | 101.18 | 62.00 | 61.65 | 0.08 |
| 29.91 | 184.00 | 174.26 | 1.74 | 104.08 | 59.80 | 59.78 | 0.00 |
| 33.60 | 139.00 | 132.87 | 1.22 | 105.00 | 59.70 | 58.98 | 0.36 |
| 37.28 | 98.40 | 97.08 | 0.11 | 106.92 | 56.90 | 57.05 | 0.02 |
| 40.94 | 67.20 | 67.47 | 0.04 | 109.71 | 54.20 | 53.67 | 0.24 |
| 44.57 | 43.50 | 44.29 | 0.83 | 110.00 | 54.30 | 53.28 | 0.87 |
| 48.19 | 26.90 | 27.30 | 0.56 | 112.43 | 49.90 | 49.89 | 0.00 |
| 51.78 | 15.80 | 16.15 | 1.22 | 115.00 | 46.90 | 46.09 | 0.74 |
| 55.34 | 10.10 | 10.14 | 0.03 | 115.09 | 44.50 | 45.95 | 2.67 |
| 58.87 | 8.49 | 8.40 | 0.25 | 117.68 | 41.70 | 42.08 | 0.21 |
| 62.37 | 10.20 | 10.07 | 0.41 | 120.00 | 39.10 | 38.76 | 0.19 |
| 65.84 | 14.30 | 14.26 | 0.01 | 120.21 | 38.30 | 38.47 | 0.05 |
| 69.28 | 20.20 | 20.17 | 0.01 | 122.68 | 34.40 | 35.28 | 1.62 |
| 72.67 | 26.90 | 27.01 | 0.05 | 125.00 | 33.10 | 32.74 | 0.30 |
| 76.03 | 34.10 | 34.21 | 0.03 | 130.00 | 30.00 | 29.35 | 1.17 |
| 79.34 | 41.30 | 41.20 | 0.01 | 135.00 | 29.60 | 29.59 | 0.00 |
| 82.61 | 48.00 | 47.59 | 0.18 | 140.43 | 33.80 | 34.50 | 0.26 |
| 85.84 | 53.90 | 53.06 | 0.60 | 145.00 | 41.90 | 42.31 | 0.06 |
| 89.01 | 58.20 | 57.37 | 0.52 | 150.00 | 52.10 | 54.02 | 0.85 |
| 92.13 | 61.10 | 60.39 | 0.34 | 155.00 | 66.40 | 67.90 | 0.32 |
| 95.20 | 62.20 | 62.08 | 0.01 | 160.00 | 79.70 | 82.44 | 0.72 |

Приведем теперь результаты расчетов для векторной поляризации $T_t$, полученные с исходными фазами из [91] и их сравнение с экспериментальными данными $T_e$ работ [204-206]. Ошибки поляризаций полагались равными 10% от их величины, хотя реально, некоторые из них доходят до 200% [204-206].

| θ | $T_{11e}$ | $T_{11t}$ | $\chi^2_i$ | θ | $T_{11e}$ | $T_{11t}$ | $\chi^2_i$ |
|---|---|---|---|---|---|---|---|
| 18.75 | +3.80E-03 | -2.64E-02 | +6.31E+03 | 101.18 | -1.65E-01 | -1.84E-01 | +1.37E+00 |
| 26.20 | +2.14E-02 | +5.38E-04 | +9.50E+01 | 104.08 | -2.10E-01 | -2.29E-01 | +8.52E-01 |
| 29.91 | +2.07E-02 | +5.72E-03 | +5.24E+01 | 105.00 | -2.27E-01 | -2.43E-01 | +5.35E-01 |
| 33.60 | +1.92E-02 | +3.53E-03 | +6.66E+01 | 106.92 | -2.60E-01 | -2.71E-01 | +1.74E-01 |
| 37.28 | +1.39E-02 | -8.27E-03 | +2.54E+02 | 109.71 | -2.96E-01 | -3.07E-01 | +1.39E-01 |
| 40.94 | -8.80E-03 | -3.34E-02 | +7.82E+02 | 110.00 | -2.91E-01 | -3.10E-01 | +4.41E-01 |
| 44.57 | -4.35E-02 | -7.78E-02 | +6.23E+01 | 112.43 | -3.31E-01 | -3.35E-01 | +1.90E-02 |
| 48.19 | -8.59E-02 | -1.50E-01 | +5.54E+01 | 115.00 | -3.52E-01 | -3.53E-01 | +1.30E-03 |
| 51.78 | -1.50E-01 | -2.50E-01 | +4.48E+01 | 115.09 | -3.58E-01 | -3.53E-01 | +1.95E-02 |
| 55.34 | -1.76E-01 | -3.20E-01 | +6.67E+01 | 117.68 | -3.67E-01 | -3.59E-01 | +5.19E-02 |
| 58.87 | -5.37E-02 | -2.13E-01 | +8.80E+02 | 120.00 | -3.71E-01 | -3.50E-01 | +3.29E-01 |
| 62.37 | +1.13E-01 | +9.80E-03 | +8.34E+01 | 120.21 | -3.74E-01 | -3.48E-01 | +4.67E-01 |
| 65.84 | +2.03E-01 | +1.46E-01 | +8.07E+00 | 122.68 | -3.40E-01 | -3.19E-01 | +3.77E-01 |
| 69.28 | +2.18E-01 | +1.91E-01 | +1.56E+00 | 125.00 | -3.09E-01 | -2.72E-01 | +1.47E+00 |
| 72.67 | +1.97E-01 | +1.91E-01 | +7.28E-02 | 130.00 | -1.26E-01 | -1.01E-01 | +3.86E+00 |
| 76.03 | +1.74E-01 | +1.70E-01 | +3.27E-02 | 135.00 | +1.20E-01 | +1.25E-01 | +1.23E-01 |
| 79.34 | +1.39E-01 | +1.38E-01 | +4.61E-03 | 140.43 | +3.46E-01 | +3.27E-01 | +3.07E-01 |
| 82.61 | +1.04E-01 | +9.94E-02 | +2.18E-01 | 145.00 | +4.40E-01 | +4.08E-01 | +5.41E-01 |
| 85.84 | +5.91E-02 | +5.59E-02 | +2.86E-01 | 150.00 | +4.20E-01 | +4.12E-01 | +3.37E-02 |
| 89.01 | +8.60E-03 | +9.60E-03 | +1.36E+00 | 155.00 | +4.05E-01 | +3.66E-01 | +9.29E-01 |
| 92.13 | -2.96E-02 | -3.86E-02 | +9.23E+00 | 160.00 | +3.20E-01 | +2.97E-01 | +5.14E-01 |
| 95.20 | -7.84E-02 | -8.76E-02 | +1.39E+00 | 165.00 | +2.48E-01 | +2.22E-01 | +1.09E+00 |
| 98.22 | -1.21E-01 | -1.37E-01 | +1.57E+00 | | | | |

$$\chi^2 = 195.37$$

Из этих результатов видно, что и поляризации в целом согласуются с экспериментальными данными. Расчеты поляризаций с измененными фазами дают примерно такие же результаты.

## 6.5 Нетождественные частицы со спином 1/2

Если не учитывать тензорное и спин - орбитальное взаимодействие, сечение рассеяния двух частиц со спином 1/2, например, в системах $p^3He$, $^3H^3He$, может быть представлено в виде [209]

$$\frac{d\sigma(\theta)}{d\Omega} = 1/4\frac{d\sigma_s(\theta)}{d\Omega} + 3/4\frac{d\sigma_t(\theta)}{d\Omega} \quad , \tag{6.26}$$

где индексы s и t относятся к синглетному (с полным спином 0) и триплетному (с полным спином 1) состоянию рассеяния и

$$\frac{d\sigma_s(\theta)}{d\Omega} = |f_s(\theta)|^2 \quad , \qquad \frac{d\sigma_t(\theta)}{d\Omega} = |f_t(\theta)|^2 \quad . \tag{6.27}$$

Амплитуды рассеяния записываются аналогично (6.2) и (6.3)

$$f_{t,s}(\theta) = f_c(\theta) + f^N_{t,s}(\theta) \, , \tag{6.28}$$

$$f_c(\theta) = -\left(\frac{\eta}{2kSin^2(\theta/2)}\right)\exp\{i\eta\ln[Sin^{-2}(\theta/2)] + 2i\sigma_0\} \quad ,$$

$$f_s^N(\theta) = \frac{1}{2ik}\sum_L (2L+1)\exp(2i\sigma_L)[S_L^s - 1]P_L(Cos\theta) \quad . \tag{6.29}$$

$$f_t^N(\theta) = \frac{1}{2ik}\sum_L (2L+1)\exp(2i\sigma_L)[S_L^t - 1]P_L(Cos\theta) \quad .$$

где $S_L^{t,s} = \eta_L^{t,s}\exp[2i\delta_L^{t,s}(k)]$ - матрица рассеяния в триплетном или синглетном состоянии.

В случае процессов рассеяния тождественных фермионов с полуцелым спином, например, NN, $^3He^3He$, без учета тензорных и спин - орбитальных взаимодействий знак плюс в (6.9) заменяется на минус для каждого спинового состояния

$$\frac{d\sigma(\theta)}{d\Omega} = \left| f(\theta) - f(\pi - \theta) \right|^2 , \tag{6.30}$$

а суммирование выполняется только по нечетным моментам, поскольку четные парциальные волны не дают вклада в суммарное сечение.

Приведем текст компьютерной программы для расчета сечений упругого рассеяния в системе частиц, каждая из которых имеет спин 1/2. Здесь использованы практически те же обозначения, что и в программе для $^4He^4He$ рассеяния [210].

**REM CALCULATE OF CROSS SECTION FOR COMPLEX PHASE SHIFTS FOR SYSTEM WITH SPIN 1/2+1/2**

```
CLS: DEFDBL A-Z: DEFINT I,J,K,L,N,M: N=200
DIM E(N), DE(N), E1(N), ETA(N), SEC(N), SECE(N), FM(N/10),
FR(N/10), ST(N), SS(N)
REM *********************************************
ISAVE=0: REM =0 - NO SAVE, =1 - SAVE IN FILE
G$="C:\BASICA\SEC\SEC-1-2.DAT"
REM ********** INPUT PARAMETERS ***************
PI=4*ATN(1):NN=0: NV=0: LN=0: LV=2: LH=1: TMI=10:
TMA=170
TH=2: AM1=1: AM2=3: Z1=1: Z2=2: A1=41.4686
PM=AM1*AM2/(AM1+AM2): B1=2*PM/A1
REM ********** ENERGY IN LAB. SYSTEM ***********
E1(0)=11.48
REM **************** PHASE SHIFTS ****************
REM *********** FOR P- 3HE ON E=11.48 *************
FT(0)=-88.8: FT(1)=55: FT(2)=2.5: FS(0)=-84.6: FS(1)=21.4: FS(2)=-
18.6
111 FOR L=LN TO LV STEP LH: FMT(L)=FMT(L)*PI/180
FT(L)=FT(L)*PI/180: FMS(L)=FMS(L)*PI/180: FS(L)=FS(L)*PI/180
ET(L)=EXP(-2*FMT(L)): ES(L)=EXP(-2*FMS(L)): NEXT
REM *********** TRANSFORM TO C.M. *************
FOR I=NN TO NV: E(I)=E1(I)*PM/AM1: NEXT I
REM ************ TOTAL CROSS SECTION *********
FOR J=NN TO NV: SK=E(J)*B1: SS=SQR(SK)
GG=3.44476E-02*Z1*Z2*PM/SS: SIGS=0: SIGT=0
```

```
 FOR L=LN TO LV STEP LH: FT=FT(L): FS=FS(L)
 SIGS=SIGS+(2*L+1)*ET(L)^2*SIN(FT)^2
 SIGT=SIGT+(2*L+1)*ES(L)^2*SIN(FS)^2: NEXT L
 SIG=1/4*SIGS+3/4*SIGT: SIGMAS=10*4*PI*SIG/SK
 PRINT "                    SIGMS-TOT=";
 PRINT USING " ####.### ";SIGMAS: NEXT J: PRINT
 REM ****** DIFFERENTIAL CROSS SECTION *******
 IUSLOV=0: REM = 0 - P - 3HE , =1 - 3HE - 3HE
 CALL SEC (FT(), FS(), GG, SS, TMI, TMA, TH, SEC(), ETT(), LN,
LV, LH, IUSLOV)
 FOR T=TMI TO TMA/(3.3) STEP TH
 PRINT USING "####.##   "; T; SEC(T); T+42; SEC(T+42); T+84;
SEC(T+84); T+126; SEC(T+126): NEXT
211 REM ************* SAVE IN FILE *************
 IF ISAVE=0 GOTO 221: OPEN "O",1,G$
 PRINT#1, "          ALPHA - ALPHA FOR LAB E=";
 PRINT#1, E1(NN): FOR T=TMI TO TMA STEP TH
 PRINT#1, USING " #.###^^^^ ";T;SEC(T): NEXT
221 END
```

**SUB SEC (FT(100), FS(100), GG, SS, TMI, TMA, TH, S(100), E(100), LMI, LMA, LH, NYS)**

```
 SHARED PI,ET(),ES(),ST(),SS(): DIM S0(20),P(20)
 RECUL1=0: AIMCUL1=0: CALL CULFAZ(GG,S0())
 FOR TT=TMI TO TMA STEP TH: T=TT*PI/180: X=COS(T): A=2/(1-
X)
 S0=2*S0(0): BB=-GG*A: ALO=GG*LOG(A)+S0
 RECUL=BB*COS(ALO): AIMCUL=BB*SIN(ALO)
 IF NYS=0 GOTO 555: X1=COS(T): A1=2/(1+X1)
 BB1=-GG*A1: ALO1=GG*LOG(A1)+S0:
 RECUL1=BB1*COS(ALO1)
 AIMCUL1=BB1*SIN(ALO1):555 RET=0: AIT=0: RES=0: AIS=0
 FOR L=LMI TO LMA STEP LH: AL=ET(L)*COS(2*FT(L))-1
 BE=ET(L)*SIN(2*FT(L)): LL=2*L+1: SL=2*S0(L)
 CALL POLLEG(X,L,P())
 RET=RET+LL*(BE*COS(SL)+AL*SIN(SL))*P(L)
 AIT=AIT+LL*(BE*SIN(SL)-AL*COS(SL))*P(L)
 AL=ES(L)*COS(2*FS(L))-1: BE=ES(L)*SIN(2*FS(L))
 RES=RES+LL*(BE*COS(SL)+AL*SIN(SL))*P(L)
 AIS=AIS+LL*(BE*SIN(SL)-AL*COS(SL))*P(L): NEXT L
 IF NYS=0 GOTO 556
 AIT=2*AIT: RET=2*RET: AIS=2*AIS: RES=2*RES
556 RETR=RECUL+RECUL1+RET:
AITR=AIMCUL+AIMCUL1+AIT
 RESI=RECUL+RECUL1+RES: AISI=AIMCUL+AIMCUL1+AIS
```

```
ST(TT)=10*(RETR^2+AITR^2)/4/SS^2
SS(TT)=10*(RESI^2+AISI^2)/4/SS^2
S(TT)=1/4*SS(TT)+3/4*ST(TT): NEXT TT: END SUB
SUB POLLEG(X,L,P(20))
P(0)=1: P(1)=X: FOR I=2 TO L: P(I)=(2*I-1)*X/I*P(I-1)-(I-1)/I*P(I-2)
NEXT: END SUB
SUB CULFAZ(G,F(20))
REM COULOMB PHASE SHIFTS
C=0.577215665: S=0: N=50: A1=1.202056903/3: A2=1.036927755/5
FOR I=1 TO N: A=G/I-ATN(G/I)-(G/I)^3/3+(G/I)^5/5: S=S+A
NEXT: FAZ=-C*G+A1*G^3-A2*G^5+S: F(0)=FAZ: FOR I=1 TO 20
F(I)=F(I-1)+ATN(G/(I)): NEXT: END SUB
```

Рассмотрим пример рассеяния в $p^3He$ системе при энергии 11.48 МэВ. В работе [211] приведены дифференциальные сечения, которые мы повторяем в таблице 6.10. Ошибки сечений для разных углов составляют 2.2 - 2.5 %.

Таблица 6.10 - Сечения рассеяния.

| $\theta$ , град | $\sigma_e$ , мб/ст | $\Delta\sigma_e$ , мб/ст |
|---|---|---|
| 27.64 | 223.1 | 5.58 |
| 31.97 | 222.0 | 5.55 |
| 36.71 | 211.9 | 5.30 |
| 82.53 | 54.27 | 1.36 |
| 90.00 | 36.76 | 0.92 |
| 96.03 | 25.70 | 0.64 |
| 102.80 | 16.78 | 0.42 |
| 110.55 | 13.21 | 0.33 |
| 116.57 | 13.21 | 0.33 |
| 125.27 | 20.26 | 0.51 |
| 133.48 | 32.21 | 0.81 |
| 140.79 | 45.95 | 1.15 |
| 147.21 | 58.82 | 1.47 |
| 153.90 | 75.46 | 1.89 |
| 162.14 | 92.72 | 2.32 |
| 165.67 | 97.70 | 2.44 |
| 166.59 | 101.10 | 2.53 |

Фазы рассеяния для этой энергии приведены в работе [212] и в таблице 6.11. Поскольку мы не учитываем спин - орбитальное расщепление, вместо трех триплетных Р фаз нужно использовать их среднее значение, для которого примем $55^0$ [210].

Таблица 6.11 - Фазы рассеяния.

| S = 0 | | | S = 1 | | | | |
|---|---|---|---|---|---|---|---|
| $^1S_0$ | $^1P_1$ | $^1D_2$ | $^3S_0$ | $^3P_0$ | $^3P_1$ | $^3P_2$ | $^3D_2$ |
| -84.6 | 21.4 | -18.6 | -88.8 | 44.3 | 49.4 | 66.7 | 2.5 |

Приведем теперь дифференциальные сечения, вычисленные по нашей программе с этими фазами

| θ | σ | θ | σ | θ | σ | θ | σ |
|---|---|---|---|---|---|---|---|
| 10.00 | 770.70 | 52.00 | 149.99 | 94.00 | 26.44 | 136.00 | 37.41 |
| 12.00 | 398.83 | 54.00 | 142.53 | 96.00 | 23.41 | 138.00 | 41.45 |
| 14.00 | 274.67 | 56.00 | 135.08 | 98.00 | 20.70 | 140.00 | 45.66 |
| 16.00 | 232.31 | 58.00 | 127.68 | 100.00 | 18.34 | 142.00 | 50.02 |
| 18.00 | 219.10 | 60.00 | 120.35 | 102.00 | 16.33 | 144.00 | 54.49 |
| 20.00 | 216.29 | 62.00 | 113.12 | 104.00 | 14.67 | 146.00 | 59.03 |
| 22.00 | 216.71 | 64.00 | 106.03 | 106.00 | 13.38 | 148.00 | 63.60 |
| 24.00 | 217.54 | 66.00 | 99.10 | 108.00 | 12.46 | 150.00 | 68.16 |
| 26.00 | 217.70 | 68.00 | 92.34 | 110.00 | 11.90 | 152.00 | 72.68 |
| 28.00 | 216.83 | 70.00 | 85.78 | 112.00 | 11.73 | 154.00 | 77.11 |
| 30.00 | 214.86 | 72.00 | 79.42 | 114.00 | 11.93 | 156.00 | 81.40 |
| 32.00 | 211.86 | 74.00 | 73.29 | 116.00 | 12.51 | 158.00 | 85.53 |
| 34.00 | 207.95 | 76.00 | 67.39 | 118.00 | 13.47 | 160.00 | 89.44 |
| 36.00 | 203.26 | 78.00 | 61.73 | 120.00 | 14.80 | 162.00 | 93.11 |
| 38.00 | 197.91 | 80.00 | 56.34 | 122.00 | 16.49 | 164.00 | 96.50 |
| 40.00 | 192.00 | 82.00 | 51.20 | 124.00 | 18.54 | 166.00 | 99.58 |
| 42.00 | 185.64 | 84.00 | 46.34 | 126.00 | 20.93 | 168.00 | 102.30 |
| 44.00 | 178.93 | 86.00 | 41.76 | 128.00 | 23.65 | 170.00 | 104.66 |
| 46.00 | 171.94 | 88.00 | 37.47 | 130.00 | 26.68 | | |
| 48.00 | 164.75 | 90.00 | 33.49 | 132.00 | 30.00 | | |
| 50.00 | 157.42 | 92.00 | 29.81 | 134.00 | 33.59 | | |

Сравнение расчета с экспериментом приведено на рис.6.1. Из этих результатов видно вполне хорошее согласие с экспериментальными данными, не смотря на то, что в расчетах использовалась средняя Р фаза, т.е. не учитывалось спин - орбитальное расщепление, которое будет рассмотрено в следующем параграфе.

## 6.6 Нетождественные частицы со спином 1/2 и спин - орбитальными силами

Если в системе частиц со спинами 1/2 учесть спин - орбитальное взаимодействие, то для триплетного сечения можно использовать формулы, которые применялись ранее для $^2H^4He$ системы, по-скольку она находится именно в триплетном состоянии

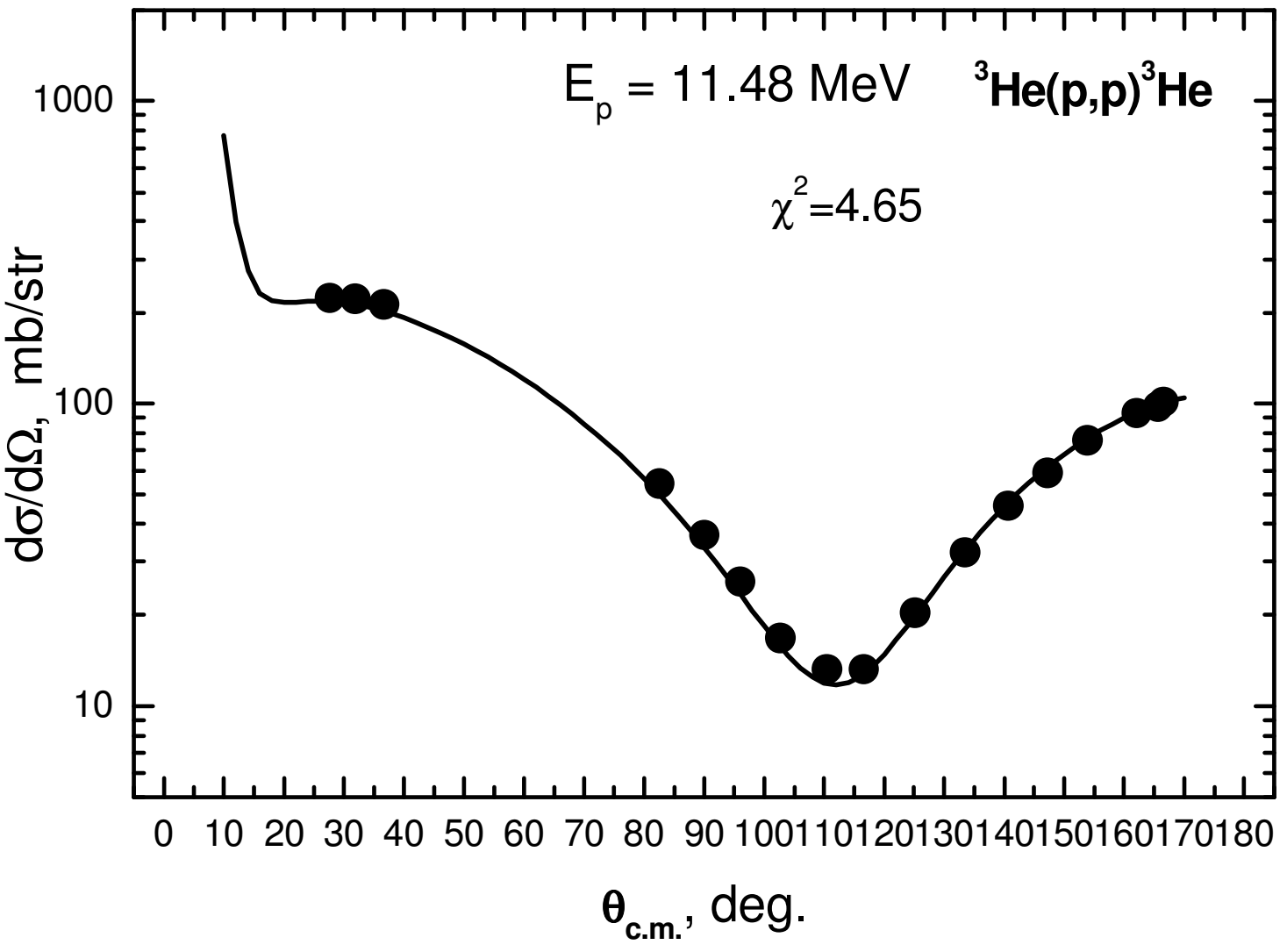

Рис.6.1 Сравнение расчета и экспериментальных данных для упругого p$^3$He рассеяния.

$$\frac{d\sigma_t(\theta)}{d\Omega} = \frac{1}{3}\left[|A|^2 + 2\left(|B|^2 + |C|^2 + |D|^2 + |E|^2\right)\right] \ .$$

Для синглетного сечения используем формулы предыдущего параграфа (6.26), (6.27) и (6.29).

Приведем теперь текст программы для расчетов упругих сечений частиц со спином 1/2 с учетом спин - орбитального взаимодействия. Обозначения практически совпадают с предыдущей программой и программой для рассеяния в $^2H^4He$ системе [203].

```
REM CALCULATE OF CROSS SECTION ON COMPLEX PHASE SHIFTS FOR SYSTEM WITH 1/2+1/2 SPIN
CLS: DEFDBL A-Z: DEFINT I,J,K,L,N,M: N=200
DIM E(N),E1(N),EP(N/10),EM(N/10),E0(N/10)
DIM SEC(N), FM(N/10), F0(N/10), FP(N/10)
REM *******************************************
ISAVE=1: REM =0 - NO SAVE, =1 - SAVE IN FILE
G$="C:\BASICA\SEC\AL-d1.DAT"
REM ************ INPUT PARAMETERS *************
PI=4*ATN(1): NN=0: NV=0: LN=0: LV=2: LH=1: TMI=10: TMA=170
```

```
TH=2: AM1=1: AM2=3: Z1=1: Z2=2: A1=41.4686
PM=AM1*AM2/(AM1+AM2): B1=2*PM/A1
REM ******** ENERGY IN LAB. SYSTEM *************
E1(0)=11.48
REM **** ECSPERIMENTAL CROSS SECTION ********
SECE(1)=223.1: SECE(2)=222: SECE(3)=211.9: SECE(4)=54.27
SECE(5)=36.76
SECE(6)=25.7: SECE(7)=16.78: SECE(8)=13.21: SECE(9)=13.21
SECE(10)=20.26: SECE(11)=32.21: SECE(12)=45.95:
SECE(13)=58.82
SECE(14)=75.46: SECE(15)=92.72: SECE(16)=97.7: SECE(17)=101.1
TT(1)=27.64: TT(2)=31.97: TT(3)=36.71: TT(4)=82.53: TT(5)=90
TT(6)=96.03: TT(7)=102.8: TT(8)=110.55: TT(9)=116.57:
TT(10)=125.27
TT(11)=133.48: TT(12)=140.79: TT(13)=147.21: TT(14)=153.9
TT(15)=162.14: TT(16)=165.67: TT(17)=166.59: NT=17
REM ***************** PHASE SHIFTS *************
REM ********** FOR P-3HE ON E=11.48 *************
FP(0)=-88.8:    FPI(0)=0:  FP(1)=66.7:     FPI(1)=0
FP(2)=2.5:      FPI(2)=0:  F0(0)=-88.8:    F0I(0)=0
F0(1)=49.4:     F0I(1)=0:  F0(2)=2.5:      F0I(2)=0
FM(0)=-88.8:    FMI(0)=0: FM(1)=44.3:     FMI(1)=0
FM(2)=2.5:      FMI(2)=0: FS(0)=-84.6:    FSI(0)=0
FS(1)=21.4:     FSI(1)=0:  FS(2)=-18.6:    FSI(2)=0
FS(3)=0:        FSI(3)=0:  FT(0)=-88.8:    FTI(0)=0
FT(1)=55:       FTI(1)=0:  FT(2)=2.5:      FTI(2)=0
111 FOR L=LN TO LV STEP LH: FM(L)=FM(L)*PI/180
FP(L)=FP(L)*PI/180: F0(L)=F0(L)*PI/180: FMI(L)=FMI(L)*PI/180
FPI(L)=FPI(L)*PI/180: F0I(L)=F0I(L)*PI/180: FT(L)=FT(L)*PI/180
FS(L)=FS(L)*PI/180: FSI(L)=FSI(L)*PI/180: FTI(L)=FTI(L)*PI/180
EP(L)=EXP(-2*FPI(L)): EM(L)=EXP(-2*FMI(L)): E0(L)=EXP(-
2*F0I(L))
ET(L)=EXP(-2*FTI(L)): ES(L)=EXP(-2*FSI(L)): NEXT
REM ************ TRANSFORM TO C.M. *************
FOR I=NN TO NV: E(I)=E1(I)*PM/AM1: NEXT I
REM ***********  TOTAL CROSS SECTION  **********
FOR J=NN TO NV: SK=E(J)*B1: SS=SQR(SK)
GG=3.44476E-02*Z1*Z2*PM/SS: SRT=0: SRS=0:   SST=0:  SSS=0
FOR L=LN TO LV STEP LH: AP=FP(L): AM=FM(L): A0=F0(L)
ASS=FS(L): L1=2*L+3: L2=2*L+1: L3=2*L-1
SRT=SRT+L1*(1-EP(L)^2)+L2*(1-E0(L)^2)+L3*(1-EM(L)^2)
SRS=SRS+L2*(1-ES(L)^2)
SST=SST+L1*EP(L)^2*SIN(AP)^2+L2*E0(L)^2*SIN(A0)^2+L3*EM(
L)^2*SIN(AM)^2: SSS=SSS+L2*ES(L)^2*SIN(ASS)^2: NEXT L
```

```
 SRT=10*PI*SRT/SK/3:                              SRS=10*PI*SRS/SK:
SIGR=1/4*SRS+3/4*SRT
 SST=10*4*PI*SST/SK/3: SSS=10*4*PI*SSS/SK
 SIGS=1/4*SSS+3/4*SST
 PRINT "                        SIGMR-TOT=";: PRINT USING " ####.###
";SIGR
 PRINT "                        SIGMS-TOT=";: PRINT USING " ####.###
";SIGS
 NEXT J: PRINT
 REM *****  DIFFERENTIAL CROSS SECTION  **********
 CALL SEC(SS,GG,SEC(),POL()): FOR T=TMI TO (TMA/3.3) STEP
TH
 PRINT USING " ### ";T;: PRINT USING " ####.##      ";SEC(T);
 PRINT T+42;: PRINT USING " ####.##      ";SEC(T+42);
 PRINT T+84;: PRINT USING " ####.##      ";SEC(T+84);
 PRINT T+126;: PRINT USING " ####.##      ";SEC(T+126): NEXT
 REM ***************** XI^2 **********************
 S1=0: FOR K=1 TO NT STEP 1
 DE(K)=((SEC(K)-SECE(K))/(SECE(K)*0.025))^2: S1=S1+DE(K)
 NEXT: XI=S1/NT: XIS=SQR(XI): PRINT
 PRINT "                  XI=";XI;XIS: PRINT
 REM ***************   SAVE IN FILE   ***************
 IF ISAVE=0 GOTO 221: OPEN "O",1,G$
 PRINT#1, "             P - 3He FOR LAB E=";: PRINT#1, E1(NN)
 FOR T=TMI TO TMA STEP TH
 PRINT#1, USING "  +#.###^^^^  "; T; SEC(T); POL(T): NEXT
221 END
```

**SUB SEC (SS, GG, S(100), POL(100))**

```
 SHARED FP(),EP(),F0(),E0(),FM(),EM(),FS(),ES(),FT(),ET()
 SHARED                PI,TMI,TMA,TH,LN,LV,LH:                  DIM
S0(20),P(20),P1(20),P2(20)
 CALL CULFAZ(GG,S0()): FOR TT=TMI TO TMA STEP TH
 T=TT*PI/180: X=COS(T): A=2/(1-X): S0=2*S0(0)
 BB=-GG*A: ALO=GG*LOG(A)+S0: RECUL=BB*COS(ALO)
 AMCUL=BB*SIN(ALO): REA=0: AMA=0: REB=0: AMB=0: REC=0
 AMC=0: RED=0: AMD=0: REE=0: AME=0
 FOR L=LN TO LV STEP LH
 CALL POLLEG(X,L,P()): FP=2*FP(L): FM=2*FM(L)
 F0=2*F0(L): SL=2*S0(L): C=COS(SL): S=SIN(SL)
 AL1P=EP(L)*COS(FP)-1: AL2P=EP(L)*SIN(FP)
 AL1M=EM(L)*COS(FM)-1: AL2M=EM(L)*SIN(FM)
 AL10=E0(L)*COS(F0)-1: AL20=E0(L)*SIN(F0)
 A1=(L+1)*AL1P+L*AL1M: A2=(L+1)*AL2P+L*AL2M
 REA=REA+(A2*C+A1*S)*P(L): AMA=AMA+(A2*S-A1*C)*P(L)
```

```
 B1=(L+2)*AL1P+(2*L+1)*AL10+(L-1)*AL1M
 B2=(L+2)*AL2P+(2*L+1)*AL20+(L-1)*AL2M
 REB=REB+(B2*C+B1*S)*P(L)/2: AMB=AMB+(B2*S-B1*C)*P(L)/2
 IF L<1 GOTO 1111: CALL FUNLEG1(X,L,P1()): C1=AL1P-AL1M
 C2=AL2P-AL2M: CC1=1/(SQR(2))
 REC=REC+(C2*C+C1*S)*P1(L)*CC1
 AMC=AMC+(C2*S-C1*C)*P1(L)*CC1: DD1=1/(SQR(2)*L*(L+1))
 D1=L*(L+2)*AL1P-(2*L+1)*AL10-(L^2-1)*AL1M
 D2=L*(L+2)*AL2P-(2*L+1)*AL20-(L^2-1)*AL2M
 RED=RED+(D2*C+D1*S)*P1(L)*DD1
 AMD=AMD+(D2*S-D1*C)*P1(L)*DD1
1111 IF L<2 GOTO 2222:CALL FUNLEG2(X,L,P2()): EE1=1/(2*L*(L+1))
 E1=L*AL1P-(2*L+1)*AL10+(L+1)*AL1M
 E2=L*AL2P-(2*L+1)*AL20+(L+1)*AL2M
 REE=REE+(E2*C+E1*S)*P2(L)*EE1
 AME=AME+(E2*S-E1*C)*P2(L)*EE1
2222 NEXT L: RET=0: AMT=0: RES=0:  AMS=0
 FOR L=LN TO LV STEP LH
 C=COS(2*S0(L)): S=SIN(2*S0(L)): ALS=ES(L)*COS(2*FS(L))-1
 BS=ES(L)*SIN(2*FS(L)): RES=RES+(2*L+1)*(BS*C+ALS*S)*P(L)
 AMS=AMS+(2*L+1)*(BS*S-ALS*C)*P(L): NEXT L
 RES=RECUL+RES: AMS=AMCUL+AMS
 SES=10*(RES^2+AMS^2)/4/SS^2: REA=RECUL+REA
 AMA=AMCUL+AMA: REB=RECUL+REB: AMB=AMCUL+AMB
 AA=REA^2+AMA^2: BB=REB^2+AMB^2: CC=REC^2+AMC^2
 DD=RED^2+AMD^2: EE=REE^2+AME^2
 SEC=10*(AA+2*(BB+CC+DD+EE))/4/SS^2/3:
S(TT)=3/4*SEC+1/4*SES
 NEXT TT: END SUB
 SUB POLLEG(X,L,P(20))
 P(0)=1: P(1)=X: FOR I=2 TO L: P(I)=(2*I-1)*X/I*P(I-1)-(I-1)/I*P(I-2)
 NEXT: END SUB
SUB FUNLEG1(X,L,P(20))
 P(0)=0: P(1)=SQR(ABS(1-X^2)): FOR I=2 TO L
 P(I)=(2*I-1)*X/(I-1)*P(I-1)-I/(I-1)*P(I-2): NEXT: END SUB
SUB FUNLEG2(X,L,P(20))
 P(0)=0: P(1)=0: P(2)=3*ABS(1-X^2): FOR I=3 TO L
 P(I)=(2*I-1)*X/(I-2)*P(I-1)-(I+1)/(I-2)*P(I-2): NEXT: END SUB
SUB CULFAZ(G,F(20))
 C=0.577215665: S=0: N=50: A1=1.202056903/3: A2=1.036927755/5
 FOR I=1 TO N: A=G/I-ATN(G/I)-(G/I)^3/3+(G/I)^5/5: S=S+A
 NEXT: FAZ=-C*G+A1*G^3-A2*G^5+S: F(0)=FAZ: FOR I=1 TO 20
 F(I)=F(I-1)+ATN(G/(I)): NEXT: END SUB
```

Приведем результаты контрольного счета для $p^3He$ системы при энергии 11.48 МэВ, которая была рассмотрена в предыдущем параграфе [212]. Теперь мы учитываем спин - орбитальное взаимодействие и используем все три P фазы из работы [212], результаты по которым также были приведены в предыдущем параграфе.

В данном случае был выполнен расчет при углах рассеяния, которые приведены в работе [212] и вычислен полный $\chi^2$ со средней экспериментальной ошибкой 2.5% на точку.

| $\theta$ | $\sigma_e$ | $\sigma_t$ |
|---|---|---|
| 27.64 | 223.10 | 229.16 |
| 31.97 | 222.00 | 222.69 |
| 36.71 | 211.90 | 211.15 |
| 82.53 | 54.27 | 53.52 |
| 90.00 | 36.76 | 36.25 |
| 96.03 | 25.70 | 25.47 |
| 103.80 | 16.78 | 16.16 |
| 110.55 | 13.21 | 12.60 |
| 116.57 | 13.21 | 13.12 |
| 125.27 | 20.26 | 19.96 |
| 133.48 | 32.21 | 32.33 |
| 140.79 | 45.95 | 46.98 |
| 147.21 | 58.82 | 61.39 |
| 153.90 | 75.46 | 76.52 |
| 162.14 | 92.72 | 93.06 |
| 165.67 | 97.70 | 98.82 |
| 166.59 | 101.10 | 100.16 |

$$\chi^2 = 0.74$$

Из этих результатов видно, что согласие с экспериментом существенно улучшилось, а $\chi^2$ вполне согласуется со значением, приведенным в [212], где получено $\chi^2 = 0.45$. Отметим, что в предыдущем случае, когда не использовалось спин - орбитальное расщепление, величина $\chi^2$ была намного больше и равнялась 4.6.

Здесь мы, по - прежнему, не учитываем возможное смешивание состояний с разным спиновым моментом, которое будет рассмотрено в следующем параграфе.

## 6.7 Нетождественные частицы со спином 1/2, спин - орбитальными силами и смешиванием триплет - синглетных состояний

При рассеянии нетождественных частиц с полуцелым спином, например, $N^3H$, $N^3He$, с учетом спин - орбитальных взаимодействий, смешивания различных орбитальных состояний за счет тензорных сил и смешивания синглет - триплетных состояний, дифференциальное сечение рассеяния имеет более сложный вид, и в формулы для сечений входят, как фазы рассеяния, так и параметры смешивания состояний с разным спином и орбитальным моментом [213]

$$\frac{d\sigma(\theta)}{d\Omega} = \frac{1}{2k^2}\{|A|^2 + |B|^2 + |C|^2 + |D|^2 + |E|^2 + |F|^2 + |G|^2 + |H|^2\} \quad . \qquad (6.31)$$

Амплитуды рассеяния записываются в наиболее полном виде

$$A = f_c' + \frac{1}{4}\sum_{L=0}^{\infty} P_L(x)\Big\{-\sqrt{L(L-1)}U^{L-1}_{L,1;L-2,1} + (L+2)U^{L+1}_{L,1;L,1} + (2L+1)U^{L}_{L,1;L,1} +$$
$$+ (L-1)U^{L-1}_{L,1;L,1} - \sqrt{(L+1)(L+2)}\,U^{L+1}_{L,1;L+2,1}\Big\} , \qquad (6.32)$$

$$B = f_c' + \frac{1}{4}\sum_{L=0}^{\infty} P_L(x)\Big\{\sqrt{L(L-1)}U^{L-1}_{L,1;L-2,1} + (L+1)U^{L+1}_{L,1;L,1} + (2L+1)U^{L}_{L,0;L,0} +$$
$$+ LU^{L-1}_{L,1;L,1} + \sqrt{(L+1)(L+2)}\,U^{L+1}_{L,1;L+2,1}\Big\} ,$$

$$C = \frac{1}{4}\sum_{L=0}^{\infty} P_L(x)\Big\{\sqrt{L(L-1)}U^{L-1}_{L,1;L-2,1} + (L+1)U^{L+1}_{L,1;L,1} - (2L+1)U^{L}_{L,0;L,0} +$$
$$+ LU^{L-1}_{L,1;L,1} + \sqrt{(L+1)(L+2)}\,U^{L+1}_{L,1;L+2,1}\Big\} ,$$

$$D = -\frac{1}{4} i\mathrm{Sin}\theta \sum_{L=1}^{\infty} P_L'(x)/\sqrt{L(L+1)}\Big\{-\sqrt{(L+1)(L-1)}U^{L-1}_{L,1;L-2,1} + \sqrt{L(L+1)}U^{L+1}_{L,1;L,1} -$$
$$- \sqrt{L(L+1)}U^{L-1}_{L,1;L,1} + \sqrt{L(L+2)}\,U^{L+1}_{L,1;L+2,1} - (2L+1)U^{L}_{L,1;L,0}\Big\},$$

$$E = -\frac{1}{4} i\mathrm{Sin}\theta \sum_{L=1}^{\infty} P_L'(x)/\sqrt{L(L+1)}\Big\{-\sqrt{(L+1)(L-1)}U^{L-1}_{L,1;L-2,1} + \sqrt{L(L+1)}U^{L+1}_{L,1;L,1} -$$
$$- \sqrt{L(L+1)}U^{L-1}_{L,1;L,1} + \sqrt{L(L+2)}U^{L+1}_{L,1;L+2,1} + (2L+1)U^{L}_{L,1;L,0}\Big\} ,$$

$$F = -\frac{1}{4} i\mathrm{Sin}^2\theta \sum_{L=2}^{\infty} P_L''(x)/\sqrt{(L-1)L(L+1)(L+2)}\Big\{-\sqrt{(L+1)(L+2)}U^{L-1}_{L,1;L-2,1} +$$

$$+\sqrt{\frac{L(L-1)(L+2)}{L+1}}U^{L+1}_{L,1;\,L,1} - (2L+1)\sqrt{\frac{(L-1)(L+2)}{L(L+1)}}U^{L}_{L,1;\,L,1} +$$

$$+\sqrt{\frac{(L-1)(L+1)(L+2)}{L}}U^{L-1}_{L,1;L,1} - \sqrt{L(L-1)}\,U^{L+1}_{L,1;L+2,1}\Big\},$$

$$G = -\frac{1}{4}i\mathrm{Sin}\theta\sum_{L=1}^{\infty}P_L^{'}(x)/\sqrt{L(L+1)}\Bigg\{\sqrt{(L-1)(L+1)}U^{L-1}_{L,1;L-2,1} + (L+2)\sqrt{\frac{L}{L+1}}U^{L+1}_{L,1;L,1} -$$

$$-\frac{(2L+1)}{\sqrt{L(L+1)}}U^{L}_{L,1;L,1} - (L-1)\sqrt{\frac{(L+1)}{L}}U^{L-1}_{L,1;L,1} - \sqrt{L(L+2)}U^{L+1}_{L,1;L+2,1} - (2L+1)U^{L}_{L,0;L,1}\Big\}\ ,$$

$$H = -\frac{1}{4}i\mathrm{Sin}\theta\sum_{L=1}^{\infty}P_L^{'}(x)/\sqrt{L(L+1)}\Bigg\{\sqrt{(L-1)(L+1)}U^{L-1}_{L,1;L-2,1} + (L+2)\sqrt{\frac{L}{L+1}}U^{L+1}_{L,1;L,1} -$$

$$-\frac{(2L+1)}{\sqrt{L(L+1)}}U^{L}_{L,1;L,1} - (L-1)\sqrt{\frac{(L+1)}{L}}U^{L-1}_{L,1;L,1} - \sqrt{L(L+2)}U^{L+1}_{L,1;L+2,1} + (2L+1)U^{L}_{L,0;L,1}\Big\}\ .$$

Здесь матрица рассеяния представляется в форме

$$U^{J}_{L,S;L',S'} = U^{J}_{L',S';L,S} = \exp[i(\alpha_L + \alpha_{L'})](S^{J}_{L',S';L,S} - \delta_{L,L'}\delta_{S,S'})$$

и, например, при L = 1 и J = 1 с учетом смешивания $\varepsilon^1_{1,0} = \varepsilon^J_{S,S'}$ синглетного и триплетного состояний записывается [212]

$$S^1_{1,0;1,0} = \mathrm{Cos}^2\varepsilon^1_{1,0}\exp(2i\delta^1_{0,1}) + \mathrm{Sin}^2\varepsilon_{1,1}\exp(2i\delta^1_{1,1})\quad , \qquad (6.33)$$

$$S^1_{1,1;1,1} = \mathrm{Sin}^2\varepsilon^1_{1,0}\exp(2i\delta^1_{0,1}) + \mathrm{Cos}^2\varepsilon_{1,1}\exp(2i\delta^1_{1,1})\quad ,$$

$$S^1_{1,0;1,1} = S^1_{1,1;1,0} = \frac{1}{2}\mathrm{Sin}(2\varepsilon^1_{1,0})\Big(\exp(2i\delta^1_{0,1}) - \exp(2i\delta^1_{1,1})\Big)\quad ,$$

где $\delta_{k,k'}$ - дельта функция, x = Cos(θ), величины без штриха обозначают начальное состояние, а со штрихом - конечное при том же полном моменте J, кулоновские фазы $\alpha_L$ были определены в (6.8), а ядерные фазы $\delta^J_{S,L}$ считаются комплексными, чтобы учесть неупругие каналы.

Смешивание $\varepsilon_1 = \varepsilon^J_{S,S'}$ триплетных (спины S и S', и полный момент равны 1) S и D состояний определяется следующими выражениями для матрицы рассеяния

$$S^1_{0,0} = \mathrm{Cos}^2\varepsilon_1\exp(2i\delta^1_0) + \mathrm{Sin}^2\varepsilon_1\exp(2i\delta^1_2)\ , \qquad (6.34)$$

$$S^1_{2,2} = \mathrm{Sin}^2\varepsilon_1 \exp(2i\delta^1_0) + \mathrm{Cos}^2\varepsilon_1 \exp(2i\delta^1_2) \ ,$$

$$S^1_{0,2} = S^1_{2,0} = \frac{1}{2}\mathrm{Sin}(2\varepsilon_1)\left[\exp(2i\delta^1_0) - \exp(2i\delta^1_2)\right] .$$

Штрихи у полиномов Лежандра обозначают производные, а кулоновская амплитуда рассеяния записана несколько в другой форме (за знак модуля вынесена величина 1/i)

$$f'_c(\theta) = -\left(\frac{i\eta}{2\ \mathrm{Sin}^2(\theta/2)}\right)\exp\left\{i\eta\ln[\mathrm{Sin}^{-2}(\theta/2)]\right\} .$$

Производные полиномов Лежандра связаны с функциями Лежандра следующим образом

$$P^m_n(x) = (1-x^2)^{m/2}\frac{d^m P_n(x)}{dx^m} = (1-x^2)^{m/2}\frac{d^{m+n}(x^2-1)^n}{dx^{m+n}} = \mathrm{Sin}^m\theta\frac{d^m P_n(\mathrm{Cos}\theta)}{(d\mathrm{Cos}\theta)^m}$$

Поляризация (например, протонов в $p^3H$ рассеянии) может быть записана в виде

$$P = -\left[\frac{2\,\mathrm{Re}\left(AE^* + BH^* + CG^* + DF^*\right)}{|A|^2+|B|^2+|C|^2+|D|^2+|E|^2+|F|^2+|G|^2+|H|^2}\right] . \qquad (6.35)$$

Если в выражениях (6.32) пренебречь тензорным взаимодействием и синглет - триплетным смешиванием, то матрица рассеяния примет обычный вид $\exp(2i\delta_{S,L})$ с комплексной фазой.

В том случае, если в выражениях (6.32) отбросить синглетное состояние (все матричные элементы с S = 0 положить равными нулю), то эти формулы преобразуются к виду, который можно использовать для расчета сечений в $^2H^4He$ системе не только с учетом спин - орбитальных, но и тензорных взаимодействий.

В случае рассеяния не тождественных частиц с полным моментом 3/2, например, $p^2H$ $p^6Li$ и т.д., формулы для сечений упругого рассеяния приведены в работах [214,215,216], а процессы рассеяния тождественных частиц с полуцелым спином, например, NN, $^3H^3H$, $^3He^3He$ и т.д. описаны в работе [217].

Во всех случаях, если оказываются открытыми неупругие процессы, фазы рассеяния становятся комплексными и мнимая часть учитывает переход сталкивающихся частиц в неупругий ка-нал.

Приведем теперь программу расчета дифференциальных сечений упругого рассеяния в системе частиц со спином 1/2 и учетом синглет - триплетного смешивания. Обозначения в программе практически совпадают с обозначениями в предыдущем случае [210].

```
REM CALCULATE OF CROSS SECTION ON COMPLEX PHASE SHIFTS FOR SYSTEM WITH 1/2+1/2 SPIN
CLS: DEFDBL A-Z: DEFINT I,J,K,L,N,M: N=200
DIM E(N), DE(N), E1(N), EP(N/10), EM(N/10), E0(N/10), TT(N), SEC(N), SECE(N), FM(N/10), F0(N/10), FP(N/10), EPSS(N), FPI(N), FPI(N), FMI(N)
REM *********************************************
ISAVE=1: REM =0 - NO SAVE, =1 - SAVE IN FILE
G$="C:\BASICA\SEC\P-3HE-SL.DAT"
REM ************* INPUT PARAMETERS ************
PI=4*ATN(1): NN=0: NV=0: LN=0: LV=2: LH=1: TMI=10: TMA=170
TH=2: AM1=1: AM2=3: Z1=1: Z2=2: A1=41.4686
PM=AM1*AM2/(AM1+AM2): B1=2*PM/A1
REM ********** ENERGY IN LAB. SYSTEM **********
IXI=1: REM IF =1 - XI^2 WILL BE CALCULATE, =0 NO CALCULATE
E1(0)=11.48
REM *** ECSPERIMENTAL CROSS SECTION 11.48 *****
SECE(1)=223.1: SECE(2)=222: SECE(3)=211.9
SECE(4)=54.27: SECE(5)=36.76: SECE(6)=25.7
SECE(7)=16.78: SECE(8)=13.21: SECE(9)=13.21
SECE(10)=20.26: SECE(11)=32.21: SECE(12)=45.95
SECE(13)=58.82: SECE(14)=75.46: SECE(15)=92.72
SECE(16)=97.7: SECE(17)=101.1: TT(1)=27.64: TT(2)=31.97
TT(3)=36.71: TT(4)=82.53: TT(5)=90: TT(6)=96.03: TT(7)=103.8
TT(8)=110.55: TT(9)=116.57: TT(10)=125.27: TT(11)=133.48
TT(12)=140.79: TT(13)=147.21: TT(14)=153.9: TT(15)=162.14
TT(16)=165.67: TT(17)=166.59: NT=17
REM ********** FOR P-3HE ON E=11.48 *************
EPSS(1)=-11.2
FP(0)=-88.8:   FPI(0)=0: FP(1)=66.7:   FPI(1)=0
FP(2)=2.5:     FPI(2)=0: FP(3)=0:      FPI(3)=0
F0(0)=0:       F0I(0)=0: F0(1)=49.4:   F0I(1)=0
F0(2)=2.5:     F0I(2)=0: F0(3)=0:      F0I(3)=0
FM(0)=0:       FMI(0)=0: FM(1)=44.3:   FMI(1)=0
FM(2)=2.5:     FMI(2)=0: FM(3)=0:      FMI(3)=0
```

```
FS(0)=-84.6:     FSI(0)=0: FS(1)=21.4:      FSI(1)=0
FS(2)=-18.6:     FSI(2)=0: FS(3)=0:         FSI(3)=0
FOR L=LN TO LV+2 STEP LH: FM(L)=FM(L)*PI/180
FP(L)=FP(L)*PI/180: F0(L)=F0(L)*PI/180: FMI(L)=FMI(L)*PI/180
FPI(L)=FPI(L)*PI/180: F0I(L)=F0I(L)*PI/180: FS(L)=FS(L)*PI/180
FSI(L)=FSI(L)*PI/180:       EP(L)=EXP(-2*FPI(L)):       EM(L)=EXP(-
2*FMI(L))
E0(L)=EXP(-2*F0I(L)):                          ES(L)=EXP(-2*FSI(L)):
EPS(L)=EPS(L)*PI/180
EPSS(L)=EPSS(L)*PI/180: NEXT
REM ************* TRANSFORM TO C.M. ************
FOR I=NN TO NV: E(I)=E1(I)*PM/AM1: NEXT I
REM ***********  TOTAL CROSS SECTION  **********
FOR J=NN TO NV: SK=E(J)*B1: SS=SQR(SK)
GG=3.44476E-02*Z1*Z2*PM/SS:   SERT=0:   SERS=0:        SEST=0:
SESS=0
FOR L=LN TO LV STEP LH: AP=FP(L): AM=FM(L): A0=F0(L):
ASS=FS(L)
L1=2*L+3: L2=2*L+1: L3=2*L-1
SERT = SERT + L1*(1 - EP(L)^2) + L2*(1 - E0(L)^2) + L3*(1 -
EM(L)^2)
SERS=SERS+L2*(1-ES(L)^2)
SEST = SEST + L1*EP(L)^2*SIN(AP)^2 + L2*E0(L)^2*SIN(A0)^2 +
L3*EM(L)^2*SIN(AM)^2:      SESS=SESS+L2*ES(L)^2*SIN(ASS)^2:
NEXT L
SIGRT=10*PI*SERT/SK/3: SIGRS=10*PI*SERS/SK
SIGR=1/4*SIGRS+3/4*SIGRT: SIGST=10*4*PI*SEST/SK/3
SIGSS=10*4*PI*SESS/SK: SIGS=1/4*SIGSS+3/4*SIGST
PRINT "                    SIGMR-TOT=";: PRINT USING " ####.###
";SIGR
PRINT "                    SIGMS-TOT=";: PRINT USING " ####.###
";SIGS
NEXT J: PRINT
REM *******  DIFFERENTIAL CROSS SECTION  *******
CALL SEC(SS,GG,SEC(),POL())
REM ******************* XI^2 *******************
IF IXI=0 GOTO 211: S1=0: FOR K=1 TO NT STEP 1
DE(K)=((SEC(K)-SECE(K))/(SECE(K)*0.025))^2: S1=S1+DE(K)
NEXT: XI=S1/NT: XIS=SQR(XI)
PRINT "    T      S-EXP        DEL-S-EXP       S-TER          DE"
211 FOR I=1 TO NT STEP 1: PRINT USING " ###.## ";TT(I);
PRINT USING " ######.##        "; SECE(I); SECE(I)*0.025; SEC(I);
DE(I)
NEXT: PRINT: PRINT "               XI=";XI;XIS: PRINT
```

```
REM *************   SAVE IN FILE   **************
IF ISAVE=0 GOTO 221: OPEN "O",1,G$
PRINT#1, "            P - 3HE FOR LAB E=";
PRINT#1, E1(NN): PRINT#1, "  T           SEC "
FOR T=TMI TO TMA STEP TH: PRINT#1, USING "   +#.###^^^^
";T;SEC(T)
NEXT: PRINT#1, "               XI=";XI;XIS
221 END
```

**SUB SEC(SS,GG,S(100),POL(100))**

```
SHARED FP(), EP(), F0(), E0(), FM(), EM(), FS(), ES(), FT(), ET(),
TT()
SHARED PI,TMI,TMA,TH,LN,LV,LH,NT,EPSS()
DIM S0(20),P(20),P1(20),P2(20)
CALL CULFAZ(GG,LV+2,S0()): FOR I=1 TO NT STEP 1
T=TT(I)*PI/180
X=COS(T): CALL AMPCUL(X,GG,S0(),RECUL,AMCUL)
CALL POLLEG(X,LV,P()): CALL FUNLEG1(X,LV,P1())
CALL FUNLEG2(X,LV,P2())
REA=0: AMA=0: REB=0: AMB=0: REC=0
AMC=0: RED=0: AMD=0: REE=0: AME=0:RRG=0: AAG=0: REH=0
AMH=0:    REF=0:   AMF=0:   FOR   L=LN   TO   LV   STEP   LH:
FP=2*FP(L)
FM=2*FM(L): F0=2*F0(L): SL=2*S0(L): C=COS(SL): S=SIN(SL)
FS=2*FS(L): SO=SIN(EPSS(L))^2: CO=COS(EPSS(L))^2
AL1P=EP(L)*COS(FP)-1: AL2P=EP(L)*SIN(FP)
AL1M=EM(L)*COS(FM)-1: AL2M=EM(L)*SIN(FM)
AL10=SO*ES(L)*COS(FS)+CO*E0(L)*COS(F0)-1
AL20=SO*ES(L)*SIN(FS)+CO*E0(L)*SIN(F0)
A1=(L+2)*AL1P+(2*L+1)*AL10+(L-1)*AL1M
A2=(L+2)*AL2P+(2*L+1)*AL20+(L-1)*AL2M
REA=REA+(A1*C-A2*S)*P(L)/2:
AMA=AMA+(A1*S+A2*C)*P(L)/2
ALS=CO*ES(L)*COS(FS)+SO*E0(L)*COS(F0)-1
BS=CO*ES(L)*SIN(FS)+SO*E0(L)*SIN(F0)
RES=(2*L+1)*(ALS*C-BS*S): AMS=(2*L+1)*(ALS*S+BS*C)
B1=(L+1)*AL1P+L*AL1M: B2=(L+1)*AL2P+L*AL2M
REB=REB+(B1*C-B2*S+RES)*P(L)/2
AMB=AMB+(B1*S+B2*C+AMS)*P(L)/2
REC=REC+(B1*C-B2*S-RES)*P(L)/2
AMC=AMC+(B1*S+B2*C-AMS)*P(L)/2
IF L<1 GOTO 1111: SI2=1/2*SIN(2*EPSS(L))
AL1=SI2*(ES(L)*COS(FS)-E0(L)*COS(F0))
AL2=SI2*(ES(L)*SIN(FS)-E0(L)*SIN(F0))
RE1=(2*L+1)*(AL2*C+AL1*S)/SQR(L*(L+1))
```

```
 AM1=(2*L+1)*(AL2*S-AL1*C)/SQR(L*(L+1))
 C1=AL1P-AL1M: C2=AL2P-AL2M
 RED=RED+(C2*C+C1*S-RE1)*P1(L)/2
 AMD=AMD+(C2*S-C1*C-AM1)*P1(L)/2
 REE=REE+(C2*C+C1*S+RE1)*P1(L)/2
 AME=AME+(C2*S-C1*C+AM1)*P1(L)/2
 D1=(L+2)/(L+1)*AL1P-(2*L+1)/(L*(L+1))*AL10-(L-1)/L*AL1M
 D2=(L+2)/(L+1)*AL2P-(2*L+1)/(L*(L+1))*AL20-(L-1)/L*AL2M
 RRG=RRG+(D2*C+D1*S-RE1)*P1(L)/2
 AAG=AAG+(D2*S-D1*C-AM1)*P1(L)/2
 REH=REH+(D2*C+D1*S+RE1)*P1(L)/2
 AMH=AMH+(D2*S-D1*C+AM1)*P1(L)/2
1111 IF L<2 GOTO 2122
 F1=1/(L+1)*AL1P-(2*L+1)/(L*(L+1))*AL10+AL1M/L
 F2=1/(L+1)*AL2P-(2*L+1)/(L*(L+1))*AL20+AL2M/L
 REF=REF+(F2*C+F1*S)*P2(L)/2: AMF=AMF+(F2*S-F1*C)*P2(L)/2
2122 NEXT L: CALL SINGL(X,P(),S0(),LN,LV,LH,RES,AMS)
 RES=RECUL+RES: AMS=AMCUL+AMS
 SES=10*(RES^2+AMS^2)/4/SS^2: REA=RECUL+REA
 AMA=AMCUL+AMA: REB=RECUL+REB: AMB=AMCUL+AMB
 AA=REA^2+AMA^2: BB=REB^2+AMB^2: CC=REC^2+AMC^2
 DD=RED^2+AMD^2: EE=REE^2+AME^2: FF=REF^2+AMF^2
 HH=REH^2+AMH^2: GGG=RRG^2+AAG^2
 SUM=AA+BB+CC+DD+EE+GGG+HH+FF: S(I)=10*SUM/2/SS^2/4
 POL(I)=-2*(REA*REE + AMA*AME + REB*REH + AMB*AMH +
REC*RRG + AMC*AAG + RED*REF + AMD*AMF)/SUM
 REM POL(I)=POL(I)*SIN(T): NEXT I: END SUB
 SUB AMPCUL(X,GG,S0(20),RECUL,AMCUL)
 A=2/(1-X): S0=2*S0(0): BB=-GG*A: AL=GG*LOG(A)+S0
 RECUL=-BB*SIN(AL): AMCUL=BB*COS(AL): END SUB
 SUB SINGL (X, P(20), S0(20), LN, LV, LH, RES, AMS)
 SHARED FS(),ES()
 RES=0:   AMS=0: FOR L=LN TO LV STEP LH: SL=2*S0(L):
C=COS(SL)
 S=SIN(SL): FT=2*FT(L): FS=2*FS(L): ALT=ET(L)*COS(FT)-1
 BT=ET(L)*SIN(FT): ALS=ES(L)*COS(FS)-1: BS=ES(L)*SIN(FS)
 RES=RES+(2*L+1)*(BS*C+ALS*S)*P(L)
 AMS=AMS+(2*L+1)*(BS*S-ALS*C)*P(L): NEXT L: END SUB
 SUB POLLEG(X,L,P(20))
 P(0)=1: P(1)=X: FOR I=2 TO L: P(I)=(2*I-1)*X/I*P(I-1)-(I-1)/I*P(I-2)
 NEXT: END SUB
 SUB FUNLEG1(X,L,P(20))
 P(0)=0: P(1)=SQR(ABS(1-X^2)): FOR I=2 TO L
 P(I)=(2*I-1)*X/(I-1)*P(I-1)-I/(I-1)*P(I-2): NEXT: END SUB
```

```
SUB FUNLEG2(X,L,P(20))
P(0)=0: P(1)=0: P(2)=3*ABS(1-X^2): FOR I=3 TO L
P(I)=(2*I-1)*X/(I-2)*P(I-1)-(I+1)/(I-2)*P(I-2): NEXT: END SUB
SUB CULFAZ(G,L,F(20))
C=0.577215665: S=0: N=50: A1=1.202056903/3: A2=1.036927755/5
FOR I=1 TO N: A=G/I-ATN(G/I)-(G/I)^3/3+(G/I)^5/5: S=S+A: NEXT
FAZ=-C*G+A1*G^3-A2*G^5+S: F(0)=FAZ: FOR I=1 TO L
F(I)=F(I-1)+ATN(G/(I)): NEXT: END SUB
```

Дадим теперь результаты счета по этой программе для $p^3He$ рассеяния при энергии 11.48 МэВ [212] с учетом синглет - триплетного смешивания и заметим, что при ε, равном нулю, все результаты совпадают с результатами, полученными по предыдущей программе. Здесь мы используем значение ε, которое определяет величину синглет - триплетного смешивания, равное $11.2^0$, как приведено в работе [212]. Смешивание S и D волн в триплетном состоянии, обусловленное тензорными силами, не учитывается, поскольку практически не приводит к каким - нибудь заметным изменениям расчетных сечений.

Экспериментальное сечение $\sigma_e$, экспериментальные ошибки $\Delta\sigma_e$ из расчета 2.5% на точку, результаты расчета по нашей программе $\sigma_t$, вычисленные $\chi^2_i$ на каждую точку и среднее $\chi^2$ по всем точкам приведены ниже [210].

| θ | $\sigma_e$ | $\Delta\sigma_e$ | $\sigma_t$ | $\chi^2_i$ |
|---|---|---|---|---|
| 27.64 | 223.10 | 5.58 | 228.04 | 0.78 |
| 31.97 | 222.00 | 5.55 | 221.73 | 0.00 |
| 36.71 | 211.90 | 5.30 | 210.38 | 0.08 |
| 82.53 | 54.27 | 1.36 | 54.22 | 0.00 |
| 90.00 | 36.76 | 0.92 | 36.97 | 0.05 |
| 96.03 | 25.70 | 0.64 | 26.15 | 0.49 |
| 103.80 | 16.78 | 0.42 | 16.74 | 0.01 |
| 110.55 | 13.21 | 0.33 | 13.03 | 0.29 |
| 116.57 | 13.21 | 0.33 | 13.39 | 0.31 |
| 125.27 | 20.26 | 0.51 | 19.97 | 0.32 |
| 133.48 | 32.21 | 0.81 | 32.07 | 0.03 |
| 140.79 | 45.95 | 1.15 | 46.48 | 0.21 |
| 147.21 | 58.82 | 1.47 | 60.70 | 1.64 |
| 153.90 | 75.46 | 1.89 | 75.66 | 0.01 |
| 162.14 | 92.72 | 2.32 | 92.03 | 0.09 |
| 165.67 | 97.70 | 2.44 | 97.73 | 0.00 |
| 166.59 | 101.10 | 2.53 | 99.05 | 0.66 |

$$\chi^2 = 0.29$$

Полученное среднее значение $\chi^2$ несколько меньше, приведенной в работе [212] величины 0.45, поскольку мы использовали среднее значение экспериментальных ошибок 2.5%, а реально, некоторые их них доходят до 2.2%, увеличивая, тем самым, среднюю величину $\chi^2$. Изменение параметра смешивания ε в любую сторону приводит к резкому скачку $\chi^2$, ухудшая описание экспериментальных данных.

# 7. МЕТОДЫ МНОГОПАРАМЕТИЧЕСКОЙ ВАРИАЦИОННОЙ ЗАДАЧИ

В этой главе рассмотрены многопараметрические вариационные методы, которые используются для решения задачи фазового анализа при упругом рассеянии ядерных частиц с различным спином. Выполняется минимизация функционала $\chi^2$ для поиска реальных фаз, описывающих процессы ядерного рассеяния, а именно, дифференциальные сечений при разных энергиях сталкивающихся частиц. Для расчетов дифференциальных сечений используются результаты предыдущей главы.

Множество задач теоретической ядерной физики требуют знания ядерных фаз упругого и неупругого рассеяния, которые могут быть определены из сечений рассеяния различных ядерных частиц. Задача определения ядерных фаз из упругих сечений обычно называется фазовым анализом, и в процессе ее решения возникает не мало специфических проблем.

Когда известны экспериментальные сечения рассеяния ядерных частиц и математические выражения, которые описывают эти сечения (они приведены в предыдущей главе) в зависимости от некоторых параметров $\delta_L$, называемых ядерными фазами рассеяния, возникает многопараметрическая вариационная задача нахождения этих параметров. В разных ядерных системах, в зависимости от энергии сталкивающихся частиц, число этих параметров может колебаться от 3-5 до 20-30.

Поскольку не существует общих методов решения многопараметрической вариационной задачи для поиска глобального минимума, мы можем надеяться найти только некоторые локальные минимумы при каждой энергии и, исходя уже из физических соображений, отобрать те из них, которые могут являться решениями исходной задачи. Основным критерием такого отбора является требование плавного поведения каждой ядерной фазы, как функции энергии в нерезонансной области.

## 7.1 Система частиц с нулевым спином

Рассмотрим методы решения такой задачи для простейшей системы ядерных частиц $^4$He$^4$He, которые имеют нулевой спин и в фазовом анализе нужно учитывать только четные парциальные волны. В случае упругого рассеяния бесспиновых частиц, сечения выражаются через амплитуды и фазы ядерного рассеяния следующим образом (более подробно методы расчета сечений приведены в предыдущей главе) [57]

$$\frac{d\sigma(\theta)}{d\Omega} = |f(\theta)|^2 \ , \qquad (7.1)$$

где амплитуда рассеяния представляется в виде суммы кулоновской и ядерной амплитуд

$$f(\theta) = f_c(\theta) + f_N(\theta) \qquad (7.2)$$

Зная дифференциальные сечения рассеяния, которые определяются экспериментальным путем, можно найти некоторый набор фаз, способный с той или иной точностью, передать поведение этих сечений. Качество описания экспериментальных данных на основе некоторой функции (функционала нескольких переменных, а именно, фаз рассеяния) можно оценить по методу $\chi$ - квадрат [57]

$$\chi^2 = \frac{1}{N}\sum_{i=1}^{N}\left[\frac{\sigma_i(\text{theor}) - \sigma_i(\exp)}{\Delta\sigma_i(\exp)}\right]^2 \ ,$$

Чем меньше величина $\chi^2$, тем лучше описание экспериментальных данных на основе выбранного теоретического представления. Обычно результаты расчетов можно считать вполне удовлетворительными, если $\chi^2$ порядка единицы, т.е. отклонение расчетных и экспериментальных величин примерно равно величине экспериментальных ошибок.

Для поиска ядерных фаз рассеяния по экспериментальны сечениям, нужно выполнить процедуру минимизации функционала $\chi^2$, как функции 2N переменных, каждая из которых является фазой $\delta_L$ определенной парциальной волны рассеяния, и неупругостью $\eta_L$ в этой волне. Чем меньше парциальных волн присутствует при минимизации функционала, тем легче решается такая задача.

Отметим, что задача поиска минимума функционала многих переменных не имеет общего решения, поэтому ищется минимум в некоторой ограниченной области переменных. В частности, величина $\eta_L$ может принимать только значения от 0 до 1, а фазы $\delta_L$ обычно ищут в области $0^0$ - $180^0$, причем в нерезонансной области энергий эти величины должны меняться плавно.

Приведем теперь текст компьютерной программы для поиска фаз рассеяния по заданным экспериментальным дифференциальным сечениям [218].

```
REM * ПРОГРАММА ФАЗОВОГО АНАЛИЗА ДЛЯ AL-AL *
CLS:DEFDBL A-Z:DEFINT I,L,J,N,M,K
```

```
DIM SE(50), ST(50), DS(50), FR(50), FM(50), ET(50), TT(50), XP(50), DE(50)
REM *********** НАЧАЛЬНЫЕ ЗНАЧЕНИЯ **************
FAIL$="C:\BASICA\FAZ-ANAL.DAT": PI=4*ATN(1.): Z1=2: Z2=2
AM1=4: AM2=4: AM=AM1+AM2: P1=3.14159265: A1=41.4686
PM=AM1*AM2/(AM1+AM2): B1=2*PM/A1: LMI=0: LH=2: LMA=8
NYS=1:  REM IF =1 THEN 4HE-4HE: EP=1.0D-03: NV=1: FH=0.1
NI=10: NP=2*LMA+LH
REM *********** НАЧАЛЬНЫЕ ПАРАМЕТРЫ *************
REM ************ CROSS SECTIONS *********************
SE(1)=900: SE(2)=800: SE(3)=580: SE(4)=320: SE(5)=250
SE(6)=170: SE(7)=120: SE(8)=75: SE(9)=47: SE(10)=30
SE(11)=18: se(12)=8: SE(13)=2.7: SE(14)=0.9: SE(15)=2.2
SE(16)=6.2: SE(17)=21: SE(18)=32: SE(19)=41: SE(20)=47
SE(21)=53: SE(22)=58: SE(23)=55: SE(24)=50: SE(25)=42
SE(26)=33: SE(27)=22: SE(28)=6: SE(29)=3.5: SE(30)=5
SE(31)=20: SE(32)=33: SE(33)=50: SE(34)=65: SE(35)=77
SE(36)=85: SE(37)=92: SE(38)=92
DE(1)=50: DE(2)=50: DE(3)=30: DE(4)=20: DE(5)=20: DE(6)=20
DE(7)=20: DE(8)=5: DE(9)=3: DE(10)=2: DE(11)=2: DE(12)=1
DE(13)=5: DE(14)=3: DE(15)=5: DE(16)=5: DE(17)=1: DE(18)=2
DE(19)=1: DE(20)=3: DE(21)=3: DE(22)=2: DE(23)=2: DE(24)=2
DE(25)=2: DE(26)=3: DE(27)=2: DE(28)=5: DE(29)=5: DE(30)=5
DE(31)=2: DE(32)=3: DE(33)=3: DE(34)=3: DE(35)=3: DE(36)=3
DE(37)=3: DE(38)=3
TT(1)=12: TT(2)=14: TT(3)=17.5: TT(4)=21: TT(5)=22.5: TT(6)=25
TT(7)=27: TT(8)=28: TT(9)=30: TT(10)=32.5: TT(11)=35
TT(12)=37: TT(13)=38: TT(14)=41: TT(15)=42.5: TT(16)=44
TT(17)=47.5: TT(18)=49: TT(19)=51: TT(20)=53: TT(21)=55
TT(22)=57: TT(23)=59: TT(24)=60: TT(25)=62: TT(26)=64
TT(27)=66: TT(28)=69: TT(29)=72: TT(30)=74: TT(31)=77
TT(32)=79: TT(33)=81: TT(34)=83: TT(35)=85: TT(36)=87
TT(37)=88:TT(38)=90
REM ************** FOR AL-AL ON E=47.1 ***************
NT=38: EL=51.1
FR(0)=105: FR(2)=48: FR(4)=138: FR(6)=28: FR(8)=2: FM(0)=12.1
FM(2)=22.1: FM(4)=16.3: FM(6)=3.2
REM *********** ENERGY IN LAB. SYSTEM **************
FOR L=LMI TO LMA STEP LH: FM(L)=FM(L)*PI/180
FR(L)=FR(L)*PI/180: NEXT: FH=FH*PI/180
FOR I=LMI TO LMA STEP LH: XP(I)=FR(I): XP(I+LMA+LH)=FM(I)
NEXT
REM ************ TRANSFORM TO C.M. *****************
```

```
EC=EL*PM/AM1: SK=EC*B1: SS=SQR(SK)
GG=3.44476E-02*Z1*Z2*PM/SS
REM ********* DIFFERENTIAL CROSSS SECTION *********
CALL VAR(ST(),FH,LMA,NI,XP(),EP,XI,NV)
PRINT: PRINT "                    XI-KV=";
PRINT USING " ####.#### ";XI;NI
REM *********** TOTAL CROSSS SECTION **************
SIGMAR=0: SIGMAS=0: FOR L=LMI TO LMA STEP LH
FR(L)=XP(L): FM(L)=XP(L+LMA+LH): A=FR(L): ETA(L)=1
IF NP=LMA GOTO 3456: ETA(L)=EXP(-2*FM(L))
3456 SIGMAR=SIGMAR+(2*L+1)*(1-(ETA(L))^2)
SIGMAS=SIGMAS+(2*L+1)*(ETA(L))^2*(SIN(A))^2: NEXT L
SIGMAR=10*4*PI*SIGMAR/SK: SIGMAS=10*4*PI*SIGMAS/SK
PRINT "                    SIGMR-TOT=";: PRINT USING " ####.###
";SIGMAR
PRINT "                    SIGMS-TOT=";: PRINT USING " ####.###
";SIGMAS
PRINT "   T    SE    ST    XI    T    SE    ST    XI"
FOR I=1 TO NT/2
PRINT USING "#####.## "; TT(I); SE(I); ST(I); DS(I), TT(I+17);
SE(I+17); ST(I+17); DS(I+17): NEXT I
FOR L=LMI TO LMA STEP LH: FM(L)=FM(L)*180/PI
FR(L)=FR(L)*180/PI: PRINT USING " ###.#### ";FR(L);FM(L);:
NEXT
OPEN "O",1,G$: PRINT#1, "            ALPHA - ALPHA FOR LAB
E=";
PRINT#1, E1(NN): FOR T=TMI TO TMA STEP TH
PRINT#1, USING " #.###^^^^ ";T;SEC(T): NEXT: END
SUB VAR(ST(50),PHN,LMA,NI,XP(50),EP,AMIN,NV)
DIM XPN(50): SHARED LH,LMI,NT,PI,DS(),NP
REM ************** ПОИСК МИНИМУМА ***************
FOR I=LMI TO NP STEP LH: XPN(I)=XP(I): NEXT: NN=LMI
PRINT USING " ### ";NN;
PRINT USING " +###.#########   ";XPN(NN)*180/PI: PH=PHN
CALL DET(XPN(),ST(),ALA): B=ALA: IF NV=0 GOTO 3012
PRINT USING "        +#.#######^^^^ ";ALA
REM -----------------------------------------------------------------------
FOR IIN=1 TO NI: NN=-LH
PRINT USING "        +#.#######^^^^ ";ALA;IIN:  GOTO 1119
1159 XPN(NN)=XPN(NN)-PH*XP(NN)
1119 NN=NN+LH: IN=0
2229 A=B: XPN(NN)=XPN(NN)+PH*XP(NN)
IF XPN(NN)<0 GOTO 1159: IN=IN+1
REM -----------------------------------------------------------------------
```

```
 CALL DET(XPN(),ST(),ALA): B=ALA
 PRINT USING " ### ";NN;
 PRINT USING " +###.#########    ";XPN(NN)*180/PI;
 PRINT USING "          +#.#######^^^^ ";ALA;: PRINT
 REM ---------------------------------------------------------------------------
 IF ABS((A-B)/(B))<EP GOTO 4449: IF B<A GOTO 2229
 C=A: XPN(NN)=XPN(NN)-PH*XP(NN)
 IF IN>1 GOTO 3339: PH=-PH: GOTO 5559
3339 IF ABS((C-B)/(B))<EP GOTO 4449:  PH=PH/2
5559 B=C: GOTO 2229
4449 PH=PHN: B=C: IF NN<NP GOTO 1119: AMIN=B:PH=PH/NI
 NEXT IIN: FOR I=LMI TO NP STEP LH: XP(I)=XPN(I): NEXT
 END SUB
SUB DET(XP(50),ST(50),XI)
 SHARED SE(),DS(),DE(),NT
 S=0: CALL SEC(XP(),ST()): FOR I=1 TO NT
 S=S+((ST(I)-SE(I))/DE(I))^2: DS(I)=((ST(I)-SE(I))/DE(I))^2
 NEXT:XI=S/NT:END SUB
 SUB SEC(XP(50),S(50))
 SHARED PI,NT,TT(),GG,SS,LMI,LMA,LH,NYS,NP
 DIM S0(20),P(20),FR(50),ET(50)
 FOR I=LMI TO LMA STEP LH: FR(I)=XP(I): ET(I)=1: IF NP=LMA
 GOTO 1234
 ET(I)=EXP(-2*XP(I+LMA+LH))
1234 NEXT: RECUL1=0: AIMCUL1=0: CALL CULFAZ(GG,S0())
 FOR I=1 TO NT: T=TT(I)*PI/180: X=COS(T): A=2/(1-X)
 S0=2*S0(0): BB=-GG*A: ALO=GG*LOG(A)+S0
 RECUL=BB*COS(ALO): AIMCUL=BB*SIN(ALO)
 IF NYS=0 GOTO 555: X1=COS(T): A1=2/(1+X1): BB1=-GG*A1
 ALO1=GG*LOG(A1)+S0: RECUL1=BB1*COS(ALO1)
 AIMCUL1=BB1*SIN(ALO1):555 RENUC=0: AIMNUC=0
 FOR L=LMI TO LMA STEP LH: AL=ET(L)*COS(2*FR(L))-1
 BE=ET(L)*SIN(2*FR(L)): LL=2*L+1: SL=2*S0(L)
 CALL POLLEG(X,L,P())
 RENUC=RENUC+LL*(BE*COS(SL)+AL*SIN(SL))*P(L)
 AIMNUC=AIMNUC+LL*(BE*SIN(SL)-AL*COS(SL))*P(L):   NEXT
L
 IF NYS=0 GOTO 556: AIMNUC=2*AIMNUC: RENUC=2*RENUC
556 RE=RECUL+RECUL1+RENUC
 AIM=AIMCUL+AIMCUL1+AIMNUC
 S(I)=10*(RE^2+AIM^2)/4/SS^2: NEXT I : END SUB
 SUB POLLEG(X,L,P(20))
 P(0)=1: P(1)=X: FOR I=2 TO L: P(I)=(2*I-1)*X/I*P(I-1)-(I-1)/I*P(I-2)
 NEXT: END SUB
```

```
SUB CULFAZ(G,F(20))
C=0.577215665: S=0: N=50: A1=1.202056903/3: A2=1.036927755/5
FOR I=1 TO N: A=G/I-ATN(G/I)-(G/I)^3/3+(G/I)^5/5
S=S+A: NEXT: FAZ=-C*G+A1*G^3-A2*G^5+S
F(0)=FAZ: FOR I=1 TO 20: F(I)=F(I-1)+ATN(G/(I)): NEXT: END SUB
```

Приведем вариант контрольного счета по этой программе, который выполнен для $^4$He$^4$He рассеяния при энергии 29.5 МэВ с данными из работы [200]. В работе приведены экспериментальные сечения при энергиях 18 - 30 МэВ и результаты фазового анализа, которые даны в первом параграфе предыдущей главы в таблице 6.1 для E = 29.5 МэВ [219].

Как видно из результатов, приведенных в предыдущей главе для $^4$He$^4$He рассеянии, наблюдается хорошее согласие вычисленных величин с данными по сечениям $\sigma_e$, приведенными в работе [200] с $\chi^2$=1.086. В этой работе для среднего $\chi^2$ была получена величина 0.68, но методы ее расчета несколько отличаются от изложенных выше (учитывался некоторый весовой множитель), поэтому значение 0.68 нельзя напрямую сравнивать с нашими результатами. Если учесть этот весовой множитель, то для $\chi^2$ можно получить величину 0.60, вполне совпадающую с результатами работы [200].

Далее, представляется интересным выяснить, насколько хорошо был выполнен фазовый анализ сечений при 29.5 МэВ, и можно ли получить меньший $\chi^2$, варьируя фазы из работы [200]. Выполняем уточнение (варьирование) фаз по нашей программе при одной итерации $N_i$=1 (без учета весовых множителей)

$\chi^2$ = 0.679 $\sigma_s$ = 1045.78

| θ | $\sigma_e$ | $\sigma_t$ | $\chi^2$ | θ | $\sigma_e$ | $\sigma_t$ | $\chi^2$ |
|---|---|---|---|---|---|---|---|
| 22.04 | 1523.00 | 1512.94 | 0.72 | 56.09 | 139.86 | 140.68 | 1.33 |
| 24.05 | 1164.00 | 1166.08 | 0.04 | 58.10 | 127.10 | 126.55 | 0.44 |
| 26.05 | 885.90 | 868.17 | 4.00 | 60.10 | 107.83 | 107.47 | 0.20 |
| 28.05 | 616.10 | 619.39 | 0.21 | 62.10 | 86.66 | 86.17 | 0.38 |
| 30.06 | 422.60 | 419.08 | 0.39 | 64.10 | 66.12 | 65.59 | 0.59 |
| 32.06 | 270.00 | 267.49 | 0.38 | 66.10 | 48.43 | 48.56 | 0.07 |
| 34.06 | 160.20 | 159.94 | 0.01 | 68.11 | 37.43 | 37.38 | 0.03 |
| 36.07 | 91.50 | 91.03 | 0.09 | 70.11 | 33.77 | 33.80 | 0.02 |
| 38.07 | 55.53 | 55.08 | 0.38 | 72.11 | 38.34 | 38.50 | 0.24 |
| 40.07 | 44.68 | 44.60 | 0.12 | 74.11 | 50.74 | 51.31 | 1.11 |
| 42.08 | 52.96 | 52.35 | 1.53 | 76.11 | 70.82 | 71.15 | 0.19 |
| 44.08 | 71.74 | 71.16 | 0.61 | 78.11 | 95.55 | 96.14 | 0.45 |
| 46.08 | 95.44 | 94.63 | 0.94 | 80.11 | 124.20 | 123.88 | 0.10 |
| 48.08 | 118.46 | 117.38 | 1.74 | 82.11 | 153.40 | 151.61 | 2.86 |
| 50.09 | 135.58 | 135.39 | 0.08 | 84.11 | 177.50 | 176.57 | 0.81 |
| 52.09 | 145.62 | 145.80 | 0.10 | 86.11 | 197.00 | 196.27 | 0.51 |
| 54.09 | 147.60 | 147.51 | 0.01 | 88.11 | 209.74 | 208.73 | 1.11 |
| 56.09 | 139.86 | 140.68 | 1.33 | 90.11 | 211.20 | 212.69 | 1.98 |

150.88 86.71 121.08 2.19 0.10 - Улучшенный вариант фаз.

Как видно, очень не большие изменения фаз позволяют заметно улучшить описание экспериментальных данных, заметно уменьшая величину $\chi^2$. Выполним теперь дополнительные расчеты фаз при десяти итерациях $N_i = 10$

$\chi^2 = 0.602$ $\sigma_s = 1046.77$

| θ | $\sigma_e$ | $\sigma_t$ | $\chi^2$ | θ | $\sigma_e$ | $\sigma_t$ | $\chi^2$ |
|---|---|---|---|---|---|---|---|
| 22.04 | 1523.00 | 1513.27 | 0.67 | 56.09 | 139.86 | 140.70 | 1.38 |
| 24.05 | 1164.00 | 1166.32 | 0.05 | 58.10 | 127.10 | 126.53 | 0.47 |
| 26.05 | 885.90 | 868.34 | 3.93 | 60.10 | 107.83 | 107.43 | 0.26 |
| 28.05 | 616.10 | 619.52 | 0.23 | 62.10 | 86.66 | 86.10 | 0.50 |
| 30.06 | 422.60 | 419.20 | 0.36 | 64.10 | 66.12 | 65.51 | 0.78 |
| 32.06 | 270.00 | 267.62 | 0.34 | 66.10 | 48.43 | 48.48 | 0.01 |
| 34.06 | 160.20 | 160.08 | 0.00 | 68.11 | 37.43 | 37.31 | 0.17 |
| 36.07 | 91.50 | 91.19 | 0.04 | 70.11 | 33.77 | 33.74 | 0.02 |
| 38.07 | 55.53 | 55.26 | 0.13 | 72.11 | 38.34 | 38.48 | 0.17 |
| 40.07 | 44.68 | 44.80 | 0.23 | 74.11 | 50.74 | 51.33 | 1.17 |
| 42.08 | 52.96 | 52.56 | 0.67 | 76.11 | 70.82 | 71.21 | 0.27 |
| 44.08 | 71.74 | 71.37 | 0.25 | 78.11 | 95.55 | 96.26 | 0.64 |
| 46.08 | 95.44 | 94.82 | 0.56 | 80.11 | 124.20 | 124.05 | 0.02 |
| 48.08 | 118.46 | 117.54 | 1.26 | 82.11 | 153.40 | 151.83 | 2.19 |
| 50.09 | 135.58 | 135.52 | 0.01 | 84.11 | 177.50 | 176.85 | 0.40 |
| 52.09 | 145.62 | 145.89 | 0.24 | 86.11 | 197.00 | 196.58 | 0.17 |
| 54.09 | 147.60 | 147.57 | 0.00 | 88.11 | 209.74 | 209.06 | 0.50 |
| 56.09 | 139.86 | 140.70 | 1.38 | 90.11 | 211.20 | 213.03 | 2.99 |

150.76 86.61 121.00 2.16 0.09 - Улучшенный вариант фаз.

И в этом случае получаем небольшое улучшение описания экспериментальных данных. Рассмотрим теперь возможность включения десятой парциальной волны в наш фазовый анализ, т.е. задаем L = 10 и $N_i = 10$

$\chi^2 = 0.574$ $\sigma_s = 1045.93$

| θ | $\sigma_e$ | $\sigma_t$ | $\chi^2$ | θ | $\sigma_e$ | $\sigma_t$ | $\chi^2$ |
|---|---|---|---|---|---|---|---|
| 22.04 | 1523.00 | 1512.85 | 0.73 | 56.09 | 139.86 | 140.89 | 2.10 |
| 24.05 | 1164.00 | 1165.69 | 0.03 | 58.10 | 127.10 | 126.77 | 0.16 |
| 26.05 | 885.90 | 867.60 | 4.27 | 60.10 | 107.83 | 107.69 | 0.03 |
| 28.05 | 616.10 | 618.78 | 0.14 | 62.10 | 86.66 | 86.37 | 0.13 |
| 30.06 | 422.60 | 418.53 | 0.52 | 64.10 | 66.12 | 65.76 | 0.28 |
| 32.06 | 270.00 | 267.09 | 0.51 | 66.10 | 48.43 | 48.66 | 0.21 |
| 34.06 | 160.20 | 159.72 | 0.03 | 68.11 | 37.43 | 37.40 | 0.01 |
| 36.07 | 91.50 | 90.99 | 0.10 | 70.11 | 33.77 | 33.72 | 0.07 |
| 38.07 | 55.53 | 55.21 | 0.19 | 72.11 | 38.34 | 38.34 | 0.00 |
| 40.07 | 44.68 | 44.86 | 0.54 | 74.11 | 50.74 | 51.10 | 0.44 |
| 42.08 | 52.96 | 52.69 | 0.31 | 76.11 | 70.82 | 70.93 | 0.02 |
| 44.08 | 71.74 | 71.53 | 0.08 | 78.11 | 95.55 | 95.97 | 0.23 |
| 46.08 | 95.44 | 94.98 | 0.30 | 80.11 | 124.20 | 123.81 | 0.15 |
| 48.08 | 118.46 | 117.70 | 0.87 | 82.11 | 153.40 | 151.67 | 2.67 |
| 50.09 | 135.58 | 135.66 | 0.01 | 84.11 | 177.50 | 176.78 | 0.49 |
| 52.09 | 145.62 | 146.03 | 0.56 | 86.11 | 197.00 | 196.61 | 0.15 |
| 54.09 | 147.60 | 147.73 | 0.03 | 88.11 | 209.74 | 209.15 | 0.38 |
| 56.09 | 139.86 | 140.89 | 2.10 | 90.11 | 211.20 | 213.14 | 3.37 |

150.93 86.70 121.04 2.22 0.15 0.06 - Полученные фазы.

Вновь получаем некоторое улучшение результатов описания экспериментальных сечений, которое показано на рис.7.1, однако, почти все эти изменения значений фаз находятся в пределах ошибок их определения, указанных в работе [200] и приведенных в табл.6.2.

Рассмотрим теперь возможность учета максимального числа парциальных волн с включением их мнимой части. Будем учитывать L до 16 и наравне с действительной частью, варьировать мнимую часть фаз рассеяния. В результате получим

$$\chi^2 = 0.4053$$

| θ | $\sigma_e$ | $\sigma_t$ | $\chi^2$ |
|---|---|---|---|
| 2.204E+01 | 1.523E+03 | 1.518E+03 | 1.570E-01 |
| 2.405E+01 | 1.164E+03 | 1.172E+03 | 5.547E-01 |
| 2.605E+01 | 8.859E+02 | 8.738E+02 | 1.865E+00 |
| 2.805E+01 | 6.161E+02 | 6.244E+02 | 1.315E+00 |
| 3.006E+01 | 4.226E+02 | 4.228E+02 | 1.482E-03 |
| 3.206E+01 | 2.700E+02 | 2.698E+02 | 3.048E-03 |
| 3.406E+01 | 1.602E+02 | 1.610E+02 | 7.600E-02 |
| 3.607E+01 | 9.150E+01 | 9.125E+01 | 2.469E-02 |

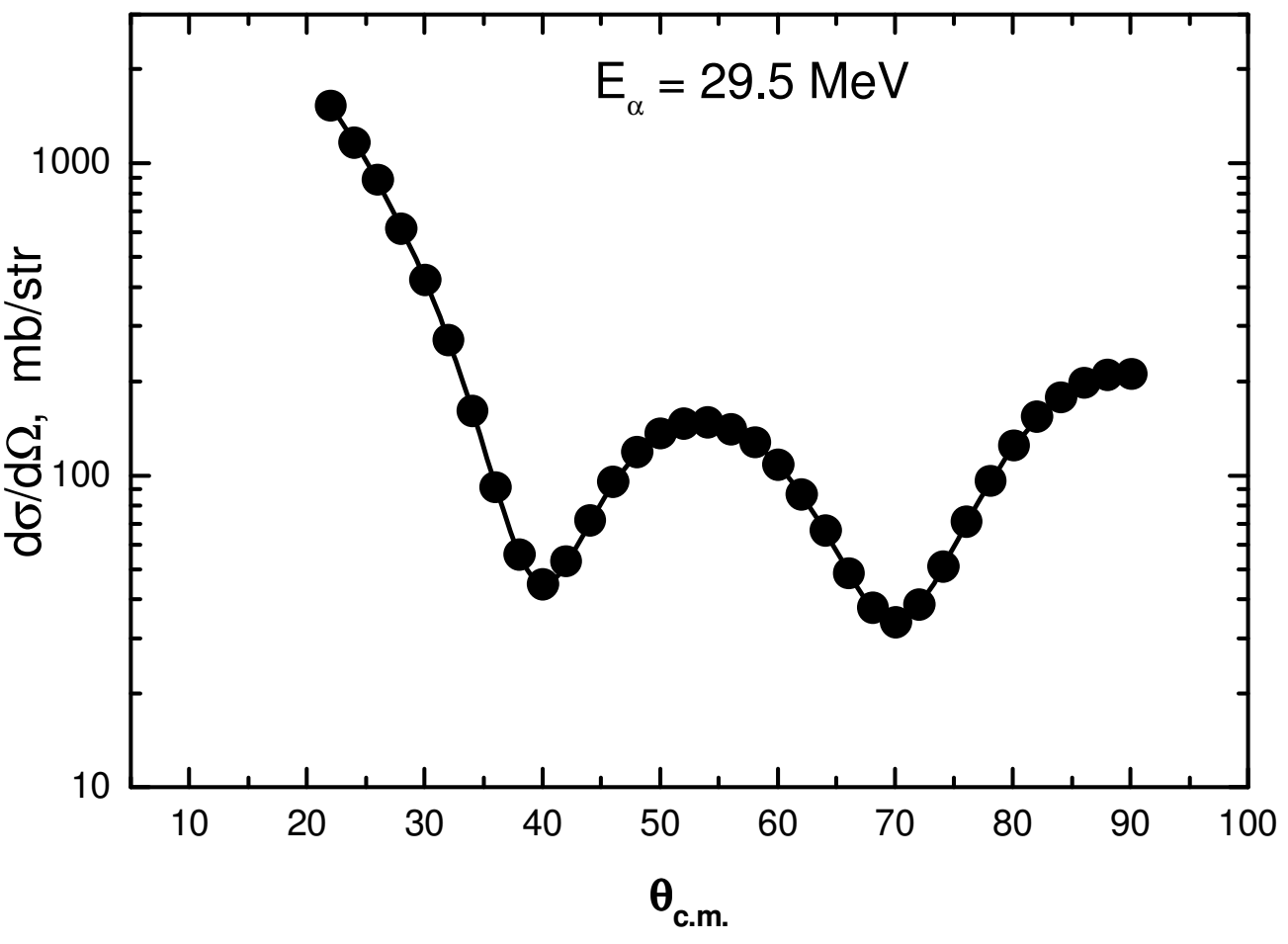

Кружки – экспериментальные данные, сплошная кривая - расчет сечений с найденными фазами.

Рисунок 7.1 - Дифференциальные сечения упругого рассеяния альфа-частиц на ядрах гелия при энергии 29.5 МэВ.

| | | | |
|---|---|---|---|
| 3.807E+01 | 5.553E+01 | 5.506E+01 | 4.083E-01 |
| 4.007E+01 | 4.468E+01 | 4.476E+01 | 9.927E-02 |
| 4.208E+01 | 5.296E+01 | 5.285E+01 | 4.948E-02 |
| 4.408E+01 | 7.174E+01 | 7.192E+01 | 5.968E-02 |
| 4.608E+01 | 9.544E+01 | 9.542E+01 | 4.368E-04 |
| 4.808E+01 | 1.185E+02 | 1.180E+02 | 3.558E-01 |
| 5.009E+01 | 1.356E+02 | 1.356E+02 | 1.039E-02 |
| 5.209E+01 | 1.456E+02 | 1.458E+02 | 5.671E-02 |
| 5.409E+01 | 1.476E+02 | 1.473E+02 | 1.727E-01 |
| 5.609E+01 | 1.399E+02 | 1.405E+02 | 7.909E-01 |
| 5.810E+01 | 1.271E+02 | 1.265E+02 | 4.671E-01 |
| 6.010E+01 | 1.078E+02 | 1.077E+02 | 3.380E-02 |
| 6.210E+01 | 8.666E+01 | 8.654E+01 | 2.201E-02 |
| 6.410E+01 | 6.612E+01 | 6.600E+01 | 3.062E-02 |
| 6.610E+01 | 4.843E+01 | 4.886E+01 | 7.058E-01 |
| 6.811E+01 | 3.743E+01 | 3.747E+01 | 2.062E-02 |
| 7.011E+01 | 3.377E+01 | 3.369E+01 | 1.831E-01 |
| 7.211E+01 | 3.834E+01 | 3.828E+01 | 3.325E-02 |
| 7.411E+01 | 5.074E+01 | 5.110E+01 | 4.388E-01 |
| 7.611E+01 | 7.082E+01 | 7.106E+01 | 1.043E-01 |
| 7.811E+01 | 9.555E+01 | 9.625E+01 | 6.168E-01 |

8.011E+01 1.242E+02 1.242E+02 1.320E-03
8.211E+01 1.534E+02 1.520E+02 1.723E+00
8.411E+01 1.775E+02 1.770E+02 2.271E-01
8.611E+01 1.970E+02 1.967E+02 9.627E-02
8.811E+01 2.097E+02 2.091E+02 4.458E-01
9.011E+01 2.112E+02 2.130E+02 3.034E+00

Действительная часть фаз - 149.7790 85.8000 119.9222 1.8941
0.0383 0.0792 0.1541 0.1233 0.0154
Мнимая часть фаз - 0.0000 0.0000 0.0660 0.0823 0.0594
0.0018 0.0071 0.0592 0.0000

Видно, что учет высших парциальных волн несколько уменьшает величину первых четырех действительных фаз рассеяния, поскольку их влияние перераспределяется на фазы более высоких L. Хотя мнимая часть фаз мала, но оказывает вполне заметное влияние на общее поведение сечений рассеяния, улучшая величину $\chi^2$ примерно в полтора раза по сравнению с вариантом фазового анализа для чисто действительных фаз. Дальнейшее увеличение числа парциальных волн, а рассматривались варианты с L=20 ($\chi^2$ = 0.400) и L=26 ($\chi^2$ = 0.275), не приводит уже к заметному улучшению описания экспериментальных данных и L=14-16 с мнимой частью фаз оказывается вполне достаточно для воспроизведения имеющихся экспериментальных результатов.

Например, для L=26 можно получить

$\chi^2$ = 0.27467
Действительная часть фаз - 149.5111 85.5608 119.6096 1.8579
0.0018 0.0882 0.2221 0.1343 0.0188 0.0853 0.0753 0.0618
0.0591 0.0106
Мнимая часть фаз - 0.0290 0.0998 0.0718 0.1123 0.0663
0.0190 0.0440 0.1371 0.0786 0.0608 0.0516 0.0475 0.0079
0.0347

Эти фазы практически не отличаются от приведенных выше, но учет более высоких парциальных волн позволяет несколько улучшить величину $\chi^2$. Здесь, как и прежде, наблюдается тенденция уменьшение величины первых трех парциальных волн и перераспределение их влияния на состояния с более высокими L.

Теперь более подробно рассмотрим результаты фазового анализа, которые получаются из упругих сечений для $^4$He$^4$He системы при разных энергиях. Остановимся вначале на низких энергиях, меньше 30 МэВ, при которых был выполнен фазовый анализ и сравним наши результаты с ранее полученными.

В работе [198] были выполнены измерения сечений при энергии 6.47 МэВ, которые приведены на рисунках (табличные данные по сечениям не приводятся). Фазовый анализ таких сечений приводит к следующим значениям фаз $\delta_0 = 79.5\pm2^0$, $\delta_2 = 80.8\pm2^0$ (фазы приведены в таблице). Используя описанную программу, найдем для таких фаз дифференциальные сечения, величину $\chi^2$ по каждой точке, среднее $\chi^2$ по всем точкам и величину полного сечения упругого рассеяния в таких процессах $\sigma_s$.

$\chi^2 = 0.262$ $\sigma_s = 2351.43$

| $\theta$ | $\sigma_e$ | $\sigma_t$ | $\chi^2$ | $\theta$ | $\sigma_e$ | $\sigma_t$ | $\chi^2$ |
|---|---|---|---|---|---|---|---|
| 30.00 | 1400.00 | 1402.28 | 0.00 | 65.00 | 80.00 | 83.69 | 0.55 |
| 35.00 | 1050.00 | 1062.02 | 0.06 | 70.00 | 200.00 | 188.60 | 0.32 |
| 40.00 | 700.00 | 704.00 | 0.01 | 75.00 | 300.00 | 313.73 | 0.47 |
| 45.00 | 400.00 | 395.75 | 0.02 | 80.00 | 430.00 | 428.93 | 0.00 |
| 50.00 | 180.00 | 174.58 | 0.07 | 85.00 | 500.00 | 509.12 | 0.09 |
| 55.00 | 60.00 | 54.17 | 1.36 | 90.00 | 530.00 | 537.77 | 0.07 |
| 60.00 | 33.00 | 29.90 | 0.38 | | | | |

79.50 80.80 - Начальные фазы.

Экспериментальные дифференциальные сечения и их ошибки определялись из рисунков работы [198], поэтому величина ошибок больше, чем было найдено в реальных измерениях сечений. Выполним теперь варьирование значений фаз, приведенных в работе [198], с 10 итерациями.

$\chi^2 = 0.175$ $\sigma_s = 2352.89$

| $\theta$ | $\sigma_e$ | $\sigma_t$ | $\chi^2$ | $\theta$ | $\sigma_e$ | $\sigma_t$ | $\chi^2$ |
|---|---|---|---|---|---|---|---|
| 30.00 | 1400.00 | 1413.91 | 0.08 | 65.00 | 80.00 | 81.02 | 0.04 |
| 35.00 | 1050.00 | 1073.05 | 0.21 | 70.00 | 200.00 | 183.85 | 0.65 |
| 40.00 | 700.00 | 713.43 | 0.07 | 75.00 | 300.00 | 307.25 | 0.13 |
| 45.00 | 400.00 | 403.01 | 0.01 | 80.00 | 430.00 | 421.16 | 0.09 |
| 50.00 | 180.00 | 179.38 | 0.00 | 85.00 | 500.00 | 500.56 | 0.00 |
| 55.00 | 60.00 | 56.39 | 0.52 | 90.00 | 530.00 | 528.95 | 0.00 |
| 60.00 | 33.00 | 29.60 | 0.46 | | | | |

80.43 80.73 - Улучшенный вариант фаз.

Видно, что удается несколько улучшить описание эксперимента при не большом изменении значений фаз, причем, уточненные фазы находятся в пределах ошибок их определения, приведенных в работе [198].

В работах [197,220] измерены упругие сечения при энергии 12.3 МэВ (данные приведены в таблице) и получены фазы $\delta_0$ = 29±4$^0$, $\delta_2$ = 103±8$^0$, $\delta_4$ = 3±1.5$^0$ (данные также приведены в таблице). С такими фазами по нашей программе получаются следующие результаты

$\chi^2$ = 3.944 $\sigma_s$ = 1060.58

| θ | $\sigma_e$ | $\sigma_t$ | $\chi^2$ | θ | $\sigma_e$ | $\sigma_t$ | $\chi^2$ |
|---|---|---|---|---|---|---|---|
| 22.00 | 1357.00 | 1254.13 | 6.96 | 48.00 | 57.00 | 57.95 | 0.40 |
| 24.00 | 1203.00 | 1117.86 | 4.53 | 50.00 | 32.50 | 30.14 | 4.61 |
| 26.00 | 1074.00 | 989.81 | 12.30 | 52.00 | 12.30 | 12.13 | 0.03 |
| 28.00 | 870.00 | 867.43 | 0.02 | 55.00 | 2.28 | 2.37 | 0.05 |
| 30.00 | 759.00 | 750.40 | 0.29 | 60.00 | 24.70 | 26.16 | 4.32 |
| 32.00 | 688.00 | 639.28 | 8.21 | 65.00 | 86.50 | 87.80 | 0.42 |
| 35.00 | 467.00 | 485.65 | 2.41 | 70.00 | 157.00 | 170.96 | 15.03 |
| 40.00 | 271.00 | 271.71 | 0.01 | 75.00 | 270.00 | 258.47 | 3.15 |
| 42.00 | 196.00 | 202.86 | 2.80 | 80.00 | 337.00 | 334.34 | 0.13 |
| 45.00 | 130.00 | 118.76 | 9.75 | 85.00 | 408.00 | 385.55 | 7.50 |
| 46.00 | 93.90 | 95.90 | 0.83 | 90.00 | 418.00 | 403.60 | 3.01 |

29.00 103.00 3.00 - Исходные фазы.

Выполним дополнительное варьирование значений фаз с 10 итерациями

$\chi^2$ = 3.432 $\sigma_s$ = 1039.73

| θ | $\sigma_e$ | $\sigma_t$ | $\chi^2$ | θ | $\sigma_e$ | $\sigma_t$ | $\chi^2$ |
|---|---|---|---|---|---|---|---|
| 22.00 | 1357.00 | 1279.60 | 3.94 | 48.00 | 57.00 | 58.89 | 1.59 |
| 24.00 | 1203.00 | 1134.61 | 2.92 | 50.00 | 32.50 | 30.77 | 2.48 |
| 26.00 | 1074.00 | 1000.78 | 9.31 | 52.00 | 12.30 | 12.40 | 0.01 |
| 28.00 | 870.00 | 874.64 | 0.05 | 55.00 | 2.28 | 2.03 | 0.39 |
| 30.00 | 759.00 | 755.23 | 0.06 | 60.00 | 24.70 | 24.77 | 0.01 |
| 32.00 | 688.00 | 642.66 | 7.11 | 65.00 | 86.50 | 85.51 | 0.24 |
| 35.00 | 467.00 | 487.93 | 3.04 | 70.00 | 157.00 | 168.00 | 9.33 |
| 40.00 | 271.00 | 273.40 | 0.12 | 75.00 | 270.00 | 255.08 | 5.27 |
| 42.00 | 196.00 | 204.43 | 4.22 | 80.00 | 337.00 | 330.73 | 0.72 |
| 45.00 | 130.00 | 120.07 | 7.60 | 85.00 | 408.00 | 381.84 | 10.18 |
| 46.00 | 93.90 | 97.11 | 2.13 | 90.00 | 418.00 | 399.87 | 4.77 |

28.37 105.03 2.62 - Улучшенный вариант фаз.

Полученные фазы также находятся в пределах ошибок, приведенных в работе [197].

В той же работе измерены сечения и фазы при энергии 17.8 МэВ. В результате фазового анализа было получено 7±2$^0$, 104±4$^0$, 16.2±2$^0$ (данные по сечениям и фазам приведены в таблице). Расчет с такими фазами по нашей программе приводит нас к результатам

$\chi^2$ = 1.322 $\sigma_s$ = 793.78

| θ | $\sigma_e$ | $\sigma_t$ | $\chi^2$ | θ | $\sigma_e$ | $\sigma_t$ | $\chi^2$ |
|---|---|---|---|---|---|---|---|
| 42.00 | 186.00 | 187.21 | 0.09 | 60.00 | 61.90 | 59.42 | 0.78 |
| 46.00 | 131.00 | 125.82 | 2.79 | 65.00 | 74.90 | 70.96 | 4.29 |
| 48.00 | 110.00 | 104.31 | 3.16 | 70.00 | 98.50 | 96.11 | 1.30 |
| 50.00 | 91.50 | 87.79 | 1.64 | 75.00 | 130.00 | 130.26 | 0.01 |
| 52.00 | 79.30 | 75.53 | 1.82 | 80.00 | 163.00 | 165.49 | 0.32 |
| 55.00 | 65.10 | 63.89 | 0.17 | 85.00 | 188.00 | 192.09 | 0.70 |

7.00 104.00 16.20 - Исходные фазы.

После варьирования с 10 итерациями получаем заметное улучшение согласия расчетных сечений с экспериментом при сравнительно не большом изменении исходных фаз

$\chi^2$ = 0.461 $\sigma_s$ = 804.40

| θ | $\sigma_e$ | $\sigma_t$ | $\chi^2$ | θ | $\sigma_e$ | $\sigma_t$ | $\chi^2$ |
|---|---|---|---|---|---|---|---|
| 42.00 | 186.00 | 186.64 | 0.03 | 60.00 | 61.90 | 63.36 | 0.27 |
| 46.00 | 131.00 | 127.45 | 1.31 | 65.00 | 74.90 | 73.32 | 0.69 |
| 48.00 | 110.00 | 106.94 | 0.91 | 70.00 | 98.50 | 96.65 | 0.78 |
| 50.00 | 91.50 | 91.25 | 0.01 | 75.00 | 130.00 | 129.39 | 0.03 |
| 52.00 | 79.30 | 79.59 | 0.01 | 80.00 | 163.00 | 163.86 | 0.04 |
| 55.00 | 65.10 | 68.37 | 1.27 | 85.00 | 188.00 | 190.19 | 0.20 |

7.25 103.93 17.00 - Улучшенный вариант фаз.

В работах [197,220] была рассмотрена и энергия 22.9 МэВ, для которой получены фазы $\delta_0$ = 169.3±2$^0$, $\delta_2$ = 94.0±2$^0$, $\delta_4$ = 59.2±2$^0$, $\delta_6$ = 1.09$^0$ (данные по сечениям и фазам приведены в таблице). Наши вычисления с такими фазами приводят к следующему результату

$\chi^2$ = 3.059 $\sigma_s$= 1326.12

| θ | $\sigma_e$ | $\sigma_t$ | $\chi^2$ | θ | $\sigma_e$ | $\sigma_t$ | $\chi^2$ |
|---|---|---|---|---|---|---|---|
| 26.00 | 1041.60 | 1053.45 | 3.65 | 48.60 | 231.80 | 229.42 | 1.57 |
| 28.00 | 748.20 | 756.29 | 0.81 | 52.00 | 283.10 | 274.93 | 16.70 |
| 30.60 | 454.20 | 453.44 | 0.02 | 54.80 | 285.70 | 285.30 | 0.04 |
| 34.00 | 202.30 | 201.18 | 0.19 | 60.00 | 241.80 | 240.44 | 0.51 |
| 36.00 | 120.20 | 121.76 | 0.96 | 65.00 | 150.20 | 150.76 | 0.16 |
| 38.00 | 84.70 | 84.68 | 0.00 | 70.20 | 60.30 | 61.14 | 1.95 |
| 40.00 | 81.10 | 81.07 | 0.00 | 76.20 | 7.20 | 7.47 | 2.62 |
| 42.00 | 101.00 | 101.50 | 0.25 | 82.00 | 6.30 | 5.79 | 5.88 |
| 45.00 | 157.00 | 157.08 | 0.00 | 86.00 | 17.20 | 15.37 | 18.05 |
| 48.60 | 231.80 | 229.42 | 1.57 | 90.00 | 21.10 | 20.10 | 4.77 |

169.30 94.00 59.20 1.09 - Исходные фазы.

При дополнительном варьировании значений фаз с 10 итерациями получим некоторое уменьшение $\chi^2$ при очень не большом изменении значений фаз

$\chi^2 = 1.457$ $\sigma_s = 1330.77$

| $\theta$ | $\sigma_e$ | $\sigma_t$ | $\chi^2$ | $\theta$ | $\sigma_e$ | $\sigma_t$ | $\chi^2$ |
|---|---|---|---|---|---|---|---|
| 26.00 | 1041.60 | 1059.08 | 7.95 | 48.60 | 231.80 | 231.07 | 0.15 |
| 28.00 | 748.20 | 760.11 | 1.75 | 52.00 | 283.10 | 277.12 | 8.93 |
| 30.60 | 454.20 | 455.41 | 0.06 | 54.80 | 285.70 | 287.64 | 0.94 |
| 34.00 | 202.30 | 201.63 | 0.07 | 60.00 | 241.80 | 242.26 | 0.06 |
| 36.00 | 120.20 | 121.76 | 0.95 | 65.00 | 150.20 | 151.53 | 0.90 |
| 38.00 | 84.70 | 84.51 | 0.05 | 70.20 | 60.30 | 61.01 | 1.39 |
| 40.00 | 81.10 | 80.98 | 0.02 | 76.20 | 7.20 | 7.28 | 0.20 |
| 42.00 | 101.00 | 101.69 | 0.47 | 82.00 | 6.30 | 6.35 | 0.06 |
| 45.00 | 157.00 | 157.91 | 0.42 | 86.00 | 17.20 | 16.47 | 2.85 |
| 48.60 | 231.80 | 231.07 | 0.15 | 90.00 | 21.10 | 21.41 | 0.46 |

169.30 94.49 59.55 1.00 - Улучшенный вариант фаз.

В работе [200] приведены данные для энергии 25.5 МэВ, для которой получены следующие фазы $\delta_0$ = 160,36±1.01$^0$, $\delta_2$ = 89.37±1.54$^0$, $\delta_4$ = 88.64±1.77$^0$, $\delta_6$ = 1.61±0.39$^0$, $\delta_8$ = 0.36±0.19$^0$ (фазы и сечения даны в таблице). Наш расчет с этими фазами приводит к следующим сечениям

$\chi^2 = 2.127$ $\sigma_s = 1442.59$

| $\theta$ | $\sigma_e$ | $\sigma_t$ | $\chi^2$ | $\theta$ | $\sigma_e$ | $\sigma_t$ | $\chi^2$ |
|---|---|---|---|---|---|---|---|
| 24.04 | 1578.00 | 1556.78 | 2.05 | 54.08 | 281.00 | 277.69 | 7.88 |
| 26.04 | 1154.00 | 1126.52 | 4.77 | 56.08 | 277.70 | 274.84 | 7.74 |
| 28.05 | 759.00 | 765.24 | 0.37 | 58.08 | 258.10 | 257.62 | 0.20 |
| 30.05 | 485.60 | 479.86 | 0.55 | 60.09 | 230.90 | 228.74 | 2.02 |
| 32.05 | 272.80 | 267.25 | 1.01 | 62.09 | 193.00 | 192.29 | 0.21 |
| 34.06 | 127.00 | 121.98 | 2.10 | 64.09 | 152.80 | 152.38 | 0.10 |
| 36.06 | 38.70 | 37.60 | 0.40 | 66.09 | 114.20 | 113.13 | 0.76 |
| 37.06 | 14.80 | 14.86 | 0.00 | 68.09 | 78.00 | 78.21 | 0.04 |
| 38.06 | 3.48 | 3.27 | 0.29 | 70.09 | 50.41 | 50.49 | 0.01 |
| 39.06 | 1.42 | 1.40 | 0.01 | 72.09 | 31.76 | 31.78 | 0.00 |
| 40.06 | 7.85 | 7.81 | 0.01 | 74.09 | 22.91 | 22.74 | 0.81 |
| 42.07 | 39.30 | 39.85 | 0.20 | 75.09 | 22.03 | 21.75 | 4.92 |
| 44.07 | 87.70 | 87.86 | 0.01 | 76.10 | 22.88 | 22.95 | 0.16 |
| 46.07 | 144.50 | 141.84 | 2.60 | 78.10 | 31.45 | 30.93 | 2.08 |
| 48.07 | 195.00 | 193.28 | 1.34 | 80.10 | 45.51 | 44.44 | 4.83 |
| 50.08 | 238.20 | 235.82 | 3.70 | 82.10 | 61.76 | 60.77 | 3.44 |
| 52.08 | 265.70 | 264.61 | 1.16 | 84.10 | 78.54 | 77.11 | 6.23 |
| 54.08 | 281.00 | 277.69 | 7.88 | 86.10 | 92.16 | 90.82 | 6.03 |

160.36 89.37 88.64 1.61 0.39 - Исходные фазы.

Выполним теперь дополнительное варьирование фаз рассеяния с 10 итерациями и получим заметное улучшение описания имеющихся данных

| | $\chi^2 = 0.886$ | | | | $\sigma_s = 1442.01$ | | |
|---|---|---|---|---|---|---|---|
| $\theta$ | $\sigma_e$ | $\sigma_t$ | $\chi^2$ | $\theta$ | $\sigma_e$ | $\sigma_t$ | $\chi^2$ |
| 24.04 | 1578.00 | 1555.27 | 2.35 | 54.08 | 281.00 | 279.67 | 1.26 |
| 26.04 | 1154.00 | 1126.00 | 4.96 | 56.08 | 277.70 | 276.56 | 1.24 |
| 28.05 | 759.00 | 765.12 | 0.36 | 58.08 | 258.10 | 258.96 | 0.63 |
| 30.05 | 485.60 | 479.75 | 0.57 | 60.09 | 230.90 | 229.67 | 0.65 |
| 32.05 | 272.80 | 267.00 | 1.11 | 62.09 | 193.00 | 192.83 | 0.01 |
| 34.06 | 127.00 | 121.56 | 2.46 | 64.09 | 152.80 | 152.58 | 0.03 |
| 36.06 | 38.70 | 37.14 | 0.80 | 66.09 | 114.20 | 113.09 | 0.81 |
| 37.06 | 14.80 | 14.44 | 0.12 | 68.09 | 78.00 | 78.06 | 0.00 |
| 38.06 | 3.48 | 2.94 | 1.86 | 70.09 | 50.41 | 50.32 | 0.01 |
| 39.06 | 1.42 | 1.21 | 1.53 | 72.09 | 31.76 | 31.69 | 0.02 |
| 40.06 | 7.85 | 7.79 | 0.01 | 74.09 | 22.91 | 22.81 | 0.27 |
| 42.07 | 39.30 | 40.28 | 0.64 | 75.09 | 22.03 | 21.91 | 0.85 |
| 44.07 | 87.70 | 88.80 | 0.48 | 76.10 | 22.88 | 23.21 | 3.60 |
| 46.07 | 144.50 | 143.27 | 0.56 | 78.10 | 31.45 | 31.39 | 0.03 |
| 48.07 | 195.00 | 195.10 | 0.00 | 80.10 | 45.51 | 45.08 | 0.77 |
| 50.08 | 238.20 | 237.87 | 0.07 | 82.10 | 61.76 | 61.57 | 0.12 |
| 52.08 | 265.70 | 266.72 | 1.02 | 84.10 | 78.54 | 78.03 | 0.80 |
| 54.08 | 281.00 | 279.67 | 1.26 | 86.10 | 92.16 | 91.83 | 0.38 |

160.49 89.00 88.60 1.41 0.18 - Улучшенный вариант фаз.

Таким образом, во всех рассмотренных случаях, наши конечные результаты, в пределах приведенных ошибок, совпадают с данными, полученными ранее, в различных работах и разными авторами.

Перейдем теперь к рассмотрению области энергий 30-38 МэВ, рассмотренной в работе [221] (сечения приведены в таблицах), где выполнен фазовый анализ этих экспериментальных данных, но фазы показаны только на рисунках.

Для первой из этих энергий 30.3 МэВ в работе [221] получены следующие фазы $135\pm5^0$, $75\pm5^0$, $110\pm5^0$,$-2\pm2^0$, которые мы будем рассматривать, как входные параметры для нашей программы. В результате нашего фазового анализа при L=16 и у чете мнимой части фаз найдено

$\chi^2 =$ 0.2449

| $\theta$ | $\sigma_e$ | $\sigma_t$ | $\chi^2$ |
|---|---|---|---|
| 3.000E+01 | 4.420E+02 | 4.441E+02 | 1.024E-02 |
| 3.250E+01 | 2.660E+02 | 2.620E+02 | 6.160E-02 |
| 3.500E+01 | 1.410E+02 | 1.420E+02 | 1.502E-02 |
| 4.000E+01 | 8.900E+01 | 8.849E+01 | 5.333E-03 |
| 4.500E+01 | 1.170E+02 | 1.168E+02 | 8.217E-04 |
| 5.000E+01 | 1.320E+02 | 1.336E+02 | 3.912E-02 |
| 5.500E+01 | 1.380E+02 | 1.344E+02 | 1.563E-01 |
| 6.000E+01 | 1.060E+02 | 1.080E+02 | 1.160E-01 |
| 6.500E+01 | 4.860E+01 | 4.799E+01 | 4.072E-02 |
| 7.000E+01 | 2.490E+01 | 2.523E+01 | 2.644E-02 |
| 7.500E+01 | 7.900E+01 | 7.672E+01 | 2.084E-01 |
| 8.000E+01 | 1.320E+02 | 1.336E+02 | 1.626E-01 |
| 8.500E+01 | 2.180E+02 | 1.971E+02 | 2.232E+00 |
| 9.000E+01 | 2.340E+02 | 2.405E+02 | 3.545E-01 |

Действительная часть фаз - 136.8425 72.6246 121.0292 0.0000
1.0747 3.5335 0.0881 4.2693 0.0000
Мнимая часть фаз - 1.8943 4.2426 0.4210 2.7232 0.0702
4.5630 0.6693 3.5361 1.5092

Качество описания дифференциальных сечений показано на рис.7.2, а полученные фазы мало отличаются от приведенных в работе [221].

Энергия 31.8 МэВ так же была рассмотрена в работе [221], где на рисунках приведены фазы рассеяния. Используя их в качестве начальных, выполним варьирование по нашей программе. В результате получим для L=8 и без учета мнимой части фаз

$\chi^2 = 2.0$

| $\theta$ | $\sigma_e$ | $\sigma_t$ | $\chi^2$ |
|---|---|---|---|
| 3.000E+01 | 3.600E+02 | 3.768E+02 | 7.844E-01 |
| 3.500E+01 | 1.640E+02 | 1.507E+02 | 2.751E+00 |
| 4.060E+01 | 8.700E+01 | 8.526E+01 | 1.896E-01 |
| 4.500E+01 | 1.060E+02 | 9.951E+01 | 1.686E+00 |
| 5.000E+01 | 1.100E+02 | 1.191E+02 | 3.298E+00 |
| 5.500E+01 | 1.000E+02 | 1.085E+02 | 2.889E+00 |
| 6.000E+01 | 7.800E+01 | 7.123E+01 | 5.099E+00 |
| 6.500E+01 | 4.300E+01 | 3.604E+01 | 5.376E+00 |
| 7.000E+01 | 3.120E+01 | 3.310E+01 | 1.600E+00 |
| 7.500E+01 | 7.600E+01 | 7.334E+01 | 7.872E-01 |
| 8.000E+01 | 1.360E+02 | 1.414E+02 | 5.864E-01 |
| 8.500E+01 | 2.120E+02 | 2.041E+02 | 6.295E-01 |
| 9.000E+01 | 2.350E+02 | 2.293E+02 | 3.943E-01 |

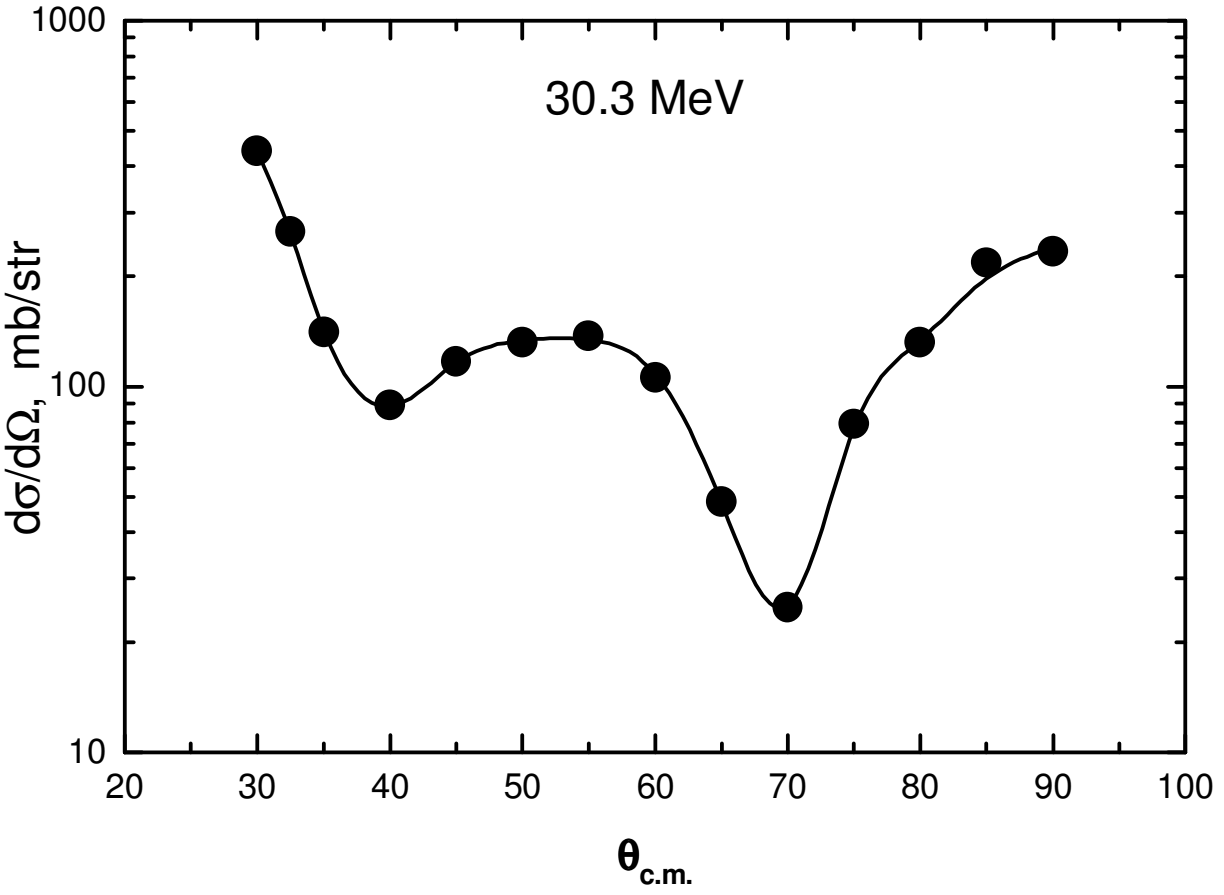

Точки – экспериментальные данные, сплошная кривая - расчет сечений с найденными фазами.

Рисунок 7.2 - Дифференциальные сечения упругого рассеяния альфа-частиц на ядрах гелия при энергии 30.3 МэВ [221].

Для фаз рассеяния найдены следующие величины 143.7759, 78.4276, 125.7295, 0.0002, 0.0002, которые практически не отличаются от результатов фазового анализа, выполненного в работе [221], где на рисунках дано $148^0 \pm 5^0$, $77^0 \pm 5^0$, $125^0 \pm 5^0$.

Рассмотрим теперь возможность улучшения описания этих экспериментальных данных при увеличении числа парциальных волн и учете мнимой части фаз рассеяния. Для L=16 получим

$$\chi^2 = 0.1089$$

| $\theta$ | $\sigma_e$ | $\sigma_t$ | $\chi^2$ |
|---|---|---|---|
| 3.000E+01 | 3.600E+02 | 3.627E+02 | 2.031E-02 |
| 3.500E+01 | 1.640E+02 | 1.626E+02 | 3.190E-02 |
| 4.060E+01 | 8.700E+01 | 8.782E+01 | 4.190E-02 |
| 4.500E+01 | 1.060E+02 | 1.042E+02 | 1.302E-01 |
| 5.000E+01 | 1.100E+02 | 1.116E+02 | 1.030E-01 |
| 5.500E+01 | 1.000E+02 | 9.910E+01 | 3.258E-02 |
| 6.000E+01 | 7.800E+01 | 7.810E+01 | 1.163E-03 |
| 6.500E+01 | 4.300E+01 | 4.286E+01 | 2.076E-03 |
| 7.000E+01 | 3.120E+01 | 3.124E+01 | 7.440E-04 |
| 7.500E+01 | 7.600E+01 | 7.547E+01 | 3.111E-02 |
| 8.000E+01 | 1.360E+02 | 1.397E+02 | 2.834E-01 |
| 8.500E+01 | 2.120E+02 | 2.041E+02 | 6.193E-01 |
| 9.000E+01 | 2.350E+02 | 2.381E+02 | 1.189E-01 |

Действительная часть фаз - 149.8762 70.1656 126.6079 0.0000 1.4289 4.3003 0.0000 2.3328 0.1076
Мнимая часть фаз - 0.0001 2.0772 4.5555 1.6315 0.4573 5.5964 1.9072 2.7722 0.4582

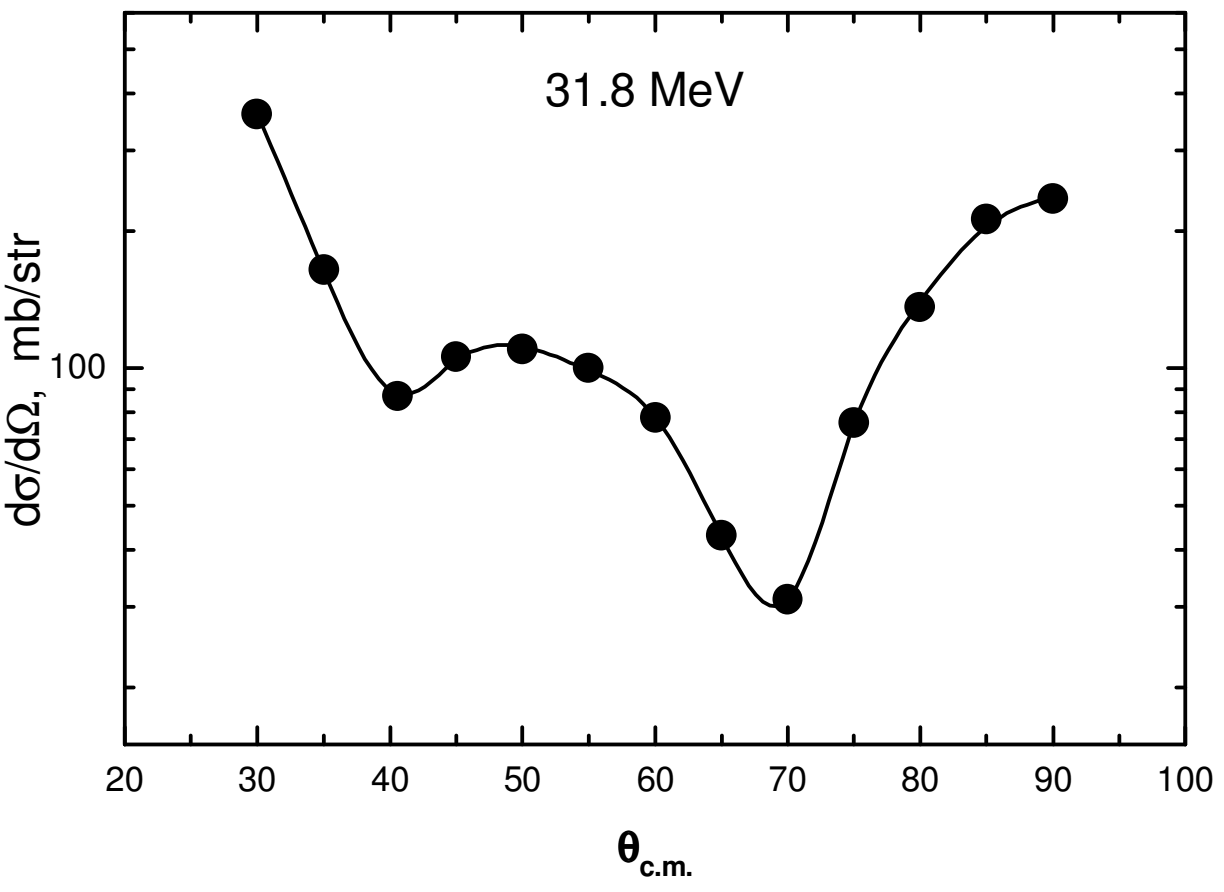

Точки – экспериментальные данные, сплошная кривая - расчет сечений с найденными фазами.
Рисунок 7.3 - Дифференциальные сечения упругого рассеяния альфа-частиц на ядрах гелия при энергии 31.8 МэВ [221].

Качество описания дифференциальных сечений с таким фазами показано на рисунке 7.3.

Далее в работе [221] была рассмотрена энергия 34.2 МэВ, где на рисунках приведены фазы рассеяния $145\pm5^0$, $65\pm5^0$, $145\pm10^0$, $5\pm2^0$. С этими фазами по нашей программе и шестью парциальны-ми волнами можно получить $\chi^2$=6.4. Выполняя далее варьирование с 10 итерациями и включая восьмую парциальную волну, получим заметное улучшение описания экспериментальных данных

$\chi^2$ = 0.970 $\sigma_s$ = 602.99

| $\theta$ | $\sigma_e$ | $\sigma_t$ | $\chi^2$ | $\theta$ | $\sigma_e$ | $\sigma_t$ | $\chi^2$ |
|---|---|---|---|---|---|---|---|
| 30.00 | 263.00 | 252.95 | 0.60 | 60.00 | 29.70 | 29.03 | 0.18 |
| 32.50 | 184.00 | 182.60 | 0.01 | 62.50 | 14.60 | 15.01 | 0.26 |
| 35.00 | 125.00 | 135.94 | 1.48 | 65.00 | 8.70 | 8.49 | 0.09 |
| 40.00 | 112.00 | 96.75 | 3.64 | 67.50 | 11.80 | 12.23 | 0.29 |
| 42.50 | 100.00 | 93.63 | 0.83 | 70.00 | 29.00 | 27.63 | 0.83 |
| 45.00 | 88.00 | 94.26 | 2.45 | 72.50 | 49.10 | 54.43 | 2.32 |
| 47.50 | 94.00 | 94.03 | 0.00 | 75.00 | 101.00 | 90.67 | 4.26 |
| 50.00 | 89.00 | 89.74 | 0.03 | 80.00 | 175.00 | 176.96 | 0.06 |
| 55.00 | 67.40 | 64.99 | 0.53 | 85.00 | 247.00 | 250.85 | 0.30 |
| 60.00 | 29.70 | 29.03 | 0.18 | 90.00 | 286.00 | 279.81 | 0.27 |

145.35 69.41 143.86 4.41 0.76 - Улучшенные фазы рассеяния.

Учтем теперь, как и для предыдущих энергий, 16 парциальных волн и мнимую часть фаз, тогда получим

$\chi^2$ = 0.9049

| $\theta$ | $\sigma_e$ | $\sigma_t$ | $\chi^2$ |
|---|---|---|---|
| 3.000E+01 | 2.630E+02 | 2.562E+02 | 2.762E-01 |
| 3.250E+01 | 1.840E+02 | 1.842E+02 | 1.965E-04 |
| 3.500E+01 | 1.250E+02 | 1.365E+02 | 1.640E+00 |
| 4.000E+01 | 1.120E+02 | 9.810E+01 | 3.021E+00 |
| 4.250E+01 | 1.000E+02 | 9.443E+01 | 6.323E-01 |
| 4.500E+01 | 8.800E+01 | 9.375E+01 | 2.063E+00 |
| 4.750E+01 | 9.400E+01 | 9.235E+01 | 5.583E-02 |
| 5.000E+01 | 8.900E+01 | 8.786E+01 | 8.168E-02 |
| 5.500E+01 | 6.740E+01 | 6.542E+01 | 3.617E-01 |
| 6.000E+01 | 2.970E+01 | 2.996E+01 | 2.592E-02 |
| 6.250E+01 | 1.460E+01 | 1.507E+01 | 3.394E-01 |
| 6.500E+01 | 8.700E+00 | 8.058E+00 | 8.418E-01 |
| 6.750E+01 | 1.180E+01 | 1.217E+01 | 2.176E-01 |
| 7.000E+01 | 2.900E+01 | 2.842E+01 | 1.477E-01 |
| 7.250E+01 | 4.910E+01 | 5.574E+01 | 3.604E+00 |
| 7.500E+01 | 1.010E+02 | 9.167E+01 | 3.484E+00 |
| 8.000E+01 | 1.750E+02 | 1.758E+02 | 1.109E-02 |
| 8.500E+01 | 2.470E+02 | 2.502E+02 | 2.047E-01 |
| 9.000E+01 | 2.860E+02 | 2.808E+02 | 1.853E-01 |

Действительная часть фаз - 142.4348 64.9232 142.1401 2.3609
0.0549 0.0577 0.7871 0.7618 0.4919
Мнимая часть фаз - 1.1109 0.1086 0.1434 0.0726 0.8405
0.0000 0.0000 0.1712 0.6064

Можно еще несколько улучшить качество описания экспериментальных данных, если принять L=20

$$\chi^2 = 0.5645$$

| θ | $\sigma_e$ | $\sigma_t$ | $\chi^2$ |
|---|---|---|---|
| 3.000E+01 | 2.630E+02 | 2.651E+02 | 2.492E-02 |
| 3.250E+01 | 1.840E+02 | 1.712E+02 | 9.769E-01 |
| 3.500E+01 | 1.250E+02 | 1.329E+02 | 7.775E-01 |
| 4.000E+01 | 1.120E+02 | 1.081E+02 | 2.382E-01 |
| 4.250E+01 | 1.000E+02 | 9.738E+01 | 1.404E-01 |
| 4.500E+01 | 8.800E+01 | 9.058E+01 | 4.173E-01 |
| 4.750E+01 | 9.400E+01 | 8.950E+01 | 4.137E-01 |
| 5.000E+01 | 8.900E+01 | 8.893E+01 | 3.022E-04 |
| 5.500E+01 | 6.740E+01 | 6.748E+01 | 5.491E-04 |
| 6.000E+01 | 2.970E+01 | 2.931E+01 | 6.032E-02 |
| 6.250E+01 | 1.460E+01 | 1.492E+01 | 1.623E-01 |
| 6.500E+01 | 8.700E+00 | 8.335E+00 | 2.721E-01 |
| 6.750E+01 | 1.180E+01 | 1.212E+01 | 1.638E-01 |
| 7.000E+01 | 2.900E+01 | 2.792E+01 | 5.220E-01 |
| 7.250E+01 | 4.910E+01 | 5.565E+01 | 3.497E+00 |
| 7.500E+01 | 1.010E+02 | 9.296E+01 | 2.586E+00 |
| 8.000E+01 | 1.750E+02 | 1.780E+02 | 1.427E-01 |
| 8.500E+01 | 2.470E+02 | 2.494E+02 | 1.200E-01 |
| 9.000E+01 | 2.860E+02 | 2.805E+02 | 2.091E-01 |

Действительная часть фаз - 142.6946 63.8742 134.1698 1.1397
0.0000 1.0921 2.0543 0.6701 1.5754 1.0617 0.6621
Мнимая часть фаз - 0.0000 0.0000 2.2654 0.0000 0.0000
0.0000 1.6313 1.1373 2.1817 1.3685 0.6630

Качество описания экспериментальных дифференциальных сечений показано на рисунке 7.4.

Далее, в работе [221] рассмотрена энергия 35.1 МэВ, для которой на рисунке приведены фазы рассеяния $147\pm5^0$, $80\pm5^0$, $150\pm5^0$, $7\pm2^0$. Используя их в качестве начальных, выполним варьирование с 10 итерациями, L=16 и учетом мнимой части. В результате получится очень хорошее описание экспериментальных данных

$$\chi^2 = 0.01495$$

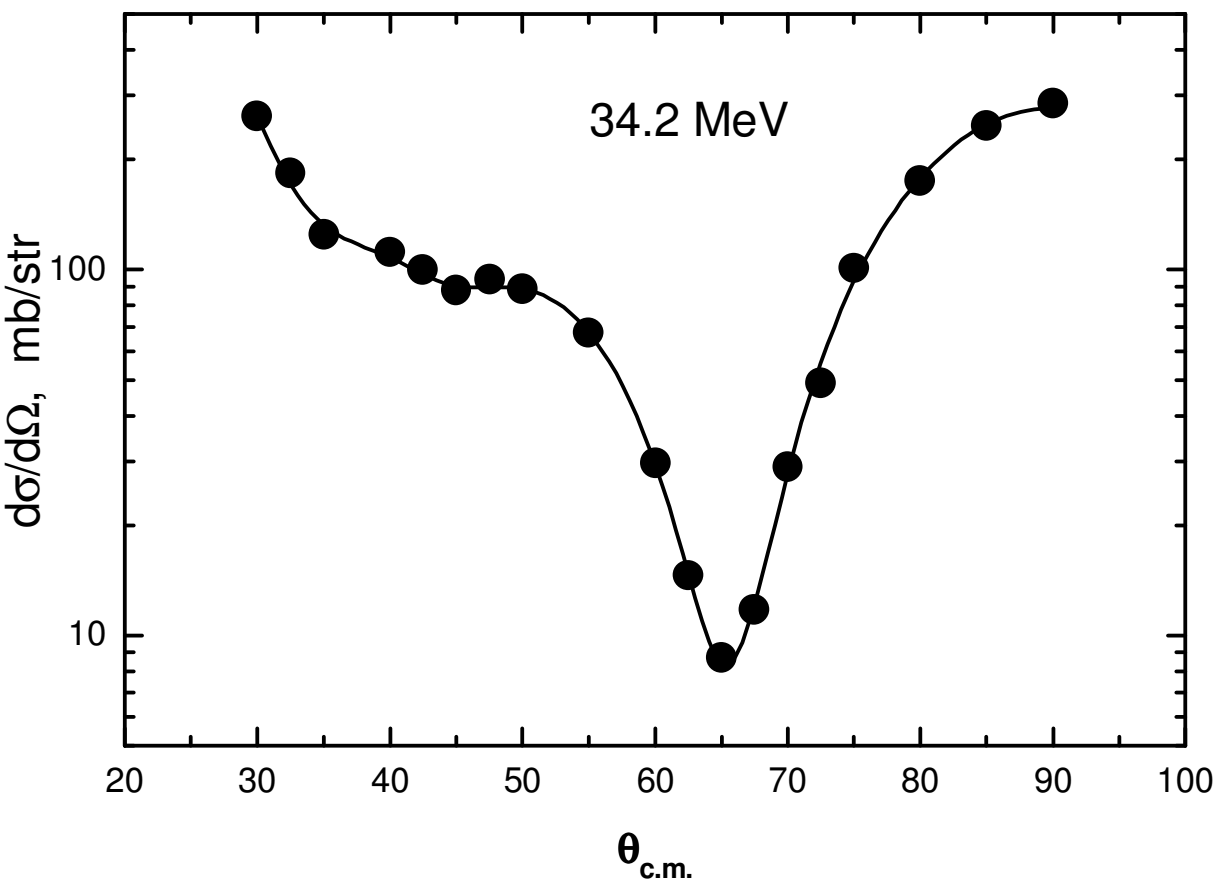

Точки – экспериментальные данные, сплошная кривая - расчет сечений с найденными фазами.

Рисунок 7.4 - Дифференциальные сечения упругого рассеяния альфа-частиц на ядрах гелия при энергии 34.2 МэВ [221].

| $\theta$ | $\sigma_e$ | $\sigma_t$ | $\chi^2$ |
|---|---|---|---|
| 3.000E+01 | 3.180E+02 | 3.189E+02 | 3.453E-03 |
| 3.500E+01 | 1.530E+02 | 1.525E+02 | 5.475E-03 |
| 4.000E+01 | 1.050E+02 | 1.054E+02 | 8.040E-03 |
| 4.500E+01 | 8.800E+01 | 8.766E+01 | 7.292E-03 |
| 5.000E+01 | 7.800E+01 | 7.845E+01 | 8.134E-03 |
| 5.500E+01 | 7.100E+01 | 7.081E+01 | 2.322E-03 |
| 6.000E+01 | 2.870E+01 | 2.873E+01 | 3.956E-04 |
| 6.500E+01 | 7.800E+00 | 7.801E+00 | 1.406E-06 |
| 7.000E+01 | 2.690E+01 | 2.689E+01 | 9.947E-05 |
| 7.500E+01 | 7.610E+01 | 7.635E+01 | 4.632E-03 |
| 8.000E+01 | 1.340E+02 | 1.325E+02 | 4.341E-02 |
| 8.500E+01 | 2.000E+02 | 2.018E+02 | 6.279E-02 |
| 9.000E+01 | 2.610E+02 | 2.581E+02 | 4.828E-02 |

Действительная часть фаз - 155.1029 70.2311 137.7721 0.0000 1.0646 3.1268 1.9988 0.4711 1.7829

Мнимая часть фаз - 3.3855 0.4609 4.6500 0.8327 1.1338 1.5133 5.1306 0.0000 3.3445

Результаты описания дифференциальных сечений представлены на рис.7.5.

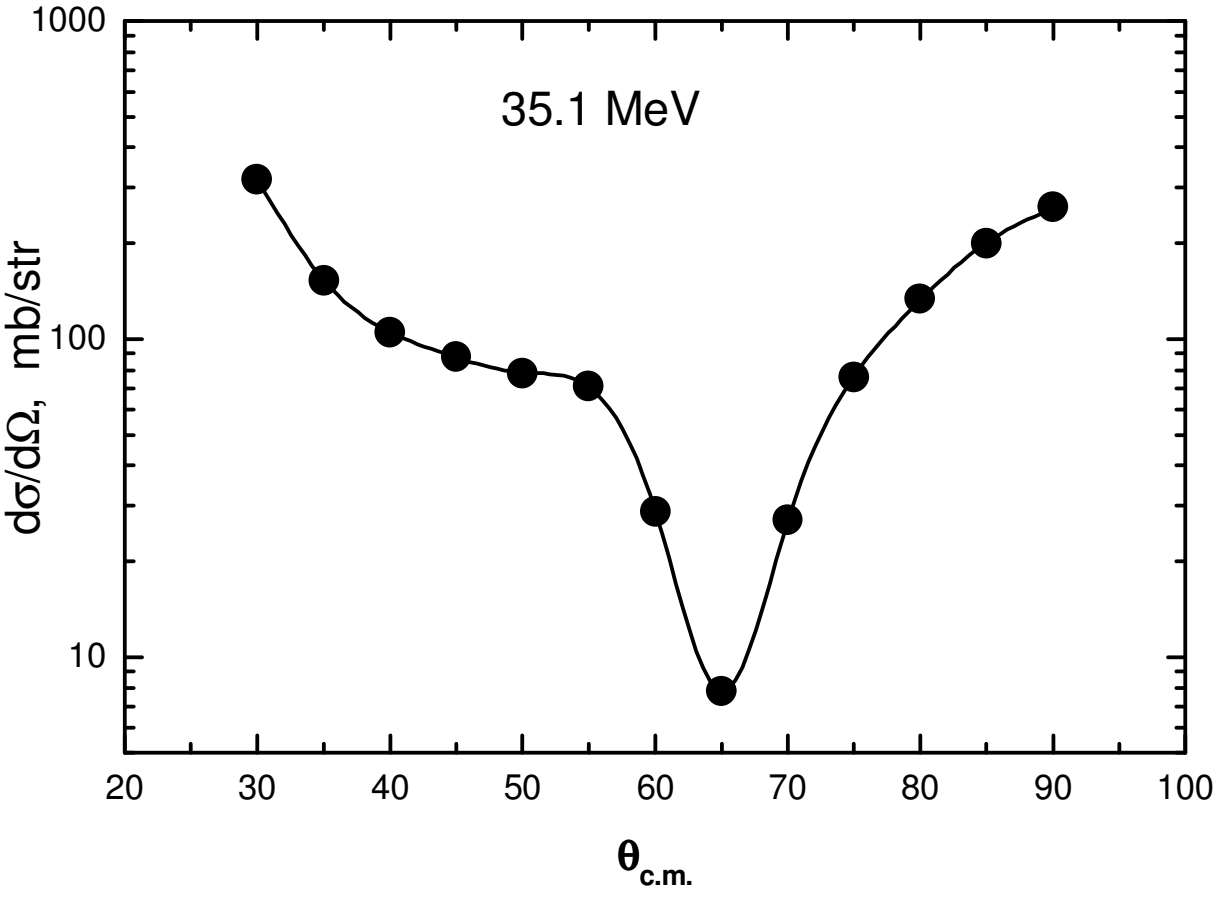

Точки – экспериментальные данные, сплошная кривая - расчет сечений с найденными фазами.

Рисунок 7.5 - Дифференциальные сечения упругого рассеяния альфа-частиц на ядрах гелия при энергии 35.1 МэВ [221].

Следующая энергия рассмотренная в работе [221] - это 37.0 МэВ, для которой на рисунках приведены фазы $137\pm5^0$, $72\pm5^0$, $145\pm5^0$, $-2\pm2^0$. Выполняя варьирование фаз с этими начальными условиями при L=20 получим

$\chi^2$ = 0.1990

| θ | $\sigma_e$ | $\sigma_t$ | $\chi^2$ |
|---|---|---|---|
| 3.000E+01 | 3.310E+02 | 3.299E+02 | 4.584E-03 |
| 3.500E+01 | 1.830E+02 | 1.830E+02 | 3.691E-05 |
| 4.000E+01 | 1.330E+02 | 1.336E+02 | 6.507E-03 |
| 4.500E+01 | 1.270E+02 | 1.258E+02 | 2.230E-02 |
| 5.000E+01 | 6.540E+01 | 6.553E+01 | 9.390E-04 |
| 5.500E+01 | 3.670E+01 | 3.687E+01 | 4.789E-03 |
| 6.000E+01 | 1.920E+01 | 1.891E+01 | 2.870E-02 |
| 6.500E+01 | 5.200E+00 | 5.207E+00 | 1.334E-03 |
| 7.000E+01 | 2.050E+01 | 2.090E+01 | 5.481E-02 |
| 7.500E+01 | 7.210E+01 | 7.138E+01 | 5.771E-02 |
| 8.000E+01 | 1.030E+02 | 1.059E+02 | 3.432E-01 |
| 8.500E+01 | 1.960E+02 | 1.862E+02 | 1.514E+00 |
| 9.000E+01 | 2.080E+02 | 2.147E+02 | 5.485E-01 |

Действительная часть фаз - 114.1774 55.9445 140.2266 7.0498

0.0000 9.8996 3.2214 7.1907 2.5893 1.7230 3.1121
Мнимая часть фаз - 5.3564 1.5250 8.9494 6.5947 0.0000
8.4460 7.8207 9.7521 1.9544 0.0000 1.3431

Качество описания экспериментальных данных показано на рис.7.6. Отметим, что при L=16 удается получить величину $\chi^2$ около 1 и только учет высших парциальных волн позволяет практически идеально описать приведенные дифференциальные сечения, как это было для предыдущих энергий.

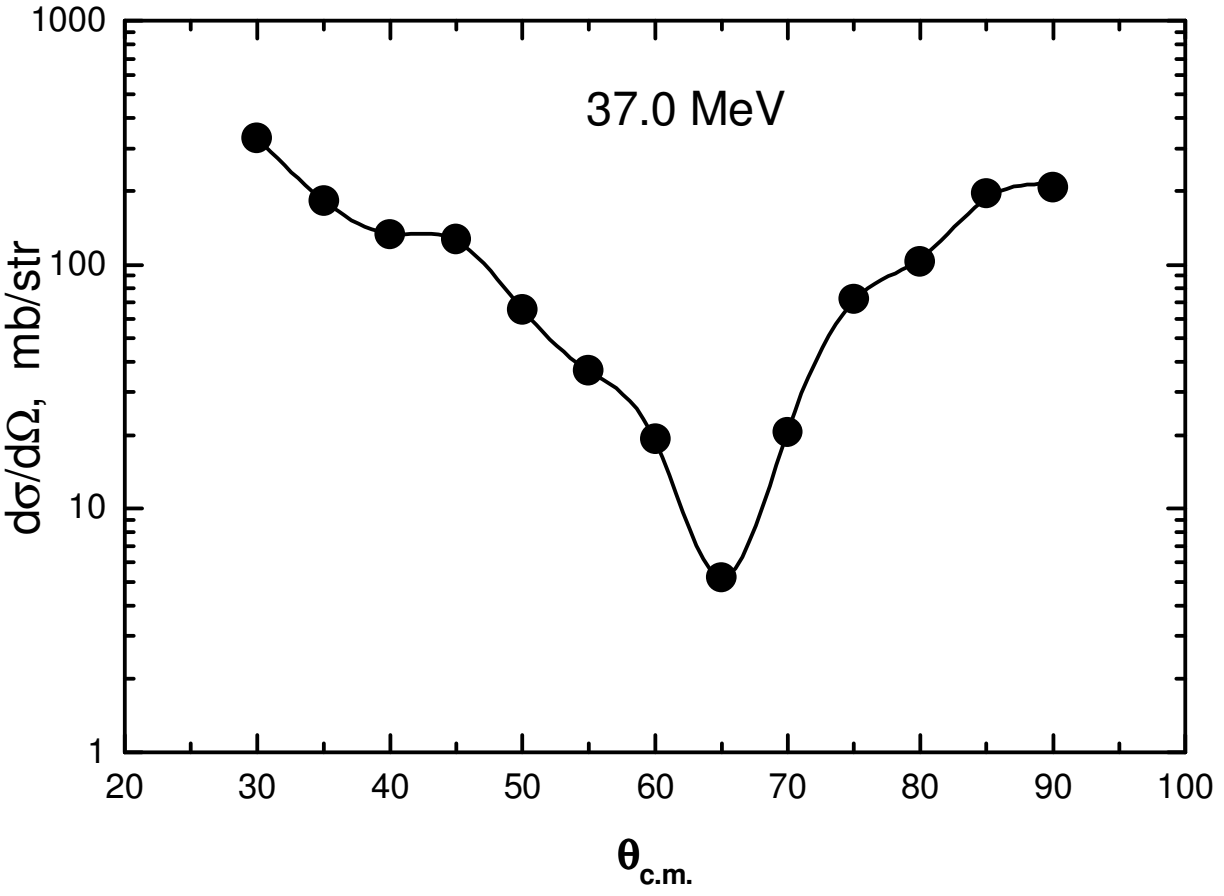

Точки – экспериментальные данные, сплошная кривая - расчет сечений с найденными фазами.

Рисунок 7.6- Дифференциальные сечения упругого рассеяния альфа-частиц на ядрах гелия при энергии 37.0 МэВ [221].

В этой же работы [221] была рассмотрена и энергия 38.4 МэВ, где на рисунках приведены следующие фазы рассеяния $135\pm5^0$, $75\pm5^0$, $170\pm10^0$, $5\pm2^0$. С такими значениями фаз по нашей программе получается $\chi^2 = 19.518$. Варьируем теперь эти значения фаз с 10 итерациями, получим существенное улучшение описания эксперимента при восьми парциальных волнах

$\chi^2$ = 1.425 $\sigma_s$ = 384.87

| θ | $\sigma_e$ | $\sigma_t$ | $\chi^2$ | θ | $\sigma_e$ | $\sigma_t$ | $\chi^2$ |
|---|---|---|---|---|---|---|---|
| 30.00 | 314.00 | 323.52 | 0.46 | 64.00 | 0.50 | 0.50 | 0.00 |
| 35.00 | 212.00 | 225.92 | 1.60 | 65.00 | 0.86 | 0.94 | 0.26 |
| 40.60 | 140.00 | 138.61 | 0.08 | 66.00 | 2.70 | 1.95 | 2.24 |
| 45.00 | 97.00 | 87.22 | 3.83 | 67.50 | 4.20 | 4.56 | 0.52 |
| 50.00 | 49.00 | 45.71 | 2.70 | 70.00 | 11.60 | 11.87 | 0.46 |
| 55.00 | 19.30 | 18.89 | 0.05 | 75.00 | 36.40 | 36.67 | 0.03 |
| 60.00 | 3.45 | 3.96 | 5.87 | 80.00 | 69.10 | 69.08 | 0.00 |
| 62.00 | 1.52 | 1.22 | 1.70 | 85.00 | 100.00 | 96.97 | 0.57 |
| 62.50 | 1.45 | 0.84 | 5.03 | 90.00 | 110.00 | 108.01 | 0.25 |

137.01 89.16 175.00 5.96 0.0002 - Полученные фазы.

Выполним теперь варьирование фаз с учетом мнимой части и L=16

$$\chi^2 = 0.5768$$

| θ | $\sigma_e$ | $\sigma_t$ | $\chi^2$ |
|---|---|---|---|
| 3.000E+01 | 3.140E+02 | 3.103E+02 | 6.828E-02 |
| 3.500E+01 | 2.120E+02 | 2.145E+02 | 5.334E-02 |
| 4.060E+01 | 1.400E+02 | 1.391E+02 | 2.913E-02 |
| 4.500E+01 | 9.700E+01 | 9.606E+01 | 3.510E-02 |
| 5.000E+01 | 4.900E+01 | 4.961E+01 | 9.314E-02 |
| 5.500E+01 | 1.930E+01 | 1.711E+01 | 1.486E+00 |
| 6.000E+01 | 3.450E+00 | 3.655E+00 | 9.576E-01 |
| 6.200E+01 | 1.520E+00 | 1.338E+00 | 6.233E-01 |
| 6.250E+01 | 1.450E+00 | 9.912E-01 | 2.887E+00 |
| 6.400E+01 | 5.000E-01 | 5.982E-01 | 6.702E-01 |
| 6.500E+01 | 8.600E-01 | 9.505E-01 | 3.198E-01 |
| 6.600E+01 | 2.700E+00 | 1.857E+00 | 2.843E+00 |
| 6.750E+01 | 4.200E+00 | 4.334E+00 | 7.219E-02 |
| 7.000E+01 | 1.160E+01 | 1.153E+01 | 3.329E-02 |
| 7.500E+01 | 3.640E+01 | 3.644E+01 | 5.043E-04 |
| 8.000E+01 | 6.910E+01 | 7.007E+01 | 1.512E-01 |
| 8.500E+01 | 1.000E+02 | 9.906E+01 | 5.517E-02 |
| 9.000E+01 | 1.100E+02 | 1.103E+02 | 4.187E-03 |

Действительная часть фаз - 135.0456 82.0880 169.1763 4.0532
1.5103 1.5318 0.9571 0.4465 0.1501
Мнимая часть фаз - 0.0000 2.9087 0.0637 0.1203 0.6567
2.1785 0.5149 1.1855 1.0905

Увеличение числа парциальных волн до 20 приводят к уменьшению $\chi^2$ до 0.47, что уже существенно не влияет на качество описания экспериментальных данных, которое показано на рисунке 7.7.

В работе [222], в таблице приведены дифференциальные сечения при энергии 38.5 МэВ, но фазовый анализ этих данных не выполнялся. Используем в качестве начальных, фазы, полученные в предыдущем случае и при L=16 находим $\chi^2$ = 0.55 со следующими

фазами

Действительная часть фаз - 130.9165 79.8475 166.1612 2.8989
1.5999 1.5393 0.5942 0.1918 0.1599
Мнимая часть фаз - 0.0000 1.6270 1.5261 1.2301 0.0000
2.3412 0.2508 1.6414 0.8989

При L=24 получаем $\chi^2$ = 0.50 и только при 26 парциальных волнах $\chi^2$ начинает резко уменьшаться и оказывается равен 0.265, а для L=30 $\chi^2$ достигает своего предела, равного 0.207.

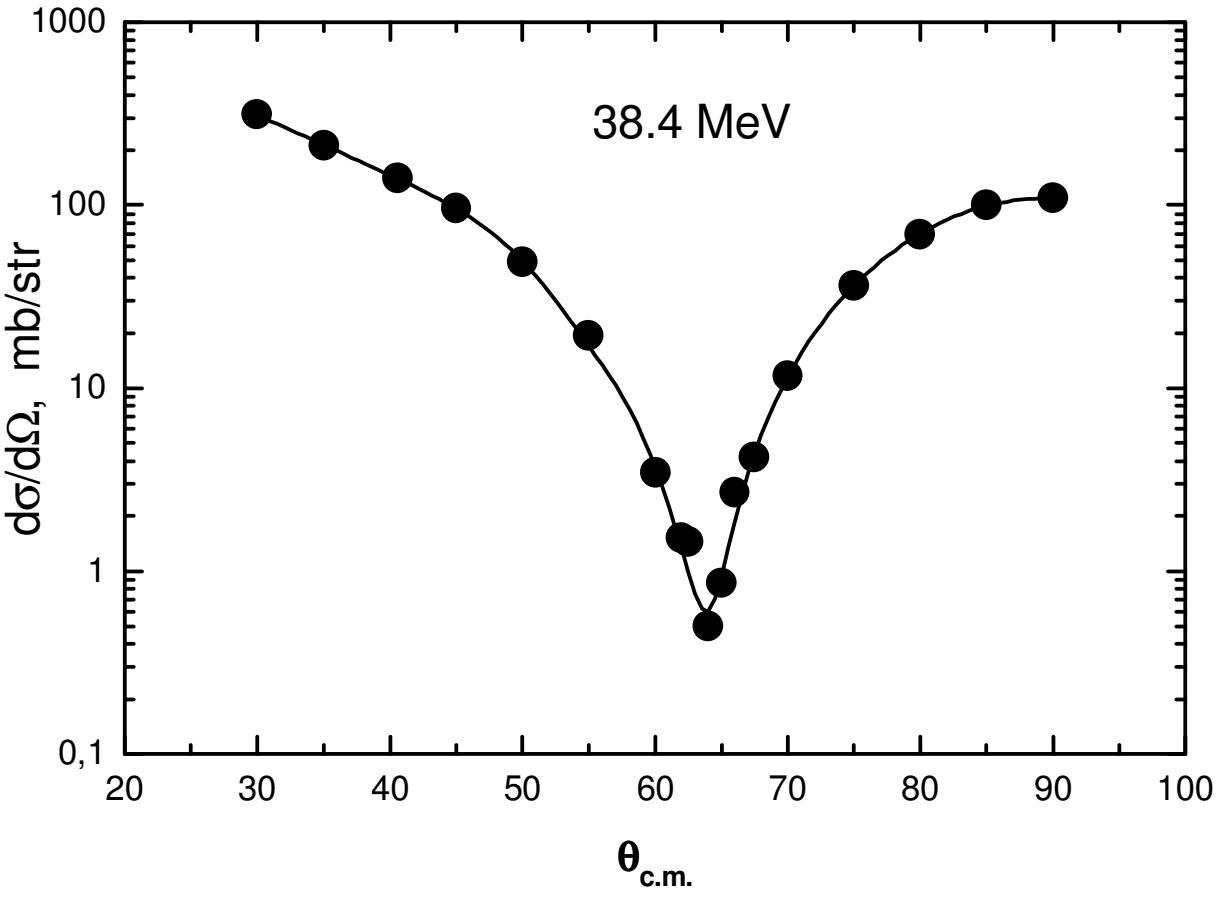

Точки – экспериментальные данные, сплошная кривая - расчет сечений с найденными фазами.

Рисунок 7.7 - Дифференциальные сечения упругого рассеяния альфа-частиц на ядрах гелия при энергии 38.4 МэВ [221].

$\chi^2 = 0.2067$

| $\theta$ | $\sigma_e$ | $\sigma_t$ | $\chi^2$ |
|---|---|---|---|
| 3.000E+01 | 3.140E+02 | 3.143E+02 | 3.215E-04 |
| 3.500E+01 | 2.120E+02 | 2.118E+02 | 4.271E-04 |
| 4.000E+01 | 1.400E+02 | 1.401E+02 | 1.825E-04 |
| 4.500E+01 | 9.700E+01 | 9.689E+01 | 5.244E-04 |
| 5.000E+01 | 4.900E+01 | 4.903E+01 | 1.909E-04 |
| 5.500E+01 | 1.930E+01 | 1.926E+01 | 3.928E-04 |
| 6.000E+01 | 3.450E+00 | 3.447E+00 | 2.058E-04 |
| 6.200E+01 | 1.520E+00 | 1.661E+00 | 3.771E-01 |
| 6.250E+01 | 1.450E+00 | 1.203E+00 | 8.374E-01 |
| 6.400E+01 | 5.000E-01 | 5.087E-01 | 5.220E-03 |
| 6.500E+01 | 8.600E-01 | 9.148E-01 | 1.175E-01 |
| 6.600E+01 | 2.700E+00 | 1.990E+00 | 2.015E+00 |
| 6.750E+01 | 4.200E+00 | 4.500E+00 | 3.594E-01 |
| 7.000E+01 | 1.160E+01 | 1.157E+01 | 5.831E-03 |
| 7.500E+01 | 3.640E+01 | 3.647E+01 | 1.788E-03 |
| 8.000E+01 | 6.910E+01 | 6.910E+01 | 3.205E-06 |
| 8.500E+01 | 1.000E+02 | 1.000E+02 | 1.017E-04 |
| 9.000E+01 | 1.100E+02 | 1.100E+02 | 7.037E-05 |

Действительная часть фаз - 129.9036 77.1998 165.6615 1.5187 0.5671 2.1129 0.0601 0.0000 0.6059 0.4907 0.0005 0.0896 0.3055 0.4272 0.1328 0.0000

Мнимая часть фаз - 3.8528 2.4699 1.1513 0.9460 1.4118 0.0000 1.7083 2.0442 0.0000 0.2196 0.8160 2.4573 0.0000 0.9449 0.4563 0.0000

На рис.7.8 приведены результаты расчетов дифференциальных сечений с полученными фазами.

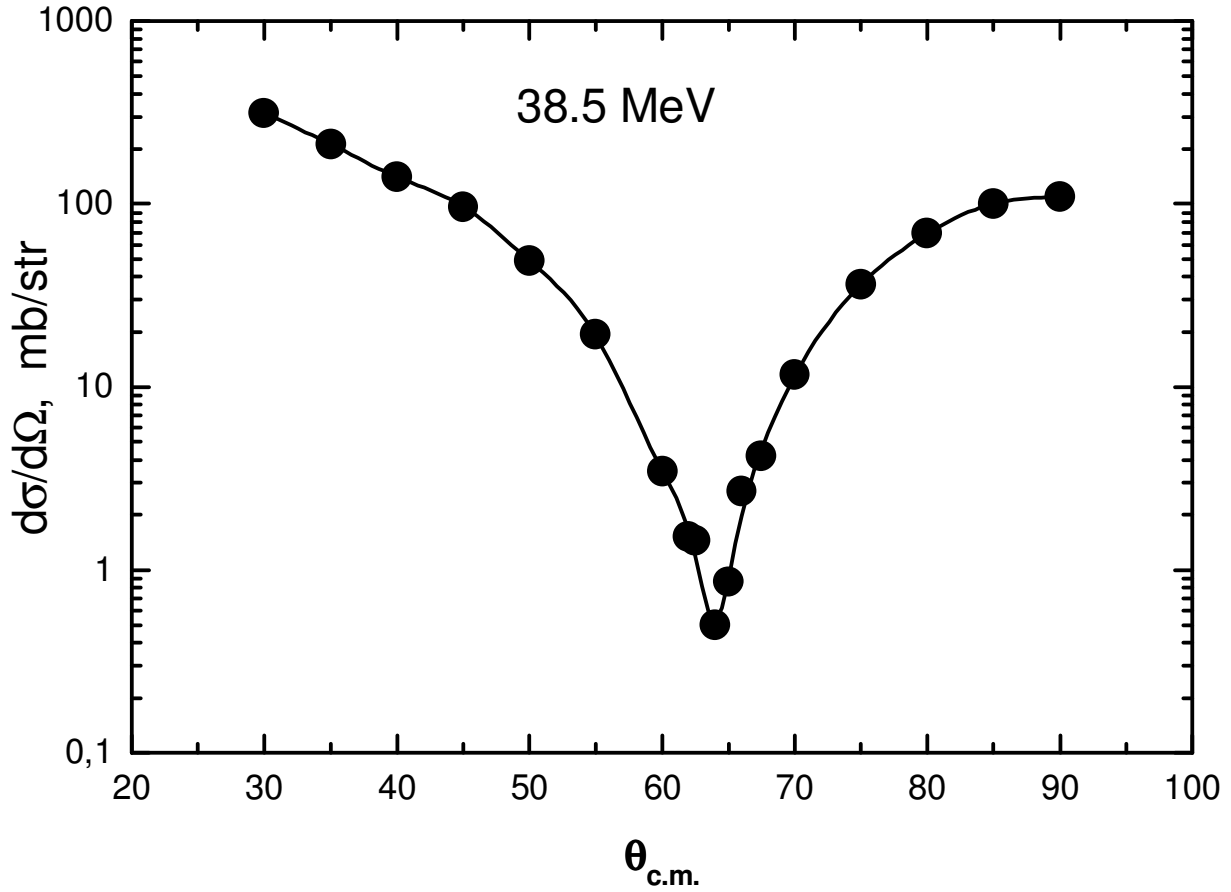

Точки – экспериментальные данные, сплошная кривая - расчет сечений с найденными фазами.

Рисунок 7.8 - Дифференциальные сечения упругого рассеяния альфа-частиц на ядрах гелия при энергии 38.5 МэВ.

В работе [223] на рисунках приведены данные для энергий 39, 40 и 41 МэВ, используем одну их них, а именно 40 МэВ, для фазового анализа. Принимая в качестве начальных фаз результаты предыдущего анализа, для L=12 получим

$\chi^2$ = 0.2234

Действительная часть фаз - 69.4967 49.5392 81.4271 1.3593 0.0000 0.9287 0.0255

Мнимая часть фаз - 0.8975 0.0000 4.5934 7.1150 1.2930 0.0000 0.1762

Качество описания дифференциальных сечений показано на рис.7.9. Дальнейшее увеличение числа парциальных волн не приводит к заметному уменьшению $\chi^2$. Отметим, что полученные в результате фазового анализа фазы рассеяния, заметно отличаются от найденных для энергий 38.4, 38.5 МэВ и далее рассмотренной энергии 40.77 МэВ.

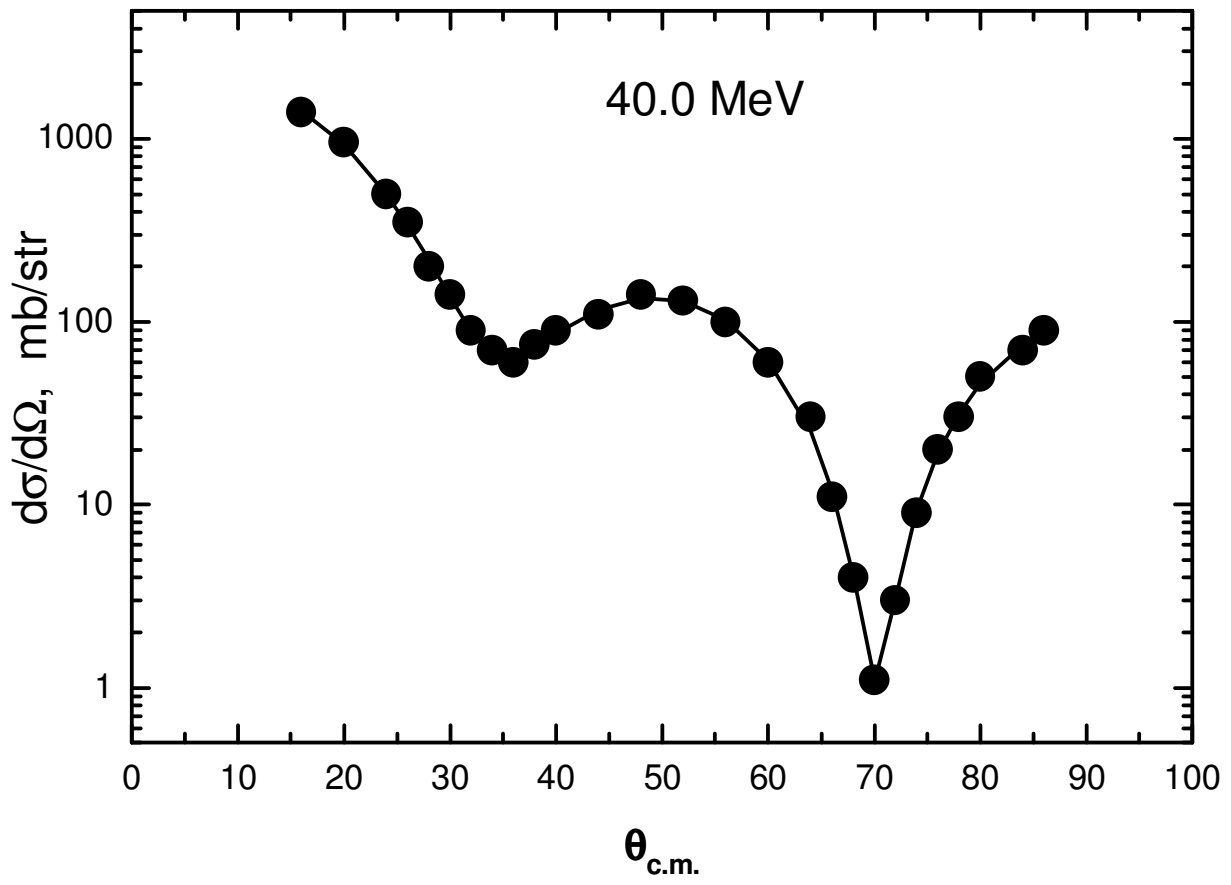

Точки – экспериментальные данные, сплошная кривая - расчет сечений с найденными фазами.

Рисунок 7.9 - Дифференциальные сечения упругого рассеяния альфа-частиц на ядрах гелия при энергии 40.0 МэВ.

Перейдем теперь к рассмотрению энергий в области 36-47 МэВ. Экспериментальные исследования дифференциальных сечений выполнены в работе [224], а фазовый анализ этих данных вообще не проводился. Эта область энергий и 23-38 МэВ рассматривалась в работе [225], где выполнена подгонка параметров оптических потенциалов, а затем, из них вычислялись фазы упругого рассеяния.

Для энергии 36.85 МэВ в [225] найдены фазы $135^0$, $78{,}7^0$, $139^0$, $2{,}0^0$, $0{,}07^0$, которые мы будем использовать в качестве начальных при варьировании фаз рассеяния по нашей программе. При L=16 можно получить $\chi^2 = 2.135$, при L=20 находим $\chi^2 = 1.40$, и только при L=30 удается уменьшить $\chi^2$

$$\chi^2 = 1.2846$$

| $\theta$ | $\sigma_e$ | $\sigma_t$ | $\chi^2$ |
|---|---|---|---|
| 1.370E+01 | 1.391E+03 | 1.389E+03 | 4.654E-02 |
| 1.570E+01 | 1.116E+03 | 1.130E+03 | 1.286E+00 |
| 1.770E+01 | 9.710E+02 | 9.588E+02 | 1.840E+00 |
| 1.970E+01 | 8.180E+02 | 8.252E+02 | 3.601E-01 |
| 2.170E+01 | 6.910E+02 | 7.101E+02 | 3.640E+00 |
| 2.370E+01 | 6.260E+02 | 6.059E+02 | 6.293E+00 |
| 2.570E+01 | 5.050E+02 | 5.104E+02 | 8.240E-01 |
| 2.770E+01 | 4.250E+02 | 4.249E+02 | 1.733E-04 |
| 2.970E+01 | 3.510E+02 | 3.515E+02 | 1.146E-02 |
| 3.170E+01 | 2.900E+02 | 2.901E+02 | 2.303E-04 |
| 3.370E+01 | 2.410E+02 | 2.381E+02 | 9.179E-01 |
| 3.570E+01 | 1.890E+02 | 1.940E+02 | 2.749E+00 |
| 3.770E+01 | 1.590E+02 | 1.577E+02 | 1.796E-01 |
| 3.970E+01 | 1.320E+02 | 1.300E+02 | 1.029E+00 |
| 4.170E+01 | 1.080E+02 | 1.103E+02 | 1.309E+00 |
| 4.370E+01 | 9.770E+01 | 9.709E+01 | 1.880E-01 |
| 4.570E+01 | 8.690E+01 | 8.785E+01 | 2.774E-01 |
| 4.770E+01 | 7.980E+01 | 7.960E+01 | 2.913E-02 |
| 4.970E+01 | 7.040E+01 | 7.040E+01 | 1.251E-06 |
| 5.170E+01 | 6.100E+01 | 6.040E+01 | 3.589E-01 |
| 5.370E+01 | 4.910E+01 | 5.065E+01 | 1.997E+00 |
| 5.570E+01 | 4.170E+01 | 4.117E+01 | 5.635E-01 |
| 5.770E+01 | 3.080E+01 | 3.111E+01 | 1.481E-01 |
| 5.970E+01 | 2.040E+01 | 2.052E+01 | 3.888E-02 |
| 6.170E+01 | 1.140E+01 | 1.110E+01 | 5.624E-01 |
| 6.370E+01 | 4.800E+00 | 5.111E+00 | 1.151E+00 |
| 6.570E+01 | 4.630E+00 | 3.981E+00 | 2.503E+00 |
| 6.770E+01 | 7.720E+00 | 8.080E+00 | 5.196E-01 |
| 6.970E+01 | 1.830E+01 | 1.758E+01 | 1.045E+00 |
| 7.170E+01 | 3.040E+01 | 3.275E+01 | 6.815E+00 |
| 7.370E+01 | 5.640E+01 | 5.301E+01 | 7.966E+00 |
| 7.570E+01 | 7.630E+01 | 7.680E+01 | 1.123E-01 |
| 7.770E+01 | 1.010E+02 | 1.027E+02 | 7.378E-01 |
| 7.970E+01 | 1.290E+02 | 1.298E+02 | 1.780E-01 |
| 8.170E+01 | 1.600E+02 | 1.564E+02 | 1.436E+00 |
| 8.370E+01 | 1.750E+02 | 1.793E+02 | 1.160E+00 |
| 8.570E+01 | 1.990E+02 | 1.959E+02 | 1.094E+00 |
| 8.770E+01 | 2.020E+02 | 2.054E+02 | 7.337E-01 |
| 8.970E+01 | 2.090E+02 | 2.089E+02 | 9.085E-04 |

Действительная часть фаз - 126.3556 62.3458 132.7907 2.5076 0.5551 7.0659 9.1342 6.3779 2.3275 1.5439 0.9287 0.0000 0.4619 0.2778 0.1248 0.0000

Мнимая часть фаз - 0.0000 0.0000 2.2830 18.8637 0.0000 0.8903 1.8405 2.9630 2.5709 2.3248 0.8069 0.0000 0.2250 0.0232 0.1390 0.0912

Качество описания экспериментальных данных показано на рис.7.10. Дальнейшее варьирование параметров или фаз рассеяния, или увеличение размерности базиса, т.е. при L>30, не приводит к существенному улучшению величины $\chi^2$. Полученные при этой энергии фазы заметно отличаются от наших результатов для 37.0 МэВ.

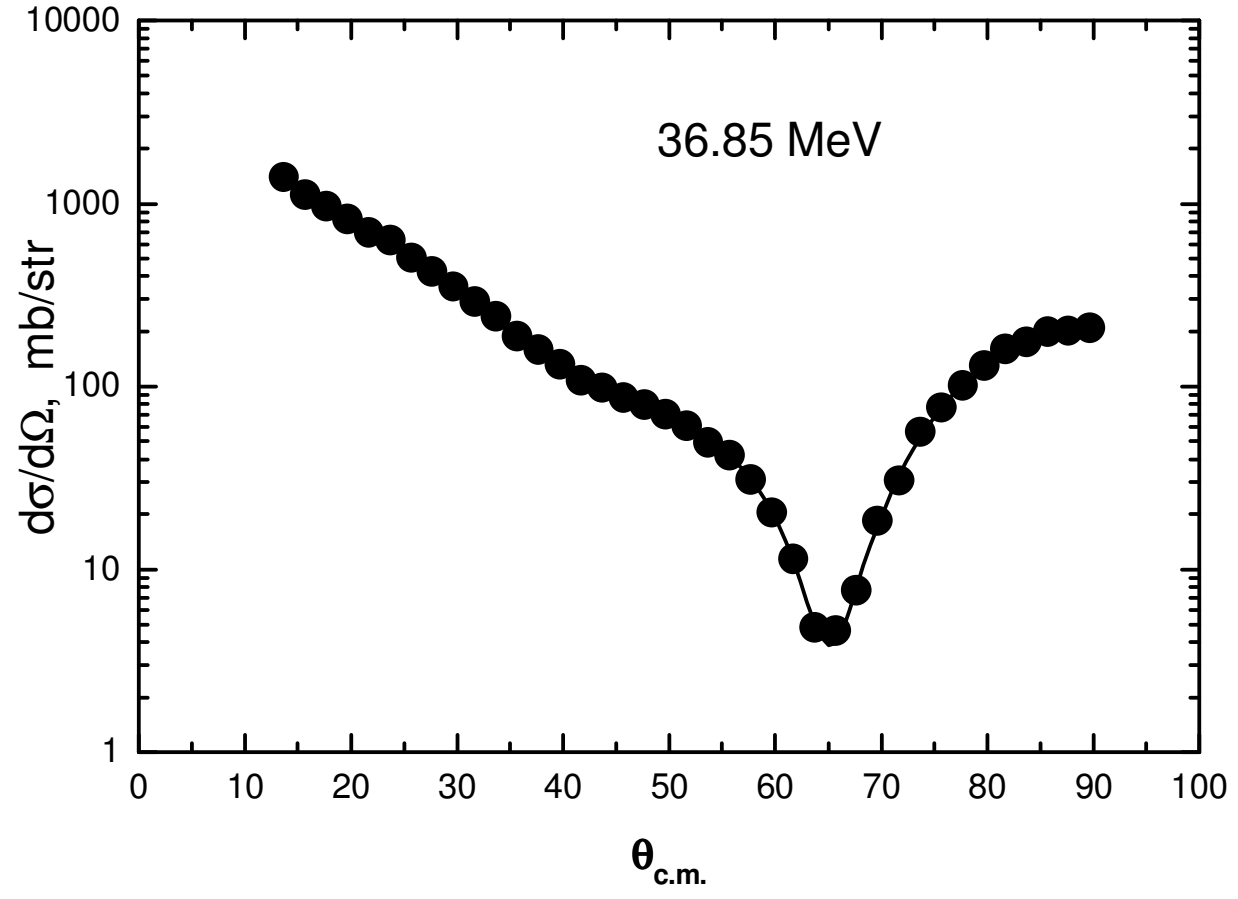

Точки – экспериментальные данные, сплошная кривая - расчет сечений с найденными фазами.

Рисунок 7.10 – Дифференциальные сечения упругого рассеяния альфа-частиц на ядрах гелия при энергии 36.85МэВ [224].

В той же работе рассмотрена энергия 38.83 МэВ, для которой в [225] получены фазы 133.4, 78.8, 154, 6.0, 0,35. Будем использовать их в качестве начальных и выполняя варьирование с L=20

$$\chi^2 = 1.900$$

| θ | $\sigma_e$ | $\sigma_t$ | $\chi^2$ |
|---|---|---|---|
| 1.170E+01 | 9.790E+02 | 1.030E+03 | 1.484E+00 |
| 1.370E+01 | 9.100E+02 | 8.563E+02 | 3.683E+00 |
| 1.770E+01 | 7.170E+02 | 7.243E+02 | 8.369E-01 |
| 2.170E+01 | 6.200E+02 | 6.107E+02 | 1.774E+00 |
| 2.570E+01 | 4.440E+02 | 4.521E+02 | 2.626E+00 |
| 2.770E+01 | 3.830E+02 | 3.795E+02 | 7.632E-01 |
| 2.970E+01 | 3.240E+02 | 3.213E+02 | 4.675E-01 |
| 3.170E+01 | 2.750E+02 | 2.772E+02 | 5.583E-01 |
| 3.370E+01 | 2.420E+02 | 2.432E+02 | 1.642E-01 |
| 3.570E+01 | 2.170E+02 | 2.142E+02 | 8.830E-01 |
| 3.770E+01 | 1.870E+02 | 1.870E+02 | 5.312E-04 |
| 3.970E+01 | 1.620E+02 | 1.615E+02 | 5.736E-02 |
| 4.170E+01 | 1.360E+02 | 1.387E+02 | 1.841E+00 |
| 4.370E+01 | 1.200E+02 | 1.188E+02 | 3.451E-01 |
| 4.570E+01 | 1.020E+02 | 1.004E+02 | 6.173E-01 |
| 4.770E+01 | 8.270E+01 | 8.204E+01 | 4.379E-01 |
| 4.970E+01 | 6.020E+01 | 6.386E+01 | 7.929E+00 |
| 5.170E+01 | 4.880E+01 | 4.752E+01 | 2.550E+00 |
| 5.370E+01 | 3.280E+01 | 3.431E+01 | 3.553E+00 |
| 5.570E+01 | 2.610E+01 | 2.417E+01 | 1.033E+01 |
| 5.770E+01 | 1.470E+01 | 1.638E+01 | 1.130E+01 |
| 5.970E+01 | 1.070E+01 | 1.051E+01 | 4.221E-01 |
| 6.170E+01 | 7.900E+00 | 6.516E+00 | 5.696E+00 |
| 6.370E+01 | 4.090E+00 | 4.380E+00 | 1.349E+00 |
| 6.570E+01 | 4.100E+00 | 3.989E+00 | 3.930E-02 |
| 6.770E+01 | 5.540E+00 | 5.355E+00 | 2.379E-01 |
| 6.970E+01 | 8.950E+00 | 8.529E+00 | 5.467E-01 |
| 7.170E+01 | 1.280E+01 | 1.327E+01 | 8.982E-01 |
| 7.370E+01 | 1.800E+01 | 1.897E+01 | 1.459E+00 |
| 7.570E+01 | 2.580E+01 | 2.477E+01 | 2.928E+00 |
| 7.770E+01 | 2.970E+01 | 2.984E+01 | 2.079E-02 |
| 7.970E+01 | 3.280E+01 | 3.354E+01 | 8.610E-01 |
| 8.370E+01 | 3.760E+01 | 3.746E+01 | 1.047E-02 |
| 8.570E+01 | 4.000E+01 | 3.919E+01 | 2.560E-01 |
| 8.770E+01 | 4.250E+01 | 4.100E+01 | 7.793E-01 |
| 8.970E+01 | 4.080E+01 | 4.197E+01 | 7.024E-01 |

Действительные фазы - 121.9545 100.4211 163.5241 3.8082
6.4402 4.7599 2.5298 1.3435 1.1103 0.7199 0.0000
Мнимые фазы - 6.7218 6.5487 0.4986 0.4164 0.0906 0.0000
2.5561 0.6655 0.4853 0.4851 0.9691

Качество описания экспериментальных данных приведено на рис.7.11. При увеличении L до 24 удается получить $\chi^2$ на уровне 1.87, при L=30 для величины $\chi^2$ найдено 1.80, а для L=34 получено $\chi^2$ = 1.73, но такое уменьшение $\chi^2$ практически не сказывается на поведении фаз рассеяния .

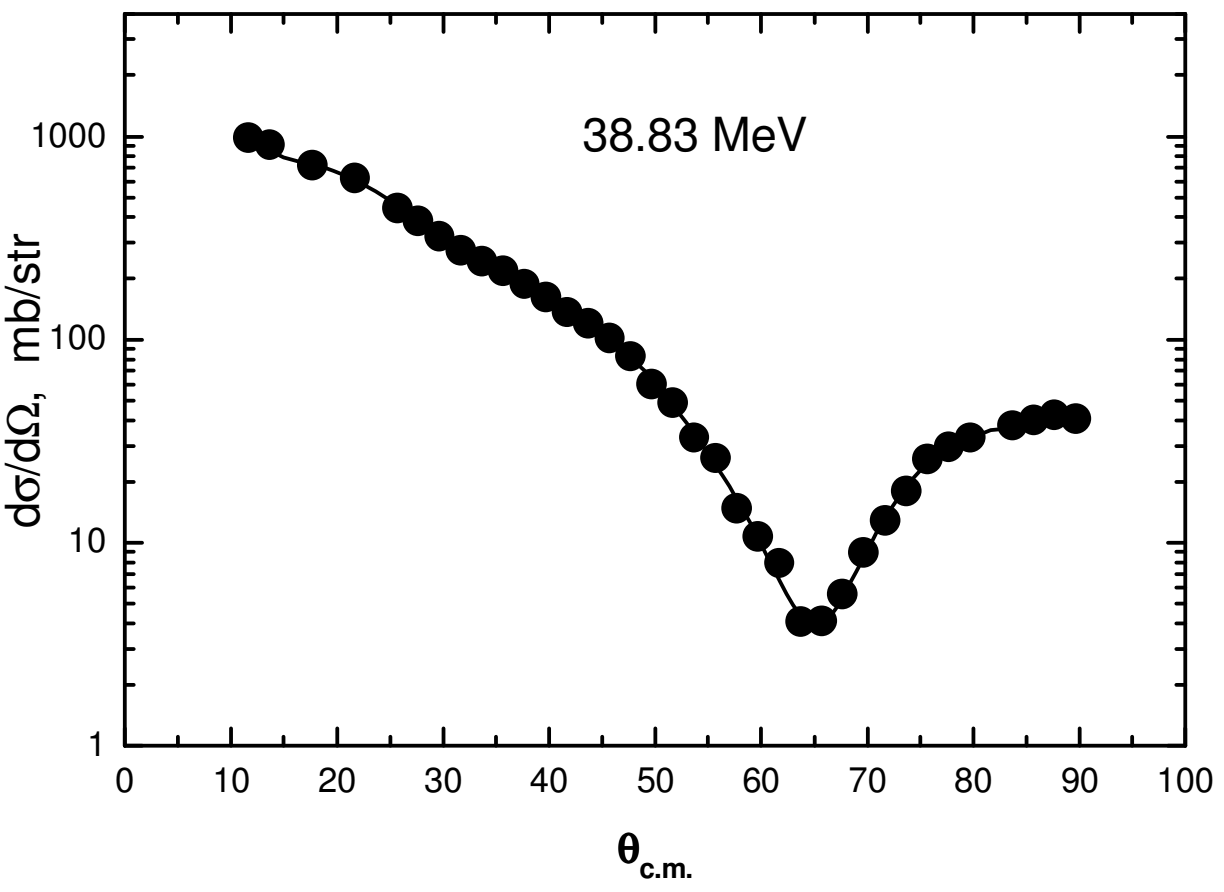

Точки – экспериментальные данные, сплошная кривая - расчет сечений с найденными фазами.

Рисунок 7.11 - Дифференциальные сечения упругого рассеяния альфа-частиц на ядрах гелия при энергии 38.83 МэВ [224].

Следующая энергия, рассмотренная в работе [224], это 40.77 МэВ, для которой в [225] получены фазы 74.4, 21.8, 86.0, 5.4, 0.38. Их мы будем использовать в качестве начальных и выполним варьирование при L=16, тогда получаем $\chi^2$ = 3.51. Изменение L до 22 практически не меняет величину $\chi^2$. Только при 24 парциальных волнах начинается уменьшение $\chi^2$ , которое оказывается равно 2.1. Приведем здесь результаты для 30 парциальных волн

$$\chi^2 = 1.5560$$

| θ | $\sigma_e$ | $\sigma_t$ | $\chi^2$ |
|---|---|---|---|
| 2.730E+01 | 1.770E+02 | 1.768E+02 | 8.316E-03 |
| 3.130E+01 | 2.950E+01 | 3.023E+01 | 4.405E-01 |
| 3.330E+01 | 1.230E+01 | 1.106E+01 | 1.888E+00 |
| 3.530E+01 | 8.820E+00 | 8.899E+00 | 1.553E-01 |
| 3.720E+01 | 3.320E+01 | 3.070E+01 | 4.353E+00 |
| 3.930E+01 | 6.660E+01 | 6.863E+01 | 2.110E+00 |
| 4.130E+01 | 1.050E+02 | 1.034E+02 | 6.080E-01 |
| 4.330E+01 | 1.370E+02 | 1.372E+02 | 3.028E-03 |
| 4.730E+01 | 1.840E+02 | 1.848E+02 | 3.848E-02 |
| 5.130E+01 | 1.820E+02 | 1.813E+02 | 2.806E-02 |
| 5.320E+01 | 1.720E+02 | 1.715E+02 | 2.442E-02 |
| 5.520E+01 | 1.530E+02 | 1.535E+02 | 1.158E-02 |
| 5.930E+01 | 8.940E+01 | 8.992E+01 | 2.248E-01 |
| 6.130E+01 | 6.030E+01 | 5.807E+01 | 1.721E+00 |
| 6.330E+01 | 3.220E+01 | 3.262E+01 | 7.671E-02 |
| 6.530E+01 | 1.400E+01 | 1.445E+01 | 4.116E-01 |
| 6.730E+01 | 5.310E+00 | 4.742E+00 | 9.600E-01 |
| 6.930E+01 | 4.390E+00 | 5.036E+00 | 1.380E+00 |
| 7.130E+01 | 1.420E+01 | 1.409E+01 | 2.869E-01 |
| 7.330E+01 | 2.690E+01 | 2.961E+01 | 6.083E+00 |
| 7.530E+01 | 5.760E+01 | 5.340E+01 | 3.338E+00 |
| 7.730E+01 | 9.120E+01 | 8.812E+01 | 1.304E+00 |
| 7.930E+01 | 1.210E+02 | 1.285E+02 | 6.218E+00 |
| 8.130E+01 | 1.760E+02 | 1.638E+02 | 9.266E+00 |
| 8.330E+01 | 1.840E+02 | 1.903E+02 | 2.507E+00 |
| 8.530E+01 | 2.130E+02 | 2.129E+02 | 1.302E-03 |
| 8.730E+01 | 2.350E+02 | 2.337E+02 | 1.119E-01 |
| 8.930E+01 | 2.460E+02 | 2.464E+02 | 1.169E-02 |

Действительные фазы - 127.3870 40.8110 87.6083 2.2358
0.0173 1.3451 1.4785 0.0000 4.2453 2.6973 4.3213 0.2210
0.0000 0.1755 0.5197 0.4039
Мнимые фазы - 18.8252 0.9065 0.3905 1.9687 0.0345 1.4207
2.1880 0.4515 0.0000 0.3531 0.3290 2.1834 1.1953 0.3142
0.0000 0.0000

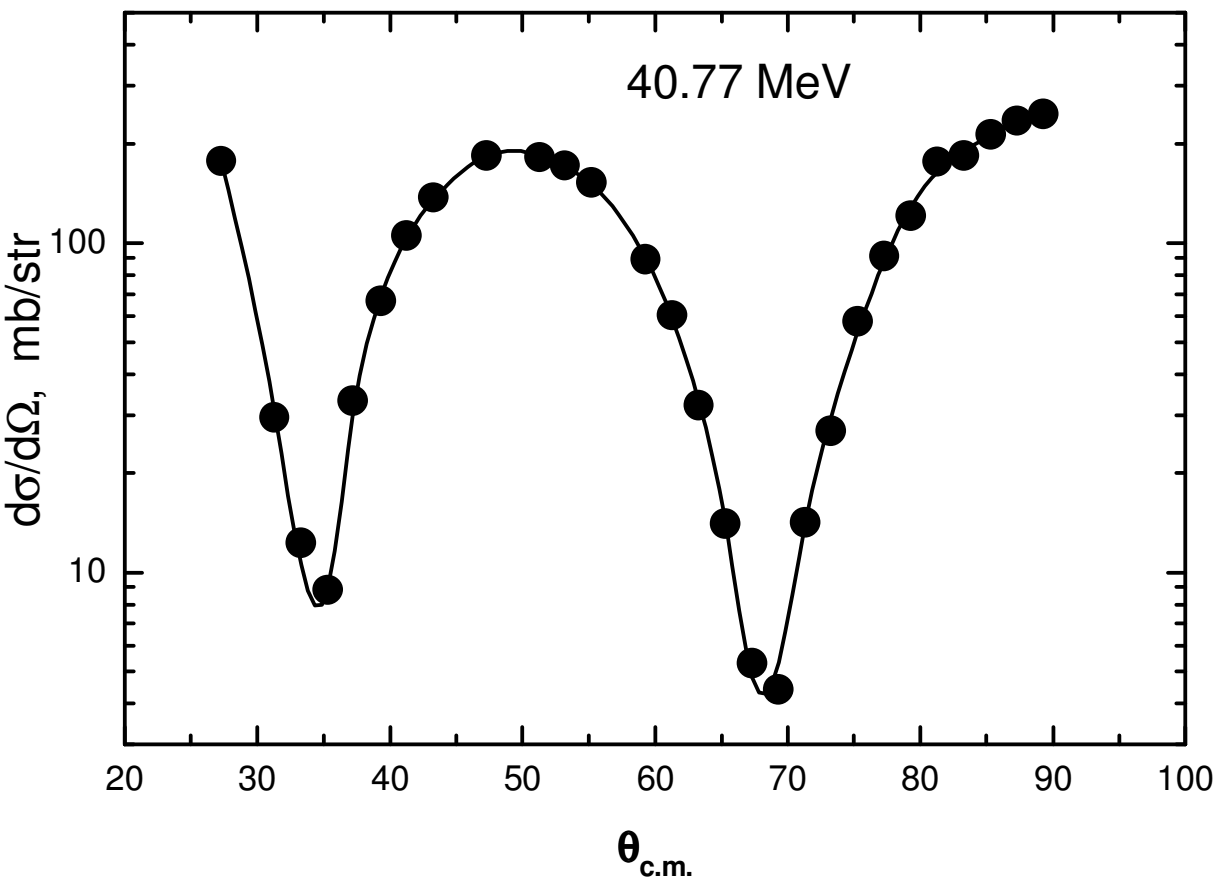

Точки – экспериментальные данные, сплошная кривая - расчет сечений с найденными фазами.

Рисунок 7.12 - Дифференциальные сечения упругого рассеяния альфа-частиц на ядрах гелия при энергии 40.77 МэВ [224].

Качество описания экспериментальных данных показано на рис.7.12, а найденные фазы заметно отличаются от результатов для энергии 40.0 МэВ. Увеличение L до 34 приводит нас к $\chi^2$ = 1.34, а при L= 40 к $\chi^2$ = 1.21085 мало изменяя фазы рассеяния, которые приведены ниже

Действительные фазы - 124.9455 37.3915 86.8874 1.8977
0.0000 1.1045 1.9122 0.0897 4.1335 2.7145 4.0382 0.3077
0.0000 0.1285 0.7936 1.1646 0.4322 0.0000 0.1584 0.0000
0.1347
Мнимые фазы - 18.7685 4.3167 0.0011 1.5199 0.0001 1.4317
1.9380 0.9304 0.0000 0.3225 0.4839 2.2038 1.0627 0.6151
0.0822 0.0000 0.5439 0.4801 0.1615 0.1930 0.0000

Далее, в работе [224] была рассмотрена энергия 41.9 МэВ. В качестве начальных принимаем фазы для предыдущей энергии и по нашей программе с 10 итерациями для L=8-10 получаем $\chi^2$ порядка 10. Увеличение L до 14 приводит к несколько лучшему результату $\chi^2$ = 5.66 с фазами

Действительные фазы - 105.8413 53.7189 107.0094 13.1469
2.1598 0.1235 1.1350 1.3931
Мнимые фазы - 22.0202 0.0000 0.0000 0.0000 0.0000 0.0000
0.0821 0.3104

Выясним, насколько надо увеличить размерность L, чтобы получить величину $\chi^2$ порядка единицы. Как показывают расчеты, нужно принять L=26, тогда мы можем получить $\chi^2$ = 4.91 с фазами

Действительные фазы - 107.8003 53.5006 105.7207 13.9290
1.6358 0.0000 1.8523 1.6216 0.0000 0.2889 0.2598 0.2007
0.0224 0.0000
Мнимые фазы - 24.6993 0.0000 0.0000 0.0000 0.0000
0.0000 0.2984 0.3034 0.0000 0.0123 0.0000 0.0114 0.1151
0.1607

При L=30 находим

$\chi^2$ = 4.0970

| $\theta$ | $\sigma_e$ | $\sigma_t$ | $\chi^2$ |
|---|---|---|---|
| 1.220E+01 | 2.197E+03 | 1.930E+03 | 3.842E+01 |
| 1.400E+01 | 1.824E+03 | 1.712E+03 | 1.028E+01 |
| 1.600E+01 | 1.466E+03 | 1.451E+03 | 4.213E-01 |
| 1.780E+01 | 1.132E+03 | 1.208E+03 | 2.000E+01 |
| 1.980E+01 | 9.540E+02 | 9.469E+02 | 1.569E-01 |
| 2.160E+01 | 7.861E+02 | 7.349E+02 | 1.952E+01 |
| 2.340E+01 | 5.694E+02 | 5.520E+02 | 9.636E+00 |
| 2.520E+01 | 3.923E+02 | 3.988E+02 | 3.488E+00 |
| 2.700E+01 | 2.683E+02 | 2.728E+02 | 2.426E+00 |
| 2.880E+01 | 1.738E+02 | 1.718E+02 | 1.391E+00 |
| 3.060E+01 | 9.440E+01 | 9.527E+01 | 6.320E-01 |
| 3.240E+01 | 4.500E+01 | 4.369E+01 | 4.802E+00 |
| 3.420E+01 | 1.620E+01 | 1.671E+01 | 2.695E+00 |
| 3.620E+01 | 1.289E+01 | 1.256E+01 | 1.198E+00 |
| 3.800E+01 | 2.679E+01 | 2.667E+01 | 1.346E-01 |
| 3.980E+01 | 5.090E+01 | 5.116E+01 | 1.375E-01 |
| 4.180E+01 | 8.300E+01 | 8.343E+01 | 2.846E-01 |
| 4.360E+01 | 1.147E+02 | 1.125E+02 | 3.330E+00 |
| 4.580E+01 | 1.430E+02 | 1.445E+02 | 1.004E+00 |
| 4.780E+01 | 1.676E+02 | 1.675E+02 | 1.216E-03 |
| 5.000E+01 | 1.844E+02 | 1.824E+02 | 1.284E+00 |
| 5.200E+01 | 1.847E+02 | 1.831E+02 | 7.668E-01 |
| 5.420E+01 | 1.712E+02 | 1.697E+02 | 8.656E-01 |
| 5.620E+01 | 1.450E+02 | 1.479E+02 | 3.346E+00 |
| 5.840E+01 | 1.203E+02 | 1.181E+02 | 2.748E+00 |
| 6.060E+01 | 8.620E+01 | 8.544E+01 | 7.212E-01 |
| 6.280E+01 | 5.150E+01 | 5.281E+01 | 4.733E+00 |
| 6.500E+01 | 2.566E+01 | 2.493E+01 | 4.613E+00 |
| 6.700E+01 | 7.770E+00 | 8.024E+00 | 1.997E+00 |
| 6.920E+01 | 2.220E+00 | 2.032E+00 | 1.563E+00 |
| 7.140E+01 | 9.830E+00 | 1.017E+01 | 1.902E+00 |
| 7.360E+01 | 3.120E+01 | 3.068E+01 | 1.099E+00 |
| 7.580E+01 | 6.180E+01 | 6.076E+01 | 1.699E+00 |
| 7.800E+01 | 9.510E+01 | 9.869E+01 | 7.614E+00 |
| 8.020E+01 | 1.448E+02 | 1.426E+02 | 1.705E+00 |
| 8.220E+01 | 1.875E+02 | 1.833E+02 | 3.025E+00 |
| 8.440E+01 | 2.234E+02 | 2.216E+02 | 4.999E-01 |
| 8.640E+01 | 2.439E+02 | 2.453E+02 | 1.681E-01 |
| 8.847E+01 | 2.567E+02 | 2.579E+02 | 1.105E-01 |
| 9.040E+01 | 2.681E+02 | 2.603E+02 | 3.459E+00 |

Действительные фазы - 112.6611 53.5342 104.2262 15.7344
0.0000 0.0000 3.9450 1.8310 0.0000 1.0262 0.4863 0.0000
0.3765 0.0000 0.0000 0.0000
Мнимые фазы - 28.4867 0.0000 0.0000 0.5922 0.0000 0.0000
0.3468 0.0000 0.0109 0.2538 0.0000 0.0000 0.2712 0.1621

0.0000 0.0978

И только увеличение L до 40 позволяет получить заметно меньшее $\chi^2$ = 0.8573, со следующими фазами рассеяния

Действительные фазы - 105.8962 53.0717 103.3716 16.0950
0.0000 0.0710 3.9094 1.5391 0.0000 0.5518 1.4615 0.0091
0.4501 0.0000 1.1544 0.0631 0.1885 0.8750 0.0000 0.4869
0.0000
19.0395 0.0005 0.0000 0.2844 0.0000 0.1207 0.4153
0.0000 0.2214 0.1117 0.0968 0.0069 0.5820 0.2484 0.2132
0.3826 0.3416 0.3118 0.0642 0.1986 0.1950

Которые несколько отличаются от, полученных при L=30, только в S – волне. Качество писания экспериментальных сечений показано на рисунке 7.13.

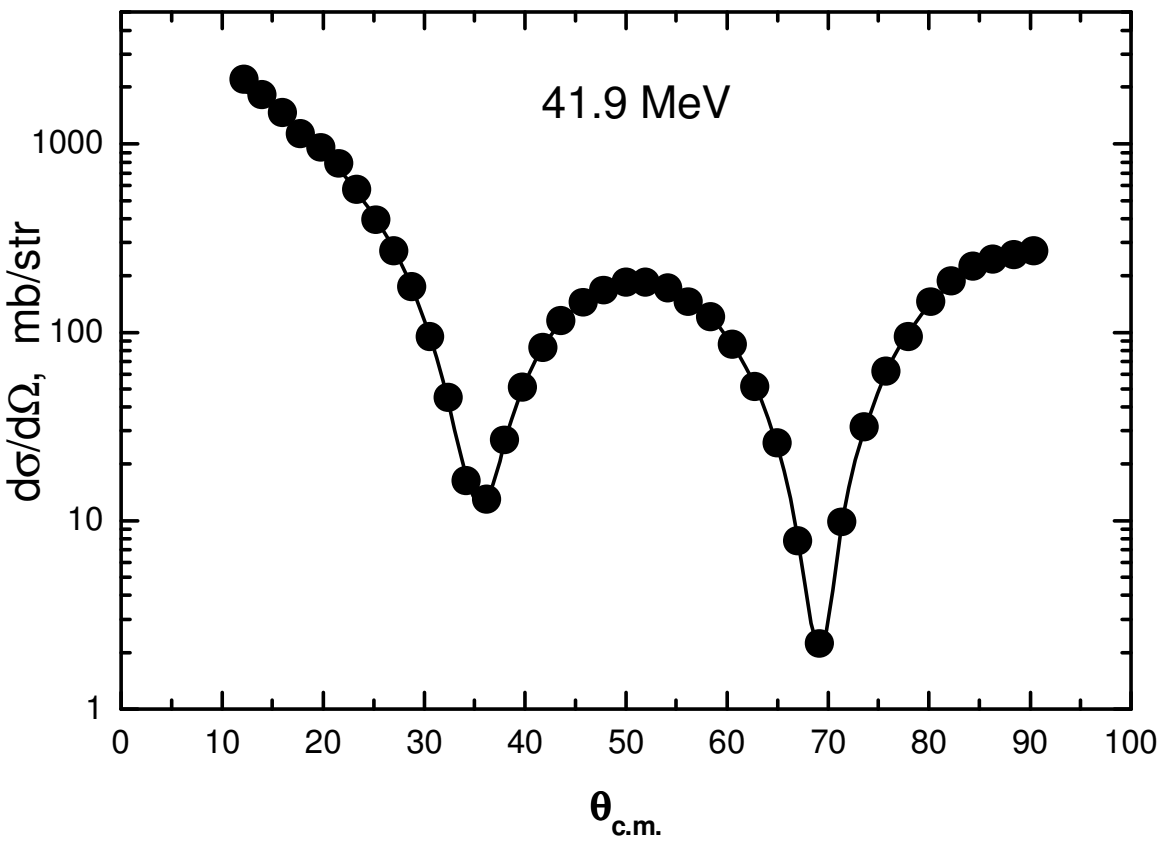

Точки – экспериментальные данные, сплошная кривая - расчет сечений с найденными фазами.

Рисунок 7.13 - Дифференциальные сечения упругого рассеяния альфа-частиц на ядрах гелия при энергии 41.9 МэВ [224].

Следующая энергия, рассмотренная в работе [224], это 44.41 МэВ. Используем фазы, полученные в предыдущем случае и выполним варьирование с L=20 и учете мнимой части фаз. Один из лучших результатов, который можно получить дает $\chi^2$ = 4.97, с фазами

Действительные фазы - 99.4301 74.5262 126.0065 18.5705
4.4035 0.7213 0.2493 0.0000 0.1380 0.5229 0.3010

Мнимые фазы - 23.6077 4.0092 4.9234 0.0626 0.8233 1.4425
0.5864 0.1271 0.0000 0.0000 0.1546

Отметим, что при меньших L, например, 12-16 величина $\chi^2$ находится на уровне 6.0-6.2. Увеличим L до 30, тогда получим $\chi^2$ = 0.97 и фазы

Действительные фазы - 119.3874 72.3364 116.1525 16.8884
3.6282 0.3916 0.0005 0.0670 0.0000 2.5644 2.2477 0.0013
0.1206 0.0000 0.0000 0.1876
Мнимые фазы - 29.7369 9.3575 3.6540 0.0717 1.4200
3.1324 1.7457 1.0906 0.0000 0.1427 0.4842 0.0000 0.2488
0.0862 0.0244 0.0019

Видно, что только учет комбинации малых вкладов высших парциальных волн позволяет, как и в предыдущем случае, получить приемлемую величину $\chi^2$ порядка единицы.

Проверим, можно ли получить лучшие результаты, если еще увеличить число парциальных волн. В частности, для L=34 можно получить некоторое улучшение качества описания экспериментальных сечений при $\chi^2$ = 0.68 и только увеличение L до 40 приводит к заметному улучшению описания дифференциальных сечений, которое приведено ниже

$\chi^2$ = 0.47672

| $\theta$ | $\sigma_e$ | $\sigma_t$ | $\chi^2$ |
|---|---|---|---|
| 1.220E+01 | 1.820E+03 | 1.869E+03 | 1.748E+00 |
| 1.400E+01 | 1.602E+03 | 1.550E+03 | 2.521E+00 |
| 1.600E+01 | 1.263E+03 | 1.272E+03 | 1.248E-01 |
| 1.780E+01 | 1.058E+03 | 1.082E+03 | 1.323E+00 |
| 1.980E+01 | 9.050E+02 | 8.935E+02 | 5.874E-01 |
| 2.160E+01 | 7.114E+02 | 7.085E+02 | 3.722E-02 |
| 2.480E+01 | 3.820E+02 | 3.858E+02 | 5.897E-01 |
| 2.680E+01 | 2.580E+02 | 2.554E+02 | 7.759E-01 |
| 2.880E+01 | 1.760E+02 | 1.779E+02 | 9.484E-01 |
| 3.080E+01 | 1.210E+02 | 1.182E+02 | 1.903E+00 |
| 3.280E+01 | 6.440E+01 | 6.476E+01 | 1.987E-01 |
| 3.480E+01 | 2.840E+01 | 2.840E+01 | 1.354E-06 |
| 3.680E+01 | 1.360E+01 | 1.356E+01 | 4.869E-02 |
| 3.880E+01 | 1.420E+01 | 1.426E+01 | 8.793E-02 |
| 4.080E+01 | 2.600E+01 | 2.592E+01 | 7.489E-02 |
| 4.480E+01 | 7.290E+01 | 7.319E+01 | 1.753E-01 |
| 4.580E+01 | 8.760E+01 | 8.749E+01 | 1.390E-02 |
| 4.760E+01 | 1.126E+02 | 1.120E+02 | 2.646E-01 |
| 5.000E+01 | 1.344E+02 | 1.353E+02 | 4.518E-01 |
| 5.200E+01 | 1.442E+02 | 1.431E+02 | 5.952E-01 |
| 5.420E+01 | 1.435E+02 | 1.442E+02 | 2.838E-01 |
| 5.620E+01 | 1.399E+02 | 1.402E+02 | 3.298E-02 |
| 5.840E+01 | 1.249E+02 | 1.240E+02 | 6.015E-01 |
| 6.080E+01 | 9.160E+01 | 9.210E+01 | 3.076E-01 |
| 6.480E+01 | 3.880E+01 | 3.868E+01 | 8.438E-02 |
| 6.680E+01 | 1.920E+01 | 1.924E+01 | 2.047E-02 |
| 7.080E+01 | 1.790E+00 | 1.791E+00 | 5.952E-04 |
| 7.480E+01 | 2.050E+01 | 2.057E+01 | 2.996E-02 |
| 7.880E+01 | 6.230E+01 | 6.205E+01 | 1.268E-01 |
| 8.220E+01 | 1.038E+02 | 1.047E+02 | 3.731E-01 |
| 8.640E+01 | 1.512E+02 | 1.495E+02 | 8.028E-01 |
| 8.840E+01 | 1.600E+02 | 1.618E+02 | 5.796E-01 |
| 9.040E+01 | 1.651E+02 | 1.648E+02 | 1.804E-02 |

Действительные фазы - 117.3916 72.1036 115.9565 16.7765
3.3375 0.3926 0.0756 0.0270 0.0000 2.6021 2.2519 0.2051
0.1094 0.0091 0.0000 0.2321 0.3424 0.0896 0.0557 0.0609
0.0279

Мнимые фазы - 30.1833 9.4258 3.4545 0.1919 1.4489
2.9905 1.8006 1.2168 0.0054 0.1728 0.3725 0.0587 0.3169
0.1322 0.0572 0.0000 0.0583 0.0817 0.0583 0.0316 0.0383

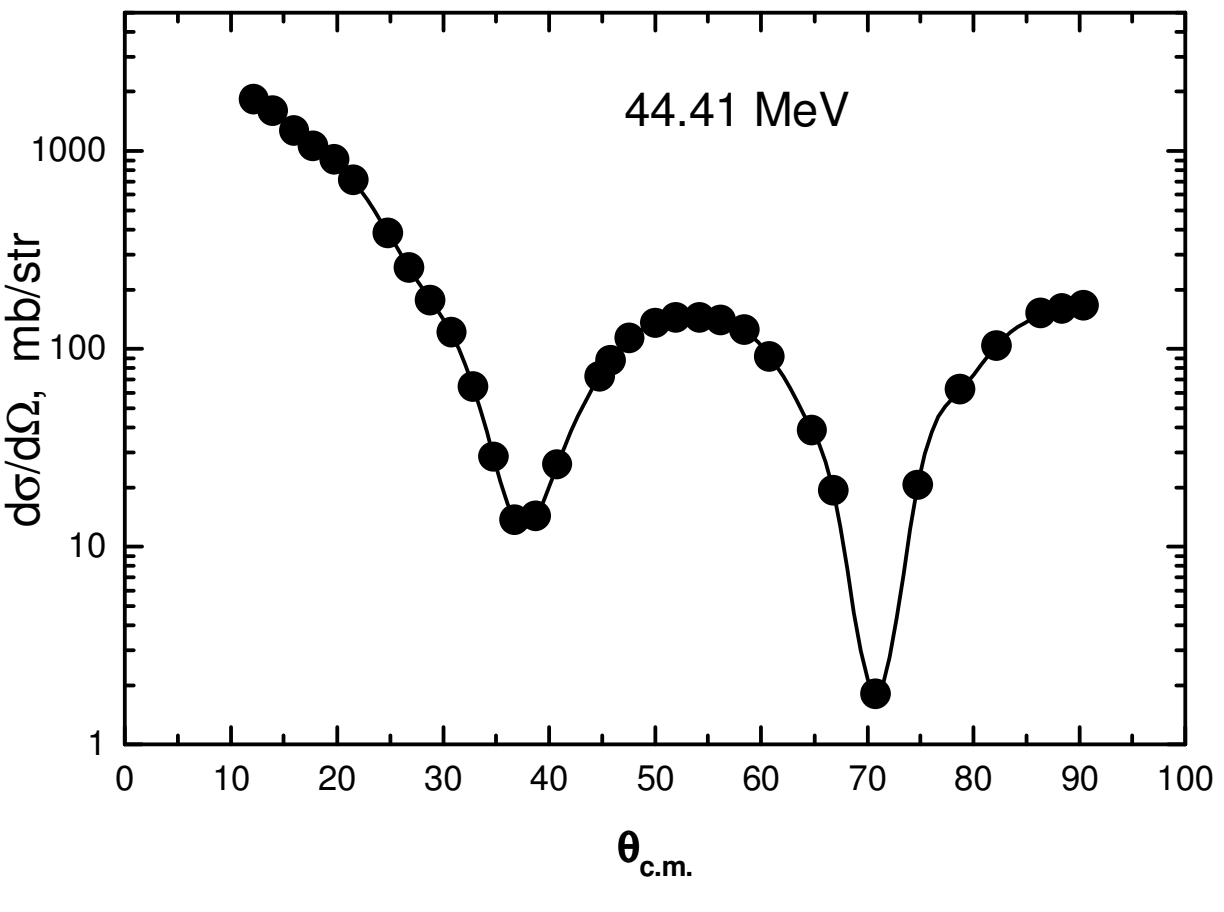

Точки – экспериментальные данные, сплошная кривая - расчет сечений с найденными фазами.

Рисунок 7.14 - Дифференциальные сечения упругого рассеяния альфа-частиц на ядрах гелия при энергии 44.41 МэВ [224].

Результаты описания дифференциальных сечений показаны на рис.7.14.

Следующая энергия, рассмотренная в работе [224], это 46.12 МэВ. Принимая в качестве начальных, фазы полученные в предыдущем случае получим для L=20 величину $\chi^2$ равную 5.08. При увеличении числа парциальных волн до 34 находим $\chi^2$ = 3.97 и только при L=40 получаем

$\chi^2$ = 2.2050

| $\theta$ | $\sigma_e$ | $\sigma_t$ | $\chi^2$ |
|---|---|---|---|
| 1.220E+01 | 1.319E+03 | 1.304E+03 | 1.798E-01 |
| 1.400E+01 | 1.012E+03 | 1.044E+03 | 3.126E+00 |
| 1.600E+01 | 8.570E+02 | 8.179E+02 | 6.802E+00 |
| 1.780E+01 | 6.660E+02 | 6.866E+02 | 2.936E+00 |
| 1.980E+01 | 5.800E+02 | 5.729E+02 | 5.025E-01 |
| 2.160E+01 | 4.561E+02 | 4.597E+02 | 2.692E-01 |
| 2.580E+01 | 2.240E+02 | 2.214E+02 | 7.735E-01 |
| 2.780E+01 | 1.470E+02 | 1.503E+02 | 2.800E+00 |
| 2.980E+01 | 9.510E+01 | 9.328E+01 | 2.307E+00 |
| 3.180E+01 | 4.880E+01 | 4.899E+01 | 9.975E-02 |
| 3.380E+01 | 2.370E+01 | 2.390E+01 | 4.445E-01 |
| 3.580E+01 | 1.530E+01 | 1.508E+01 | 1.231E+00 |
| 3.780E+01 | 1.740E+01 | 1.766E+01 | 1.753E+00 |
| 3.980E+01 | 3.170E+01 | 3.081E+01 | 4.991E+00 |
| 4.180E+01 | 5.080E+01 | 5.207E+01 | 4.478E+00 |
| 4.380E+01 | 7.520E+01 | 7.398E+01 | 2.338E+00 |
| 4.580E+01 | 9.340E+01 | 9.435E+01 | 9.021E-01 |
| 4.780E+01 | 1.155E+02 | 1.146E+02 | 4.679E-01 |
| 5.000E+01 | 1.313E+02 | 1.318E+02 | 1.365E-01 |
| 5.200E+01 | 1.379E+02 | 1.378E+02 | 3.423E-03 |
| 5.420E+01 | 1.375E+02 | 1.360E+02 | 1.049E+00 |
| 5.620E+01 | 1.254E+02 | 1.287E+02 | 5.504E+00 |
| 5.840E+01 | 1.101E+02 | 1.070E+02 | 6.479E+00 |
| 5.980E+01 | 8.560E+01 | 8.573E+01 | 1.793E-02 |
| 6.180E+01 | 5.370E+01 | 5.536E+01 | 5.601E+00 |
| 6.380E+01 | 3.440E+01 | 3.284E+01 | 9.715E+00 |
| 6.580E+01 | 1.550E+01 | 1.581E+01 | 2.472E+00 |
| 6.780E+01 | 4.130E+00 | 4.077E+00 | 4.421E-01 |
| 6.980E+01 | 2.070E+00 | 2.101E+00 | 2.619E-01 |
| 7.180E+01 | 9.700E+00 | 9.462E+00 | 1.413E+00 |
| 7.380E+01 | 2.570E+01 | 2.610E+01 | 9.809E-01 |
| 7.580E+01 | 5.290E+01 | 5.299E+01 | 2.041E-02 |
| 7.780E+01 | 8.470E+01 | 8.435E+01 | 1.207E-01 |
| 7.980E+01 | 1.180E+02 | 1.141E+02 | 3.874E+00 |
| 8.180E+01 | 1.360E+02 | 1.421E+02 | 9.351E+00 |
| 8.440E+01 | 1.817E+02 | 1.790E+02 | 1.682E+00 |
| 8.640E+01 | 2.039E+02 | 2.028E+02 | 2.506E-01 |
| 8.840E+01 | 2.162E+02 | 2.169E+02 | 1.034E-01 |
| 9.040E+01 | 2.194E+02 | 2.202E+02 | 1.157E-01 |

Действительные фазы - 134.8902 56.3774 126.0122 12.2513
1.8657 0.0000 0.7520 2.6370 0.0000 0.4870 1.6064 0.0000
0.6390 0.9047 0.0000 0.5808 0.4622 0.0000 0.4953 0.3246
0.2539

Мнимые фазы - 34.2819 5.3999 1.4059 1.8203 0.9929
2.1676 0.8192 1.3869 0.4775 0.2946 0.7548 0.2448 0.0286
0.0773 0.1094 0.1046 0.3101 0.0000 0.0000 0.0359 0.0545

Качество описания экспериментальных сечений показано на рис.7.15. Как видно, при данной энергии даже при 40 парциальных волнах не удается получить приемлемое значение $\chi^2$.

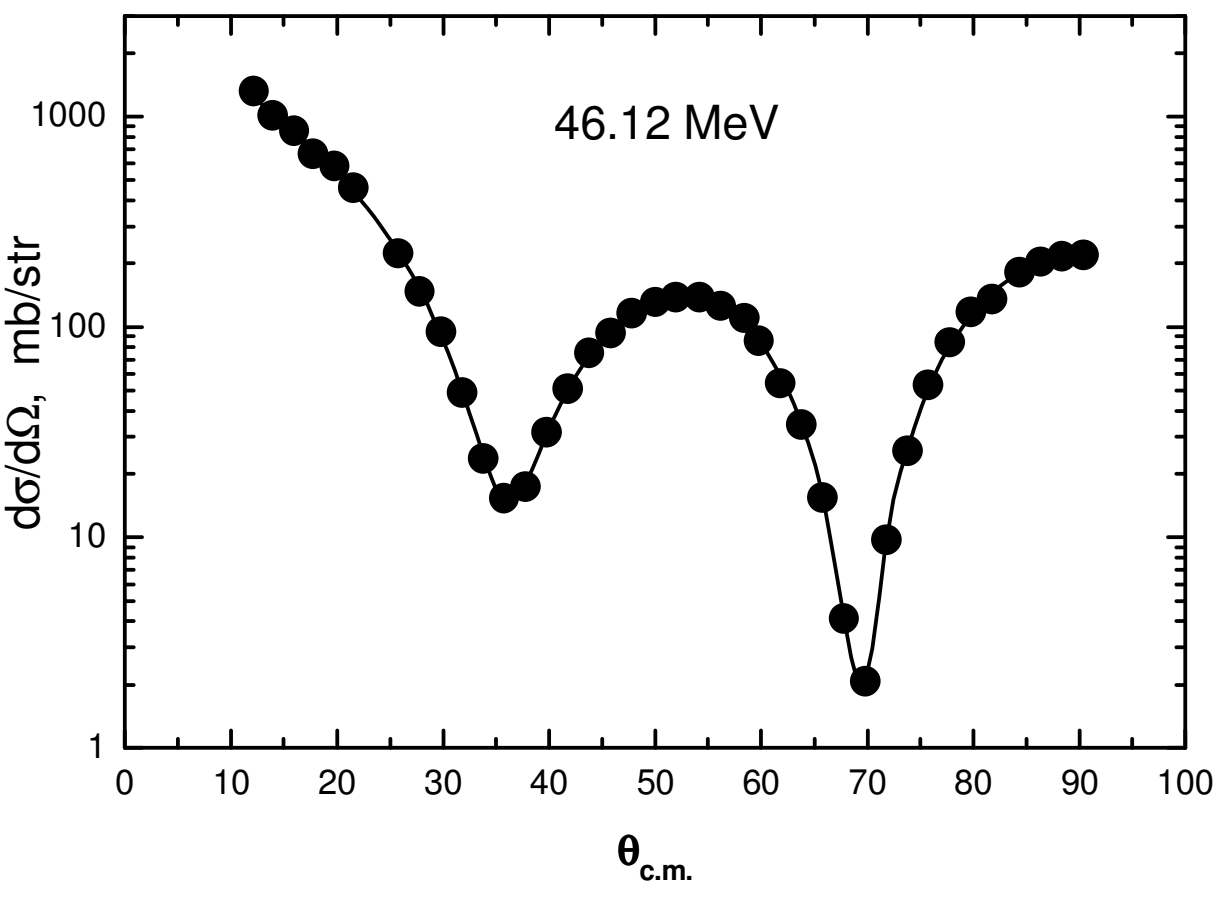

Точки – экспериментальные данные, сплошная кривая - расчет сечений с найденными фазами.

Рисунок 7.15 - Дифференциальные сечения упругого рассеяния альфа-частиц на ядрах гелия при энергии 46.12 МэВ [224].

В работе [224] приведены и данные по дифференциальным сечениям и для энергии 47.1 МэВ. Измеренные сечения также даны в таблицах, а анализ [225] приводит к следующим действительным фазам рассеяния $99^0$, $51.8^0$, $145.5^0$, $18.7^0$, $2.8^0$. Наши вычисления с такими фазами дают следующий результат для $\chi^2$ = 156.

Видно, что согласие с экспериментом сравнительно плохое, потому что при такой энергии, как и в предыдущем случае, уже возможны неупругие каналы, а мнимая часть фаз в работе [225] не приводится. Используем варьирование фаз по нашей программе с включением их дополнительной мнимой части и 10 итерациями. В результате находим для L=8

$\chi^2$ = 2.63

Действительная часть - 105.3432 55.0540 140.6032 18.8508 2.8351

Мнимая часть - 5.4793 0.5167 0.7782 0.8478 0.0251

Увеличим теперь L и посмотрим, сколько парциальных волн нужно учитывать для получения $\chi^2$ порядка единицы. При четырнадцати парциальных волнах получаем $\chi^2$ = 1.33, и только для L=20 находим

$\chi^2$ = 1.0181

| $\theta$ | $\sigma_e$ | $\sigma_t$ | $\chi^2$ |
|---|---|---|---|
| 2.150E+01 | 3.680E+02 | 3.693E+02 | 1.949E-01 |
| 2.350E+01 | 2.750E+02 | 2.755E+02 | 2.807E-02 |
| 2.550E+01 | 1.966E+02 | 1.941E+02 | 1.953E+00 |
| 2.750E+01 | 1.275E+02 | 1.280E+02 | 1.252E-01 |
| 2.950E+01 | 7.740E+01 | 7.844E+01 | 2.199E+00 |
| 3.150E+01 | 4.600E+01 | 4.480E+01 | 2.933E+00 |
| 3.350E+01 | 2.550E+01 | 2.548E+01 | 2.196E-03 |
| 3.550E+01 | 1.860E+01 | 1.865E+01 | 2.393E-02 |
| 3.750E+01 | 2.220E+01 | 2.249E+01 | 5.202E-01 |
| 3.950E+01 | 3.550E+01 | 3.506E+01 | 7.805E-01 |
| 4.150E+01 | 5.420E+01 | 5.397E+01 | 6.503E-02 |
| 4.350E+01 | 7.510E+01 | 7.627E+01 | 8.126E-01 |
| 4.550E+01 | 9.890E+01 | 9.868E+01 | 1.898E-02 |
| 4.750E+01 | 1.177E+02 | 1.181E+02 | 5.638E-02 |
| 4.950E+01 | 1.316E+02 | 1.323E+02 | 1.080E-01 |
| 5.150E+01 | 1.425E+02 | 1.395E+02 | 1.986E+00 |
| 5.350E+01 | 1.383E+02 | 1.391E+02 | 1.526E-01 |
| 5.550E+01 | 1.303E+02 | 1.307E+02 | 4.182E-02 |
| 5.750E+01 | 1.129E+02 | 1.149E+02 | 1.076E+00 |
| 5.950E+01 | 9.320E+01 | 9.307E+01 | 6.900E-03 |
| 6.150E+01 | 6.940E+01 | 6.790E+01 | 1.571E+00 |
| 6.350E+01 | 4.240E+01 | 4.272E+01 | 2.052E-01 |
| 6.550E+01 | 2.140E+01 | 2.118E+01 | 3.026E-01 |
| 6.750E+01 | 6.530E+00 | 6.643E+00 | 4.988E-01 |
| 6.950E+01 | 1.650E+00 | 1.639E+00 | 2.656E-02 |
| 7.150E+01 | 7.630E+00 | 7.512E+00 | 3.496E-01 |
| 7.350E+01 | 2.360E+01 | 2.427E+01 | 2.835E+00 |
| 7.550E+01 | 5.120E+01 | 5.059E+01 | 4.531E-01 |
| 7.750E+01 | 8.350E+01 | 8.390E+01 | 7.011E-02 |
| 7.950E+01 | 1.229E+02 | 1.207E+02 | 2.976E+00 |
| 8.150E+01 | 1.567E+02 | 1.570E+02 | 1.061E-02 |
| 8.350E+01 | 1.783E+02 | 1.893E+02 | 1.192E+01 |
| 8.550E+01 | 2.167E+02 | 2.150E+02 | 2.345E-01 |
| 8.750E+01 | 2.342E+02 | 2.321E+02 | 9.678E-01 |
| 8.950E+01 | 2.389E+02 | 2.397E+02 | 1.310E-01 |

Действительная часть - 105.3332 52.3046 135.3000 17.4462
2.5823 0.6332 0.7840 0.0600 0.2724 0.1426 0.2977
Мнимая часть - 11.8524 2.4806 1.1357 0.1707 0.0803 0.8866
0.7752 0.4579 0.3014 0.1366 0.2743

Поскольку фазы при такой энергии становятся комплексными, то появляется сечение неупругих процессов или реакций $\sigma_r$ и учет мнимой части фаз позволяет улучшить согласие расчета с экспериментом, как показывает рисунок 7.16.

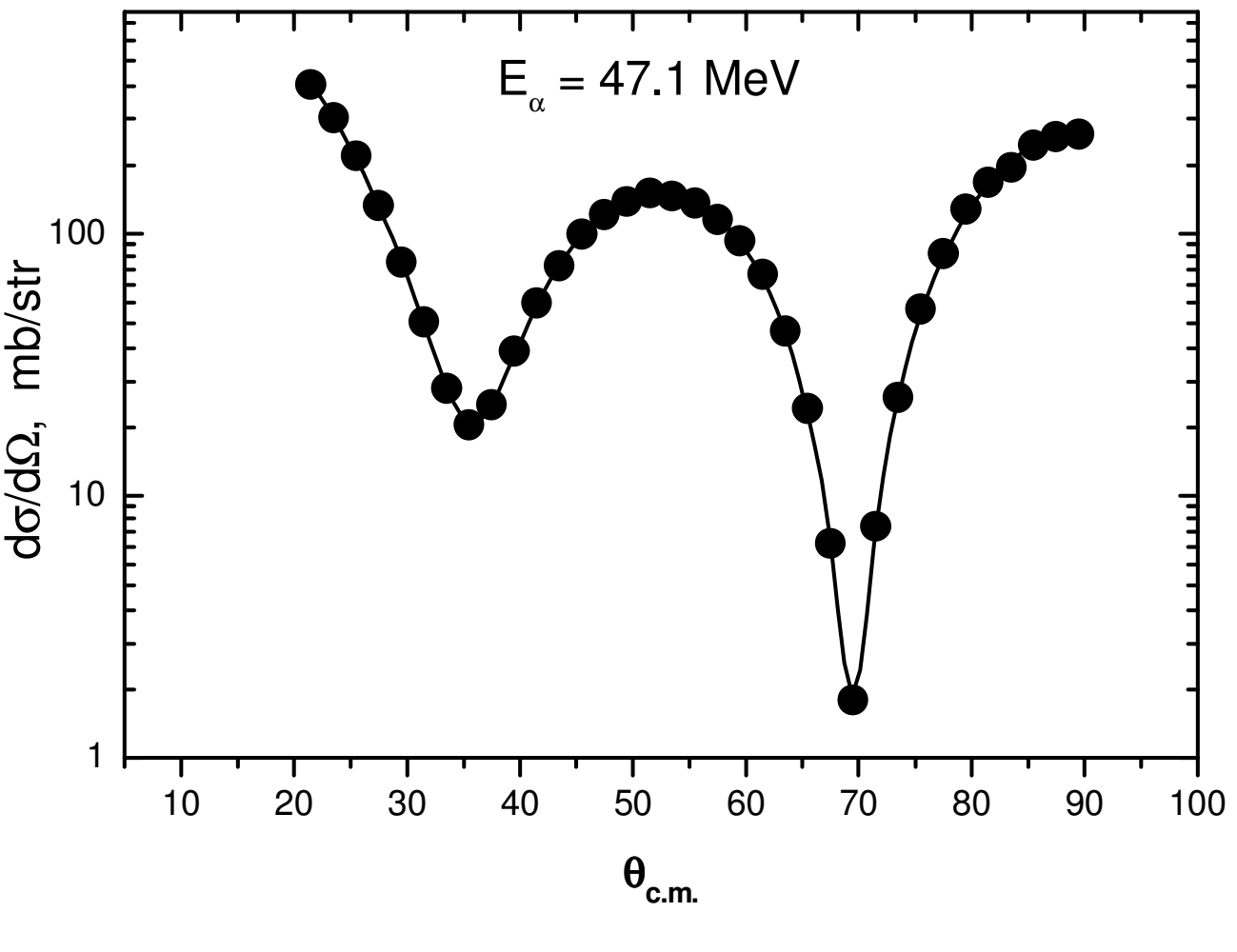

Кружки – экспериментальные данные, сплошная кривая - расчет сечений с найденными фазами.

Рисунок 7.16 - Дифференциальные сечения упругого рассеяния альфа-частиц на ядрах гелия при энергии 47.1 МэВ.

При увеличении L до 30 получает $\chi^2$ = 0.700 со следующими фазами

Действительная часть - 106.1704 51.6346 134.1084 17.2203
2.3269 0.3964 0.6347 0.0000 0.4220 0.2338 0.4530 0.0000
0.0020 0.0000 0.1307 0.3588
Мнимая часть - 11.9789 3.4758 1.2031 0.0000 0.0544 1.0305
1.1103 0.5890 0.2763 0.1363 0.4951 0.2665 0.2795 0.0000
0.0000 0.1021

Такие фазы мало отличаются от приведенных выше для случая с L=20. Это говорит о том, что в процессе поиска ядерных фаз рассеяния достигнуто насыщение, т.е. при увеличении числа парциальных волн каждая фаза стремиться к некоторому пределу, который и является ее истинным значением.

Рассмотрим теперь энергию 51.1 МэВ. Сечения были измерены в работе [102] (данные приведены на рисунках), а фазовый анализ вообще не проводился. Поэтому используем в качестве начальных фаз результаты работы [226] при 53.4 МэВ, где для реальной части фаз получено $\delta_0$ = 104,8±2.4$^0$, $\delta_2$ = 47.9±1.7$^0$, $\delta_4$ = 137.9±1.3$^0$, $\delta_6$ = 27.5±0.6$^0$, $\delta_8$ = 2.0±0.5$^0$. Для мнимой части найдено 12.1±3.1$^0$, 22.1±1.7$^0$, 16.3±1.1$^0$, 3.2±0.5$^0$, 0±0.4$^0$. Выполняя по нашей программе варьирование начальных фаз с десятью итерациями и L=20, получим $\chi^2$ = 0.97 и фазы рассеяния

Действительная часть фаз - 110.8739 55.0885 151.8536 24.9089
3.2213 0.0379 0.0000 0.2313 0.1331 0.2534 0.2644
Мнимая часть фаз - 14.9625 20.3880 23.6627 1.8434 0.3412
0.1910 0.0009 0.0000 0.0000 0.0000 0.1214

Увеличение L до 30 приводит нас к следующим фазам при $\chi^2$ = 0.565

Действительная часть фаз - 110.4496 55.9184 151.6671 24.5942
2.9661 0.0000 0.0540 0.3384 0.4010 0.3354 0.2822 0.0000
0.0987 0.0696 0.2609 0.3980
Мнимая часть фаз - 16.3963 20.2218 23.6944 1.6822 0.4220
0.2718 0.0000 0.0000 0.1552 0.0598 0.1692 0.1597 0.0271
0.0604 0.0002 0.1763

Найденные, таким образом, фазы вполне согласуются с общим ходом фаз в этой области энергий, а экспериментальные и вычис-ленные, с найденными фазами, сечения рассеяния показаны на рисунке 7.17.

В качестве экспериментальных ошибок, использовались ошибки определения сечений из рисунка, которые составляют примерно 5-10%.

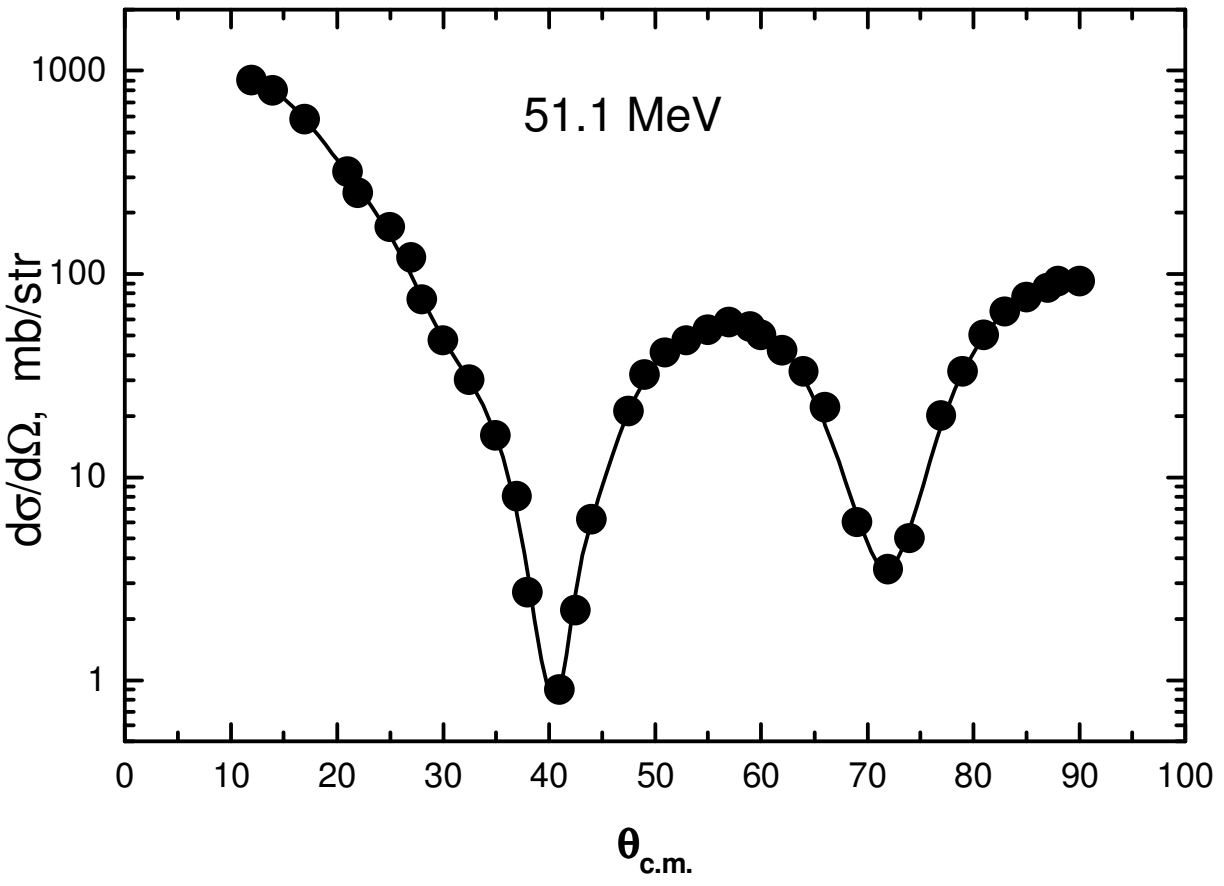

Кружки – экспериментальные данные, сплошная кривая - расчет сечений с найденными фазами.

Рисунок 7.17 - Сечения упругого $^4He^4He$ рассеяния при энергии 51.1 МэВ.

И в заключение выполним фазовый анализ экспериментальных данных при энергии 49.9 МэВ [227], для которой такие расчеты не проводился, принимая в качестве начальных фаз предыдущие результаты. Ранее было получено, что при L=8 величина $\chi^2$ оказывается порядка 130 [219]. Столь большое значение обусловлено малостью экспериментальных ошибок, полученные в работе [227].

Рассмотрим теперь более высокие парциальные волна. Так при L=20 для $\chi^2$ получается еще довольно большая величина 20.2 со следующими фазами

Действительная часть фаз - 128.4722 26.0885 138.4599 6.8897 0.0000 0.9080 3.3861 1.2455 4.9559 2.7870 0.0000
Мнимая часть фаз - 20.8384 21.9793 5.6456 0.0000 4.7497 6.2505 4.7320 0.0000 1.4833 1.1623 0.2628

И только при 30 парциальных волн удается хорошо воспроизвести экспериментальные результаты по дифференциальным сечениям

$\chi^2 = 0.0189$

| θ | $\sigma_e$ | $\sigma_t$ | $\chi^2$ |
|---|---|---|---|
| 2.000E+01 | 1.539E+02 | 1.539E+02 | 1.493E-04 |
| 2.400E+01 | 1.211E+02 | 1.211E+02 | 1.918E-03 |
| 2.800E+01 | 6.790E+01 | 6.792E+01 | 8.305E-04 |
| 3.200E+01 | 2.600E+01 | 2.599E+01 | 5.691E-04 |
| 3.600E+01 | 1.350E+01 | 1.351E+01 | 1.308E-03 |
| 4.000E+01 | 1.970E+01 | 1.969E+01 | 4.197E-03 |
| 4.400E+01 | 4.160E+01 | 4.163E+01 | 8.151E-03 |
| 4.800E+01 | 6.520E+01 | 6.515E+01 | 1.474E-02 |
| 5.200E+01 | 7.470E+01 | 7.474E+01 | 9.153E-03 |
| 5.600E+01 | 7.060E+01 | 7.055E+01 | 1.293E-02 |
| 6.000E+01 | 3.950E+01 | 3.951E+01 | 8.638E-04 |
| 6.400E+01 | 2.600E+01 | 2.599E+01 | 1.315E-03 |
| 6.800E+01 | 5.000E+00 | 5.003E+00 | 2.404E-03 |
| 7.200E+01 | 5.000E+00 | 5.006E+00 | 8.882E-03 |
| 7.600E+01 | 3.070E+01 | 3.068E+01 | 1.911E-02 |
| 8.000E+01 | 6.600E+01 | 6.606E+01 | 3.758E-02 |
| 8.400E+01 | 1.038E+02 | 1.038E+02 | 1.204E-02 |
| 9.000E+01 | 1.300E+02 | 1.318E+02 | 2.042E-01 |

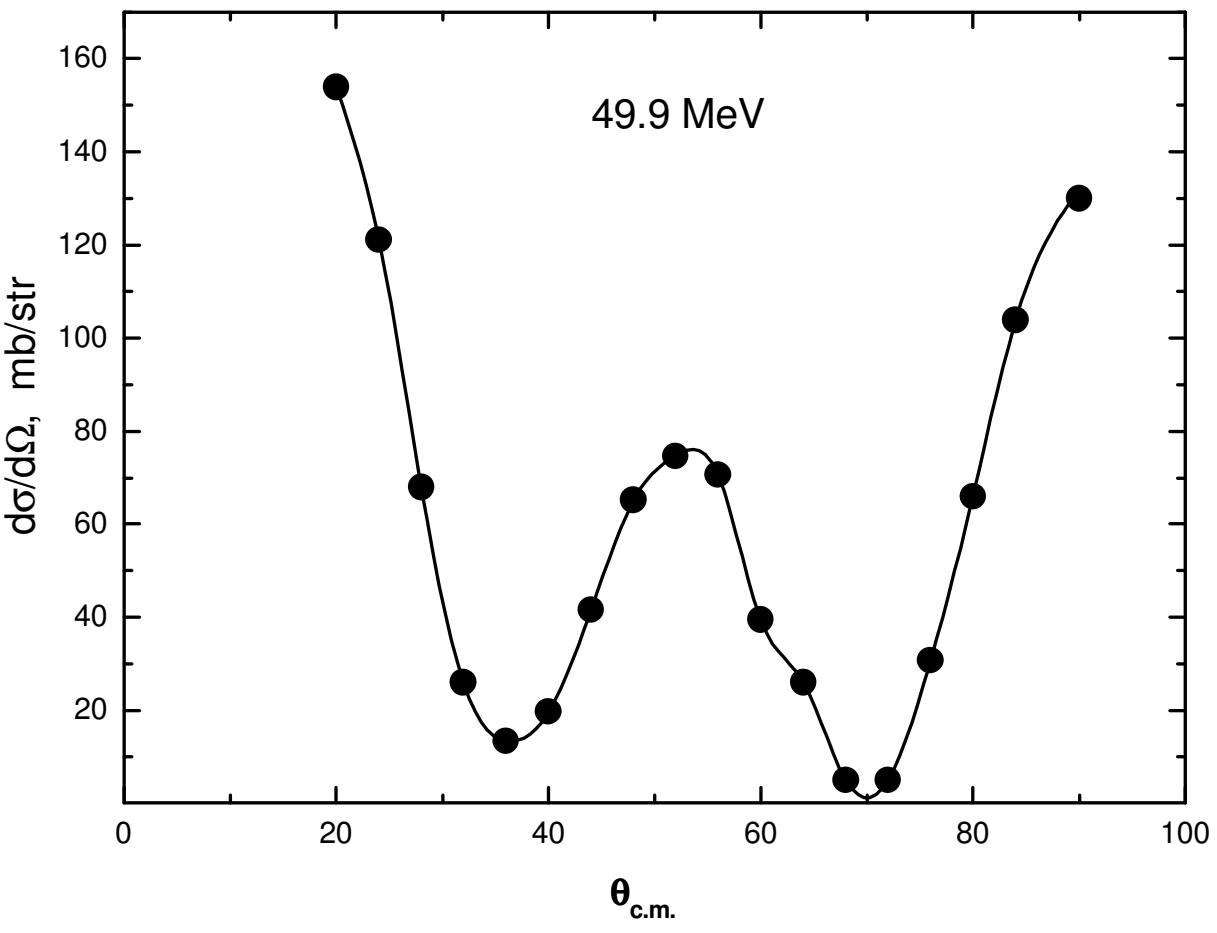

Кружки – экспериментальные данные, сплошная кривая - расчет сечений с найденными фазами.

Рисунок 7.18 - Сечения упругого $^4He^4He$ рассеяния при энергии 49.9 МэВ.

Действительная часть фаз - 127.5003 34.3877 141.3168 8.0940

0.0092 0.3824 1.8194 0.7321 4.4653 2.4917 0.0005 0.9836
0.1437 0.0018 0.5503 0.3922

Мнимая часть фаз - 31.2338 17.9109 6.1097 0.2166 3.4990
5.1733 4.1288 0.0000 1.8539 1.1013 0.0491 0.9386 0.3148
0.0746 0.2505 0.1446

На рисунке 7.18 показаны расчетные и экспериментальные сечения рассеяния.

Приведем теперь сводную таблицу 7.2 фаз рассеяния в области энергий 30-50 МэВ, полученных в наших расчетах и результаты при более низких энергиях 6-30 МэВ.

Таблица 7.2 - Сводная таблица фаз $^4He^4He$ рассеяния.

| E, МэВ | | $\delta_0$ | $\delta_2$ | $\delta_4$ | $\delta_6$ | $\delta_8$ | $\chi^2$ |
|---|---|---|---|---|---|---|---|
| 6.47 | Re $\delta_L$ | 80.4 | 80.7 | --- | --- | --- | 0.2 |
| 12.3 | Re $\delta_L$ | 28.4 | 105.0 | 2.6 | --- | --- | 3.4 |
| 17.8 | Re $\delta_L$ | 7.2 | 103.9 | 17.0 | --- | --- | 0.5 |
| 22.9 | Re $\delta_L$ | 169.3 | 94.5 | 59.5 | 1.0 | --- | 1.5 |
| 25.5 | Re $\delta_L$ | 160.5 | 89.0 | 88.6 | 1.4 | 0.2 | 0.9 |
| 29.5 | Re $\delta_L$ | 150.9 | 86.7 | 121.0 | 2.2 | 0.15 | 0.6 |
| 30.3 | Re $\delta_L$ | 136.8 | 72.6 | 121.0 | 0.0 | 1.1 | 0.2 |
| | Im $\delta_L$ | 1.9 | 4.2 | 0.4 | 2.7 | 0.1 | |
| 31.8 | Re $\delta_L$ | 149.9 | 70.1 | 126.6 | 1.4 | 4.3 | 0.1 |
| | Im $\delta_L$ | 0.0 | 2.1 | 4.6 | 1.6 | 0.5 | |
| 34.2 | Re $\delta_L$ | 142.7 | 63.9 | 134.2 | 1.1 | 0.0 | 0.6 |
| | Im $\delta_L$ | 0.0 | 0.0 | 2.3 | 0.0 | 0.0 | |
| 35.1 | Re $\delta_L$ | 155.1 | 70.2 | 137.8 | 0.0 | 1.0 | 0.01 |
| | Im $\delta_L$ | 3.4 | 0.5 | 4.6 | 0.8 | 1.0 | |
| 37.0 | Re $\delta_L$ | 114.1 | 55.9 | 140.2 | 7.0 | 0.0 | 0.2 |
| | Im $\delta_L$ | 5.3 | 1.5 | 8.9 | 6.6 | 0.0 | |
| 38.4 | Re $\delta_L$ | 135.0 | 82.1 | 169.2 | 4.0 | 1.5 | 0.6 |
| | Im $\delta_L$ | 0.0 | 2.9 | 0.1 | 0.1 | 0.7 | |
| 38.5 | Re $\delta_L$ | 129.9 | 77.2 | 165.7 | 1.5 | 0.6 | 0.2 |
| | Im $\delta_L$ | 3.8 | 2.5 | 1.1 | 0.9 | 1.4 | |
| 40.0 | Re $\delta_L$ | 69.5 | 49.5 | 81.4 | 1.4 | 0.0 | 0.2 |
| | Im $\delta_L$ | 0.9 | 0.0 | 4.6 | 7.1 | 1.3 | |
| 36.85 | Re $\delta_L$ | 126.3 | 62.3 | 132.8 | 2.5 | 0.6 | 1.3 |
| | Im $\delta_L$ | 0.0 | 0.0 | 2.3 | 18.9 | 0.0 | |
| 38.83 | Re $\delta_L$ | 121.9 | 100.4 | 163.5 | 3.8 | 6.4 | 1.9 |
| | Im $\delta_L$ | 6.7 | 6.5 | 0.5 | 0.4 | 0.1 | |
| 40.77 | Re $\delta_L$ | 127.4 | 40.8 | 87.6 | 2.2 | 0.0 | 1.6 |
| | Im $\delta_L$ | 18.8 | 0.9 | 0.4 | 2.0 | 0.0 | |
| 41.9 | Re $\delta_L$ | 105.9 | 53.1 | 103.4 | 16.1 | 0.0 | 0.9 |
| | Im $\delta_L$ | 19.0 | 0.0 | 0.0 | 0.3 | 0.0 | |
| 44.41 | Re $\delta_L$ | 117.4 | 72.1 | 116.0 | 16.8 | 3.3 | 0.5 |
| | Im $\delta_L$ | 30.2 | 9.4 | 3.4 | 0.2 | 1.4 | |
| 46.12 | Re $\delta_L$ | 134.8 | 56.4 | 126.0 | 12.2 | 1.9 | 2.2 |
| | Im $\delta_L$ | 34.3 | 5.4 | 1.4 | 1.8 | 1.0 | |
| 47.1 | Re $\delta_L$ | 106.2 | 51.6 | 134.1 | 17.2 | 2.3 | 0.7 |
| | Im $\delta_L$ | 12.0 | 3.5 | 1.2 | 0.0 | 0.0 | |
| 49.9 | Re $\delta_L$ | 127.5 | 34.4 | 141.3 | 8.1 | 0.0 | 0.02 |
| | Im $\delta_L$ | 31.2 | 17.9 | 6.1 | 0.2 | 3.5 | |
| 51.1 | Re $\delta_L$ | 110.4 | 55.9 | 151.7 | 24.6 | 3.0 | 0.6 |
| | Im $\delta_L$ | 16.4 | 20.2 | 23.7 | 1.7 | 0.4 | |

На рисунках 7.19-7.22 приведено сравнение, полученных в наших расчетах фаз рассеяния (треугольниками) в тех областях энергий, где отсутствовал фазовый анализ или был представлен ранее только на рисунках, с имеющимися на сегодняшний день результатами фазовых анализов при других энергиях.

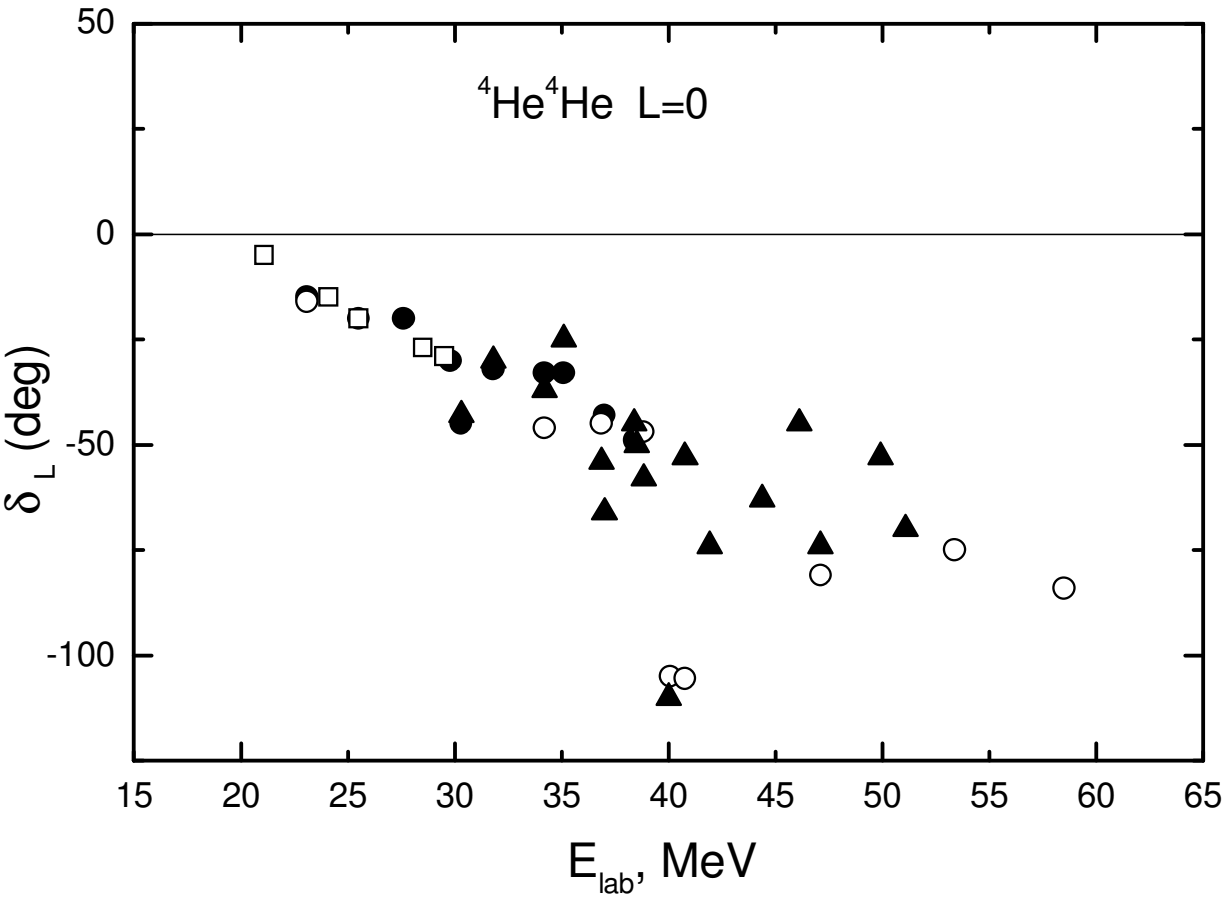

Кружки – данные работ [225] и [226] при энергии больше 50 МэВ, точки – [221], квадраты – [200], треугольники – наши результаты.
Рисунок 7.19 - Фазы упругого $^4$He$^4$He рассеяния при L=0.

Следует отметить, что найденные наборы параметров рассматриваемой вариационной задачи, т.е. фазы рассеяния при каж-дой энергии, могут лишь претендовать на то, что каждый из них определяет глобальный минимум $\chi^2$. В частности, при энергии 46.12 МэВ, по-видимому, так и не удалось найти нужный набор фаз рассеяния, определяющий истинный минимум величины $\chi^2$.

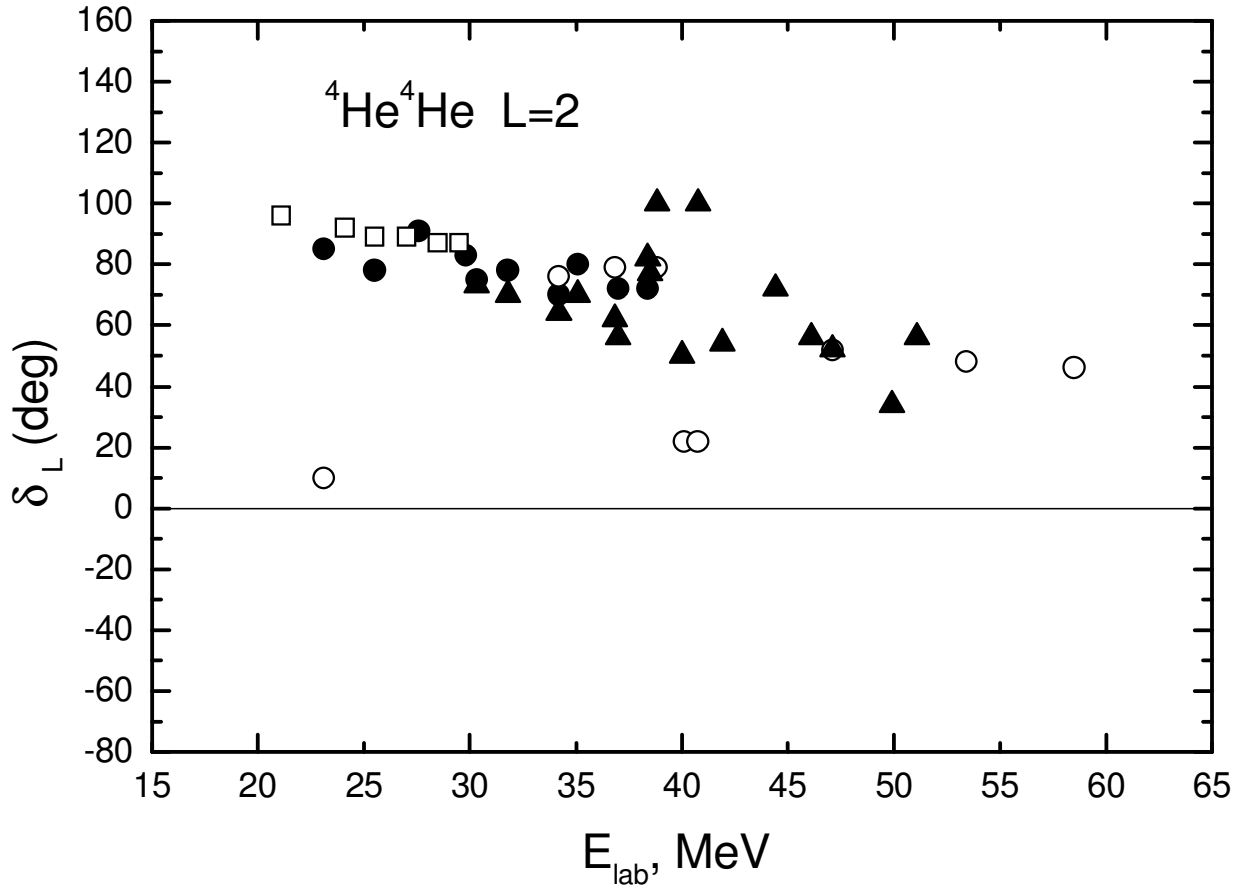

Кружки – данные работ [225] и [226] при энергии больше 50 МэВ, точки – [221], квадраты – [200], треугольники – наши результаты.
Рисунок 7.20 - Фазы упругого $^4He^4He$ рассеяния при L=2.

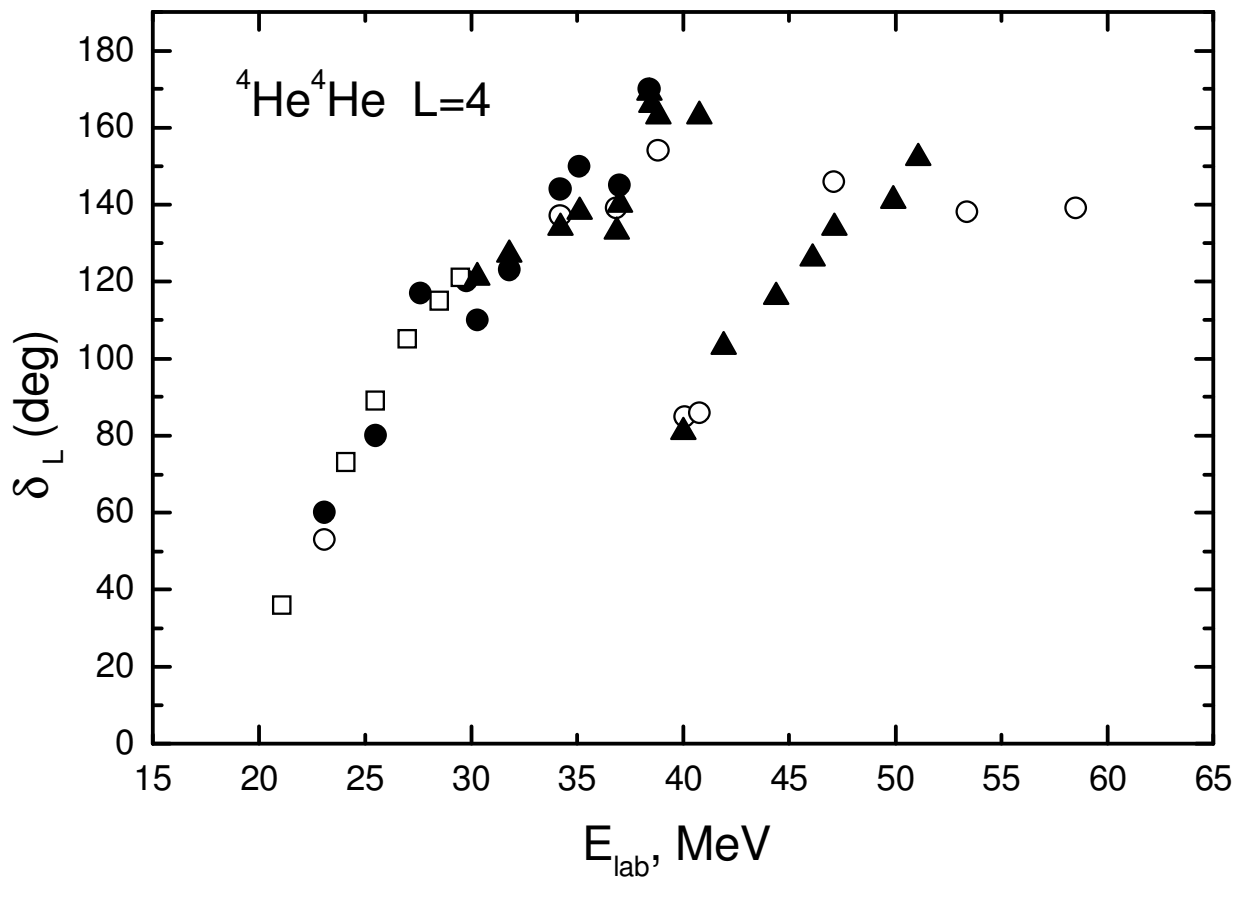

Кружки – данные работ [225] и [226] при энергии больше 50 МэВ, точки – [221], квадраты – [200], треугольники – наши результаты.
Рисунок 7.21 - Фазы упругого $^4He^4He$ рассеяния при L=4.

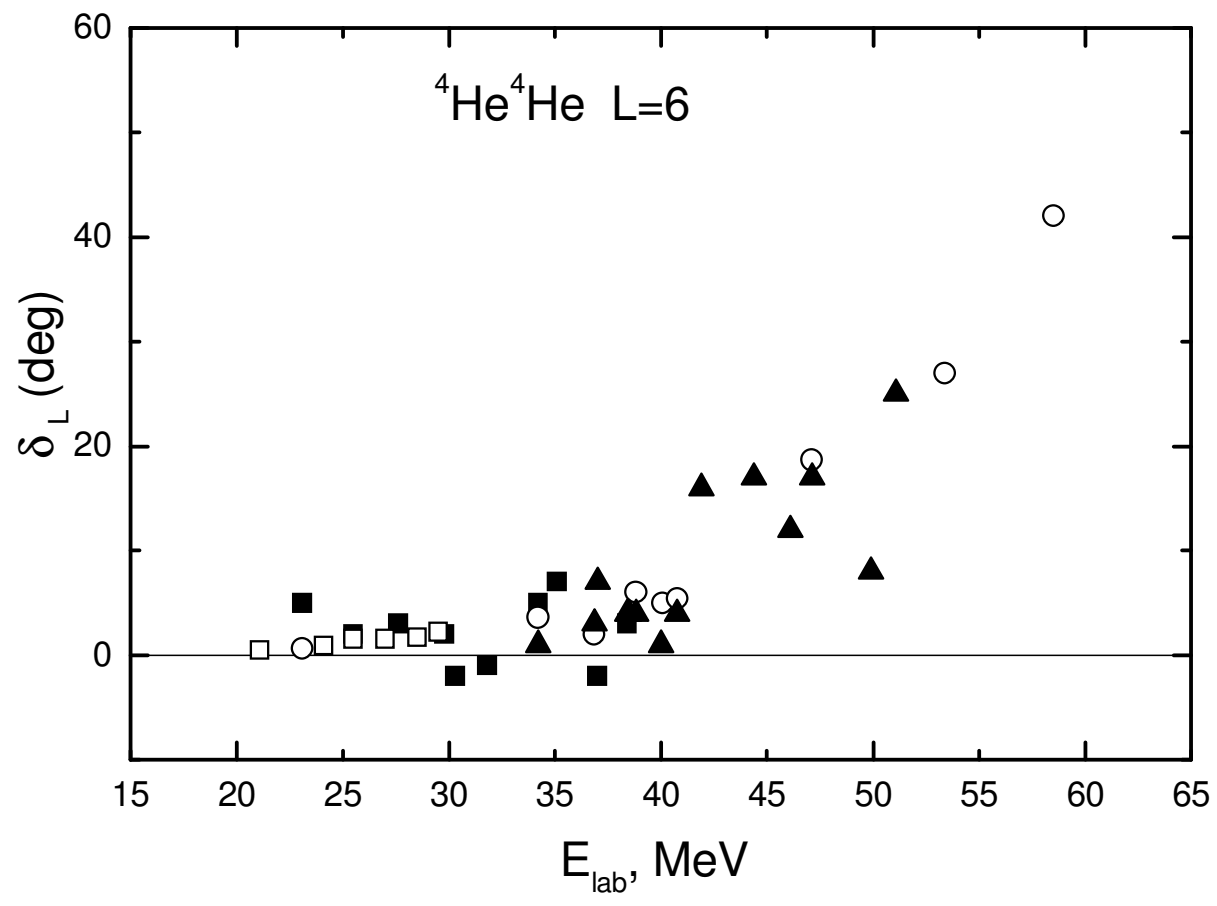

Кружки – данные работ [225] и [226] при энергии больше 50 МэВ, точки – [221], квадраты – [200], треугольники – наши результаты.
7.22 - Фазы упругого $^4$He$^4$He рассеяния при L=6.

Однако, при многих других энергиях, можно найти такие наборы параметров или фаз рассеяния, которые с увеличением числа парциальных волн, стремятся к некоторому пределу, определяющему их истинное значение.

Из этих результатов видно, что предложенная программа позволяет вполне успешно выполнять минимизацию функционала $\chi^2$ в поле многих вариационных параметров (действительных и комплексных фаз рассеяния) при всех рассмотренных энергиях сталкивающихся ядерных частиц.

В широкой энергетической области выполнено уточнение известных значений ядерных фаз $^4$He$^4$He рассеяния [228]. Фазовый анализ при других рассмотренных энергиях в области 30-50 МэВ приводит нас к вполне разумным результатам и, в целом, согласуется с данными других работ.

## 7.2 Система частиц с полным спином 1/2

Перейдем теперь к рассмотрению процессов упругого рассеяния в ядерной системе p$^4$He, полный спин которой равен 1/2. В случае упругого рассеяния таких частиц сечения выражаются через фазы ядерного рассеяния следующим образом (эти формулы были приведены в предыдущей главе, но мы снова приведем их для большей наглядности) [57]

$$\frac{d\sigma(\theta)}{d\Omega} = |A(\theta)|^2 + |B(\theta)|^2 \ , \tag{7.3}$$

где

$$A(\theta) = f_c(\theta) + \frac{1}{2ik}\sum_{L=0}^{\infty}\{(L+1)S_L^+ + LS_L^- - (2L+1)\}\exp(2i\sigma_L)P_L(Cos\theta) \ ,$$

$$B(\theta) = \frac{1}{2ik}\sum_{L=0}^{\infty}(S_L^+ - S_L^-)\exp(2i\sigma_L)P_L^1(Cos\theta) \ . \tag{7.4}$$

Здесь $S_L^\pm = \eta_L^\pm \exp(2i\delta_L^\pm)$ - матрица рассеяния, $\eta^\pm_L$ - параметры неупругости, а знаки “±“ соответствуют полному моменту системы J = L ± 1/2, $f_c$ - кулоновская амплитуда, описанная выше (6.3), $P_n^m(x)$ - присоединенные полиномы или функции Лежандра.

Приведем теперь текст компьютерной программы на языке Turbo Basic для поиска фаз рассеяния в системе частиц со спином 1/2 по заданным экспериментальным сечениям [229].

```
REM * ПРОГРАММА ФАЗОВОГО АНАЛИЗА AL-N 9.954 *
CLS: DEFDBL A-Z: DEFINT I,L,J,N,M,K
DIM SE(50), ST(50), DS(50), FP(50), FM(50), TT(50), XP(50),
DE(50), POL(50)
REM ********* НАЧАЛЬНЫЕ ЗНАЧЕНИЯ *****************
FAIL$="C:\BASICA\FAZ-ALN.DAT": PI=4*ATN(1.): Z1=1: Z2=2
AM1=1: AM2=4: AM=AM1+AM2: P1=3.14159265: A1=41.4686
PM=AM1*AM2/(AM1+AM2): B1=2*PM/A1: LMI=0: LH=1: LMA=3
EP=1.0D-05: NV=1: FH=1.1: NI=10: NPP=2*LMA
REM ********* НАЧАЛЬНЫЕ ПАРАМЕТРЫ АЛЬФА ********
REM ********* PHASE SHIFTS FOR P-AL ON E=9.954 ********
FP(0)=107.2:  FP(1)=103.3:  FP(2)=3.2:  FP(3)=2.3:   FP(4)=0.
FM(0)=FP(0):  FM(1)=54.7:   FM(2)=-2.9:  FM(3)=2.9:   FM(4)=0.
FPI(0)=.1:    FPI(1)=.1:   FPI(2)=.1:  FPI(3)=.1:  FPI(4)=.1
FMI(0)=FPI(0):  FMI(1)=.1:   FMI(2)=.1:  FMI(3)=.1:  FMI(4)=.1
REM ******** ECSPERIMENTAL CROSS SECTION 9.954 *****
SE(1)=371:   SE(2)=339:   SE(3)=305:   SE(4)=232:    SE(5)=205
SE(6)=176:   SE(7)=124:   SE(8)=82.0:  SE(9)=49.2:   SE(10)=39.1
SE(11)=26.2:  SE(12)=22.5:  SE(13)=21:    SE(14)=23: SE(15)=24.5
SE(16)=31.9: SE(17)=33.2:  SE(18)=37.8: SE(19)=47.3
SE(20)=54.0:  SE(21)=61.6:  SE(22)=70.4: SE(23)=78.4: SE(24)=84.9
TT(1)= 25.1: TT(2)=30.89:   TT(3)=35.07:   TT(4)=49.03
```

```
TT(5)=54.7: TT(6)= 60.1: TT(7)=70.10: TT(8)=80: TT(9)=90
TT(10)=94.07: TT(11)=102.17: TT(12)=106.9: TT(13)=109.9
TT(14)=120.6: TT(15)=122.8: TT(16)=130.13: TT(17)=130.9
TT(18)=134.87: TT(19)=140.8: TT(20)=145: TT(21)=149.4
TT(22)=154.9: TT(23)=160: TT(24)=164.4
REM *************** FOR AL-N ON E=9.954 ***************
NT=24: EL=9.954
REM ************ ENERGY IN LAB. SYSTEM **************
FOR L=LMI TO LMA STEP LH: FM(L)=FM(L)*PI/180
FP(L)=FP(L)*PI/180
FMI(L)=FMI(L)*PI/180: FPI(L)=FPI(L)*PI/180: NEXT:
FH=FH*PI/180
REM NP=2*NPP+1: REM - ПРИ УЧЕТЕ МНИМОЙ ЧАСТИ ФАЗ
NP=NPP: REM - ПРИ УЧЕТЕ ТОЛЬКО ДЕЙСТВИТЕЛЬНОЙ
ЧАСТИ ФАЗ
FOR I=LMI TO LMA STEP LH: XP(I)=FP(I): NEXT
FOR I=LMI TO LMA-1 STEP LH: XP(I+LMA+1)=FM(I+1): NEXT
FOR I=LMI TO LMA STEP LH: XP(I+2*LMA+1)=FPI(I): NEXT
FOR I=LMI TO LMA-1 STEP LH: XP(I+3*LMA+2)=FMI(I+1):
NEXT
REM *********** TRANSFORM TO C.M. *****************
EC=EL*PM/AM1: SK=EC*B1: SS=SQR(SK)
GG=3.44476E-02*Z1*Z2*PM/SS
CALL VAR(ST(),FH,LMA,NI,XP(),EP,XI,NV)
CLS: PRINT: PRINT "                XI-KV=";
PRINT USING " ####.### ";XI;NI
REM ************** TOTAL CROSSS SECTION *************
FOR I=LMI TO LMA STEP LH: FP(I)=XP(I): NEXT
FOR I=LMI TO LMA-1 STEP LH: FM(I+1)=XP(I+LMA+1): NEXT
FOR I=LMI TO LMA STEP LH: FPI(I)=XP(I+2*LMA+1): NEXT
FOR I=LMI TO LMA-1 STEP LH: FMI(I+1)=XP(I+3*LMA+2):
NEXT
FMI(0)=FPI(0): FM(0)=FP(0): SIGMAR=0: SIGMAS=0
FOR L=LMI TO LMA STEP LH: AP=FP(L): AM=FM(L)
135 ETP(L)=EXP(-2*FPI(L)): ETM(L)=EXP(-2*FMI(L))
SIGMAR=SIGMAR+((L+1)*(1-(ETP(L))^2)+L*(1-(ETM(L))^2))
SIGMAS = SIGMAS + ((L + 1)*(ETP(L))^2*(SIN(AP))^2 +
L*(ETM(L))^2*(SIN(AM))^2): NEXT L
SIGMAR=10*4*PI*SIGMAR/SK: SIGMAS=10*4*PI*SIGMAS/SK
PRINT "                SIGMR-TOT=";
PRINT USING " ####.## ";SIGMAR
PRINT "                SIGMS-TOT=";
PRINT USING " ####.## ";SIGMAS
PRINT "   T    SE     ST     XI     T     SE     ST     XI"
```

```
 FOR I=1 TO NT/2
 PRINT USING " ####.## ";TT(I); SE(I); ST(I); DS(I), TT(I+12);
SE(I+12); ST(I+12); DS(I+12): NEXT I: CLS
 PRINT "     T      POL       T       POL ": FOR I=1 TO NT/2
 PRINT USING " ####.## ";TT(I); POL(I)*100; TT(I+12);
POL(I+12)*100
 NEXT I: PRINT "   FP      FPI      FM      FMI"
 FOR L=LMI TO LMA STEP LH: FM(L)=FM(L)*180/PI
 FP(L)=FP(L)*180/PI: FMI(L)=FMI(L)*180/PI: FPI(L)=FPI(L)*180/PI
 PRINT USING " ###.### ";FP(L);FPI(L);FM(L);FMI(L): NEXT
 GOTO 1111: OPEN "O",1,G$
 PRINT#1, "             ALPHA - ALPHA FOR LAB E=";
 PRINT#1, E1(NN): FOR T=TMI TO TMA STEP TH
 PRINT#1, USING " #.####^^^^ ";T;SEC(T): NEXT
1111 END
 SUB VAR(ST(50),PHN,LMA,NI,XP(50),EP,AMIN,NV)
 DIM XPN(50): SHARED LH,LMI,NT,PI,DS(),NP,NPP
 REM ************* ПОИСК МИНИМУМА *****************
 FOR I=LMI TO NP STEP LH: XPN(I)=XP(I): NEXT: NN=LMI
 PRINT USING " ### ";NN;
 PRINT USING " +###.#########   ";XPN(NN)*180/PI: PH=PHN
 CALL DET(XPN(),ST(),ALA): B=ALA: IF NV=0 GOTO 3012
 PRINT USING "         +#.#######^^^^ ";ALA
 REM ------------------------------------------------------------------------
 FOR IIN=1 TO NI: NN=-LH
 PRINT USING "         +#.#######^^^^ ";ALA;IIN
 GOTO 1119
 1159 XPN(NN)=XPN(NN)-PH*XP(NN)
1119  NN=NN+LH:REM IF NN>NP GOTO 3012: IN=0
2229  A=B:  XPN(NN)=XPN(NN)+PH*XP(NN):  IF  NP=NPP  GOTO
7777
 IF NN<(NP/2) GOTO 7777: IF XPN(NN)<0 GOTO 1159
7777 IN=IN+1
 REM ------------------------------------------------------------------------
 CALL DET(XPN(),ST(),ALA): B=ALA: GOTO 5678
 PRINT USING " ### ";NN;
 PRINT USING " +###.#########   ";XPN(NN)*180/PI;
 PRINT USING "         +#.#######^^^^ ";ALA;
 PRINT
5678 REM ---------------------------------------------------------------------
  IF B<A GOTO 2229:  C=A:  XPN(NN)=XPN(NN)-PH*XP(NN)
  IF IN>1 GOTO 3339:  PH=-PH:  GOTO 5559
3339 IF ABS((C-B)/(B))<EP GOTO 4449:  PH=PH/2
5559 B=C: GOTO 2229
```

```
4449 PH=PHN:  B=C:  IF NN<NP GOTO 1119
3012  AMIN=B:  PH=PHN: NEXT IIN: FOR I=LMI TO NP STEP LH
 XP(I)=XPN(I): NEXT: END SUB
 SUB DET(XP(50),ST(50),XI)
 SHARED SE(),DS(),DE(),NT
 S=0: CALL SEC(XP(),ST()): FOR I=1 TO NT
 DS(I)=((ST(I)-SE(I))/0.01/SE(I))^2
 S=S+DS(I): NEXT: XI=S/NT: END SUB
 SUB SEC(XP(50),S(50))
 SHARED PI,NT,GG,SS,LMI,LMA,LH,POL(),TT(),NP,NPP
 DIM S0(20),P(20),PP(20),FP(20),FM(20),EP(20),EM(20)
 FOR I=LMI TO LMA STEP LH: FP(I)=XP(I): NEXT
 FOR I=LMI TO LMA-1 STEP LH: FM(I+1)=XP(I+LMA+1): NEXT
 FOR I=LMI TO LMA STEP LH: FPI(I)=XP(I+2*LMA+1): NEXT
 FOR  I=LMI  TO  LMA-1  STEP  LH:  FMI(I+1)=XP(I+3*LMA+2):
NEXT
 FMI(0)=FPI(0): FM(0)=FP(0): CALL CULFAZ(GG,S0())
 FOR I=1 TO NT: TT=TT(I): T=TT*PI/180: X=COS(T): A=2/(1-X)
 S0=2*S0(0): BB=-GG*A
 ALO=GG*LOG(A)+S0: REC=BB*COS(ALO): AMC=BB*SIN(ALO)
 REA=0: AMA=0:  REB=0:   AMB=0: FOR L=LMI TO LMA STEP
LH
 FP=2*FP(L): FM=2*FM(L): EP(L)=EXP(-2*FPI(L))
 EM(L)=EXP(-2*FMI(L)): A=EP(L)*COS(FP)-EM(L)*COS(FM)
 B=EP(L)*SIN(FP)-EM(L)*SIN(FM): SL=2*S0(L)
 CALL FUNLEG(X,L,PP())
 REB=REB+(B*COS(SL)+A*SIN(SL))*PP(L)
 AMB=AMB+(B*SIN(SL)-A*COS(SL))*PP(L): LL=2*L+1: JJ=L+1
 A=JJ*EP(L)*COS(FP)+L*EM(L)*COS(FM)-LL
 B=JJ*EP(L)*SIN(FP)+L*EM(L)*SIN(FM): CALL POLLEG(X,L,P())
 REA=REA+(B*COS(SL)+A*SIN(SL))*P(L)
 AMA=AMA+(B*SIN(SL)-A*COS(SL))*P(L): NEXT L
 REA=REC+REA: AMA=AMC+AMA
 RE=REA^2+AMA^2: AM=REB^2+AMB^2
 S(I)=10*(RE+AM)/4/SS^2
 POL(I)=2*(REB*AMA-REA*AMB)/(RE+AM)
 NEXT I: END SUB
 SUB POLLEG(X,L,P(20))
 P(0)=1: P(1)=X: FOR I=2 TO L: P(I)=(2*I-1)*X/I*P(I-1)-(I-1)/I*P(I-2)
 NEXT: END SUB
 SUB FUNLEG(X,L,P(20))
 P(0)=0: P(1)=SQR(ABS(1-X^2)): P(2)=3*X*P(1)
 FOR I=2 TO L: P(I+1)=(2*I+1)*X/I*P(I)-(I+1)/I*P(I-1)
 NEXT: END SUB
```

```
SUB CULFAZ(G,F(20))
REM COULOMB PHASE SHIFTS
C=0.577215665: S=0: N=50: A1=1.202056903/3: A2=1.036927755/5
FOR I=1 TO N: A=G/I-ATN(G/I)-(G/I)^3/3+(G/I)^5/5: S=S+A
NEXT: FAZ=-C*G+A1*G^3-A2*G^5+S: F(0)=FAZ: FOR I=1 TO 20
F(I)=F(I-1)+ATN(G/(I)): NEXT: END SUB
```

Приведем теперь вариант контрольного счета по этой программе, который выполнен для $p^4He$ рассеяния при энергии 9.954 МэВ с данными из работы [202]. В работе приведены экспериментальные сечения при энергиях 2 - 11 МэВ и результаты фазового анализа, которые приведены в предыдущей главе и в таблице 7.3 для E = 9.954 МэВ.

Таблица 7.3 - Фазы $p^4He$ рассеяния.

| E, МэВ | $S_0$, град | $P_{3/2}$, град | $P_{1/2}$, град | $D_{5/2}$, град | $D_{3/2}$, град |
|---|---|---|---|---|---|
| 9.954 | 111,0 | 103,0 | 52,0 | -2,0 | -4,0 |

Результаты расчета сечений $\sigma_t$ при этой энергии с точными значениями углов рассеяния и табличными фазами приведены в предыдущей главе, где было получено $\chi^2 = 0.96$.

Представляется интересным выяснить, насколько хорошо был выполнен фазовый анализ сечений, и можно ли получить меньший $\chi^2$ варьируя фазы из работы [202]. Выполняем уточнение (варьирование) фаз по нашей программе при 10 итерациях (экспериментальные ошибки по сечениям приняты равными 2%)

$\chi^2 = 0.401$ $\sigma_s = 1402.41$

| θ | $\sigma_e$ | $\sigma_t$ | $\chi^2$ | θ | $\sigma_e$ | $\sigma_t$ | $\chi^2$ |
|---|---|---|---|---|---|---|---|
| 25.10 | 371.00 | 371.09 | 0.00 | 109.90 | 21.00 | 20.71 | 0.47 |
| 30.89 | 339.00 | 333.69 | 0.61 | 120.60 | 23.00 | 22.61 | 0.72 |
| 35.07 | 305.00 | 310.01 | 0.67 | 122.80 | 24.50 | 24.23 | 0.30 |
| 49.03 | 232.00 | 232.29 | 0.00 | 130.13 | 31.90 | 32.01 | 0.03 |
| 54.70 | 205.00 | 200.96 | 0.97 | 130.90 | 33.20 | 33.01 | 0.08 |
| 60.00 | 176.00 | 172.49 | 0.99 | 134.87 | 37.80 | 38.58 | 1.05 |
| 70.10 | 124.00 | 122.35 | 0.45 | 140.80 | 47.30 | 47.84 | 0.32 |
| 80.00 | 82.00 | 81.03 | 0.35 | 145.00 | 54.00 | 54.74 | 0.47 |
| 90.00 | 49.20 | 49.49 | 0.09 | 149.40 | 61.60 | 61.96 | 0.08 |
| 94.07 | 39.10 | 39.90 | 1.04 | 154.90 | 70.60 | 70.53 | 0.00 |
| 102.17 | 26.20 | 26.54 | 0.42 | 160.00 | 78.40 | 77.61 | 0.26 |
| 106.90 | 22.00 | 22.20 | 0.21 | 164.40 | 83.00 | 82.76 | 0.02 |

$S_0 = 109.14$, $P_{3/2} = 101.92$, $P_{1/2} = 50.74$, $D_{5/2} = -2.21$,
$D_{3/2} = -5.40$ - Улучшенный вариант фаз.

Как видно, очень не большие изменения фаз позволяют заметно улучшить описание экспериментальных сечений. Отметим, что такие изменения фаз заметно меняют расчетные поляризации, поэтому минимизацию $\chi^2$ нужно проводить при совместном анализе сечений и поляризаций.

В работе [230] был приведен вариант фазового анализа для энергии 9.89 МэВ, в котором получены положительные D фазы и среднее $\chi^2 = 0.60$. В этом анализе использованы 22 точки по сечениям из [202] при энергии 9.954 МэВ (в [230] не указано, какие именно 22 точки были взяты из 24 - х, приведенных в работе [202]) и несколько точек по поляризациям из работ [230, 231]. В последнем случае, по - видимому, использованы данные при углах $46.5^0$, $55.9^0$, $56.2^0$, $73.5^0$, $89.7^0$, $99.8^0$, $114.3^0$, $128.3^0$ и энергиях 9.89, 9.84 и 9.82 МэВ.

Фазы из работы [230] приведены в таблице 7.4, а $\chi^2$ по нашей программе с учетом 24 точек только по сечениям из [202] (при энергии 9.954 МэВ) и этими фазами получается равным 0.59, как показано на распечатке ниже.

Таблица 7.4 - Фазы рассеяния из работы [230].

| E, МэВ | $S_0$, град | $P_{3/2}$, град | $P_{1/2}$, град | $D_{5/2}$, град | $D_{3/2}$, град |
|---|---|---|---|---|---|
| 9.954 | 119,3 $^{+2.0}_{-1.8}$ | 112,4 $^{+3.5}_{-5.2}$ | 65,7 $^{+2.7}_{-3.2}$ | 5,3 $^{+1.6}_{-2.5}$ | 3,7 $^{+1.6}_{-2.8}$ |

$\chi^2_s = 0.586$ $\sigma_s = 1363.87$

| θ | $\sigma_e$ | $\sigma_t$ | $\chi^2$ | θ | $\sigma_e$ | $\sigma_t$ | $\chi^2$ |
|---|---|---|---|---|---|---|---|
| 25.10 | 371.00 | 366.85 | 0.31 | 109.90 | 21.00 | 20.70 | 0.51 |
| 30.89 | 339.00 | 331.54 | 1.21 | 120.60 | 23.00 | 22.59 | 0.79 |
| 35.07 | 305.00 | 308.40 | 0.31 | 122.80 | 24.50 | 24.19 | 0.40 |
| 49.03 | 232.00 | 230.61 | 0.09 | 130.13 | 31.90 | 31.91 | 0.00 |
| 54.70 | 205.00 | 199.10 | 2.07 | 130.90 | 33.20 | 32.90 | 0.21 |
| 60.00 | 176.00 | 170.56 | 2.39 | 134.87 | 37.80 | 38.44 | 0.71 |
| 70.10 | 124.00 | 120.59 | 1.89 | 140.80 | 47.30 | 47.69 | 0.17 |
| 80.00 | 82.00 | 79.77 | 1.85 | 145.00 | 54.00 | 54.62 | 0.33 |
| 90.00 | 49.20 | 48.82 | 0.15 | 149.40 | 61.60 | 61.88 | 0.05 |
| 94.07 | 39.10 | 39.44 | 0.19 | 154.90 | 70.60 | 70.54 | 0.00 |
| 102.17 | 26.20 | 26.39 | 0.13 | 160.00 | 78.40 | 77.71 | 0.19 |
| 106.90 | 22.00 | 22.15 | 0.12 | 164.40 | 83.00 | 82.94 | 0.00 |

Для данных [230,231] по поляризациям при энергиях 9.82-9.89 МэВ и восьми углах, с фазами из [230], можно получить следующие результаты (углы в даны градусах, поляризации в процентах, т.е. поляризация умножена на 100%, а энергия, по-прежнему, задается равной 9.954 МэВ)

$\chi^2_p = 0.589$

| $\theta$ | $P_e$ | $\Delta P_e$ | $P_t$ | $\chi^2_i$ |
|---|---|---|---|---|
| 46.50 | -32.30 | 2.10 | -33.11 | 0.15 |
| 55.90 | -41.30 | 2.20 | -42.50 | 0.30 |
| 56.20 | -44.40 | 0.90 | -42.81 | 3.11 |
| 73.50 | -62.60 | 3.00 | -62.84 | 0.01 |
| 73.50 | -64.80 | 1.90 | -62.84 | 1.06 |
| 89.70 | -76.10 | 3.60 | -76.33 | 0.01 |
| 89.70 | -75.50 | 2.40 | -76.33 | 0.12 |
| 99.80 | -59.30 | 2.50 | -58.55 | 0.09 |
| 114.30 | 48.20 | 3.20 | 51.03 | 0.78 |
| 128.30 | 99.40 | 3.30 | 97.66 | 0.28 |

Здесь $P_e$ - экспериментальные поляризации, $\Delta P_e$ - экспериментальные ошибки для поляризаций и $P_t$ - расчетные поляризации. Если усреднить $\chi^2$ по всем точкам (24+10=34), результаты для которых приведены выше, т.е. использовать выражение

$$\chi^2 = \frac{1}{(N_s + N_P)}\left\{\sum_{i=1}^{N}\left[\frac{\sigma_i^{\ t} - \sigma_i^{\ e}}{\Delta\sigma_i^{\ e}}\right]^2 + \sum_{i=1}^{N}\left[\frac{P_i^{\ t} - P_i^{\ e}}{\Delta P_i^{\ e}}\right]^2\right\}$$

то получается величина $\chi^2 = 0.5875 \approx 0.59$ в полном соответствии с результатами работы [230], где для значения $\chi^2$ приведено 0.60. Здесь $N_s$ и $N_P$ - число данных по сечениям (24 точки) и поляризациям (10 точек), $s^e$, $P^e$, $s^t$, $P^t$ – экспериментальные и теоретические значения сечений и поляризаций.

Если выполнить дополнительную минимизацию $\chi^2$ по приведенной программе, то для $\chi^2_s$ по сечениям получим 0.576, для поляризаций $\chi^2_p = 0.561$ и полное $\chi^2 = 0.572$ при следующих значениях фаз

$S_0 = 119.01$ , $P_{3/2} = 112.25$ , $P_{1/2} = 65.39$ , $D_{5/2} = 5.24$ , $D_{3/2} = 3.63$

которые полностью ложатся в полосу ошибок, приведенных в работе [230].

Учтем теперь данные по поляризациям для энергии 10 МэВ

при девяти углах из работы [232], тогда, для 19 - ти точек по поляризациям, получим (здесь использованы фазы из работы [230] и энергия 9.954 МэВ)

$\chi^2_p = 0.825$

| $\theta$ | $P_e$ | $\Delta P_e$ | $P_t$ | $\chi^2_i$ |
|---|---|---|---|---|
| 46.50 | -32.30 | 2.10 | -33.11 | 0.15 |
| 55.90 | -41.30 | 2.20 | -42.50 | 0.30 |
| 56.20 | -44.40 | 0.90 | -42.81 | 3.11 |
| 73.50 | -62.60 | 3.00 | -62.84 | 0.01 |
| 73.50 | -64.80 | 1.90 | -62.84 | 1.06 |
| 89.70 | -76.10 | 3.60 | -76.33 | 0.00 |
| 89.70 | -75.50 | 2.40 | -76.33 | 0.12 |
| 99.80 | -59.30 | 2.50 | -58.55 | 0.09 |
| 114.30 | 48.20 | 3.20 | 51.03 | 0.78 |
| 128.30 | 99.40 | 3.30 | 97.66 | 0.28 |
| 123.30 | 96.20 | 1.70 | 95.61 | 0.12 |
| 124.70 | 95.40 | 1.80 | 97.46 | 1.30 |
| 126.90 | 99.20 | 1.60 | 98.25 | 0.35 |
| 127.40 | 98.00 | 1.80 | 98.13 | 0.01 |
| 128.70 | 96.80 | 1.50 | 97.37 | 0.14 |
| 129.60 | 97.20 | 1.60 | 96.51 | 0.18 |
| 130.00 | 94.30 | 1.50 | 96.06 | 1.38 |
| 131.40 | 92.30 | 1.60 | 94.16 | 1.35 |
| 133.50 | 93.70 | 1.40 | 90.58 | 4.96 |

Для полного $\chi^2$, найденного по сечениям (24 точки по сечениям из [202] и фазы из [230]) и поляризациям (19 точек из [230,231,232] и фазы из [230]), получим 0.692, т.е. всего учитывались 43-и точки с экспериментальными данными.

Используем теперь фазы, найденные в работе [232], которые приведены в таблице 7.5. Тогда для сечений и поляризаций (43 точки с данными) получим следующие результаты (энергия рассеяния, по - прежнему, задается 9.954 МэВ)

Таблица 7.5 - Фазы рассеяния из работы [232].

| E, МэВ | $S_0$, град | $P_{3/2}$, град | $P_{1/2}$, град | $D_{5/2}$, град | $D_{3/2}$, град |
|---|---|---|---|---|---|
| 9.954 | 119,75±0.69 | 112,99±0.92 | 66.18±1.26 | 5,39±0.57 | 3,78±0.57 |

$\chi^2_s = 0.678$ $\sigma_s = 1358.18$

| $\theta$ | $\sigma_e$ | $\sigma_t$ | $\chi^2$ | $\theta$ | $\sigma_e$ | $\sigma_t$ | $\chi^2$ |
|---|---|---|---|---|---|---|---|
| 25.10 | 371.00 | 366.04 | 0.45 | 109.90 | 21.00 | 20.80 | 0.22 |
| 30.89 | 339.00 | 330.43 | 1.60 | 120.60 | 23.00 | 22.65 | 0.57 |
| 35.07 | 305.00 | 307.24 | 0.13 | 122.80 | 24.50 | 24.24 | 0.27 |
| 49.03 | 232.00 | 229.64 | 0.26 | 130.13 | 31.90 | 31.93 | 0.00 |
| 54.70 | 205.00 | 198.30 | 2.67 | 130.90 | 33.20 | 32.92 | 0.17 |
| 60.00 | 176.00 | 169.91 | 2.99 | 134.87 | 37.80 | 38.46 | 0.75 |
| 70.10 | 124.00 | 120.24 | 2.30 | 140.80 | 47.30 | 47.71 | 0.19 |
| 80.00 | 82.00 | 79.66 | 2.04 | 145.00 | 54.00 | 54.64 | 0.35 |
| 90.00 | 49.20 | 48.86 | 0.12 | 149.40 | 61.60 | 61.91 | 0.07 |
| 94.07 | 39.10 | 39.51 | 0.28 | 154.90 | 70.60 | 70.59 | 0.00 |
| 102.17 | 26.20 | 26.50 | 0.32 | 160.00 | 78.40 | 77.78 | 0.16 |
| 106.90 | 22.00 | 22.26 | 0.35 | 164.40 | 83.00 | 83.03 | 0.00 |

$\chi^2_p = 0.740$

| θ | $P_e$ | $\Delta P_e$ | $P_t$ | $\chi^2_i$ |
|---|---|---|---|---|
| 46.50 | -32.30 | 2.10 | -33.35 | 0.25 |
| 55.90 | -41.30 | 2.20 | -42.80 | 0.46 |
| 56.20 | -44.40 | 0.90 | -43.12 | 2.02 |
| 73.50 | -62.60 | 3.00 | -63.26 | 0.05 |
| 89.70 | -76.10 | 3.60 | -76.80 | 0.04 |
| 99.80 | -59.30 | 2.50 | -59.05 | 0.01 |
| 114.30 | 48.20 | 3.20 | 50.36 | 0.45 |
| 123.30 | 96.20 | 1.70 | 95.35 | 0.25 |
| 124.70 | 95.40 | 1.80 | 97.27 | 1.08 |
| 126.90 | 99.20 | 1.60 | 98.17 | 0.42 |
| 127.40 | 98.00 | 1.80 | 98.06 | 0.00 |
| 128.30 | 99.40 | 3.30 | 97.63 | 0.29 |
| 128.70 | 96.80 | 1.50 | 97.35 | 0.14 |
| 129.60 | 97.20 | 1.60 | 96.53 | 0.18 |
| 130.00 | 94.30 | 1.50 | 96.09 | 1.42 |
| 131.40 | 92.30 | 1.60 | 94.22 | 1.44 |
| 133.50 | 93.70 | 1.40 | 90.69 | 4.61 |

Для полного $\chi^2$ теперь получим 0.705, а в работе [232] приводится значение 0.6. Если отбросить последнюю точку по поляризациям, которая дает наибольший $\chi^2_i$ (т.е. использовать 18 точек), то получим $\chi^2_p = 0.52$ и полный $\chi^2 = 0.6$ в полном согласии с результатами работы [232].

Выполним теперь варьирование фаз при десяти итерациях $N_i = 10$ с этими начальными фазами (при 42 - х экспериментальных точках) и одновременно минимизируя $\chi^2$ по сечениям и поляризациям

$\chi^2_s = 0.605$ $\sigma_s = 1363.85$

| $\theta$ | $\sigma_e$ | $\sigma_t$ | $\chi^2$ | $\theta$ | $\sigma_e$ | $\sigma_t$ | $\chi^2$ |
|---|---|---|---|---|---|---|---|
| 25.10 | 371.00 | 366.53 | 0.36 | 109.90 | 21.00 | 20.67 | 0.62 |
| 30.89 | 339.00 | 331.37 | 1.27 | 120.60 | 23.00 | 22.49 | 1.23 |
| 35.07 | 305.00 | 308.30 | 0.29 | 122.80 | 24.50 | 24.07 | 0.74 |
| 49.03 | 232.00 | 230.62 | 0.09 | 130.13 | 31.90 | 31.76 | 0.04 |
| 54.70 | 205.00 | 199.14 | 2.04 | 130.90 | 33.20 | 32.75 | 0.44 |
| 60.00 | 176.00 | 170.62 | 2.34 | 134.87 | 37.80 | 38.29 | 0.43 |
| 70.10 | 124.00 | 120.69 | 1.78 | 140.80 | 47.30 | 47.55 | 0.07 |
| 80.00 | 82.00 | 79.89 | 1.66 | 145.00 | 54.00 | 54.49 | 0.21 |
| 90.00 | 49.20 | 48.92 | 0.08 | 149.40 | 61.60 | 61.77 | 0.02 |
| 94.07 | 39.10 | 39.51 | 0.28 | 154.90 | 70.60 | 70.46 | 0.01 |
| 102.17 | 26.20 | 26.41 | 0.17 | 160.00 | 78.40 | 77.66 | 0.22 |
| 106.90 | 22.00 | 22.14 | 0.11 | 164.40 | 83.00 | 82.92 | 0.00 |

$\chi^2_p = 0.538$

| $\theta$ | $P_e$ | $\Delta P_e$ | $P_t$ | $\chi^2_i$ |
|---|---|---|---|---|
| 46.50 | -32.30 | 2.10 | -33.43 | 0.29 |
| 55.90 | -41.30 | 2.20 | -42.81 | 0.47 |
| 56.20 | -44.40 | 0.90 | -43.13 | 2.02 |
| 73.50 | -62.60 | 3.00 | -62.96 | 0.01 |
| 73.50 | -64.80 | 1.90 | -62.96 | 0.94 |
| 89.70 | -76.10 | 3.60 | -76.02 | 0.00 |
| 89.70 | -75.50 | 2.40 | -76.02 | 0.05 |
| 99.80 | -59.30 | 2.50 | -58.03 | 0.26 |
| 114.30 | 48.20 | 3.20 | 51.34 | 0.96 |
| 128.30 | 99.40 | 3.30 | 97.56 | 0.31 |
| 123.30 | 96.20 | 1.70 | 95.69 | 0.09 |
| 124.70 | 95.40 | 1.80 | 97.48 | 1.34 |
| 126.90 | 99.20 | 1.60 | 98.20 | 0.39 |
| 127.40 | 98.00 | 1.80 | 98.05 | 0.00 |
| 128.70 | 96.80 | 1.50 | 97.24 | 0.09 |
| 129.60 | 97.20 | 1.60 | 96.36 | 0.27 |
| 130.00 | 94.30 | 1.50 | 95.89 | 1.13 |
| 131.40 | 92.30 | 1.60 | 93.94 | 1.06 |

$\chi^2 = 0.572 \approx 0.57$ – Среднее значение.

$S_0 = 119.36$, $P_{3/2} = 112.40$, $P_{1/2} = 65.67$, $D_{5/2} = 5.32$, $D_{3/2} = 3.89$

В этом случае для среднего $\chi^2$ по всем этим точкам получим 0.57, т.е. процессы минимизации такого функционала приводят к дальнейшему улучшению описания экспериментальных данных. Как и в предыдущем случае, мы снова получаем для $\chi^2$ такую же величину 0.57, которая является, по-видимому, минимально возможной для такого типа экспериментальных данных.

## 7.3 Нетождественные частицы с полуцелым спином

Рассмотрим теперь процессы упругого рассеяния в системе нетождественных частиц, типа $p^3$He, когда спины каждой частицы равны 1/2. В случае упругого рассеяния таких частиц сечения выражаются через фазы ядерного рассеяния следующим образом [57]

$$\frac{d\sigma(\theta)}{d\Omega} = 1/4\frac{d\sigma_s(\theta)}{d\Omega} + 3/4\frac{d\sigma_t(\theta)}{d\Omega} \quad , \tag{7.5}$$

где индексы s и t относятся к синглетному и триплетному состоянию рассеяния и

$$\frac{d\sigma_s(\theta)}{d\Omega} = |f_s(\theta)|^2 \quad , \qquad \frac{d\sigma_t(\theta)}{d\Omega} = |f_t(\theta)|^2 \quad . \tag{7.6}$$

Для триплетного состояния, при учете спин - орбитального расщепления, можно использовать формулы, которые совпадают с выражениями для $^2H^4$He рассеяния [57]

$$\frac{d\sigma_t(\theta)}{d\Omega} = \frac{1}{3}\left[|A|^2 + 2\left(|B|^2 + |C|^2 + |D|^2 + |E|^2\right)\right] \quad ,$$

а для синглетного сечения используем формулы предыдущей главы (6.26), (6.27) и (6.29).

Приведем теперь программу для поиска ядерных фаз на основе дифференциальных сечений упругого рассеяния в такой системе [233].

```
REM ***** ПРОГРАММА ФАЗОВОГО АНАЛИЗА ДЛЯ 3He-P
ПРИ 11.48 МЭВ***
CLS: DEFDBL A-Z: DEFINT I,L,J,N,M,K
DIM SE(50), ST(50), DS(50), FP(50), F0(50), FM(50), FS(50), FT(50),
TT(50), T20(50), T22(50), T21(50)
DIM XP(50), DE(50), POL(50), FPI(50), F0I(50), FMI(50), FSI(50)
REM ********* НАЧАЛЬНЫЕ ЗНАЧЕНИЯ **********
FAIL$="C:\BASICA\FAZ-ALN.DAT": PI=4*ATN(1.): Z1=1: Z2=2
AM1=1: AM2=3: AM=AM1+AM2: P1=3.14159265: A1=41.4686
PM=AM1*AM2/(AM1+AM2): B1=2*PM/A1: LMI=0: LH=1: LMA=2
LN=LMI: LV=LMA: EP=1.0D-05: NV=1: FH=.1: NI=10:
```

```
NPP=2*LMA
 REM ****** НАЧАЛЬНЫЕ ПАРАМЕТРЫ АЛЬФА ******
 REM *** ECSPERIMENTAL CROSS SECTION 11.48 *****
 SE(1)=223.1: SE(2)=222: SE(3)=211.9: SE(4)=54.27
 SE(5)=36.76: SE(6)=25.7: SE(7)=16.78: SE(8)=13.21
 SE(9)=13.21: SE(10)=20.26: SE(11)=32.21: SE(12)=45.95
 SE(13)=58.82: SE(14)=75.46: SE(15)=92.72: SE(16)=97.7:
SE(17)=101.1
 TT(1)=27.64: TT(2)=31.97: TT(3)=36.71: TT(4)=82.53
 TT(5)=90: TT(6)=96.03: TT(7)=103.8: TT(8)=110.55
 TT(9)=116.57: TT(10)=125.27: TT(11)=133.48
 TT(12)=140.79: TT(13)=147.21: TT(14)=153.9
 TT(15)=162.14: TT(16)=165.67: TT(17)=166.59: NT=17
 REM *********** FOR P-3HE ON E=11.48 ************
 FP(0)=-88.8:   FPI(0)=1: FP(1)=66.7:    FPI(1)=1
 FP(2)=2.5:     FPI(2)=1: FP(3)=1:       FPI(3)=1
 F0(0)=-88.8:   F0I(0)=1: F0(1)=49.4:    F0I(1)=1
 F0(2)=2.5:     F0I(2)=1: F0(3)=1:       F0I(3)=1
 FM(0)=-88.8:   FMI(0)=1: FM(1)=44.3:    FMI(1)=1
 FM(2)=2.5:     FMI(2)=1: FM(3)=1:       FMI(3)=1
 FS(0)=-84.6:   FSI(0)=1: FS(1)=21.4:    FSI(1)=1
 FS(2)=-18.6:   FSI(2)=1: FS(3)=1:       FSI(3)=1
 REM ********* TRANSFORM TO RADIANS **********
 FOR L=LN TO LV STEP LH
 FM(L)=FM(L)*PI/180: FP(L)=FP(L)*PI/180
 F0(L)=F0(L)*PI/180: FMI(L)=FMI(L)*PI/180
 FPI(L)=FPI(L)*PI/180: F0I(L)=F0I(L)*PI/180
 FT(L)=FT(L)*PI/180: FS(L)=FS(L)*PI/180
 FSI(L)=FSI(L)*PI/180: EP(L)=EXP(-2*FPI(L))
 EM(L)=EXP(-2*FMI(L)): E0(L)=EXP(-2*F0I(L))
 ES(L)=EXP(-2*FSI(L)): NEXT: FH=FH*PI/180
 REM ********************************************
 REM NP=4*NPP+3: REM ПРИ УЧЕТЕ КОМПЛЕКСНЫХ ФАЗ
 NP=2*NPP+1: REM ПРИ УЧЕТЕ ТОЛЬКО ДЕЙСТВИТЕЛЬНЫХ
ФАЗ
 IF NP<>(2*NPP+1) GOTO 9988: FOR L=LN TO LV STEP LH:
FMI(L)=0
 FPI(L)=0: F0I(L)=0: FSI(L)=0: NEXT
9988 FOR I=LMI TO LMA STEP LH: XP(I)=FP(I): NEXT
 FOR I=LMI TO LMA-1 STEP LH: XP(I+LMA+1)=F0(I+1): NEXT
 FOR I=LMI TO LMA-1 STEP LH: XP(I+2*LMA+1)=FM(I+1): NEXT
 FOR I=LMI TO LMA STEP LH: XP(I+3*LMA+1)=FS(I): NEXT
 FOR I=LMI TO LMA STEP LH: XP(I+4*LMA+2)=FPI(I): NEXT
 FOR I=LMI TO LMA-1 STEP LH: XP(I+5*LMA+3)=F0I(I+1): NEXT
```

```
 FOR I=LMI TO LMA-1 STEP LH: XP(I+6*LMA+3)=FMI(I+1):
NEXT
 FOR I=LMI TO LMA STEP LH: XP(I+7*LMA+3)=FSI(I): NEXT
 REM ********** TRANSFORM TO C.M. ***************
 EL=11.48: EC=EL*PM/AM1: SK=EC*B1: SS=SQR(SK)
 GG=3.44476E-02*Z1*Z2*PM/SS
 CALL VAR(ST(),FH,LMA,NI,XP(),EP,XI,NV)
 CLS: PRINT: PRINT "                    XI-KV=";
 PRINT USING " ####.### ";XI
 REM *********** TOTAL CROSSS SECTION ***********
 FOR I=LMI TO LMA STEP LH: FP(I)=XP(I): NEXT
 FOR I=LMI TO LMA-1 STEP LH: F0(I+1)=XP(I+LMA+1): NEXT
 FOR I=LMI TO LMA-1 STEP LH: FM(I+1)=XP(I+2*LMA+1): NEXT
 FOR I=LMI TO LMA STEP LH: FS(I)=XP(I+3*LMA+1): NEXT
 F0(0)=FP(0): FM(0)=FP(0): FOR I=LMI TO LMA STEP LH
 FPI(I)=XP(I+4*LMA+2): NEXT: FOR I=LMI TO LMA-1 STEP LH
 F0I(I+1)=XP(I+5*LMA+3): NEXT: FOR I=LMI TO LMA-1 STEP LH
 FMI(I+1)=XP(I+6*LMA+3): NEXT: FOR I=LMI TO LMA STEP LH
 FSI(I)=XP(I+7*LMA+3): NEXT: F0I(0)=FPI(0): FMI(0)=FPI(0)
 FOR L=LN TO LV STEP LH: EP(L)=EXP(-2*FPI(L))
 EM(L)=EXP(-2*FMI(L))
 E0(L)=EXP(-2*F0I(L)): ES(L)=EXP(-2*FSI(L)): NEXT
 SRT=0: SRS=0:  SST=0:  SSS=0: FOR L=LN TO LV STEP LH
AP=FP(L): AM=FM(L): A0=F0(L): ASS=FS(L): L1=2*L+3
 L2=2*L+1: L3=2*L-1
 SRT = SRT + L1*(1  - EP(L)^2) + L2*(1 - E0(L)^2) + L3*(1 -
EM(L)^2)
 SRS=SRS + L2*(1 - ES(L)^2)
 SST = SST + L1*EP(L)^2*SIN(AP)^2 + L2*E0(L)^2*SIN(A0)^2 +
L3*EM(L)^2*SIN(AM)^2: SSS = SSS + L2*ES(L)^2*SIN(ASS)^2
 NEXT L
 SRT=10*PI*SRT/SK/3: SRS=10*PI*SRS/SK
 SIGR=1/4*SRS+3/4*SRT: SST=10*4*PI*SST/SK/3
 SSS=10*4*PI*SSS/SK: SIGS=1/4*SSS+3/4*SST
 PRINT "                 SIGMR-TOT=";
 PRINT USING " ####.## ";SIGR: PRINT "                    SIGMS-
TOT=";
 PRINT USING " ####.## ";SIGS
 PRINT "   T     SE      ST      XI      T      SE      ST      XI"
 FOR I=1 TO NT/2: PRINT USING " ####.##
";TT(I);SE(I);ST(I);DS(I),TT(I+9);SE(I+9);ST(I+9);DS(I+9): NEXT I
 PRINT
 PRINT "   FP     FPI     F0     F0I     FM     FMI     FS     FSI"
 FOR L=LMI TO LMA STEP LH: FM(L)=FM(L)*180/PI
```

```
 FP(L)=FP(L)*180/PI
 FMI(L)=FMI(L)*180/PI: FPI(L)=FPI(L)*180/PI
 F0(L)=F0(L)*180/PI: F0I(L)=F0I(L)*180/PI
 FS(L)=FS(L)*180/PI: FSI(L)=FSI(L)*180/PI
 PRINT USING "+###.### "; FP(L); FPI(L), F0(L); F0I(L), FM(L);
FMI(L), FS(L); FSI(L): NEXT: GOTO 1111: OPEN "O",1,G$
 PRINT#1, "             P - 3He FOR LAB E=";
 PRINT#1, EC: FOR I=1 TO NP: PRINT#1, USING " #.####^^^^
";TT(I);ST(I): NEXT
1111 END
 SUB VAR(ST(50),PHN,LMA,NI,XP(50),EP,AMIN,NV)
 DIM XPN(50): SHARED LH,LMI,NT,PI,DS(),NP,NPP
 REM ************ ПОИСК МИНИМУМА *************
 FOR I=LMI TO NP STEP LH: XPN(I)=XP(I): NEXT: NN=LMI
 PRINT USING " ### ";NN;
 PRINT USING " +###.#########   ";XPN(NN)*180/PI: PH=PHN
 CALL DET(XPN(),ST(),ALA): B=ALA: IF NV=0 GOTO 3012
 PRINT USING "          +#.#######^^^^ ";ALA
 PRINT "---------------------------------------------------------------"
 REM -----------------------------------------------------------------
 FOR IIN=1 TO NI: NN=-LH: PRINT USING "          +#.#######^^^^
";ALA;IIN
 GOTO 1119
 1159 XPN(NN)=XPN(NN)-PH*XP(NN)
1119  NN=NN+LH: REM IF NN>NP GOTO 3012: IN=0
2229 A=B: XPN(NN)=XPN(NN)+PH*XP(NN)
 IF NP=2*NPP+1 GOTO 7777: IF NN<(NP/2) GOTO 7777
 IF XPN(NN)<0 GOTO 1159
7777 IN=IN+1
 REM ----------------------------------------------------------------
 CALL DET(XPN(),ST(),ALA): B=ALA: GOTO 5678
 PRINT USING " ### ";NN;
 PRINT USING " +###.#########   ";XPN(NN)*180/PI;
 PRINT USING "          +#.#######^^^^ ";ALA;: PRINT
5678 REM -------------------------------------------------------------
  IF B<A GOTO 2229:  C=A:  XPN(NN)=XPN(NN)-PH*XP(NN)
  IF IN>1 GOTO 3339:  PH=-PH:  GOTO 5559
3339 IF ABS((C-B)/(B))<EP GOTO 4449:  PH=PH/2
5559 B=C: GOTO 2229
4449 PH=PHN: B=C: IF NN<NP GOTO 1119
3012  AMIN=B: PH=PHN: NEXT IIN: FOR I=LMI TO NP STEP LH
 XP(I)=XPN(I): NEXT: END SUB
 SUB DET(XP(50),ST(50),XI)
 SHARED SE(),DS(),DE(),NT
```

```
S=0: CALL SEC(XP(),ST()): FOR I=1 TO NT
DS(I)=((ST(I)-SE(I))/0.025/SE(I))^2: S=S+DS(I)
NEXT: XI=S/NT: END SUB
SUB SEC(XP(50),S(50))
SHARED FP(), FPI(), EP(), F0(), FOI(), E0(), FM(), FMI(), EM(),
FS(), FSI(), ES(), T20(), T22(), T21()
SHARED SS,GG,PI,LN,LV,LH,NT,POL(),TT(),NP
DIM S0(20),P(20),P1(20),P2(20)
FOR I=LN TO LV STEP LH: FP(I)=XP(I): NEXT
FOR I=LN TO LV-1 STEP LH: F0(I+1)=XP(I+LV+1): NEXT
FOR I=LN TO LV-1 STEP LH: FM(I+1)=XP(I+2*LV+1): NEXT
FOR I=LN TO LV STEP LH: FS(I)=XP(I+3*LV+1): NEXT
F0(0)=FP(0): FM(0)=FP(0): FOR I=LN TO LV STEP LH
FPI(I)=XP(I+4*LV+2)
NEXT: FOR I=LN TO LV-1 STEP LH: F0I(I+1)=XP(I+5*LV+3)
NEXT: FOR I=LN TO LV-1 STEP LH: FMI(I+1)=XP(I+6*LV+3)
NEXT: FOR I=LN TO LV STEP LH: FSI(I)=XP(I+7*LV+3)
NEXT: F0I(0)=FPI(0): FMI(0)=FPI(0): FOR L=LN TO LV STEP LH
EP(L)=EXP(-2*FPI(L)): EM(L)=EXP(-2*FMI(L))
E0(L)=EXP(-2*F0I(L)): ES(L)=EXP(-2*FSI(L)): NEXT
CALL CULFAZ(GG,S0()): FOR I=1 TO NT STEP 1
T=TT(I)*PI/180: X=COS(T)
CALL CULAMP(X,GG,S0(),RECUL,AMCUL)
CALL POLLEG(X,LV,P()): CALL FUNLEG1(X,LV,P1())
CALL FUNLEG2(X,LV,P2()): RES=0: AMS=0: REA=0: AMA=0
REB=0: AMB=0: REC=0: AMC=0: RED=0: AMD=0: REE=0:
AME=0
FOR L=LN TO LV STEP LH: FP=2*FP(L): FM=2*FM(L):
F0=2*F0(L)
SL=2*S0(L): C=COS(SL): S=SIN(SL): FS=2*FS(L)
ALS=ES(L)*COS(FS)-1: BS=ES(L)*SIN(FS)
RES=RES+(2*L+1)*(BS*C+ALS*S)*P(L)
AMS=AMS+(2*L+1)*(BS*S-ALS*C)*P(L)
AL1P=EP(L)*COS(FP)-1: AL2P=EP(L)*SIN(FP)
AL1M=EM(L)*COS(FM)-1: AL2M=EM(L)*SIN(FM)
AL10=E0(L)*COS(F0)-1: AL20=E0(L)*SIN(F0)
A1=(L+1)*AL1P+L*AL1M: A2=(L+1)*AL2P+L*AL2M
REA=REA+(A2*C+A1*S)*P(L): AMA=AMA+(A2*S-A1*C)*P(L)
B1=(L+2)*AL1P+(2*L+1)*AL10+(L-1)*AL1M
B2=(L+2)*AL2P+(2*L+1)*AL20+(L-1)*AL2M
REB=REB+(B2*C+B1*S)*P(L)/2: AMB=AMB+(B2*S-B1*C)*P(L)/2
IF L<1 GOTO 2111: C1=AL1P-AL1M: C2=AL2P-AL2M
CC1=1/(SQR(2)): REC=REC+(C2*C+C1*S)*P1(L)*CC1
AMC=AMC+(C2*S-C1*C)*P1(L)*CC1: DD1=1/(SQR(2)*L*(L+1))
```

```
D1=L*(L+2)*AL1P-(2*L+1)*AL10-(L^2-1)*AL1M
D2=L*(L+2)*AL2P-(2*L+1)*AL20-(L^2-1)*AL2M
RED=RED+(D2*C+D1*S)*P1(L)*DD1
AMD=AMD+(D2*S-D1*C)*P1(L)*DD1
2111 IF L<2 GOTO 2222: EE1=1/(2*L*(L+1))
E1=L*AL1P-(2*L+1)*AL10+(L+1)*AL1M
E2=L*AL2P-(2*L+1)*AL20+(L+1)*AL2M
REE=REE+(E2*C+E1*S)*P2(L)*EE1
AME=AME+(E2*S-E1*C)*P2(L)*EE1
2222 NEXT L: RES=RECUL+RES: AMS=AMCUL+AMS
SES=10*(RES^2+AMS^2)/4/SS^2
REA=RECUL+REA: AMA=AMCUL+AMA: REB=RECUL+REB
AMB=AMCUL+AMB: AA=REA^2+AMA^2: BB=REB^2+AMB^2
CC=REC^2+AMC^2: DD=RED^2+AMD^2: EE=REE^2+AME^2
SET=10*(AA+2*(BB+CC+DD+EE))/4/SS^2/3:
S(I)=3/4*SET+1/4*SES
POL(TT)=SQR(2/3)*SQR(3/2)*2*SQR(2)/3*(AMA*REC-
REA*AMC+AMB*RED-REB*AMD+AMD*REE-RED*AME)/SEC
T20(TT)=1/SQR(2)*(1-(AA+2*DD)/SEC)
T22(TT)=1/SQR(3)*(2*(REB*REE+AMB*AME)-CC)/SEC
T21(TT)=-SQR(2/3)*(REA*REC+AMA*AMC-REB*RED-
AMB*AMD+RED*REE+AMD*AME)/SEC: NEXT I: END SUB
 SUB CULAMP(X,GG,S0(20),RECUL,AMCUL)
 A=2/(1-X): S0=2*S0(0): BB=-GG*A: AL=GG*LOG(A)+S0
 RECUL=BB*COS(AL):  AMCUL=BB*SIN(AL): END SUB
 SUB POLLEG(X,L,P(20))
 P(0)=1: P(1)=X: FOR I=2 TO L: P(I)=(2*I-1)*X/I*P(I-1)-(I-1)/I*P(I-2)
 NEXT: END SUB
 SUB FUNLEG1(X,L,P(20))
 P(0)=0: P(1)=SQR(ABS(1-X^2)): FOR I=2 TO L
 P(I)=(2*I-1)*X/(I-1)*P(I-1)-I/(I-1)*P(I-2): NEXT: END SUB
 SUB FUNLEG2(X,L,P(20))
 P(0)=0: P(1)=0: P(2)=3*ABS(1-X^2): FOR I=3 TO L
 P(I)=(2*I-1)*X/(I-2)*P(I-1)-(I+1)/(I-2)*P(I-2): NEXT: END SUB
 SUB CULFAZ(G,F(20))
 C=0.577215665: S=0: N=50: A1=1.202056903/3: A2=1.036927755/5
 FOR I=1 TO N: A=G/I-ATN(G/I)-(G/I)^3/3+(G/I)^5/5: S=S+A
 NEXT: FAZ=-C*G+A1*G^3-A2*G^5+S: F(0)=FAZ
 FOR I=1 TO 20: F(I)=F(I-1)+ATN(G/(I)): NEXT: END SUB
```

Приведем результаты контрольного счета по этой программе для рассеяния в системе $p^3$He при энергии 11.48 МэВ, которая рассматривалась в предыдущей главе. Экспериментальные сечения определялись в работе [211], а фазовый анализ выполнен в работе [212] (фазы приводились в шестой главе).

В последней работе для $\chi^2$ приведена величина 0.45, полученная для найденных фаз рассеяния с учетом спин - орбитального взаимодействия и синглет - триплетного смешивания состояний. В предыдущей главе, с этими фазами, мы получили для $\chi^2$ значение 0.74 при учете только спин - орбитального взаимодействия.

Посмотрим теперь насколько можно улучшить величину $\chi^2$, используя только спин - орбитальное расщепление фаз рассеяния. Выполняя варьирование исходных фаз из [212] по нашей программе с одной итерацией, получим

$\chi^2 = 0.316$ $\sigma_s = 1148.53$

| $\theta$ | $\sigma_e$ | $\sigma_t$ | $\chi^2$ | $\theta$ | $\sigma_e$ | $\sigma_t$ | $\chi^2$ |
|---|---|---|---|---|---|---|---|
| 27.64 | 223.10 | 228.88 | 1.07 | 125.27 | 20.26 | 20.12 | 0.08 |
| 31.97 | 222.00 | 222.71 | 0.02 | 133.48 | 32.21 | 32.19 | 0.00 |
| 36.71 | 211.90 | 211.32 | 0.01 | 140.79 | 45.95 | 46.51 | 0.24 |
| 82.53 | 54.27 | 53.82 | 0.11 | 147.21 | 58.82 | 60.64 | 1.53 |
| 90.00 | 36.76 | 36.65 | 0.01 | 153.90 | 75.46 | 75.48 | 0.00 |
| 96.03 | 25.70 | 25.93 | 0.13 | 162.14 | 92.72 | 91.71 | 0.19 |
| 103.80 | 16.78 | 16.66 | 0.08 | 165.67 | 97.70 | 97.37 | 0.02 |
| 110.55 | 13.21 | 13.07 | 0.19 | 166.59 | 101.10 | 98.68 | 0.92 |

| | $\delta^+$ | $\delta^0$ | $\delta^-$ | $\delta_S$ |
|---|---|---|---|---|
| L = 0 | -87.948 | -87.948 | -87.948 | -86.224 |
| L = 1 | +66.540 | +48.926 | +44.300 | +22.838 |
| L = 2 | +3.220 | +2.452 | +2.716 | -18.511 |

Видно, что удается заметно улучшить описание экспериментальных данных даже при не большом изменении исходных фаз рассеяния. Выполним теперь варьирование фаз с 10 итерациями.

$\chi^2 = 0.291$ $\sigma_s = 1141.38$

| $\theta$ | $\sigma_e$ | $\sigma_t$ | $\chi^2$ | $\theta$ | $\sigma_e$ | $\sigma_t$ | $\chi^2$ |
|---|---|---|---|---|---|---|---|
| 27.64 | 223.10 | 226.09 | 0.29 | 125.27 | 20.26 | 20.10 | 0.10 |
| 31.97 | 222.00 | 220.17 | 0.11 | 133.48 | 32.21 | 32.18 | 0.00 |
| 36.71 | 211.90 | 209.07 | 0.29 | 140.79 | 45.95 | 46.52 | 0.24 |
| 82.53 | 54.27 | 53.69 | 0.19 | 147.21 | 58.82 | 60.66 | 1.56 |
| 90.00 | 36.76 | 36.61 | 0.03 | 153.90 | 75.46 | 75.51 | 0.00 |
| 96.03 | 25.70 | 25.93 | 0.12 | 162.14 | 92.72 | 91.78 | 0.16 |
| 103.80 | 16.78 | 16.67 | 0.07 | 165.67 | 97.70 | 97.45 | 0.01 |
| 110.55 | 13.21 | 13.07 | 0.19 | 166.59 | 101.10 | 98.76 | 0.86 |

| | $\delta^+$ | $\delta^0$ | $\delta^-$ | $\delta_S$ |
|---|---|---|---|---|
| L = 0 | -87.948 | -87.948 | -87.948 | -86.224 |
| L = 1 | +65.900 | +48.503 | +44.087 | +22.838 |
| L = 2 | +2.602 | +2.788 | +2.896 | -18.511 |

И в этом случае небольшое изменение фаз приводит к уменьшению величины $\chi^2$.

## 7.4 Частицы с полуцелым спином и синглет - триплетным смешиванием

Система с полуцелым спином и триплет - синглетным смешиванием была рассмотрена в предыдущей главе, и здесь мы приведем только программу фазового анализа с учетом синглет - триплетного смешивания состояний [234].

```
REM *** ПРОГРАММА ФАЗОВОГО АНАЛИЗА ДЛЯ 3He-P
11.48 ****
CLS:DEFDBL A-Z:DEFINT I,L,J,N,M,K
 DIM SE(50), ST(50), DS(50), FP(50), F0(50), FM(50), FS(50), FT(50),
TT(50)
 DIM  XP(50), DE(50), POL(50), FPI(50), F0I(50), FMI(50), FSI(50),
EPS(50)
 REM ************* НАЧАЛЬНЫЕ ЗНАЧЕНИЯ *************
 G$="C:\BASICA\FAZ-ALN.DAT": PI=4*ATN(1.): Z1=1: Z2=2
 AM1=1: AM2=3: AM=AM1+AM2: P1=3.14159265: A1=41.4686
 PM=AM1*AM2/(AM1+AM2): B1=2*PM/A1: LMI=0: LH=1: LMA=2
 LN=LMI: LV=LMA: EPP=1.0D-05: NV=1: FH=.1: NI=10:
NPP=2*LMA
 REM ******** НАЧАЛЬНЫЕ ПАРАМЕТРЫ АЛЬФА *********
 REM ******** ECSPERIMENTAL CROSS SECTION 11.48 *****
 SE(1)=223.1: SE(2)=222: SE(3)=211.9: SE(4)=54.27: SE(5)=36.76
 SE(6)=25.7: SE(7)=16.78: SE(8)=13.21: SE(9)=13.21: SE(10)=20.26
 SE(11)=32.21: SE(12)=45.95: SE(13)=58.82: SE(14)=75.46
 SE(15)=92.72: SE(16)=97.7: SE(17)=101.1
 TT(1)=27.64: TT(2)=31.97: TT(3)=36.71: TT(4)=82.53
 TT(5)=90: TT(6)=96.03: TT(7)=103.8: TT(8)=110.55
 TT(9)=116.57: TT(10)=125.27: TT(11)=133.48
 TT(12)=140.79: TT(13)=147.21: TT(14)=153.9
 TT(15)=162.14: TT(16)=165.67: TT(17)=166.59: NT=17
 REM ************* FOR P-3HE ON E=11.48 ****************
 FP(0)=-88.8:   FPI(0)=1: FP(1)=66.7:    FPI(1)=1
 FP(2)=3.0:     FPI(2)=1: FP(3)=.0:      FPI(3)=1
 F0(0)=FP(0):   F0I(0)=1: F0(1)=49.4:    F0I(1)=1
```

```
 F0(2)=2.5:       F0I(2)=1: F0(3)=.0:        F0I(3)=1
 FM(0)=FP(0):  FMI(0)=1: FM(1)=44.3:    FMI(1)=1
 FM(2)=2.:        FMI(2)=1: FM(3)=.0:       FMI(3)=1
 FS(0)=-84.6:   FSI(0)=1: FS(1)=21.4:    FSI(1)=1
 FS(2)=-18.6:   FSI(2)=1: FS(3)=.0:        FSI(3)=1: EPS(1)=-11.2
 REM ************ TRANSFORM TO RADIANS *************
 FOR L=LN TO LV STEP LH
 FM(L)=FM(L)*PI/180: FP(L)=FP(L)*PI/180
 F0(L)=F0(L)*PI/180: EPS(L)=EPS(L)*PI/180
 FMI(L)=FMI(L)*PI/180: FPI(L)=FPI(L)*PI/180
 F0I(L)=F0I(L)*PI/180: FS(L)=FS(L)*PI/180
 FSI(L)=FSI(L)*PI/180: EP(L)=EXP(-2*FPI(L))
 EM(L)=EXP(-2*FMI(L)): E0(L)=EXP(-2*F0I(L))
 ES(L)=EXP(-2*FSI(L)): NEXT
 REM ****************************************************
 FH=FH*PI/180: NP=9*LMA+4: REM Для комплексных фаз рассея-
ния
 NP=5*LMA+2: REM Для действительных  фаз рассеяния
 IF NP>(5*LMA+2) GOTO 9988: FOR L=LN TO LV STEP LH
 FMI(L)=0: FPI(L)=0: F0I(L)=0: FSI(L)=0: NEXT
9988 FOR I=LMI TO LMA STEP LH: XP(I)=FP(I): NEXT
 FOR I=LMI TO LMA-1 STEP LH: XP(I+LMA+1)=F0(I+1)
 NEXT: FOR I=LMI TO LMA-1 STEP LH: XP(I+2*LMA+1)=FM(I+1)
 NEXT: FOR I=LMI TO LMA STEP LH: XP(I+3*LMA+1)=FS(I)
 NEXT: FOR I=LMI TO LMA STEP LH: XP(I+4*LMA+2)=EPS(I)
 NEXT: FOR I=LMI TO LMA STEP LH: XP(I+5*LMA+3)=FPI(I)
 NEXT: FOR I=LMI TO LMA-1 STEP LH: XP(I+6*LMA+4)=F0I(I+1)
 NEXT:     FOR     I=LMI     TO     LMA-1     STEP     LH:
XP(I+7*LMA+4)=FMI(I+1)
 NEXT:  FOR  I=LMI  TO  LMA  STEP  LH:  XP(I+8*LMA+4)=FSI(I):
NEXT
 REM   ************** TRANSFORM TO C.M. *************
 EL=11.48: EC=EL*PM/AM1
 SK=EC*B1: SS=SQR(SK): GG=3.44476E-02*Z1*Z2*PM/SS
 CALL VAR(ST(),FH,LMA,NI,XP(),EPP,XI,NV): CLS
 PRINT: PRINT "                    XI-KV=";: PRINT USING " ####.###
";XI
 REM ************* TOTAL CROSSS SECTION *************
 FOR I=LMI TO LMA STEP LH: FP(I)=XP(I): NEXT
 FOR I=LMI TO LMA-1 STEP LH: F0(I+1)=XP(I+LMA+1): NEXT
 FOR I=LMI TO LMA-1 STEP LH: FM(I+1)=XP(I+2*LMA+1): NEXT
 FOR I=LMI TO LMA STEP LH: FS(I)=XP(I+3*LMA+1): NEXT
 FOR I=LMI TO LMA STEP LH: EPS(I)=XP(I+4*LMA+2): NEXT
 F0(0)=FP(0): FM(0)=FP(0): FOR I=LMI TO LMA STEP LH
```

```
FPI(I)=XP(I+5*LMA+3): NEXT: FOR I=LMI TO LMA-1 STEP LH
F0I(I+1)=XP(I+6*LMA+4): NEXT: FOR I=LMI TO LMA-1 STEP LH
FMI(I+1)=XP(I+7*LMA+4): NEXT: FOR I=LMI TO LMA STEP LH
FSI(I)=XP(I+8*LMA+4): NEXT: F0I(0)=FPI(0): FMI(0)=FPI(0)
FOR L=LN TO LV STEP LH: EP(L)=EXP(-2*FPI(L))
EM(L)=EXP(-2*FMI(L))
E0(L)=EXP(-2*F0I(L)):      ET(L)=EXP(-2*FTI(L)):      ES(L)=EXP(-
2*FSI(L))
NEXT: SRT=0: SRS=0:   SST=0:  SSS=0: FOR L=LN TO LV STEP
LH
AP=FP(L): AM=FM(L): A0=F0(L): ASS=FS(L)
L1=2*L+3: L2=2*L+1: L3=2*L-1
SRT=SRT+L1*(1-EP(L)^2)+L2*(1-E0(L)^2)+L3*(1-EM(L)^2)
SRS=SRS+L2*(1-ES(L)^2)
SST=SST+L1*EP(L)^2*SIN(AP)^2+L2*E0(L)^2*SIN(A0)^2+L3*EM(
L)^2*SIN(AM)^2: SSS=SSS+L2*ES(L)^2*SIN(ASS)^2: NEXT L
SRT=10*PI*SRT/SK/3: SRS=10*PI*SRS/SK
SIGR=1/4*SRS+3/4*SRT: SST=10*4*PI*SST/SK/3
SSS=10*4*PI*SSS/SK: SIGS=1/4*SSS+3/4*SST
PRINT "                    SIGMR-TOT=";: PRINT USING " ####.##
";SIGR
PRINT "                    SIGMS-TOT=";: PRINT USING " ####.##
";SIGS
PRINT "   T    SE     ST     XI     T     SE     ST     XI"
FOR I=1 TO NT/2
PRINT  USING  "  ####.##  ";TT(I);  SE(I);  ST(I);  DS(I),  TT(I+8);
SE(I+8); ST(I+8); DS(I+8): NEXT I: PRINT "     L      FP      F0
FM    FS    EPS"
FOR L=LMI TO LMA STEP LH
FM(L)=FM(L)*180/PI: FP(L)=FP(L)*180/PI
FMI(L)=FMI(L)*180/PI: FPI(L)=FPI(L)*180/PI
F0(L)=F0(L)*180/PI: F0I(L)=F0I(L)*180/PI
FS(L)=FS(L)*180/PI: FSI(L)=FSI(L)*180/PI
EPS(L)=EPS(L)*180/PI
PRINT USING "+###.### ";L;FP(L);F0(L);FM(L);FS(L);EPS(L)
PRINT USING "+###.### ";L;FPI(L);F0I(L);FMI(L);FSI(L): NEXT
INPUT A: IF A=0 GOTO 1111: CLS: PRINT "      T       POL       T
POL "
FOR   I=1   TO   NT/2:   PRINT   USING   "   ####.###
";TT(I);POL(I);TT(I+8);POL(I+8)
NEXT I: GOTO 1111: OPEN "O",1,G$
PRINT#1, "           P - 3He FOR LAB E=";
PRINT#1, EC: FOR T=TMI TO TMA STEP TH
PRINT#1, USING " #.####^^^^ ";T;ST(T): NEXT
```

```
1111 END
 SUB VAR(ST(50),PHN,LMA,NI,XP(50),EP,AMIN,NV)
 DIM XPN(50): SHARED LH,LMI,NT,PI,DS(),NP,NPP
 REM ************* ПОИСК МИНИМУМА *****************
 FOR I=LMI TO NP STEP LH: XPN(I)=XP(I): NEXT
 NN=LMI: PRINT USING " ### ";NN;
 PRINT USING " +###.#########   ";XPN(NN)*180/PI
 PH=PHN: CALL DET(XPN(),ST(),ALA): B=ALA: IF NV=0 GOTO
3012
 PRINT USING "         +#.#######^^^^ ";ALA
 PRINT "--------------------------------------------------------------"
 REM --------------------------------------------------------------
 FOR IIN=1 TO NI: NN=-LH: PRINT USING "          +#.#######^^^^
";ALA;IIN
 GOTO 1119
 1159 XPN(NN)=XPN(NN)-PH*XP(NN)
1119  NN=NN+LH: IN=0
2229 A=B: XPN(NN)=XPN(NN)+PH*XP(NN)
 IF NN<(5*LMA+3) GOTO 7777
 IF XPN(NN)<0 GOTO 1159
7777 IN=IN+1
 REM --------------------------------------------------------------
 CALL DET(XPN(),ST(),ALA): B=ALA: GOTO 5678
 PRINT USING " ### ";NN;
 PRINT USING " +###.#########   ";XPN(NN)*180/PI;
 PRINT USING "         +#.#######^^^^ ";ALA;: PRINT
5678 REM --------------------------------------------------------------
 IF B<A GOTO 2229: C=A:  XPN(NN)=XPN(NN)-PH*XP(NN)
 IF IN>1 GOTO 3339: PH=-PH:  GOTO 5559
3339 IF ABS((C-B)/ABS(B))<EP GOTO 4449: PH=PH/2
5559 B=C: GOTO 2229
4449 PH=PHN: B=C: IF NN<NP GOTO 1119
3012  AMIN=B: PH=PHN: NEXT IIN: FOR I=LMI TO NP STEP LH
 XP(I)=XPN(I): NEXT: END SUB
 SUB DET(XP(50),ST(50),XI)
 SHARED SE(),DS(),DE(),NT
 S=0: CALL SEC(XP(),ST()): FOR I=1 TO NT
 DS(I)=((ST(I)-SE(I))/0.025/SE(I))^2: S=S+DS(I): NEXT
 XI=S/NT: END SUB
 SUB SEC(XP(50),S(50))
 SHARED SS,GG,PI,LN,LV,LH,NT,POL(),TT()
 DIM S0(20),P(20),P1(20),P2(20)
 FOR I=LN TO LV STEP LH: FP(I)=XP(I): NEXT
 FOR I=LN TO LV-1 STEP LH: F0(I+1)=XP(I+LV+1): NEXT
```

```
FOR I=LN TO LV-1 STEP LH: FM(I+1)=XP(I+2*LV+1): NEXT
FOR I=LN TO LV STEP LH: FS(I)=XP(I+3*LV+1): NEXT
FOR I=LN TO LV STEP LH: EPS(I)=XP(I+4*LV+2): NEXT
F0(0)=FP(0): FM(0)=FP(0): FOR I=LN TO LV STEP LH
FPI(I)=XP(I+5*LV+3): NEXT: FOR I=LN TO LV-1 STEP LH
F0I(I+1)=XP(I+6*LV+4): NEXT: FOR I=LN TO LV-1 STEP LH
FMI(I+1)=XP(I+7*LV+4): NEXT: FOR I=LN TO LV STEP LH
FSI(I)=XP(I+8*LV+4): NEXT: F0I(0)=FPI(0): FMI(0)=FPI(0)
FOR L=LN TO LV STEP LH? EP(L)=EXP(-2*FPI(L))
EM(L)=EXP(-2*FMI(L)): E0(L)=EXP(-2*F0I(L))
ES(L)=EXP(-2*FSI(L)): NEXT: CALL CULFAZ(GG,S0())
FOR I=1 TO NT STEP 1: T=TT(I)*PI/180: X=COS(T)
CALL CULAMP(X,GG,S0(),RECUL,AMCUL)
CALL POLLEG(X,LV,P()): CALL FUNLEG1(X,LV,P1())
CALL FUNLEG2(X,LV,P2()): REA=0: AMA=0: REB=0: AMB=0
REC=0: AMC=0: RED=0: AMD=0: REE=0
AME=0: RRG=0: AAG=0: REH=0: AMH=0: REF=0: AMF=0
FOR L=LN TO LV STEP LH: FP=2*FP(L): FM=2*FM(L): F0=2*F0(L)
SL=2*S0(L): C=COS(SL): S=SIN(SL): FS=2*FS(L)
SO=SIN(EPS(L))^2: CO=COS(EPS(L))^2
AL1P=EP(L)*COS(FP)-1: AL2P=EP(L)*SIN(FP)
AL1M=EM(L)*COS(FM)-1: AL2M=EM(L)*SIN(FM)
AL10=SO*ES(L)*COS(FS)+CO*E0(L)*COS(F0)-1
AL20=SO*ES(L)*SIN(FS)+CO*E0(L)*SIN(F0)
A1=(L+2)*AL1P+(2*L+1)*AL10+(L-1)*AL1M
A2=(L+2)*AL2P+(2*L+1)*AL20+(L-1)*AL2M
REA=REA+(A1*C-A2*S)*P(L)/2: AMA=AMA+(A1*S+A2*C)*P(L)/2
ALS=CO*ES(L)*COS(FS)+SO*E0(L)*COS(F0)-1
BS=CO*ES(L)*SIN(FS)+SO*E0(L)*SIN(F0)
RES=(2*L+1)*(ALS*C-BS*S): AMS=(2*L+1)*(ALS*S+BS*C)
B1=(L+1)*AL1P+L*AL1M: B2=(L+1)*AL2P+L*AL2M
REB=REB+(B1*C-B2*S+RES)*P(L)/2
AMB=AMB+(B1*S+B2*C+AMS)*P(L)/2
REC=REC+(B1*C-B2*S-RES)*P(L)/2
AMC=AMC+(B1*S+B2*C-AMS)*P(L)/2: IF L<1 GOTO 1211
SI2=1/2*SIN(2*EPS(L)): AL1=SI2*(ES(L)*COS(FS)-E0(L)*COS(F0))
AL2=SI2*(ES(L)*SIN(FS)-E0(L)*SIN(F0))
RE1=(2*L+1)*(AL2*C+AL1*S)/SQR(L*(L+1))
AM1=(2*L+1)*(AL2*S-AL1*C)/SQR(L*(L+1))
C1=AL1P-AL1M: C2=AL2P-AL2M
RED=RED+(C2*C+C1*S-RE1)*P1(L)/2
```

```
 AMD=AMD+(C2*S-C1*C-AM1)*P1(L)/2
 REE=REE+(C2*C+C1*S+RE1)*P1(L)/2
 AME=AME+(C2*S-C1*C+AM1)*P1(L)/2
 D1=(L+2)/(L+1)*AL1P-(2*L+1)/(L*(L+1))*AL10-(L-1)/L*AL1M
 D2=(L+2)/(L+1)*AL2P-(2*L+1)/(L*(L+1))*AL20-(L-1)/L*AL2M
 RRG=RRG+(D2*C+D1*S-RE1)*P1(L)/2
 AAG=AAG+(D2*S-D1*C-AM1)*P1(L)/2
 REH=REH+(D2*C+D1*S+RE1)*P1(L)/2
 AMH=AMH+(D2*S-D1*C+AM1)*P1(L)/2
1211 IF L<2 GOTO 2122
 F1=1/(L+1)*AL1P-(2*L+1)/(L*(L+1))*AL10+AL1M/L
 F2=1/(L+1)*AL2P-(2*L+1)/(L*(L+1))*AL20+AL2M/L
 REF=REF+(F2*C+F1*S)*P2(L)/2
 AMF=AMF+(F2*S-F1*C)*P2(L)/2
2122 NEXT L: REA=RECUL+REA: AMA=AMCUL+AMA
 REB=RECUL+REB: AMB=AMCUL+AMB
 AA=REA^2+AMA^2: BB=REB^2+AMB^2
 CC=REC^2+AMC^2: DD=RED^2+AMD^2
 EE=REE^2+AME^2: FF=REF^2+AMF^2
 HH=REH^2+AMH^2: GGG=RRG^2+AAG^2
 SUM=AA+BB+CC+DD+EE+GGG+HH+FF
 S(I)=10*SUM/2/SS^2/4: POL(I)= - 2*(REA*REE + AMA*AME +
REB*REH + AMB*AMH + REC*RRG + AMC*AAG + RED*REF +
AMD*AMF)/SUM
 NEXT I: END SUB
 SUB CULAMP(X,GG,S0(20),RECUL,AMCUL)
 A=2/(1-X): S0=2*S0(0): BB=-GG*A: AL=GG*LOG(A)+S0
 RECUL=-BB*SIN(AL): AMCUL=BB*COS(AL): END SUB
 SUB POLLEG(X,L,P(20))
 P(0)=1: P(1)=X: FOR I=2 TO L
 P(I)=(2*I-1)*X/I*P(I-1)-(I-1)/I*P(I-2): NEXT: END SUB
 SUB FUNLEG1(X,L,P(20))
 P(0)=0: P(1)=SQR(ABS(1-X^2)): FOR I=2 TO L
 P(I)=(2*I-1)*X/(I-1)*P(I-1)-I/(I-1)*P(I-2): NEXT: END SUB
 SUB FUNLEG2(X,L,P(20))
 P(0)=0: P(1)=0: P(2)=3*ABS(1-X^2)
 FOR I=3 TO L: P(I)=(2*I-1)*X/(I-2)*P(I-1)-(I+1)/(I-2)*P(I-2): NEXT
 END SUB
 SUB CULFAZ(G,F(20))
 C=0.577215665: S=0: N=50: A1=1.202056903/3: A2=1.036927755/5
 FOR I=1 TO N: A=G/I-ATN(G/I)-(G/I)^3/3+(G/I)^5/5: S=S+A: NEXT
 FAZ=-C*G+A1*G^3-A2*G^5+S: F(0)=FAZ
 FOR I=1 TO 20: F(I)=F(I-1)+ATN(G/(I)): NEXT: END SUB
```

Приведем результаты контрольного счета по этой программе для рассеяния в системе $p^3He$ при энергии 11.48 МэВ. Экспериментальные сечения определялись в работе [211], а фазовый анализ выполнен в работе [212]. В последней работе для $\chi^2$ приведена величина 0.45, полученная для найденных фаз рассеяния с учетом спин - орбитального взаимодействия и синглет - триплетного смешивания состояний. Для параметра смешивания было получено $11.2^0$.

По нашей программе [234] с такими фазами и триплет - синглетным смешиванием можно получить следующие результаты (они приведены в предыдущей главе, но для большей наглядности мы приводим из снова)

$\chi^2 = 0.294$ $\sigma_s = 1146.06$

| $\theta$ | $\sigma_e$ | $\sigma_t$ | $\chi^2$ | $\theta$ | $\sigma_e$ | $\sigma_t$ | $\chi^2$ |
|---|---|---|---|---|---|---|---|
| 27.64 | 223.10 | 228.04 | 0.78 | 116.57 | 13.21 | 13.39 | 0.31 |
| 31.97 | 222.00 | 221.73 | 0.00 | 125.27 | 20.26 | 19.97 | 0.32 |
| 36.71 | 211.90 | 210.38 | 0.08 | 133.48 | 32.21 | 32.07 | 0.03 |
| 82.53 | 54.27 | 54.22 | 0.00 | 140.79 | 45.95 | 46.48 | 0.21 |
| 90.00 | 36.76 | 36.97 | 0.05 | 147.21 | 58.82 | 60.70 | 1.64 |
| 96.03 | 25.70 | 26.15 | 0.49 | 153.90 | 75.46 | 75.66 | 0.01 |
| 103.80 | 16.78 | 16.74 | 0.01 | 162.14 | 92.72 | 92.03 | 0.09 |
| 110.55 | 13.21 | 13.03 | 0.29 | 165.67 | 97.70 | 97.73 | 0.00 |

| | $\delta^+$ | $\delta^0$ | $\delta^-$ | $\delta_S$ | $\varepsilon$ |
|---|---|---|---|---|---|
| L = 0 | -88.800 | -88.800 | -88.800 | -84.600 | +0.000 |
| L = 1 | +66.700 | +49.400 | +44.300 | +21.400 | -11.200 |
| L = 2 | +2.500 | +2.500 | +2.500 | -18.600 | +0.000 |

Найденное значение $\chi^2$ меньше, приведенной в работе [212] величины 0.45, поскольку мы использовали среднее значение экспериментальных ошибок, приняв их равными 2.5%, а реально, некоторые их них доходят до 2.2%, увеличивая, тем самым, среднюю величину $\chi^2$. Если принять экспериментальную ошибку равной 2.2%, то для величины $\chi^2$ получается 0.6 и среднее между этими значениями дает нам 0.45, в полном соответствии с результатами работы [212].

Посмотрим теперь насколько можно улучшить величину $\chi^2$, используя спин - орбитальное расщепление фаз рассеяния и синглет - триплетное смешивание. Выполняя варьирование исходных фаз из [212] по нашей программе с одной итерацией, получим

$\chi^2 = 0.276$ $\sigma_s = 1141.87$

| $\theta$ | $\sigma_e$ | $\sigma_t$ | $\chi^2$ | $\theta$ | $\sigma_e$ | $\sigma_t$ | $\chi^2$ |
|---|---|---|---|---|---|---|---|
| 27.64 | 223.10 | 226.64 | 0.40 | 116.57 | 13.21 | 13.42 | 0.39 |
| 31.97 | 222.00 | 220.49 | 0.07 | 125.27 | 20.26 | 19.99 | 0.29 |
| 36.71 | 211.90 | 209.27 | 0.25 | 133.48 | 32.21 | 32.06 | 0.03 |
| 82.53 | 54.27 | 53.95 | 0.06 | 140.79 | 45.95 | 46.43 | 0.17 |
| 90.00 | 36.76 | 36.80 | 0.00 | 147.21 | 58.82 | 60.60 | 1.47 |
| 96.03 | 25.70 | 26.06 | 0.31 | 153.90 | 75.46 | 75.51 | 0.00 |
| 103.80 | 16.78 | 16.71 | 0.03 | 162.14 | 92.72 | 91.83 | 0.15 |
| 110.55 | 13.21 | 13.04 | 0.27 | 165.67 | 97.70 | 97.52 | 0.01 |

| . | $\delta^+$ | $\delta^0$ | $\delta^-$ | $\delta_S$ | $\varepsilon$ |
|---|---|---|---|---|---|
| L = 0 | -88.800 | -88.800 | -88.800 | -84.600 | +0.000 |
| L = 1 | +66.060 | +49.357 | +44.300 | +21.437 | -11.317 |
| L = 2 | +2.548 | +2.478 | +2.500 | -18.600 | +0.000 |

Видно, что удается несколько улучшить описание экспериментальных данных при небольшом изменении исходных фаз рассеяния. Выполним теперь варьирование исходных фаз с 10 итерациями

$\chi^2 = 0.274$ $\sigma_s = 1142.47$

| $\theta$ | $\sigma_e$ | $\sigma_t$ | $\chi^2$ | $\theta$ | $\sigma_e$ | $\sigma_t$ | $\chi^2$ |
|---|---|---|---|---|---|---|---|
| 27.64 | 223.10 | 226.61 | 0.40 | 116.57 | 13.21 | 13.41 | 0.38 |
| 31.97 | 222.00 | 220.52 | 0.07 | 125.27 | 20.26 | 19.98 | 0.31 |
| 36.71 | 211.90 | 209.32 | 0.24 | 133.48 | 32.21 | 32.05 | 0.04 |
| 82.53 | 54.27 | 53.95 | 0.06 | 140.79 | 45.95 | 46.42 | 0.17 |
| 90.00 | 36.76 | 36.81 | 0.00 | 147.21 | 58.82 | 60.61 | 1.47 |
| 96.03 | 25.70 | 26.06 | 0.32 | 153.90 | 75.46 | 75.52 | 0.00 |
| 103.80 | 16.78 | 16.72 | 0.02 | 162.14 | 92.72 | 91.86 | 0.14 |
| 110.55 | 13.21 | 13.04 | 0.26 | 165.67 | 97.70 | 97.55 | 0.00 |

| | $\delta^+$ | $\delta^0$ | $\delta^-$ | $\delta_S$ | $\varepsilon$ |
|---|---|---|---|---|---|
| L = 0 | -88.800 | -88.800 | -88.800 | -84.563 | +0.000 |
| L = 1 | +65.900 | +49.357 | +44.455 | +21.736 | -12.080 |
| L = 2 | +2.618 | +2.504 | +2.579 | -18.632 | +0.000 |

Из приведенных результатов видно, что дополнительное варьирование фаз позволяет выполнять минимизацию функционала $\chi^2$ в поле нескольких параметров - фаз ядерного рассеяния и, найдя другие варианты для фаз, получить несколько лучшее согласие расчетных дифференциальных сечений с экспериментальными данными.

Таким образом, были протестированы все программы для фазового анализа и получено хорошее согласие с более ранними ре-зультатами других авторов.

На их основе получены новые физические результаты по фазовому анализу в упругом рассеянии $^4$He$^4$He при энергиях 49.9 МэВ и некоторых других энергиях, которые хорошо описывают экспериментальные сечения упругого рассеяния и в целом согласуются с общим ходом фаз в этой области энергий.

# 8. МАТЕМАТИЧЕСКИЕ МЕТОДЫ РАСЧЕТА СЕЧЕНИЙ ФОТОЯДЕРНЫХ ПРОЦЕССОВ

В настоящей главе изложены математические и численные методы, используемые для решения задач ядерных фотопроцессов. В этом случае решается одно уравнение Шредингера, на основе этого решения находятся волновые функции системы и вычисляются матричные элементы фотопроцессов, которые определяют полные сечения фотоядерных реакций.

## 8.1 Векторные соотношения

Двухкластерная модель предполагает наличие только двух обособленных фрагментов - кластеров, между которыми перераспределены все нуклоны ядра. Первый кластер содержит $M_1$ нуклонов с зарядом $Z_1$, второй $M_2$ с зарядом $Z_2$. Векторная схема кластерной модели приведена на рисунке 8.1. Межкластерное расстояние R определяет относительное положение центров масс фрагментов. Радиусы $\rho_i$ и $\rho_j$ задают положение каждого нуклона в первом и втором кластерах относительно их центров масс. Радиусы $r_i$ и $r_j$ указывают положение каждого нуклона в обоих кластерах относительно общего центра масс ядра. Векторы $R_1$ и $R_2$ определяют положение центров масс кластеров относительно их общего центра масс.

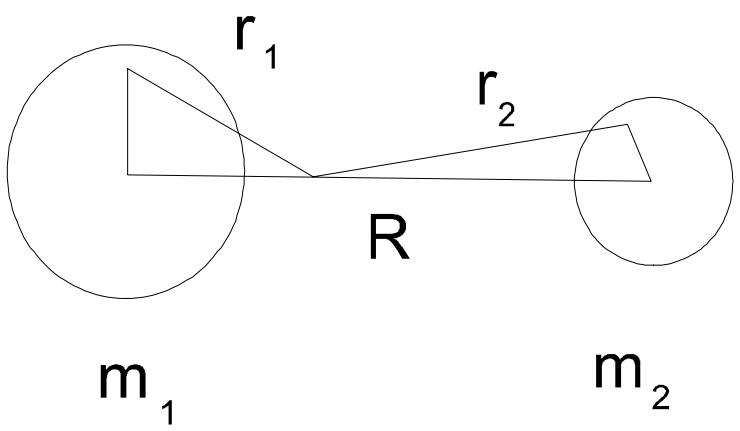

Рисунок 8.1 - Векторная схема кластерной модели.

При таком определении радиус - векторов, между ними существуют простые соотношения:

$$\Sigma\, r_k = \Sigma\, r_i + \Sigma\, r_j = 0\ ,\ \Sigma\, \rho_i = \Sigma\, \rho_j = 0\ ,$$

$$r_i = R_1 + \rho_i = RM_2/M + \rho_i, \qquad r_j = R_2 + \rho_j + RM_1/M - \rho_j,$$

$$R = R_1 - R_2\ ,$$

$$R_1 = 1/M\ \sum_i r_i = \frac{M_2}{M} R\ ,$$

$$R_2 = 1/M\ \sum_j r_j = -\frac{M_1}{M} R\ , \tag{8.1}$$

$$1 < i < M_1\ , \qquad M_1 + 1 < j < M\ , \qquad 1 < k < M\ ,$$

$M = M_1 + M_2$ , $Z = Z_1 + Z_2$ , $\mu = M_1M_2/M$ .

Эти векторные соотношения будут использоваться в дальнейшем для вычисления различных ядерных характеристик в двухкластерной модели.

Рассмотрим, например, вывод формулы для среднеквадратичного зарядового радиуса ядра, который определяется следующим образом

$$\langle r^2 \rangle = \langle \Psi | r^2 | \Psi \rangle \cdot$$

В кластерной модели квадрат радиус - вектора может быть представлен в виде

$$r^2 = 1/M \sum_k r_k^2 \cdot$$

Таким же образом определим зарядовые радиусы кластеров

$$\langle r^2 \rangle_{1,2} = \langle \Psi(1,2) | \frac{1}{M_{1,2}} \sum_n \rho_n^2(1,2) | \Psi(1,2) \rangle ,$$

где 1,2 - первый или второй кластер, а индекс n определяет суммирование по i или j. Используя теперь выражения (8.1), устанавливающие связь между межкластерным расстоянием и векторами $r_k$

$$r^2 = 1/M \, \Sigma \, r_i^2 + 1/M \, \Sigma \, r_j^2 = 1/M \, \Sigma \, \rho_i^2 + 1/M \, \Sigma \, \rho_j^2 + \mu/M \, R^2,$$

для радиуса ядра в кластерной модели с волновыми функциями (2.3) получим окончательное выражение

$$R_r^2 = \frac{M_1}{M} \langle r^2 \rangle_1 + \frac{M_2}{M} \langle r^2 \rangle_2 + \frac{M_1 M_2}{M^2} I_2 , \qquad (8.2)$$

где

$$I_2 = \langle \Phi(R) | R^2 | \Phi(R) \rangle$$

матричный элемент по радиальным волновым функциям относительного движения кластеров от квадрата межкластерного расстояния.

Таким образом, радиус ядра в кластерной модели может быть легко выражен через радиусы кластеров и эффективное межкластерное расстояние. Аналогичным образом можно использовать векторные соотношения кластерной модели при выводе формул для формфакторов, квадрупольных, магнитных моментов ядер, матричных элементов ядерных реакций, в частности процессов фоторазвала или радиационного захвата ассоциаций и т.д.

Рассмотрим далее методы вычисления полных сечений ядерных фотопроцессов, а также характеристик связанных состояний кластеров в ядре для чисто центральных межкластерных потенциалов. Затем перейдем к учету тех эффектов, которые дают тензорные взаимодействия в двухчастичной системе, и приведем некоторые основные формулы для рассмотрения сечений рассеяния и реакций в супермультиплетном приближении, которое используется для анализа взаимодействий легчайших кластерных систем.

## 8.2 Фоторазвал и радиационный захват

Одной из самых интересных ядерных реакций является процесс ядерного фоторазвала или обратная ему реакция - радиационного захвата. Налетающая частица - фотон не вступает в сильные ядерные взаимодействия с ядром мишенью. Происходит только электромагнитное взаимодействие, операторы которого точно известны. Поэтому можно учитывать только ядерные взаимодействия связанных кластеров, что существенно упрощает рассмотрение по сравнению с трехтельной задачей, когда, наряду с межкластерными силами, нужно включать и ядерное взаимодействие налетающей частицы [235,236, 237,238,239,240,241,242,243].

Общие методы расчета сечений подобных процессов подробно изложены в прекрасной монографии [244]. Поэтому далее будем исходить из уже известных определений дифференциальных сечений радиационных и фотоядерных процессов. Для расчетов сечений радиационного захвата в длинноволновом приближении будем использовать известное выражение [244,245]

$$\frac{d\sigma_c(N)}{d\Omega} = \frac{K\mu}{2\pi\ \hbar^2 q}\ \frac{1}{(2S_1+1)(2S_2+1)} \sum_{m_i,m_f,\lambda} |M_{J\lambda}(N)|^2 , \tag{8.3}$$

где N = E - электрические или M - магнитные переходы и

$$M_{J\lambda}(N) = \sum_J i^J \sqrt{2\pi\ (2J+1)}\ \frac{K^J}{(2J+1)!!}\left[\frac{J+1}{J}\right]^{1/2} \times$$

$$\lambda \sum_m D^J_{m\lambda} \langle f \left| H_{Jm}(N) \right| i \rangle,$$

$$H_{Jm}(E) = Q_{Jm}(L) + Q_{Jm}(S),$$

$$H_{Jm}(M) = W_{Jm}(L) + W_{Jm}(S) ,$$

$$Q_{Jm}(L) = e \sum_i Z_i r_i^J Y_{Jm}(\Omega_i) ,$$

$$Q_{Jm}(S) = -\frac{e\hbar}{m_0 c} K \left[\frac{J}{J+1}\right]^{1/2} \sum_i \mu_i \hat{S}_i r_i^J Y_{Jm}(\Omega_i) ,$$

$$W_{Jm}(L) = i \frac{e\hbar}{m_0 c} \frac{1}{(J+1)} \sum_i \frac{Z_i}{M_i} \hat{L}_i \nabla_i (r_i^J Y_{Jm}(\Omega_i)) ,$$

$$W_{Jm}(S) = i \frac{e\hbar}{m_0 c} \sum_i \mu_i \hat{S}_i \nabla_i (r_i^J Y_{Jm}(\Omega_i)) .$$

Здесь J - мультипольность, q - волновое число относительного движения кластеров, $D^J_{m\lambda}$ - функция Вигнера, μ - приведенная масса, $M_i$, $Z_i$, $S_i$ и $S_i$ - массы, заряды, спины и орбитальные моменты i - го кластера, $\mu_i$ - магнитные моменты кластеров, K - волновое число фотона, $m_0$ - масса нуклона.

Знак оператора $Q_{Jm}(S)$ выбран отрицательным, как приведено в работе [114]. Интегрируя по углам и суммируя это выражение по λ, для полного сечения захвата получаем [19,213,246]

$$\sigma_c(J) = \frac{8\pi K^{2J+1}}{\hbar^2 q} \frac{\mu}{(2S_1+1)(2S_2+1)} \frac{J+1}{J[(2J+1)!!]^2} \sum_{m, m_i, m_f} \left| M_{Jm}(N) \right|^2 ,$$

$$M_{Jm}(N) = i^J \langle f \left| H_{Jm}(N) \right| i \rangle . \qquad (8.4)$$

В кластерной модели электромагнитные операторы принимают простой вид

$$Q_{Jm}(L) = e\mu^J \left[ \frac{Z_1}{M_1^J} + (-1)^J \frac{Z_2}{M_2^J} \right] R^J Y_{Jm} = A_J R^J Y_{Jm},$$

$$Q_{J\,m}(S) = -\frac{e\hbar}{m_0\,c} K\left[\frac{J}{J+1}\right]^{1/2}\left[\mu_1 \hat{S}_1 \frac{M_2^J}{M^J} + (-1)^J \mu_2 \hat{S}_2 \frac{M_1^J}{M^J}\right] R^J Y_{J\,m} =$$

$$= (B_{1J}\ \hat{S}_1 + B_{2J}\ \hat{S}_2) R^J Y_{J\,m},$$

$$W_{J\,m}(L) = i\frac{e\hbar}{m_0 c}\frac{\sqrt{J(2J+1)}}{J+1}\left[\frac{Z_1}{M_1}\frac{M_2^J}{M^J} + (-1)^{J-1}\ \frac{Z_2}{M_2}\frac{M_1^J}{M^J}\right] R^{J-1}\ \hat{L}\, Y_{J\,m}^{J-1} =$$

$$= C_J\ R^{J-1} \hat{L}\ \ Y_{J\,m}^{J-1},$$

$$W_{J\,m}(S) = i\frac{e\hbar}{m_0 c}\sqrt{J(2J+1)}\left[\mu_1\ \hat{S}_1 \frac{M_2^{J-1}}{M^{J-1}} + (-1)^{J-1}\ \mu_2 \hat{S}_2 \frac{M_1^{J-1}}{M^{J-1}}\right] R^{J-1}\ \ Y_{J\,m}^{J-1} =$$

$$= (D_{1J}\ \hat{S}_1 + D_{2J}\ \hat{S}_2) R^{J-1}\ Y_{J\,m}^{J-1}.$$

Здесь R - межкластерное расстояние и M - масса ядра. Используем в дальнейшем волновые функции связанных состояний кластеров в обычной форме [19,213]

$$|f\rangle = \Psi_f\ = \sum_{L_f J_f} R_{L_f\ J_f}\ \Phi_{J_f m_f}^{L_f S},\ R_{LJ} = \frac{U_{LJ}}{r}. \tag{8.5}$$

Функцию рассеяния запишем в виде разложения по спин - угловым функциям [19,213]

$$|i\rangle = \Psi_i\ = \frac{1}{q}\sum_{L_i J_i} i^{L_i}\sqrt{4\pi(2L_i+1)}(L_i\, 0 S m_i\ |\ J_i\, m_i) e^{i\delta_{L_i J_i}} R_{L_i J_i}\ \Phi_{J_i m_i}^{L_i S}\ . \tag{8.6}$$

Здесь $R_{LJ}$ - радиальная волновая функция рассеяния, получаемая из решения уравнения Шредингера (В.4) с заданными межкластерными потенциалами, $\Phi_{Jm}^{LS}$ - спин - угловая функция начального i состояния системы, $\delta_{LJ}$ - фазы упругого рассеяния [247].

Используя, приведенные в [120], известные формулы матричных элементов различных операторов, для полного сечения захвата можно получить окончательное выражение [19,213]

$$\sigma_c(J) = \frac{8\pi\ K^{2J+1}}{\hbar^2 q^3}\frac{\mu}{(2S_1+1)(2S_2+1)}\frac{J+1}{J[(2J+1)!!]^2}\sum_{\substack{L_i,L_f,\\ J_i,J_f}}\left|T_J\ (N)\right|^2, \quad (8.7)$$

где матричные элементы приобретают вид

$$T_J(E) = A_J\, I_J\, P_J + (\, B_{1J}\, N_{1J} + B_{2J}\, N_{2J}\,)\, I_J\,,$$

$$T_J(M) = C_J\, I_{J-1}\, G_J + (\, D_{1J}\, N_{1J} + D_{2J}\, N_{2J}\,)\, I_{J-1}\,,$$

$$P_J = \sqrt{4\pi}\left\langle J_f L_f S\right\|Y_J\left\|J_i L_i S\right\rangle = (-1)^{J+S+L_i+L_f}(L_i 0 J 0 \,|\, L_f 0)\times$$

$$\times\sqrt{(2J_i+1)(2J_f+1)(2J+1)(2L_i+1)}\begin{Bmatrix} L_i & S & J_i \\ J_f & J & L_f \end{Bmatrix},$$

$$G_J = \sqrt{4\pi}\left\langle J_f L_f S\right\|\hat{L}\, Y_J^k\left\|J_i L_i S\right\rangle = (-1)^{S+L_i+J_i}(L_i 0 k 0 \,|\, L_f 0)(2L_i+1)\times$$

$$\times\sqrt{L_i(2L_i+1)(2k+1)(2J_i+1)(2J_f+1)(2J+1)}\begin{Bmatrix} L_i & 1 & L_i \\ k & L_f & J \end{Bmatrix}\begin{Bmatrix} S & L_i & J_i \\ J & J_f & L_f \end{Bmatrix},$$

$$N_J = \sqrt{4\pi}\left\langle J_f L_f S\right\|\hat{S}\, Y_J^k\left\|J_i L_i S\right\rangle = (-1)^{k+1-J+L_i+L_f+2S-J_i-J_f}(L_i 0 k 0 \,|\, L_f 0)\times$$

$$\times\begin{Bmatrix} S & 1 & S \\ L_i & k & L_f \\ J_i & J & J_f \end{Bmatrix}\sqrt{S(S+1)(2S+1)(2k+1)(2L_i+1)(2J_i+1)(2J_f+1)}\,,$$

а $I_J$ - радиальные интегралы от волновых функций

$$I_J = \langle J_f L_f \,|\, R^J \,|\, J_i\, L_i \rangle\, .$$

Сечение обратного процесса - фоторазвала можно получить из принципа детального равновесия [244]

$$\sigma_d(J) = \frac{q^2(2S_1+1)(2S_2+1)}{K^2\, 2(2J_0+1)}\sigma_c(J), \quad (8.8)$$

где $J_0$ - полный момент ядра в основном состоянии.

В полученных выражениях аналитически вычисляются все величины, кроме радиальных интегралов, которые находятся численно по определенным из решения уравнения Шредингера волновым функциям связанных состояний и рассеяния. Асимптотика радиальной волновой функции рассеяния обычно представляется в виде суперпозиции кулоновских $F_L$ и $G_L$ функций на границе области ядерного взаимодействия при r = R

$$R_{LJ} \rightarrow N[\ F_L(qr)\ Cos(\delta_{LJ}) + G_L(qr)\ Sin(\delta_{LJ})\ ]\ , \qquad (8.9)$$

где $\delta_{LJ}$ - фазы рассеяния, N - нормировочная константа. Получив численную радиальную волновую функцию, из этого соотношения можно определить фазы рассеяния и нормировочную константу с данным орбитальным L и полным J моментами кластерной системы.

Приведенные выше выражения позволяют выполнять расчеты полных сечений ядерных фотопроцессов в кластерной модели ядра, когда известно межкластерное взаимодействие. Однако, ядерные потенциалы, как правило, неизвестны и приходится использовать различные дополнительные методы и предположения для их определения. Одним из таких методов является анализ фаз упругого рассеяния кластеров, который позволяет определять приближенный вид кластерных потенциалов. Обычно считается, что если потенциалы способны правильно передать экспериментальные фазы рассеяния, то они могут быть использованы для рассмотрения ядерных характеристик связанных состояний кластеров в ядре [248].

В дальнейшем, результаты таких расчетов, будут целиком зависеть от степени кластеризации ядра в рассматриваемый кластерный канал. Это предположение вытекает из общего принципа квантовой механики, который утверждает, что квантовая система должна иметь единый гамильтониан взаимодействия в дискретном и непрерывном спектре. Поэтому перейдем теперь к рассмотрению различных характеристик для связанных состояний, считая, что межкластерная волновая функция и потенциал взаимодействия в принципе известны или могут быть определены теми или иными методами [249].

### 8.3. Программа расчетов фотоядерных процессов

Приведем теперь саму программу [250] для расчета сечений электрических E1 процессов $n^6Li$ модели ядра $^7Li$. Здесь использованы следующие обозначения: AM1, AM2, Z1, Z2 - Массы и заряды частиц, V0, R0, V00, R00 - Параметры ядерных потенциалов, RCU, L, L0, L2 - Кулоновский радиус и орбитальные моменты, NN, NV, NH, EN, EH - Задание энергии для расчета фотосечений, SKN, HC - Интервал для поиска энергии связанного состояния, EP - Точность вычислений.

**REM РАСЧЕТ СЕЧЕНИЙ ФОТОРАЗВАЛА ЯДРА 7Li B N6Li**

**КАНАЛ**

```
 DEFDBL A - Z: DEFNT I,J,K,L,N,M
 DIM EL(50),F32(50),EG(50),F52(50),F72(50),SZ(50)
 DIM S0(50),S2(50),SR(50),ECM(50): NN=4000
 DIM V(NN+1),U(NN+1),U0(NN+1)
 REM *************************************************
 A$=" ВЫЧИСЛЕНИЕ СЕЧЕНИЯ ФОТОЗАХВАТА "
 B$="E(GAM) E(LAB) S(R) S(Z) S0 S2 "
 C$=" Q R F FP G GP W": H$=" Z1 Z2 M1 M2 H N NM"
 G$="C:\BASICA\FOTLI7N8.DAT":
GG$="C:\BASICA\IMPALT.DAT"
 GGG$="C:\BASICA\WF - ALT.DAT": F$=" V R": F1$=" V3/2 R"
 F2$=" V5/2 R": F0$=" VS R"
 REM ***********************************************
 PRINT: PRINT A$: PRINT
 REM ******* ВХОДНЫЕ ПАРАМЕТРЫ ************
 NN=0: NV=50: NH=1: EN=0.1: EH=.5: AM1=1: AM2=6: Z1=0: Z2=3
 PI=3.1415926535899: PM=AM1*AM2/(AM1+AM2): A1=41.4686
 B1=2*PM/A1: AK1=1.439975*Z1*Z2*B1
 GK=3.44476E - 02*Z1*Z2*PM
 REM ************************************************
 N=1000: N3=2*N: H=0.02: SKN= - 8: HC=1: SKN=SKN*B1
 HC=HC*B1: EP=1.E - 05
 REM *********** ПАРАМЕТРЫ ПОТЕНЦИАЛОВ ******
 V0=252.6: R0=0.25: V00=140: R00=0.15: V22=V00: R22=R00
 A2= - V0*B1: A00= - V00*B1: A22= - V22*B1: RCU=0
 L=1: L0=0: L2=2
 REM ************ БЛОК ПЕЧАТИ *************
 PRINT F$: PRINT USING " +#.####^^^^"; - V0,R0
 PRINT F0$: PRINT USING " +#.####^^^^"; - V00,R0
 PRINT F1$: PRINT USING " +#.####^^^^"; - V32,R32
 PRINT F2$: PRINT USING " +#.####^^^^"; - V52,R52
 PRINT H$: PRINT USING ".^^^^ ";Z1,Z2,AM1,AM2,H,N3,NV
 REM ******** ПОИСК ЭНЕРГИИ СОСОЯНИЯ ********
 CALL MIN(EP, B1, PM, SKN, HC, H, N, L, A2, R0, AK1, RCU, GK,
E32, SKS)
 REM **** РЕШЕНИЕ УРАВНЕНИЯ ШРЕДИНГЕРА ****
 U(0)=0: U(1)=0.001: HK=H*H: FOR K=1 TO N - 1: X=K*H
 Q1=A2*EXP( - R0*X*X)+L*(L+1)/(X*X): IF X>RCU GOTO 157
 Q1=Q1+(3 - (X/RCU)^2)*AK1/(2*RCU): GOTO 158
157 Q1=Q1+AK1/X
158 Q2= - Q1*HK - 2+SKS*HK: U(K+1)= - Q2*U(K) - U(K - 1):
NEXT K
 REM * * * * * * ВЫЧИСЛЕНИЕ НОРМИРОВКИ * * * *
```

```
 XX=H*N": SS=SQR(ABS(SKS)): SQQ=SQR(2.*SS)
 WW=EXP( - XX*SS): GG=GK/SS
 CCC=U(N)/(WW*SQQ)*(SS*2.*XX)^GG: FOR I1=N+1 TO N3
 XX=I1*H: SXS=XX*SS: WW=EXP( - SXS)*SQQ/(2.*SXS)^GG
 U(I1)=CCC*WW: NEXT I1: FOR I1=0 TO N3: V(I1)=U(I1)^2
 NEXT I1: A=0: B=0: FOR II=1 TO N3 - 1 STEP 2: B=B+V(II)
 NEXT II: FOR JJ=2 TO N3 - 2 STEP 2: A=A+V(JJ): NEXT JJ
 SIM=H*(V(0)+V(N3)+2*A+4*B)/3: HN=1/SQR(SIM): FOR I1=0 TO
N3
 U(I1)=U(I1)*HN: NEXT I1
 REM * * * * * * ВЫЧИСЛЕНИЕ РАДИУСА * * * * * * *
 FOR I1=0 TO N3: X=I1*H: V(I1)=X^2*U(I1)^2: NEXT I1
 A=0: B=0: FOR II=1 TO N3 - 1 STEP 2: B=B+V(II): NEXT II
 FOR JJ=2 TO N3 - 2 STEP 2: A=A+V(JJ): NEXT JJ
 RKV=H*(V(0)+V(N3)+2*A+4*B)/3: AM=AM1+AM2
 RK= AM1 / AM * (0) ^ 2 + AM2 / AM * (2.50) ^ 2 + AM1 * AM2 /
AM ^ 2 * RKV: PRINT: PRINT " (R^2)^1/2=";
 PRINT USING " #.####^^^^ ";SQR(RK);SQR(RKV); RKV: PRINT
 REM ВЫЧИСЛЕНИЕ ФУНКЦИЙ РАССЕЯНИЯ И МАТРИЧНЫХ
ЭЛЕМЕНТОВ
 PRINT B$: FOR I=NN TO NV STEP NH: ECM(I)=EN+EH*I
 EG(I)=ECM(I)+ABS(E32): EG1=ECM(I)+ER: SK=ECM(I)*B1
 SS1=SQR(SK): G=GK/SS1: X1=H*SS1*(N3 - 4): X2=H*SS1*(N3)
 REM ** ВЫЧИСЛЕНИЕ КУЛОНОВСКИХ ФУНКЦИЙ **
 CALL CULFUN(L0,X1,G,F10,G10,W0)
 CALL CULFUN(L0,X2,G,F20,G20,W0)
 CALL CULFUN(L2,X1,G,F12,G12,W0)
 CALL CULFUN(L2,X2,G,F22,G22,W0)
 REM * * ВЫЧИСЛЕНИЕ ФУНКЦИЙ РАССЕЯНИЯ * * *
 U0(0)=0: U0(1)=0.001: HK=H*H: FOR K=1 TO N3 - 1: X=K*H
 Q1=A00*EXP( - R00*X*X)+L0*(L0+1)/(X*X): IF X>RCU GOTO
1157
 Q1=Q1+(3 - (X/RCU)^2)*AK1/(2*RCU): GOTO 1158
1157 Q1=Q1+AK1/X
1158 Q2= - Q1*HK - 2.+SK*HK: U0(K+1)= - Q2*U0(K) - U0(K - 1)
 NEXT K
 REM ********* ВЫЧИСЛЕНИЕ ФАЗ ************
 U10=U0(N3 - 4): U20=U0(N3): F1=F10: G1=G10: F2=F20: G2=G20
 AF= - (F1 - F2*U10/U20)/(G1 - G2*U10/U20): F00=ATN(AF)
 XH0=(COS(F00)*F2+SIN(F00)*G2)/U20: IF F00>0 GOTO 90
 F00=F00+PI
90 F32(I)=F00*180/PI: FOR J=0 TO N3: U0(J)=U0(J)*XH0: NEXT J
 REM ** ВЫЧИСЛЕНИЕ МАТРИЧНЫХ ЭЛЕМЕНТОВ **
 FOR II=0 TO N3: X=H*II: V(II)=U(II)*X*U0(II): NEXT II
```

```
 A=0: B=0: FOR II=1 TO N3 - 1 STEP 2: B=B+V(II): NEXT II
 FOR JJ=2 TO N3 - 2 STEP 2: A=A+V(JJ): NEXT JJ
 S=H*(V(0)+V(N3)+2*A+4*B)/3: AI0=S
 REM * * * ВЫЧИСЛЕНИЕ ФУНКЦИЙ РАССЕЯНИЯ * *
 FOR K=1 TO N3 - 1: X=K*H
 Q1=A22*EXP( - R22*X*X)+L2*(L2+1)/(X*X)
 IF X>RCU GOTO 2157
 Q1=Q1+(3 - (X/RCU)^2)*AK1/(2*RCU)
 GOTO 2158
2157 Q1=Q1+AK1/X
2158 Q2= - Q1*HK - 2.+SK*HK: U0(K+1)= - Q2*U0(K) - U0(K - 1)
 NEXT K
 REM ********** ВЫЧИСЛЕНИЕ ФАЗ ************
 U10=U0(N3 - 4): U20=U0(N3): F1=F12: G1=G12: F2=F22: G2=G22
 AF= - (F1 - F2*U10/U20)/(G1 - G2*U10/U20): F00=ATN(AF)
 XH0=(COS(F00)*F2+SIN(F00)*G2)/U20: IF F00>0. GOTO 902
 F00=F00+PI
902 F52(I)=F00*180/PI: FOR J=0 TO N3: U0(J)=U0(J)*XH0: NEXT J
 REM ** ВЫЧИСЛЕНИЕ МАТРИЧНЫХ ЭЛЕМЕНТОВ **
 FOR II=0 TO N3: X=H*II: V(II)=U(II)*X*U0(II): NEXT II: A=0: B=0
 FOR II=1 TO N3 - 1 STEP 2: B=B+V(II): NEXT II
 FOR JJ=2 TO N3 - 2 STEP 2
 A=A+V(JJ): NEXT JJ: S=H*(V(0)+V(N3)+2*A+4*B)/3: AI2=S
 REM * * * * * * * ВЫЧИСЛЕНИЕ СЕЧЕНИЙ * * * ** * *
 M0=4*AI0^2: M2=14*AI2^2: KP=SS1: KG=(EG(I))/197.331
 BBB=344.46*8*PI*2/(2*3*9)*PM^3*(Z1/AM1-Z2/AM2)^2/1000
 S0(I)=BBB*(KG/KP)^3*M0: S2(I)=BBB*(KG/KP)^3*M2
 SZ(I)=S0(I)+S2(I)
 SR(I)=SZ(I)*(KP/KG)^2*2*3/(2*4):
S0(I)=S0(I)*(KP/KG)^2*2*3/(2*4)
 S2(I)=S2(I)*(KP/KG)^2*2*3/(2*4): EL(I)=ECM(I)*AM1/PM
 PRINT USING " #.####^^^^"; EG(I); EL(I); SR(I); SZ(I); S0(I); S2(I)
 NEXT I
 REM *********** ЗАПИСЬ В ФАЙЛ ************
 OPEN "O",1,G$
 PRINT1,: PRINT1, A$: PRINT1,: PRINT1, F$
 PRINT1, USING " +#.####^^^^"; - V0,R0
 PRINT1,: PRINT1, F0$: PRINT1, USING " +#.####^^^^"; - V00,R00
 PRINT1,: PRINT1, F1$: PRINT1, USING " +#.####^^^^"; - V22,R22
 PRINT1,: PRINT1, H$: PRINT1,
 PRINT1, USING "#.####^^^^ ";Z1,Z2,AM1,AM2,H,N3,NM
 PRINT1,: PRINT1, " ES=";E32: PRINT1,: PRINT1, B$
 PRINT1,: FOR I=NN TO NV STEP NH
 PRINT1, USING " #.####^^^^"; EG(I); ECM(I); EL(I); SR(I); SZ(I);
```

```
S0(I); S2(I)
 NEXT I: PRINT1,: PRINT1, " R U": PRINT1,: FOR L=1 TO N3
 R=L*H: PRINT1, USING " +#.####^^^^ ";R,U(L)
 NEXT L
 END
```

**SUB CULFUN(L0,X1,G,F10,G10,W)**

```
 REM ВЫЧИСЛЕНИЕ КУЛОНОВСКИХ ФУНКЦИЙ
 Q=G: R=X1: F0=1: GK=Q*Q: GR=Q*R: RK=R*R
 B01=(L0+1)/R+Q/(L0+1): K=1: BK=(2*L0+3)*((L0+1)*(L0+2)+GR)
 AK= - R*((L0+1)^2+GK)/(L0+1)*(L0+2): DK=1/BK: DEHK=AK*DK
 S=B01+DEHK
15 K=K+1: AK= - RK*((L0+K)^2 - 1)*((L0+K)^2+GK)
 BK=(2*L0+2*K+1)*((L0+K)*(L0+K+1)+GR): DK=1/(DK*AK+BK)
 IF DK>0 GOTO 35
25 F0= - F0
35 DEHK=(BK*DK-1)*DEHK: S=S+DEHK
 IF (ABS(DEHK)-1E-6)>0 GOTO 15
 FL=S: K=1: RMG=R - Q: LL=L0*(L0+1): CK= - GK - LL: DK=Q
 GKK=2*RMG: HK=2: AA1=GKK*GKK+HK*HK: PBK=GKK/AA1
 RBK= - HK/AA1: OMEK=CK*PBK - DK*RBK
 EPSK=CK*RBK+DK*PBK
 PB=RMG+OMEK: QB=EPSK
52 K=K+1: CK= - GK - LL+K*(K - 1): DK=Q*(2*K - 1): HK=2*K
 FI=CK*PBK - DK*RBK+GKK: PSI=PBK*DK+RBK*CK+HK
 AA2=FI*FI+PSI*PSI: PBK=FI/AA2: RBK= - PSI/AA2
 VK=GKK*PBK - HK*RBK: WK=GKK*RBK+HK*PBK
 OM=OMEK: EPK=EPSK: OMEK=VK*OM - WK*EPK - OM
 EPSK=VK*EPK+WK*OM - EPK: PB=PB+OMEK
 QB=QB+EPSK: IF (ABS(OMEK)+ABS(EPSK) - 1E - 06)>0 GOTO
52
 PL= - QB/R: QL=PB/R: G0=(FL - PL)*F0/QL
 G0P=(PL*(FL - PL)/QL - QL)*F0: F0P=FL*F0
 ALFA=1/(SQR(ABS(F0P*G0 - F0*G0P))): G10=ALFA*G0
 GP10=ALFA*G0P: F10=ALFA*F0: FP10=ALFA*F0P
 W=1 - FP10*G10+F10*GP10
 END SUB
```

**SUB MIN(EP, B1, PM, SKN, HC, H, N3, L, A22, R0, AK1, RCU, GK, E, SKS)**

```
 REM ВЫЧИСЛЕНИЕ ЭНЕРГИИ СОСТОЯНИЙ
 I1=0
777 A2=SKN: DK=A2
 CALL DET1(DK,GK,N3,A22,R0,L,AK1,RCU,H,DD)
 D12=DD: B2=A2+HC
51 DK=B2: CALL DET1(DK,GK,N3,A22,R0,L,AK1,RCU,H,DD)
```

```
 D11=DD: IF D12*D11>0 GOTO 4
3 I1=I1+1: A3=A2: B3=B2
11 C3=(A3+B3)/2: IF (ABS(A3 - B3))<1D - 10 GOTO 151: DK=C3
 CALL DET1(DK,GK,N3,A22,R0,L,AK1,RCU,H,DD): F2=DD
 IF D12*F2>0 GOTO 14: B3=C3: D11=F2: GOTO 155
14 A3=C3: D12=F2
155 IF ABS(F2)>EP GOTO 11
151 CO=C3: IF I1<NC GOTO 777: GOTO 7
4 IF ABS(D11*D12)<1D - 10 GOTO 3: A2=A2+HC: B2=B2+HC
 D12=D11: IF B2 - SKV<0.1 GOTO 51: YS=SKV: GOTO 88
7 YS=NC
88 E=CO/B1
 SKS=CO
 END SUB
```

**SUB DET1(DK,GK,N3,A22,R0,L,AK1,RCU,H,DD)**

```
 REM ВЫЧИСЛЕНИЕ ДЕТЕРМИНАНТА
 HK=H^2: S1=SQR(ABS(DK)): G2=GK/S1: D1=0: D=1: N1=N3 - 1
 FOR II=1 TO N1: X=II*H: F=A22*EXP( - X*X*R0)+L*(L+1)/(X*X)
 IF X>RCU GOTO 67: F=F+AK1/(2*RCU)*(3 - (X/RCU)^2): GOTO
66
67 F=F+AK1/X
66 D2=D1: D1=D: OM=DK*HK - F*HK - 2: D=D1*OM - D2: NEXT
II
 X=H*N3: F=A22*EXP( - X*X*R0)+L*(L+1)/(X*X)
 IF X>RCU GOTO 68
 F=F+AK1/(2*RCU)*(3 - (X/RCU)^2): GOTO 69
68 F=F+AK1/X
69 Z=2*X*S1: OM=DK*HK - F*HK - 2
 W5= - S1 - 2*S1*G2/Z - 2*S1*(L - G2)/(Z*Z)
 OM=OM+2*H*W5
 DD=OM*D - 2*D1
 END SUB
```

**SUB WW(SK,L,GK,R,N,H,WH,V(5000))**

```
 REM ВЫЧИСЛЕНИЕ ФУНКЦИЙ УИТТЕКЕРА
 SS=SQR(ABS(SK)): AA=GK/SS: BB=L: NN=N: HH=0.02
 ZZ=1+AA+BB: AAA=1/ZZ: NNN=100000: FOR I2=1 TO NNN
 AAA=AAA*I2/(ZZ+I2): NEXT I2: GAM=AAA*NNN^ZZ: RR=R
 CC=RR*SS*2: FOR I=0 TO NN: TT=HH*I
 V(I)=TT^(AA+BB)*(1+TT/CC)^(BB - AA)*EXP( - TT): NEXT I
 A=0: B=0: FOR II=1 TO NN - 1 STEP 2: B=B+V(II): NEXT II
 FOR JJ=2 TO NN - 2 STEP 2: A=A+V(JJ)
 NEXT JJ
 SIM=HH*(V(0)+V(NN)+2*A+4*B)/3
 WH=SIM*EXP( - CC/2)/(CC^AA*GAM)
```

END SUB

В качестве контрольного счета, описанной программы, можно привести результаты расчета полных сечений фоторазвала дейтрона и сравнение их с классической формулой фоторазвала [59]

$$\sigma_c(E1) = \frac{8\pi e^2 h^2 W^{1/2} E^{3/2}}{3hcm(E+W)(1-k_0 r_{0t})}$$

Здесь E - энергия нуклонов в непрерывном спектре, W - энергия связи дейтрона. На рис 8.2 показаны результаты расчета. Пунктир – приведенная выше формула, непрерывная линия - один из NN потенциалов, который правильно передает низкоэнергетические параметры рассеяния [59].

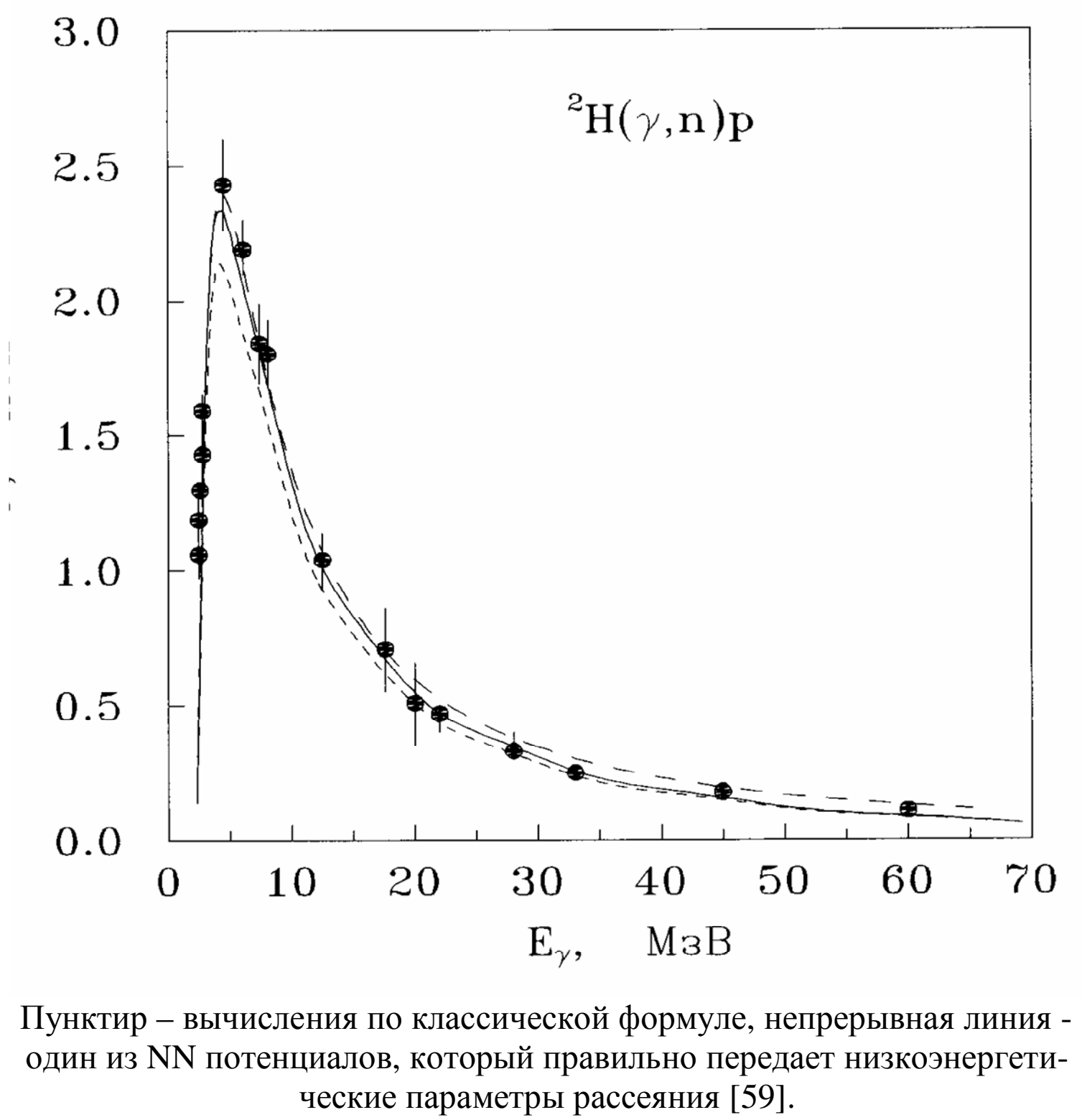

Пунктир – вычисления по классической формуле, непрерывная линия - один из NN потенциалов, который правильно передает низкоэнергетические параметры рассеяния [59].

Рисунок 8.2. Полные сечения фоторазвала дейтрона.

Приведем теперь текст программы для вычисления сечений E2 процессов в системе $^2$H$^4$He ядра $^6$Li [251,252].

**REM РАСЧЕТ СЕЧЕНИЯ ФОТОЗАХВАТА АЛЬФА ДЕЙТОН В 6Li ДЛЯ E2 ПЕРЕХОДОВ**

```
 DEFDBL A - Z: DEFINT I,J,K,L,N,M
 DIM EG(50),SE(50),FA1(50),ECM(50),FA2(50),SEC(50)
 DIM DEE(50),SR(50),S1(50),SZ(50),EL(50),FA3(50): N=4000
 DIM U(N),V(N),U1(N)
 REM ********************************************
 A$=" ВЫЧИСЛЕНИЕ СЕЧЕНИЯ ФОТОЗАХВАТА АЛЬФА ДЕЙ-
ТОН"
 B$=" E(G) E(CM) E(L) SR(NB) SZ(NB)": C$=" Q R F FP G GP W"
 H$=" Z1 Z2 M1 M2 H N NM":
G$="C:\BASICA\FOT\RAZALDE2.DAT"
 GGG$="C:\BASICA\FOT\IMPULSN.DAT": F$=" V R": F1$=" V1
R1"
 F2$=" V2 R2": F3$=" V3 R3"
 REM *********************************************
 PRINT: PRINT A$: PRINT
 REM ******* ВХОДНЫЕ ПАРАМЕТРЫ ***********
 NN=1: NV=20: NH=1: HE=.5: EN=.01: MAG=0.8574: AM1=4:
AM2=2
 Z1=2: Z2=1: PI=3.1415926535899: PM=AM1*AM2/(AM1+AM2)
 A1=41.4686: B1=2*PM/A1: AK1=1.439975*Z1*Z2*B1
 GK=3.44476E - 02*Z1*Z2*PM
 REM *********************************************
 N=1000: N3=2*N: H=0.02: H1=H: SKN= - 2: HC=.1: SKN=SKN*B1
 HC=HC*B1: EP=1.E - 07
 REM ****** ПАРАМЕТРЫ ПОТЕНЦИАЛОВ *******
 V0=76.12: R0=0.2: V11=53: R11=0.164: V22=62.9: R22=0.174
 V33=80.88: R33=0.19: A2= - V0*B1: A11= - V11*B1: A22= -
V22*B1
 A33= - V33*B1: RCU=0: L=0: L2=2
 REM *********** БЛОК ПЕЧАТИ *************
 PRINT F$: PRINT USING " +#.###^^^^"; - V0,R0
 PRINT F1$: PRINT USING " +#.###^^^^"; - V11,R11
 PRINT F2$: PRINT USING " +#.###^^^^"; - V22,R22
 PRINT F3$: PRINT USING " +#.###^^^^"; - V33,R33
 PRINT H$: PRINT USING "#.###^^^^ ";Z1,Z2,AM1,AM2,H,N3,NV
 REM **** РЕШЕНИЕ УРАВНЕНИЯ ШРЕДИНГЕРА **
 CALL MIN(EP, B1, PM, SKN, HC, H1, N, L, A2, R0, AK1, RCU, GK,
ES, SKS)
 CALL FUN(N,H,U(),L,A2,AK1,SKS,R0,RCU)
 REM * * * * * * ВЫЧИСЛЕНИЕ НОРМИРОВКИ * * * *
 XX=H*N: SS=SQR(ABS(SKS)): SQQ=SQR(2.*SS)
 WW=EXP( - XX*SS): GGG=GK/SS
```

```
 CCC=U(N)/(WW*SQQ)*(SS*XX*2.)^(GGG): SQQ=SQR(2.*SS)
 FOR I1=N+1 TO N3: XX=I1*H: SXS=XX*SS
 REM CALL WW(SKS,L,GK,XX,N,H,WW)
 WW=EXP( - SXS)*SQQ/(SXS*2.)^(GGG): U(I1)=CCC*WW: NEXT
I1
 FOR I1=0 TO N3: V(I1)=U(I1)^2: NEXT I1: CALL SIM(V(),N3,H,S)
 HN=1/SQR(S): FOR I1=0 TO N3: U(I1)=U(I1)*HN: NEXT I1
 REM * * * * ВЫЧИСЛЕНИЕ РАДИУСА * * * * * * *
 FOR I1=1 TO N3: X=I1*H: V(I1)=X^2*U(I1)^2: NEXT I1
 CALL SIM(V(),N3,H,S): AM=AM1+AM2
 RK = AM1/AM * (1.67) ^ 2 + AM2 / AM * (1.96) ^ 2 + AM1 * AM2 /
AM ^ 2*S
 PRINT " (R^2)^1/2=";: PRINT USING " #.####^^^^ ";SQR(RK):
PRINT
 REM ВЫЧИСЛЕНИЕ ФУНКЦИЙ РАССЕЯНИЯ ФАЗ И МАТ-
РИЧНЫХ ЭЛЕМЕНТОВ
 PRINT B$: FOR I=NN TO NV STEP NH: EG(I)=EN+I*HE+ABS(ES)
 ECM(I)=EG(I) - ABS(ES): SK=ECM(I)*B1: SS1=SQR(SK):
G=GK/SS1
 X1=H*SS1*(N3 - 4): X2=H*SS1*(N3)
 REM **** ВЫЧИСЛЕНИЕ КУЛОНОВСКИХ ФУНКЦИЙ
 CALL CULFUN(L2,X1,G,F11,G11,W0)
 CALL CULFUN(L2,X2,G,F22,G22,W0)
 REM * * ВЫЧИСЛЕНИЕ ФУНКЦИЙ РАССЕЯНИЯ * *
 CALL FUN(N3,H,U1(),L2,A11,AK1,SK,R11,RCU)
 REM ******** ВЫЧИСЛЕНИЕ ФАЗ *************
 U10=U1(N3 - 4): U20=U1(N3): F1=F11: G1=G11: F2=F22: G2=G22
 AF= - (F1 - F2*U10/U20)/(G1 - G2*U10/U20): FA1=ATN(AF)
 XH1=(COS(FA1)*F2+SIN(FA1)*G2)/U20: IF FA1>0. GOTO 90
 FA1=FA1+PI
90 FA1(I)=FA1*180/PI: FOR J=0 TO N3: U1(J)=U1(J)*XH1: NEXT J
 REM ВЫЧИСЛЕНИЕ МАТРИЧНЫХ ЭЛЕМЕНТОВ
 FOR II=0 TO N3: X=H*II: V(II)=U1(II)*X^2*U(II): NEXT II
 CALL SIM(V(),N3,H,S): AI11=S
 REM * * ВЫЧИСЛЕНИЕ ФУНКЦИЙ РАССЕЯНИЯ * * *
 CALL FUN(N3,H,U1(),L2,A22,AK1,SK,R22,RCU)
 REM ********* ВЫЧИСЛЕНИЕ ФАЗ *************
 U10=U1(N3 - 4): U20=U1(N3): F1=F11: G1=G11: F2=F22: G2=G22
 AF= - (F1 - F2*U10/U20)/(G1 - G2*U10/U20): FA1=ATN(AF)
 XH1=(COS(FA1)*F2+SIN(FA1)*G2)/U20: IF FA1>0 GOTO 901
 FA1=FA1+PI
901 FA2(I)=FA1*180/PI: FOR J=0 TO N3: U1(J)=U1(J)*XH1: NEXT J
 REM ВЫЧИСЛЕНИЕ МАТРИЧНЫХ ЭЛЕМЕНТОВ
 FOR II=0 TO N3: X=H*II: V(II)=U1(II)*X^2*U(II): NEXT II
```

```
CALL SIM(V(),N3,H,S): AI22=S
REM * * * ВЫЧИСЛЕНИЕ ФУНКЦИЙ РАССЕЯНИЯ * *
CALL FUN(N3,H,U1(),L2,A33,AK1,SK,R33,RCU)
REM ********** ВЫЧИСЛЕНИЕ ФАЗ ************
U10=U1(N3 - 4): U20=U1(N3): F1=F11: G1=G11: F2=F22: G2=G22
AF= - (F1 - F2*U10/U20)/(G1 - G2*U10/U20): FA1=ATN(AF)
XH1=(COS(FA1)*F2+SIN(FA1)*G2)/U20: IF FA1>0. GOTO 902
FA1=FA1+PI
902 FA3(I)=FA1*180/PI: FOR J=0 TO N3: U1(J)=U1(J)*XH1: NEXT J
REM * ВЫЧИСЛЕНИЕ МАТРИЧНЫХ ЭЛЕМЕНТОВ **
FOR II=0 TO N3: X=H*II: V(II)=U1(II)*X^2*U(II): NEXT II
CALL SIM(V(),N3,H,S): AI33=S
REM ******** ВЫЧИСЛЕНИЕ СЕЧЕНИЙ ********
ME1=3*AI11^2: ME2=5*AI22^2: ME3=7*AI33^2:
ME=ME1+ME2+ME3
KG=(ECM(I)+ABS(ES))/197.331: KP=SS1
BBB= 344.46 * 4 * PI / 225 * PM ^ 5 * (Z1 / AM1 ^ 2 + Z2 / AM2 ^ 2)
^ 2 * 1000
SZ(I)=BBB*(KG/KP)^3*ME*KG^2: SR(I)=SZ(I)/2*(KP/KG)^2
EL(I)=ECM(I)*AM2/PM: PRINT USING " #.##^^^^"; EG(I); ECM(I);
EL(I); SR(I); SZ(I): NEXT I
REM ********* ЗАПИСЬ В ФАЙЛ ************
OPEN "O",1,G$
PRINT#1,: PRINT#1, A$: PRINT#1,: PRINT#1, F$
PRINT#1, USING " +#.###^^^^"; - V0,R0
PRINT#1,: PRINT#1, F1$: PRINT#1, USING " +#.###^^^^"; -
V11,R11
PRINT#1,: PRINT#1, F2$: PRINT#1, USING " +#.###^^^^"; -
V22,R22
PRINT#1,: PRINT#1, F3$: PRINT#1, USING " +#.###^^^^"; -
V33,R33
PRINT#1,: PRINT#1, H$: PRINT#1,
PRINT#1, USING "#.###^^^^ "; Z1, Z2, AM1, AM2, H, N3, NM
PRINT#1,: PRINT#1, " E=";ES: PRINT#1,: PRINT#1, B$: PRINT#1,
FOR LL=NN TO NV STEP NH
PRINT#1, USING " #.##^^^^"; ECM(LL); FA1(LL); FA2(LL);
FA3(LL); SR(LL); SZ(LL): NEXT LL
PRINT#1,: PRINT#1, " R U": PRINT#1,: FOR L=1 TO N3: R=L*H
PRINT#1, USING " +#.###^^^^ ";R,U(L): NEXT L: PRINT#1,
PRINT#1," КОНЕЦ "
REM *********** КОНЕЦ ПРОГРАММЫ ************
PRINT: PRINT " КОНЕЦ ПРОГРАММЫ": END
SUB CULFUN(L0,X1,G,F10,G10,W)
Q=G: R=X1: F0=1: GK=Q*Q: GR=Q*R: RK=R*R
```

```
 B01=(L0+1)/R+Q/(L0+1): K=1: BK=(2*L0+3)*((L0+1)*(L0+2)+GR)
 AK= - R*((L0+1)^2+GK)/(L0+1)*(L0+2): DK=1/BK: DEHK=AK*DK
 S=B01+DEHK
15 K=K+1: AK= - RK*((L0+K)^2 - 1)*((L0+K)^2+GK)
 BK=(2*L0+2*K+1)*((L0+K)*(L0+K+1)+GR): DK=1/(DK*AK+BK)
 IF DK>0 GOTO 35
25 F0= - F0
35 DEHK=(BK*DK-1)*DEHK: S=S+DEHK
 IF (ABS(DEHK)-1E- 6)>0 GOTO 15
 FL=S: K=1: RMG=R - Q: LL=L0*(L0+1): CK= - GK - LL: DK=Q
 GKK=2*RMG: HK=2: AA1=GKK*GKK+HK*HK: PBK=GKK/AA1
 RBK= - HK/AA1: OMEK=CK*PBK - DK*RBK
 EPSK=CK*RBK+DK*PBK: PB=RMG+OMEK: QB=EPSK
52 K=K+1: CK= - GK - LL+K*(K - 1): DK=Q*(2*K - 1): HK=2*K
 FI=CK*PBK - DK*RBK+GKK: PSI=PBK*DK+RBK*CK+HK
 AA2=FI*FI+PSI*PSI: PBK=FI/AA2: RBK= - PSI/AA2
 VK=GKK*PBK - HK*RBK: WK=GKK*RBK+HK*PBK:
OM=OMEK
 EPK=EPSK: OMEK=VK*OM - WK*EPK - OM
 EPSK=VK*EPK+WK*OM - EPK: PB=PB+OMEK: QB=QB+EPSK
 IF (ABS(OMEK)+ABS(EPSK) - 1E - 06)>0 GOTO 52: PL= - QB/R
 QL=PB/R: G0=(FL - PL)*F0/QL: G0P=(PL*(FL - PL)/QL - QL)*F0
 F0P=FL*F0: ALFA=1/(SQR(ABS(F0P*G0 - F0*G0P))):
G10=ALFA*G0
 GP10=ALFA*G0P: F10=ALFA*F0: FP10=ALFA*F0P
 W=1 - FP10*G10+F10*GP10: END SUB
```

**SUB MIN(EP, B1, PM, SKN, HC, H, N3, L, A22, R0, AK1, RCU, GK, E, SKS)**

```
 I1=0
777 A2=SKN: DK=A2
 CALL DET1(DK,GK,N3,A22,R0,L,AK1,RCU,H,DD)
 D12=DD: B2=A2+HC
51 DK=B2: CALL DET1(DK,GK,N3,A22,R0,L,AK1,RCU,H,DD)
 D11=DD
 IF D12*D11>0 GOTO 4
3 I1=I1+1: A3=A2: B3=B2
11 C3=(A3+B3)/2: IF (ABS(A3 - B3))<1D - 10 GOTO 151: DK=C3
 CALL DET1(DK,GK,N3,A22,R0,L,AK1,RCU,H,DD): F2=DD
 IF D12*F2>0 GOTO 14: B3=C3: D11=F2: GOTO 155
14 A3=C3: D12=F2
155 IF ABS(F2)>EP GOTO 11
151 CO=C3: IF I1<NC GOTO 777: GOTO 7
4 IF ABS(D11*D12)<1D - 10 GOTO 3: A2=A2+HC: B2=B2+HC
 D12=D11: IF B2 - SKV<+0.1 GOTO 51: YS=SKV: GOTO 88
```

```
7 YS=NC
88 E=CO/B1:SKS=CO:END SUB
 SUB DET1(DK,GK,N3,A22,R0,L,AK1,RCU,H,DD)
 HK=H^2: S1=SQR(ABS(DK)): G2=GK/S1: D1=0: D=1: N1=N3 - 1
 FOR II=1 TO N1: X=II*H: F=A22*EXP( - X*X*R0)+L*(L+1)/(X*X)
 IF X>RCU GOTO 67: F=F+AK1/(2*RCU)*(3 - (X/RCU)^2): GOTO
66
67 F=F+AK1/X
66 D2=D1: D1=D: OM=DK*HK - F*HK - 2: D=D1*OM - D2: NEXT
II
 X=H*N3: F=A22*EXP( - X*X*R0)+L*(L+1)/(X*X)
 IF X>RCU GOTO 68
 F=F+AK1/(2*RCU)*(3 - (X/RCU)^2): GOTO 69
68 F=F+AK1/X
69 Z=2*X*S1: OM=DK*HK - F*HK - 2
 W5= - S1 - 2*S1*G2/Z - 2*S1*(L - G2)/(Z*Z): OM=OM+2*H*W5
 DD=OM*D - 2*D1: END SUB
 SUB WW(SK,L,GK,R,N,H,WH,V(5000))
 SS=SQR(ABS(SK)): AA=GK/SS: BB=L: NN=2000: HH=0.01
 ZZ=1+AA+BB: AAA=1/ZZ: NNN=30000: FOR I2=1 TO NNN
 AAA=AAA*I2/(ZZ+I2): NEXT I2: GAM=AAA*NNN^ZZ: RR=R
 CC=RR*SS*2: FOR I=0 TO NN: TT=HH*I
 V(I)=TT^(AA+BB)*(1+TT/CC)^(BB - AA)*EXP( - TT): NEXT I
 A=0: B=0: FOR II=1 TO NN - 1 STEP 2: B=B+V(II): NEXT II
 FOR JJ=2 TO NN - 2 STEP 2: A=A+V(JJ): NEXT JJ
 SIM=HH*(V(0)+V(NN)+2*A+4*B)/3
 WH=SIM*EXP( - CC/2)/(CC^AA*GAM)
 END SUB
 SUB FUN(N,H,U(4000),L,AV,AK,SK,R0,RCU)
 U(0)=0: U(1)=0.001: HK=H*H: FOR K=1 TO N - 1: X=K*H
 Q1=AV*EXP( - R0*X*X)+L*(L+1)/(X*X): IF X>RCU GOTO 1571
 Q1=Q1+(3 - (X/RCU)^2)*AK/(2*RCU): GOTO 1581
1571 Q1=Q1+AK/X
1581 Q2= - Q1*HK - 2+SK*HK: U(K+1)= - Q2*U(K) - U(K - 1)
 NEXT K: END SUB
  SUB SIM(V(4000),N,H,S)
 A=0: B=0: FOR II=1 TO N - 1 STEP 2: B=B+V(II): NEXT II
 FOR JJ=2 TO N - 2 STEP 2: A=A+V(JJ): NEXT JJ
 S=H*(V(0)+V(N)+2*A+4*B)/3: END SUB
```

Приведем результаты расчета для фоторазвала ядра $^7$Li в n$^6$Li канал, выполненные по первой из этих программ. Здесь E(GAM) - энергия гамма кванта, S(R) - полные сечения процесса фоторазвала (непрерывная линия на рисунке 8.3), S0 - сечение обусловленное S

- волной (штриховая линия внизу на рисунке 8.3) и S2 - сечение обусловленное D - волной (штриховая линия вверху на рисунке 8.3) [45].

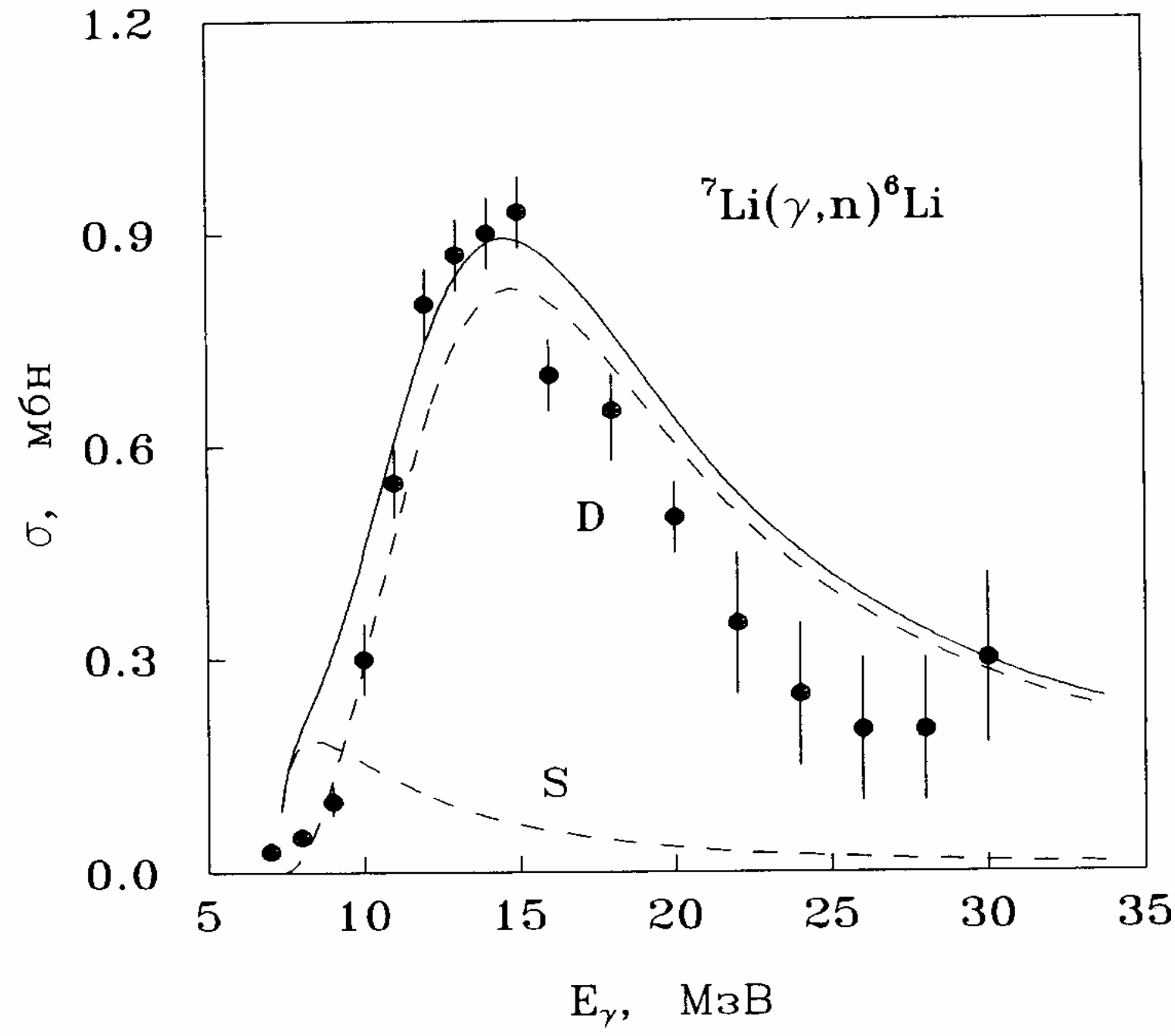

Непрерывная линия - расчет полных E1 сечений с полученным потенциалом. Штриховые линии – вклады фотопроцессов от $^2$S и $^2$D волн рассеяния. Точки – экспериментальные данные [253].

Рисунок 8.3 - Полные сечения фоторазвала ядра $^7$Li в n$^6$Li канал [254].

| E(GAM) | E(LAB) | S(R) | S(Z) | S0 | S2 |
|---|---|---|---|---|---|
| 7.34E+00 | 1.17E-01 | 8.74E-02 | 3.90E-02 | 8.72E-02 | 1.52E-04 |
| 7.84E+00 | 7.00E-01 | 1.84E-01 | 1.56E-02 | 1.71E-01 | 1.26E-02 |
| 8.34E+00 | 1.28E+00 | 2.36E-01 | 1.24E-02 | 1.85E-01 | 5.02E-02 |
| 8.84E+00 | 1.87E+00 | 2.90E-01 | 1.17E-02 | 1.81E-01 | 1.09E-01 |
| 9.34E+00 | 2.45E+00 | 3.52E-01 | 1.21E-02 | 1.71E-01 | 1.81E-01 |
| 9.84E+00 | 3.03E+00 | 4.22E-01 | 1.30E-02 | 1.59E-01 | 2.63E-01 |
| 1.03E+01 | 3.62E+00 | 4.97E-01 | 1.42E-02 | 1.47E-01 | 3.50E-01 |
| 1.08E+01 | 4.20E+00 | 5.74E-01 | 1.55E-02 | 1.35E-01 | 4.39E-01 |
| 1.13E+01 | 4.78E+00 | 6.49E-01 | 1.69E-02 | 1.24E-01 | 5.25E-01 |
| 1.18E+01 | 5.37E+00 | 7.18E-01 | 1.81E-02 | 1.13E-01 | 6.04E-01 |
| 1.23E+01 | 5.95E+00 | 7.77E-01 | 1.92E-02 | 1.04E-01 | 6.73E-01 |
| 1.28E+01 | 6.53E+00 | 8.25E-01 | 2.01E-02 | 9.52E-02 | 7.29E-01 |
| 1.33E+01 | 7.12E+00 | 8.59E-01 | 2.08E-02 | 8.74E-02 | 7.72E-01 |
| 1.38E+01 | 7.70E+00 | 8.81E-01 | 2.12E-02 | 8.06E-02 | 8.00E-01 |
| 1.43E+01 | 8.28E+00 | 8.91E-01 | 2.14E-02 | 7.45E-02 | 8.16E-01 |
| 1.48E+01 | 8.87E+00 | 8.91E-01 | 2.14E-02 | 6.92E-02 | 8.22E-01 |
| 1.53E+01 | 9.45E+00 | 8.83E-01 | 2.13E-02 | 6.44E-02 | 8.19E-01 |
| 1.58E+01 | 1.00E+01 | 8.68E-01 | 2.10E-02 | 6.01E-02 | 8.08E-01 |
| 1.63E+01 | 1.06E+01 | 8.49E-01 | 2.06E-02 | 5.62E-02 | 7.93E-01 |
| 1.68E+01 | 1.12E+01 | 8.25E-01 | 2.02E-02 | 5.26E-02 | 7.73E-01 |
| 1.73E+01 | 1.18E+01 | 7.99E-01 | 1.97E-02 | 4.94E-02 | 7.50E-01 |
| 1.78E+01 | 1.24E+01 | 7.71E-01 | 1.92E-02 | 4.63E-02 | 7.25E-01 |
| 1.83E+01 | 1.30E+01 | 7.42E-01 | 1.86E-02 | 4.35E-02 | 6.98E-01 |
| 1.88E+01 | 1.35E+01 | 7.12E-01 | 1.80E-02 | 4.09E-02 | 6.71E-01 |
| 1.93E+01 | 1.41E+01 | 6.81E-01 | 1.75E-02 | 3.85E-02 | 6.43E-01 |
| 1.98E+01 | 1.47E+01 | 6.52E-01 | 1.69E-02 | 3.64E-02 | 6.15E-01 |
| 2.03E+01 | 1.53E+01 | 6.23E-01 | 1.63E-02 | 3.44E-02 | 5.89E-01 |
| 2.08E+01 | 1.59E+01 | 5.95E-01 | 1.58E-02 | 3.26E-02 | 5.63E-01 |
| 2.13E+01 | 1.65E+01 | 5.69E-01 | 1.52E-02 | 3.09E-02 | 5.38E-01 |
| 2.18E+01 | 1.70E+01 | 5.44E-01 | 1.47E-02 | 2.95E-02 | 5.15E-01 |
| 2.23E+01 | 1.76E+01 | 5.21E-01 | 1.43E-02 | 2.81E-02 | 4.93E-01 |
| 2.28E+01 | 1.82E+01 | 4.99E-01 | 1.38E-02 | 2.69E-02 | 4.72E-01 |
| 2.33E+01 | 1.88E+01 | 4.79E-01 | 1.34E-02 | 2.58E-02 | 4.53E-01 |
| 2.38E+01 | 1.94E+01 | 4.60E-01 | 1.30E-02 | 2.47E-02 | 4.35E-01 |
| 2.43E+01 | 2.00E+01 | 4.42E-01 | 1.27E-02 | 2.38E-02 | 4.18E-01 |
| 2.48E+01 | 2.05E+01 | 4.26E-01 | 1.24E-02 | 2.29E-02 | 4.03E-01 |
| 2.53E+01 | 2.11E+01 | 4.10E-01 | 1.21E-02 | 2.20E-02 | 3.88E-01 |
| 2.58E+01 | 2.17E+01 | 3.95E-01 | 1.18E-02 | 2.12E-02 | 3.74E-01 |
| 2.63E+01 | 2.23E+01 | 3.81E-01 | 1.15E-02 | 2.04E-02 | 3.61E-01 |
| 2.68E+01 | 2.29E+01 | 3.68E-01 | 1.12E-02 | 1.96E-02 | 3.48E-01 |
| 2.73E+01 | 2.35E+01 | 3.55E-01 | 1.10E-02 | 1.89E-02 | 3.37E-01 |
| 2.77E+01 | 2.39E+01 | 3.46E-01 | 1.08E-02 | 1.84E-02 | 3.27E-01 |
| 2.82E+01 | 2.45E+01 | 3.34E-01 | 1.05E-02 | 1.78E-02 | 3.17E-01 |
| 2.87E+01 | 2.51E+01 | 3.23E-01 | 1.03E-02 | 1.71E-02 | 3.06E-01 |
| 2.92E+01 | 2.57E+01 | 3.13E-01 | 1.01E-02 | 1.66E-02 | 2.97E-01 |
| 2.97E+01 | 2.63E+01 | 3.03E-01 | 9.88E-03 | 1.60E-02 | 2.87E-01 |
| 3.02E+01 | 2.68E+01 | 2.94E-01 | 9.69E-03 | 1.56E-02 | 2.79E-01 |
| 3.07E+01 | 2.74E+01 | 2.86E-01 | 9.51E-03 | 1.51E-02 | 2.70E-01 |
| 3.12E+01 | 2.80E+01 | 2.77E-01 | 9.35E-03 | 1.47E-02 | 2.63E-01 |
| 3.17E+01 | 2.86E+01 | 2.70E-01 | 9.19E-03 | 1.43E-02 | 2.55E-01 |
| 3.22E+01 | 2.92E+01 | 2.63E-01 | 9.05E-03 | 1.40E-02 | 2.49E-01 |
| 3.27E+01 | 2.98E+01 | 2.56E-01 | 8.92E-03 | 1.36E-02 | 2.42E-01 |
| 3.32E+01 | 3.03E+01 | 2.50E-01 | 8.80E-03 | 1.33E-02 | 2.36E-01 |
| 3.37E+01 | 3.09E+01 | 2.44E-01 | 8.68E-03 | 1.30E-02 | 2.31E-01 |
| 3.42E+01 | 3.15E+01 | 2.39E-01 | 8.58E-03 | 1.28E-02 | 2.26E-01 |

Приведем теперь E2 сечения, вычисленные по второй программе, для $^2H^4He$ захвата на основное состояние ядра $^6Li$ [18]. Здесь E(G), E(CM), E(L) - энергия гамма кванта, частиц в системе центра масс и лабораторной системе в МэВ, SR и SZ сечение развала и захвата в нанобарн (nb). На рисунке 8.4 эти результаты показаны верхней непрерывной линией.

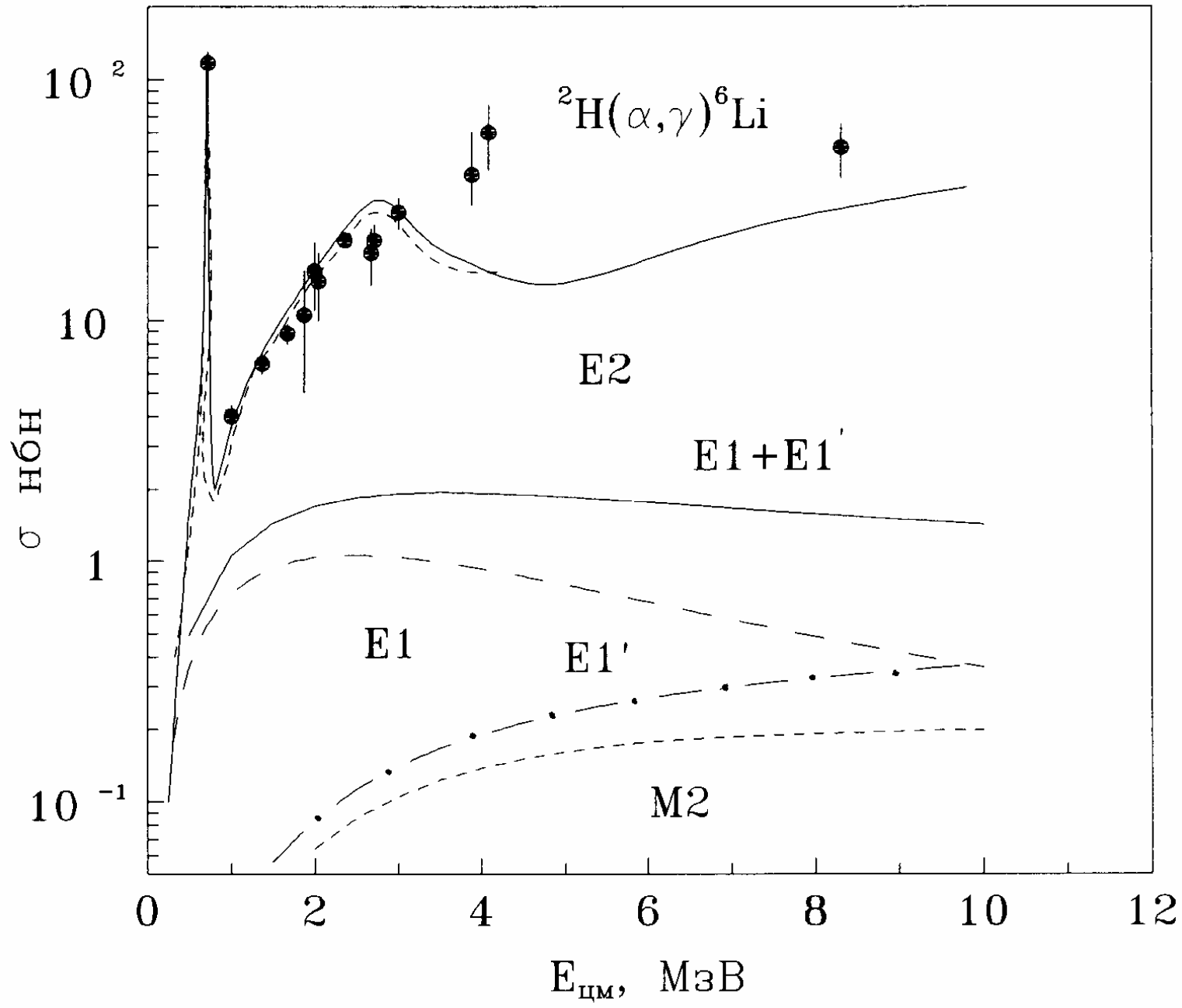

Непрерывная линия вверху - расчетные E2 сечения, полученные из приведенной программы. Точечная линия внизу - сечения M2 процесса, штриховая линия - E1 сечения, обусловленное орбитальной частью электрического оператора, штрих - пунктир - E1' сечение, обусловленное спиновой частью электрического оператора, непрерывная линия внизу - E1 сечение, полученное с учетом орбитальной и спиновой частей электрического оператора. Точки - экспериментальные данные [255].

Рисунок 8.4 - полные сечения процесса радиационного захвата в $^4He^2H$ системе с образованием ядра $^6Li$ в основном состоянии.

| E(G) | E(CM) | E(L) | SR(nb) | SZ(nb) |
|---|---|---|---|---|
| 1.58E+00 | 1.10E-01 | 1.65E-01 | 1.02E+00 | 1.85E-02 |
| 1.68E+00 | 2.10E-01 | 3.15E-01 | 1.43E+01 | 1.54E-01 |
| 1.78E+00 | 3.10E-01 | 4.65E-01 | 5.67E+01 | 4.64E-01 |
| 1.88E+00 | 4.10E-01 | 6.15E-01 | 1.43E+02 | 9.89E-01 |
| 1.98E+00 | 5.10E-01 | 7.65E-01 | 3.03E+02 | 1.87E+00 |
| 2.08E+00 | 6.10E-01 | 9.15E-01 | 7.15E+02 | 4.06E+00 |
| 2.18E+00 | 7.10E-01 | 1.07E+00 | 2.36E+04 | 1.27E+02 |
| 2.28E+00 | 8.10E-01 | 1.22E+00 | 3.89E+02 | 2.00E+00 |
| 2.38E+00 | 9.10E-01 | 1.37E+00 | 5.51E+02 | 2.74E+00 |
| 2.48E+00 | 1.01E+00 | 1.52E+00 | 7.58E+02 | 3.69E+00 |
| 2.58E+00 | 1.11E+00 | 1.67E+00 | 9.72E+02 | 4.66E+00 |
| 2.68E+00 | 1.21E+00 | 1.82E+00 | 1.19E+03 | 5.67E+00 |
| 2.78E+00 | 1.31E+00 | 1.97E+00 | 1.43E+03 | 6.73E+00 |
| 2.88E+00 | 1.41E+00 | 2.12E+00 | 1.67E+03 | 7.86E+00 |
| 2.98E+00 | 1.51E+00 | 2.27E+00 | 1.92E+03 | 9.05E+00 |
| 3.08E+00 | 1.61E+00 | 2.42E+00 | 2.19E+03 | 1.03E+01 |
| 3.18E+00 | 1.71E+00 | 2.57E+00 | 2.47E+03 | 1.17E+01 |
| 3.28E+00 | 1.81E+00 | 2.72E+00 | 2.77E+03 | 1.32E+01 |
| 3.38E+00 | 1.91E+00 | 2.87E+00 | 3.09E+03 | 1.48E+01 |
| 3.48E+00 | 2.01E+00 | 3.02E+00 | 3.43E+03 | 1.65E+01 |
| 3.58E+00 | 2.11E+00 | 3.17E+00 | 3.80E+03 | 1.85E+01 |
| 3.68E+00 | 2.21E+00 | 3.32E+00 | 4.20E+03 | 2.06E+01 |
| 3.78E+00 | 2.31E+00 | 3.47E+00 | 4.61E+03 | 2.28E+01 |
| 3.88E+00 | 2.41E+00 | 3.62E+00 | 5.04E+03 | 2.52E+01 |
| 3.98E+00 | 2.51E+00 | 3.77E+00 | 5.46E+03 | 2.75E+01 |
| 4.08E+00 | 2.61E+00 | 3.92E+00 | 5.80E+03 | 2.96E+01 |
| 4.18E+00 | 2.71E+00 | 4.07E+00 | 6.01E+03 | 3.10E+01 |
| 4.28E+00 | 2.81E+00 | 4.22E+00 | 5.99E+03 | 3.12E+01 |
| 4.38E+00 | 2.91E+00 | 4.37E+00 | 5.74E+03 | 3.03E+01 |
| 4.48E+00 | 3.01E+00 | 4.52E+00 | 5.30E+03 | 2.83E+01 |
| 4.58E+00 | 3.11E+00 | 4.67E+00 | 4.81E+03 | 2.59E+01 |
| 4.68E+00 | 3.21E+00 | 4.82E+00 | 4.35E+03 | 2.37E+01 |
| 4.78E+00 | 3.31E+00 | 4.97E+00 | 3.98E+03 | 2.19E+01 |
| 4.88E+00 | 3.41E+00 | 5.12E+00 | 3.69E+03 | 2.06E+01 |
| 4.98E+00 | 3.51E+00 | 5.27E+00 | 3.47E+03 | 1.96E+01 |
| 5.08E+00 | 3.61E+00 | 5.42E+00 | 3.30E+03 | 1.88E+01 |
| 5.18E+00 | 3.71E+00 | 5.57E+00 | 3.14E+03 | 1.82E+01 |
| 5.28E+00 | 3.81E+00 | 5.72E+00 | 3.00E+03 | 1.76E+01 |
| 5.38E+00 | 3.91E+00 | 5.87E+00 | 2.87E+03 | 1.70E+01 |
| 5.48E+00 | 4.01E+00 | 6.02E+00 | 2.74E+03 | 1.64E+01 |
| 5.58E+00 | 4.11E+00 | 6.17E+00 | 2.62E+03 | 1.59E+01 |
| 5.68E+00 | 4.21E+00 | 6.32E+00 | 2.51E+03 | 1.54E+01 |
| 5.78E+00 | 4.31E+00 | 6.47E+00 | 2.42E+03 | 1.50E+01 |
| 5.88E+00 | 4.41E+00 | 6.62E+00 | 2.34E+03 | 1.47E+01 |
| 5.98E+00 | 4.51E+00 | 6.77E+00 | 2.28E+03 | 1.44E+01 |
| 6.08E+00 | 4.61E+00 | 6.92E+00 | 2.23E+03 | 1.43E+01 |
| 6.18E+00 | 4.71E+00 | 7.07E+00 | 2.19E+03 | 1.42E+01 |
| 6.28E+00 | 4.81E+00 | 7.22E+00 | 2.17E+03 | 1.42E+01 |
| 6.38E+00 | 4.91E+00 | 7.37E+00 | 2.16E+03 | 1.43E+01 |

В этой главе, в качестве примера, были рассмотрены фотопроцессы на ядре $^6$Li в $^2$H$^4$He и на ядре $^7$Li в n$^6$Li моделях с запрещенными состояниями для потенциалов согласованными с фазами рассеяния этих ядерных частиц и характеристиками их связанных состояний.

# ЛИТЕРАТУРА